\documentclass[graybox,envcountchap,sectrefs]{svmono-arxiv}

\usepackage{type1cm}         

\usepackage{makeidx}         
\usepackage{graphicx}        
\usepackage{multicol}        
\usepackage[bottom]{footmisc}

\usepackage{newtxtext}       %
\usepackage[varvw]{newtxmath}       

\usepackage{amsmath}
\usepackage[bookmarks=true,hyperfigures=true]{hyperref}
\usepackage{graphicx}
\usepackage{slashed}
\usepackage{color}
\usepackage{stackrel}
\usepackage{array} 
\usepackage{nicefrac}  
\usepackage{enumerate}
\usepackage{dsfont}
\usepackage{booktabs} 
\usepackage{bbm}

\usepackage{empheq}
\usepackage{caption}
\usepackage{subcaption}

\def\shuffle{\sqcup\mathchoice{\mkern-7mu}{\mkern-7mu}{\mkern-3.2mu}{\mkern-3.8mu}\sqcup}

\usepackage{tikz}
\usepackage{tikz-feynman}
\usetikzlibrary{decorations.pathmorphing}
\usetikzlibrary{decorations.markings}
\usetikzlibrary{positioning, shapes, snakes, arrows}

\tikzset{
	graviton/.style={decorate, double, double distance=.8pt,
        decoration={snake,amplitude=2pt, segment length=8pt}},
		worldline/.style={gray, line width=1pt},
	worldlineBold/.style={black, line width=.6pt},
	zUndirected/.style={line width=1pt},
	zParticle/.style={line width=1pt,postaction={decorate},decoration={markings,mark=at position .6 with {\arrow[#1]{latex}}}},
	zParticleF/.style={line width=1pt,postaction={decorate}},
	cscalar/.style={line width=1pt,postaction={decorate},decoration={markings,mark=at position .6 with {\arrow[#1]{latex}}}},
	cscalar2/.style={line width=1pt,postaction={decorate},decoration={markings,mark=at position .8 with {\arrow[#1]{latex}}}},
	photon/.style={line width =.8pt, decorate, decoration={snake, segment length=4pt, amplitude=1.8pt,  pre length=.1cm, post length=.1cm}},
	blob/.style={circle, minimum size =1.5cm, draw=black, thick, fill=lightgray, text=black}
}

\makeindex             

\allowdisplaybreaks[3]

\usepackage[font=small,labelfont=bf,width=0.85\textwidth]{caption}

\makeatletter
\let\old@startsection=\@startsection
\renewcommand{\@startsection}[6]{\old@startsection{#1}{#2}{#3}{#4}{#5}{#6\mathversion{bold}}}
\makeatother

\makeatletter
\newlength{\apb@width}
\newcommand{\autoparbox}[2][c]{\settowidth{\apb@width}{#2}\parbox[#1]{\apb@width}{#2}}

\makeatother

\let\oldPhi=\Phi
\let\oldPsi=\Psi
\let\oldGamma=\Gamma
\let\oldDelta=\Delta
\let\oldSigma=\Sigma
\let\oldLambda=\Lambda
\let\oldTheta=\Theta
\let\oldPi=\Pi
\let\oldXi=\Xi
\let\oldUpsilon=\Upsilon
\let\oldOmega=\Omega
\renewcommand{\Phi}{\mathnormal{\oldPhi}}
\renewcommand{\Psi}{\mathnormal{\oldPsi}}
\renewcommand{\Gamma}{\mathnormal{\oldGamma}}
\renewcommand{\Sigma}{\mathnormal{\oldSigma}}
\renewcommand{\Delta}{\mathnormal{\oldDelta}}
\renewcommand{\Theta}{\mathnormal{\oldTheta}}
\renewcommand{\Lambda}{\mathnormal{\oldLambda}}
\renewcommand{\Pi}{\mathnormal{\oldPi}}
\renewcommand{\Xi}{\mathnormal{\oldXi}}
\renewcommand{\Upsilon}{\mathnormal{\oldUpsilon}}
\renewcommand{\Omega}{\mathnormal{\oldOmega}}

\ifx\genfrac\sdflkaj\else\fi
\newcommand{\sfrac}[2]{{\textstyle\frac{#1}{#2}}}
\newcommand{\half}{\sfrac{1}{2}}

\newcommand{\alg}[1]{\mathfrak{#1}}

\newcommand{\vev}[1]{\langle#1\rangle}
\newcommand{\bev}[1]{ [#1]}

\newcommand{\be}{\begin{equation}}
\newcommand{\ee}{\end{equation}}

\newcommand{\Tr}{\mathop{\mathrm{Tr}}}

\renewcommand{\Re}{\mathop{\mathrm{Re}}}

\newcommand{\euler}{\ensuremath{\gamma_{\text{E}}}}

\newcommand{\nn}{\nonumber}

\makeatletter
\def\mr@ignsp#1 {\ifx\:#1\@empty\else #1\expandafter\mr@ignsp\fi}%
\newcommand{\multiref}[1]{\begingroup
\xdef\mr@no@sparg{\expandafter\mr@ignsp#1 \: }%
\def\mr@comma{}%
\@for\mr@refs:=\mr@no@sparg\do{\mr@comma\def\mr@comma{,}\ref{\mr@refs}}%
\endgroup}
\makeatother

\ifx\href\asklfhas\newcommand{\href}[2]{#2}\fi

\newcommand{\hypref}[2]{\ifx\href\asklfhas #2\else\href{#1}{#2}\fi}

\renewcommand{\eqref}[1]{(\multiref{#1})}
\newcommand{\eqn}[1]{eq.~\eqref{#1}}

\newcommand{\R}{\ensuremath{\mathbb{R }}}

\newcommand{\1}{\ensuremath{\mathds{1}}}
\renewcommand{\i}{\ensuremath{\mathrm{i}}}
\renewcommand{\d}{\ensuremath{\mathrm{d}}}
\renewcommand{\Re}{\ensuremath{\text{Re}\,}}

\newcommand{\tr}[1]{\mathrm{Tr}\left(#1\right)}

\newcommand{\cO}{\mathcal{O}}

\newcommand{\cU}{\mathcal{U}}

\ifx\hypersetup\sadfkjashdfkxja\newcommand\hypersetup[1]{}\fi

\hypersetup{plainpages=false}
\hypersetup{pdfpagemode=UseNone}
\hypersetup{bookmarksnumbered=true}
\hypersetup{pdfstartview=FitH}
\hypersetup{colorlinks=false}
\hypersetup{citebordercolor={.5 1 .5}}
\hypersetup{urlbordercolor={.5 1 1}}
\hypersetup{linkbordercolor={1 .7 .7}}

\allowdisplaybreaks

\renewcommand{\a}{\alpha}
\newcommand{\da}{{\dot{\alpha}}}
\newcommand{\db}{{\dot{\beta}}}
\renewcommand{\b}{\beta}
\newcommand{\la}{\lambda}
\newcommand{\tla}{\tilde\lambda}

\renewcommand \widebar [1] {\overline{#1}}

\newcommand{\ft}[2]{{\textstyle\frac{#1}{#2}}}
\newcommand{\cL}{\mathcal{L}}
\newcommand{\cA}{\mathcal{A}}

\newcommand{\nicehalf}{\nicefrac{1}{2}}

\def\l<{\langle}
\def\r>{\rangle}
\newcommand{\lan}{\langle}
\newcommand{\ran}{\rangle}

\newcommand{\spA}[2]{\langle #1 #2 \rangle}
\newcommand{\spB}[2]{[ #1 #2 ]}
\newcommand{\spAB}[3]{\langle #1 | #2 | #3 ]}
\newcommand{\spBA}[3]{[ #1 | #2 | #3 \rangle}

\newcommand{\spBB}[3]{[ #1 | #2 | #3 ]}
\newcommand{\trm}[4]{ {\rm tr}_-( \slashed{#1} \slashed{#2} \slashed{#3} \slashed{#4} )}
\newcommand{\trmgamtwo}[4]{ {\rm tr}_-( \slashed{#1} #2 \slashed{#3} \slashed{#4} )}

\newcommand{\MHVb}{$\overline{\rm MHV}$}

\newcommand{\eps}{\epsilon}
\newcommand{\epsdot}[2]{(\epsilon_{#1}\cdot\epsilon_{#2})}

\newcommand{\bal}[1]{\setlength{\jot}{1em}\begin{align}\begin{aligned}#1\end{aligned}\end{align}}

\tikzset{
  c/.style={every coordinate/.try}}

\newcounter{exernumber}[chapter]
\renewcommand{\theexernumber}{\thechapter.\arabic{exernumber}}

\newenvironment{exer}[1]
    {\refstepcounter{exernumber}
     \begin{question}{Exercise \theexernumber : #1}
    }
    { 
    \end{question}
    }

\begin{document}

\author{Simon Badger, Johannes Henn, Jan Plefka and Simone Zoia}
\title{Scattering Amplitudes in Quantum Field Theory}
\subtitle{-- Monograph --}
\maketitle

\frontmatter

\include{dedication}
\preface

Scattering amplitudes have been termed the ``most perfect microscopic structures in the Universe''.\footnote{L.~J.~Dixon,
{\it ``Scattering amplitudes: the most perfect microscopic structures in the universe,''}
J. Phys. A \textbf{44} (2011), 454001.} 
Scattering amplitudes are elementary building blocks in quantum field theory that allow us to predict probabilities for the outcome of particle collisions. 
Therefore they are a crucial link connecting theory and experiment. For example, they allow us to test predictions from the Standard Model of particle physics against the collider data being collected at the Large Hadron Collider (LHC) at CERN.

\smallskip
The study of scattering amplitudes in quantum field theory has a long history,
dating back to the analytic $S$-matrix program of the 1960's. The modern field can trace its
origins to the 1980's. At that time, the state of the art of amplitude
computation was five-gluon scattering at tree level, i.e.\ the lowest order in
perturbation theory. Parke and Taylor famously managed to simplify what were
page-long results to a beautiful single-line formula, thus providing the first
of many hints for the underlying simplicity of scattering amplitudes in gauge
theory. To date, the state of the art has advanced by several loop orders,
i.e.\ to higher orders in the perturbative expansion in the coupling constant.
This was made possible by many conceptual advances. Modern approaches often use
gauge-invariant building blocks, as opposed to the traditional Feynman
diagrams, to organise calculations.
Perturbative unitarity made it possible to write down relations that recycle
these gauge-invariant building blocks, effectively leading to a recursion, both
in the number of particles and in the loop order. This is closely related to a
deeper understanding of the analytic structure of scattering amplitudes at loop level.

 \smallskip 
 
Our lecture notes provide an introduction for anyone wishing to learn more about this fascinating subject.
They are suitable for students at M.Sc.\ or Ph.D.\ levels.
We have made a specific selection of topics, so that they can be used for a one-semester university lecture series. This is based on our teaching experience, both at M.Sc.\ level at the universities of Berlin and Turin, as well as at Ph.D.-level summer schools worldwide. Additionally, they may be a useful primer for graduate students wishing to do research in this area. 
The chapters of the book cover necessary quantum field theory background, on-shell techniques for tree-level and one-loop scattering amplitudes, as well as dedicated techniques for evaluating Feynman loop integrals. 
Many exercises complement the text, and we provide fully worked-out solutions, as well as examples of how to implement the material using computer algebra codes. 
Supplementary material, \textsc{Mathematica} notebooks, corrections and further information are provided and maintained at the following 
dedicated website:
\begin{center}
\href{https://scattering-amplitudes.mpp.mpg.de/scattering-amplitudes-in-qft/}{https://scattering-amplitudes.mpp.mpg.de/scattering-amplitudes-in-qft/} \,.
\end{center}

\smallskip
Our aim is to provide a useful starting point to enable readers to access and contribute to current research in this thriving field.
Indeed, all the advancements mentioned above have significantly contributed to the community's ability to make more accurate predictions for particle scattering processes relevant to collider physics.
Beyond that, recent applications use these multi-loop techniques in Einstein's theory of gravity to provide efficient, high-precision predictions for the gravitational waveforms emitted in the encounter of black holes and
neutron stars. Therefore they provide the basis for data analysis in present and future gravitational wave detectors.
Furthermore, some of the properties that were revealed in the recent studies of scattering amplitudes in quantum field theories, 
such as hidden symmetries, dualities, and conceptually new approaches, 
have taught us entirely novel ways of thinking about quantum field theory. 
This may ultimately lead to a reformulation of the theory, where Feynman diagrams no longer play a fundamental role.
Since 2009, progress in this fast-growing field of theoretical physics is evident via the ``Amplitudes'' conferences held annually, which bring together researchers 
interested in both formal and phenomenological aspects of scattering amplitudes, with a wide range of applications from pure mathematics to collider and gravitational wave physics.

\smallskip

Some elements of modern on-shell methods for scattering amplitudes are also discussed in the quantum field theory textbooks
of Srednicki~{\it ``Quantum field theory''},
  Zee~{\it ``Quantum field theory in a nutshell''}
  and Schwartz~{\it ``Quantum Field Theory and the Standard Model''}, as well as in the dedicated reviews by Mangano and Parke~{\it ``Multiparton amplitudes in gauge theories''}, and by Dixon~{\it ``Calculating scattering amplitudes efficiently''}.
Other recent reviews on topics covered in these lecture notes include those by Weinzierl~{\it ``Tales of 1001 Gluons''}, and {\it ``Feynman Integrals''}. 
These offer a comprehensive account which can be especially interesting to mathematically-inclined readers.
For advanced topics in scattering amplitudes, focusing in particular on developments in supersymmetric theories, we highly recommend the textbook by Elvang and Huang~{\it ``Scattering Amplitudes in Gauge Theory and Gravity''}
 as well as the research monograph 
by Arkani-Hamed, Bourjaily, Cachazo, Goncharov, Postnikov and Trnka~{\it ``Grassmannian Geometry of Scattering Amplitudes''}.

\vspace{\baselineskip}
\begin{flushright}\noindent
Berlin, Munich and Turin\hfill {\it Simon Badger}\\
June 2023\hfill {\it Johannes M.\ Henn}\\
{\it Jan C.\ Plefka}\\
{\it Simone Zoia}\\
\end{flushright}
%
%

\extrachap{Acknowledgements}

We thank Sorana Scholtes for coordination, Ryan Moodie for help with the website, and Elena Mendoza and Ekta Chaubey for 
helpful comments on the manuscript.

This project received funding from the European Research Council (ERC) under the European Union's Horizon 2020 research and innovation programmes ``{\it Novel structures in scattering amplitudes}'' (grant agreement No.~725110) and ``{\it High precision multi-jet dynamics at the LHC}'' (grant agreement No.~772099).

\tableofcontents

%
%

\extrachap{Acronyms}

\begin{description}[CABR]
\item[BCFW]{Britto, Cachazo, Feng, Witten}
\item[BCJ]{Bern, Carrasco, Johansson}
\item[CERN]{Conseil européen pour la recherche nucléaire (European Organization for Nuclear Research)}
\item[DDM]{Del Duca, Dixon, Maltoni}
\item[DE]{Differential equation}
\item[EH]{Einstein, Hilbert}
\item[IBP]{Integration by parts}
\item[IR]{Infrared}
\item[ISP]{Irreducible scalar product}
\item[KLT]{Kawai, Lewellen, Tye}
\item[LHC]{Large hadron collider}
\item[LHS]{Left-hand side}
\item[LO]{Leading order}
\item[MHV]{Maximally helicity violating}
\item[MI]{Master integral}
\item[NMHV]{Next-to maximally helicity violating}
\item[OPP]{Ossola, Papadopoulos, Pittau}
\item[QCD]{Quantum chromodynamics}
\item[QED]{Quantum electrodynamics}
\item[QFT]{Quantum field theory}
\item[RHS]{Right-hand side}
\item[sYM]{super Yang-Mills}
\item[UV]{Ultraviolet}
\item[YM]{Yang-Mills}
\end{description}

\mainmatter

\chapter{Introduction and foundations}
\label{ch:intro}

\abstract{
In this introductory chapter we review the foundations
of perturbative, relativistic quantum field theory.
We focus on space-time and internal symmetries that are a highly successful guiding
principle in the construction and classification of relativistic quantum field theories.
We begin with the Poincar\'e group---the fundamental space-time symmetry of 
nature---that achieves the classification of elementary particles in terms of their masses and spins.
We review scalars, fermions, gauge fields and gravity, and expose their
perturbative quantisation leading to their Feynman rules.  Helicity
spinors are introduced that capture the polarisation and
momentum degrees of freedom of the scattered particles.
The internal non-Abelian gauge symmetry is reviewed and two methods for an efficient management of the colour degrees of freedom are discussed. They lead to
 the central concept of colour-ordered amplitudes. In the final section, we employ 
 this colour-ordered formalism to 
evaluate tree-level three- and four-gluon amplitudes, and depict general classes of vanishing tree-amplitudes of gluons and gravitons.
}

\section{Poincar\'e group and its representations}
\label{sec:1.1}

Quantum field theory unifies 
quantum mechanics and special relativity, and 
as such is a fundamental cornerstone of theoretical physics.\footnote{See 
\cite{ch1_Ramond:1981pw}, \cite{ch1_Weinberg:1995mt}, \cite{ch1_Schwartz:2014sze} for introductory
textbooks reviewing in particular symmetry aspects.} The underlying symmetry group of special relativity is the
Poincar\'e group, given as the semi-direct product of the Lorentz group and the Abelian group of space-time translations.
The four dimensional Lorentz group $\text{SO}(1,3)$ is a linear homogeneous coordinate transformation that leaves 
the relativistic length $x^{2}$ invariant,
\be
x^{\prime\mu} = \Lambda^{\mu}{}_{\nu}\, x^{\nu}\, ,
\qquad \text{with} \quad x^{\prime 2} = x^{2}=\eta_{\mu\nu}\, x^{\mu}\, x^{\nu} \, ,
\label{I.1}
\ee
where $\eta_{\mu\nu} = \text{diag}(+,-,-,-)$ denotes the flat space-time Minkowski metric. The Lorentz transformation matrices
$\Lambda^{\mu}{}_{\nu}$ depend on six parameters: three for spatial rotations, and three for Lorentz boosts.
This can be seen as follows. Demanding invariance of the relativistic length implies the following defining condition 
\be
\eta_{\mu\nu}\,\Lambda^{\mu}{}_{\rho}\, 
\Lambda^{\nu}{}_{\kappa} = \eta_{\rho\kappa}\, .
\label{I.2}
\ee 
Infinitesimally, we write
$\Lambda^{\mu}{}_{\nu}= \delta^{\mu}{}_{\nu}+\omega^{\mu}{}_{\nu} +\cO(\omega^{2})$ and 
find from 
eq. (\ref{I.2})
that $\omega_{\mu\nu}=\eta_{\mu\rho}\omega^{\rho}{}_{\nu}$ must be
antisymmetric, i.e., $\omega_{\mu\nu}=-\omega_{\nu\mu}$. Hence  $\omega_{\mu\nu}$ has 
six degrees of freedom, matching the above counting of rotations and boosts. 

In quantum theory the symmetry generators are represented by unitary operators, which we denote by $\cU(\Lambda)$. 
These furnish a representation of the Lorentz group and hence must obey the composition property
\be
\cU(\Lambda)\, \cU(\Lambda') = \cU(\Lambda\, \Lambda') \, .
\ee
Infinitesimally close to the identity we have
\begin{align}
\cU(\1+\omega) = \1 + \frac{\i}{2}\, \omega_{\mu\nu}\, M^{\mu\nu} \, ,
\end{align}
with the Hermitian operators $M^{\mu\nu}=-M^{\nu\mu}$ acting on the Hilbert space of the
quantum theory in question. The $M^{\mu\nu}$ are known as the generators of the Lorentz 
group. We would now like to derive the Lorentz algebra, i.e.~the commutation relations of the generators
$M^{\mu\nu}$. For this, consider 
\begin{align}
\cU(\Lambda)^{-1}\, \cU(\Lambda')\, \cU(\Lambda) = \cU(\Lambda^{-1}\, \Lambda' \, \Lambda)\,,
\end{align}
in the case of an infinitesimal $\Lambda'=\1+\omega'$. Expanding to linear order in $\omega'$
on both sides of the equation for arbitrary anti-symmetric $\omega_{\mu\nu}'$ yields
the transformation property of the Lorentz generator:
\be
\cU(\Lambda)^{-1}\, M^{\mu\nu}\, \cU(\Lambda) = \Lambda^{\mu}{}_{\rho}\,  
\Lambda^{\nu}{}_{\kappa}\,
M^{\rho\kappa}\, .
\label{I.3}
\ee
We see that each space-time index of $M^{\mu\nu}$ transforms with a $\Lambda^{\mu}{}_{\nu}$ matrix. Therefore a space-time vector such as $P^{\mu}$ should transform as
\be
\cU(\Lambda)^{-1}\, P^{\mu}\, \cU(\Lambda) = \Lambda^{\mu}{}_{\nu}\, P^{\nu}\, ,
\label{I.3b}
\ee
which holds in particular for the generator of space-time translations, the momentum operator
$P^{\mu}$ considered here. We now take the remaining Lorentz transformation $\Lambda$ in \eqn{I.3} and
\eqn{I.3b} to be infinitesimal as well,
$\Lambda=\1+\omega$. Stripping off the arbitrary anti-symmetric parameter $\omega_{\rho\kappa}$ on both sides of the resulting linearised equations yields the Poincar\'e algebra.

\newpage

\begin{svgraybox}{\bf Poincar\'e algebra.}
The central space-time symmetry group of nature is the Poincar\'e algebra.
From \eqn{I.3} and
\eqn{I.3b} we deduce the commutation relations
\begin{align}
\!\!\!\! [M^{\mu\nu} , M^{\rho\kappa}] &= \i \,(\, \eta^{\nu\rho}\, M^{\mu\kappa} +
 \eta^{\mu\kappa}\, M^{\nu\rho} - \eta^{\nu\kappa}\, M^{\mu\rho} - 
 \eta^{\mu\rho}\, M^{\nu\kappa} \, )\, , \label{I.4}\\
 [M^{\mu\nu} , P^{\rho}] &= -\i \, \eta^{\mu\rho}\, P^{\nu} +
 \i \, \eta^{\nu\rho}\, P^{\mu} \, ,
 \label{I.5}
 \end{align}
 where $[A,B]=AB-BA$ is the commutator of $A$ and $B$. It is of fundamental importance for relativistic
 quantum field theory.
 \end{svgraybox}

In quantum field theory the Lorentz generators act not only on the space-time
coordinates but also on the fields. A general representation of the Lorentz generators
of \eqn{I.4} then takes the form
\be
\label{Mrepgen}
(M^{\mu\nu})^{i}{}_{j}= \i\,  \left (x^{\mu}\,\frac{\partial}{\partial x_{\nu}} - 
x^{\nu}\,\frac{\partial}{\partial x_{\mu}} \right )\, \delta^{i}{}_{j} + (S^{\mu\nu})^{i}{}_{j}\, ,
\ee
with the $x^{\mu}$-independent $d_{R}\times d_{R}$ representation
matrices $(S^{\mu\nu})^{i}{}_{j}$ obeying the commutation relations of \eqn{I.4}.

We now wish to classify the possible representations of the Lorentz group. For this 
we drop for the moment 
the covariant notation and
define the rotation and boost generators
\be
J_{i} \coloneqq \ft{1}{2}\, \epsilon_{ijk}\, M_{jk}\, , \qquad K_{i} \coloneqq M_{0i}\, ,
\ee
with $i,j,k=1,2,3$ running over the spatial indices only. The $J_{i}$  obey
the $su(2)$ Lie algebra relations known from the angular momentum or spin commutation relations
in quantum mechanics. Introducing the following complex combinations of Hermitian 
generators,
\be
N_{i} \coloneqq \ft{1}{2}(J_{i} + \i\, K_{i} ) \, , \qquad
N_{i}^{\dagger} \coloneqq \ft{1}{2}(J_{i} - \i\, K_{i} ) \, ,
\ee
we see that the $so(1,3)$ algebra of \eqn{I.4} may be mapped to two commuting copies of
$su(2)$, 
\be
[N_{i},N_{j}]=\i\, \epsilon_{ijk}\, N_{k}\, , \qquad
[N_{i}^{\dagger},N_{j}^{\dagger}]=\i\, \epsilon_{ijk}\, N^{\dagger}_{k}\, , \qquad
[N_{i},N^{\dagger}_{j}]=0 \, .
\ee
Based on our knowledge of the representation theory of $\text{SU}(2)$ from the study of angular momentum
in quantum mechanics, we conclude that the
representations of the $\text{SO}(1,3)$ Lorentz group may be labeled by a doublet of half-integers
$(m,n)$ related to the eigenvalues $m(m+1)$  and $n(n+1)$ 
of the Casimir operators $N_{i}\, N_{i}$ and $N^{\dagger}_{i}\, N^{\dagger}_{i}$,
respectively. Moreover, since $J_{3}=N_{3}+N_{3}^{\dagger}$, we identify $m+n$ as
the spin of the representation $(m,n)$. We give a classification of the lower spin representations of the four-dimensional Lorentz group in table~\ref{tab:reps}.
\begin{table}[t]
   \centering
   \begin{tabular}{@{} lcll @{}} 
      \toprule
      Rep.    & Spin  & Field & Lorentz transformation property\\
      \midrule
      $(0,0)$      & 0   & scalar $\phi(x)$ & 
      $\cU(\Lambda^{-1})\,\phi(x)\,\cU(\Lambda) =\phi(\Lambda^{-1}x)  $\\
      $(\nicefrac{1}{2},0)$      & $\nicefrac{1}{2}$   & left-handed Weyl spinor $\chi_{\alpha}(x)$ 
      &
      $\cU(\Lambda^{-1})\,\chi_{\a}(x)\, \cU(\Lambda) =L(\Lambda)_{\a}{}^{\b}\chi_{\b}(\Lambda^{-1}x)$\\
      $(0,\nicefrac{1}{2})$      & $\nicefrac{1}{2}$   & right-handed Weyl spinor 
      $\bar\xi_{\da}(x)$ &
       $\cU(\Lambda^{-1})\,\bar\xi_{\da}(x)\,\cU(\Lambda) =R(\Lambda)_{\da}{}^{\db}\bar\xi_{\db}(\Lambda^{-1}x)$
      \\
           $(\nicefrac{1}{2},\nicefrac{1}{2})$      & $1$   & vector 
      $A_{\mu}(x)$ &
      $\cU(\Lambda^{-1})\,A_{\mu}(x)\,\cU(\Lambda) =\Lambda_{\mu}{}^{\nu}A_{\nu}(\Lambda^{-1}x)$ \\
     $(1,0)$      & $1$   & self-dual rank 2 tensor
      $B_{\mu\nu}(x)$ & \\
          $(0,1)$      & $1$   & anti-self-dual rank 2 tensor
      $\tilde B_{\mu\nu}(x)$ &  \\
       $(1,\nicefrac{1}{2})$      & $\nicefrac{3}{2}$   & gravitino  $\psi^{\mu}_{\alpha}(x)$ & \\
       $(1,1)$      & $2$   & graviton $h_{\mu\nu}(x)$ & \\
      \bottomrule
   \end{tabular}
   \caption{Lower spin representations of the four-dimensional Lorentz group. For the considerations in this text only the first
   four will be of importance. We have $\a=1,2$, $\da=1,2$, and $\mu=0,1,2,3$.}
   \label{tab:reps}
\end{table}

\section{Weyl and Dirac spinors}
\label{sec:1.2}

We now wish to construct a Lagrangian for the $(\nicefrac{1}{2},0)$ representation: the left-handed Weyl
spinor of table~\ref{tab:reps}. 
The relevant $2\times 2$ representation matrix $S^{\mu\nu}_{\rm L}$ arising in the corresponding 
representation of $M^{\mu\nu}$ in \eqn{Mrepgen} takes the form
\be \label{eq:S_L}
(S^{\mu\nu}_{\rm L})_{\a}{}^{\b}= \ft{\i}{4}\, (\sigma^{\mu}\,\bar\sigma^{\nu}-
\sigma^{\nu}\,\bar\sigma^{\mu})_{\a}{}^{\b} \, , \qquad \a,\b=1,2\, , 
\ee
with $(\bar\sigma^{\mu})^{\da\a}=(\1,-\vec\sigma)$ and $(\sigma^{\mu})_{\a\da}=\epsilon_{\a\b}
\, \epsilon_{\da\db}\, (\bar\sigma^{\mu})^{\db\b}= (\1,\vec\sigma)$, where $\vec\sigma$ is the list of Pauli matrices, and $\epsilon_{\a\b}$ is the Levi-Civita tensor.\footnote{Our conventions are summarised in appendix~\ref{sec:Conventions}.}
The free Lagrangian for a massive Weyl spinor field reads
\be
\cL_{\rm W}=\i\tilde\chi_{\da}\, (\bar\sigma^{\mu})^{\da\a}\, \partial_{\mu}\chi_{\a}
-\ft{1}{2}\, m\, \chi^{\a}\,\chi_{\a} -\ft{1}{2}\, m^{\ast}\, \tilde\chi_{\da}\,\tilde\chi^{\da}\,  ,
\label{I.8}
\ee
where we denote $(\chi_{\a})^{\dagger}=\tilde\chi_{\da}$ and $\partial_{\mu}=\frac{\partial}{\partial x^{\mu}}$. 
It is invariant under Poincar\'e transformations.
Recall that the half-integer spin fields are  anti-commuting (Gra{\ss}mann odd)
quantities. The equations of motion follow from the action $S=\int \d^{4}x \, \cL_{\rm W}$ by
variation of the action w.r.t.~$\chi$ and $\tilde\chi$. We find
\begin{align}
-\frac{\delta S}{\delta \tilde\chi_{\da}}&=-\i(\bar\sigma^{\mu})^{\da\a}\,\partial_{\mu}\chi_{\a}
+m^{\ast}\, \tilde\chi^{\da}=0 \, , \label{I.9} \\
-\frac{\delta S}{\delta \chi^{\a}}&=-\i(\sigma^{\mu})_{\a\da}\,\partial_{\mu}\tilde\chi^{\da}
+m\, \chi_{\a}=0 \, .
\label{I.10}
\end{align}
Note that the second equation follows from complex conjugation of the first and is therefore
spurious. In general the mass may be taken to be complex $m=|m|\,\mathrm{e}^{\i\alpha}$. However, the phase of a complex mass
may be absorbed in a redefinition of the spinor fields, so that in the end we may set $m=m^{\ast}$. These equations of
motion may be unified into a four-component notation as
\be
0=   \begin{pmatrix} 
      m\,\delta_{\a}{}^{\b} & -\i(\sigma^{\mu})_{\a\db}\,\partial_{\mu} \\
      -\i(\bar\sigma^{\mu})^{\da\b}\,\partial_{\mu}  & m\, \delta^{\da}{}_{\db} \\
   \end{pmatrix}
   \,
   \begin{pmatrix} \chi_{\b} \\ \tilde\chi^{\db}
   \end{pmatrix}\, .
\ee
Introducing the $4\times 4$ Dirac matrices in the chiral representation\footnote{See exercise~\ref{Ex:1.2} for an analysis of the chiral and Dirac representations of the Dirac matrices.}
\be
\gamma^{\mu} \coloneqq
\begin{pmatrix} 
   0 & (\sigma^{\mu})_{\a\db} \\
       (\bar\sigma^{\mu})^{\da\b}  &  0 \\
   \end{pmatrix}
   \, ,
   \label{diracchiral}
\ee
which obey the Clifford algebra $\{\gamma^{\mu},\gamma^{\nu}\}=2\,\eta^{\mu\nu}$,
calls for the introduction of a four-component spinor field 
\be
\psi_{M}= 
 \begin{pmatrix} \chi_{\b} \\ \tilde\chi^{\db}
   \end{pmatrix}\, ,
\ee
known as the Majorana field $\psi_{M}(x)$.
Using this, the equation of motion
may be cast in the form of a Dirac equation:
\be
(-\i\, \gamma^{\mu}\, \partial_{\mu} +m)\, \psi_{M} =0 \, .
\ee

\begin{svgraybox}{\bf Dirac spinor and equation.}
Generalising this, we may combine a left-handed Weyl spinor $\chi_{\a}$ and an 
independent right-handed Weyl spinor
$\bar\xi^{\da}$ into a four component Dirac spinor,
\be
\psi= \begin{pmatrix} \chi_{\a} \\ \bar\xi^{\da}
   \end{pmatrix} \, .
\ee
We can then write down the Dirac equation $(-\i\, \gamma^{\mu}\partial_{\mu} + m)\, \psi =0$,
and the associated Lagrangian---in an index free notation---reads
\be
\cL_{\rm D}= \i\,\bar\psi\, \gamma^{\mu}\,\partial_{\mu}\psi - m \, \bar\psi\psi\, ,
\quad \text{with} \quad \bar\psi \coloneqq \psi^{\dagger}\,\gamma^{0}\, .
\label{I.11}
\ee
This is the reducible field theory of the sum of a $(\nicefrac{1}{2},0)$ left-handed Weyl and a $(0,\nicefrac{1}{2})$ right-handed
Weyl spinor. 
\end{svgraybox}

\begin{exer}{Manipulating spinor indices}
\label{Ex:1.1}
The Levi-Civita symbols are used to raise and lower Weyl indices according to
$\bar\xi_{\dot\alpha}=\epsilon_{\dot\alpha\dot\beta} \, \bar\xi^{\dot\beta}$ and
$\chi^{\alpha}=\epsilon^{\alpha\beta}\, \chi_{\beta}$. We have
$$
\epsilon_{12}=\epsilon_{\dot 1\dot 2}=\epsilon^{21}=\epsilon^{\dot 2\dot 1}=1\, ,\qquad
\epsilon_{21}=\epsilon_{\dot 2\dot 1}=\epsilon^{12}=\epsilon^{\dot 1\dot 2}=-1\, .
$$
The sigma-matrix four-vector is defined by $(\bar\sigma^{\mu})^{\dot\alpha\alpha}=(\1,-\vec\sigma)$.
Moreover we have $(\sigma^{\mu})_{\alpha\dot\alpha}\coloneqq
\epsilon_{\alpha\beta}\, \epsilon_{\dot\alpha\dot\beta}\, (\bar\sigma^{\mu})^{\dot\beta\beta}$.
Prove the relations
\begin{align*}
\begin{array}{ll}
(1) \ (\sigma^{\mu})_{\alpha\dot\alpha}=(\1,\vec\sigma)\,, &
(2) \ (\sigma_{\mu})_{\alpha\dot\alpha}=(\1,-\vec\sigma)\,, \\
(3) \ \tr{\sigma^{\mu} \bar\sigma^{\nu}} = 2 \, \eta^{\mu \nu}\,, \quad \quad \quad & 
(4) \ (\sigma^{\mu})_{\alpha\dot\alpha}\, (\sigma_{\mu})_{\beta\dot\beta} = 2\,\epsilon_{\alpha\beta}\,
 \epsilon_{\dot\alpha\dot\beta}\  \,. 
 \end{array}
\end{align*}
For the solution see \hyperref[Prob:1.1]{chapter~5}.
\end{exer}

\vspace{-0.4cm}

\begin{exer}{Massless Dirac equation and Weyl spinors}
\label{Ex:1.2}
Consider the Dirac representation of the Dirac matrices:
\begin{align} \label{eq:gamma_dirac}
\gamma^{0}= \left ( \begin{matrix} {\1}_{2\times 2}& 0\cr 0& -{\1}_{2\times 2} 
\end{matrix}\right )\, ,\quad
\gamma^{i}= \left ( \begin{matrix} 0& {\sigma^{i}}\cr -{\sigma^{i}}
& 0 \end{matrix}\right )\, ,\quad
\gamma^{5}= \i\gamma^{0}\gamma^{1}\gamma^{2}\gamma^{3}=
\left ( \begin{matrix} 0 &{\1}_{2\times 2}\cr {\1}_{2\times 2} &0 
\end{matrix}\right )\, .
\end{align}
\begin{enumerate}[a)]
\item Show that the solutions of the massless Dirac equation $\gamma^{\mu} k_{\mu}\psi =0$
may be chosen~as
\begin{align}
u_{+}(k) =v_{-}(k) = \frac{1}{\sqrt{2}}\, \left (\begin{matrix} \sqrt{k^{+}}\cr
\sqrt{k^{-}}\, \mathrm{e}^{\i\phi(k)}\cr \sqrt{k^{+}}\cr
\sqrt{k^{-}}\, \mathrm{e}^{\i\phi(k)}\cr 
\end{matrix} \right )\, , \quad 
u_{-}(k) =v_{+}(k) = \frac{1}{\sqrt{2}}\, \left (\begin{matrix} \sqrt{k^{-}}\, \mathrm{e}^{-\i\phi(k)}\cr
-\sqrt{k^{+}}\cr -\sqrt{k^{-}}\, \mathrm{e}^{-\i\phi(k)}\cr
\sqrt{k^{+}}\cr 
\end{matrix} \right ) \,,
\end{align}
where 
\begin{align}
\mathrm{e}^{\pm \i\phi(k)}\coloneqq \frac{k^{1}\pm \i k^{2}}{\sqrt{k^{+}\,k^{-}}}\,, \qquad
k^{\pm}\coloneqq k^{0}\pm k^{3}\, ,
\end{align}
and that the spinors $u_{\pm}(k)$ and $v_{\pm}(k)$ obey the helicity
relations
\begin{align}
P_{\pm}u_{\pm}=u_{\pm}\, ,
\qquad 
P_{\pm}u_{\mp}=0\, , \qquad P_{\pm}v_{\pm}=0\, , \qquad
P_{\pm}v_{\mp}=v_{\mp}\, ,
\end{align}
with
\begin{align}
P_{\pm} \coloneqq\frac{1}{2}\, (\1 \pm \gamma^{5})\,.
\end{align}

\item What helicity relations hold for the conjugate expressions $\bar u_{\pm}(k)$ and
$\bar v_{\pm}(k)$, where we define $\bar\psi \coloneqq \psi^{\dagger}\gamma^{0}$?
\item Now consider the unitary transformation to the chiral representation of the Dirac matrices,
$$
\psi\to U\,  \psi\, \, ,\qquad \gamma^{\mu}\to U\, \gamma^{\mu}\, U^{\dagger} \, ,
$$
with $U= (\1_4 -\i \gamma^{1}\,\gamma^{2}\,\gamma^{3})/\sqrt{2}$.
Determine $\gamma^{\mu}$, $\gamma^{5}$, and the spinors $u_{\pm}(k)$ and $v_{\pm}(k)$ in this chiral basis.

\item Using the chiral representation of the Dirac matrices, prove that
\begin{align} \label{eq:trSigma2gamma}
\Tr\left( \sigma^{\mu} \bar{\sigma}^{\nu} \sigma^{\rho} \bar{\sigma}^{\tau} \right) = \frac{1}{2} \Tr\left(\gamma^{\mu} \gamma^{\nu} \gamma^{\rho} \gamma^{\tau} ({\1}-\gamma_5)\right) \,.
\end{align}
\end{enumerate}
For the solution see \hyperref[Prob:1.2]{chapter~5}.
\end{exer}

\section{Non-Abelian gauge theories}
\label{sec:1.3}

We now discuss the  
principle of local gauge invariance due to Yang and Mills~\cite{ch1_Yang:1954ek}, which is central
to the theory of the fundamental non-gravitational interactions in nature. This will lead us 
to non-Abelian gauge theories that are constitutional for the standard model of elementary particles and beyond.
The spin $\nicehalf$ Lagrangians for a Weyl spinor $\cL_{\rm W}$ of \eqn{I.8} and a Dirac spinor $\cL_{\rm D}$ of \eqn{I.11} are invariant under
global phase transformations:
\be
\chi\to \mathrm{e}^{\i\alpha}\, \chi\,  , \qquad \qquad \psi\to \mathrm{e}^{\i\alpha}\, \psi \,,
\ee
with $\a\in\R$, respectively.
We now wish to elevate this global symmetry to a \emph{local} symmetry, i.e.~to allow for a space-time dependent phase transformation
$\alpha(x)$. The kinetic terms in the actions $\cL_{\rm W}$ and $\cL_{\rm D}$ are then
no longer invariant, as the space-time derivative $\partial_{\mu}$ may hit the $\alpha(x)$.
This can be cured by introducing a gauge field $A_{\mu}(x)$ to cancel the unwanted terms.
For this purpose, 
the local gauge transformations of the Dirac field (we specialise to this case from now on) 
and of the novel gauge field $A_{\mu}(x)$ are given by
\be
\psi\to \mathrm{e}^{\i\, e\,\alpha(x)}\, \psi\, , \qquad \qquad A_{\mu} \to A_{\mu} + \partial_{\mu}\alpha(x) \,,
\label{u(1)transformation}
\ee
where $e$ denotes the coupling constant.
Moreover, the derivative $\partial_{\mu}$ in the Dirac action of \eqn{I.11} is replaced 
by the covariant derivative $D_{\mu}=\partial_{\mu}-\i e A_{\mu}$.

\begin{svgraybox}{\bf Quantum electrodynamics.}
Introducing the covariant derivative in the Dirac Lagrangian we are led to consider
the theory
\be
\cL_{\rm QED}= \i\,\bar\psi\, \gamma^{\mu}\,D_{\mu}\psi - m \, \bar\psi\psi
-\ft{1}{4}\, F_{\mu\nu}\, F^{\mu\nu}\, ,
\label{I.12}
\ee
where we also added the field-strength tensor term $\ft 14 F_{\mu\nu}F^{\mu\nu}$
known from Maxwell's theory of electromagnetism. It generates the kinetic term for the gauge field $A_{\mu}$. Recall that  $F_{\mu\nu}$ is defined as
\be
F_{\mu\nu} = \frac{\i}{e}\, [D_{\mu},D_{\nu}] = \partial_{\mu}A_{\nu} -\partial_{\nu}A_{\mu}\, .
\label{I.FieldStrength}
\ee
\end{svgraybox}

This is an Abelian gauge theory invariant under the $\text{U}(1)$
transformations~\eqref{u(1)transformation}.
For the case of $e$ and $m$ being the charge and mass of the electron, this is 
the theory of quantum electrodynamics (QED).

Let us now formalise this construction slightly by associating to the local Abelian gauge transformation of \eqn{u(1)transformation}
an $x$-dependent $\text{U}(1)$ group element $U(x)=\mathrm{e}^{\i e \alpha(x)}$ obeying  $U(x)^{\dagger}\, U(x)=1$. It 
generates the transformations
\be
\psi \to U(x)\, \psi \, , \qquad D_{\mu}\to U(x)\,D_{\mu}\, U^{\dagger}(x)\, ,
\label{U1transf}
\ee
which leave $\bar\psi\gamma^{\mu}D_{\mu}\psi$ and $\bar\psi\psi$ manifestly invariant, as
$\bar\psi\to\bar\psi \,U^{\dagger}(x)$. The transformation
rule for $D_{\mu}=\partial_{\mu}- \i eA_{\mu}$ above implies the transformation of the gauge field
\be
A_{\mu}(x) \to U(x)\, A_{\mu}(x)\, U^{\dagger}(x) + \frac{\i}{e}\, U(x) \, \partial_{\mu}U^{\dagger}(x)\, .
\ee
One indeed easily verifies the equivalence to \eqn{u(1)transformation}.

We now wish to lift this construction to a \emph{non-Abelian gauge symmetry}, i.e.~we want to find 
matrix-valued generalisations of $U(x)$. To this end, consider a
set of $N_{c}$ Dirac spinor fields $\psi_{i,A}$ with spinor index $A=(\a,\da)$ and  an
additional  index $i=1,\ldots, N_{c}$. The number of components $N_{c}$ is referred to as
the number of ``colours'' for actually no good reason. 
The associated Dirac Lagrangian
\be
\cL_{D_{N}}= \sum_{i=1}^{N_{c}} \i \, \bar\psi^{i} \,\slashed{\partial}\psi_{i} - m\,\bar\psi^{i}\psi_{i}
\ee
is again invariant under the global unitary transformation
\be
\psi_{i}(x)\to U_{i}{}^{j}\, \psi_{j}(x)\,,
\label{I.14}
\ee
with the $N_{c}\times N_{c}$ matrices $U_{i}{}^{j}$ obeying $U^{\dagger}\,U=U\, U^{\dagger}=\1$.
These matrices $U_{i}{}^{j}$ span the Lie group of unitary transformations $\text{U}(N_{c})$. 
We shall further specialise to the case of special unitary transformations $\text{SU}(N_{c})$ by imposing
 the additional condition $\text{det}(U)=1$. 

We will focus here on gauge theories built from
$\text{SU}(N_c)$, although all other compact semi-simple Lie groups $\text{SO}(N_c)$, $\text{Sp}(2N_c)$, and the five 
exceptional $\text{G}_{2}$, 
$\text{F}_{4}$, $\text{E}_{6}$, $\text{E}_{7}$ and $\text{E}_{8}$ may in principle be also 
considered.

The global symmetry of $\cL_{D_{N}}$ under \eqn{I.14} may now be turned into a \emph{local}
 non-Abelian symmetry
$U_{i}{}^{j}\to U_{i}{}^{j}(x)$ with arbitrary space-time dependence. 
We introduce the covariant, matrix-valued derivative
\be
(D_{\mu})_{i}{}^{j} \coloneqq \delta_{i}{}^{j}\, \partial_{\mu} - \i \, g\, (A_{\mu})_{i}{}^{j}(x)\,,
\ee
with the  $\text{SU}(N_c)$ gauge field $(A_{\mu})_{i}{}^{j}(x)$. 
The coupling constant is now denoted by $g$, generalising the electric charge $e$ of the Abelian case. We then generalise the Abelian \eqn{U1transf}~to a covariant 
non-Abelian transformation 
\be
(D_{\mu})_{i}{}^{j}\to U_{i}{}^{k}(x)\,(D_{\mu})_{k}{}^{l}\, (U^{\dagger})_{l}{}^{j}(x)
\ee
that leads to the transformation rule (now in matrix notation)
\be
\label{1.32}
A_{\mu}(x) \to U(x)\, A_{\mu}(x)\, U^{\dagger}(x) + \frac{\i}{g}\, U(x) \, \partial_{\mu}U^{\dagger}(x)\, .
\ee

\newpage

\begin{svgraybox}{\bf $\text{\bf SU}(N_c)$ gauge theory.}
With the help of this construction the ``gauged'' Lagrangian
(here and in the following we omit the sum over colour indices)
\be
\cL_{D_{n}}'= \i\,\bar\psi^{i}\,\slashed{D}_{i}{}^{j} \psi_{j} - m\,\bar\psi^{i}\psi_{i}
\ee
is invariant under \emph{local} $\text{SU}(N_c)$ gauge transformations. Next, we need to construct 
the kinetic term for the
non-Abelian gauge field $(A_{\mu})_{i}{}^{j}$. In generalisation of the Abelian construction above, cf.~\eqn{I.FieldStrength}, 
the natural quantity to take is (again in matrix notation)
\be
F_{\mu\nu} = \frac{\i}{g}\, [D_{\mu},D_{\nu}] = \partial_{\mu}A_{\nu} -\partial_{\nu}A_{\mu}
- \i \, g \, [A_{\mu},A_{\nu}]\, .
\label{I.NAF}
\ee
This is the non-Abelian field strength tensor. Note that it is not invariant under gauge transformation, but
transforms as 
\begin{align}
F_{\mu\nu}\to U(x)\, F_{\mu\nu}\, U^{\dagger}(x)\,.
\end{align}
One says it transforms
covariantly under gauge transformations.
The kinetic term for the gauge field then is
\be
\label{FmnFmn}
\cL_{\rm YM}= -\ft{1}{4}\, \Tr(F_{\mu\nu}\, F^{\mu\nu}) \,,
\ee
which is both gauge invariant, thanks to the colour trace, and Lorentz invariant, as
all indices are properly contracted. We note that, as opposed to the Abelian $\text{U}(1)$ case, the 
non-Abelian gauge field is self-interacting due to the
 commutator term
in \eqn{I.NAF}. The interaction strength is controlled by the coupling constant $g$. 
The complete Lagrangian of $\text{SU}(N_c)$ gauge theory interacting with a Dirac ``matter'' field is then
given by the sum of $\cL_{D_{n}}'$ and $\cL_{\rm YM}$.
\end{svgraybox}

In order to better understand the structure of this gauge theory it is useful to look at 
gauge transformations infinitesimally close to the identity
\be
\label{1.36}
U_{i}{}^{j}(x)=\delta_{i}{}^{j}- \i \, g\, \theta^{a}(x)\, (T^{a})_{i}{}^{j}\, ,
\ee
where we introduced the Lie algebra generators $(T^{a})_{i}{}^{j}$ of $\text{SU}(N_c)$ with $a=1,\ldots, N_c^{2}-1$ and
$i,j=1,\ldots, N_c$. In the above, the $\theta^{a}(x)$ serve as the local transformation parameters
generalising the phase $\a(x)$ of the Abelian
case ($N_c=1$). As a consequence of the $\text{SU}(N_c)$ group properties, i.e.~$U^{\dagger}U=\1$,
the $(T^{a})_{i}{}^{j}$ are Hermitian traceless $N_c\times N_c$ matrices. They obey the commutation relations
\be
[ \, T^{a}, T^{b}\, ] = \i\sqrt{2}\, f^{abc}\, T^{c}\,,
\label{I.15}
\ee
with the \emph{structure constants} $f^{abc}$. The factor of $\sqrt{2}$ is our normalisation
convention. We can choose a diagonal basis for the generators $T^{a}$ such that
\be
\Tr(T^{a}\, T^{b}) =\delta^{ab}\, .
\label{I.15.5}
\ee
Based on this we may write
\be
f^{abc}=-\frac{\i}{\sqrt{2}}\, \Tr( \, T^{a}\, [T^{b},T^{c}]\, ) \, ,
\label{I.17}
\ee
which renders the structure constants totally anti-symmetric in all 
indices.\footnote{Concrete expressions for the $\text{SU}(N_c)$ generators
may be found in appendix~\ref{sec:Conventions}.}
The Jacobi identity for the generators
\be \label{eq:JacobiGenerators}
\bigl[T^{a}, [ T^{b}, T^{c}]\bigr] + \bigl[T^{b}, [ T^{c}, T^{a}]\bigr] +\bigl[T^{c}, [ T^{a}, T^{b}]\bigr]  = 0 
\ee
then directly translates into the relation
\be
\label{Jacobif}
f^{abe} f^{ceg} + f^{bce} f^{aeg} + f^{cae} f^{beg} =0 \,,
\ee
known as the Jacobi relation for the structure constants.
Furthermore we note the important $\text{SU}(N_c)$ identity
\be \label{eq:completenessSUN}
(T^{a})_{i_{1}}{}^{j_{1}}\, (T^{a})_{i_{2}}{}^{j_{2}} = \delta_{i_{1}}{}^{j_{2}}\,  
\delta_{i_{2}}{}^{j_{1}}
-\frac{1}{N_c}\, \delta_{i_{1}}{}^{j_{1}}\,  \delta_{i_{2}}{}^{j_{2}}\,,
\ee
which is nothing but a completeness relation for $N_c\times N_c$ Hermitian matrices and where
the last term ensures the tracelessness of the $(T^{a})_{i}{}^{j}$.

\begin{exer}{$\text{SU}(N_c)$ identities}
\label{Ex:1.3}
\vspace{-0.6cm}
\begin{enumerate}[a)]
\item Prove the Jacobi identity~\eqref{eq:JacobiGenerators} for the generators. Hint: expand all commutators.
\smallskip
\item Prove the Jacobi identity~\eqref{Jacobif} for the structure constants. Hint: use \eqn{I.15} to trade commutators for structure constants in the Jacobi identity for the generators.
\smallskip
\item Prove the $\text{SU}(N_c)$ completeness relation given in \eqn{eq:completenessSUN}. Hint: consider an arbitrary $N_c \times N_c$ complex matrix, and expand it in the basis given by the identity $\1_{N_c}$ and the $su(N_c)$ generators $T^a$.
\end{enumerate}
For the solution see \hyperref[Sol:1.3]{chapter~5}.
\end{exer}

As we took the gauge fields to be traceless Hermitian matrices we can expand them in the 
basis of $\text{SU}(N_c)$ generators $T^{a}$, as
\be
(A_{\mu})_{i}{}^{j}(x) = A_{\mu}^{a}(x)\, (T^{a})_{i}{}^{j} \quad
\Leftrightarrow \quad A^{a}_{\mu}(x) = \Tr\Bigl(T^{a}\, A_{\mu}(x)\Bigr)\, .
\ee
Similarly, the field strength may be decomposed as $F_{\mu\nu}(x)
= F^{a}_{\mu\nu}(x)\, T^{a}$ and
\be
F^{c}_{\mu\nu}= \partial_{\mu} A^{c}_{\nu} -\partial_{\nu} A^{c}_{\mu}
+\sqrt{2}\, g\, f^{abc}\, A^{a}_{\mu}\, A^{b}_{\nu}\, ,
\ee
yielding $\cL_{\rm YM}= -\frac{1}{4}\, F^{c}_{\mu\nu}\, F^{c\, \mu\nu}$.

\begin{table}[t]
   \centering
   \begin{tabular}{@{} lccccccc @{}}
      \toprule
      quarks:    & up  & down & charm & strange & top & bottom  & gluon \\
      \midrule
      $\widebar{\text{MS}}$ mass:    & 2.16 MeV  &4.67 MeV &  1.27 GeV & 93.4 MeV & 172.7 GeV & 4.18 GeV & 0\\
      symbol: & $\Psi_{1,i}$ & $\Psi_{2,i}$ & $\Psi_{3,i}$ & $\Psi_{4,i}$ & $\Psi_{5,i}$ & 
      $\Psi_{6,i}$ & $A_{\mu}^{a}$ \\
      spin: & $\nicehalf$  & $\nicehalf$& $\nicehalf$& $\nicehalf$& $\nicehalf$& $\nicehalf$ & 1 \\
      SU(3) rep.: & {\bf 3 } & {\bf 3}  & {\bf 3}  & {\bf 3}  & {\bf 3}  & {\bf 3} & {\bf 8}\\
      \bottomrule
   \end{tabular}
   \caption{The fields of QCD. Note the enormous spread in quark masses: $m_{t}/m_{u}=7.9\times10^{5}$}
   \label{tab:qcd}
\end{table}

\begin{svgraybox}{\bf Quantum Chromodynamics.}
The most important realisation of non-Abelian gauge field theory is Quantum Chromodynamics (QCD), the theory of the strong interactions, that describes the interactions of quarks and gluons in nature. It is built on the gauge group $\text{SU}(3)$. The field content  consists of 8 gauge fields known as gluons, 
$A^{a}_{\mu}(x)$ $(a=1,\ldots, 8)$,
together with 6 flavours of quark fields, $\psi_{I,i}$ ($i=1,2,3$, and $I=1,\ldots, 6$),
see~fig.\ref{tab:qcd}.
The QCD Lagrangian reads
\be
\label{QCDLagrangian}
\cL_{\rm QCD} = \i\,\bar\psi_{I}^{i}\, \,\slashed{D}_{i}{}^{j} \psi_{I\, j} 
- m_{I}\,\bar\psi_{I}^{i}\psi_{I\, i} -\ft{1}{4} \, F^{c}_{\mu\nu}\, F^{c\, \mu\nu}\, ,
\ee
where the masses $m_{I}$ span a range from 2 MeV for the up-quark to
172 GeV for the top-quark.
\end{svgraybox}

Given the structure constants $f^{abc}$ of a non-Abelian gauge group as in \eqn{I.15},
one can search for representations of the group in terms of $d_{R}\times d_{R}$ dimensional
matrices $(T^{a}_{R})_{I}{}^{J}$ with $I,J=1,\ldots , d_{R}$ obeying
\be
[T_{R}^{a}, T^{b}_{R}] = \i\sqrt{2}\, f^{abc}\, T^{c}_{R} \, ,
\label{I.20}
\ee
with $d_{R}$ denoting the dimension of the representation. 
So far we discussed the
fundamental or defining representation of $\text{SU}(N_c)$ in the form of $N_c\times N_c$ Hermitian
matrices. 
As $f^{abc}$ is real, we see by  complex conjugating \eqn{I.20} that for a given representation $T^{a}_{R}$ there always exists a complex conjugate representation  $T^{a}_{\bar R} \coloneqq -T^{a\, \ast}_{R}$.
Another important representation is the 
$(N_{c}^{2}-1)$-dimensional adjoint representation
induced by the structure constants $f^{abc}$ themselves. Its generators $(T^{a}_{A})^{bc}$ are defined as
\be \label{eq:GeneratorsAdjoint}
(T^{a}_{A})^{bc} \coloneqq -\i\sqrt{2}\, f^{abc} \, .
\ee
These indeed furnish a representation of the algebra due to the Jacobi identity for the structure constants~\eqref{Jacobif}.
The matrix indices are raised and lowered freely in this representation thanks to the diagonal 
metric in \eqn{I.15.5}. 
(We also note that $T^{a}_{\bar A}= T^{a}_{A}$, as $f^{abc}$ is real.)
The infinitesimal transformation of the gauge fields following from \eqn{1.32} and
\eqn{1.36} reads
\be \begin{aligned}
\delta A^{a}_{\mu} & = \partial_{\mu}\Theta^{a} +\sqrt{2}\, g\, f^{abc}\, \Theta^{b}\,
A^{c}_{\mu}  \\
& = \partial_{\mu}\Theta^{a} + \i\, g\, \Theta^{b}(T^{b}_{A})^{ac}\, 
A^{c}_{\mu} \, .
\end{aligned} \ee
Quarks, on the other hand, transform in the $N_c$-dimensional fundamental representation
\be
\delta \psi_{i} = \Theta^{a}\, (T^{a}_{F})_{i}{}^{j}\, \psi_{j}\, .
\ee
Comparing the two, we see that gauge fields transform in the adjoint
representation in their homogenous part.
As a final comment, representations are often denoted by their dimensionality in 
boldfaced letters, e.g.~for
QCD the quarks are in the {\bf 3}, the anti-quarks in the ${\bf \bar 3}$,
 whereas the gluons are in the {\bf 8} of $\text{SU}(3)$.
 See \cite{ch1_Cvitanovic:1976am} for further reading on group theoretical aspects
 of gauge theories.

 \begin{exer}{Casimir operators}
\label{Ex:1.4}
A \emph{Casimir operator} is an element of a Lie algebra which commutes with all generators.
\begin{enumerate}[a)]
\item Prove that the quadratic operator $T^a T^a$ is a Casimir operator of $su(N_c)$.
\item By Schur's lemma, the Casimir operator of an irreducible representation $R$ must be proportional to the identity,
\begin{align}
T^a_R T^a_R = C_R \, \1_{d_R} \,,
\end{align}
where $d_R$ is the dimension of the representation $R$, and $C_R$ is called (quadratic) Casimir invariant. Prove that the Casimir invariants of the fundamental and the adjoint representations are given by
\begin{align}
C_F = \frac{N_c^2-1}{N_c} \,, \qquad \quad C_A = 2 \, N_c \,.
\end{align}
\end{enumerate}
For the solution see \hyperref[Sol:1.4]{chapter~5}.
\end{exer}

 \section{Feynman rules for non-Abelian gauge theories}
 \label{chap:1.4}

Let us now discuss the Feynman rules for non-Abelian gauge theories. In the pure Yang-Mills theory
we have the following explicit form of the Lagrangian~\eqref{FmnFmn}:
\begin{align} \label{YMexp} \begin{aligned}
\cL_{\rm YM} = \ & 
- \sfrac{1}{2} \partial_{\mu} A_{\nu}^{a}  \partial^{\mu} A^{a\, \nu} + \sfrac{1}{2} \partial_{\mu} A_{\nu}^{a}  \partial^{\nu} A^{\mu\, a} \\
& -g f^{abc} A^{a\, \mu} A^{b\, \nu} \partial_{\mu}A_{\nu}^{c} -\sfrac{1}{4} \, g^{2} f^{abe}f^{cde}A^{a\, \mu}A^{b\, \nu}
A^{c}_{\mu}A^{d}_{\nu} \,.
\end{aligned} \end{align}
In principle, the first line of eq. (\ref{YMexp}) yields the kinetic terms of the theory. 
However, due to the local gauge symmetry, we need
to first fix a gauge in the path integral quantisation in order to not ``overcount'' physically-equivalent field configurations.
A popular covariant gauge fixing function is $G^{a}= \partial^{\mu}A^{a}_{\mu}$. In the Fadeev-Popov procedure this is
implemented by adding a gauge-fixing and a ghost term,
\begin{align} \label{FadeevPopovYM} \begin{aligned}
& \mathcal{L}_{\rm GF} = - \frac{1}{2\xi} G^{a} G^{a}= -\frac{1}{2\xi} (\partial^{\mu}A_{\mu}^{a})^{2} \,, \\
& \mathcal{L}_{\rm Ghost} = - \bar c^{a} \frac{\delta G^{a}}{\delta \theta^{b}} c^{b}= -\bar c^{a}(\partial^{\mu}D_{\mu}c)^{a} \,.
\end{aligned} \end{align}
where the anti-commuting field $c^{a}(x)$ is referred to as the ghost and $\bar c^{a}(x)$ as the
anti-ghost field.
The ghost term arises from the gauge transformation of the gauge-fixing function: $\delta A^{a}_{\mu}= (D_{\mu}\theta)^{a}$
and $\frac{\delta G^{a}}{\delta \theta^{b}}=\frac{\partial^{\mu}\delta A_{\mu}^{a}}{\delta \theta^{b}}=\delta^{a}_{b}\partial^{\mu}D_{\mu}$. Adding the gauge-fixing term 
$\mathcal{L}_{\rm GF}$ to the kinetic terms of \eqn{YMexp} yields an invertible kinetic operator. The full Lagrangian then takes the form
\begin{align}
\label{fullQCDLagrangian}
\cL_{\rm full QCD } = &\, \i\,\bar\psi_{I}^{i}\, \,\slashed{D}_{i}{}^{j} \psi_{I\, j} 
- m_{I}\,\bar\psi_{I}^{i}\psi_{I\, i} -\bar c^{a}(\partial^{\mu}D_{\mu}c)^{a} \nn\\ &
- \sfrac{1}{2} \partial_{\mu} A_{\nu}^{a}  \partial^{\mu} A^{a\, \nu} +\sfrac{\xi-1}{2\xi} (\partial^{\mu}A_{\mu}^{a})^{2}\\
& -g f^{abc} A^{a\, \mu} A^{b\, \nu} \partial_{\mu}A_{\nu}^{c} -\sfrac{1}{4} \, g^{2} f^{abe}f^{cde}A^{a\, \mu}A^{b\, \nu}
A^{c}_{\mu}A^{d}_{\nu} \,.\nn
\end{align}

\begin{svgraybox}{\bf QCD Feynman rules.}
 In momentum space
we therefore find the propagator for the gluon field
\be
\begin{tikzpicture}[baseline={(current bounding box.center)}]
\begin{feynman}
\vertex (a);
\vertex [right of = a] (b);
\draw (a) node [above] {$\mu$} node [below] {$a$};
\draw (b) node [above] {$\nu$} node [below] {$b$};
\diagram*{(a) -- [gluon, edge label'={$p$}]  (b)};
\end{feynman}
\end{tikzpicture}
=\frac{-\i\delta^{ab}}{p^2+\i 0}\left(\eta_{\mu\nu}+ (\xi-1)\frac{p_{\mu}p_{\nu}}{p^{2}}\right)\,,
	\ee
which simplifies for the convenient choice of the gauge fixing parameter $\xi=1$ (Feynman gauge). The $\i 0$ factor provides the Feynman prescription for the propagator.

The interaction parts of the Lagrangian
give rise to the three-gluon vertex
\begin{align}
\label{V3YM}
\raisebox{0.25cm}{\begin{tikzpicture}[baseline={(current bounding box.center)}]
\begin{feynman}
\coordinate (a) at (-0.7,0);
\coordinate (b) at (0.7,0);
\coordinate (c) at (0,-1.4);
\coordinate (o) at (0,-0.5);
\diagram*{ (o) -- [gluon, momentum'={$p_{1}$}] (a), (o) --  [gluon, momentum'={$p_{2}$}] (b),
(o) --  [gluon, momentum'={$p_{3}$}] (c)};
\draw [fill] (a) circle (.0) node [above] {$\mu_{1}$} node [left] {$a_{1}$};
\draw [fill] (b) circle (.0) node [above] {$\mu_{2}$} node [right] {$a_{2}$};
\draw [fill] (c) circle (.0) node [left] {$\mu_{3}$} node [right] {$a_{3}$};
\draw [fill] (o) circle (.06);
\end{feynman}
\end{tikzpicture}}
\begin{aligned}
= \ &
	-g f^{a_{1}a_{2}a_{3}} \bigl[ (p_{2}-p_{3})_{\mu_{1}} \eta_{\mu_{2}\mu_{3}} + (p_{1}-p_{2})_{\mu_{3}} \eta_{\mu_{1}\mu_{2}} \\
	& \phantom{-g f^{a_{1}a_{2}a_{3}} \bigl[} + 
	(p_{3}-p_{1})_{\mu_{2}} \eta_{\mu_{3}\mu_{1}} 	\bigr] \,.
	\end{aligned} 
\end{align}
Here all momenta are taken to be outgoing, as indicated by the arrows. 
\end{svgraybox}

\newpage

\begin{svgraybox}
For the four-gluon vertex we have
\begin{align}
\label{V4YM}
	\raisebox{0.25cm}{\begin{tikzpicture}[baseline={(current bounding box.center)}]
	\begin{feynman}
	\coordinate (a) at (-0.5,0);
	\coordinate (b) at (0.5,0);
\coordinate (c) at (0.5,-1);
\coordinate (d) at (-0.5,-1);
\coordinate (o) at (0,-0.5);
\diagram*{(a) -- [gluon] (o) -- [gluon] (b), (c) -- [gluon] (o) -- [gluon] (d)};
\draw [fill] (a) circle (.0) node [above] {$\mu_{1}$} node [left] {$a_{1}$};
\draw [fill] (b) circle (.0) node [above] {$\mu_{2}$} node [right] {$a_{2}$};
\draw [fill] (c) circle (.0) node [below] {$\mu_3$} node [right] {$a_{3}$};
\draw [fill] (d) circle (.0) node [below] {$\mu_4$} node [left] {$a_{4}$};
\draw [fill] (o) circle (.06);
\end{feynman}
	\end{tikzpicture}}
\begin{aligned} 
  = \ & -\i g^{2} \bigl[  f^{a_{1}a_{2}e} f^{a_{3}a_{4}e} (\eta_{\mu_{1}\mu_{3}}\eta_{\mu_{2}\mu_{4}}-  \eta_{\mu_{1}\mu_{4}}\eta_{\mu_2\mu_3})\\
  & \phantom{-\i g^{2} \bigl[ } 
	+  f^{a_{1}a_{3}e} f^{a_{4}a_{2}e} (\eta_{\mu_{1}\mu_4}\eta_{\mu_3\mu_2}-  \eta_{\mu_{1}\mu_2}\eta_{\mu_3\mu_4}) \\
  & \phantom{-\i g^{2} \bigl[ } +  f^{a_{1}a_{4}e} f^{a_{2}a_{3}e} (\eta_{\mu_{1}\mu_2}\eta_{\mu_3\mu_4}-  \eta_{\mu_{1}\mu_3}\eta_{\mu_2\mu_4})
		\bigr] \,. 
\end{aligned} 
\end{align}
The ghost Vertex takes the form
\be
\label{VgYM}
\begin{tikzpicture}[baseline={(current bounding box.center)}]
\begin{feynman}
	\coordinate (x) at (-0.75,0);
	\coordinate (o) at (0,0);
	\coordinate (y) at (0.75,0);
		\coordinate (z) at (0,-0.9);
		\diagram*{(x) -- [ghost] (o) -- [ghost, momentum =$p$] (y), (o) -- [gluon] (z)};
		\draw [fill] (x) circle (.0) node [left] {$c^{a}$} ;
	\draw [fill] (y) circle (.0) node [right] {$\bar c^{b}$} ;
	\draw [fill] (o) circle (.06);
		\draw [fill] (z) circle (.0) node [right] {$c$} node [left] {$\mu$} ;
    \end{feynman}
	\end{tikzpicture}=
	 -g f^{abc} p^{\mu}
	\, . 
	\ee
Including matter in the form of Dirac fermions as in \eqn{QCDLagrangian} we augment these rules with the fermion propagator
\be
\begin{tikzpicture}[baseline={(current bounding box.center)}]
\begin{feynman}
	\coordinate (x) at (-.7,0);
	\coordinate (y) at (0.5,0);
	\diagram*{(x) -- [fermion, momentum' =$p$] (y)};
		\draw [fill] (x) circle (.0) node [left] {$i$} ;
	\draw [fill] (y) circle (.0) node [right] {$j$} ;
	\end{feynman}
	\end{tikzpicture}=\delta_{j}{}^{i}\frac{\i (\slashed{p}+m)}{p^2-m^{2}+\i 0}\,, 
	\ee
and the interaction vertex
\be
\label{VqYM}
\begin{tikzpicture}[baseline={(current bounding box.center)}]
\begin{feynman}
	\coordinate (x) at (-0.7,0);
	\coordinate (o) at (0,0);
	\coordinate (y) at (0.75,0);
		\coordinate (z) at (0,-0.7);
		\diagram*{(x) -- [fermion] (o) -- [fermion] (y), (o) -- [gluon] (z)};
		\draw [fill] (x) circle (.0) node [left] {$i$} ;
	\draw [fill] (y) circle (.0) node [right] {$j$} ;
	\draw [fill] (o) circle (.06);
		\draw [fill] (z) circle (.0) node [right] {$a$} node [left] {$\mu$} ;
    \end{feynman}
	\end{tikzpicture}=
	\i g \gamma^{\mu} (T_{R}^{a})_{j}{}^{i}
	\, . 
	\ee
\end{svgraybox}

\section{Scalar QCD}
\label{Sec:sQCD}

One may also couple massive scalar fields to gauge theory, a model known as
scalar QCD. If we take the complex scalar $\phi_{i}(x)$ in the representation $R$ we
have the matter action
\be
\label{sQCD}
\cL_{\rm sQCD}= [D_{\mu}\phi(x)]^{\dagger}_{i}\, [D_{\mu}\phi(x)]_{i}
- m^{2} \phi^{\dagger}_{i}\phi_{i}\, ,
\ee
where $[D_{\mu}\phi(x)]_{i} = \partial_{\mu}\phi_{i}(x) - \i g A_{\mu}^{a}(x)\, (T_{R}^{a})_{i}{}^{j}
\phi_{j}(x)$. The associated Feynman rules are for the scalar propagator
\be
\begin{tikzpicture}[baseline={(current bounding box.center)}]
\begin{feynman}
	\coordinate (x) at (-.7,0);
	\coordinate (y) at (0.5,0);
	\diagram*{(x) -- [charged scalar, momentum' =$p$] (y)};
		\draw [fill] (x) circle (.0) node [left] {$i$} ;
	\draw [fill] (y) circle (.0) node [right] {$j$} ;
	\end{feynman}
	\end{tikzpicture}=\delta_{j}{}^{i}\frac{\i}{p^2-m^{2}+\i 0}\,, 
\ee
while we have two interaction vertices: a three-point interaction, 
\be
\label{VsqYM}
\begin{tikzpicture}[baseline={(current bounding box.center)}]
\begin{feynman}
	\coordinate (x) at (-0.7,0);
	\coordinate (o) at (0,0);
	\coordinate (y) at (0.75,0);
		\coordinate (z) at (0,-0.7);
		\diagram*{(x) -- [charged scalar] (o) -- [charged scalar] (y), (o) -- [gluon] (z)};
		\draw [fill] (x) circle (.0) node [above] {$1$} node [below] {$i$} ;
	\draw [fill] (y) circle (.0) node [above] {$2$} node [below] {$j$}  ;
	\draw [fill] (o) circle (.06);
		\draw [fill] (z) circle (.0) node [right] {$a$} node [left] {$\mu$} ;
    \end{feynman}
	\end{tikzpicture}=
	\i g  (T_{R}^{a})_{j}{}^{i}(p_{2}-p_{1})^{\mu}\, .
	\ee
with all momenta outgoing, and a four-point interaction originating from the term quartic in the fields in \eqref{sQCD},
\be
	\mbox{\begin{tikzpicture}[baseline={(current bounding box.center)}]
	\begin{feynman}
	\coordinate (a) at (-0.5,0);
	\coordinate (b) at (0.5,0);
\coordinate (c) at (0.5,-1);
\coordinate (d) at (-0.5,-1);
\coordinate (o) at (0,-0.5);
\diagram*{(a) -- [gluon] (o) -- [gluon] (b), (d) -- [charged scalar] (o) -- [charged scalar] (c)};
\draw [fill] (a) circle (.0) node [above] {$\mu$} node [left] {$a$};
\draw [fill] (b) circle (.0) node [above] {$\nu$} node [right] {$b$};
\draw [fill] (c) circle (.0) node [right] {$i$};
\draw [fill] (d) circle (.0) node [left] {$j$};
\draw [fill] (o) circle (.06);
\end{feynman}
	\end{tikzpicture}}=
	\i  g^{2} (T^{b}_{R})_{i}{}^{k}\, (T^{a}_{R})_{k}{}^{j}  \eta^{\mu\nu}
	\,. 
\ee
Note that in the case of an adjoint scalar field ($R=A$) one replaces $(T^{a}_{A})_{b}{}^{c}=-\i\sqrt{2}\, f^{abc}$ in the above expressions.

\section{Perturbative quantum gravity}
\label{sec:1.4}

The second fundamental theory of nature is Einstein's theory of gravity. Here we want to
discuss its perturbative quantisation.
It is famously known to be a non-renormalisable theory, which excludes it as a 
fundamental \emph{quantum} field theory of nature in its present form. Yet,
the modern viewpoint on the non-renormalisability of Einstein's gravity is to  
understand it as an effective quantum field theory valid
for energy scales below the Planck mass, see e.g.~\cite{ch1_Donoghue:1994dn}. 
 In this setting graviton scattering amplitudes  
 amplitudes can be computed, one needs to include counter
  terms order by order in the loop expansion.
Doing so physical quantities may be extracted. In this fashion systematic 
quantum corrections to Newton's potential, studies 
in a perturbative weak gravity (post-Newtonian or post-Minkowskian)
approach to the gravitational two body problem for bound and scattering scenarios,
or cosmological scenarios have been addressed.
Moreover, the study of graviton scattering
amplitudes has led to many surprising connections to scattering
amplitudes in non-Abelian gauge theory, which we will address later in this chapter. 

Let us now discuss the perturbative quantisation of Einstein's theory. We assume the
reader to be familiar with classical general relativity.
The gravitational field is given by the metric $g_{\mu\nu}(x)$. The minimal coupling of
gravity to matter emerges by replacing the flat-space Minkowski metric $\eta_{\mu\nu}$
by $g_{\mu\nu}(x)$ in the Lagrangians. This works fine for bosonic fields, while fermions
need a special treatment.

\begin{important}{Einstein-Hilbert Lagrangian.}
The dynamics of gravity is dictated by the Einstein-Hilbert Lagrangian
\be
\label{EinsteinHilbert}
\mathcal{L}_{\text{EH}}= \frac{2}{\kappa^{2}} \sqrt{-g}\, R \,,
\ee
where $g=\text{det}(g_{\mu\nu})$ and $R=g^{\mu\nu}R_{\mu\nu}$ is the Ricci scalar
built from the Ricci tensor $R_{\mu\nu}$ that describes the
curvature of space-time. The gravitational coupling constant $\kappa$ has inverse mass dimension one in four dimensions 
(in general $D$ we have $[\kappa]=(D-2)/2$). It is related to Newton's gravitational constant $G$ via $\kappa^{2}=32\pi G$. 
\end{important}

The Ricci tensor is defined by
\begin{align} \begin{aligned}
R_{\mu\nu}& = \partial_{\mu}\Gamma^{\rho}{}_{\rho\nu} - \partial_{\rho}\Gamma^{\rho}{}_{\mu\nu}
+\Gamma^{\rho}{}_{\mu\lambda}\Gamma^{\lambda}{}_{\rho\nu} - \Gamma^{\rho}{}_{\rho\lambda}\Gamma^{\lambda}{}_{\mu\nu} \,, \\
\Gamma^{\rho}{}_{\mu\nu}&= \sfrac{1}{2} g^{\rho\kappa} \left ( \partial_{\mu}g_{\nu\kappa} + \partial_{\nu}g_{\mu\kappa}
-\partial_{\kappa}g_{\mu\nu}
\right ) \, ,
\end{aligned} \end{align}
with the affine connection $\Gamma^{\rho}{}_{\mu\nu}$. In perturbative quantum gravity we assume a weak gravitational
field: the metric is flat on which small fluctuations propagate. These are
given by the \emph{graviton field} $h_{\mu\nu}(x)$. Therefore, we write the metric as
\be
\label{gravexpand}
g_{\mu\nu}(x) = \eta_{\mu\nu} + \kappa\, h_{\mu\nu}(x) \, .
\ee
In the classical theory the graviton field $h_{\mu\nu}$ represents gravitational waves.\footnote{In fact the quantum field theory methods to be discussed may also be applied to this case in their classical limit. This has proven to be a very
efficient approach, see e.g.~\cite{ch1_Goldberger:2004jt,ch1_Bern:2019crd,ch1_Mogull:2020sak}.} We now insert this expression 
for the metric into the
Einstein-Hilbert action, and perform a power series expansion in powers of $\kappa$ and the graviton field.
This is a weak field expansion. Let us gather the various building blocks in this expansion.
For the inverse metric one has
\be
g^{\mu\nu}(x) = \eta^{\mu\nu} - \kappa \, h^{\mu\nu} + \kappa^{2} h^{\mu\alpha}h_{\alpha}{}^{\nu} + \cO\bigl(\kappa^{3}\bigr)\, .
\ee
From now on we raise and lower indices with the flat Minkowski metric $\eta_{\mu\nu}$. The further quantities entering
$\mathcal{L}_{\text{EH}}$ take the following forms up to cubic order in $\kappa$:
\begin{align}
\begin{aligned}
\sqrt{-g} = \ & 1 + \frac{\kappa}{2} h + \frac{\kappa^{2}}{8} (h^{2}-2 \, h^{\alpha\beta}h_{\alpha\beta}) + \cO\bigl(\kappa^{3}\bigr) \,, \\
\Gamma^{\rho}{}_{\mu\nu} = \ & \frac{\kappa}{2} ( \partial_{\mu}h^{\rho}{}_{\nu}+\partial_{\nu}h^{\rho}{}_{\mu}- \partial^{\rho}h_{\mu\nu}) -\frac{\kappa^{2}}{2} h^{\rho\sigma}( \partial_{\mu}h_{\nu\sigma}+ \partial_{\nu}h_{\mu\sigma}
-\partial_{\sigma}h_{\mu\nu}) + \cO\bigl(\kappa^{3}\bigr) \,, \\
R  = \ & \kappa (\partial^{2}h - \partial^{\alpha}\partial^{\beta}h_{\alpha\beta}) - \frac{\kappa^{2}}{2} \Bigl [
h^{\alpha\beta} ( \partial^{2}h_{\alpha\beta}+ \partial_{\alpha}\partial_{\beta}h - 2 \, \partial_{\rho}\partial_{\alpha}
h^{\rho}{}_{\beta}) + \partial_{\alpha}h \, \partial_{\beta}h^{\alpha\beta} \\
& - (\partial_{\alpha}h)^{2} + \sfrac{1}{2}
\partial_{\gamma}h_{\alpha\beta} \, \partial^{\gamma}h^{\a\b}- \partial_{\a}h_{\gamma\b} \,
\partial^{\b}h^{\gamma\a} + \text{total derivatives} \Bigr ] +\cO\bigl(\kappa^{3}\bigr)  \,,
\end{aligned}
\end{align}
where $h \coloneqq h^{\alpha}{}_{\alpha}$.
Inserting these expansions into the Einstein-Hilbert Lagrangian~\eqref{EinsteinHilbert} yields to leading order in
$\kappa$ the expression 
\begin{align} 
\label{LEHquadr}
\begin{aligned}
\mathcal{L}_{\text{EH}} = \ & \partial_{\alpha}h \, \partial_{\beta}h^{\a\b}
- \partial_{\a}h_{\b\gamma}\, \partial^{\b}h^{\a\gamma} - \sfrac{1}{2}(\partial_{\a}h)^{2} +\sfrac 12 (\partial_{\gamma}h_{\a\b})^{2} \\ 
& + \text{total derivatives} +\cO\bigl(\kappa, h^{3} \bigr) \,.
\end{aligned}
\end{align}
These quadratic terms in $h_{\mu\nu}$ give rise to the kinetic term for the graviton. The omitted infinite series of higher powers
in $\kappa$ gives rise to the graviton self interactions. They take the schematic form 
\be
\mathcal{L}_{\text{EH,int}}=\sum_{n=1}^{\infty} \kappa^{n} \left[\partial^{2} h^{n+2} \right] \,,
\ee
where the term in brackets simply denotes the order in derivatives and fields encountered
in this expansion. In general one finds
all possible tensor structures. Hence, the Feynman rules for perturbative quantum gravity have vertices  of \emph{all} multiplicities.
Yet, in a computation to a given order in $\kappa$ only a finite number of vertices enter,
as the power of $\kappa$ of a vertex grows with its multiplicity.

Gravity is invariant under general coordinate transformations, which take the infinitesimal form
\be
\label{diffeos}
x^{\mu}\to x^{\mu}+ \xi^{\mu}(x) \,,
\ee
with an arbitrary space-time dependent vector $\xi^{\mu}(x)$. Under these coordinate transformations the graviton field
transforms as\footnote{Recall that we symmetrise with unit weight $a^{(\mu}b^{\nu)} \coloneqq (a^{\mu}b^{\nu}+  
a^{\nu}b^{\mu})/2$.}
\be
\delta h_{\mu\nu} = 2 \, h_{\sigma ( \mu}\partial_{\nu)} \xi^{\sigma}+ \xi^{\sigma}\partial_{\sigma}h_{\mu
\nu}+ \frac{2}{\kappa} \, \partial_{(\mu}\xi_{\nu)}\, .
\ee
Just as in Yang-Mills theory, this local invariance necessitates a gauge fixing in oder not to ``overcount'' in
the path-integral over $h_{\mu\nu}$ through the Fadeev-Popov procedure. As our transformation freedom lies
in an arbitrary space-time vector $\xi^{\mu}(x)$, we need to gauge fix four components of $h_{\mu\nu}$. A popular and convenient choice is the de~Donder gauge: 
\be
\label{deDonder}
G_{\mu}= \partial^{\nu}h_{\mu\nu}- \ft{1}{2}\partial_{\mu}h = 0 \,,
\ee
that we shall also employ. Note that this is the linearised version (in $\kappa$) of the harmonic coordinate
choice $g^{\mu\nu}\Gamma^{\rho}{}_{\mu\nu}=0$, frequently used in general relativity. The gauge fixing term to be added
to the Lagrangian takes the form\footnote{In analogy to the gauge theory discussion around eqs.~\eqref{FadeevPopovYM}, with suitable choice of gauge-fixing parameter $\xi=-1/2$.}
\be
\mathcal{L}_{\text{GF}}=  G_{\mu}G^{\mu} = \partial^{\nu}h_{\mu\nu} \, \partial^{\rho}h^{\mu}{}_{\rho} + \ft{1}{4} (\partial_{\mu}h)^{2}-\partial^{\nu}h_{\mu\nu} \, \partial^{\mu}h \,.
\ee
Adding this to $\mathcal{L}_{\text{EH}}$ then cancels the first two terms in \eqn{LEHquadr} and yields a nice, invertible
quadratic term:
\begin{align}\label{EH+GF}
\begin{aligned}
\mathcal{L}_{\text{EH}}|_{h^{2}} + \mathcal{L}_{\text{GF}} & = -\ft{1}{2} h_{\alpha\beta} \, \partial^{2} h_{\alpha\beta}
+\ft{1}{4} \, h \, \partial^{2}h \\ 
& = - \ft 12 h_{\alpha\beta}\underbrace{ \left[  \eta^{\alpha(\gamma}\eta^{\delta)\beta}-\ft 12 \eta^{\alpha\beta}\eta^{\gamma\delta} \right]}_{= \, I^{\alpha\beta,\gamma\delta}}
\partial^{2} h_{\gamma\delta}\, .
\end{aligned}
\end{align}

\section{Feynman rules for perturbative quantum gravity}
\label{sec:1.6}

Going to momentum space and inverting the differential operator $I^{\alpha\beta\gamma\delta}\partial^{2}$ of \eqn{EH+GF} leads us to the graviton propagator
\be
\begin{tikzpicture}[baseline={(current bounding box.center)}]
\begin{feynman}
\vertex (a) {$\alpha\beta$};
\vertex [right of = a] (b) {$\gamma\delta$};
\diagram*{(a) -- [graviton] (b)};
\end{feynman}
\end{tikzpicture}
= \frac{\i \, P_{\alpha\beta,\gamma\delta}}{k^{2}+\i 0}\,  \qquad \text{with} \quad
P_{\alpha\beta,\gamma\delta} =  \eta_{\alpha(\gamma}\eta_{\delta)\beta} - \ft{1}{D-2}\eta_{\alpha\beta}\eta_{\gamma\delta}\, .
\ee
One indeed verifies that $I^{\alpha\beta,\gamma\delta}\, P_{\gamma\delta,\rho\kappa}=\delta^{\alpha}_{(\rho}\delta^{\beta}_{\kappa)}$. The graviton self-interaction vertices take an involved structure due to a proliferation
of indices. For example, we exhibit the three-graviton vertex~\cite{ch1_Sannan:1986tz}
\begin{align}
\begin{tikzpicture}[baseline={(current bounding box.center)}]
\begin{feynman}
\coordinate (a1) at (-0.5,0);
\coordinate (a2) at (0.5,0);
\coordinate (a3) at (0,-1.0);
\coordinate (o) at (0,-0.5);
\diagram*{ (a1) -- [graviton] (o), (a2) -- [graviton] (o), (a3) -- [graviton] (o)};
\draw [fill] (a1) circle (.0) node [below] {$\mu \quad $} node [left] {${}^1$} node [above] {$\alpha$};
\draw [fill] (a2) circle (.0) node [above] {$\nu$} node [right] {${}^{2}$} node [below] {$\quad  \beta$};
\draw [fill] (a3) circle (.0) node [left] {$\rho$} node [below] {${}_{3}$} node [right] {$\gamma$};
\draw [fill] (o) circle (.06);
\end{feynman} 
\end{tikzpicture} &= \i\kappa\,  \text{sym}[ \ft 12 P_{3}( k_{1}\cdot k_{2} \eta_{\mu\nu}\eta_{\alpha\beta}\eta_{\rho\gamma}) \nn \\[-0.8cm]& 
-\ft 12 P_{3}( k_{1}\cdot k_{2} \eta_{\mu\alpha}\eta_{\nu\beta}\eta_{\rho\gamma})
-\ft 12 P_{6}( k_{1\nu} k_{1\beta} \eta_{\mu\alpha}\eta_{\rho\gamma})
\nn \\ &
+       P_{6}( k_{1}\cdot k_{2} \eta_{\mu\alpha}\eta_{\nu\rho}\eta_{\beta\gamma}) 
+      2P_{3}( k_{1\nu} k_{1\gamma} \eta_{\mu\alpha}\eta_{\beta\rho})
-       P_{3}( k_{1\beta} k_{2\mu} \eta_{\alpha\nu}\eta_{\rho\gamma}) \nn \\ &
+       P_{3}( k_{1\rho} k_{2\gamma} \eta_{\mu\nu}\eta_{\alpha\beta}) 
+       P_{6}( k_{1\rho} k_{1\gamma} \eta_{\mu\nu}\eta_{\alpha\beta})
+      2P_{6}( k_{1\nu} k_{2\gamma} \eta_{\beta\mu}\eta_{\alpha\rho}) \nn \\ &
+      2P_{3}( k_{1\nu} k_{2\mu} \eta_{\beta\rho}\eta_{\gamma\alpha})
-      2P_{3}( k_{1}\cdot k_{2} \eta_{\nu\alpha}\eta_{\rho\beta}\eta_{\mu\gamma}) 
]
 \, , 
 \label{3gravitonvertex}
\end{align}
where ``$\text{sym}$'' means symmetrisation in the index pairs $(\mu\alpha)$, $(\nu\beta)$
and $(\rho\gamma)$. The symbol $P_{n}$ denotes the symmetrisation in the momentum-index 
combinations ${(k_1\mu\alpha, k_2\nu\beta, k_{3} \rho\gamma})$ associated with the three legs and results in $n$ distinct terms. For example the first term above evaluates to
\begin{align}
\text{sym}\ft 12 P_{3}( k_{1}\cdot k_{2} \eta_{\mu\nu}\eta_{\alpha\beta}\eta_{\rho\gamma})
=&\ft 12 ( k_{1}\cdot k_{2} \eta_{\mu(\nu}\eta_{\beta)\alpha}\eta_{\rho\gamma}
+k_{2}\cdot k_{3} \eta_{\nu(\rho}\eta_{\gamma)\beta}\eta_{\mu\alpha}\nn\\ &
+k_{3}\cdot k_{1} \eta_{\rho(\mu}\eta_{\alpha)\gamma}\eta_{\nu\beta})\, .
\end{align}

The higher-point vertices take the schematic structure
\be
\begin{tikzpicture}[baseline={(current bounding box.center)}]
\label{gravitonvertexscaling}
\begin{feynman}
	\coordinate (a1) at (-0.5,0);
	\coordinate (a2) at (0.5,0);
\coordinate (a3) at (0.5,-1);
\coordinate (a4) at (-0.5,-1);
\coordinate (o) at (0,-0.5);
\diagram*{ (a1) -- [graviton] (o), (a2) -- [graviton] (o), (a3) -- [graviton] (o), (a4) -- [graviton] (o)};
\draw [fill] (o) circle (.06);
\end{feynman} 
\end{tikzpicture} \sim \kappa^{2} k^{2}  \, , \qquad
\begin{tikzpicture}[baseline={(current bounding box.center)}]
\begin{feynman}
	\coordinate (a1) at (-0.5,-0.2);
	\coordinate (a2) at (0.5,-0.2);
\coordinate (a3) at (0.5,-1);
\coordinate (a4) at (-0.5,-1);
\coordinate (a5) at (0,0);
\coordinate (o) at (0,-0.5);
\diagram*{ (a1) -- [graviton] (o), (a2) -- [graviton] (o), (a3) -- [graviton] (o), (a4) -- [graviton] (o), (a5) -- [graviton] (o)};
\draw [fill] (o) circle (.06);
\end{feynman} 
\end{tikzpicture} \sim \kappa^{3} k^{2}  \,, \qquad
\begin{tikzpicture}[baseline={(current bounding box.center)}]
\begin{feynman}
	\coordinate (a1) at (-0.5,-0.2);
	\coordinate (a2) at (0.5,-0.2);
\coordinate (a3) at (0.5,-0.8);
\coordinate (a4) at (-0.5,-0.8);
\coordinate (a5) at (0,0);
\coordinate (a6) at (0,-1);
\coordinate (o) at (0,-0.5);
\diagram*{ (a1) -- [graviton] (o), (a2) -- [graviton] (o), (a3) -- [graviton] (o), (a4) -- [graviton] (o), (a5) -- [graviton] (o), (a6) -- [graviton] (o)};
\draw [fill] (o) circle (.06);
\end{feynman} 
\end{tikzpicture} \sim \kappa^{4} k^{2}  \,, \qquad \ldots
\ee
and grow considerably in size. E.g.~the four-graviton vertex consists of 60 distinct terms,
see~\cite{ch1_Sannan:1986tz} for its explicit form.
Through the Fadeev-Popov procedure one also picks up a ghost sector. The local symmetry transformations are now the general coordinate transformations given in \eqn{diffeos}. 
Hence, the gravity ghosts carry a vector index: $b^{\nu}(x)$ and $\bar b^{\mu}(x)$. The ghost contribution to the Lagrangian takes the form
\be
\mathcal{L}_{\text{GH}}= -\bar b^{\mu} \left( \kappa \frac{\delta G_{\mu}}{\delta \xi^{\nu}} \right)  b^{\nu}\, .
\ee
From the de~Donder gauge-fixing function of \eqn{deDonder} one deduces the differential operator in the ghost sector
\be
\kappa\frac{\delta G_{\mu}}{\delta \xi^{\nu}} =  \eta_{\mu\nu} \partial^{2}+
\kappa \bigl[ \partial^{\rho}h_{\mu\nu}\partial_{\rho}
+ \partial^{\rho}h_{\nu\rho}\partial_{\mu}+ \partial^{\rho}(\partial_{\nu}h_{\mu\rho}) - \partial_{\mu}h_{\nu\rho}
\partial^{\rho} - \ft 1 2 \partial_{\mu}(\partial_{\nu} h) \bigr] \,,
\ee
where the first term gives rise to the kinetic term of the ghost fields, yielding the propagator
\be
\begin{tikzpicture}[baseline={(current bounding box.center)}]
\begin{feynman}
\vertex (a) {$\alpha$};
\vertex [right of = a] (b) {$\beta$};
\diagram*{(a) -- [ghost] (b)};
\end{feynman}
\end{tikzpicture}
= \frac{\i \,\eta_{\alpha\beta}}{p^{2}+\i 0}\, .
\ee
The remaining terms yield a graviton-ghost-anti-ghost interaction vertex,
\be
\begin{tikzpicture}[baseline={(current bounding box.center)}]
\begin{feynman}
	\coordinate (x) at (-0.7,0);
	\coordinate (o) at (0,0);
	\coordinate (y) at (0.75,0);
		\coordinate (z) at (0,-0.7);
		\diagram*{(x) -- [ghost] (o) -- [ghost] (y), (o) -- [graviton] (z)};
		\draw [fill] (x) circle (.0) node [left] {$\alpha$} ;
	\draw [fill] (y) circle (.0) node [right] {$\beta$} ;
	\draw [fill] (o) circle (.06);
		\draw [fill] (z) circle (.0) node [right] {$\nu$} node [left] {$\mu$} ;
    \end{feynman}
	\end{tikzpicture} 	\ee
However, ghosts will play no role in the modern approaches to scattering amplitudes 
developed in these lecture notes. Therefore we do not need to spell out this involved
vertex here.

\section{Spinor-helicity formalism for massless particles}
\label{sec:1.7}

In this section we will introduce a formalism that efficiently captures the kinematical data
of the scattering states in the $S$-matrix: the momenta and polarisations of the scattered particles.
The spinor-helicity variables allow one to express this data (momenta and helicities) for a massless particle
in a uniform object thereby guaranteeing the on-shell conditions. In fact, 
the scattering amplitudes involving massless  scalars, fermions, gluons, photons and gravitons expressed in these variables take very compact
forms.

The starting point is to rewrite the four-momentum $p^{\mu}$ as a bi-spinor $p^{\da\a}$ that we
may represent as a $2\times 2$ matrix
\be
p^{\mu} \to p^{\da\a} = \bar\sigma^{\da\a}_{\mu}\, p^{\mu} =
\begin{pmatrix}
p^{0}+p^{3} & p^{1}-\i p^{2}\\ p^{1}+\i p^{2} & p^{0}-p^{3}
\end{pmatrix}\, ,
\ee
where $\bar\sigma^{\da\a}_{\mu}=(\1, +\vec\sigma)$, cf.~exercise~\ref{Ex:1.1}. The determinant of this matrix is given by
\be
\text{det}\bigl(p^{\da\a}\bigr) = \bigl(p^{0}\bigr)^{2}-{\vec p}^{2}=p^{2} \,,
\ee
where $\vec{p}=(p^{1},p^{2},p^{3})$ is the spatial momentum vector. If we put the
momentum $p^{\mu}$ on the mass shell, i.e.~$p^{2}=m^{2}$, we see that the determinant
equals $m^{2}$. Hence, there is a distinction between the massive and massless case.
 In the massive case the hermitian $2\times 2$ matrix $p^{\da\a}$ has rank 2 and may be decomposed into the sum of two outer products of commuting Weyl spinors,
\be
\label{masssivep}
p^{\da\a} = \tilde \la^{\da}\la^{\a} + \tilde \mu^{\da}\mu^{\a} \,,
\ee
with complex conjugates $\tilde\la= \pm\la^{*}$ and  $\tilde\mu= \pm\mu^{*}$.
In the massless case---which we are mostly interested in---the determinant of $p^{\da\a}$ vanishes, and hence the matrix $p^{\da\a}$ has rank one. 
\begin{svgraybox}{\bf Helicity spinors.}
In the massless case, we may then write the light-like momentum as a direct product of two conjugate spinors:
\be
\label{spinheldef}
p^{\da\a} = \tilde \la^{\da}\la^{\a} \, .
\ee
These are the \emph{helicity spinors} associated to the light-like momentum $p^{\mu}$. They are commuting (Gra{\ss}mann even) Weyl spinors in the $(\nicefrac{1}{2},0)$ and $(0,\nicefrac{1}{2})$ representations,
respectively. 
The reality of $p^{\mu}$ implies the hermiticity of $p^{\da\a}$, which in turn implies the conjugation
property $(\lambda^{\a})^{\ast}= \pm \tilde\la^{\da}$. In fact, the sign may be  related to the
sign of the energy $p^{0}$. An explicit realisation of the helicity spinors associated to the
real momentum $p^{\mu}$ is given by
\be
\label{helspinexpl}
\la^{\a}= \frac{1}{\sqrt{p^{0}+p^{3}}}\, \begin{pmatrix}
p^{0}+ p^{3} \\ p^{1} - \i p^{2}
\end{pmatrix}\, , \qquad
\tla^{\da}= \frac{1}{\sqrt{p^{0}+p^{3}}}\, \begin{pmatrix}
p^{0}+ p^{3} \\ p^{1} + \i p^{2}
\end{pmatrix}\, .
\ee
Note that indeed $\sqrt{p^{0}+p^{3}}$ is real (imaginary) for positive (negative) $p^{0}$ as claimed 
above due to the constraint $|p^{0}|\ge |p^{3}|$, which follows from the on-shell condition.
\end{svgraybox}

The spinor indices are raised and lowered with the antisymmetric Levi-Cevita tensors $\epsilon_{\a\b}$ and $\epsilon_{\da\db}$:
\be
\la_{\a} \coloneqq \epsilon_{\a\b}\la^{\b}\, , \qquad
\tla_{\da} \coloneqq \epsilon_{\da\db}\tla^{\db}\, .
\ee
In a scattering amplitude involving $n$ massless particles we have $n$ momenta $p_{i}^{\mu}$ ($i=1,\ldots, n$), and hence $n$ pairs
of helicity spinors $\{\la_{i}^{\a}, \tla_{i}^{\da}\}$. Out of these we may assemble Lorentz invariant
quantities:
\begin{align} \begin{aligned} \label{bevvev}
& \vev{\la_{i},\la_{j}} \coloneqq \la_{i}^{\a}\, \la_{j\, \a}=\phantom{+}\epsilon_{\a\b}\la_{i}^{\a}\la_{j}^{\b}=
-\vev{\la_{j},\la_{i}}= \vev{ij} \,, \\ 
& \bev{\tla_{i},\tla_{j}}\coloneqq \tla_{i\,\da}\, \tla_{j}^{\da}=-\epsilon_{\da\db}\tla_{i}^{\da}\tla_{j}^{\db}=-\bev{\tla_{j},\tla_{i}}= \bev{ij} \, .
\end{aligned} \end{align}
Note the opposite index contraction convention between the un-dotted (NW-SE) and dotted (SW-NE) spinors.
The Mandelstam invariants $s_{ij}=2 \, p_{i}\cdot p_{j}$ may then be written as
\be
\vev{ij}\bev{ji} =  p_{i}^{\da\a}p_{j\, \a\da}= \epsilon_{\da\db}\epsilon_{\a\b}
 p_{i}^{\da\a}p_{j}^{\db\b}=
\underbrace{\epsilon_{\da\db}\epsilon_{\a\b}\bar\sigma^{\da\a}_{\mu}\bar\sigma^{\db\b}_{\nu}}_{2\eta_{\mu\nu}}p_{i}^{\mu} p_{j}^{\nu}
=2 \, p_{i}\cdot p_{j}=s_{ij}\, ,
 \ee
 where we used that $\tr{\sigma^{\mu} \bar\sigma^{\nu}} = 2 \, \eta^{\mu \nu}$ (see exercise~\ref{Ex:1.1}).

\smallskip

The representation theory of the Poincar\'e group teaches us that massless particles with
spin carry a Lorentz-invariant quantity, the helicity $h$, which is the projection of the particle's spin onto their direction of motion. It takes values $h=\pm s$ for a particle of spin $s$. The Dirac equation for positive energy $u(p)$ and negative energy $v(p)$ solutions,
\be
(\slashed{p}-m)u(p)=0 \, , \qquad (\slashed{p}+m)v(p) \,,
\ee
degenerates in the massless limit to $\slashed{p}u(p)=0=\slashed{p}v(p)$, in which $u(p)$
and $v(p)$ may be identified. Projecting onto definite helicity states
\be
\label{uvpmfd}
u_{\pm}=\ft 12 (1 \pm \gamma_{5})u(p)\, , \qquad
v_{\mp}=\ft 12 (1 \pm \gamma_{5})v(p)\, ,
\ee
allows for the identification $u_{\pm}(p)=v_{\mp}(p)$. On-shell massless spin-$\nicefrac{1}{2}$ states may
then be labeled as $|\la_{i},\tla_{i},\pm\nicefrac{1}{2}\rangle$, reflecting both momentum and helicity. Using the helicity spinors and the chiral representation of the Dirac matrices,\footnote{Note that $(\sigma^{\mu})_{\a\db}=
\epsilon_{\a\b}\epsilon_{\db\da}(\bar\sigma^{\mu})^{\da\b}$.} the massless Dirac operator in momentum space reads
\be
\label{pslashed}
\slashed{p} = p_{\mu}\gamma^{\mu} = p_{\mu}
\begin{pmatrix} 0 & (\sigma^{\mu})_{\a\db}\\ (\bar\sigma^{\mu})^{\da\b} & 0 
\end{pmatrix} = 
\begin{pmatrix} 0 & p_{\a\db}\\ p^{\da\b} & 0 
\end{pmatrix} = \begin{pmatrix} 0 & \la_{\a}\tla_{\db}\\ \tla^{\da}\la^{\b} & 0 
\end{pmatrix} \,.
\ee
The helicity spinors may then be identified with the solutions to the massless Dirac equation as
\be
\label{uvpmdef}
u_{+}(p)= v_{-}(p)=\begin{pmatrix} \la_{\a} \\ 0 \end{pmatrix} \eqqcolon |\la\rangle\, \qquad
u_{-}(p)= v_{+}(p)=\begin{pmatrix} 0 \\ \tla^{\da} \end{pmatrix} \eqqcolon |\tla]\, ,
\ee
which obey $\slashed{p} u_{\pm}(p)=0$ as $\la^{\beta}\lambda_{\beta}=0=\tla_{\db}\tla^{\db}$. 
The conjugate spinors $\bar u = u^{\dagger}\gamma^{0}$ then take the form 
\begin{align}
\label{uvpmdef2}
\begin{aligned}
\bar u_{+}(p)= \bar v_{-}(p)&=\begin{pmatrix} \la_{\a}^{\ast} & 0 \end{pmatrix} \,
\begin{pmatrix} 0 & \1 \\ \1 & 0 \end{pmatrix} =
\begin{pmatrix} 0& \tla_{\da} \end{pmatrix}
\eqqcolon [\tla| \,, \\
\bar u_{-}(p)= \bar v_{+}(p) &=
\begin{pmatrix} 0 & \tla^{\da\,\ast} \end{pmatrix} \,
\begin{pmatrix} 0 &\1 \\ \1 & 0 \end{pmatrix} =
\begin{pmatrix} \la^{\a} & 0 \end{pmatrix}
 \eqqcolon \langle \la |\, .
\end{aligned}
\end{align}
In the above we introduced a crafty bra-ket notation that we shall use frequently from now on. 
In this way we may rewrite \eqn{pslashed} in an index-free notation as
\be
\slashed{p} = |\la\rangle [\tla| + |\tla]\langle \la |\, ,
\ee
for an on-shell light-like momentum $p^{\mu}$. 
This makes the relations $\slashed{p}|\la\rangle =0 $ and  $\slashed{p}|\tla] =0 $ 
manifest, i.e.~the
helicity spinors $\la_{\a}$ and $\tla^{\da}$ solve the massless Dirac equation.

Were we to consider a massive or off-shell momentum---represented by two helicity spinors as done in \eqn{masssivep}---we could write
\be
\label{offshellpsslash}
\slashed{p} = |\la\rangle [\tla| + |\tla]\langle \la | + 
 |\mu\rangle [\tilde\mu| + |\tilde\mu]\langle \mu |\, .
 \ee
Notation wise, in the context of scattering amplitudes we often just write the particle number label in the brackets and drop the $\lambda$'s, e.g.~for the $i$th particle
\be
|\la_{i}\rangle = |i\rangle\, , \qquad |\la_{i}] = |i] \, , 
\ee
or use the on-shell momentum itself as the label, e.g.
\be
\label{pspinorde}
\slashed{p} = |p\rangle [p| + |p]\langle p |\, .
\ee
Note that the quantities $\langle i|j]$ and $[i|j\rangle$ vanish. We
then deduce the relations
\be
\langle i | \slashed{p} | j\rangle = 0 = [i | \slashed{p} | j]\, \quad \text{but} \quad
[i|\slashed{p} |j\rangle = \langle j|\slashed{p} |i] \, ,
\ee
that also hold true for an off-shell $p^{\mu}$ using \eqn{offshellpsslash}.
Stripping off the $p_{\mu}$ in the second relation allows us to define a four-vector object
$\langle i| \gamma^{\mu} |j] = [ j| \gamma^{\mu} |i\rangle $. Using the notation of
\eqn{pspinorde} we have
\be \label{eq:pmufromspinors}
\langle p| \gamma^{\mu} |p] = 2p^{\mu}= [ p| \gamma^{\mu} |p\rangle \, .
\ee
For redirecting momenta we define
\be
\label{negpspin}
|-p\rangle = \i |p\rangle\, , \qquad |-p]=\i |p]\, ,
\ee
which is also consistent with \eqn{helspinexpl}.

We note that the four-vector $p^{\mu}$ in \eqn{spinheldef} does not completely fix the spinor-helicity variables $\la^{\a}$ and $\tla^{\da}$, as the rescaling 
\be \label{eq:little_group}
\la^{\a} \to \mathrm{e}^{-\i\varphi}\la^{\a}\, , \qquad
\tla^{\da} \to \mathrm{e}^{+\i\varphi}\tla^{\da}\, , 
\ee
leaves $p^{\da\a}= \tla^{\da}\la^{\a}$ and therefore also $p^{\mu}$ invariant. 
This rescaling freedom is known as the \emph{little group}: a rotation that leaves the
momentum invariant.

There exist a number of important identities for a set of helicity spinors
$\{\la_{i},\tla_{i}\}$ ($i=1,\ldots,n$) parametrising an $n$-point scattering amplitude that we would like to collect:
\begin{enumerate}[i)]
\item Mandelstam invariants: \be 2 \, p_{i}\cdot p_{j} = \bev{ij}\vev{ji}\, . \ee

\smallskip
\item Schouten identity: 
\be \label{eq:Schouten}
\vev{\la_{1}\la_{2}}\la_{3}^{\a} + \vev{\la_{2}\la_{3}}\la_{1}^{\a}
+\vev{\la_{3}\la_{1}}\la_{2}^{\a}=0\, ,\ee
and similarly for the $\tilde{\lambda}_i$'s.
This identity simply reflects the fact that one cannot have a completely anti-symmetric three-tensor in two-dimensions. 

\smallskip
\item Total momentum conservation \be\sum_{i=1}^{n}\vev{ai}\bev{ib} =0\, , \ee
for any $a$, $b$. This holds if the $n$ helicity spinors parameterise the external states of a scattering amplitude  with total momentum conservation $\sum_{i=1}^{n}p_{i}^{\mu}=0$. 
Note that in our convention we take the momenta of all legs in a scattering amplitude to be \emph{outgoing}.

\smallskip
\item Fierz rearrangement:
\be \label{eq:Fierz}
[i|\gamma^{\mu}|j\rangle \, [ k | \gamma_{\mu}| l \rangle = 2 \, \bev{ik}\, \vev{lj}\, .
\ee
\item Complex conjugation: due to the definition in \eqn{bevvev} we have
\be
\vev{ij}^{\ast} = \bev{ji}\, , \qquad 
\bev{ij}^{\ast} = \vev{ji}\, .
\ee
\end{enumerate}

\begin{exer}{Spinor identities}
\label{Ex:1.6}
Prove the following identities:
\begin{multicols}{2}
\begin{enumerate}[a)]
\item $\spBA{i}{\gamma^{\mu}}{j} = (\tilde{\lambda}_i)_{\dot{\alpha}} (\bar{\sigma}^{\mu})^{\dot{\alpha} \alpha} (\lambda_j)_{\alpha}$,
\item $\spAB{i}{\gamma^{\mu}}{j} = \lambda_i^{\alpha} (\sigma^{\mu})_{\alpha\dot{\alpha}} \tilde{\lambda}_j^{\dot{\alpha}}$,
\item $\spBA{i}{\gamma^{\mu}}{i} = \spAB{i}{\gamma^{\mu}}{i} $,
\item $\spAB{i}{\gamma^{\mu}}{i} = 2 p_i^{\mu}$,
\item the Schouten identity~\eqref{eq:Schouten},
\item the Fierz rearrangement~\eqref{eq:Fierz}.
\end{enumerate}
\end{multicols}
For the solution see \hyperref[Sol:1.6]{chapter~5}.
\end{exer}

\begin{exer}{Lorentz generators in the spinor-helicity formalism}
\label{Ex:1.5}
\vspace{-0.6cm}
\begin{enumerate}[a)]
\item Prove that the Lorentz generators in the scalar representation take the following form in momentum space, 
\begin{align}
\tilde{M}^{\mu\nu} = \i \left( p^{\mu} \frac{\partial}{\partial p_{\nu}} - p^{\nu} \frac{\partial}{\partial p_{\mu}} \right) \,.
\end{align}

\smallskip
\item The Lorentz generators in the helicity-spinor formalism come in two pairs of symmetric tensors $m_{\alpha \beta}$ and $\overline{m}_{\dot\alpha \dot\beta}$ originating from the projections $m_{\alpha \beta}=\left(S^{\mu\nu}_{\rm L}\right)_{\alpha \beta} \tilde{M}_{\mu\nu}$ and  $\overline{m}_{\dot\alpha \dot\beta}=\left(S^{\mu \nu}_{\rm R}\right)_{\dot\alpha \dot\beta} \tilde{M}_{\mu \nu}$, where $S^{\mu\nu}_{\rm L}$ and $S^{\mu\nu}_{\rm R}$ are the $2 \times 2$ representation matrices for the $(\nicefrac{1}{2},0)$ and $(0,\nicefrac{1}{2})$ representations, respectively. $S^{\mu\nu}_L$ is given by eq.~\eqref{eq:S_L}, and $S^{\mu \nu}_{\rm R} = \i\left(\bar{\sigma}^{\mu} \sigma^{\nu}-\bar{\sigma}^{\nu} \sigma^{\mu} \right)/4 $.
Show that
\begin{align}
m_{\alpha \beta} = \lambda_{\alpha} \frac{\partial}{\partial \lambda^{\beta}}+ \lambda_{\beta} \frac{\partial}{\partial \lambda^{\alpha}} \,, \qquad \overline{m}_{\dot\alpha \dot\beta} = \tilde{\lambda}_{\dot\alpha} \frac{\partial}{\partial \tilde{\lambda}^{\dot\beta}}+ \tilde{\lambda}_{\dot\beta} \frac{\partial}{\partial \tilde{\lambda}^{\dot{\alpha}}} \,,
\end{align}
where $p^{\dot\alpha\alpha} = \tilde{\lambda}^{\dot\alpha} \lambda^{\alpha}$.

\smallskip
\item The representation of the Lorentz generators on a function of $n$ momenta $p_i$ ($p_i^{\dot\alpha\alpha} = \tilde{\lambda}_i^{\dot\alpha} \lambda_i^{\alpha}$) is obtained by summing over all  single-momentum generators. Use the appropriate Lorentz generators to show that the following quantities are Lorentz invariant: $\vev{ij}$, $\bev{ij}$, $s_{ij}$.

\end{enumerate}
For the solution see \hyperref[Sol:1.5]{chapter~5}.
\end{exer}

\section{Polarisations of massless particles of spin $\half$, 1 and 2}
\label{sec:1.8}

External states in scattering amplitudes are parameterised by their momenta and polarisations.
We now analyse the polarisations of massless particles of spin $\half$, $1$ and $2$, and how they may be expressed via the helicity spinors. We recall that scalar states do not have polarisations. 

\begin{important}{Spin $\mathbf \half$.}
The polarisations for massive fermions and anti-fermions are captured by the
positive $u(p)$ and negative $v(p)$ energy Dirac-spinors, which we saw coincide in the
massless case. Spin-$\half$ states of definite helicities $\pm \half$ are obtained by
the projections $u_{\pm}$ of \eqn{uvpmfd}, which  coincide with $|\la\rangle$ and $|\tla]$
as we saw in \eqn{uvpmdef}.
In our convention all momenta are outgoing, therefore $|\la\rangle$
and $\langle \la|$ represent $-\half$ helicity states (or outgoing anti-fermions)
whereas $|\tla]$
and $[ \tla|$ represent $+_{}\half$ helicity states (or outgoing fermions).
This observation allows us to introduce the \emph{helicity generator}: 
\be \label{eq:hel_gen}
h= \frac 1 2 \sum_{i=1}^{n} \left[
-\la_{i}^{\a}\frac{\partial}{\partial \la_{i}^{\a}} + \tla_{i}^{\da}\frac{\partial}{\partial \tla_{i}^{\da}}
\right]\, .
\ee
In fact $h\, \la_{i}^{\a}= -\half \la_{i}^{\a}$ and $h \, \tla_{i}^{\da}= \half \tla_{i}^{\da}$, so $h$ measures the helicity. This is the reason why we call $\la$ and $\tla$  
\emph{helicity spinors}. They capture 
\emph{both} the momentum and polarisation of an external scattering state thereby guaranteeing
the on-shell conditions.
\end{important}

\begin{important}{Spin $\mathbf{s=1}$.} 
Gauge fields may have the helicities $h=\pm 1$. Their polarisation
vectors are denoted by $\epsilon_{\pm}^{\mu}(p)$ and obey the relations
\begin{align}
\label{s1polarprop}
\begin{aligned}
& \epsilon_{\pm}^{\mu}(p)^{\ast} = \epsilon_{\mp}^{\mu}(p) \, ,
& & p\cdot \epsilon_{\pm}(p) = 0\, , \\
& \epsilon_{+}(p)\cdot \epsilon_{-}(p)  = -1\, ,  \qquad
& & \epsilon_{+}(p)^{2} = \epsilon_{-}(p)^{2}= 0\, .
\end{aligned}
\end{align}
The polarisation vectors may be expressed as bi-spinors using the spinor-helicity variables associated with the momentum $p$ ($p^{\a\da}=\la^{\a}\tla^{\da}$) as
\be
\label{s1polarrel}
\epsilon_{+,i}^{\a\da}= -\sqrt{2} \, \frac{\tla^{\da}_{i}\, \mu_{i}^{\a}}{\vev{\la_{i}\mu_{i}}}\,
,\qquad
\epsilon_{-,i}^{\a\da}= \sqrt{2} \, \frac{\tilde\mu^{\da}_{i}\, \la_{i}^{\a}}{\bev{\tla_{i}\tilde\mu_{i}}}\, .
\ee
One directly checks the properties of \eqn{s1polarrel} in this representation. 
We also find $h\, \epsilon_{\pm,i}^{\a\da}= \pm \epsilon_{\pm,i}^{\a\da}$, hence the
helicity assignments check.
The
spinor pair $\mu_{i}$, $\tilde\mu_{i}$ appearing in the above are arbitrary reference
spinors needed to define the polarisations. They parameterise a light-like reference momentum
$r_{i}=\mu_{i}\tilde\mu_{i}$ associated to every leg $i=1,\ldots, n$ in a scattering
amplitude. 
The only condition on the reference spinors is that they are nor parallel to $\la_{i}$ an $\tla_{i}$:
\be
\mu_{i}\neq c \la_{i}\, , \qquad \tilde\mu_{i}\neq c^{\ast} \tilde\la_{i}\, .
\ee
Moreover, we see from \eqn{s1polarrel} that $\epsilon_{\pm,i} \cdot r_i = 0$.
\end{important}

\begin{exer}{Gluon polarisations}
\label{Ex:1.7}
\vspace{-0.6cm}
\begin{enumerate}[a)]
\item Show that the properties in \eqn{s1polarprop}, together with the gauge choice $\epsilon_{\pm,i} \cdot r_i = 0$ (with $r_i^2=0$), lead to the representation of \eqn{s1polarrel}. Hint: expand the polarisation vector in a basis constructed from the spinors associated with $p_i^{\dot{\alpha}\alpha}$ and $r_i$.
\smallskip
\item Verify that the representation~\eqn{s1polarrel} fulfils the polarisation sum rule
\begin{align}
\sum_{h=\pm} \epsilon_{h,i}^{\mu} \epsilon_{h,i}^{*\nu} = - \eta^{\mu\nu} + \frac{p_i^{\mu} r_i^{\nu}+p_i^{\nu} r_i^{\mu}}{p_i \cdot r_i} \,.
\end{align}
\end{enumerate}
For the solution see \hyperref[Sol:1.7]{chapter~5}.
\end{exer}

The appearance of the reference spinors $\mu_{i}$ and $\tilde\mu_{i}$ with associated
reference momentum $r_{i}^{\a\da} \coloneqq  \mu_{i}^{\a}\, \tilde\mu_{i}^{\da}$ in the 
polarisation vectors corresponds 
to the freedom of performing local gauge transformations of the gauge field. To see this let us  compute the infinitesimal change
of the polarisation $\epsilon_{+}^{\a\da}$ induced by an infinitesimal shift of the reference
spinor $\mu\to\mu+ \delta\mu$:
\begin{align}
\label{gtargument}
\delta \epsilon_{+}^{\a\da} &= - \sqrt{2}\, \left( \frac{\tla^{\da}\, \delta\mu^{\a}}{\vev{\la\mu}}
-\tla^{\da}\,\mu^{\a}\, \frac{\vev{\la\, \delta\mu}}{\vev{\la\, \mu}^{2}} \right) =
-\sqrt{2}\, \frac{\tla^{\da}}{\vev{\la\, \mu}^{2}}\, \underbrace{(\delta\mu^{\a}\, \vev{\la\, \mu} -
\mu^{\da}\, \vev{\la\, \delta\mu} )}_{=\la^{\a}\, \vev{\mu\, \delta\mu}} \nn\\
&= p^{\a\da}\, \left(\sqrt{2}\, \frac{\vev{ \delta\mu \mu}}{\vev{\la\, \mu}^{2}}  \right)
= p^{\a\da}\,\xi(p,\mu,\delta\mu)\, ,
\end{align}
where we used Schouten's identity \eqref{eq:Schouten} in step two. We see that the induced
change of the
polarisation vector  $\epsilon_{+}^{\mu}$
is a gauge transformation $\delta \epsilon^{\mu}= p^{\mu}\xi$.
By transversality of the amplitude this implies the invariance under the shift $\mu\to\mu+ \delta\mu$. Therefore the amplitude depends only on $\{\lambda_{i}, \tilde\lambda_{i}, h_{i}=\pm 1\}$,
and we may freely choose the reference spinors $\mu_{i}$ and $\tilde\mu_{i}$ for \emph{every}
leg at our convenience---reflecting the local gauge invariance of the theory.

\begin{important}{Spin $\mathbf{s=2}$.} The polarisation of a graviton has the helicities 
$h=\pm 2$ and  is captured by a symmetric rank-two tensor
$\epsilon_{++/--}^{\mu\nu}(p)$. It is transverse
\be
p_{\mu} \epsilon_{++/--}^{\mu\nu}(p) = 0 \,,
\ee
and may be chosen to be traceless $\eta_{\mu \nu}\epsilon_{++/--}^{\mu \nu}(p)=0$. A very convenient
parametrisation is given by doubling the gauge field polarisations:
\be
\label{gravitonpolarize}
\epsilon_{++}^{\mu\nu}(p)= \epsilon_{+}^{\mu}(p) \, \epsilon_{+}^{\nu}(p)\, , \qquad
\epsilon_{--}^{\mu\nu}(p)= \epsilon_{-}^{\mu}(p) \, \epsilon_{-}^{\nu}(p)\, .
\ee
In this way the above properties of transversality and tracelessness are consequences
of the properties of the vector polarisations $\epsilon_{\pm}^{\mu}(p)$ in \eqn{s1polarprop}.

It is then straightforward to translate this into helicity spinor variables. We now have 
the spin-tensors 
\be
\epsilon_{++}^{\a\b\da\db} = \epsilon_{+}^{\a\da}\, \epsilon_{+}^{\b\db}\, , \qquad \quad
\epsilon_{--}^{\a\b\da\db} = \epsilon_{-}^{\a\da}\, \epsilon_{-}^{\b\db}\, ,
\ee
using the polarisation bi-spinors of the gauge fields \eqn{s1polarrel}.  
\end{important}
Again we have reference spinors $\mu_{i}$ and $\tilde\mu_{i}$ for each leg. The amplitude does not depend on 
this choice, and one may show that under a shift $\mu \to \mu + \delta\mu$ the graviton
polarisation transforms as
\be
\epsilon_{\mu\nu}\to \epsilon_{\mu\nu}'=\epsilon_{\mu\nu} + 2 p_{(\mu}\xi_{\nu)} \,,
\ee
by virtue of the same argument as in \eqn{gtargument}. This change 
of course leaves the amplitude invariant.

\section{Colour decompositions for gluon amplitudes}
\label{sec:1.9}

Let us now focus on non-Abelian gauge field theories and the management of colour. The tools
to be developed will allow us to disentangle the colour degrees of freedom from the kinematic ones. There are two such formalisms for an efficient colour management which we shall discuss: the trace based, and the structure constant based formalism. Focussing on $\text{SU}(N_c)$ gauge theories coupled to matter, one mostly 
encounters two representations of
the gauge group: 
\begin{itemize}
    \item {Adjoint representation:} gluons $A^a_\mu$  carry
    adjoint indices $a-1,\ldots,N^2_c-1$;
 \smallskip
    \item {Fundamental \& anti-fundamental representation:} fermions and scalars carry fundamental indices
    $i=1,\ldots, N_c$ and $\bar i=1,\ldots , N_c$.
\end{itemize}
As discussed above, the generators of the $\text{SU}(N_c)$ algebra in the fundamental 
representation are $N_c\!\times\!N_c$ hermitian, traceless
 matrices $(T^a)_i{}^{{j}}$ . 
 We recall \eqn{I.17}
\begin{equation}
\label{fasTr}
    f^{abc}=-\frac{\i}{\sqrt{2}} \, \Tr \bigl(T^a [T^b, T^c] \bigr)\, , 
\end{equation}
or $[T^a, T^b] {=} \i\sqrt{2} f^{abc} T^c$, with $\Tr(T^a T^b){=}\delta^{ab}$. 
Moreover, we have the $\text{SU}(N_c)$ identity of \eqn{eq:completenessSUN},
\begin{equation}
\label{SUNFierz}
 (T^a)_{i_1}{}^{{j}_1}\,  (T^a)_{i_2}{}^{{j}_2}= \delta_{i_1}{}^{{j}_2}\,
\delta_{i_2}{}^{{j}_1} - \frac{1}{N_c}\delta_{i_1}{}^{{j}_1}\,
\delta_{i_2}{}^{{j}_2} \, ,
\end{equation}
which will be important for the trace based decomposition.
It can be understood as a completeness relation for a basis of Hermitian matrices spanned
by $\{\1, T^a\}$. Introducing a graphical representation for the structure constants
and generators via
\be
\label{Tagraphical}
f^{abc}=
   \begin{tikzpicture}[baseline={(current bounding box.center)}]
	\begin{feynman}
	\coordinate (a) at (-.4,-0.8);
	\coordinate (b) at (-.4,0.8);
	\coordinate (o) at (00,0);
\coordinate (c) at (0.9,0);
 \draw [fill] (o) circle (.08);
\diagram*{(a)  -- [gluon] (o), (b) -- [gluon] (o),(c) -- [gluon] (o)};
\draw (a) node [below] {a};\draw (b) node [above] {b};\draw (c) node [right] {c};
\end{feynman}\end{tikzpicture}\, ,
\qquad
(T^{a})_{i_1}{}^{ j_{1}} \, = \,
\begin{tikzpicture}[baseline={(current bounding box.center)}]
	\begin{feynman}
	\coordinate (j1) at (-1,0.4);
	\coordinate (i1) at (-1,-0.4);
	\coordinate (l) at (-0.5,0);
     \coordinate (r) at (0.3,0);
\diagram*{(j1) -- [fermion,quarter left, looseness=1.5] (l) -- [fermion, quarter left] (i1)};
\diagram*{(l) -- [gluon] (r)};
\draw (r) node [right] {$a$};
\draw (i1) node [left] {$i_{1}$};\draw (j1) node [left] {$ j_{1}$};
 \draw [fill=lightgray,thick] (l) circle (.08);
\end{feynman}
	\end{tikzpicture} \,,
\ee
one may illustrate these two relations as in figure~\ref{fig:colour-decomposition}.

Moreover, we recall the Jacobi identity for the structure constants
of \eqn{Jacobif},
\be
\label{Jacobif2}
f^{abe} f^{ceg} + f^{bce} f^{aeg} + f^{cae} f^{beg} =0 \, ,
\ee
which we shall put to use in the structure constant based formalism.
\begin{figure}[tbp]
   \centering
   $$
   \begin{tikzpicture}[baseline={(current bounding box.center)}]
	\begin{feynman}
	\coordinate (a) at (-.4,-0.8);
	\coordinate (b) at (-.4,0.8);
	\coordinate (o) at (00,0);
\coordinate (c) at (0.9,0);
 \draw [fill] (o) circle (.08);
\diagram*{(a)  -- [gluon] (o), (b) -- [gluon] (o),(c) -- [gluon] (o)};
\draw (a) node [below] {a};\draw (b) node [above] {b};\draw (c) node [right] {c};
\end{feynman}\end{tikzpicture}
=
-\frac{\i}{\sqrt 2}\left (\,
\begin{tikzpicture}[baseline={(current bounding box.center)}]
	\begin{feynman}
	\coordinate (a) at  (-.4,-0.8);
	\coordinate (u) at (0,-.3);
	\coordinate (b) at  (-.4,0.8);
	\coordinate (o) at (0,.3);
	\coordinate (r) at (0.4,0);
\coordinate (c) at (1.2,0);
\diagram*{(a) -- [gluon] (u) -- [fermion,quarter right, looseness=1.5] (r) -- [fermion, quarter right] (o) -- [fermion,half right] (u), (b) -- [gluon] (o),(c) -- [gluon] (r)};
\draw (a) node [below] {a};\draw (b) node [above] {b};\draw (c) node [right] {c};
 \draw [fill=lightgray,thick] (r) circle (.08);\draw [fill=lightgray,thick] (u) circle (.08);\draw [fill=lightgray,thick] (o) circle (.08);
\end{feynman}
	\end{tikzpicture}
	-
\begin{tikzpicture}[baseline={(current bounding box.center)}]
	\begin{feynman}
	\coordinate (a) at  (-.4,-0.8);
	\coordinate (u) at (0,-.3);
	\coordinate (b) at  (-.4,0.8);
	\coordinate (o) at (0,.3);
	\coordinate (r) at (0.4,0);
\coordinate (c) at (1.2,0);
\diagram*{(a) -- [gluon] (u) -- [fermion,half left, looseness=1.5] (o) -- [fermion, quarter left] (r) -- [fermion,quarter left] (u), (b) -- [gluon] (o),(c) -- [gluon] (r)};
\draw (a) node [below] {a};\draw (b) node [above] {b};\draw (c) node [right] {c};
 \draw [fill=lightgray,thick] (r) circle (.08);\draw [fill=lightgray,thick] (u) circle (.08);\draw [fill=lightgray,thick] (o) circle (.08);
\end{feynman}
	\end{tikzpicture}
		\right )
$$
$$
\begin{tikzpicture}[baseline={(current bounding box.center)}]
	\begin{feynman}
	\coordinate (j1) at (-1,0.4);
	\coordinate (i1) at (-1,-0.4);
	\coordinate (j2) at (1,0.4);
	\coordinate (i2) at (1,-0.4);
	\coordinate (l) at (-0.5,0);
     \coordinate (r) at (0.5,0);
\diagram*{(j1) -- [fermion,quarter left, looseness=1.5] (l) -- [fermion, quarter left] (i1)};
\diagram*{(i2) -- [fermion,quarter left, looseness=1.5] (r)  -- [fermion, quarter left]  (j2)};
\diagram*{(l) -- [gluon] (r)};
\draw (i1) node [left] {$i_{1}$};\draw (j1) node [left] {$ j_{1}$};
\draw (i2) node [right] {$ j_{2}$};\draw (j2) node [right] {$i_{2}$};
 \draw [fill=lightgray,thick] (r) circle (.08);\draw [fill=lightgray,thick] (l) circle (.08);
\end{feynman}
	\end{tikzpicture}
	=
\begin{tikzpicture}[baseline={(current bounding box.center)}]
	\begin{feynman}
	\coordinate (j1) at (-0.75,0.4);
	\coordinate (i1) at (-0.75,-0.4);
	\coordinate (j2) at (0.75,-0.4);
	\coordinate (i2) at (0.75,0.4);
\diagram*{(j1) -- [fermion] (i2)};\diagram*{(j2) -- [fermion] (i1)};
\draw (i1) node [left] {$i_{1}$};\draw (j1) node [left] {$ j_{1}$};
\draw (i2) node [right] {$i_{2}$};\draw (j2) node [right] {$ j_{2}$};
\end{feynman}
	\end{tikzpicture}	
	-
	\frac{1}{N_{c}}
	\begin{tikzpicture}[baseline={(current bounding box.center)}]
	\begin{feynman}
	\coordinate (j1) at (-0.75,0.4);
	\coordinate (i1) at (-0.75,-0.4);
	\coordinate (j2) at (0.75,0.4);
	\coordinate (i2) at (0.75,-0.4);
	\coordinate (l) at (-0.25,0);
     \coordinate (r) at (0.25,0);
\diagram*{(j1) -- [fermion,quarter left, looseness=1.5] (l) -- [fermion, quarter left] (i1)};
\diagram*{(i2) -- [fermion,quarter left, looseness=1.5] (r) -- [fermion, quarter left] (j2)};
\draw (i1) node [left] {$i_{1}$};\draw (j1) node [left] {$ j_{1}$};
\draw (i2) node [right] {$ j_{2}$};\draw (j2) node [right] {$i_{2}$};
\end{feynman}
	\end{tikzpicture}
$$
     \caption{\it Graphical representation of eqs.~(\ref{fasTr}) and (\ref{SUNFierz})
     using (\ref{Tagraphical}) for the generators. }
   \label{fig:colour-decomposition}
\end{figure}

\subsection{Trace basis}
\label{sec:trace_basis}

A look at the QCD Feynman rules of section~\ref{chap:1.4} tells us that the colour dependence of a given Feynman graph arises from its vertices. The three-gluon vertex in \eqn{V3YM} carries one structure constant $f^{abc}$, whereas the four-gluon vertex in \eqn{V4YM} contains 
two $f^{abc}$'s.
Coupling to matter, the quark-anti-quark-gluon
vertex \eqn{VqYM} comes with a generator $(T^a)_i{}^{ j}$. 
In order to work out the colour dependence of a given Feynman diagram in the trace basis, we replace all structure constants appearing
in it by the trace formula~\eqref{fasTr}. 
This transforms the expression to products
of generators  $(T^{a})_{i}{}^{ j}$ with contracted and open indices. 
Open fundamental indices $(i, j)$ correspond to external quark lines in the diagram, open adjoint indices $(a)$
to the external gluon states.
Contracted adjoint indices can be used to merge traces and products of generators 
by repeatedly applying the $\text{SU}(N_{c})$ identity~\eqref{SUNFierz}:
\begin{align}
\label{SUNFierz2}
(A\, T^{a}  B)_{i}{}^{ j}\, 
(C\, T^{a}  D)_{k}{}^{ l}
= (A \, D)_{i}{}^{l}\, (C\, B)_{k}{}^{ j}
-\frac{1}{N_{c}} (A\, B)_{i}{}^{j}\, 
(C\, D)_{k}{}^{ l} \, ,
\end{align}
where $A, B,C$ and $D$ are arbitrary $N_{c}\times N_{c}$ matrices made of products of generators.
By iterating this procedure we arrive at a final expression of traces and strings 
of generators $T^a$'s with only open adjoint and fundamental indices corresponding to the external states.
They take the generic form
\begin{align}
\Tr(T^{a_{1}}\cdots T^{a_{n}})\ldots
\Tr(T^{b_{1}}\cdots T^{b_{m}})\, (T^{c_{1}}
\cdots T^{c_{p}})_{i_{1}}{}^{ j_{1}}
\ldots
 (T^{d_{1}} \cdots
T^{d_{p}})_{i_{s}}{}^{j_{s}}\, .
\end{align}
For pure gluon amplitudes there is a further simplification: in pure Yang-Mills theory the interaction
vertices of the $\text{SU}(N_{c})$ and $\text{U}(N_{c})$ gauge groups are identical, as $f^{0bc}\!=\!0$ due to \eqn{fasTr}
where 
$T^{0}= \1/\sqrt{N_{c}}$ is the $\text{U}(1)$ generator. Hence, the ${1}/{N_{c}}$ part in
the relation~\eqref{SUNFierz2}, responsible for tearing apart traces, is not active here.
In conclusion, tree-level gluon amplitudes reduce to a \emph{single}-trace structure, and can
be brought into the \emph{colour}-decomposed form 
\begin{align}
\label{I.26}
\cA_{n}^{\text{tree}}(\{ a_{i}, h_{i},p_{i}\}) 
=  
\sum_{\sigma\in S_{n}/\mathbb{Z}_{n}}\, 
\Tr(T^{a_{\sigma_{1}}}\,  T^{a_{\sigma_{2}}}\cdots T^{a_{\sigma_{n}}} )\, 
A_{n}^{\text{tree}}({\sigma_{1}}, {\sigma_{2}}, 
\ldots , \sigma_{n}) \, .
\end{align}
Here $h_{i}$ denote the helicities and $a_{i}$ the adjoint colour indices of the external states, 
and we use the compact notation $\sigma = \{p_{\sigma},h_{\sigma}\}$ in the argument of the $A_{n}^{\text{tree}}$. 
The latter are called \emph{partial} or \emph{colour-ordered} amplitudes
and carry all kinematic information. 
Moreover, $S_{n}/\mathbb{Z}_{n}$ is the set of all non-cyclic permutations of $n$ elements, which is
equivalent to $S_{n-1}$. Therefore for an $n$-particle amplitude
we have $(n-1)!$ distinct colour-ordered amplitudes in the
trace basis.
For example in the four-point case we find
\begin{align}
\begin{aligned}
\cA_{4}^{\text{tree}}
= \ & 
\Tr(T^{a_{1}} T^{a_{2}} T^{a_{3}} T^{a_{4}} )\, A_{4}^{\text{tree}}(1,2,3,4) +
\Tr(T^{a_{1}} T^{a_{3}} T^{a_{4}} T^{a_{2}} )\, A_{4}^{\text{tree}}(1,3,4,2)  \\ 
& + \Tr(T^{a_{1}} T^{a_{4}} T^{a_{2}} T^{a_{3}} )\, A_{4}^{\text{tree}}(1,4,2,3)+
\Tr(T^{a_{1}}  T^{a_{2}} T^{a_{4}} T^{a_{3}} )\, A_{4}^{\text{tree}}(1,2,4,3)  \\ 
& + \Tr(T^{a_{1}} T^{a_{4}} T^{a_{3}} T^{a_{2}} )\, A_{4}^{\text{tree}}(1,4,3,2) +
\Tr(T^{a_{1}} T^{a_{3}} T^{a_{2}} T^{a_{4}} )\, A_{4}^{\text{tree}}(1,3,2,4) \, .
\end{aligned}
\end{align}
This is the promised separation of colour and kinematical degrees of freedom.
The colour ordered amplitudes $A_{n}$ are simpler than the full amplitudes $\cA_{n}$, 
as they are individually gauge invariant. This is due to the fact that any shift 
of a polarisation $\epsilon_{i}\to p_{i}$ in \eqref{I.26} will lead to a vanishing
left-hand side for $\mathcal{A}_{n}$. On the right-hand side the colour factors form a linearly independent basis, hence the individual factors of $A_{n}$ need to vanish
individually. In addition, they
exhibit poles only whenever cyclically adjacent momenta go on-shell,
$(p_{i}+p_{i+1}+\cdots + p_{i+s})^{2}\to 0$, see 
Figure~\ref{Fig:regionmomenta}. This property will be exploited in chapter~\ref{ch:trees}.
\begin{figure}[tt]
\begin{center}
$$
  \begin{tikzpicture}[baseline={(current bounding box.center)}]
  \begin{feynman}
    \coordinate (1) at (-3.5,-2);
    \coordinate (2) at (-4,-0.5);
    \coordinate (3) at (-3,1);
    \coordinate (4) at (-1,1.4);
    \coordinate (5) at (0,1.4);
    \coordinate (6) at (2,1.4);
    \coordinate (7) at (2,-.6);
    \coordinate (9) at (-0.8,-2);
    \coordinate (8) at (0.8,-2);
        \coordinate (10) at (-1.8,-2);
    \coordinate (a) at (-3.2,-0.7);
    \coordinate (b) at (-2,-0);
    \coordinate (c) at (-1,0);
    \coordinate (d) at (0,0);
    \coordinate (e) at (1,0);
    \coordinate (f) at (1,0.9);
     \coordinate (g) at (-0.2,-1);
 \diagram{ (1)-- [gluon] (a)  -- [gluon] (2), (10)  -- [gluon] (b),
     (10)  -- [gluon] (b),
   (a)  -- [gluon] (b) ,
    (3)  -- [gluon] (b),
      (4)  -- [gluon] (c),     (5)  -- [gluon] (f),
      (6)  -- [gluon] (f),    (7)  -- [gluon] (e),
      (8)  -- [gluon] (g),   (9)  -- [gluon] (g),
        (b)  -- [gluon] (c),
          (c)  -- [gluon] (d),
               (d)  -- [gluon] (e),
                      (e)  -- [gluon] (f),
                             (d)  -- [gluon] (g)};
  \draw (1) node [right] {$1$};
  \draw (2) node [left] {$2$};
  \draw (3) node [left] {$3$};
  \draw (4) node [left] {$4$};
  \draw (5) node [left] {$5$};
  \draw (6) node [right] {$6$};
  \draw (7) node [right] {$7$};  
  \draw (8) node [right] {$8$};
    \draw (9) node [left] {$9$};
        \draw (10) node [left] {$10$};
        \draw (-2,-0.7)  node [left] {$(a)$};
         \draw (-1.1,0.4)  node [left] {$(b)$};
         \draw (-0.1,0.4)  node [left] {$(c)$};
             \draw (0.9,-0.4)  node [left] {$(d)$};
       \draw [fill] (a) circle (.08);
       \draw [fill] (b) circle (.08);
      \draw [fill] (c) circle (.08);
      \draw [fill] (d) circle (.08);
       \draw [fill] (e) circle (.08);
      \draw [fill] (f) circle (.08);       
      \draw [fill] (g) circle (.08);
      \end{feynman}
              \end{tikzpicture}
$$
\begin{align*}
&(a):  (p_{1}+p_{2})^{2}\to 0\, , \qquad & \qquad & (b):  (p_{10}+ p_{1}+p_{2}+p_{3})^{2}\to 0\,, \\
&(c):  (p_{5}+p_{6}+p_{7}+p_{8}+p_{9})^{2}\to 0\, , & \qquad & (d):  (p_{5}+p_{6}+p_{7})^{2}\to 0 \,.
\end{align*}
\end{center}
\caption{
Some of the possible poles in a colour-ordered Feynman diagram. 
}
\label{Fig:regionmomenta}
\end{figure}

For tree-level gluon-quark-anti-quark amplitudes with a single quark line one has
\begin{align} \begin{aligned}
\cA_{n,q\bar q}^{\text{tree}} \Bigl(\{ a_{i}, h_{i},p_{i}\}|\Bigl\{i,q_{1}^{h_{q_{1}}},
 j,\bar{q}_{2}^{h_{q_{2}}}\Bigr\}\Bigr) & =
 \sum_{\sigma\in S_{n-2}}   
(T^{a_{\sigma_{1}}}\, \cdots T^{a_{\sigma_{n-2}}})_{i}{}^{j}\, \\ 
& \times A_{n,q\bar q}^{\text{tree}}\Bigl(\sigma_{1},
\ldots, {\sigma_{n-2}}| q_{1}^{h_{q_{1}}} , \bar{q}_{2}^{h_{q_{2}}}\Bigr) \, . 
\end{aligned} \end{align}
Increasing the number of quark lines to $m>1$ 
yields a more involved structure, as more strings and factors of $1/N_c$ appear. Here the $m$ quark lines will yield $m$ products of strings in $T$-matrices,
$(T^{a_1}\, T^{a_2} \cdots T^{a_r})_{i_s}{}^{{j}_s}$, where the adjoint indices are
either contracted with an outer gluon leg or connect to another quark line. Using the $\text{SU}(N_{c})$ 
identity~\eqref{SUNFierz2} for these internal contractions leads to a final basis of $m$ products
of open $T$-matrix strings with only external adjoint indices. The general construction
of this colour-decomposition is rather involved and we shall not discuss it here. We refer to~\cite{ch1_Mangano:1990by,ch1_Ita:2011ar,ch1_Badger:2012pg,ch1_Reuschle:2013qna} for a detailed
analysis.
In this case, some of the colour factors also include explicit factors of 
$1/N_c$ stemming from the last term in \eqn{SUNFierz2}. Nevertheless, all of the
kinematical dependences can still be constructed from
suitable linear combinations of the partial amplitudes for
external quarks, anti-quarks and external gluons generated by the colour-ordered
Feynman diagrams. Hence, the partial amplitudes 
are the atoms of gauge-theory scattering amplitudes.

At loop level, pure gluon amplitudes contain also multi-trace contributions arising from 
the merging performed using \eqn{SUNFierz}. For example, at one loop one has
\begin{align}
\begin{split}
\label{leading-colour-one-loop}
\cA_{n}^{\text{1-loop}}(\{ a_{i}, h_{i},p_{i}\}) &=  N_c\,
 \sum_{\sigma\in S_{n}/\mathbb{Z}_{n}}\,
\Tr(T^{a_{\sigma_{1}}}\, T^{a_{\sigma_{2}}}\cdots T^{a_{\sigma_{n}}} )\, 
A_{n;1}^{(1)}(\sigma_{1}, 
\ldots , \sigma_{n} ) \\
+ \sum_{i=2}^{[n/2]+1}\, \sum_{\sigma\in S_n/\mathbb{Z}_n} &
\Tr(T^{a_{\sigma_{1}}} \cdots T^{a_{\sigma_{i-1}}} )\,
\Tr(T^{a_{\sigma_{i}}}\cdots T^{a_{\sigma_{n}}})\, 
A_{n;i}^{(1)}(\sigma_{1}, \ldots , \sigma_{n}  )\, ,
\end{split}
\end{align}
where the $A_{n;1}^{(1)}$ are called the primitive (colour-ordered) amplitudes, and the 
$A_{n;c>1}^{(1)}$ are the higher primitive amplitudes, $[n]$ is the lower integer part of $n$. The latter can be expressed as linear
combinations of the primitive ones~\cite{ch1_Bern:1990ux}. 
In the large-$N_{c}$ limit the single-trace 
contributions are enhanced: one speaks of the \emph{leading-colour} contributions. In colour-summed cross sections, which are of interest for computing 
collider physics observables, the contribution of the higher primitive amplitudes is suppressed by a factor of $1/N_{c}^2$.

\subsection{Structure constant basis}

\begin{figure}[tt]
\begin{center}
$$
  \begin{tikzpicture}[baseline={(current bounding box.center)}]
  \begin{feynman}
    \coordinate (1) at (120:2);
    \coordinate (2) at (75:2);
    \coordinate (3) at (30:2);
    \coordinate (4) at (-15:2);
    \coordinate (5) at (-60:2);
    \coordinate (6) at (-105:2);
    \coordinate (7) at (-150:2);
    \coordinate (8) at (-195:2);
    \coordinate (a) at (-1.0,0.75);
    \coordinate (b) at (-0.0,0.5);
    \coordinate (c) at (0.0,-0.5);
    \coordinate (d) at (1.0,-0.75);    
    \coordinate (e) at (0.75,1.15);    
    \coordinate (f) at (-0.75,-1.0);    
\diagram*{  (1) -- [gluon](a)};
\diagram*{  (8) -- [gluon](a)};
\diagram*{  (2) -- [gluon](e)};
\diagram*{  (3) -- [gluon](e)} ;
\diagram*{  (4) -- [gluon](d)};
\diagram*{  (5) -- [gluon](d)};
\diagram*{  (6) -- [gluon](f)};
\diagram*{  (7) -- [gluon](f)};
\diagram*{  (b) -- [gluon](a)};
\diagram*{  (b) -- [gluon](e)};
\diagram*{  (b) -- [gluon](c)};
\diagram*{  (c) -- [gluon](d)};
\diagram*{  (c) -- [gluon](f)};
  \draw (1) node [left] {$a_1$};
  \draw (2) node [left] {$a_2$};
  \draw (3) node [right] {$a_3$};
  \draw (4) node [right] {$a_4$};
  \draw (5) node [right] {$a_5$};
  \draw (6) node [right] {$a_6$};
  \draw (7) node [left] {$a_7$};  
  \draw (8) node [left] {$a_8$};
       \draw [fill] (a) circle (.08);
       \draw [fill] (b) circle (.08);
      \draw [fill] (c) circle (.08);
      \draw [fill] (d) circle (.08);
       \draw [fill] (e) circle (.08);
      \draw [fill] (f) circle (.08); 
      \end{feynman}      
              \end{tikzpicture}
$$
\end{center}
\caption{ 
Typical colour tree in a structure constant based expansion. 
}
\label{Fig:tree}
\end{figure}
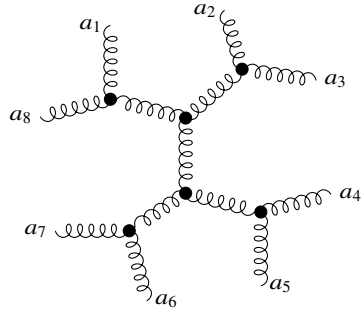

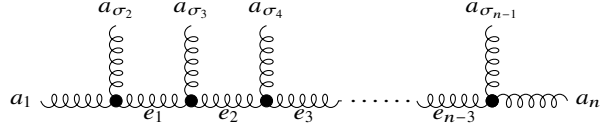
\begin{figure}[tt]
\begin{center}
$$
\begin{tikzpicture}[baseline={(current bounding box.center)}]
\begin{feynman}
\coordinate (1) at (-2,0);
\coordinate (2) at (-1,1);
\coordinate (a2) at (-1,0);
\coordinate (3) at (0,1);
\coordinate (a3) at (0,0);
\coordinate (4) at (1,1);
\coordinate (a4) at (1,0);
\coordinate (a5) at (2,0);
\coordinate (a6) at (3,0);
\coordinate (n-1) at (4,1);
\coordinate (an-1) at (4,0);
\coordinate (n) at (5,0);
\draw [gluon] (1) -- (a2);
\draw [gluon] (a2) to node[below]{$e_{1}$} (a3);
\draw [gluon] (a3) to node[below]{$e_{2}$} (a4);
\draw [gluon] (a6) to node[below]{$e_{n-3}$} (an-1);
\draw [gluon] (a4) to node[below]{$e_{3}$} (a5);
\draw [gluon] (2) -- (a2);
\draw [gluon] (3) -- (a3) ;
\draw [gluon] (4) -- (a4) ;above
\draw [gluon] (n-1) -- (an-1) ;
\draw [gluon] (n) -- (an-1) ;
  \draw (a5) node [right] {$\ldots\ldots$};
  \draw (1) node [left] {$a_{1}$};
    \draw (2) node [above] {$a_{\sigma_{2}}$};
      \draw (3) node [above] {$a_{\sigma_{3}}$};
        \draw (4) node [above] {$a_{\sigma_{4}}$};
          \draw (n-1) node [above] {$a_{\sigma_{n-1}}$};
            \draw (n) node [right] {$a_{n}$};  
        \draw [fill] (an-1) circle (.08);
       \draw [fill] (a2) circle (.08);
         \draw [fill] (a3) circle (.08);
       \draw [fill] (a4) circle (.08);  
             \draw [fill] (an-1) circle (.08);  
             \end{feynman}
\end{tikzpicture}
$$
\end{center}
\caption{
Half-ladder form.
}
\label{Fig:halfladder}
\end{figure}

An alternative basis for the colour decomposition of pure-gluon  
amplitudes employs the structure constants $f^{abc}$ and is due to Del Duca, Dixon and Maltoni (DDM)~\cite{ch1_DelDuca:1999rs}. To begin with, we consider the colour dependence of an $n$-gluon tree amplitude. This may
be represented as a sum over only tri-valent graphs with vertices linear in $f^{abc}$.
In order to reach this tri-valent representation we artificially ``blow'' up a 
four-valent gluon vertex to sums of products of tri-valent vertices. This is done
 by multiplying it by $1=q^{2}/q^{2}$ where $\i/q^{2}$ is the ``blown up'' propagator.
Concretely, if we go back to the Feynman rule of \eqn{V4YM} for the four-gluon vertex, we
rewrite
\be
f^{abe}f^{cde} = f^{abe}\, \frac{\delta_{ef} (p_{a}+p_{b})^{2}}{(p_{a}+p_{b})^{2}}f^{cdf} \,,
\ee
where $p_{a}$($p_{b}$) is the momentum flowing into leg $a$ ($b$), and analogously for the
other two $ff$-terms. The resulting structure will then be a trivalent tree as in Figure~\ref{Fig:tree}.
Now we use the Jacobi identity~\eqref{Jacobif} for the structure constants,
\begin{align} \begin{aligned}
f^{dae}f^{bce}\qquad - \qquad  f^{dbe}f^{ace}\qquad  &= \qquad f^{abe} f^{cde} \,, \\
 \begin{tikzpicture}[baseline={(current bounding box.center)}]
  \begin{feynman}
 \coordinate (a) at (-0.75,0.6);
    \coordinate (b) at (0.75,.6);
    \coordinate (c) at (0.75,-0.6);
    \coordinate (d) at (-0.75,-0.6);
    \coordinate (l) at (-0.3,0); 
    \coordinate (r) at (0.3,0);      
  \diagram*{  (a) -- [gluon](l) -- [gluon] (r) -- [gluon] (b), (d) -- [gluon] (l), (c) -- [gluon] (r)};
    \draw [fill] (l) circle (.08);
       \draw [fill] (r) circle (.08);
        \draw (a) node [left] {$a$};
  \draw (d) node [left] {$d$};
  \draw (b) node [right] {$b$};
  \draw (c) node [right] {$c$};
    \end{feynman}     
  \end{tikzpicture}  
  -
   \begin{tikzpicture}[baseline={(current bounding box.center)}]
  \begin{feynman}
 \coordinate (a) at (-0.75,0.6);
    \coordinate (b) at (0.75,.6);
    \coordinate (c) at (0.75,-0.6);
    \coordinate (d) at (-0.75,-0.6);
    \coordinate (l) at (-0.5,-0.25); 
    \coordinate (r) at (0.5,-0.25);      
  \diagram*{  (a) -- [gluon](r) -- [gluon] (l) -- [gluon] (b), (d) -- [gluon] (l), (c) -- [gluon] (r)};
    \draw [fill] (l) circle (.08);
       \draw [fill] (r) circle (.08);
        \draw (a) node [left] {$a$};
  \draw (d) node [left] {$d$};
  \draw (b) node [right] {$b$};
  \draw (c) node [right] {$c$};
    \end{feynman}     
  \end{tikzpicture}  
  &=
    \begin{tikzpicture}[baseline={(current bounding box.center)}]
  \begin{feynman}
 \coordinate (a) at (-0.75,0.6);
    \coordinate (b) at (0.75,.6);
    \coordinate (c) at (0.75,-0.6);
    \coordinate (d) at (-0.75,-0.6);
    \coordinate (l) at (0,0.4); 
    \coordinate (r) at (00,-0.4);      
  \diagram*{  (a) -- [gluon](l) -- [gluon] (b), (r) -- [gluon] (l), (d) -- [gluon] (r), (c) -- [gluon] (r)};
    \draw [fill] (l) circle (.08);
       \draw [fill] (r) circle (.08);
        \draw (a) node [left] {$a$};
  \draw (d) node [left] {$d$};
  \draw (b) node [right] {$b$};
  \draw (c) node [right] {$c$};
    \end{feynman}     
  \end{tikzpicture}  \,,
\end{aligned}  \end{align}
in order to successively shrink branched trees to branchless ones, resulting in
a final ``half-ladder'' expression. The shrinking of a branched tree is thereby performed via the operation
\be
\begin{tikzpicture}[baseline={(current bounding box.center)}]
 \begin{feynman}   
  \tikzstyle{blob}=[
    circle,
    minimum size =.15cm,
    draw=black,
    thick,
    fill=lightgray,
    text=black
]
\coordinate (in) at (-1.,0.5);
\coordinate (out) at (1.,0.5);
\coordinate (a) at (0,0.5);
\coordinate (b) at (0,1.2);
\coordinate (1) at (-0.75,1.5);
\coordinate (2) at (0.75,1.5);
\diagram*{ (in) -- [gluon] (a)};
\diagram*{(out) -- [gluon] (a)};
\diagram*{ (a) -- [gluon] (b)};
\diagram*{ (1) -- [gluon] (b)} ;
\diagram*{ (2) -- [gluon] (b) };
      \draw [fill] (a) circle (.08);
       \draw [fill] (b) circle (.08);
        \node [blob] (1) at (-.75,1.5){$1$};
                \node [blob] (2) at (.75,1.5){$2$};
       \draw [fill] (a) circle (.08);
       \draw [fill] (b) circle (.08);  
        \end{feynman}  
\end{tikzpicture} \quad
=  \quad
\begin{tikzpicture}[baseline={(current bounding box.center)}]
 \begin{feynman}  
  \tikzstyle{blob}=[
    circle,
    minimum size =.15cm,
    draw=black,
    thick,
    fill=lightgray,
    text=black
]
\coordinate (in) at (-1.,0.5);
\coordinate (out) at (1.,0.5);
\coordinate (a) at (-0.3,0.5);
\coordinate (b) at (0.3,0.5);
\coordinate (1) at (-0.75,1.5);
\coordinate (2) at (0.75,1.5);
\diagram*{ (in) -- [gluon] (a)};
\diagram*{ (out) -- [gluon] (b)};
\diagram*{ (a) -- [gluon] (b)};
\diagram*{ (1) -- [gluon] (a) };
\diagram*{ (2) -- [gluon] (b) };
      \draw [fill] (a) circle (.08);
       \draw [fill] (b) circle (.08);
        \node [blob] (1) at (-0.75,1.5){$1$};
                \node [blob] (2) at (0.75,1.5){$2$};
       \draw [fill] (a) circle (.08);
       \draw [fill] (b) circle (.08);  
        \end{feynman}  
\end{tikzpicture}\quad
- \quad 
\begin{tikzpicture}[baseline={(current bounding box.center)}]
\begin{feynman} 
  \tikzstyle{blob}=[
    circle,
    minimum size =.15cm,
    draw=black,
    thick,
    fill=lightgray,
    text=black
]
\coordinate (in) at (-1.,0.5);
\coordinate (out) at (1.,0.5);
\coordinate (a) at (-0.3,0.5);
\coordinate (b) at (0.3,0.5);
\coordinate (1) at (-0.75,1.5);
\coordinate (2) at (0.75,1.5);
\diagram*{ (in) -- [gluon] (a)};
\diagram*{ (out) -- [gluon] (b)};
\diagram*{ (a) -- [gluon] (b)};
\diagram*{ (1) -- [gluon] (a) };
\diagram*{ (2) -- [gluon] (b) };
      \draw [fill] (a) circle (.08);
       \draw [fill] (b) circle (.08);
        \node [blob] (1) at (-0.75,1.5){$2$};
                \node [blob] (2) at (0.75,1.5){$1$};
       \draw [fill] (a) circle (.08);
       \draw [fill] (b) circle (.08);          
    \end{feynman}   
\end{tikzpicture} \ .
\ee
In summary, one may colour-reduce any diagram, e.g.\ as depicted in fig.~\ref{Fig:tree},
to the ``half-ladder'' structure shown in fig.~\ref{Fig:halfladder}.
In this way we can completely reduce an amplitude to a sum of colour-ordered amplitudes
in the half-ladder basis in colour space:
\begin{align}
\label{DDM}
\begin{split}
\cA_{n}^{\text{tree}}(\{ a_{i}, h_{i},p_{i}\}) 
\!=\!  \\
\sum_{\sigma\in S_{n-2}}\, 
f^{a_{1}a_{\sigma_{2}}e_{1}} & \, f^{e_{1}a_{\sigma_{3}}e_{2}} \, f^{e_{2}a_{\sigma_{4}}e_{3}} 
\cdots f^{e_{n-3}a_{\sigma_{n-1}}a_{n}} 
A_{n}^{\text{tree}}(1, {\sigma_{2}}, 
\ldots , {\sigma_{n-1}}, n) \, , 
\end{split}
\end{align}
where we now sum over the permutations $\sigma$ of the $n{-}2$ elements $\{2,3,\ldots, n-1\}$.
The half-ladder colour basis fixes two (arbitrary) legs, here 1 and  $n$, see Figure~\ref{Fig:halfladder}.

In consequence, the DDM basis consists of $(n-2)!$ independent partial amplitudes.
This is to be contrasted with the  $(n-1)!$ partial amplitudes that we found in the trace basis.
Hence, there must exist non-trivial identities between  colour-ordered amplitudes
allowing one to reduce the basis accordingly. These are known as  Kleiss-Kuijf
relations \cite{ch1_Kleiss:1988ne}, and take the form
\be
\label{KleissKruif}
A_{n}^{\text{tree}}(1,\{\alpha\}, n, \{\beta\}) = (-1)^{n_{\beta}}
\sum_{\sigma\in \alpha \shuffle \beta^{\top}} A_{n}^{\text{tree}}(1,\sigma, n) \, , 
\ee
where $n_{\beta}$ denotes the number of elements in the set $\beta$, and $\beta^{\top}$ is the
set $\beta$ with reversed ordering. The shuffle or ordered permutation
${\displaystyle \alpha \shuffle \beta^{\top}}$ of the two sets merges $\alpha$ and $\beta^{\top}$ while preserving the individual orderings of $\alpha$ and $\beta^{\top}$. 
An example illustrates~this:
$$
\{12\} \shuffle \{34\}^{\top}= \{1243\}+\{1423\}+\{1432\}+\{4123\}+\{4312\}+\{4132\} \,.
$$
In fact, one can prove the Kleiss-Kuijf relations~\eqref{KleissKruif}  by
rewriting the DDM basis in terms of the trace basis discussed above.

It turns out that there exists a further non-trivial identity between colour-ordered amplitudes,
allowing one to further reduce the basis of colour-ordered (or partial) amplitudes
to $(n-3)!$ independent elements. This is due to the  Bern-Carrasco-Johansson relation~\cite{ch1_Bern:2008qj,ch1_Bern:2010ue}, to be discussed in Chapter~\ref{sec:2.5}. It takes the schematic form
\be
\label{BCJ}
A_{n}^{\text{tree}}(\sigma_{1},\ldots,\sigma_{n}) = 
\sum_{\rho\in S_{n-3}} 
K^{(\sigma)}_{\rho}
A_{n}^{\text{tree}}(1,2,{\rho_{3}},\ldots, \rho_{n-1},n)\, , 
\ee
with coefficients $K^{(\sigma)}_{\rho}$ depending on the external momenta.
Finally, we note that there is also a useful generalisation of the DDM basis to include fundamental matter that we will not discuss here, see \cite{ch1_Johansson:2015oia,ch1_Melia:2015ika}.

\begin{svgraybox}{\bf Colour-ordered Feynman rules.}
One may write down colour-ordered Feynman rules that generate the colour-ordered (partial) 
amplitudes upon stripping off
the colour factors from the usual Feynman rules. This is trivial for the gluon
and quark propagators (working in Feynman gauge):
\be
\begin{tikzpicture}[baseline={(current bounding box.center)}]
\begin{feynman}
\vertex (a);
\vertex [right of = a] (b);
\draw (a) node [above] {$\mu$} ;
\draw (b) node [above] {$\nu$};
\diagram*{(a) -- [gluon, edge label'={$k$}]  (b)};
\end{feynman}
\end{tikzpicture}
= -\frac{\i}{k^{2}+\i 0}\eta_{\mu\nu}\, ,\qquad \qquad
\begin{tikzpicture}[baseline={(current bounding box.center)}]
\begin{feynman}
	\coordinate (x) at (-.7,0);
	\coordinate (y) at (0.5,0);
	\diagram*{(x) -- [fermion, momentum' =$k$] (y)};
		\draw [fill] (x) circle (.0) ;
	\draw [fill] (y) circle (.0)  ;
	\end{feynman}
	\end{tikzpicture}
= \frac{\i \slashed{k}}{k^{2}+\i 0}\, .
\ee
\end{svgraybox}

\begin{svgraybox}
To obtain the colour-ordered vertex rules one inserts into the standard
Feynman rules of sections \ref{chap:1.4} and \ref{Sec:sQCD} the
trace expression \eqref{fasTr} for $f^{abc}$, and
together with the identity \eqref{SUNFierz} reduces everything to a
string of generators. Extracting only a single ordering of the $T^{a}$'s then 
yields the colour-ordered vertices:
\begin{align}
\label{Feynrulveert}
\begin{tikzpicture}[baseline={(current bounding box.center)}]
\begin{feynman}
\coordinate (a) at (-0.5,0);
\coordinate (b) at (0.5,0);
\coordinate (c) at (0,-1.2);
\coordinate (o) at (0,-0.5);
\diagram*{ (o) -- [gluon, momentum'={$p_{3}$}] (a), (o) --  [gluon, momentum'={$p_{1}$}] (b),
(o) --  [gluon, momentum'={$p_{2}$}] (c)};
\draw [fill] (a) circle (.0) node [left] {$\mu_{3}$} ;
\draw [fill] (b) circle (.0) node [right] {$\mu_{1}$};
\draw [fill] (c) circle (.0) node [right] {$\mu_{2}$} ;
\draw [fill] (o) circle (.06);
\end{feynman}
\end{tikzpicture}
&= \i \frac{g}{\sqrt{2}}\Big[ (p_1-p_2)^{\mu_3}\eta^{\mu_1\mu_2}+(p_2-p_3)^{\mu_1}\eta^{\mu_2\mu_3}+(p_3-p_1)^{\mu_2}\eta^{\mu_3\mu_1}\Big]\nn \\
\begin{tikzpicture}[baseline={(current bounding box.center)}]
	\begin{feynman}
	\coordinate (a) at (-0.5,0);
	\coordinate (b) at (0.5,0);
\coordinate (c) at (0.5,-1);
\coordinate (d) at (-0.5,-1);
\coordinate (o) at (0,-0.5);
\diagram*{(a) -- [gluon] (o) -- [gluon] (b), (c) -- [gluon] (o) -- [gluon] (d)};
\draw [fill] (a) circle (.0) node [above] {$\mu_{1}$} ;
\draw [fill] (b) circle (.0) node [above] {$\mu_{2}$} ;
\draw [fill] (c) circle (.0) node [below] {$\mu_{3}$} ;
\draw [fill] (d) circle (.0) node [below] {$\mu_{4}$} ;
\draw [fill] (o) circle (.06);
\end{feynman}
	\end{tikzpicture}
&= \frac{\i}{2}g^{2} \Bigl [ 
2\eta_{\mu_{1}\mu_{3}}\eta_{\mu_{2}\mu_{4}} -\eta_{\mu_{1}\mu_{2}}\eta_{\mu_{3}\mu_{4}} -
\eta_{\mu_{2}\mu_{3}}\eta_{\mu_{4}\mu_{1}}\Big ] \\
\begin{tikzpicture}[baseline={(current bounding box.center)}]
\begin{feynman}
	\coordinate (x) at (-0.7,0);
	\coordinate (o) at (0,0);
	\coordinate (y) at (0.75,0);
		\coordinate (z) at (0,-0.7);
		\diagram*{(x) -- [fermion] (o) -- [fermion] (y), (o) -- [gluon] (z)};
		\draw [fill] (x) circle (.0) ;
	\draw [fill] (y) circle (.0) ;
	\draw [fill] (o) circle (.06);
		\draw [fill] (z) circle (.0)  node [below] {$\mu$} ;
    \end{feynman}
	\end{tikzpicture}
&= \i \frac{g}{\sqrt{2}}\, \gamma^{\mu} \nn
\end{align}
\end{svgraybox}

Similarly, in the scalar QCD sector we have the propagator
\be
\begin{tikzpicture}[baseline={(current bounding box.center)}]
\begin{feynman}
	\coordinate (x) at (-1,0);
	\coordinate (y) at (0.5,0);
	\diagram*{(x) -- [charged scalar, momentum' =$p$] (y)};
		\draw [fill] (x) circle (.0);
	\draw [fill] (y) circle (.0);
	\end{feynman}
	\end{tikzpicture}=\frac{\i}{p^2-m^{2}+\i 0}\,, 
\ee
and the vertices
\be
\label{cOsQCD}
\begin{tikzpicture}[baseline={(current bounding box.center)}]
\begin{feynman}
	\coordinate (x) at (-0.7,0);
	\coordinate (o) at (0,0);
	\coordinate (y) at (0.75,0);
		\coordinate (z) at (0,-0.7);
		\diagram*{(x) -- [charged scalar] (o) -- [charged scalar] (y), (o) -- [gluon] (z)};
		\draw [fill] (x) circle (.0) node [left] {$1$}  ;
	\draw [fill] (y) circle (.0) node [right] {$2$}   ;
	\draw [fill] (o) circle (.06);
		\draw [fill] (z) circle (.0) node [below] {$\mu$} ;
    \end{feynman}
	\end{tikzpicture}=
	\i \frac{g}{\sqrt{2}} \, (p_{2}-p_{1})^{\mu}\, ,\qquad
	\mbox{\begin{tikzpicture}[baseline={(current bounding box.center)}]
	\begin{feynman}
	\coordinate (a) at (-0.5,0);
	\coordinate (b) at (0.5,0);
\coordinate (c) at (0.5,-1);
\coordinate (d) at (-0.5,-1);
\coordinate (o) at (0,-0.5);
\diagram*{(a) -- [gluon] (o) -- [gluon] (b), (d) -- [charged scalar] (o) -- [charged scalar] (c)};
\draw [fill] (a) circle (.0) node [above] {$\mu$};
\draw [fill] (b) circle (.0) node [above] {$\nu$};
\draw [fill] (o) circle (.06);
\end{feynman}
	\end{tikzpicture}}=
	\i  g^{2}  \eta^{\mu\nu}
	\,. 
\ee

\begin{exer}{Colour-ordered Feynman rules}
\label{Ex:1.8}
Derive the form of the colour-ordered four-gluon vertex in eq.~\ref{Feynrulveert} from the Feynman rules of \eqn{V4YM}.
For the solution see \hyperref[Sol:1.8]{chapter~5}.
\end{exer}

\begin{important}{General properties of colour-ordered amplitudes.}
\label{ampprop}
Due to the factorisation of the colour degrees of freedom, the partial or colour-ordered
amplitudes are individually gauge invariant. Next to the Kleiss-Kruif~\eqref{KleissKruif} and Bern-Carrasco-Johansson~\eqref{BCJ} relations, they
obey further general properties which reduce considerably the number of independent structures. We list them below, denoting by $A(1,2,\ldots,n)$ the colour-ordered amplitudes, where the argument $i$ refers to a colour-ordered gluon, while a quark (anti-quark) leg is denoted by $i_{q}$ ($i_{\bar{q}}$).
\begin{itemize}
\item[{\bf 1.}]~Cyclicity: 
\be A(1,2,\ldots, n) = A(2,\ldots, n,1)
\, , 
\ee
which follows from the cyclicity of the trace and the definition of \eqn{I.26}. 

\smallskip
\item[{\bf 2.}]~Parity: 
\be A(\bar 1, \bar 2,\ldots,\bar n)= A(1,2,\ldots, n) \Big|_{\vev{ij}\to \bev{ji}, \bev{ij} \to\vev{ji} }\, . 
\ee
Here the bar over the particle number denotes the inversion of the particle's helicity. 
Note the flip in the helicity spinor brackets under parity.
\item[{\bf 3.}]~Charge conjugation: 
\be A(1_q,2_{\bar q},3,\ldots, n) = -
A(1_{\bar q},2_{q},3,\ldots, n)\ , 
\ee
that is, flipping the helicity of a quark line changes the sign of the amplitude. This descends from 
the colour-ordered gluon-quark-anti-quark vertex above.

\smallskip
\item[{\bf 4.}]~Reflection: \be A(1,2,\ldots, n) = (-1)^n\, A(n,n-1,\ldots,1)\, . 
\ee
This relation follows from the anti-symmetry of the colour-ordered gluon vertices
under reflection of all legs. It also holds in the presence of quark lines but only at tree~level.

\smallskip
\item[{\bf 5.}]~Photon or $\text{U}(1)$ decoupling identity:
\begin{align}
\label{U1decoupling}
\sum_{\sigma\in \mathbb{Z}_{n-1}} A(\sigma_1, \ldots , \sigma_{n-1}, n) =0\, , 
\end{align}
where $\sigma = \{\sigma_1, \ldots , \sigma_{n-1}\}$ are cyclic permutations of $\{1, 2,  \ldots, n-1\}$. 
This powerful identity follows from \eqn{I.26} and the fact that a gluon amplitude with a single photon vanishes since~$f^{0bc}{=}0$. Here $0$ is the colour index of the $\text{U}(1)$ generator $T^{0}=\1/\sqrt{N_c}$.

\smallskip
\item[{\bf 6.}]~We restate the Kleiss-Kuijf relations~\cite{ch1_Kleiss:1988ne} of \eqn{KleissKruif},
\be
\label{KleissKruif2}
A_{n}^{\text{tree}}(1,\{\alpha\}, n, \{\beta\}) = (-1)^{n_{\beta}}
\sum_{\sigma\in \alpha \shuffle \beta^{\top}} A_{n}^{\text{tree}}(1,\sigma, n) \, , 
\ee
which may be derived by the transition from the DDM basis to the trace basis. 
\smallskip
\item[{\bf 7.}]~Finally, there is a final set of relations emerging from the double copy or Bern-Carrasco-Johansson~\cite{ch1_Bern:2008qj} duality between graviton and gluon amplitudes:
\be
\label{BCJid}
\sum_{i=2}^{n-1}p_{1}\cdot (p_{2} + \ldots + p_{i})\, A^{\text{tree}}_{n}(2,\ldots, i,1, i+1,\ldots,n) =0 \, .
\ee
This we will discuss in the later section~\ref{sec:2.5} but quote already for completeness.
\end{itemize}
In summary there are $(n-1)!$ independent colour-ordered gluon amplitudes of multiplicity $n$.
\end{important}

\begin{exer}{Independent gluon partial amplitudes}
\label{Ex:1.9}
Use the above relations amongst the colour-ordered amplitudes to determine the independent set
of colour-ordered amplitudes for four- and five-gluon scattering.
For the solution see \hyperref[Sol:1.9]{chapter~5}.
\end{exer}

\section{Colour-ordered amplitudes}
\label{sec:1.10}

Let us now begin with the actual evaluation of the first pure gluon tree-amplitudes
using the colour-ordered Feynman rules and the spinor-helicity formalism.
In fact, we will see that large classes of gluon
helicity amplitudes vanish!

\subsection{Vanishing tree amplitudes}
\label{sec:mostlyplus}

Let us restrict to tree-level amplitudes with multiplicities $n>3$ here, as there are subtleties for the three-point gluon amplitudes to be discussed later.
Our freedom to choose an arbitrary light-like reference momentum $r_{i}^{\da\a}=\mu_{i}^{\a}\, \tilde\mu_{i}^{\da}$
in the definition of the gluon polarisation vectors $\epsilon_{\pm, i}$ in \eqn{s1polarrel} 
for every leg may be used to show that entire classes of helicity gluon amplitudes vanish.

Using \eqn{s1polarrel} we find the polarisation vector products of legs $i$ and $j$ to be
\begin{align}
\epsilon_{+, i}\cdot\epsilon_{+, j}  & = \phantom{-}
\frac{\vev{\mu_{i}\, \mu_{j}}\, \bev{\lambda_{j}\, \lambda_{i}}}
{\vev{\la_{i}\, \mu_{i}}\, \vev{\lambda_{j}\, \mu_{j}}}\, , \qquad 
\epsilon_{+, i}\cdot\epsilon_{-, j}   = -
\frac{\vev{\mu_{i}\, \lambda_{j}}\, \bev{\mu_{j}\, \lambda_{i}}}
{\vev{\la_{i}\, \mu_{i}}\, \bev{\lambda_{j}\, \mu_{j}}}\, , \nn\\
\epsilon_{-, i}\cdot\epsilon_{-, j}  & = \phantom{-}
\frac{\vev{\lambda_{i}\, \lambda_{j}}\, \bev{\mu_{j}\, \mu_{i}}}
{\bev{\la_{i}\, \mu_{i}}\, \bev{\lambda_{j}\, \mu_{j}}}\, ,
\label{epscontr}
\end{align}
with the only restriction on the reference spinors of leg $i$ being distinct
to the outflowing momentum of that leg, i.e.~$\mu_{i}\neq \lambda_{i}$ and $\tilde\mu_{i}\neq \tilde\lambda_{i}$.  Clearly, if we choose the reference spinors of legs $i$ and $j$
to be identical, we have that
\be
\epsilon_{+, i}\cdot \epsilon_{+, j}= 0 =\epsilon_{-, i}\cdot \epsilon_{-, j}\qquad
\forall\, i,j\, ,
\label{uniformchoice}
\ee
due to $\vev{\mu_{i}\mu_{j}}=0=\bev{\mu_{i}\mu_{j}}$ for that choice. 
Let us see how to use this in order to identify vanishing trees.

An $n$-gluon tree amplitude must depend on the $n$-polarisation vectors involved, which have to be
contracted either with themselves (as $\epsilon_{i}\cdot\epsilon_{j}$) or with the external momenta (as $p_{i}\cdot\epsilon_{j}$). 
Now, what is the minimal number of polarisation vector contractions $\epsilon_{i}\cdot\epsilon_{j}$
arising in the terms that
constitute an $n$-gluon tree-amplitude? To find this number, we need to look at graphs
which maximise the number of momentum-polarisation contractions, i.e.~$p_{i}\cdot\epsilon_{j}$. This implies looking at graphs built entirely out of
three-point vertices. Pure three-point vertex $n$-gluon trees are made of $(n-2)$-vertices.
This follows immediately from the half-ladder form in the DDM basis, cf.~figure~\ref{Fig:halfladder}.

As an $n$-leg graph contains $n$ distinct polarisation vectors, we conclude that any 
$n$-gluon amplitude will consist of terms containing $\emph{at least}$ one polarisation contraction 
$\epsilon_{i}\cdot\epsilon_{j}$ (as this maximum configuration is attained by purely
trivalent trees). Armed with this insight, we now prove the vanishing of
three classes of tree-amplitudes.
\begin{enumerate}[i)]
\item 
Choosing the reference momenta $r_{i}$ of an $n$-gluon tree-amplitude
uniformly as in \eqn{uniformchoice} implies that
\be
A^{\text{tree}}_{n}(1^{+},2^{+},\ldots, n^{+})=0\, ,
\label{I.29}
\ee
as at least one $\epsilon_{+, i}\cdot \epsilon_{+, j}$ contraction must arise. 

\smallskip
\item Similarly, the gluon tree-amplitude
with one flipped helicity state vanishes:
\be
A^{\text{tree}}_{n}(1^{-},2^{+},\ldots, n^{+})=0\, .
\label{I.30}
\ee
This follows from the reference momenta choice
\be
r_{1}=r\neq p_{1}\, \qquad \text{and} \qquad r_{2}=\ldots= r_{n}=p_{1}\, ,
\ee
as then all terms containing a $\epsilon_{+, i}\cdot \epsilon_{+, j}=0$ contraction
with $i,j\in\{2,\ldots ,n\}$ vanish, and $\epsilon_{+, i}\cdot \epsilon_{-, 1}=0$ 
due to \eqn{epscontr} by
the specific choice above.

\smallskip
\item Finally, the $q\bar q g^{n-2}$ amplitudes vanish if all gluons have identical helicity:
\be
A^{\text{tree}}_{n}(1^{-}_{\bar q},2^{+}_{q},3^{+}\ldots, n^{+})=0\, .
\label{I.32}
\ee
Due to the presence of a quark line there is at least one contraction of the form
\be
[2|\slashed{\epsilon}_{+, i}\,|1\rangle = \tla_{2\, \da}\, \epsilon_{+, i}^{\da\a} \, \la_{1\, \a}
= -\sqrt{2}\, \frac{\bev{\lambda_{2}\, \lambda_{i}}\,\vev{\mu_{i}\, \lambda_{1}}}{\vev{\lambda_{i}\mu_{i}}}
\ee
in every term constituting the amplitude. Choosing the gluon-polarisation reference momenta uniformly
as $r_{i}=\mu_{i}\, \tilde\mu_{i} = \lambda_{1}\, \tilde\lambda_{1}$ for all $i\in\{3,\ldots, n\}$ yields
$[2|\slashed{\epsilon}_{+, i}\, |1\rangle = 0$, and hence the vanishing of \eqn{I.32}.
\end{enumerate}

By parity the vanishing of \eqn{I.29} and \eqn{I.30} implies
\be
A^{\text{tree}}_{n}(1^{\pm},2^{-},\ldots, n^{-})=0\, .
\label{I.31}
\ee
Hence, the first non-trivial class of pure gluon tree-amplitudes is the one with two flipped helicities,
$A^{\text{tree}}_{n}(1^{-},2^{+},\ldots, (i-1)^{+},i^{-},(i+1)^{+},\ldots, n^{+})$, known as
\emph{maximally helicity violating} (MHV) amplitudes. 

To understand this name recall that in our
convention all momenta are out-going. The MHV amplitudes describes, for example,
a process in which all incoming gluons have one helicity and all but two outgoing gluons---the maximal
allowed number---have the opposite helicity: flipping the momentum entails a flip in helicities. Hence, helicity is not conserved and this
process is maximally helicity violating.

Similarly, we have for the single-quark-line-gluon amplitude that
\be
A^{\text{tree}}_{n}(1^{-}_{\bar q},2^{+}_{q},3^{-}\ldots, n^{-})=0\, ,
\label{I.33}
\ee
by choosing $q_{i}=\mu_{i}\, \tilde\mu_{i} = \lambda_{2}\, \tilde\lambda_{2}$ for all $i\in\{3,\ldots, n\}$. Alternatively, this may be seen
by a parity and charge conjugation transformation of \eqn{I.32}.

As a matter of fact, the vanishing of these mostly-plus (or minus) amplitudes 
can be understood from a hidden supersymmetry in tree-level quark-gluon tree-amplitudes,
see \cite{ch1_Dixon:1996wi,ch1_Henn:2014yza} for a discussion. 

\begin{figure}[tt]
\  \centering
   \includegraphics[width=8cm]{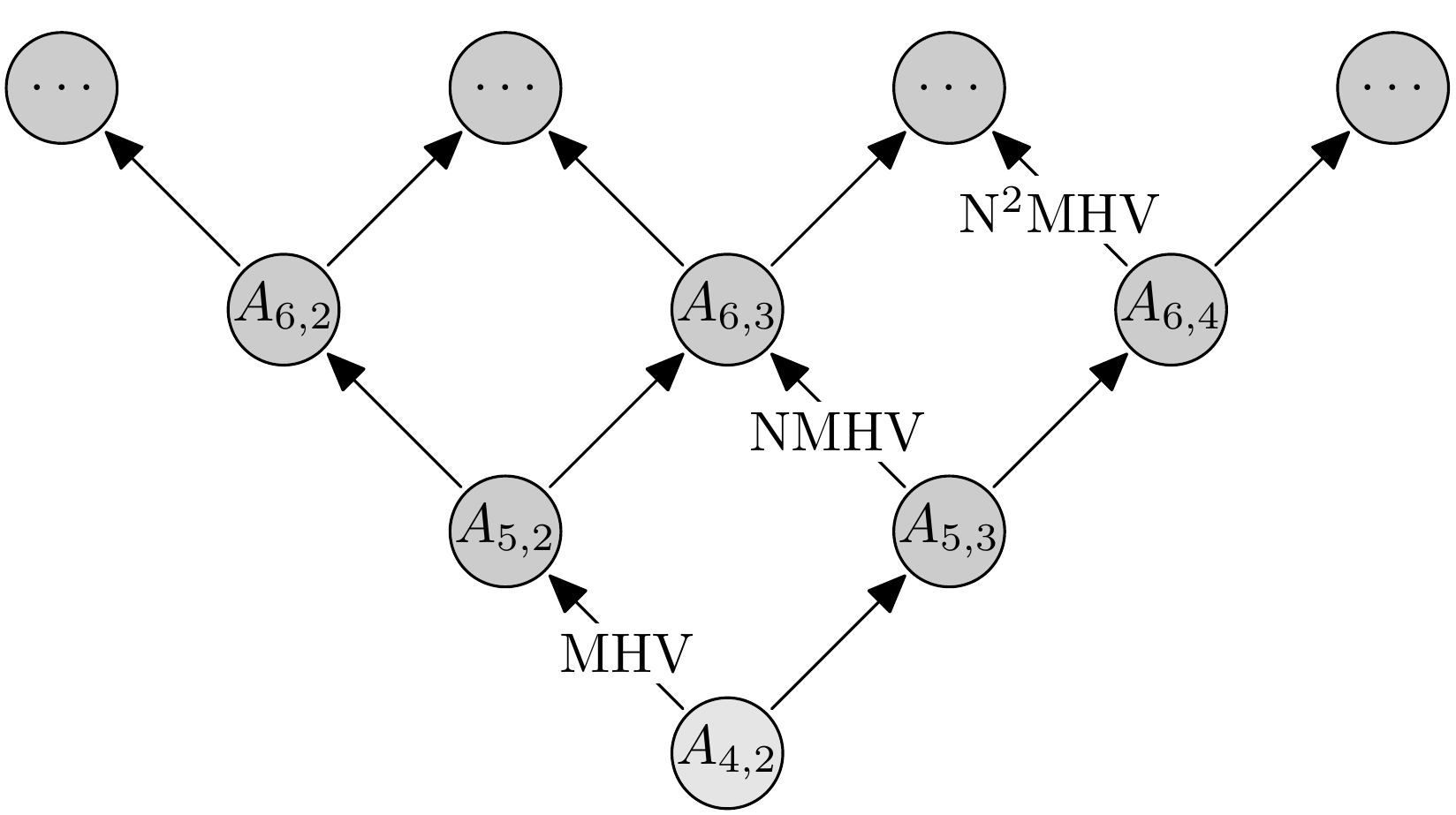}
\caption{The MHV classification of gluon amplitudes: $A_{n,m}$ denotes an $n$-gluon amplitude
with $m$ positive helicity states. Parity acts as a mirror across the vertical axis as
$A_{n,m} \leftrightarrow A_{n,n-m}$. For example the NMHV $A_{5,3}$ amplitude may be obtained
from the parity mirror of the MHV $A_{5,2}$ one. Hence, the lowest multiplicity non-trivial
NMHV amplitude is the $A_{6,3}$.
 } 
\label{Fig:MHV-Pattern}
\end{figure}

Amplitudes comprised of 3 positive helicity and $(n-3)$ negative helicity gluons
are known as \emph{next-to maximally helicity violating} (NMHV) amplitudes and
so on, see figure \ref{Fig:MHV-Pattern}

\subsection{The three-gluon tree-amplitudes}

We now want to establish the smallest amplitudes in gluon scattering. As a matter of
fact, three-point amplitudes of massless particles are very special objects. Due to
kinematics all Mandelstam invariants vanish, $p_{i}\cdot p_{j}=0$ for all $i,j$. This follows from 
momentum conservation
\begin{align}
p_{1}^{\mu}+p_{2}^{\mu}+p_{3}^{\mu}=0 \,,
\end{align}
which, with $p_i^2=0$ for all $i$, implies that 
\be
p_{i}\cdot p_{j}=0 \quad \forall \, i,j\in \{1,2,3\} \, .
\ee
Hence for real momenta this implies the vanishing of the spinor brackets
$\vev{ij}$ and $\bev{ij}$, which are the building blocks of the amplitudes for massless
particles. There is thus no Lorentz invariant object one could write down, and therefore
the three-particle amplitude must vanish.
The situation is different if one allows for {\it complex momenta} $p_{i} \in {\mathbb{C}^{4}}$.
In this case the helicity spinors
 $\lambda_{i}$ and $\tilde{\lambda}_{i}$ are independent, and the conditions $p_{i} \cdot p_{j} =0 $
can be solved either by $\bev{i j} = 0$
\emph{or} by $\vev{i j} =0$.
Hence either $\tilde\lambda_{1}^{\alpha} \propto \tilde\lambda_{2}^{\alpha} \propto 
\tilde\lambda_{3}^{\alpha}$ (collinear right-handed spinors) or
$\lambda_{1}^{\alpha} \propto \lambda_{2}^{\alpha} \propto 
\lambda_{3}^{\alpha}$ (collinear left-handed spinors) solve the constraints
$p_{i}\cdot p_{j}=0$.
The two choices correspond to the three-gluon MHV${}_{3}$ 
amplitude
\begin{eqnarray}\label{3pointmhv}
A_{3}^{\text{tree}}(1^{-},2^{-},3^{+}) = \i g\frac{ \vev{12}^3 }{\vev{23}\vev{31}}\,,\qquad \lbrack 12\rbrack=\lbrack2 3\rbrack = \lbrack 3 1\rbrack = 0\,, 
\end{eqnarray}
and the dual $\widebar{\text{MHV}}_{3}$ amplitude
\begin{eqnarray} \label{3pointmhvbar}
A_{3}^{\text{tree}}(1^{+},2^{+},3^{-}) = - \i g\frac{ \bev{12}^3 }{\bev{23}\bev{31}}\,,\qquad \vev{1 2}= \vev{2 3} = \vev{3 1} = 0\,, 
\end{eqnarray}
respectively. The two are related by a parity transformation, which flips the helicity weights and exchanges $\spA ij \leftrightarrow [ji]$.

As a useful exercise in spinor gymnastics, we will now 
derive these amplitudes from the colour-ordered Feynman rules.
Using the three-gluon vertex in \eqn{Feynrulveert} we~find  
\begin{align}
   \hspace{-0.2cm} A_{3}^\text{tree}(1^-,2^-,3^+) 
   \!=\!
   \i\frac{g}{\sqrt{2}}\Big[ & (p_1{-}p_2)\!\cdot\!\eps_{+,3} \,  \eps_{-,1}\!\cdot\!\eps_{-,2} 
   \nn\\ & +(p_2{-}p_3)\!\cdot\!\eps_{-,1} \, \eps_{-,2}\!\cdot\!\eps_{+,3} + (p_3{-}p_1)\!\cdot\!
   \eps_{-,2} \,  \eps_{+,3}\!\cdot\!\eps_{-,1}\Big] , 
\end{align}
where the polarisation vector contractions 
are given in \eqn{epscontr}. Choosing the same reference-momentum spinor $\mu_{1}=\mu_{2}=\mu_{3}=\mu$ for the gluons we have $\eps_{-,1} {\cdot} \eps_{-,2} =0$. Then, by using momentum conservation and transversality $p_i\cdot \eps_i=0$, 
we arrive at
\begin{align}
   \hspace{-0.3cm} A_{3}^{\text{tree}}(1^-, 2^-, 3^+) = \i g \sqrt{2} 
   \Big[ p_2\!\cdot\!\eps_{-,1} \, \eps_{-,2}\!\cdot\!\eps_{+,3} -p_1\!\cdot\!\eps_{-,2}\,  \eps_{+,3}\!\cdot\!\eps_{-,1}\Big]\, . 
\end{align}
From \eqn{epscontr} we derive the following expressions,
\begin{align}
\begin{split}
   \eps_{-,2}\!\cdot\!\eps_{+,3}&= -\frac{\lan 2 \mu\ran[3\mu ] }{[ 2 \mu] \lan 3 \mu\ran} \, , \qquad \qquad\, 
   \eps_{-,1}\!\cdot\!\eps_{+,3}= -\frac{\lan 1 \mu\ran[3\mu ] }{[ 1 \mu] \lan 3 \mu\ran} \, , 
   \\
     p_2\cdot \eps_{-,1}&= \frac{1}{\sqrt{2}}\frac{\lan 12 \ran[2 \mu]}{[1 \mu]}\, , \qquad \quad\ \ \,
     p_1\cdot \eps_{-,2}= -\frac{1}{\sqrt{2}}\frac{\lan 12 \ran[1 \mu]}{[2 \mu]}\, ,
     \end{split}
\end{align}
and therefore
\begin{align}
\begin{split}
    A_{3}^{\text{tree}}(1^-, 2^-, 3^+) &= -\i g \lan 12\ran \frac{[3\mu]}{\lan 3\mu\ran}
   \left(  \frac{\lan 2\mu\ran}{[1\mu]} +\frac{\lan 1\mu\ran}{[2\mu]}  \right) \\ & = 
    \i g \lan 12\ran \frac{[3\mu]}{\lan 3\mu\ran}
   \frac{\lan \mu | \slashed{p}_1 + \slashed{p}_2 |\mu] 
  }{[1\mu][2\mu]}
  =
  \i g \frac{\lan 12\ran [3\mu]^2 }{[1\mu][2\mu]}
   \, . 
   \end{split}
\end{align}
Finally, we use three-point momentum conservation to simplify
\begin{align}
\frac{[3\mu] }{[1\mu]}=\frac{\lan 23\ran [3\mu] }{\lan 23\ran[1\mu]}=\frac{\lan12\ran}{\lan23\ran}\, , \qquad \quad 
\frac{[3\mu] }{[2\mu]}=\frac{\lan 13\ran [3\mu] }{\lan 13\ran[2\mu]}=\frac{\lan12\ran}{\lan31\ran}\, ,
    \end{align}
    thus arriving at the result in \eqn{3pointmhv}. One could repeat this calculation for the scattering of three gravitons, this time using the three-graviton vertex of \eqn{3gravitonvertex}, arriving at a result proportional to  $\big[ A_{3}^{\text{tree}}(1^{-}, 2^{-}, 3^{+})\big]^2$. The involved expression for the vertex \eqn{3gravitonvertex} 
    gives no hints of such a remarkable squaring relation! 
    We shall take up this discussion in chapter~\ref{ch:trees} again.
    
\subsection{Helicity weight}
\label{sec:helscale}

There is an important consistency requirement for scattering amplitudes based on
checking their correct helicity weights.
This  is encoded in the following relation~\cite{ch1_Witten:2003nn}:
\begin{align}
\label{lg}
   \hat h_{i}\, A=  -\frac{1}{2} \left( \lambda_i^\alpha \ \frac{\partial}{\partial \lambda_i^\alpha} - \tilde{\lambda}_i^{\dot{\alpha}}\frac{\partial}{\partial \tilde{\lambda}_i^{\dot{\alpha}}}\right) \, A \, = \, h_i \, A\, , 
\end{align}
where $h_i$ is the helicity of particle $i$. 
When combined with Lorentz invariance, this relation can be used to determine the functional form of the three-point amplitudes of particles of any spin. As we argued, above a massless
three-particle amplitude in complexified momentum space can either only depend on $\langle i\, j\rangle$ with $[ i\, j] {=}0$ for all particles or vice versa.
If we choose the MHV situation $[ i\, j] {=}0$ for the helicity assignment $1^{-s}, 2^{-s}, 3^{+s}$, one can immediately see, using \eqref{lg}, that the answer must have the form  
\begin{align}
\label{generalsMHV}
        A(1^{-s}, 2^{-s}, 3^{+s}) \sim  \big[ A(1^{-}, 2^{-}, 3^{+})\big]^s\, .
\end{align}
In fact, for the gluon-amplitude with $s{=}1$, the conditions
\be
\hat h_{1} A_{3}^{\text{MHV}}= - A_{3}^{\text{MHV}}\, , \quad
\hat h_{2} A_{3}^{\text{MHV}}= - A_{3}^{\text{MHV}}\, , \quad
\hat h_{3} A_{3}^{\text{MHV}}= + A_{3}^{\text{MHV}}\,  \quad
\ee
uniquely fix the amplitude to take the form
 $A(1^-, 2^-, 3^+) \!\sim\!{\langle 1\, 2\rangle^3}/ ({\langle 2\, 3\rangle\langle 3\, 1\rangle})$.
This may be seen as an independent derivation of both the 3-point gluon and graviton
amplitudes---without referring to any Lagrangian!

\begin{exer}{The $\widebar{\text{MHV}}_{3}$ amplitude}
\label{Ex:MHV3}
Derive the anti-MHV three-gluon amplitude~\eqref{3pointmhvbar} using the colour-ordered Feynman rules.
For the solution see \hyperref[Sol:MHV3]{chapter~5}.
\end{exer}

\begin{example}{Example: A four-gluon tree-amplitude}

We shall now compute the simplest non-trivial tree-level colour-ordered gluon amplitude, namely the 
four-gluon MHV-amplitude with a split helicity distribution $A_{4}^{\text{tree}}(1^{-},2^{-},3^{+},4^{+})$.
Employing the colour-ordered Feynman rules of \eqn{Feynrulveert} we see
that the three diagrams of figure~\ref{fig:4gluon} contribute to the amplitude.
\begin{figure}[t]
   \centering
    \begin{equation*}
    A_{4}^{\text{tree}}(1^-,2^-,3^+,4^+)=
 \begin{tikzpicture}[baseline={(current bounding box.center)}]
  \begin{feynman}
    \coordinate (t) at (-0.8,0.8);
    \coordinate (d) at (-0.8,-0.8);
        \coordinate (a) at (-0.4,0);
                \coordinate (b) at (0.4,0);
           \coordinate (ot) at (0.8,0.8);
                      \coordinate (od) at (0.8,-0.8);
                       \coordinate (T) at (0,-1); \draw (T) node [below] {(I)};
         \draw [gluon] (t) -- (a);
          \draw [gluon] (a) -- (d);
         \draw [gluon] (ot) -- (b);
          \draw [gluon] (a) -- (b);
          \draw [gluon] (b) -- (od);          
               \draw [fill] (a) circle (.08); \draw [fill] (b) circle (.08);
    \draw (ot) node [right] {$3^{+}$};
          \draw (t) node [left] {$2^{-}$};
               \draw (d) node [left] {$1^{-}$};
                      \draw (od) node [right] {$4^{+}$};
                      \end{feynman}
      \end{tikzpicture}
 +    
  \begin{tikzpicture}[baseline={(current bounding box.center)}]
  \begin{feynman}
    \coordinate (t) at (-0.8,0.8);
    \coordinate (d) at (-0.8,-0.8);
        \coordinate (a) at (0,0.4);
                \coordinate (b) at (0,-0.4);
           \coordinate (ot) at (0.8,0.8);
                      \coordinate (od) at (0.8,-0.8);
                       \coordinate (T) at (0,-1); \draw (T) node [below] {(II)};
         \draw [gluon] (t) -- (a);
          \draw [gluon] (b) -- (d);
         \draw [gluon] (ot) -- (a);
          \draw [gluon] (a) -- (b);
          \draw [gluon] (b) -- (od);          
               \draw [fill] (a) circle (.08); \draw [fill] (b) circle (.08);
    \draw (ot) node [right] {$3^{+}$};
          \draw (t) node [left] {$2^{-}$};
               \draw (d) node [left] {$1^{-}$};
                      \draw (od) node [right] {$4^{+}$};
                      \end{feynman}
      \end{tikzpicture}
+
  \begin{tikzpicture}[baseline={(current bounding box.center)}]
  \begin{feynman}
    \coordinate (t) at (-0.8,0.8);
    \coordinate (d) at (-0.8,-0.8);
        \coordinate (a) at (0,0);
           \coordinate (ot) at (0.8,0.8);
                      \coordinate (od) at (0.8,-0.8);
                       \coordinate (T) at (0,-1); \draw (T) node [below] {(III)};
         \draw [gluon] (t) -- (a);
          \draw [gluon] (a) -- (d);
         \draw [gluon] (ot) -- (a);
          \draw [gluon] (a) -- (od);          
               \draw [fill] (a) circle (.08);
     \draw (ot) node [right] {$3^{+}$};
          \draw (t) node [left] {$2^{-}$};
               \draw (d) node [left] {$1^{-}$};
                      \draw (od) node [right] {$4^{+}$};
                      \end{feynman}
      \end{tikzpicture}
   \end{equation*}
   \caption{The three colour-ordered graphs contributing to the four-gluon split helicity
    MHV amplitude.}
   \label{fig:4gluon}
\end{figure}
Its computation is again considerably simplified by a clever choice of the reference momenta $r_i=\mu_{i}\,
\tilde\mu_{i}$ of the gluon polarisations,
\be
r_{1}=r_{2}=p_{4}\, , \qquad r_{3}=r_{4}=p_{1} \, ,
\label{polchoice}
\ee
where $p_{i}$ denote the physical external momenta. Then one sees, using~\eqn{epscontr}, that the polarisation-vector products
\be
\epsilon_{-, 1}\cdot\epsilon_{-, 2}   = \epsilon_{+, 3}\cdot\epsilon_{+, 4} =
\epsilon_{-, 1}\cdot\epsilon_{+, 4}  =  \epsilon_{-, 2}\cdot\epsilon_{+, 4} =
\epsilon_{-, 1}\cdot\epsilon_{+, 3}  = 0 
\label{eezero}
\ee
all vanish, and that the \emph{only} non-vanishing contraction is 
\be
\label{e2e3}
\epsilon_{-, 2}\cdot\epsilon_{+, 3}   =  - \frac{\vev{\mu_{3}\,\lambda_{2}} \, \bev{\mu_{2}\, \lambda_{3}}}
{\vev{\la_{3}\, \mu_{3}}\, \bev{\la_{2}\, \mu_{2}}} = - \frac{\vev{12}\, \bev{34}}{\vev{13}\,\bev{24}}\, .
\ee
Of course it is mandatory to use the same choice of reference momenta for all~graphs. 

\newpage

\begin{trailer}{Diagram I:}
\begin{align}
\begin{aligned}
(\text{I}) = \left(\frac{\i g}{\sqrt{2}}\right )^{2}\, \frac{-\i\,\eta_{\mu\nu}}{s_{12}}\, & \Bigl [
\epsilon_{-,2}^{\mu} \, (p_{2q}\cdot \epsilon_{-,1}) + \epsilon_{-,1}^{\mu} \, (p_{q1}\cdot \epsilon_{-,2})\, 
\Bigr ] \\
\times \Bigl [
\epsilon_{+,4}^{\nu} \, & (p_{4(-q)}\cdot \epsilon_{+,3}) + \epsilon_{+,3}^{\nu} \, (p_{(-q)3}\cdot \epsilon_{+,4})\, 
\Bigr ] \, ,
\end{aligned}
\end{align}
with $q=-p_{1}-p_{2}=p_{3}+p_{4}$, $s_{12}=(p_{1}+p_{2})^{2}=\vev{12}\bev{21}$ and $p_{ij}=p_{i}-p_{j}$. 
Due to only $\epsilon_{-,2}\cdot\epsilon_{+,3}$ surviving in the contraction, we have
\begin{align} \begin{aligned}
(\text{I}) & = \frac{\i g^{2}}{2s_{12}}\, (\epsilon_{-,2}\cdot\epsilon_{+,3})\, (p_{2}+p_{1}+p_{2})\cdot\epsilon_{-,1}\,
(-p_{3}-p_{4}-p_{3})\cdot\epsilon_{+,4} \\ 
 & = -\frac{2\i g^{2}}{s_{12}}\,(\epsilon_{-, 2}\cdot\epsilon_{+, 3})\,(p_{2}\cdot\epsilon_{-, 1})\, 
 (p_{3}\cdot\epsilon_{+, 4}) \\ 
 & = - \frac{2\i g^{2}}{s_{12}}\, \left ( - \frac{\vev{12}\, \bev{34}}{\vev{13}\, \bev{24}}\, \right )\,
  \left ( \frac{1}{\sqrt{2}}\, \frac{\vev{12}\, \bev{24}}{\bev{14}}\, \right )\,
   \left ( \frac{1}{\sqrt{2}}\, \frac{\vev{13}\, \bev{34}}{\vev{14}}\, \right ) \\
& = \i g^{2}\, \frac{\vev{12}\, \bev{34}^{2}}{\bev{12}\, \vev{14}\, \bev{41}}\, ,
\end{aligned} \end{align} 
where in the second line we used  $\displaystyle p_{2}\cdot \epsilon_{-,1}= \ft{1}{\sqrt{2}}\, \ft{\vev{12}\, 
\bev{2\,\mu_{1}}}{\bev{1\mu_{1}}} \stackrel{\mu_{1}=4}{=} \ft{1}{\sqrt{2}}\, \ft{\vev{12}\, 
\bev{24}}{\bev{14}}$, and similarly for $p_{3}\cdot \epsilon_{+,4}$. The last expression can be rewritten
exclusively in terms of $\lambda_{i}$ spinors~as
\begin{align} \begin{aligned}
-\ft{\i}{g^{2}}\, (\text{I})&=
\frac{\vev{12}\, \bev{34}^{2}}{\bev{12}\, \vev{14}\, \bev{41}} \frac{\vev{43}}{\vev{43}} =
\frac{\vev{12}\, \bev{34}}{\bev{12}\, \vev{14}}\,  \frac{ \overbrace{ \bev{34}\, \vev{43}}^{\vev{12}\,\bev{21}}}
{ \underbrace{\vev{43}\, \bev{41}}_{\hspace{-0.5cm} -\vev{34}\,\bev{41}=\vev{32}\,\bev{21} \hspace{-0.5cm}}} =
\frac{-\vev{12}^{2}}{\vev{23}\, \vev{41}}\, \underbrace{\frac{\bev{34}}{\bev{21}}}_{\frac{\vev{12}}{\vev{43}}} \\ 
& = \frac{\vev{12}^{3}}{\vev{23}\, \vev{34}\, \vev{41}} \, .
\end{aligned} \end{align}
Here in the first step the four-point kinematical relation $s_{12}=\vev{12}\,\bev{21}=s_{34}=\vev{34}\,\bev{43}$
was used.
\end{trailer}

\newpage 

\begin{trailer}{Diagram II:}
Using again the fact that the only non-vanishing contraction is that of \eqn{e2e3} we
find a vanishing result:
\begin{align}
(\text{II})  & \propto \frac{-\i}{2s_{14}}\, 
(\epsilon_{2}\cdot\epsilon_{3})\, \Bigl [\, (p_{23}\cdot\epsilon_{1})\, (p_{1(-q)}\cdot\epsilon_{4})
+ (p_{23}\cdot\epsilon_{4})\, (p_{(-q)4}\cdot\epsilon_{1}) \, \Bigr ] =0\, .
\end{align}
Here with $q=p_{1}+p_{4}$ we have $p_{1(-q)}\cdot\epsilon_{4}=2\, p_{1}\cdot\epsilon_{4}=0$, which vanishes
by virtue of our choice $q_{4}=p_{1}$ of \eqn{polchoice}. Similarly, 
$p_{(-q)4}\cdot\epsilon=-2\, p_{4}\cdot\epsilon_1 =0$ as $q_{1}=p_{4}$. 
Hence diagram II gives no contribution. 
\end{trailer}

\begin{trailer}{Diagram III:}
The same holds true for the third graph as
\begin{align}
(\text{III})\propto 2\, (\epsilon_{2}\cdot\epsilon_{4})\, (\epsilon_{1}\cdot\epsilon_{3})
- (\epsilon_{2}\cdot\epsilon_{3})\, (\epsilon_{1}\cdot\epsilon_{4})-
 (\epsilon_{2}\cdot\epsilon_{1})\, (\epsilon_{3}\cdot\epsilon_{4}) = 0 \, ,
\end{align}
where at least one of the vanishing contractions in~\eqn{eezero} appears in each term.
\end{trailer}

In summary, we have established the following compact result for the split-helicity MHV four-point amplitude:
\be
A_{4}^{\text{tree}}(1^{-},2^{-},3^{+},4^{+}) = \i g^{2}\frac{\vev{12}^{4}}{\vev{12}\,\vev{23}\, \vev{34}\, \vev{41}} \, .
\label{MHV4g1}
\ee
At this point it is also instructive to check the helicity weights of our final result 
for every leg using the helicity generator of \eqn{lg}. To wit,
$$
\hat{h}_{1}\,A_{4}=\ft 1 2\,(-4+2)\, A_{4}= -A_{4}\, , \quad \ \hat{h}_{2}\, A_{4}=-A_{4} \, ,\quad \ h_{3}\, A_{4}=+A_{4}\, ,\quad
\ \hat{h}_{4}\, A_{4}=+A_{4}\, ,
\label{helicitycheck}
$$
so everything is in order.

\smallskip

By cyclicity there is only one more independent four-gluon amplitude left to compute: the 
case $A_{4}^{\text{tree}}(1^{-},2^{+},3^{-},4^{+})$ with an alternating
helicity distribution. All other possible helicity distributions can be
related to this or  $A_{4}^{\text{tree}}(1^{-},2^{-},3^{+},4^{+})$ of \eqn{MHV4g1} by cyclicity. 
It turns out that we do not need to do another Feynman diagrammatic
computation, as the missing amplitude follows from the $\text{U}(1)$ decoupling theorem of 
\eqn{U1decoupling}:
\begin{align}
\begin{aligned}
A_{4}^{\text{tree}}(1^{-},2^{+},3^{-},4^{+}) & = - A_{4}^{\text{tree}}(1^{-},2^{+},4^{+},3^{-}) - A_{4}^{\text{tree}}(1^{-},4^{+},2^{+},3^{-})  \\
& = -\i g^{2}\left ( \frac{\vev{31}^{4}}{\vev{12}\,\vev{24}\,\vev{43}\, \vev{31}} + 
 \frac{\vev{31}^{4}}{\vev{14}\,\vev{42}\,\vev{23}\, \vev{31}} \right ) \\
 & = \i g^{2} \frac{\vev{31}^{4}}{\vev{12}\,\vev{23}\,\vev{34}\, \vev{41}} \, .
\end{aligned} \end{align}
Comparing this result to \eqn{MHV4g1} we may express all four-gluon MHV amplitudes in a single
and crafty formula,
\be
A^{\text{tree}}_{4}(\ldots i^{-}\ldots j^{-}\ldots) = \i g^{2} 
\, \frac{\vev{ij}^{4}}{\vev{12}\,\vev{23}\,\vev{34}\, \vev{41}}
\, ,
\label{I.35}
\ee
where the dots stand for positive-helicity gluon states.

In fact we shall show in chapter~\ref{ch:trees} that this formula has a straightforward generalisation to the $n$-gluon MHV case, in which only the denominator is modified~to
\be
A^{\text{tree}}_{n}(\ldots i^{-}\ldots j^{-}\ldots) =  \i g^{n-2} 
\, \frac{\vev{ij}^{4}}{\vev{12}\,\vev{23}\ldots\,\vev{(n-1)n}\, \vev{n1}}
\, ,
\label{ParkeTaylor1}
\ee
known as the Parke-Taylor amplitude~\cite{ch1_Parke:1986gb}. It is a remarkably simple and closed
expression for a tree-level amplitude with an arbitrary number $n$ of gluons.
We shall derive it in the next chapter.

By parity this results implies the $n$-gluon $\widebar{\text{MHV}}$ amplitudes 
\be
A^{\text{tree}}_{n}(\ldots i^{+}\ldots j^{+}\ldots) = (-1)^{n} \i g^{n-2} 
\, \frac{\bev{ij}^{4}}{\bev{12}\,\bev{23}\ldots\,\bev{(n-1)n}\, \bev{n1}}
\, ,
\label{ParkeTaylor1MHVbar}
\ee
in agreement with the result of exercise~\ref{Ex:MHV3} for the $n=3$ case.

\end{example}

\begin{exer}{Four-point quark-gluon scattering}
\label{Ex:1.11}
Show, by using the colour-ordered Feynman rules and a suitable choice for the gluon-polarisation 
reference vector
$r_{3}^{\a\da}=\mu_{3}^{\a}\, \tilde\mu_{3}^{\da}$,
that the first non-trivial $\bar q q g g$ scattering amplitude is given by
\be
\label{qqgg-amp}
A_{\bar q q  gg}^{\text{tree}}(1^{-}_{\bar q}, 2^{+}_{q}, 3^{-},4^{+}) = -\i g^{2}
\frac{\vev{13}^{3}\, \vev{23}}
{\vev{12}\,\vev{23}\,\vev{34}\, \vev{41}}\, .
\ee
Also convince yourself that the result for this amplitude has the correct helicity assignments, as we did in
\eqn{helicitycheck} for the pure-gluon case.
For the solution see \hyperref[Sol:1.11]{chapter~5}.
\end{exer}

\subsection{Vanishing graviton tree-amplitudes}

The observation of section~\ref{sec:mostlyplus} that the mostly-plus gluon amplitudes
vanish, $A_{n}(1^{+},\ldots, n^{+})=0=A_{n}(1^{-},2^{+},\ldots, n^{+})$, carries over
to the graviton case as well:
\be
\label{Mmostlyplus}
M_{n}(1^{++},\ldots, n^{++})=0=M_{n}(1^{--},2^{++},\ldots, n^{++}) \,.
\ee
The proof is completely analogous. Recall the representation of the graviton
polarisation tensor
as a product of two spin-1 polarisations in \eqn{gravitonpolarize}:
\be
\epsilon_{\mu\nu, \, i}^{\pm\pm}=\epsilon_{\mu, \, i}^{\pm}\, \epsilon_{\nu, \, i}^{\pm}
\qquad i=1,\ldots , n \, .
\ee
Again suitable choices for the reference momenta $r_{i}$ allow us to set to zero all possible $\epsilon^{\mu\nu}_{i} \epsilon_{j, \, \mu\nu}$
contractions 
for the two mostly-plus helicity configurations of \eqn{Mmostlyplus}.
\smallskip
\begin{enumerate}[i)]
\item $M_{n}(1^{++},\ldots, n^{++})$: 
Here we choose all reference momenta uniformly $r_{i}=r$ $\forall i$ with $r\neq p_{i}$
$\forall i\in \{1,\ldots, n\}$. This implies that all contractions
\be
\epsilon^{\mu\nu\, ++}_{i} \epsilon_{j, \, \mu\nu}^{++}=(\epsilon^{+}_{i}\cdot 
\epsilon^{+}_{j})^{2}=0
\ee
vanish by virtue of \eqref{s1polarrel}.

\smallskip
\item $M_{n}(1^{--},2^{++},\ldots, n^{++})$: 
Here we choose $r_{1}\neq p_{1}$ arbitrarily, and the remaining $r_{i}=p_{1}$ uniformly $\forall 
i\in \{2,\ldots, n\}$. This entails $\epsilon^{-}_{1}\cdot \epsilon^{+}_{i}=0$ as well as
$\epsilon^{+}_{i}\cdot \epsilon^{+}_{j}=0$, and the analogue for the graviton polarisation
contractions,
\be
\epsilon^{--}_{1}\cdot \epsilon^{++}_{i}=0=\epsilon^{++}_{i}\cdot \epsilon^{++}_{j}=0\, ,
\qquad \forall 
i,j\in \{2,\ldots, n\}\,.
\ee 
\end{enumerate}
In order to show the vanishing of these amplitudes we need to ensure that in every term comprising a
mostly-plus graviton tree-level amplitude there is at least one polarisation-tensor contraction
$\epsilon^{\mu\nu}_{i} \epsilon_{j, \, \mu\nu}$. This is easy to show: as we saw in the discussion
of section~\ref{sec:1.6}, \emph{all} multi-graviton vertices scale homogeneously quadratically in the momenta
due to $\sqrt{g} \, R \sim \sum_{n} \partial^{2} h^{n}$. Therefore, the maximum number of momenta in the numerator
of an $n$-graviton tree-level scattering amplitude arises from pure three-graviton vertex graphs.
These graphs contain $n-2$ three-vertices, hence the counting for these yields a total of $2(n-1)$ momenta
and $n$ graviton polarisation tensors, each one splitting up into a product
of two polarisation vectors.
Looking at the possible contractions of this set of $2(n-1)$ momentum vectors and
$2n$ polarisation vectors we see that at least one contraction of the type
$\epsilon_{i}\cdot\epsilon_{j}$ must happen. As these may be set to zero through
 a suitable choice of
 reference momenta, this proves the vanishing graviton trees of
 \eqn{Mmostlyplus}.

\chapter{On-shell techniques for tree-level amplitudes}
\label{ch:trees}

\abstract{
In this chapter we focus on the pole structure of tree-level amplitudes. We argue
that amplitudes factorise on these poles into lower-point amplitudes. Moreover, universal
factorisation structures emerge when two momenta become collinear as well as in the 
limit of low energy of a single particle---the soft limit. These factorisation properties
are the basis of an efficient technique for computing 
tree-level scattering amplitudes in gauge theories and gravity recursively---without
ever referring to Feynman rules or even a Lagrangian.
These recursion relations use as input lower-point amplitudes, so that 
the gauge redundancy, which is partly responsible for the complexity
of conventional Feynman graph calculations, is absent in this entirely on-shell based
formalism.
We then show the invariance of scattering amplitudes under Poincar\'e transformations, 
and introduce the conformal symmetry of gauge-theory tree-level amplitudes.
Finally, we highlight a surprising double-copy relation between gluon and graviton amplitudes.
}

\section{Factorisation properties of tree-level amplitudes}
\label{sec:2.1}

\begin{figure}[t]
   \centering
  \begin{tikzpicture}[baseline={(current bounding box.center)}, scale =1.0]
  \coordinate (i) at (70:2);
  \coordinate (n-4) at (-30:2.3);
  \coordinate (n-3) at (-18:2.5);
  \coordinate (n-2) at (-5:2.6);
  \coordinate (i) at (30:3);
  \coordinate (i+1) at (10:3.25);
  \coordinate (n) at (-35:3);
  \coordinate (1) at (-120:1.9);
  \coordinate (2) at (-150:3);
  \coordinate (3) at (-170:2.5);
  \coordinate (4) at (-190:2.5);
  \coordinate (5) at (-180:2.5);
  \coordinate (i-1) at (-210:3);
  \coordinate (x) at (-1.5,0);
  \coordinate (y) at (1.5,0);
  \draw [dashed, line width=0.5pt, color = darkgray] (0,1.2) to (0,-0.2); 
  \draw [dashed, line width=0.5pt, color = darkgray] (0,-0.9) to (0,-1.2); 
  \draw [solid, line width=2pt] (y) to (x);
  \draw [->, line width=1pt] (0.5, -0.3) to (-0.5, -0.3);
  \draw (0,-0.6) node {$P_{ij}$};
  \draw [solid, line width=2pt] (x) -- (1);
  \draw [solid, line width=2pt] (x) -- (2);
  \draw [solid, line width=2pt] (y) -- (i+1); 
  \draw [solid, line width=2pt] (y) -- (n); 
  \draw [solid, line width=2pt] (y) -- (i); 
  \draw [solid, line width=2pt] (x) -- (i-1); 
  \draw [loosely dotted, line width=1.5pt] (-160:2.5) to[bend left=30] ((-200:2.5);;
  \draw [loosely dotted, line width=1.5pt] (5:2.75) to[bend left=30] (-25:2.5);
  \draw (1) node [below] {$p_{i}$};
  \draw (2) node [left] {$p_{i+1}$};
  \draw (n) node [below] {$p_{i-1}$};
  \draw (i+1) node [right] {$p_{j+2}$};
  \draw (i) node [above] {$p_{j+1}$};
  \draw (i-1) node [above] {$p_{j}$};
  \draw (-0.9,0.3) node [right] {$-h$};            
  \draw (0.8,0.3) node [left] {$h$};
  \draw [fill=lightgray,thick, draw=black] (x) circle (0.7);
  \draw [fill=lightgray,thick, draw=black] (y) circle (0.7);
  \draw (x) node {\Large $A_L$};
  \draw (y) node {\Large $A_R$};                       
\end{tikzpicture}
   \caption{The factorisation of a colour-ordered amplitude on the multi-particle
   pole $P_{ij}=p_{i}+p_{i+1} + \ldots + p_{j}$ when $P_{ij}^{2}\to 0$. Here $h$ denotes the
   helicity of the particles crossing the pole.}
   \label{fig:factorization}
\end{figure}
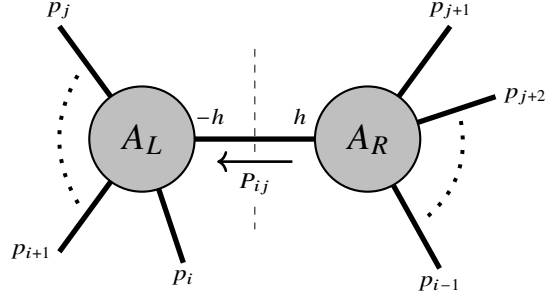

Important insights and constraints on tree-level scattering amplitudes may be gained by thinking about them
as analytic functions of the external momenta. 
In this section we will restrict ourselves to the case of massless particles.
As we already argued in chapter \ref{ch:intro} with figure~\ref{Fig:regionmomenta}, 
tree-level amplitudes have simple poles in multi-particle channels. 
This can be seen from the Feynman diagrammatic expansion. Take all diagrams which have a propagator $\sim 1 / P_{ij}^2$, where  $P_{ij}=p_{i}+\ldots +p_{j}$ is a sum of external momenta (which will be adjacent for colour-ordered amplitudes, or an arbitrary subset in gravity).
We call $P_{ij}$ the \emph{region momentum}, as it is the total momentum associated to the
 region of momenta $\{p_{i},\ldots, p_{j}\}$. %
As $P_{ij}$ goes on shell, $P_{ij}^{2}\to 0$, these singular diagrams
will collect into a product of two on-shell amplitudes, to the left and right hand
sides of the divergent propagator, in a mechanism known as
\emph{factorisation}. This process is illustrated in
figure~\ref{fig:factorization}. We can understand the details of the procedure
by studying the Feynman rules for the propagator that goes on-shell directly, each of which has
the generic form
\be
\frac{\i \, N(P_{ij})}{P_{ij}^2} \,,
\ee
where the numerator $N(P_{ij})$ depends on the type of particle.
For example, for the specific case of gluons in the axial gauge ($n^\mu A_\mu=0$) 
we would have
$N(P)\to N_g^{\mu\nu}(P,n) = -\eta^{\mu\nu}+(P^\mu n^\nu+P^\nu
n^\mu)/(P\cdot n)$. In the limit $P_{ij}^{2}\to 0$ the numerator $N$ can be
rewritten in terms of the spin sum over external polarisation vectors or
wave-functions. Again for the case of gluons, following the results of exercise~\ref{Ex:1.7}, we have that
\be
N_g^{\mu\nu}(P,n) \overset{P^2\to 0}{\to} \sum_{h=\pm} \epsilon_{h}^{\mu}(P) \epsilon_{h}^{*\nu}(P) =  \sum_{h=\pm} \epsilon_{h}^{\mu}(P) \epsilon_{-h}^{\nu}(-P) \, .
\ee
The polarisation vectors combine with the Feynman diagram components on either
side of the divergent propagator to form on-shell amplitudes. A schematic form
for a general particle type can be written as
\be
\label{eq:multiparticlepole}
A_{n}^{\text{tree}}(1,\ldots, n) \stackrel{{}_{P_{ij}^{2}\to 0}}{\longrightarrow}  \sum_{s\in s_{\rm P}}
A_{L}^{\text{tree}}\bigl(i,i+1,\ldots,j, -P_{ij}^{\bar{s}}\bigr) \frac{n_{\rm P}}{P_{ij}^{2}} A_{R}^{\text{tree}}\bigl(P_{ij}^{s},j+1,\ldots, i-1\bigr) \, ,
\ee
where P indicates the particle type of the propagator with momentum $P_{ij}$,
$s_{\rm P}$ are its possible spin states and $n_{\rm P}$ is a particle-dependent constant.
As we see from the discussion above, for gluons $s_{\rm gluon} = \pm$ are the two
helicities and $n_{\rm gluon} = \i$, for spin-$\half$ fermions one can show that $s_{\rm fermion} =
\pm \nicehalf$ and $n_{\rm fermion} = 1$, while spin $0$ has $s_{\rm scalar} = 0$ and $n_{\rm scalar}=\i$. Other particle
types are simple to determine following the same argument. Factorisation will
also occur in the case of massive propagators where $P_{ij}^2\to m_{ij}^2$
where $m_{ij}$ is the mass of propagating particle.

\subsection{Collinear limits}
\label{sect:collinear}

A special case of factorisation is the two-particle pole also known as \emph{collinear} singularity. Without loss of generality
we take the two collinear particles to be 1 and 2. 
We then have $(p_{1}+p_{2})^{2}=0$,
which implies $p_{1}\cdot p_{2}=0$ or \emph{collinearity} $p_{1}\parallel p_{2}$. We
again concentrate on the massless case.
In fact, since the factorisation now involves a three-particle amplitude, 
such a pole can only occur for collinear external momenta.
We already know from the discussion in the previous chapter that three-point amplitudes are subtle.
In the strict collinear limit $p_{1} \parallel p_{2}$ we may parametrise the collinear momenta $p_1$ and $p_2$ as
\be
p_{1}^{\mu}= x\, P^{\mu}\, ,  \qquad \qquad p_{2}^{\mu}= (1-x)\, P^{\mu} \,,
\ee
with the total collinear momentum $P=p_{1} + p_{2}$, and $x$ parametrising the amount of $P$ distribution over $p_{1}$
and $p_{2}$.
Tree-level amplitudes have a universal (singular) behaviour in the collinear limit governed by the
\emph{splitting functions},
\begin{align} \label{eq:collinear_limit}
A_{n}^{\rm tree}\bigl(1^{h_{1}}, 2^{h_{2}}, \ldots \bigr) 
\stackrel{1 \parallel 2}{\longrightarrow} \sum_{h= \pm} {\rm Split}^{\rm tree}_{h}\bigl(x,1^{h_{1}},2^{h_{2}}\bigr) \, A_{n-1}^{\rm tree}\bigl(P^{-h}, \ldots \bigr)\,,
\end{align}
as a special case of \eqn{eq:multiparticlepole}. In fact, the splitting functions are related to the three-particle amplitudes as
\be
\label{SplitfromFact}
 {\rm Split}^{\rm tree}_{h}\bigl(x,1^{h_{1}},2^{h_{2}}\bigr) = \lim_{P^2\to 0}\, A_{3}^{\text{tree}}\bigl(1^{h_{1}},2^{h_{2}},-P^{h}\bigr)\, \frac{\i}{P^{2}} \,.
\ee
The splitting functions depend on the helicities of the collinear gluons but are independent
of the helicities of the other legs $\{3,\ldots,n\}$ not participating in the collinear limit. This is
known as the universality of the splitting functions.
\begin{svgraybox}{\bf Gluon splitting functions.}
For collinear gluons the splitting functions are given by
\begin{align}
\label{splittingfunctions}
\begin{aligned}
&{\rm Split}_{+}^{\rm tree}(x, a^{+},b^{+}) = 0 \,,  &  
& {\rm Split}_{-}^{\rm tree}(x, a^{+},b^{+}) = \frac{g}{\sqrt{x (1-x)} \vev{a b}} \,, \\
&{\rm Split}_{+}^{\rm tree}(x, a^{+},b^{-}) = -\frac{(1-x)^2\, g}{\sqrt{x(1-x )} \vev{a b}} \,, 
& 
& {\rm Split}_{-}^{\rm tree}(x, a^{+},b^{-}) = - \frac{x^2\, g}{\sqrt{x (1-x)} [{a b}]}  \,.
\end{aligned} \end{align}
The remaining splitting functions may be obtained by parity
\begin{align}
\label{Splitparity}
{\rm Split}_{h}^{\rm tree}(x, a^{-\lambda_a},b^{-\lambda_b}) =   {\rm Split}_{-h}^{\rm tree}(x, a^{\lambda_a},b^{\lambda_b}) \Bigl|_{\vev{a b} \leftrightarrow [ba]} \,.
\end{align}
\end{svgraybox}
We shall now derive these expressions from our knowledge of the three-point MHV amplitude~\eqref{3pointmhv,3pointmhvbar}.
As the collinear kinematics is subtle, it is advantageous to systematically approach the collinear configuration as 
\bal{
& |1\rangle = \cos\phi\, |P\rangle -z\, \sin\phi\, |r\rangle \,, \qquad
& |1] = \cos\phi\, |P] -z\, \sin\phi\, |r] \,, \\
& |2\rangle = \sin\phi\, |P\rangle +z\, \cos\phi\, |r\rangle \,, \qquad
& |2] = \sin\phi\, |P] +z\, \cos\phi\, |r] \,
}
in the limit $z\to 0$.
Here $P^{\mu}=p_{1}^{\mu}+p_{2}^{\mu}+ \cO(z^{2})$ is the limiting collinear momentum vector,
and $r^{\mu}$ is a null reference momentum not parallel to $P^{\mu}$. Moreover, the collinear parameter above is $x=\cos^{2}\phi$. 
This parametrises the four-momenta of the collinear particles as
\bal{ \label{eq:momentadef}
p_{1}&= \cos^{2}\phi\, P - z\, \cos\phi\sin\phi\, \bigl(|P\rangle [r| + |r\rangle [P| \bigr) + z^{2} \, \sin^{2}\phi\,r \,,
\\
p_{2}&= \sin^{2}\phi\, P + z\, \cos\phi\sin\phi\, \bigl(|P\rangle [r| + |r\rangle [P| \bigr) + z^{2} \, \cos^{2}\phi\,r \,,
}
implying that $p_{1}+p_{2}=P+z^{2}\,r$, as claimed. One then has
\begin{align}
\vev{12} =& z \vev{Pr}\,,  \qquad  \vev{1P}=z \sin\phi\, \vev{Pr}\,, \qquad \vev{2P}=-z \cos\phi\, \vev{Pr}\,, \nn \\
 \bev{12} =& z \bev{Pr}\,,  \qquad  \bev{1P}=z \sin\phi\, \bev{Pr}\,, \qquad \bev{2P}=-z \cos\phi\, \bev{Pr} \, , \\
(p_{1}+p_{2})^{2}=&z^{2} \vev{Pr}\bev{rP} \,. \nn
\end{align}
The splitting functions~\eqref{splittingfunctions}
follow immediately from \eqn{SplitfromFact} and the two MHV three-point amplitudes~\eqref{3pointmhv,3pointmhvbar} upon using the above identities. 
The vanishing of ${\rm Split}_{+}^{\rm tree}(x,1^{+},2^{+})$
follows from the vanishing all-plus amplitude. Using the $\text{MHV}_{3}$ amplitude
of \eqn{3pointmhv}\footnote{Recall our convention~\eqref{negpspin} that $|-P\rangle= \i |P\rangle$ and  $|-P]= \i |P]$.} we find
\be
{\rm Split}_{-}^{\rm tree}(x,1^{+},2^{-}) =\frac{-\i g\vev{2(-P)}^{3}}{\vev{21}\,\vev{1(-P)}}\, \frac{\i}{(p_{1}+p_{2})^{2}}
= -\frac{\cos\phi^{3}g}{z\sin\phi \bev{Pr}} \,,
\ee
as well as using the $\widebar{\text{MHV}}_{3}$ amplitude of \eqn{3pointmhvbar},
\bal{
{\rm Split}_{-}^{\rm tree}(x,1^{+},2^{+}) &=  \frac{\i g[12]^{3}}{[1(-P)]\,[(-P)2]}\, \frac{\i}{(p_{1}+p_{2})^{2}}
= \frac{g}{z\cos\phi\sin\phi \vev{Pr}} \,, \\
{\rm Split}_{+}^{\rm tree}(x,1^{+},2^{-}) & =\frac{\i g[1(-P)]^{3}}{[12]\,[2(-P)]}\, \frac{\i}{(p_{1}+p_{2})^{2}}
= -\frac{\sin\phi^{3}g}{z\cos\phi \vev{Pr}}\, ,\nn
}
which prove the relations \eqref{splittingfunctions}. 

\newpage

\begin{example}{Example: Collinear limits of the five-point MHV amplitude} 
Let us use this result to test our conjectured five-point MHV amplitude~\eqref{ParkeTaylor1} from chapter \ref{ch:intro} for consistency under collinear limits. We set $g=1$ for convenience. We~have
\begin{align} \begin{aligned}
\i A_{5}^{\rm tree}(1^{-}, 2^{-}, 3^{+}, 4^{+}, 5^{+} ) & = \frac{ \vev{12}^4}{\vev{12}\vev{23}\vev{34}\vev{45}\vev{51}}  \\
& \stackrel{4 \parallel 5}{\longrightarrow} \frac{1}{\sqrt{x (1-x)} \vev{45}} \times \frac{ \vev{12}^4}{\vev{12}\vev{23}\vev{3 P} \vev{P 1}} \\
& = {\rm Split}_{-}^{\rm tree}(x, 4^{+}, 5^{+}) \times \i A_{4}^{\rm tree}(1^{-}, 2^{-}, 3^{+}, P^{+} ) \,,
\end{aligned} \end{align}
where we parametrised the collinear limit by
$|4\rangle= \sqrt{x} |P\rangle$ and $|5\rangle = \sqrt{1-x}|P\rangle$.
Indeed we find agreement with the second function in \eqn{splittingfunctions}.
Next, we take the collinear limit in a $(-+)$ channel.
We have
\begin{align} \begin{aligned}
\i A_{5}^{\rm tree}(1^{-}, 2^{-}, 3^{+}, 4^{+}, 5^{+} ) & \stackrel{2 \parallel 3}{\longrightarrow} \frac{x^2}{\sqrt{x (1-x)}} \frac{1}{\vev{23}} \frac{ \vev{1P}^4}{\vev{1 P}\vev{P 4}\vev{45}\vev{51}}  \\
& = {\rm Split}_{+}^{\rm tree}(z,2^{-},3^{+}) \times \i A_{4}^{\rm tree}(1^{-}, P^{-}, 4^{+}, 5^{+}) \,,
\end{aligned} \end{align}
from which we deduce \be{\rm Split}_{+}^{\rm tree}(x,a^{-},b^{+})= \frac{x^{2}}{\sqrt{x(1-x)}\,\vev{ab}}\,,
\ee 
in agreement with the third expression in \eqn{splittingfunctions} whilst swapping $a$ and $b$ (and $x\to 1-x$).
In order to check the fourth function we consider the collinear limit in a $(+-)$-channel,
\be
A_{5}^{\rm tree}(1^{-}, 2^{-}, 3^{+}, 4^{+}, 5^{+} )\stackrel{5 \parallel 1}{\longrightarrow}
\underbrace{\frac{(1-x)^{2}}{\sqrt{x(1-x)}\, \vev{51}}}_{{\rm Split}_{+}^{\rm tree}(x,5^{+},1^{-})}\,  A_{4}^{\rm tree}(P^{-}, 2^{-}, 3^{+}, 4^{+} ) \,,
\ee
yielding the desired result via parity~\eqref{Splitparity}. In order to check the vanishing of the uniform
helicity splitting function in \eqn{splittingfunctions} we 
have to study the collinear factorisation of the 6-point MHV amplitude with the helicity
distributions
$A_{6}^{\rm tree}(1^{-}, 2^{-}, 3^{+}, 4^{+}, 5^{+},6^{+} )$ along the legs 5 and 6.
\end{example}

\newpage

\begin{exer}{The vanishing splitting function ${\rm Split}_{+}^{\rm tree}(x, a^{+},b^{+})$}
\label{Exer:2.1}
Show that ${\rm Split}_{+}^{\rm tree}(x, a^{+},b^{+}) = 0$ by studying the factorisation properties of the six-gluon MHV tree amplitude $A_{6}^{\rm tree}(1^{-}, 2^{-}, 3^{+}, 4^{+}, 5^{+}, 6^{+} )$ from \eqn{ParkeTaylor1} in the collinear limit $5 \parallel 6$. 
For the solution see \hyperref[Sol:2.1]{chapter~5}.
\end{exer}

\begin{svgraybox}{\bf Absence of multi-particle poles in MHV amplitudes.}
General $n$-gluon scattering amplitudes have multi-particle poles when region momenta 
go on-shell. However, MHV tree-amplitudes are special, and in fact \emph{only}
 have two-particle poles or collinear singularities. 
The reason is the following. 
Consider the general factorisation formula~\eqref{eq:multiparticlepole}.
In a factorisation of an MHV amplitude there
are only three negative-helicity legs (corresponding to the two external negative helicities, and one for the internal
on-shell propagator) that are distributed over two partial amplitudes. 
However, we saw in the previous chapter that  $A_{n}(1^{\pm},2^{+},\ldots, n^{+}) = 0$ (for $n > 3$).
Therefore, either $A_{L}$ or $A_{R}$ in \eqn{eq:multiparticlepole} is always zero unless one 
partial amplitude is a three-particle amplitude. The latter case corresponds to a two-particle pole or collinear
singularity, as discussed above.
\end{svgraybox}

\subsection{Soft theorems}

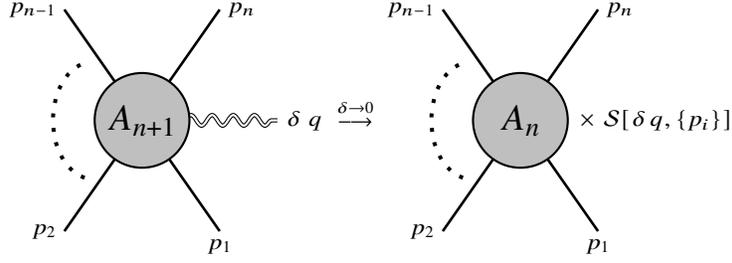
\begin{figure}[t]
   $$
  \begin{tikzpicture}[baseline={(current bounding box.center)}, scale = 0.9]
 \coordinate (q) at (0:2);
\coordinate (1) at (-55:2);
\coordinate (2) at (-125:2);
\coordinate (3) at (125:2);
\coordinate (n) at (55:2);
\coordinate (0) at (0,0);
\draw [solid,line width=1pt] (0) -- (1);
\draw [solid, line width=1pt] (0) -- (2);
\draw [solid, line width=1pt] (0) -- (3); 
\draw [solid,line width=1pt] (0) -- (n); 
\draw [graviton] (0) -- (q); 
\draw [loosely dotted, line width=1.5pt] (-135:1.2) to[bend left=70] (135:1.2);

  \draw (1) node [below] {$p_{1}$};
  \draw (2) node [left] {$p_{2}$};
    \draw (3) node [left] {$p_{n-1}$};
  \draw (n) node [right] {$p_{n}$};
    \draw (q) node [right] {$\mathbf{ \delta\,}\, q$};
    \draw [fill=lightgray,thick, draw=black] (0) circle (0.7);
 \draw (0) node {\Large $A_{n+1}$};                       
 \end{tikzpicture}
 \stackrel{\delta\to 0}{\longrightarrow}
  \begin{tikzpicture}[baseline={(current bounding box.center)}, scale = 0.9]
\coordinate (1) at (-55:2);
\coordinate (2) at (-125:2);
\coordinate (3) at (125:2);
\coordinate (n) at (55:2);
\coordinate (0) at (0,0);
\coordinate (here) at (2,0);
\draw [solid,line width=1pt] (0) -- (1);
\draw [solid, line width=1pt] (0) -- (2);
\draw [solid, line width=1pt] (0) -- (3); 
\draw [solid,line width=1pt] (0) -- (n); 
\draw [loosely dotted, line width=1.5pt] (-135:1.2) to[bend left=70] (135:1.2);

  \draw (1) node [below] {$p_{1}$};
  \draw (2) node [left] {$p_{2}$};
    \draw (3) node [left] {$p_{n-1}$};
  \draw (n) node [right] {$p_{n}$};
    \draw [fill=lightgray,thick, draw=black] (0) circle (0.7);
 \draw (0) node {\Large $A_{n}$};      
 \draw  (here) node {$\times\,  \, \mathcal{S}[ \delta\, q, \{p_{i}\} ]$};                 
 \end{tikzpicture}
  $$
   \caption{The soft factorisation of a generic $(n+1)$-particle amplitude. The soft function $\mathcal{S}[ \delta\, q, \{p_{i}\} ]$
   only depends on the momenta (not the polarisations) of the hard legs.}
   \label{fig:softthm}
\end{figure}

We continue our quest in the factorisations of scattering amplitudes with a kinematical limit
subject to just a \emph{single} leg: the soft limit. Here one particle involved in the scattering process 
has a very low energy---it is soft. Specifically, it refers to the limit where the four-momentum of the particle 
goes to zero. Again the tree-amplitude displays a universal factorisation property 
for photon, gluon and graviton amplitudes into a lower-point amplitude and a soft function. 
In order to take the limit, we parametrise the soft momentum of leg $s$ as
$p_{s}^{\mu} =  \delta\,  q^{\mu}$ and take $ \delta\, \to 0$ (do not confuse $ \delta\,  q^{\mu}$ 
with a variation). The soft theorems state that
\be
\label{centralsoft}
A_{n+1}^{\text{tree}} ( \delta\,  q, 1,\ldots , n) \stackrel{ \delta\,\to 0}{\longrightarrow}
S[ \delta\,   q,  \{p_1,\ldots,p_{n}\}]\, A_{n}^{\text{tree}}(1,\ldots , n)\, ,
\ee
which is illustrated in figure~\ref{fig:softthm}. 
The factorised soft function $S[ \delta\,   q,  \{p_1,\ldots,p_{n}\}]$
 depends on the momentum $ \delta\, \, q$ and helicity of the soft particle, as well as the
momenta of the remaining hard legs $\{p_{a}\}$. It is however independent  of 
the helicities and even particle types of the remaining hard legs, which may be massless or massive. 
The soft function diverges as $1/\delta$ at leading order, and also contains universal sub-leading~pieces:
\be
\label{softtheoremgeneral}
 S[ \delta\,   q, \{p_{a}\}] = \frac{1}{\delta} \mathcal{S}^{[0]}(q) + \mathcal{S}^{[1]}(q)
 + \delta\, \mathcal{S}^{[2]}(q)  + \cO\bigl(\delta^{2}\bigr)\,.
 \ee
 It takes a universal form for photons~\cite{ch2_Low:1958sn}, gluons and gravity~\cite{ch2_Weinberg:1964ew}, remarkably not only
 to leading order, but also to sub-leading order $\mathcal{S}^{[1]}(q)$ for gauge theories, and even to
sub-sub-leading order $\mathcal{S}^{[2]}(q)$ for gravity~\cite{ch2_Cachazo:2014fwa}. 

\begin{svgraybox}{\bf Leading soft theorems.}
The leading soft divergences take the universal form 
\begin{equation}
 \frac{1}{\delta} \mathcal{S}^{[0]}(\delta  q)= \begin{cases} 
  \displaystyle \frac{1}{\delta} {\cal S}^{[0]}_\text{EM} = \frac{1}{\delta} \,
    \sum_{a=1}^{n} e_{a}\frac{\epsilon\cdot p_{a}}{p_{a}\cdot q} \,, & \text{photon, }\\[12pt]  

  \displaystyle \frac{1}{\delta} {\cal S}^{[0]}_\text{YM} = \frac{g}{\delta\sqrt{2}} \,
    \bigg(\frac{\epsilon\cdot p_{1}}{p_{1}\cdot q}-\frac{\epsilon\cdot p_{n}}{p_{n}\cdot q}\bigg)\,, & \text{colour-ordered gluon, }\\[12pt]  
    \displaystyle \frac{1}{\delta} {\cal S}^{[0]}_\text{GR} = \frac{\kappa}{\delta} \, 
    \sum_{a=1}^{n}  \frac{\epsilon_{\mu\nu}\, p_{a}^{\mu}\, p_{a}^{\nu}}{p_{a} \cdot q} \,,
    &\text{graviton. } \end{cases}
  \label{eq:S0explicit}
\end{equation}
Here $\epsilon$ denotes the polarisation vector (tensor) of the soft particle,  $e_{a}$ denotes the $\text{U}(1)$  (electromagnetic) charge of the hard particle~$a$. 
\end{svgraybox}
These results can be made plausible through the following argument based on Feynman diagrams. For a tree-level amplitude the soft leg $\delta \, q$ is attached
either via a three-point coupling to an outgoing hard leg $a$, or to the ``bulk'' of the remaining~amplitude:
\be
\raisebox{0.6cm}{$A_{n}^{\text{tree}}(\delta\, q, \{p_{a}\})=$} 
 \begin{tikzpicture}[baseline={(current bounding box.center)}]
 \coordinate (0) at (-1,0);
\coordinate (b) at (0.25,0);
\coordinate (a) at (45:1);
\coordinate (q) at (-45:1);
\coordinate (n) at (145:1.6);
\coordinate (1) at (-145:1.6);
\coordinate (expl) at (0,-1.7);
\coordinate (mid) at (-0.2,-0.1);
\draw [solid,line width=1pt] (0) -- (1);
\draw [solid, line width=1pt] (0) -- (n);
\draw [solid, line width=1pt] (0) -- (b);
\draw [solid, line width=1pt] (b) -- (a);
\draw [graviton,] (b) -- (q); 
\draw [loosely dotted, line width=1.5pt] (-160:1.5) to[bend left=70] (160:1.5);
 \draw [fill=lightgray,thick, draw=black] (0) circle (0.4);
  \draw [fill] (b) circle (.08);
   \draw (a) node [right] {$a$};
      \draw (q) node [right] {$\delta \, q$};
      \draw [dashed, ->] (expl) to [bend left=10]  (mid);
        \draw (expl) node [right] {$\displaystyle \frac{1}{(p_{a}+\delta \,q)^{2}-m_{a}^{2}}\stackrel{\delta\to 0}{=}
        \frac{1}{2\delta} \frac{1}{p_{a}\cdot q}$};
\draw (1.5,0) node [right] {$+$} ;
\draw (-2.75,0) node [right] {$\displaystyle \sum_{a=1}^{n}$} ;
 \coordinate (0p) at (-1,0);
\coordinate (ap) at (90:.5);
\coordinate (qp) at (-90:.5);
\coordinate (np) at (145:1.6);
\coordinate (1p) at (-145:1.6);
\begin{scope}[every coordinate/.style={shift={(4,0)}}]
\draw [solid,line width=1pt] ([c] 0p) -- ([c] 1p);
\draw [solid, line width=1pt] ([c] 0p) -- ([c] np);
\draw [solid, line width=1pt] ([c] 0p) -- ([c] ap);
\draw [graviton] ([c] 0p) -- ([c] qp); 
\draw [loosely dotted, line width=1.5pt] ([c] -160:1.5) to[bend left=70] ([c] 160:1.5);
 \draw [fill=lightgray,thick, draw=black] ([c] 0p) circle (0.4);
  \draw ([c] ap) node [right] {$a$};
      \draw ([c] qp) node [right] {$\delta \, q$}; 
\end{scope}
\end{tikzpicture}
\ee
We see that the divergence of the leading order soft function 
solely arises from the three-point coupling of the gauge or graviton field
to the hard leg $a$. In the case of an amplitude with $n$-scalars and a soft photon or graviton these couplings
take the form
\be
\begin{tikzpicture}[baseline={(current bounding box.center)}]
 \coordinate (0) at (-1,0);
\coordinate (b) at (0.25,0);
\coordinate (a) at (45:1);
\coordinate (q) at (-45:1);
\draw [solid,line width=1pt] (0) -- (b);
\draw [solid, line width=1pt] (b) -- (a);
\draw [graviton,] (b) -- (q); 
  \draw [fill] (b) circle (.08);
   \draw (a) node [right] {$a$};
      \draw (q) node [right] {$\delta \, q$};
\end{tikzpicture}  
\propto  \quad
\begin{cases} 
  \displaystyle  e_{a}\, \epsilon\cdot p_{a} \,,  & \text{photon, }\\[12pt]  
    \displaystyle \kappa \,\epsilon_{\mu\nu}\, p_{a}^{\mu}\, p_{a}^{\nu} \,,
    &\text{graviton, } \end{cases}
\ee
which fixes the soft functions in \eqn{eq:S0explicit} up to an overall factor. In the case of a colour-ordered
pure gluon amplitude with one soft gluon leg, $A_{n}^{\text{tree}}(1,\ldots,n,\delta \, q)$, the soft leg can couple
either  to the hard gluon leg $1$ or $n$ due to colour ordering.
A detailed look at the Feynman rules~\eqref{Feynrulveert} then reveals the coupling $\pm g\, p_{a}\cdot\epsilon$ with
$a=1,n$ at leading order in $\delta$ with a relative sign factor reproducing the form of~\eqn{eq:S0explicit}.

It is instructive to study the gauge invariance of the leading soft functions~\eqref{eq:S0explicit}. Gauge
invariance requires the full amplitude $A_{n+1}(\delta \, q, p_{1},\ldots , p_{n})$ to vanish under the
transformation $\epsilon^{\mu}\to q^{\mu}$ (in the graviton case we again take 
$\epsilon_{\mu\nu}=\epsilon_\mu\epsilon_{\nu}$). Hence, by consistency, the soft functions in \eqn{eq:S0explicit}  
need to vanish:
\begin{equation}
  \mathcal{S}^{[0]}(\delta \, q)\Bigr |_{\epsilon\to q}= \begin{cases} 
  \displaystyle \quad \,
    \sum_{a=1}^{n} e_{a}=0 \,, & \text{photon, }\\[12pt]  
  \displaystyle  \,\,\, \frac{g}{\sqrt{2}} (1-1)=0 \,, & \text{gluon, } \\[12pt]  
    \displaystyle \kappa \, q_{\mu}
    \sum_{a=1}^{n}   p_{a}^{\mu}=0  \,,
    &\text{graviton. } \end{cases}
\end{equation}
They indeed vanish due to total charge conservation in the electromagnetic or total momentum conservation
in the gravitational case (provided the gravitational coupling is universal to all matter). Hence these soft theorems
are intimately connected to fundamental symmetries of space-time-matter.

\subsection{Spinor-helicity formulation of soft theorems}
It is instructive to translate the leading soft theorems to spinor-helicity language for the colour-ordered gluon
and graviton cases. The leading order soft factorisation states~that
\begin{align}
A_{n}^{\text{tree}}(\ldots, a, \delta q^{\pm}, b,\ldots) \stackrel{\delta\to 0}{\longrightarrow}
  \mathcal{S}^{[0]}(a,q^{\pm},b)\, A_{n-1}^{\text{tree}}(\ldots, a,  b,\ldots) \, .
\label{cosoftfactorization}
\end{align}
The factorised soft function depends on the momenta and helicities of the soft gluon and the
momenta of the colour-ordered neighbours $a$ and $b$. It is independent, however,  of 
the helicities and particle types of the neighbouring legs.
From considering the soft limit of an MHV amplitude one easily establishes that
\be
 \mathcal{S}_{\text{YM}}^{[0]}(a,q^{+},b) = g\frac{\vev{ab}}{\vev{aq}\vev{qb}} \, .
\ee
Via parity, in analogy to \eqn{Splitparity}, we find
\be \label{eq:SoftYM_minus}
\mathcal{S}_{\text{YM}}^{[0]}(a,q^{-},b) = -g\frac{\bev{ab}}{\bev{aq}\bev{qb}} \, .
\ee
Both results directly follow from \eqn{eq:S0explicit} as well. In the graviton case we deduce
\be
\mathcal{S}_{\text{GR}}^{[0]}(q^{++},1,\ldots, n) = \kappa \sum_{a=1}^{n}
\frac{\vev{xa}\vev{ya}\bev{qa}}{\vev{xq}\vev{yq}\vev{aq}}\, ,
\ee
where $x$ and $y$ are arbitrary reference spinors associated to the polarisation vectors of the soft~leg.

\begin{exer}{Soft functions in the spinor-helicity formalism}
\label{Ex:2.2}
Starting from eq.~\eqref{eq:S0explicit}, derive the leading soft functions for a colour-ordered gluon and a graviton in the spinor-helicity formalism.
For the solution see \hyperref[Sol:2.2]{chapter~5}.
\end{exer}

\subsection{Subleading soft theorems}
Remarkably, the universal soft factorisation survives to subleading order ($\cO\bigl(\delta^{0}\bigr)$) in the gauge theory, and even
sub-sub-leading order ($\cO\bigl(\delta\bigr)$) in the gravitational case, 
cf.~\eqn{softtheoremgeneral}. Here we just state the results. They may again
be derived from a careful study of gauge invariance and consistency arguments~\cite{ch2_Broedel:2014fsa,ch2_Bern:2014vva}.
The novelty is that the sub-leading soft functions are (necessarily) no longer true functions but rather  \emph{differential operators} in the external kinematical data. These then
act on the factorised hard scattering-amplitude.
Explicitly, one has the following sub-leading soft operator for a soft photon $q^{\mu}$ with polarisation 
$\epsilon^{\mu}$,
\be
\mathcal{S}_{\text{EM}}^{[1]}(q) = \sum_{a=1}^{n} e_{a} \frac{\epsilon_{\mu} q_{\nu}\, J_{a}^{\mu\nu}}{p_{a}\cdot q} \,,
\ee
with the local angular momentum operator
\be
J_{a}^{\mu\nu} = 2\, p_{a}^{[\mu} \frac{\partial}{\partial p_{a,\, \nu]}} 
+2\, \epsilon_{a}^{[\mu} \frac{\partial}{\partial \epsilon_{a,\, \nu]}} \, .
\ee
In the above $[\mu\nu]$ denotes anti-symmetrisation with unit weight, see Appendix \ref{sec:Conventions}.
Similarly, for a colour-ordered soft gluon we have
\be
\mathcal{S}_{\text{YM}}^{[1]}(n,q,1) = -\i \frac{g}{\sqrt{2}} \left (\frac{\epsilon_{\mu} q_{\nu}\, J_{1}^{\mu\nu}}{p_{1}\cdot q}
- \frac{\epsilon_{\mu} q_{\nu}\, J_{n}^{\mu\nu}}{p_{n}\cdot q} \right )\,,
\ee
while for a soft gravitons we have a sub-leading and sub-sub-leading soft operators of the forms
\bal{
\label{softnnGR}
\mathcal{S}_{\text{GR}}^{[1]}(q) &= -\i \kappa \sum_{a=1}^{n} \frac{\epsilon\cdot p_{a}\, \epsilon_{\mu} q_{\nu}\, J_{a}^{\mu\nu}}{p_{a}\cdot q} \,, \qquad 
\mathcal{S}_{\text{GR}}^{[2]}(q) = -\frac{\kappa}{2} \sum_{a=1}^{n} \frac{\bigl(\epsilon_{\mu} q_{\nu}\, J_{a}^{\mu\nu}\bigr)^{2}}{p_{a}\cdot q} \,.
}
Again, one may show that these sub-leading soft operators are consistent with gauge invariance. While this is trivial for
the gauge-theory operators by virtue of the anti-symmetry of $J^{\mu\nu}$, the gauge invariance of the gravitational $\mathcal{S}_{\text{EM}}^{[1]}(q)$ leads us to total
angular momentum conservation, 
\be
\mathcal{S}_{\text{GR}}^{[1]}(q)\Bigr |_{\epsilon\to q} = -\i \kappa \, \epsilon_{\mu} q_{\nu}\, 
\sum_{a=1}^{n} J_{a}^{\mu\nu} =0 \,,
\ee
nicely teaming up with the total momentum conservation in the leading case. Hence, the
gravitational soft theorems are directly connected to the Poincar\'e invariance of scattering amplitudes.

Expressed in spinor-helicity variables the sub-leading soft operators take the form
\bal{
\label{spinsubleadingsoft}
\mathcal{S}_{\text{YM}}^{[1]}(n,q^{+},1) &= g\left( \frac{\bev{q\partial_{n}}}{\vev{qn}}-
 \frac{\bev{q\partial_{1}}}{\vev{q1}} \right) \,, \\
 \mathcal{S}_{\text{GR}}^{[1]}(q^{+},1,\ldots, n) &= \frac{\kappa}{2}
 \sum_{a=1}^{n} \frac{\bev{aq}}{\vev{aq}}\left (\frac{\vev{ax}}{\vev{qx}} + 
  \frac{\vev{ay}}{\vev{qy}}\right)\, \bev{q\partial_{a}} \,,
  }
where as before $x$ and $y$ are arbitrary reference spinors. The sub-sub-leading gravity
operator of \eqn{softnnGR} in spinor-helicity variables simply reads
\be
 \mathcal{S}_{\text{GR}}^{[2]}(q^{+},1,\ldots, n) = \frac{\kappa}{2}
 \sum_{a=1}^{n} \frac{\bev{aq}}{\vev{aq}} \bev{q\partial_{a}}^{2}  \,,
\ee
and was found in~\cite{ch2_Cachazo:2014fwa}. 

An important subtlety for the sub-leading soft
theorems lies in balancing the total momentum conservation of the $(n+1)$-leg amplitude
with the soft leg and the factorised $n$-point hard amplitude. The
soft factorisation~\eqref{centralsoft} is really a distributional identity involving
delta-functions:
\begin{align}\begin{aligned}
\delta^{(D)}(\delta \, q + P) \, \mathcal{A}_{n+1}^{\text{tree}} (\delta \, q,\{p_{a}\}) \stackrel{\delta\to 0}{\longrightarrow}& \\
S[\delta \,  q,  \{p_{a}\}]\, & \delta^{(D)}(P)\,  \mathcal{A}_{n}^{\text{tree}}(1,\ldots , n)+ \cO\bigl(\delta^{j}\bigr)\, ,
\end{aligned}\end{align}
with $P=p_{1}+\ldots + p_{n}$ and $j=1$ or $2$ for gauge theory or gravity, respectively. This
in fact implies~that
\be
S[\delta \,  q,  \{p_{a}\}]\,  \delta^{(D)}(P) = \delta^{(D)}(\delta \, q + P)\, 
S[\delta \,  q,  \{p_{a}\}]\, ,
\ee
from which one may strongly constrain the $\delta$ expansion  of the soft
function $S[\delta \,  q,  \{p_{a}\}]$ using \eqn{softtheoremgeneral}.
It necessitates the sub-leading terms to be differential operators, and
even the functional forms of \eqn{spinsubleadingsoft} are fixed by the knowledge of the leading soft functions~\cite{ch2_Broedel:2014fsa}.

\begin{exer}{A $\bar{q}qggg$ amplitude from collinear and soft limits}
\label{Ex:qqggg}
In chapter~\ref{ch:intro} we established the following colour-ordered $\bar{q}qgg$ amplitudes involving a massless quark and anti-quark using colour-ordered Feynman rules:
\begin{align} \label{eq:Aqqgg1}
& A_{\bar q q  gg}^{\text{tree}}(1^{-}_{\bar q}, 2^{+}_{q}, 3^{+},4^{+}) = 0 \,, \\
 \label{eq:Aqqgg2}
& A_{\bar q q  gg}^{\text{tree}}(1^{-}_{\bar q}, 2^{+}_{q}, 3^{-},4^{+}) = -\i g^{2}
\frac{\vev{13}^{3}\, \vev{23}}
{\vev{12}\,\vev{23}\,\vev{34}\, \vev{41}}\, .
\end{align}
Use these and the discussed splitting and soft factorisation properties for gluonic legs to make a guess for the five-point single quark-line tree amplitude $A^{\text{tree}}_{\bar{q}qggg}(1^{-}_{\bar{q}}, 2^+_q, 3^-, 4^+, 5^+)$. Check your guess against all known factorisation properties.
Can you generalise your guess to the $n$-particle partial amplitudes \break $A^{\text{tree}}_{\bar{q}qg\ldots g}(1^{-}_{\bar{q}}, 2^+_q, 3^+, \ldots, n^+)$ and 
$A^{\text{tree}}_{\bar{q}qg\ldots g}(1^{-}_{\bar{q}}, 2^+_q, 3^-, 4^+, \ldots, n^+)$?
For the solution see \hyperref[Sol:qqggg]{chapter~5}.
\end{exer}

\section{BCFW recursion for gluon amplitudes}
\label{sec:2.2}

The Britto-Cachazo-Feng-Witten (BCFW) recursion relation~\cite{ch2_Britto:2004ap,ch2_Britto:2005fq} 
is an efficient way
to compute higher-point tree-level amplitudes
from lower-point ones. It does not make use of Feynman rules but builds upon
unitarity by artfully exploiting the
factorisation property of scattering amplitudes~\eqref{eq:multiparticlepole}
when region momenta go on-shell.
As we will see, our knowledge of the gluon (and graviton) three-point amplitudes 
of eqs.~\eqref{3pointmhv} and~\eqref{3pointmhvbar} allows for the construction of {\it all} higher-point tree-level amplitudes in a recursive fashion.
To begin with, let us concentrate on the colour-ordered case and leave the discussion of
gravitons for later. 
The central idea of the recursion is to consider a deformation 
in a {\it  single} complex variable $z$ of two adjacent momenta in a colour-ordered
 amplitude that maps
the  singularities of the amplitude into poles in  $z \in {\mathbb{C}}$.
For the tree-level $n$-gluon amplitude $A_{n}( p_{1}, \ldots p_{n})$ we introduce
the following complex shift of the helicity spinors
of two arbitrary adjacent particles, taken to be $1$ and $n$ without loss of generality:
\begin{align} \label{bcfw-n1}
\begin{split}
&\lambda_1 \rightarrow \hat{\lambda}_{1}(z) = \lambda_{1} - z \lambda_{n} \,, \qquad  \tla_{1}\rightarrow
\tla_{1} \, ,  \\
&\la_{n}\rightarrow
\la_{n} \, , \qquad\qquad \qquad \qquad 
\tilde{\lambda}_n \rightarrow \hat{\tilde\lambda}_{n}(z) = \tilde{\lambda}_{n} + z \tilde{\lambda}_{1} \, . 
\end{split}
\end{align}
We denote the shifted, $z$-dependent quantities by a hat.
This shift is often termed an $[n\, 1\ran$ shift.
It results in a  deformation of the  momenta, 
\begin{align}
\label{complexmamienta}
p_{1}^{\dot\alpha {\alpha}} \to \hat{p}_{1}^{\dot\alpha {\alpha}} (z) = 
\tilde{\lambda}_{1}^{\dot\alpha}\, (\lambda_{1} -z \lambda_{n})^{\alpha}  \,,\qquad
p_{n}^{\dot\alpha {\alpha}} \to \hat{p}_{n}^{\dot\alpha {\alpha}} (z) = (\tilde{\lambda}_{n} + z  \tilde{\lambda}_{1} )^{\dot\alpha}\, \lambda_{n}^{\alpha}  \,.
\end{align}
Importantly, the shift preserves both overall
 momentum conservation and the on-shell conditions:
\begin{equation}
\label{preservedpids}
\hat{p}_{1}(z) + \hat{p}_{n}(z) = p_{1} + p_{n}\, , \qquad 
\hat{p}_{1}^2(z) = 0\,,\qquad \hat{p}_{n}^{2}(z) = 0\, . 
\end{equation}
The $[n\, 1\ran$  shift generates
 a  one-parameter family of amplitudes,
 \be
 A_n(z) \coloneqq A_n \big(\hat{p}_1 (z), p_2, \ldots, p_{n-1}, \hat{p}_n(z)\big)\, .
 \ee 
Note that $\hat{p}_1$ and $\hat{p}_n$ in \eqn{complexmamienta} are now complex, 
as the underlying helicity spinors $\la_{1,n}$ and $\tla_{1,n}$ are
no-longer complex conjugates of each other.  This  makes  the three-point amplitudes 
involving these states of eqs.~\eqref{3pointmhv,3pointmhvbar}  non-vanishing. They will become the seeds of the recursion relation. 
What are the analytic properties of  the deformed amplitude $A_{n}(z)$?
Factorisation implies   that the deformed amplitude $A_{n}(z)$  has precisely
$n{-}3$ \emph{simple} poles in $z$. Using the region momenta  $P_{i}\coloneqq\sum_{j=1}^{i-1}p_{j}$, these $n{-}3$ poles take the form 
\begin{align}
\label{poleargumentbcfw}
\begin{split}
\frac{\i}{\hat{P}_{i}^2(z)} &
\coloneqq 
\frac{\i}{P_{i}^2 - z \, \langle n | P_{i} | 1 \rbrack} = -\frac{1}{\langle n | P_{i} | 1 \rbrack} 
\frac{\i}{z - z_{P_{i}} }
\, , 
\end{split}
\end{align}
where $\hat{P}_{i}(z) = \hat{p}_{1}(z) + p_{2} + \cdots p_{i-1}$, and 
\be
\label{zPidef}
z_{P_{i}}=\frac{P_{i}^{2}}{\lan n|P_{i}|1]}\, , \qquad \quad \forall i\in \{3,
\ldots ,n-1\}\, .
\ee
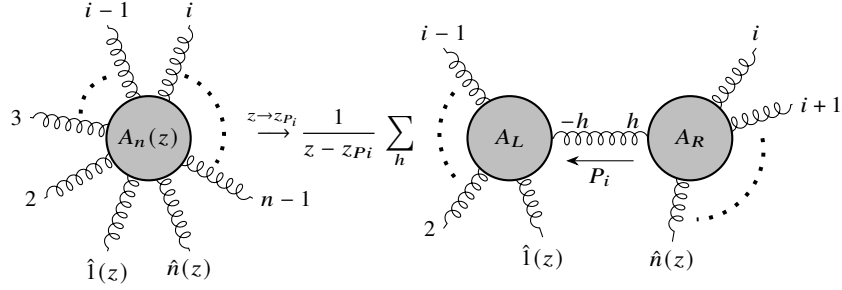
\begin{figure}[tt]
\begin{center}
\begin{equation*}
  \begin{tikzpicture}[baseline={(current bounding box.center)}, scale = 0.8]
  \tikzstyle{blob}=[
    circle,
    minimum size =1.5cm,
    draw=black,
    thick,
    fill=lightgray,
    text=black
]
\begin{feynman}
\coordinate (i) at (70:2);
\coordinate (n-1) at (-30:2);
\coordinate (n) at (-70:2);
\coordinate (1) at (-110:2);
\coordinate (2) at (-150:2);
\coordinate (3) at (-190:2);
\coordinate (i-1) at (-250:2);
\coordinate (x) at (0,0);
\draw [gluon] (x) -- (1);
\draw [gluon] (x) -- (2);
\draw [gluon] (x) -- (3); 
\draw [gluon] (x) -- (n-1); 
\draw [gluon] (x) -- (n); 
\draw [gluon] (x) -- (i); 
\draw [gluon] (x) -- (i-1); 
  \draw (1) node [below] {$\hat{1}(z)$};
  \draw (2) node [left] {$2$};
  \draw (3) node [left] {$3$};
  \draw (n) node [below] {$\hat{n}(z)$};
    \draw (n-1) node [right] {$n-1$};
        \draw (i) node [above] {$i$};
            \draw (i-1) node [above] {$i-1$};
   \draw [loosely dotted, line width=1.5pt] (-200:1.2) to[bend left=40] (-240:1.2);         
   \draw [loosely dotted, line width=1.5pt] (60:1.2) to[bend left=50] (-20:1.2); 
   \draw [fill=lightgray,thick, draw=black] (x) circle (0.7);
 \draw (x) node {$A_{n}(z)$};    
\draw (1.5,0) node [right] {$\displaystyle 
{\buildrel z\to z_{P_i} \over
{\relbar\mskip-1mu\joinrel\rightarrow}}
\frac{1}{z-z_{Pi}}\, \sum_h $};
\coordinate (ip) at (30:3);
\coordinate (i+1p) at (10:3.25);
\coordinate (np) at (-55:2.1);
\coordinate (1p) at (-120:1.9);
\coordinate (2p) at (-150:3);
\coordinate (i-1p) at (-210:3);
\coordinate (xp) at (-1.5,0);
\coordinate (yp) at (1.5,0);
\begin{scope}[every coordinate/.style={shift={(7.5,0)}}]
\draw [gluon, momentum=$P_{i}$] ([c] 0.8,0) to  ([c] -0.8,0);
\draw [gluon] ([c] xp) -- ([c] 1p);
\draw [gluon] ([c] xp) -- ([c] 2p);
\draw [gluon] ([c] yp) -- ([c] i+1p); 
\draw [gluon] ([c] yp) -- ([c] np); 
\draw [gluon] ([c] yp) -- ([c] ip); 
\draw [gluon] ([c] xp) -- ([c] i-1p); 
  \draw ([c] 1p) node [below] {$\hat{1}(z)$};
  \draw ([c] 2p) node [left] {$2$};
  \draw ([c] np) node [below] {$\hat{n}(z)$};
    \draw ([c] i+1p) node [right] {$i+1$};
        \draw ([c] ip) node [above] {$i$};
            \draw ([c] i-1p) node [above] {$i-1$};
  \draw ([c] -0.8,0.3) node [right] {$-h$};            
   \draw ([c] 0.8,0.3) node [left] {$h$};
    \draw [loosely dotted, line width=1.5pt] ([c] -165:2.5) to[bend left=40] ([c] -200:2.5);         
   \draw [loosely dotted, line width=1.5pt] ([c] 0:2.7) to[bend left=50] ([c] -40:2.1);                        
 \draw [fill=lightgray,thick, draw=black] ([c] xp) circle (0.7);
 \draw ([c] xp) node {$A_L$};    
\draw [fill=lightgray,thick, draw=black] ([c] yp) circle (0.7);
 \draw ([c] yp) node {$A_R$};    
\end{scope}
\end{feynman}
\end{tikzpicture}
\end{equation*}
\end{center}
\vspace*{-2mm}
\caption{\it 
Factorisation of the $z$-deformed amplitude $A_{n}(z)$.
}
\label{Fig:BCFW}
\end{figure}
Note that any region momentum containing \emph{both} $\hat p_{1}(z)$ and $\hat p_{n}(z)$
is independent of $z$ by virtue of \eqn{preservedpids}, and hence cannot contribute to a $z$-pole. This is why we find $n{-}3$ poles.
It follows that, as  $z\to z_{P_{i}}$,  the amplitude $A_{n}(z)$ factorises as 
\begin{align}\label{limitpolezPi}
\begin{aligned}
A_{n}(z) &{\buildrel z\to z_{P_i} \over
{\relbar\mskip-1mu\joinrel\rightarrow}}  \\ \frac{\i}{\hat{P}_{i}^2(z)} &  \!\sum_{h} A_{L}\bigl[\hat{1}(z_{P_{i}}),2,\ldots, i-1,-\hat{P}^{-h}(z_{P_{i}}) \bigr] \, A_{R} \bigl[ \hat{P}^{h}(z_{P_{i}}) ,i ,\ldots,n-1, \hat{n}(z_{P_{i}}) \bigr]  \,, 
\end{aligned} \end{align} 
as depicted in figure~\ref{Fig:BCFW}. 
The sum on $h$  runs over all possible
 helicity states  $h$ propagating between $A_{L}$ and $A_{R}$, and depends on the
 field content of the theory considered.  For gluons it is a sum over  $h= \{+1,-1\}$.

In the end we are  only interested in the undeformed amplitude, i.e.~$A_n (z{=}0)$, and  we can use complex analysis to construct it  from the knowledge of the residues of~$A_{n}(z)$:
\begin{align}
\label{BCFWrec}
\begin{aligned}
A_{n}(z{=}0) &= \frac{1}{2\pi \i}\oint_{C_{0}}\!\frac{\d z}{z } \, A_n(z) \\
&= \sum_{i=2}^{n-1} \sum_{h=\pm} A_{L}^{-h}(z_{P_{i}}) \frac{\i}{P_{i}^2} A_{R}^h(z_{P_{i}}) + {\rm Res}(z=\infty)\,.
\end{aligned}
\end{align}
\begin{figure}[tbp]
   \centering
   \includegraphics{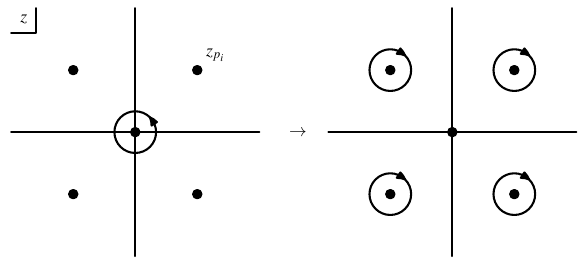}
  \caption{
 Using Cauchy's theorem to obtain  
  eq.~(\ref{BCFWrec}) we may 
   pull the initial circle $C_{0}$ off to infinity thereby encircling the other poles
   clock-wise.} 
   \label{Fig:II.1}
\end{figure}
Here $C_{0}$ is a  small circle around  $z{=}0$ that only contains  the pole around the origin. To obtain \eqn{BCFWrec} we have deformed  this into a large circle at  infinity, 
now   encircling all the poles $z_{P_{i}}$ in the complex plane but  with  an opposite
orientation. See figure \ref{Fig:II.1}.
 If $A_{n}(z) \to 0$ as $z\to \infty$ we can drop the boundary term
${\rm Res}(z=\infty)$.
As we shall argue in a moment, this is the case for gauge theories under certain conditions.

\pagebreak
\begin{svgraybox}{\bf BCFW recursion relation for gluon amplitudes.}
With this assumption, we arrive at the BCFW recursion relation~\cite{ch2_Britto:2005fq}:
\begin{align}\begin{split}\label{eq-bcfw}
A_{n}(1,\ldots, n) =&
 \sum_{i=3}^{n-1} \sum_{h=\pm} A_{i}\Bigl(\hat{1}(z_{P_{i}}),2,\ldots, 
-\hat{P}_{i}^{-h}(z_{P_{i}})\Bigr) \\
& \quad \frac{\i}{P_{i}^2}\,  A_{n+2-i}\Bigl (\hat{P}^{h}_{i}(z_{P_{i}}),i,\ldots, n-1, \hat{n}(z_{P_{i}}) \Bigr ) \, , 
\end{split}
\end{align}
with $z_{P_{i}}$  defined in \eqref{zPidef}, and $P_{i}=p_{1}+p_{2}+\ldots + p_{i-1}$.
This relation is constructive: the amplitudes appearing on the right-hand side have lower 
multiplicity than the initial $A_n$. Hence, with the seed three-gluon amplitudes
\eqref{3pointmhv} and \eqref{3pointmhvbar}  we can bootstrap this relation to construct \emph{all} $n$-gluon
trees without using Feynman diagrams at all! 
Even more, our nuclei---the three-point amplitudes---were obtained from helicity
scaling arguments alone, as discussed in section~\ref{sec:helscale}. 
In this derivation we chose to shift two neighbouring legs
$\hat{1}$ and $\hat{n}$. In fact, one can also shift non-neighbouring legs or even more than two legs
to obtain alternative recursion relations, see e.g.~\cite{ch2_Risager:2005vk,ch2_Elvang:2008vz}.
\end{svgraybox}

An open issue is the vanishing of the boundary term in \eqref{BCFWrec}. For this we need to
have that 
\be
\frac{1}{2\pi \i}\oint_{\infty}\frac{dz}{z} \, {A_{n}(z)} =0\, , \ee
which in turns requires  a large-$z$ falloff of the amplitude as $A_{n}(z) {\sim} z^{-1}$.
In fact, the large-$z$ behaviour depends on the helicities of the shifted legs, and one can show that
\begin{align}
A\bigl(\hat{1}^{+},\hat{n}^{-}\bigr)\stackrel{z\to \infty }{\sim} \frac{1}{z} \, , \qquad
A\bigl(\hat{1}^{+},\hat{n}^{+}\bigr)\stackrel{z\to \infty }{\sim} \frac{1}{z} \, , \qquad
A\bigl(\hat{1}^{-},\hat{n}^{-}\bigr)\stackrel{z\to \infty }{\sim} \frac{1}{z} \, , \qquad
\end{align}
yet $A\bigl(\hat{1}^{-},\hat{n}^{+}\bigr)\!\stackrel{z\to \infty }{\sim}\!z^{3}$, which is then a forbidden $[n1\rangle$ shift.
In the following we show the first  relation; the other scalings are more technical to derive,
and may be found in~\cite{ch2_Arkani-Hamed:2008bsc}.

\subsection{Large $z$ falloff}

\begin{figure}[tt]
\begin{center}
$$
  \begin{tikzpicture}[baseline={(current bounding box.center)}]
  \begin{feynman}
    \coordinate (1) at (120:2);
    \coordinate (2) at (75:2);
    \coordinate (3) at (30:2);
    \coordinate (3b) at (5:1.8);
    \coordinate (4) at (-15:2);
    \coordinate (5) at (-60:2);
    \coordinate (6) at (-105:2);
    \coordinate (7) at (-150:2);
    \coordinate (8) at (-195:2);
    \coordinate (a) at (-1.0,0.75);
    \coordinate (b) at (-0.0,0.5);
    \coordinate (c) at (0.0,-0.5);
    \coordinate (d) at (1.0,-0.75);    
    \coordinate (e) at (0.75,1.15);    
    \coordinate (f) at (-0.75,-1.0);    
\diagram*{  (1) -- [gluon](a)};
\diagram*{  (8) -- [gluon](a)};
\diagram*{  (2) -- [gluon](e)};
\diagram*{  (3) -- [gluon](e)};
\diagram*{  (3b) -- [gluon](b)};
\diagram*{  (4) -- [gluon](d)};
\diagram*{  (5) -- [gluon](d)};
\diagram*{  (6) -- [gluon](f)};
\diagram*{  (7) -- [gluon](f)};
\draw [gluon]  (b) -- node [above] {$\frac{1}{z}$} (a);
\diagram*{  (b) -- [gluon](e)};
\draw [gluon] (b) -- node [left] {$\frac{1}{z}$}  (c);
\draw [gluon] (c) --  node [below] {$\frac{1}{z}$} (f);
\draw (f) node [above] {$z$};\draw (b) node [above] {$1$};
\draw (c) node [below] {$z$};
\draw (a) node [below] {$z$};
\diagram*{  (c) -- [gluon](d)};
  \draw (1) node [above] {$n-1$};
  \draw (2) node [above] {$n-2$};
  \draw (5) node [below] {$3$};
  \draw (6) node [below] {$2$};
  \draw (7) node [left] {$\hat{1}$};  
  \draw (8) node [left] {$\hat n$};
    \draw [loosely dotted, line width=1.5pt] (25:2.2) to[bend left=40] (-10:2.2); 
       \draw [fill] (a) circle (.08);
       \draw [fill] (b) circle (.08);
      \draw [fill] (c) circle (.08);
      \draw [fill] (d) circle (.08);
       \draw [fill] (e) circle (.08);
      \draw [fill] (f) circle (.08); 
      \end{feynman}      
              \end{tikzpicture}
$$
\end{center}
\caption{
The $z$ scaling of a generic graph: along the path from $\hat{1}$ to $\hat n$ the propagators
scale as $1/z$, the three-point vertices as $z$, while four-point vertices do not scale. 
This sample graph scales as $1$. However, if we
would replace the four-point vertex by a three-point one, it would scale as $z$.
On top of that we have to consider the $z$ scaling of the polarisation vectors.
}
\label{fig:zscaling}
\end{figure}
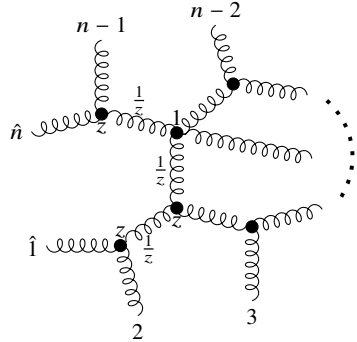

In order to estimate the large $z$ behaviour of generic tree-level amplitudes we
 perform an analysis based on Feynman graphs.
There are three sources for $z$ dependence in a generic colour-ordered amplitude: the propagators, the interaction vertices,
and the polarisation vectors. 
Consider a generic graph contributing to the tree-level $n$-gluon amplitude ($\hat{1}$ and $\hat{n}$ are assumed to be neighbours).
The $z$ dependence occurs only along the path from $\hat{1}$ to $\hat{n}$, see figure~\ref{fig:zscaling}.
Along this path, each three-gluon vertex, being linear in the momenta, maximally contributes a factor of $z$, while four-gluon vertices do not contribute, and all propagators along the path contribute a factor of $1/z$.
We may derive an upper bound for the $z$-scaling by considering the diagrams with maximal powers of $z$.
This happens when the path from $\hat{1}$ to $\hat{n}$ contains \emph{only} three-valent vertices. 
In that case it is easy to see that the graph scales as $z$:
\be
\begin{tikzpicture}[baseline={(current bounding box.center)}, scale = 0.9]
\begin{feynman}
\coordinate (1) at (-2,0);
\coordinate (2) at (-1,1);
\coordinate (a2) at (-1,0);
\coordinate (3) at (0,1);
\coordinate (a3) at (0,0);
\coordinate (4) at (1,1);
\coordinate (a4) at (1,0);
\coordinate (a5) at (2,0);
\coordinate (a6) at (3,0);
\coordinate (n-1) at (4,1);
\coordinate (an-1) at (4,0);
\coordinate (n) at (5,0);
\draw [gluon] (1) -- (a2);
\draw [gluon] (a2) to node[below]{$\frac{1}{z}$}  (a3);
\draw [gluon]  (a3) to node[below]{$\frac{1}{z}$} (a4);
\draw [gluon] (a6) to node[below]{$\frac{1}{z}$} (an-1);
\draw [gluon] (a4) to node[below]{$\frac{1}{z}$} (a5);
\draw [gluon] (2) -- (a2);
\draw [gluon] (3) -- (a3) ;
\draw [gluon] (4) -- (a4) ;
\draw [gluon] (n-1) -- (an-1) ;
\draw [gluon] (n) -- (an-1) ;
\draw (-1,0) node [below] {z};
\draw (0,0) node [below] {z};
\draw (1,0) node [below] {z};
\draw (2,0) node [below] {z};
  \draw (a5) node [right] {$\ldots\ldots$};
  \draw (1) node [left] {$\hat 1$};
            \draw (n) node [right] {$\hat{n}$};  
        \draw [fill] (an-1) circle (.08);
       \draw [fill] (a2) circle (.08);
         \draw [fill] (a3) circle (.08);
       \draw [fill] (a4) circle (.08);  
             \draw [fill] (an-1) circle (.08);  
             \end{feynman}
\end{tikzpicture}
\sim z \,.
\ee
Finally, there is an additional $z$ dependence arising from the polarisation vectors at legs $1$ and $n$:
\begin{eqnarray}
\epsilon_{1}^{+ \alpha \dot{\alpha}} &=& - \sqrt{2} \, \frac{ \tilde{\lambda}_{1}^{\dot\alpha} \mu_{1}^{\alpha}}{ \vev{ \hat{\lambda}_{1}(z) \mu_{1}}}  \sim \frac{1}{z} \,, \qquad 
\epsilon_{1}^{- \alpha \dot{\alpha}} = \sqrt{2} \, \frac{ {\hat{\lambda}_{1}}^{\alpha}(z) \tilde\mu_{1}^{\dot\alpha}}{ \lbrack \lambda_{1} \mu_{1} \rbrack } \sim z \,,  \\
\epsilon_{n}^{+ \alpha \dot{\alpha}} &=& - \sqrt{2} \, \frac{ 
\hat{\tilde{\lambda}}_{n}^{\dot\alpha}(z)
 \mu_{n}^{\alpha}}{ \vev{ \lambda_{n} \mu_{n}}}  \sim z \,, \qquad 
\epsilon_{n}^{- \alpha \dot{\alpha}} = \sqrt{2} \, \frac{ {\lambda}_{n}^{\alpha} \tilde\mu_{n}^{\dot\alpha}}{ \lbrack \hat{\lambda}_{n}(z) \mu_{n} \rbrack } \sim \frac{1}{z} \,.
\end{eqnarray}
Therefore, taking all sources of $z$ dependence into account, 
we conclude that individual graphs scale at worst as
\begin{align}
\label{zshiftscaling}
\begin{alignedat}{2}
& A\bigl( \hat{1}^{+}, \hat{n}^{-}\bigr) \stackrel{z\to \infty }{\sim} \frac{1}{z} \,, \qquad \qquad && A\bigl( \hat{1}^{+}, \hat{n}^{+} \bigr) \stackrel{z\to \infty }{\sim} z\,,\\
& A\bigl( \hat{1}^{-}, \hat{n}^{-}\bigr) \stackrel{z\to \infty }{\sim} z \,, \qquad \qquad  && A\bigl( \hat{1}^{-},\hat{n}^{+} \bigr) \stackrel{z\to \infty }{\sim} z^3\,. \\
\end{alignedat}
\end{align}
This shows that the $[-+\rangle$-shift has the desired falloff properties that allow us to drop the boundary term
at infinity in the BCFW formula~\eqref{eq-bcfw}. By cyclicity, it is always possible to find a $\{ \hat{1}^{+} , \hat{n}^{-} \}$ 
pair.
In fact, the above bound is too conservative. It was shown in~\cite{ch2_Arkani-Hamed:2008bsc}
that the $[++\rangle$ and $[--\rangle$-shifts also lead to an overall ${1}/{z}$ scaling once the sum over all Feynman graphs is performed, as the terms scaling as $z$ or $1$ cancel out.
Only the $[+-\rangle$-shift gives a non-vanishing $\text{Res}(z={\infty})$ in general, and may not be used as basis for a BCFW recursion.

\section{BCFW recursion for gravity and other theories}

Can we generalise the BCFW recursion to other \emph{massless} quantum field theories? If we 
analyse its
derivation in section \ref{sec:2.2}, we see that only two ingredients were  needed to establish it:
\begin{enumerate}
\item Tree-level amplitudes factorise on simple poles
whenever the square of the sum of a subset of external momenta vanishes. While for colour-ordered amplitudes we only need to consider adjacent channels, this is not essential for
the derivation of the BCFW recursion: factorisation is a completely general property, and that
is all that is needed.
\smallskip
\item The deformed amplitude $A_{n}(z)$ falls off as $1/z$ at infinity. This depends on the theory and is related to its ultraviolet  behaviour.
\end{enumerate}
Therefore, in order to construct 
tree-level amplitudes recursively without colour ordering from their factorisation properties, we need to
consider all multi-particle channels that may occur. We thus generalise the region momenta to include any subset $I$ of the momenta $\{p_1,\ldots,p_n\}$,
\be
P_{I}^{\mu}\coloneqq\sum_{i\in I}p_{i}^{\mu} \,.
\ee
Whenever  $P_{I}^{2}=0$ we have a pole, and, if a two-particle BCFW shift is used, the set $I$ must contain only one of the shifted momenta so that $P_{I}^{2}$ becomes $z$-dependent.
Concretely, the BCFW recursion
for a shift of legs $1$ and $n$ as in \eqn{bcfw-n1}
in gravity takes the form~\cite{ch2_Bedford:2005yy, ch2_Cachazo:2005ca}
\begin{align}\label{eq-bcfw-grav}
M_{n} =  
 \sum_Q \sum_{h=\pm\pm} 
 M_L\Bigl(\hat{1}(z_{P_{Q}}),Q, 
-\hat{P}_{Q}^{-h}(z_{P_{Q}})\Bigr)
\frac{\i}{P_{Q}^2}\,  M_{R}\Bigl (\hat{P}^{h}_{Q}(z_{P_{Q}}),\bar{Q}, \hat{n}(z_{P_{Q}}) \Bigr ) \,,
\end{align}
where the first sum runs over \emph{all} subsets $Q$ of momenta in  $\{p_{2},\ldots, p_{n-1}\}$, $\bar{Q}$ is the complement of $Q$, and
$P_{Q}=p_{1}+\sum_{i\in Q}P_{i}$.
Again, the recursion is only valid for the $[n1\rangle$ shifts
\be
\label{thisonep}
|\hat 1\rangle = |1\rangle - z |n\rangle\, , \qquad
|\hat n] = |n] + z |1]\, ,
\ee
of the types $[-+\rangle$, $[++\rangle$, and $[--\rangle$. For a derivation
see~\cite{ch2_Arkani-Hamed:2008bsc}.

Finally, we note that the BCFW recursion can be  generalised to massive theories~\cite{ch2_Badger:2005zh,ch2_Badger:2005jv} to be discussed in section~\ref{sec:2.6}, to 
the rational parts of one-loop amplitudes in QCD and gravity~\cite{ch2_Bern:2005hs,ch2_Bern:2005ji,ch2_Brandhuber:2007up,ch2_Dunbar:2010xk,ch2_Alston:2015gea} and form factors~\cite{ch2_Brandhuber:2010ad,ch2_Brandhuber:2011tv}. In supersymmetric Yang-Mills theory a
supersymmetric  version of the BCFW recursion may be formulated~\cite{ch2_Brandhuber:2008pf,ch2_Arkani-Hamed:2008owk}. In fact this recursion could be solved analytically~\cite{ch2_Drummond:2008cr}. 

\section{MHV amplitudes from the BCFW recursion relation}
\label{sect:MHVfrom BCFW}

\subsection{Proof of the Parke-Taylor formula}

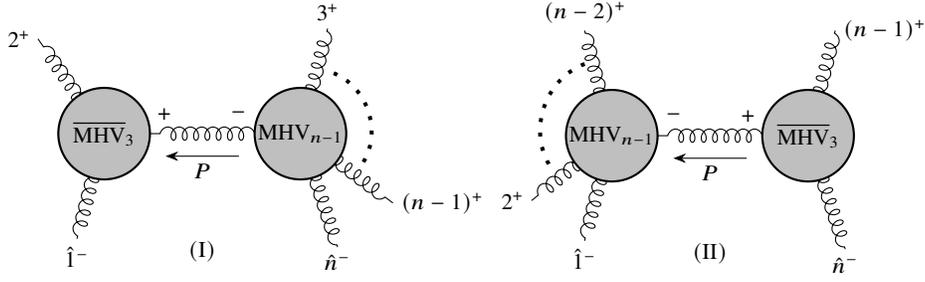
\begin{figure}[t]
\begin{center}
\begin{equation*}
  \begin{tikzpicture}[baseline={(current bounding box.center)}, scale = 0.87]
  \begin{feynman}
  \coordinate (i) at (40:2.5);
\coordinate (n-1) at (-20:3.1);
\coordinate (n) at (-40:2.7);
\coordinate (1) at (-140:2.5);
\coordinate (2) at (150:2.9);
\coordinate (x) at (-1.5,0);
\coordinate (y) at (1.5,0);
\draw [gluon, momentum=$P$] ( 0.8,0) to  ( -0.8,0);
\draw [gluon] (x) -- ( 1);
\draw [gluon] (x) -- (2);
\draw [gluon] (y) -- (i); 
\draw [gluon] (y) -- (n); 
\draw [gluon] (y) -- (n-1); 
  \draw (1) node [below] {$\hat{1}^{-}$};
  \draw (2) node [left] {$2^{+}$};
  \draw (n) node [below] {$\hat{n}^{-}$};
    \draw (n-1) node [right] {$(n-1)^{+}$};
        \draw (i) node [above] {$3^{+}$};
         \draw ( -0.8,0.3) node [right] {$+$};            
   \draw ( 0.8,0.3) node [left] {$-$};
    \draw [loosely dotted, line width=1.5pt] (-10:2.5) to[bend right=50] (25:2.2);                        
 \draw [fill=lightgray,thick, draw=black] (x) circle (0.7);
 \draw (x) node {$\widebar{\text{MHV}}_{3}$};    
\draw [fill=lightgray,thick, draw=black] (y) circle (0.7);
 \draw (y) node {$\text{MHV}_{n-1}$};  
 \draw (0,-1.8) node {$(\text{I})$};  
\end{feynman}
\end{tikzpicture}
\, 
\begin{tikzpicture}[baseline={(current bounding box.center)}, scale = 0.87]
  \begin{feynman}
  \coordinate (i) at (140:2.5);
\coordinate (n-1) at (40:2.5);
\coordinate (n) at (-40:2.7);
\coordinate (1) at (-140:2.5);
\coordinate (2) at (-160:2.9);
\coordinate (x) at (-1.5,0);
\coordinate (y) at (1.5,0);
\draw [gluon, momentum=$P$] ( 0.8,0) to  ( -0.8,0);
\draw [gluon] (x) -- ( 1);
\draw [gluon] (x) -- (2);
\draw [gluon] (x) -- (i); 
\draw [gluon] (y) -- (n); 
\draw [gluon] (y) -- (n-1); 
  \draw (1) node [below] {$\hat{1}^{-}$};
  \draw (2) node [left] {$2^{+}$};
  \draw (n) node [below] {$\hat{n}^{-}$};
    \draw (n-1) node [right] {$(n-1)^{+}$};
        \draw (i) node [above] {$(n-2)^{+}$};
         \draw ( -0.8,0.3) node [right] {$-$};            
   \draw ( 0.8,0.3) node [left] {$+$};
    \draw [loosely dotted, line width=1.5pt] (-170:2.5) to[bend left=50] (150:2.2);                        
 \draw [fill=lightgray,thick, draw=black] (x) circle (0.7);
 \draw [fill=lightgray,thick, draw=black] (y) circle (0.7);
 \draw (y) node {$\widebar{\text{MHV}}_{3}$};    
 \draw (x) node {$\text{MHV}_{n-1}$};  
 \draw (0,-1.8) node {$(\text{II})$};  
\end{feynman}
\end{tikzpicture}
 \end{equation*}
\end{center}
\vspace*{-2mm}
\caption{
The two BCFW diagrams contributing to the $\text{MHV}_{n}$ amplitude. In fact, diagram $(\text{II})$ does
not contribute.
}
\label{Fig:MHVproof}
\end{figure}

As an application of the colour-ordered BCFW recursion~\eqref{eq-bcfw}, we  now derive  the   Parke-Taylor formula~\eqref{ParkeTaylor1} for MHV amplitudes.  
We already know from  section~\ref{sec:1.10} that it  is true for $n=3$ and $n=4$ through an explicit computation.
Therefore we shall prove by induction that the formula is correct. 
We focus here on the case where particles $n$ and $1$ have negative helicity, and 
perform  the $[n1\rangle$ shifts of \eqn{bcfw-n1}. The MHV amplitude has no multi-particle factorisation, as was discussed in section~\ref{sect:collinear}. Hence, only the two BCFW diagrams of
figure~\ref{Fig:MHVproof} contribute to the BCFW recursion of \eqn{eq-bcfw}.
We recall the $[n1\rangle$ shift,
\bal{
\label{n1shiftrecap}
|\hat 1 \rangle &= |1\rangle - z |n\rangle\, , \quad
|\hat n ] = |n] + z |1]\, , \quad \hat P = P -z \, |n\rangle [1|\, ,
}
whereas $|\hat 1] = |1]$ and $|\hat n \rangle = |n\rangle$ are left inert.

In fact, diagram $(\text{II})$ does not contribute. Here, the right diagram is of $\widebar{\text{MHV}}_{3}$ type.
Its numerator reads $[\hat P | n-1]^{3}$, cf.~\eqn{3pointmhvbar}, which vanishes:
\be
[n-1|\hat P ] = \frac{[n-1|\hat P |n\rangle}{\vev{\hat P \, n}} =
\frac{[n-1|(P - z \, |1]\, \langle n|)|n\rangle}{\vev{\hat P \, n}} =
\frac{[n-1|P|n\rangle}{\vev{\hat P \, n}} \stackrel{P=-p_{n-1}-p_{n}}{=}
0 \,.
\ee
In fact, this vanishing is consistent with the observation that the
$\widebar{\text{MHV}}_{3}$ kinematical assumption of collinear left-handed spinors, i.e.~$\vev{(n-1)|\hat n}=\vev{(n-1)|n}=0$, 
forces the two momenta $p_{n-1}$ and $p_{n}$ to be collinear, $p_{n-1}\parallel p_{n}$.
This is an inconsistent assumption on the $n$-particle kinematics. This is not a problem
for diagram $(\text{I})$, where the analogue condition reads $\vev{2|\hat 1}=0$, which does not
imply collinearity of $p_{1}$ and $p_{2}$ as $|\hat 1\rangle \neq |1\rangle$. 

Therefore only the BCFW diagram $(\text{I})$ contributes, where $A_L$ is a three-point $\overline{\rm MHV}$ amplitude, and $A_R$ is an $(n-1)$-point MHV amplitude. 
From \eqn{zPidef}, the position of the pole is 
\begin{align}
z_{P} = \frac{ (p_1 + p_2 )^2 }{ \langle n | P|1 \rbrack} = \frac{ \vev{12}\lbrack2 1 \rbrack}{\vev{n 2} \lbrack2 1\rbrack} = \frac{\vev{12}}{\vev{n 2}} \, .
\end{align}
The amplitudes $A_{L}$ and $A_{R}$ are then given by
\begin{align} \begin{aligned}
& A_{L} = A^{\overline{\text{MHV}}}_{3}\bigl( \hat{1}^{-},2^{+},-\hat{P}^{+} \bigr) =  \i g  
\frac{ \lbrack 2 (-
\hat{P}) \rbrack^3}{\lbrack \hat{1}2 \rbrack \lbrack (-\hat{P}) \hat{1} \rbrack} \,, \\
& A_{R} = A^{\text{MHV}}_{n-1}\bigl(\hat{P}^- , 3^+ , 4^+, \ldots (n-1)^{+} , \hat{n}^{-}\bigr) =   -\i g^{n-3}  \frac{ \vev{\hat{n} \hat{P}}^3 }{\vev{\hat{P} 3} \vev{34} \cdots \vev{(n-1) \hat{n}}}\,.
\end{aligned} \end{align}
Using \eqref{negpspin}, the fact that $\lambda_n$ and $\tilde{\lambda}_1$ are not shifted in  our $[n1\rangle$ shift of \eqn{thisonep}, as well~as 
\begin{align}
\vev{\hat{n} \hat{P}} \bev{\hat P 2} &= \vev{n\hat 1}\, \bev{12} = \vev{n1}\, \bev{12}\, , \qquad
\vev{3\hat P}  \,\bev{\hat P 1} 
= \vev{32}\, \bev{2\hat 1} = \vev{32}\, \bev{21}\, , 
\end{align}
we find
\begin{align} \begin{aligned}
A_{L}  \frac{\i}{(p_1 + p_2)^2} A_{R} &=
\i g^{n-2}\frac{\vev{n1}^{3}
\, \bev{12}^{3}}{\bev{12}\bev{21}\vev{32}\bev{21}\, \vev{12}\vev{34}\cdots
\vev{(n-1)\, n}} \\
&
=- \i g^{n-2} \frac{ \vev{n 1}^4}{\vev{12} \cdots \vev{n1}} \ , 
\end{aligned} \end{align}
in agreement with the conjecture~\eqref{ParkeTaylor1} for the chosen helicities. This proves
the Parke-Taylor formula for adjacent negative-helicity states.

\subsection{The four-graviton MHV amplitude}

As a second example using the BCFW recursion for non-colour-ordered amplitudes~\eqref{eq-bcfw-grav} we compute the four-graviton amplitude $M_{4}(1^{--},2^{--},3^{++},4^{++})$. For
this we perform a
$[2^{--}1^{--}\rangle$ shift,
\bal{
\label{n1shiftrecap}
|\hat 1 \rangle &= |1\rangle - z |2\rangle\, , \quad
|\hat 2 ] = |2] + z |1]\, .
}
As $M_{n}(--,++,\ldots,++)=0$ for $n>3$, we again only find two channels in the BCFW
recursion, as shown in fig.~\ref{Fig:MHVproof} with the hatted leg $\hat n$ now replaced by $\hat 2$,
while the positive-helicity legs are summed over. Still, the same argument for the vanishing of the
type-$(\text{II})$ diagrams applies. For the special case of the four-point graviton amplitude we therefore
have
\be
M_{4}(1^{--},2^{--},3^{++},4^{++})=
\begin{tikzpicture}[baseline={(current bounding box.center)}, scale = 0.95]
  \begin{feynman}
\coordinate (3) at (-1,-.7);
\coordinate (4) at (1,-.7);
\coordinate (1) at (-1,.7);
\coordinate (2) at (1,.7);
\coordinate (x) at (-0.7,0);
\coordinate (y) at (0.7,0);
\draw [graviton] (x) to node [below] {$P$} (y);
\draw [graviton] (x) to (1);
\draw [graviton] (x) to (3);
\draw [graviton] (y) to (2);
\draw [graviton] (y) to (4);
\draw [ghost, momentum',white] (-0.5,-.1) to (0.5,-0.1);
\draw (1) node [left] {${\hat 1}^{--}$};
\draw (2) node [right] {${\hat 2}^{--}$};
\draw (3) node [left] {$3^{++}$};
\draw (4) node [right] {$4^{++}$};
 \draw [fill] (y) circle (.08);
 \draw [fill] (x) circle (.08);
  \draw ( -0.7,0.3) node [right] {$++$};            
   \draw (0.7,0.3) node [left] {$--$};
\end{feynman}
\end{tikzpicture} + (3\leftrightarrow 4) \,,
\ee
with $P=-p_{1}-p_{3}$. Inserting the three-graviton amplitudes~\eqref{generalsMHV} this becomes (setting $\kappa = 1$)
\bal{
M_{4}(1^{--},2^{--},3^{++},4^{++}) &= \left ( \frac{\bev{\hat P 3}^{3}}{\bev{31}\bev{1\hat P}}\right ) ^{2} \frac{\i}{(p_{1}+p_{3})^{2}}
\left ( \frac{\vev{\hat P 2}^{3}}{\vev{24}\vev{4\hat P}}\right ) ^{2} + (3\leftrightarrow 4) \nn \\
&= \frac{\i}{s_{13}} \frac{\langle 2 | \hat P | 3]^{6}}{\vev{24}^{2}\bev{31}^{2}\langle 4| \hat P|1]^{2}}
+ (3\leftrightarrow 4) \,.
}
We now use $\hat P=p_{2}+p_{4} + z |2\rangle [1|$ to find $\langle 2 | \hat P | 3]=\vev{24}\bev{43}$ and
$\langle 4| \hat P|1]=\vev{42}\bev{21}$. Hence, the $z$ dependance drops out! Inserting these two relations
one finds
\be
M_{4}(1^{--},2^{--},3^{++},4^{++}) = \i \frac{\bev{34}^{6}}{\bev{12}^{2}}
\left( \frac{\vev{24}^{2}}{\vev{13}\bev{31}^{3}} + \frac{\vev{23}^{2}}{\vev{14}\bev{41}^{3}} \right ) \, ,
\ee
where $s_{ij} = (p_i+p_j)^2$.
While this is the final result, one may write it using the Mandelstam variables
\be
s=(p_{1}+p_{2})^{2}\, , \quad  t=(p_{2}+p_{3})^{2}\, ,  \quad u=(p_{1}+p_{3})^{2}\, .
\ee
Doing this one
finally arrives at the compact result (reinstating the coupling)
\be
\label{M4MHVfinal}
M_{4}(1^{--},2^{--},3^{++},4^{++})=\i \kappa^{2} \, \frac{\vev{12}^{4}\bev{34}^{4}}{stu} \,,
\ee
with a peculiar pole structure. The correct helicity scaling is easily checked.
Similarly to the gluon case, a closed expression for the MHV $n$-graviton tree-level amplitude may also be 
conjectured and proven via BCFW recursion. Yet, it is more involved and may be found in~\cite{ch2_Berends:1988zp}.

\begin{exer}{The six-gluon split-helicity NMHV amplitude}
\label{Ex:6gNMHV}
Determine the first non-trivial next-to-maximally-helicity-violating (NMHV) amplitude 
$$A_{6}^{\rm tree}(1^{+}, 2^{+}, 3^{+}, 4^{-}, 5^{-}, 6^{-} )$$
from a BCFW recursion relation and our knowledge of the
MHV amplitudes. Consider a shift of the two helicity states $1^{+}$ and $6^{-}$, and show that
\begin{align} \label{eq:A6NMHV}
\begin{aligned}
A_{6}^{\text{tree}}(1^{+},2^{+},3^{+}, 4^{-},& 5^{-},6^{-}) = 
\i g^4 \biggl( \frac{\langle6|P_{12}|3]^{3}}{\vev{61}\vev{12}\bev{34}\bev{45}[5|P_{16}|2\rangle }
 \frac{1}{ (p_{6}+p_{1}+p_{2})^{2}} \\
& \phantom{=} + \frac{\langle 4 | P_{56}|1]^{3}}{ \vev{23}\vev{34}\bev{16}\bev{65}[5|P_{16}|2\rangle} 
\frac{1}{ (p_{5}+p_{6}+p_{1})^{2}} \biggl) \, , 
\end{aligned}
\end{align}
where $P_{ij}=p_{i}+p_{j}$. 
For the solution see \hyperref[Sol:6gNMHV]{chapter~5}.
\end{exer}

\begin{exer}{Soft limit of the six-gluon split-helicity amplitude}
\label{Ex:6gSoft}
Check the consistency of the six-gluon split-helicity amplitude of \eqn{eq:A6NMHV} with the soft limit of leg $5$.
For the solution see \hyperref[Sol:6gSoft]{chapter~5}.
\end{exer}

\section{BCFW recursion with massive particles}
\label{sec:2.6}

So far we restricted our attention to amplitudes involving massless particles,
i.e.~gluons, gravitons, and massless fermions and scalars. 
Yet, scattering amplitudes involving massive particles are of great relevance
in physics, and consequently on-shell recursions have been devised for this case as well.
Let us focus here on colour-ordered gauge-theory amplitudes involving also massive
coloured matter fields (for the colour-ordered amplitudes the representation of the matter
field is irrelevant). Concretely, we consider amplitudes involving at least two massless gluons,
which we take to be neighbours for simplicity of the discussion, say at positions $1$ and $n$,
together with $n-2$ massive states:
\be
A_{n}(p_{1},p_{2},\ldots, p_{n})\, , \qquad  p_{1}^{2}=0=p_{n}^{2}\, ,
\qquad p_{i}^{2}=m_{i}^{2} \, .
\ee
Let us now see how the BCFW on-shell recursion
derived in section \ref{sec:2.2} may be generalised to gauge theory
amplitudes with massive particles. We closely follow reference~\cite{ch2_Badger:2005zh}
in our exposition.

As before in \eqn{bcfw-n1}, we consider a complex shift  of the null gluon  momenta 
at positions $1$ and $n$  by a parameter $z\in \mathbb{C}$,
\be
{\hat p}_{1}(z)   = p_{1} - z |n\rangle [1| \, , \qquad
{\hat p}_{n}(z)   = p_{n} + z |n\rangle [1| \, .
\ee
This entails a $z$-shift of the region momentum $P_{i} \coloneqq p_{1}+\ldots p_{i-1}$,
\be
{\hat P}_{i}(z)  = P_{i} - z |n\rangle [1| \, ,  \qquad i\in \{3,\ldots , n-1\}\, ,
\ee
which also featured in the BCFW recursion
relation of \eqn{poleargumentbcfw}. 
Importantly, the on-shellness of the deformed legs $\hat 1$ and $\hat n$ as well as total momentum conservation is preserved under the shift.
As the arguments leading to the BCFW recursion 
relation are purely based on factorisation, they are applicable to a generic quantum field theory involving massive particles as well.
The BCFW recursion was obtained by thinking about the deformed amplitude 
$A_{n}(z)$ as an analytic function in $z$. Its poles arise whenever an internal propagator
associated to the $z$-shifted region momentum ${\hat P}_{i}(z)$ goes on-shell. 
This reasoning does not change in the massive case, i.e.~we will have a pole whenever 
a $z$-shifted region momentum goes on-shell, i.e.~$\hat P^{2}_{i}(z)=m^{2}$ with $i\in\{3,\ldots , n-1\}$ . 
The pole then reads in generalisation of \eqn{poleargumentbcfw} as
\begin{align}
\label{massivepoleargumentbcfw}
\frac{1}{\hat{P}_{i}(z)^{2}-m_{P_{i}}^{2}} & = \frac{1}{(\hat{p}_{1}(z) + p_{2} + \ldots p_{i-1})^2
-m_{P_{i}}^{2}} \nn \\ 
& = \frac{1}{P_{i}^2 -m_{P_{i}}^{2} - z \langle n | P_{i} | 1 \rbrack}\,, 
\end{align}
where $m_{P_{i}}$ is the mass of the associated intermediate particle going on-shell. 
Generalising \eqn{zPidef}, the location of the pole is shifted to
\be
\label{mzpole}
z_{P_{i}} =\frac{P_{i}^2-m_{P_{i}}^{2}}{\langle n | P_{i} | 1 \rbrack}\, , \qquad
\forall i \in \{3,\ldots , n-1\}\, . 
\ee
\begin{svgraybox}{\bf BCFW recursion relation with massive particles.}
Using again the complex analysis arguments of figure \ref{Fig:II.1}, one immediately arrives
at the on-shell recursion relation for amplitudes including massive particles:
\begin{align}
\begin{split}\label{massiveBCFWrec}
A_{n}(1,\ldots, n)  
= \sum_{i=3}^{n-1} \sum_{s \in s_{\rm P}}  &\, 
A_{L}\bigl(\hat 1(z_{P_{i}}),2,\ldots,i-1,-{\hat P}^{\bar{s}}(z_{P_{i}}) \bigr) 
\frac{\i \, n_{\rm P}}{P_{i}^2-m_{P_{i}}^{2}} \\ 
& \times A_{R}\bigl({\hat P}^{s}(z_{P_{i}}),i,\ldots, n-1, \hat n(z_{P_{i}})\bigr)
+ {\rm Res}(z=\infty)\,,
\end{split}
\end{align}
where the sum now is over the spins $s_{\rm P}$ of the intermediate particle $\rm P$ and $n_{\rm P}$ is the particle-dependent constant appearing in the factorisation as described below \eqn{eq:multiparticlepole}. We recall
that the legs $1$ and $n$ are assumed to be massless. 
\end{svgraybox}

Again, this formula is only of use if 
the residue at infinity, ${\rm Res}(z=\infty)$, vanishes. This turns out to be the case if the
gluon helicities of the shifted legs are not of the $[n^{+}1^{-}\rangle$ type, just as in 
\eqn{zshiftscaling}. Hence, the
statement is
\be
{\rm Res}(z=\infty)=0 \quad \text{iff}\quad 
(h_{1},h_{n})= (+,-)\, , \, (+,+)\, , \, (-,-)\, .
\ee
See~\cite{ch2_Badger:2005zh} for a derivation. This renders the massive BCFW
recursion relation~\eqref{massiveBCFWrec} very useful.

\subsection{Four-point amplitudes with gluons and massive scalars}

Let us construct an explicit example. We consider a theory of a massive complex 
scalar field coupled to gauge theory.
Concretely, we want to evaluate the four-point amplitude involving two neighbouring
gluons of positive helicity and two scalars, 
\be
\label{ggpp}
A_{4}(1^{+},2_{\phi},3_{\bar\phi},4^{+})\, .
\ee
The scalars have mass $m^{2}$. This amplitude vanishes
in the massless limit $m=0$, similarly to the vanishing of
the above amplitude when the scalars are replaced by massless fermions, as was
shown in \eqn{I.32}.  In fact, amplitudes of the above type are of interest even
in massless theories at the one-loop level. There, the need to 
regulate divergences leads one to consider internal particles propagating
in $D=4-2\epsilon$ dimensions which may be modelled using masses, as we shall discuss in detail in the next chapter.

Returning to our concrete example we employ the massive on-shell recursion 
of \eqn{massiveBCFWrec}. Only the scalar
channel contributes, 
\be
\label{massive4rec}
A_{4}(1^{+},2_{\phi},3_{\bar\phi},4^{+})=
A_{3}\bigl(\hat 1^{+}, 2_{\phi}, -{\hat P}_{\bar\phi}\bigr)\, \frac{\i}{P^{2}-m^{2}}\,
A_{3}\bigl({\hat P}_{\phi},3_{\bar\phi},\hat{4}^{+}\bigr) \,,
\ee
as an amplitude with a single scalar vanishes, $A_{3}(\hat 1^{+}, 2_{\phi},3^{\pm})=0$.
All that is needed are the $(\phi g \bar\phi)$-amplitudes.
These follow from the colour-ordered Feynman
vertices of two charged scalars and a gluon of \eqn{cOsQCD},
\be
V_{3}(l_{1},p^{\mu},l_{2})=
\raisebox{-0.6cm}{\begin{tikzpicture}
\begin{feynman}
	\coordinate (x) at (-0.7,0);
	\coordinate (o) at (0,0);
	\coordinate (y) at (0.75,0);
		\coordinate (z) at (0,-0.7);
		\diagram*{(x) -- [charged scalar] (o) -- [charged scalar] (y), (o) -- [gluon] (z)};
		\draw [fill] (x) circle (.0) node [left] {$l_{1}$} ;
	\draw [fill] (y) circle (.0) node [right] {$l_{2}$} ;
	\draw [fill] (o) circle (.06);
		\draw [fill] (z) circle (.0) node [right] {$\mu$} node [left] {$p$};
    \end{feynman}
	\end{tikzpicture}}
= \i \frac{g}{\sqrt{2}}\, (l_{2}^{\mu}-l_{1}^{\mu})\, ,
\ee
where $1$ ($2$) denotes a $\bar\phi$ ($\phi$) leg, respectively.
Contracting this with the positive-helicity gluon polarisation of \eqn{s1polarrel} one
obtains the on-shell three-point amplitudes (setting $g=1$)
\be
\label{ssp}
A_{3}(l_{1\, \bar{\phi}},p^+,l_{2 \, \phi})= -\i\frac{\langle r|l_{1}|p]}{\vev{rp}} =
A_{3}(l_{1\, {\phi}},p^+,l_{2 \, \bar\phi})\, ,
\ee
where the last relation follows by reflection.
Note that here $r$ is the arbitrary null reference 
momentum of the gluon leg related to the local gauge invariance of the theory.
By similar arguments one establishes the three-point amplitudes involving a negative
helicity gluon:
\be
\label{ssm}
A_{3}(l_{1\, \bar{\phi}},p^-,l_{2 \, \phi})=-\i \frac{\langle p|l_{1}|r]}{[pr]} =
A_{3}(l_{1\, {\phi}},p^-,l_{2 \, \bar\phi})\, .
\ee
Before moving on with the recursion, let us address a seemingly dramatic problem: 
the amplitudes of \eqn{ssp} and \eqn{ssm} apparently depend on the
reference momentum $r$---how can that be?
It turns out that, despite their representation, the amplitudes in \eqn{ssp} and \eqn{ssm}
are actually \emph{independent} of the choice of $r$. Taking the initial reference spinor
$|r\rangle$ and $|p\rangle$ as a basis in Weyl spinor space, we may parametrise 
an arbitrary reference spinor different from $|r\rangle$ as $|r'\rangle = \alpha |r\rangle + \beta |p\rangle$. Clearly, \eqn{ssp} is invariant under rescaling of the reference spinor $|r\rangle \to 
\Lambda\, |r\rangle$, thus without loss of generality
we may parameterise $|r'\rangle = |r\rangle + \gamma |p\rangle$, or  infinitesimally write
$\delta_{r}|r\rangle \propto |p\rangle$. This entails that the amplitude \eqn{ssp}  changes under
a variation of the reference spinor $|r\rangle$  as
\be
\delta_{r}A_{3}(l_{1}^{+},p^+,l_{2}^{-}) \propto \frac{\langle p|l_{1}|p]}{\vev{rp}}
=0 \, ,
\ee
where the vanishing follows from $\langle p|l_{1}|p]=2p\cdot l_{1}=0$, which is a consequence of 
the three-point kinematics:
\be
(l_{1}+p)^{2} = l_{2}^{2} \quad \rightarrow\quad  l_{1}\cdot p =0 \quad \text{as}
\quad l_{1}^{2}=l_{2}^{2}=m^{2}\, ,\,  p^{2}=0\, .
\ee
A similar argument applies to \eqn{ssm}. Again we see the subtleties in three-point amplitudes: the expressions in eqs.~\eqref{ssp} and~\eqref{ssm} are actually independent 
of $r$.
\begin{figure}[t]
   \centering
    \begin{tikzpicture}[baseline={(current bounding box.center)}]
  \begin{feynman}
  \coordinate (i) at (30:2.5);
\coordinate (n) at (-30:2.5);
\coordinate (1) at (-150:2.5);
\coordinate (2) at (150:2.5);
\coordinate (x) at (-1.2,0);
\coordinate (y) at (1.2,0);
\draw [charged scalar, momentum=$P$] ( 0.8,0) to  ( -0.8,0);
\draw [gluon] (x) -- ( 1);
\draw [charged scalar] (x) -- (2);
\draw [charged scalar] (i) -- (y); 
\draw [gluon] (y) -- (n); 
  \draw (1) node [below] {$\hat{1}^{+}$};
  \draw (2) node [above] {$2_{\phi}$};
  \draw (n) node [below] {$\hat{4}^{+}$};
           \draw (i) node [above] {$3_{\bar\phi}$};
 \draw [fill=lightgray,thick, draw=black] (x) circle (0.4);
 \draw (x) node {$A_{3}$};    
\draw [fill=lightgray,thick, draw=black] (y) circle (0.4);
 \draw (y) node {$A_{3}$};  
\end{feynman}
\end{tikzpicture}
   \caption{On-shell recursion for the massive 
   $A_{4}(1^{+},2_{\phi},3_{\bar\phi},4^{+})$ amplitude. All external momenta are
   outgoing, $P$ runs from right to left.}
   \label{fig:mss2}
\end{figure}
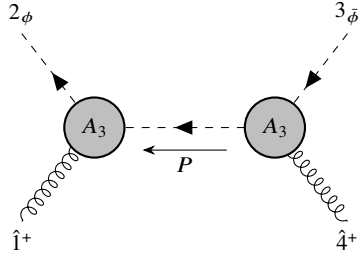

Coming back to the recursive construction of the amplitude $A_{4}(1^{+},2_{\phi},3_{\bar\phi},4^{+})$, we have
(cf.\ figure~\ref{fig:mss2})
\begin{align} \begin{aligned}
A_{4}(1^{+},2_{\phi},3_{\bar\phi},4^{+})&= A_{3}\bigl(-{\hat P}_{\bar\phi},\hat 1^{+},2_{\phi}\bigr)\, \frac{\i}{P^{2}-m^{2}}
\, A_{3}\bigl(3_{\bar\phi},\hat 4^{+},{\hat P}_{\phi} \bigr)\, \\
&= \left( \i \, \frac{\langle r_{1}|\hat P |\hat 1]}{\vev{r_{1}\hat 1}} \right) \frac{\i}{P^{2}-m^{2}} \left( - \i \,
 \frac{\langle r_{4}|p_{3}|\hat 4]}{\vev{r_{4}\hat 4}} \right) \,,
\end{aligned} \end{align}
with $P=p_{1}+p_{2}$.
Here $r_{1/4}$ denote the reference momenta of the gluon legs 1 and 4. Things are simplified
considerably with the gauge choice
\be
r_{1}=\hat p_{4}\, , \qquad r_{4}=\hat p_{1}\, .
\ee
Noting that $|\hat 1]=|1]$ and $|\hat 4 \rangle = |4\rangle$, we then have
\begin{align}
\begin{alignedat}{2}
& \langle r_{1}|\hat P|\hat 1] =  \langle 4|\hat P |1] = \langle 4|P|1] = - \langle 4|p_{3}|1] \, , \quad \quad
  && \vev{r_{1}\hat 1}=\vev{4\hat 1}=\vev{41}\, ,  \\
& \langle r_{4}|p_{3}|\hat 4] =  \langle \hat 1|p_{3} |\hat 4]\, , 
  && \vev{r_{4}\hat 4}=\vev{\hat 1 4 }=\vev{14}\, . \\
\end{alignedat} \end{align}
Plugging these into the above we find
\be
A_{4}= \i\frac{\langle 4 |p_{3}|1]\, \langle \hat 1 | p_{3}|\hat 4]}{\vev{14}^{2}\, 
[(p_{1}+p_{2})^{2} -m^{2}]}\, .
\ee
The numerator may be simplified with a trace identity (see \eqn{eq:trSigma2gamma}) to
\be
\langle 4 |p_{3}|1]\, \langle \hat 1 | p_{3}|\hat 4] = \frac{1}{2} \, \Tr \bigl(
\hat{\slashed{p}}_{4}\, \slashed{p}_{3} \,
\hat{\slashed{p}}_{1}\, \slashed{p}_{3} \,
 \bigr) = 2 \, \bigl ( \, 2 (p_{3}\cdot \hat p_{4})\, (p_{3}\cdot \hat p_{1})
 - p_{3}^{2}\, (\hat p_{1}\cdot \hat p_{4})\, \bigr )\, .
\ee
In fact $p_{3}\cdot \hat p_{4}=0$, which follows from momentum conservation:
\be
\hat P^{2} = (p_{3}+\hat p_{4})^{2}\quad \Rightarrow \quad
m^{2}= 2p_{3}\cdot \hat p_{4} + p_{3}^{2}\quad \Rightarrow \quad
p_{3}\cdot \hat p_{4}=0\, .
\ee
Moreover, we find $2\hat p_{1}\cdot \hat p_{4}=\vev{\hat 14}\bev{\hat 41}= \vev{14}\bev{41}$.
Putting everything together we arrive at the compact expression for the 
four-point amplitude
\be
\label{gpgpss}
A_{4}(1^{+},2_{\phi},3_{\bar\phi},4^{+})= \i \, \frac{\bev{14} \, m^{2}}{\vev{14}\,
[(p_{1}+p_{2})^{2} -m^{2}]}\, ,
\ee
which indeed vanishes in the massless limit, as claimed. 

\vspace{-0.1cm}

\begin{exer}{Mixed-helicity four-point scalar-gluon amplitude}
\label{Ex:2.helicity}
Compute the four-point massive-scalar-gluon amplitude with one positive and one negative gluon,
\begin{align}
A_{4}(1^{+},2_{\phi},3_{\bar\phi},4^{-}) =  \i \,
\frac{\langle 4| p_{3}| 1]^{2}}{ (p_1+p_4)^2 \, [(p_1+p_2)^2-m^2] } \,,
\end{align}
using the above recursive techniques. 
For the solution see \hyperref[Sol:2.helicity]{chapter~5}.
\end{exer}

\section{Symmetries of scattering amplitudes}
\label{sec:2.4}

We now turn to a more conceptual yet very  important subject: the question of how the
space-time symmetries of the Poincar\'e group (and beyond) that we discussed in section~\ref{sec:1.1} manifest themselves at the level of
scattering amplitudes. This has proven to be a very rich subject
in particular at tree level. The space-time symmetries of scattering amplitudes may be
grouped into obvious and less obvious symmetries.

The obvious symmetries are the Poincar\'e 
transformations of section~\ref{sec:1.1}, under which scattering amplitudes should be invariant. 
Sticking to massless amplitudes,  working in the spinor-helicity formulation of momentum space 
is highly advantageous. Here
the momentum generator $p^{\a\da}$ is represented by a multiplicative operator,
\begin{align}
p^{\a\da} & = \sum_{i=1}^n \la_i^\a \, \tla_i^\da\, , \quad
\end{align}
and any amplitude $\cA_n$ must obey
\be
p^{\a\da}\, \cA_n(\{\la_i,\tla_i\}) = 0\, .
\ee
This is in fact true in the distributional sense of the relation
$
p\, \delta(p)=0\,,
$
thanks to the total momentum conserving
delta-function present in each amplitude (that we often drop):
\be
 \cA_n(\la_i,\tla_i)=\delta^{(4)}\biggl(\sum_i p_i\biggr) \, A_n \left(\la_i,\tla_i \right) \,.
\ee
In our notation we distinguish the full amplitude $\cA_{n}$ and the delta-function ``stripped'' amplitude~$A_{n}$.

The Lorentz generators in the helicity spinor basis
come in two pairs of symmetric rank-two tensors
$m_{\a\b}$ and ${\overline m}_{\da\db}$, originating from the projections
$M^{\mu\nu}\, (\sigma_{\mu\nu})_{\a\b} = m_{\a\b}$ and 
$M^{\mu\nu}\, (\bar\sigma_{\mu\nu})_{\da\db} ={{\overline m}}_{\da\db}$.
They are first-order differential operators
in helicity spinor space,
\be
m_{\a\b} = \sum_{i=1}^n \la_{i\, (\a}\, \partial_{i\,\b)}\, , \quad
{{\overline m}}_{\da\db} = \sum_{i=1}^n \tla_{i\, (\da}\, \partial_{i\, \db)}\, ,
\ee
where $\partial_{i\a} \coloneqq \frac{\partial}{\partial \la_i^\a}$, 
$\partial_{i\da} \coloneqq \frac{\partial}{\partial \tla_i^\da}$ and
$r_{(\a\b)} \coloneqq (r_{\a\b} + r_{\b\a})/2$ denotes symmetrisation
with unit weight, cf.~exercise \ref{Ex:1.5}.
The invariance of $A_n(\la_i,\tla_i)$ under Lorentz-transformations,
\be
m_{\a\b}\, A_n(\la_i,\tla_i) = 0 = {{\overline m}}_{\da\db}\, A_n(\la_i,\tla_i)\,,
\ee
is manifest, as it is an immediate consequence of the proper 
contraction of all Weyl indices within $A_n$, i.e.~the fact
that the spinor brackets $\vev{ij}$ and $\bev{ij}$ are invariant under $m_{\a\b}$
and ${{\overline m}}_{\da\db}$. See the solution of exercise~\ref{Ex:1.5} for an explicit
calculation.

Let us now discuss a set of less obvious symmetries of $\cA_n(\la_i,\tla_i)$ in the
case of pure colour-ordered gluon amplitudes. Classical
Yang-Mills theory is in fact invariant under a larger symmetry group than Poincar\'e: 
due to the absence of dimensionful parameters in the theory (the coupling
$g$ is dimensionless) pure
Yang-Mills theory (as well as massless QCD or scalar QCD) is invariant under a scale transformation,
\be
x^\mu \to \Lambda^{-1}\, x^\mu\, , \quad \text{respectively} \quad
p^\mu \to \Lambda\, p^\mu\, .
\ee
The scale transformations of the momenta are generated by the dilatation operator $d$,
whose representation in spinor-helicity variables acting on amplitudes is
\be
d= \sum_{i=1}^n\, \left( \frac{1}{2} \la^\a_i\, \partial_{i\,\a} + 
 \frac{1}{2} \tla^\da_i\, \partial_{i\,\da} + d_0 \right) \,, \qquad d_0 \in \R \,,
\label{d_def}
\ee
reflecting the mass dimension $\nicehalf$ of the $\la_i$ and $\tla_i$ helicity spinors, 
i.e.~$[d,\lambda_i]=\lambda_i/2$ and $[d,\tla_i]= \tla_i/2$.
The constant $d_0$ is undetermined at this point. It may be fixed by requiring invariance
of the MHV gluon amplitudes,
\be
\cA_n^\text{MHV} =  \delta^{(4)}(p)\, \frac{\vev{st}^4}
{\vev{12}\ldots\vev{n1}}\, ,
\ee
where $p = \sum_{i=1}^n p_i$.
The dilatation operator $d$ of \eqn{d_def} simply measures the weight in units of mass
of the amplitude it acts on adding a factor of $n\, d_0$, namely
\be
d\, \cA_n = \bigl([\cA_n] + n\, d_0 \bigr)\, \cA_n \, ,
\ee
where $[\mathcal{O}]$ indicates the weight of $\mathcal{O}$ in dimensions of mass.
We note the weights $[\delta^{(4)}(p)]=-4$, $[\vev{st}^4]=4$ and
$[\vev{12}\ldots\vev{n1}]=n$, hence
\be
d\, \cA_n^\text{MHV} = (-4 + 4-n +n\, d_0) \, \cA_n^\text{MHV} \, ,
\ee
which vanishes for the choice $d_0=1$. 
One easily checks the invariance under dilatations of the $q\bar q gg$-amplitude of 
\eqn{qqgg-amp} and of the $\widebar{\text{MHV}}{}_n$ amplitudes as~well.

The scaling symmetry comes with a further less obvious symmetry of vectorial nature
known as special conformal transformations. Their generators, denoted by $k_{\a\da}$, are
realised in terms of a second-order differential operator in the spinor variables,
\be
k_{\a\da}= \sum_{i=1}^n \partial_{i\, \a}\, \partial_{i\, \da} \, .
\ee
Together with the Poincar\'e and dilatation
generators, the set $\{p_{\a\da}, k_{\a\da}, m_{\a\b}, {{\overline m}}_{\da\db}, d\}$ generates
the \emph{conformal group} in four dimensions, $\text{SO}(2,4)$.

Let us now prove the invariance of the MHV gluon amplitudes under special conformal
transformations. As the only dependence of $\cA^\text{MHV}_n$ on the conjugate
spinors $\tla_i$ resides in the momentum conserving delta-function, we have
\begin{align}
k_{\a\da} \, \cA^\text{MHV}_n &= 
\sum_{i=1}^n \frac{\partial}{\partial \la^\a_i}\,\frac{\partial}{\partial \tla^\da_i}\,
\left (\delta^{(4)}(p)\, A^\text{MHV}_n \right) \nn \\ &= 
\sum_{i=1}^n \frac{\partial}{\partial \la^\a_i}\,
\left[
\frac{\partial p^{\b\db}}{\partial\tla_i^\da}\, \left( \frac{\partial}{\partial p^{\b\db}}
\,\delta^{(4)}(p)\, \right) \, A_n^\text{MHV}\right]  \nn\\
&= \biggl [ \biggl( n\, \frac{\partial}{\partial p^{\a\da}} + p^{\b\db}\, 
\frac{\partial}{\partial p^{\b\da}} \, \frac{\partial}{\partial p^{\a\db}} \biggr)\, \delta^{(4)}(p)
\,   \biggr ] \, A^\text{MHV}_n \nn\\ &\qquad  + \biggl(\frac{\partial \,\delta^{(4)}(p)}{\partial p^{\b\da}}\biggr)\,
\sum_{i=1}^n\, \la_i^\b\, \frac{\partial}{\partial\la^\a_i}\, A_n^\text{MHV} \, .
\label{kAMHV}
\end{align}
The last term may be rewritten as follows. First, we note the relation
\be
\label{symlapa}
\sum_{i=1}^n\la_{i\, \a}\, \partial_{i\, \b} = \sum_{i=1}^n
\la_{i\, (\a}\, \partial_{i\, \beta)} + \frac{1}{2} \, \epsilon_{\a\b}\, \sum_i\la^\gamma_i\, 
\partial_{i\, \gamma} \, ,
\ee
which follows from decomposing the left-hand side in a symmetric and anti-symmetric piece.
The first term on the right-hand side is the Lorentz generator $m_{\a\b}$, which we already know annihilates
$A_n^\text{MHV}$. The remaining term yields
\be
\sum_{i=1}^n \la^\b_i\, \frac{\partial}{\partial\la^\a_i}\, A_n^\text{MHV}=
\frac{1}{2} \, \delta_\a^\b\, \sum_i \la^\delta_i\, \partial_{i\, \delta}\,  A_n^\text{MHV}=
(4-n)\, A_n^\text{MHV}\, .
\ee
Hence \eqn{kAMHV} turns into
\be
k_{\a\da} \, \cA^\text{MHV}_n =
\biggl [ \biggl( 4\, \frac{\partial}{\partial p^{\a\da}} + p^{\b\db}\, 
\frac{\partial}{\partial p^{\b\da}} \, \frac{\partial}{\partial p^{\a\db}} \biggr)\, \delta^{(4)}(p)
\,   \biggr] \, A^\text{MHV}_n \, .
\ee
Indeed in a distributional sense we have
\be
p^{\b\db}\, 
\frac{\partial}{\partial p^{\b\da}} \, \frac{\partial}{\partial p^{\a\db}}\, \delta^{(4)}(p)
=-4 \, \frac{\partial}{\partial p^{\a\da}}\, \delta^{(4)}(p)\, ,
\ee
which one may see by integrating the second derivative expression against a test function $F(p)$:
\begin{align} \begin{aligned}
\int \d^4 p &\, F(p)\, p^{\b\db}\, 
\frac{\partial}{\partial p^{\b\da}} \, \frac{\partial}{\partial p^{\a\db}}\, \delta^{(4)}(p)
= \\
&
= \int \d^4 p \left[ \left( \frac{\partial}{\partial p^{\b\da}}\, F(p)\right) \, 2\, \delta^{\beta}_{\alpha}
+ \left( \frac{\partial}{\partial p^{\a\db}}\, F(p)\right) 2\, \delta^{\db}_{\da} 
\right] \\
&= 4\, \int \d^4 p  \, \left(\frac{\partial}{\partial p^{\a\da}}\, F(p) \right)\, \delta^{(4)}(p)
\, .
\end{aligned} \end{align}
This proves the vanishing of $k_{\a\da} \, \cA^\text{MHV}_n$, as claimed.

Summarising, we have constructed a representation of the conformal group whose generators are represented
by differential operators of degree one ($m_{\a\b}$, ${{\overline m}}_{\da\db}$, $d$), of degree
two ($k_{\a\da}$), and as a multiplicative operator ($p_{\a\da}$) in an $n$-particle
 helicity spinor
space. This representation is natural, as amplitudes are functions in this
space. All these generators
annihilate the scattering amplitudes. 
We have verified this explicitly for the MHV amplitudes.
The representation obeys the
 commutation relations of the conformal algebra $\alg{so}(2,4)$, 
\begin{align} \label{confalgebra}
\begin{gathered} 
[d,p^{\a\da}]  = p^{\a\da} \,, \quad \ 
  [d,k_{\a\da}] = -k_{\a\da} \,, \quad \ 
  [d,m_{\a\b}]=0=[d,{{\overline m}}_{\da\db}] \,, \\
[k_{\a\da}, p^{\b\db}] = \delta_\a{}^\b \, \delta_\da{}^\db \, d + m_\a{}^\b\, \delta_\da{}^\db 
  +{{\overline m}}_\da{}^\db\, \delta_\a{}^\b \,,
\end{gathered}
\end{align}
plus the Poincar\'e commutators discussed in section~\ref{sec:1.1}.
 
The origin of this helicity spinor space representation becomes clear if one looks at the
more familiar representation of the conformal group in configuration space $x^\mu$. 
 For scalar fields, it reads
\begin{align}
\begin{alignedat}{2}
M_{\mu\nu} & = \i \, \bigl(x_\mu\, \partial_\nu -x_\nu\, \partial_\mu\bigr) \,, \qquad \qquad 
&& P_\mu = - \i \, \partial_\mu \,, \\
\mathcal{D} & =- \i \, (x^\mu\, \partial_\mu  + \Delta)\,,
&& K_\mu = \i \left[ x^2\, \partial_\mu - 2 \, x_\mu \, (x^\nu \partial_\nu 
+\Delta) \right] \,, 
\end{alignedat}
\label{confalgebra-config}
\end{align}
where $\Delta$ is the scaling dimension and $\partial_\mu \coloneqq
\partial/\partial x^\mu$.
In quantum field theory the conformal symmetry is distinguished by the 
Haag-Lopuszanski-Sohnius theorem~\cite{ch2_Haag:1974qh} as the maximal bosonic 
extension of the space-time symmetry of the $S$-matrix---generalising the
Poincar\'e algebra.

A Fourier transform $\int \d^4x \, \mathrm{e}^{\i p\cdot x} \cO(x,\partial_x)$ brings this 
representation into momentum space. From this
point of view, it is clear why $p^{\a\da}$ becomes a multiplication operator and
$k_{\a\da}$ a second-order derivative operator in momentum space as seen above.
The momentum-space representation of the conformal symmetry as it applies to
scattering amplitudes then takes the form
\begin{align}
\begin{alignedat}{2}
m^{\mu\nu} & = p^\mu\, \partial^\nu_{p} -p^{\nu}\, \partial^{\mu}_{p} \,, \qquad \qquad 
&& p^{\mu} = p^{\mu} \,, \\
d & = p_\mu\, \partial^{\mu}_{p}  + \bar\Delta\,,
&& k^{\mu} = p^{\mu} \partial_{p}^{2}- 2\bigl(p_{\nu}\partial^{\nu}_{p}+\bar{\Delta} \bigr)\partial^{\mu}_{p} \,,
\end{alignedat}
\end{align}
where $\bar{\Delta}=4-\Delta$.
This representation may be mapped to the helicity-spinor one discussed above.

\begin{important}{The conformal generators in spinor helicity space.}
Here we collect the generators of the conformal algebra in their single-particle action (with $\bar\Delta=1$, which is relevant for gauge bosons and scalar fields):
\begin{align}
\label{spinhelconfalg}
\begin{alignedat}{2}
p^{\alpha \dot{\alpha} } &= \lambda^{\alpha} {\tilde\lambda}^{\dot{\alpha}} \,, \qquad \qquad \qquad \quad
  k_{\alpha \dot{\alpha}} && = \partial_{\alpha} \partial_{\dot{\alpha}} \,, \\
m_{\alpha \beta} &= \lambda_{( \alpha} \partial_{\beta ) } \,,\qquad \qquad \qquad \quad
  \overline{m}_{ \dot{\alpha} \dot{\beta} } && = \tilde{\lambda}_{( \dot{\alpha}}  \partial_{\dot{\beta})} \,, \\
d & = \ft{1}{2} \lambda^{\alpha} \partial_\alpha + \ft{1}{2} \tilde{\lambda}^{\dot{\alpha}} \partial_{\dot{\alpha}} + 1 \,, && \\
\end{alignedat} 
\end{align} 
where $r_{(\a\b)} \coloneqq (r_{\a\b} + r_{\b\a})/2$ denotes symmetrisation of the indices.
The helicity generator is given by $h= -\frac{1}{2} \lambda^{\alpha} \partial_\alpha + \frac{1}{2} \tilde{\lambda}^{\dot{\alpha}} \partial_{\dot{\alpha}}$. It commutes with all generators of the conformal algebra.
\end{important}
Scaleless quantum field theories---such as pure Yang-Mills or massless QCD---enjoy conformal 
symmetry at the tree-level. Yet, this symmetry is usually broken at the loop level, as the need to regularise divergencies introduces a scale into the quantum theory. 
It is manifested by a non-vanishing $\beta$-function of the coupling $g$.
In fact,  
understanding the implications of broken conformal symmetry for loop amplitudes is an area of ongoing research. In the maximally supersymmetric generalisations of Yang-Mills theory,
$\mathcal{N}=4$ super Yang-Mills theory, this property is prominently absent. It is a quantum conformal
field theory, with far-reaching consequences, including a hidden infinite dimensional
symmetry known as the Yangian symmetry. In fact, tree-level amplitudes are invariant
under this extension of the super-conformal group \cite{ch2_Drummond:2009fd} and the
hidden integrability of the leading colour limit of the theory allows for exact
non-perturbative results, see~\cite{ch2_Beisert:2010jr} for a review.

\newpage
\begin{exer}{Conformal algebra}
\label{Ex:2.confalg}
Show that the representation constructed in the above \eqn{spinhelconfalg} indeed obeys
the commutation relations of the conformal algebra given in \eqn{confalgebra}.
For the solution see \hyperref[Sol:2.confalg]{chapter~5}.
\end{exer}

\begin{exer}{Inversion and special conformal transformations}
\label{Ex:2.invspecialconf}
The generator $K_\mu$ of \eqn{confalgebra-config} generates infinitesimal special conformal transformations. 
A finite special conformal transformation is given by
\be
\quad x^\mu \to x^{\prime\, \mu} = \frac{x^\mu-a^\mu\, x^2}{1-2 \, a\cdot x + a^2\, x^2} \,,
\label{finiteKaction}
\ee
where $a^\mu$ is the transformation parameter. 
\begin{enumerate}[a)]

\item An intuition on the character of these transformations may be found by noting that
the action of $K^\mu$ may be also written as $K^\mu= I\,P^\mu \, I$, i.e.~as the composition of an inversion
$I\, x^\mu = x^\mu/x^2$, followed by a translation $P^{\mu} x = x^{\mu} - a^{\mu}$, followed by
another inversion. Show that $K^\mu= I\, P^\mu \, I$ is equivalent to 
\eqn{finiteKaction}.

\smallskip
\item  A scalar field $\Phi(x)$ transforms under special conformal transformations $x\to x'$~as 
\begin{align} \label{eq:conformal_scalar}
\Phi(x) \longrightarrow \Phi'(x') = \left| \frac{\partial x'}{\partial x} \right|^{-\Delta/4} \Phi(x) \,,
\end{align}
where $|\partial x'/\partial x|$ is the Jacobian of the transformation, and $\Delta$ is the scaling dimension.
Show that the generator $K_\mu$ of \eqn{confalgebra-config} indeed generates the transformation of \eqn{finiteKaction} for a scalar field.
In other words, prove that
\begin{align} \label{eq:fxprime}
\Phi^{\prime}\left( x \right ) = \left[ 1 - \i \, a^{\mu} \, K_{\mu} +  \mathcal{O}\bigl(a^2\bigr) \right] \, \Phi(x) \,.
\end{align}
Hint: in order to compute the Jacobian factor, treat the finite special conformal transformation as the composition of inversion, translation and inversion.

\end{enumerate} 
\noindent
For the solution see \hyperref[Sol:2.invspecialconf]{chapter~5}.
\end{exer}

\section{Double-copy relations for gluon and graviton amplitudes}
\label{sec:2.5}

So far we have discussed gauge theories and gravity
rather in parallel. While their Lagrangians and Feynman rules look very different,
there exist intriguing relationships between gluon amplitudes and graviton amplitudes that suggest
a deeper relationship of these two theories---at least in their perturbative domain. In a nutshell, they express
gravity as the square of Yang-Mills theory, a property we already saw at the level of the polarisations and of the three-point amplitudes.

\subsection{Lower-point examples}

The squaring relation between gravity and gluon amplitudes
 is manifest at the level of three-point amplitudes:
\bal{
M_{3}^{\text{tree}}(1^{--},2^{--},3^{++}) &= \frac{\vev{12}^{6}}{\vev{23}^{2}\vev{31}^{2}} = \bigl[ A_{3}^{\text{tree}}(1^{-},2^{-},3^{+}) \bigr]^{2} \,,
\\
M_{3}^{\text{tree}}(1^{--},2^{++},3^{++}) &= \frac{\bev{23}^{6}}{\bev{12}^{2}\bev{31}^{2}} = \bigl[ A_{3}^{\text{tree}}(1^{-},2^{+},3^{+}) \bigr]^{2}
\, .}
For simplicity we set all couplings to unity here and absorbed a factor of $\i$ in the amplitudes. 
Hence, for any choice of polarisations we find
\be
\label{KLT1}
M_{3}^{\text{tree}}(1,2,3) = A_{3}^{\text{tree}}(1,2,3)^{2}\, .
\ee
Turning to the four-point case, we need to compare the MHV${}_{4}$ gluon amplitude
and the  MHV${}_{4}$ four-graviton amplitude of \eqn{M4MHVfinal}.
We first expose the $s$, $t$, $u$ poles in the MHV colour-ordered
gluon amplitude
explicitly. For $A_{4}(1^{-},2^{-},3^{+},4^{+})$ we have
$$
\frac{\vev{12}^{3}}{\vev{23}\vev{34}\vev{41}} = -\frac{1}{s_{12}}\frac{\vev{12}^{3} \bev{34}}{\vev{23}\vev{41}} =
-\frac{1}{s_{12}s_{23}}\frac{\vev{12}^{2} \overbrace{\vev{12}\bev{32}}^{\vev{14}\bev{43}}\bev{34}}{\vev{41}} =
-\frac{\vev{12}^{2}\bev{34}^{2}}{s_{12}s_{23}}\, .
$$
This gets close to the four-graviton amplitude~\eqref{M4MHVfinal} but is not just a simple
square. Looking at $A_{4}(1^{-},2^{-},4^{+},3^{+})$, that is obtained by swapping $3\leftrightarrow 4$ in the above, we arrive at
\be
\label{KLT2}
M_{4}^{\text{tree}}(1^{--},2^{--},3^{++},4^{++}) = s_{12}\,  A_{4}^{\text{tree}}(1^{-},2^{-},3^{+},4^{+})\, A_{4}^{\text{tree}}(1^{-},2^{-},4^{+},3^{+})\, .
\ee 
Again resulting in a squaring relation between the two. In fact, such squaring relations turn out to
be generally true for all multiplicities. 

\subsection{Colour-kinematics duality: a four-point example}

The relations \eqref{KLT1} and \eqref{KLT2} suggest a general squaring relation of
the structure $M_{n}\sim A_{n}^{2}$. This can be made precise in the context of the colour-kinematic duality, which we now want to discuss in a four-point example.

For this, let us look at the \emph{coloured} tree-level four-gluon amplitude in $D$ dimensions
using the polarisation $\epsilon_{i}^{\mu}$ and momentum $p_{i}^{\mu}$ vectors 
with $i=1,\ldots,4$ to describe the kinematics. 
It may be written as 
\be
\label{fullcolour4gluon}
\mathbf{A}_{4}^{\text{tree}}= -\i g^{2} \left ( \frac{n_{s}c_{s}}{s} + \frac{n_{t}c_{t}}{t}
+\frac{n_{u}c_{s}}{u}\right ) \,,
\ee
split into the $s$, $t$ and $u$ channels, with
\begin{align}
\label{cs}
& c_{s} = -2 f^{a_{1}a_{2}e}\, f^{ea_{3}a_{4}} \,, \\ 
\label{ns}
& \begin{aligned}
  n_{s} = -\frac{1}{2}\Bigl \{ & \bigl[ (\epsilon_{1}\cdot\epsilon_{2})\, p_{1}^{\mu} +2 (\epsilon_{1}\cdot p_{2})\, \epsilon^{\mu}_{2} - (1\leftrightarrow 2) \bigr] \\
  & \times \bigl[ (\epsilon_{3}\cdot\epsilon_{4})\, p_{3,\, \mu} +2 (\epsilon_{3}\cdot p_{4})\, \epsilon_{4,\,\mu} - (3\leftrightarrow 4) \bigr] \\ 
  & + s \, \bigl[ (\epsilon_{1}\cdot\epsilon_{3})(\epsilon_{2}\cdot\epsilon_{4}) - (\epsilon_{1}\cdot\epsilon_{4})(\epsilon_{2}\cdot\epsilon_{3})
\bigr]\Bigr \}\,, 
\end{aligned}
\end{align}
and 
\be
c_{t}\, n_{t} = c_{s}\, n_{s}\Bigr |_{1\to 2\to 3\to 1}\, , \qquad
c_{u}\, n_{u} = c_{s}\, n_{s}\Bigr |_{1\to 3\to 2\to 1}\, .
\ee
In writing the amplitude in this fashion we have split up 
contact terms emerging from the four-gluon vertex~\eqref{V4YM} into the $s$, $t$ and $u$
channels by multiplying the corresponding colour factors $n_{s}$ by $\frac{s}{s}$, and
so on. These are the terms proportional to $s$ in the last line of \eqn{ns}. It is 
instructive to study the gauge invariance of leg $4$. Replacing $\epsilon_{4}\to p_{4}$ yields
\be
\label{gfns}
 n_{s}\Bigr |_{\epsilon_{4}\to p_{4}} = \frac{s}{2} \, \Bigl [\,
  (\epsilon_{1}\cdot\epsilon_{2})  (\epsilon_{3}\cdot p_{12}) + \text{cyclic}(1,2,3)\, \Bigr ]
\eqqcolon s\, \alpha(\{\epsilon_{i},p_{i}\}) \,,
\ee 
where $p_{12}=p_{1}-p_{2}$. Crucially, the function $\alpha$ is cyclically invariant.  Therefore the gauge transformations of the other kinematical numerators read
\be
\label{gfntu}
n_{t}\Bigr |_{\epsilon_{4}\to p_{4}} = t\, \alpha(\{\epsilon_{i},p_{i}\})\, , \qquad
n_{u}\Bigr |_{\epsilon_{4}\to p_{4}} = u\, \alpha(\{\epsilon_{i},p_{i}\})\, .
\ee
Hence, the numerators are not gauge invariant individually. This is to be expected, as only the full amplitude and not individual graphs (or parts thereof) are gauge invariant. How does
$\mathbf{A}_{4}^{\text{tree}}$ then become gauge invariant? We have
\be
\mathbf{A}_{4}^{\text{tree}}\Bigr |_{\epsilon_{4}\to p_{4}} = (c_{s}+c_{t} + c_{u})\,
\alpha(\{\epsilon_{i},p_{i}\})\, ,
\ee
which is zero by virtue of Jacobi's identity~\eqref{Jacobif2},
\be
c_{s}+c_{t} + c_{u}= -2 \, \bigl(f^{a_{1}a_{2}e}\, f^{ea_{3}a_{4}}+ 
f^{a_{2}a_{3}e}\, f^{ea_{1}a_{4}}+ f^{a_{3}a_{1}e}\, f^{ea_{2}a_{4}} \bigr) =0\, .
\ee
Remarkably, one also sees that the kinematical numerators $n_{i}$ obey a Jacobi-like 
identity, 
\be
\label{kinematicJacobi}
n_{s}+n_{t}+n_{u}=0\, ,
\ee 
upon using the on-shell identities. This is known as the \emph{kinematical Jacobi identity}.
This property allows us to construct a gauge invariant object that has all the required properties
to be the four-graviton amplitude: we simply replace the colour numerators $c_{i}$ by the 
kinematical ones $n_{i}$ in \eqn{fullcolour4gluon}, obtaining
\be
\label{4gravitonamp}
M_{4}^{\text{tree}} =\mathbf{A}_{4}^{\text{tree}}\biggr|\raisebox{-0.2cm}{{\footnotesize $\begin{aligned} & c_{i}\to n_{i} \\[-12pt] & g\to\kappa/2\end{aligned}$}}= - \i \left (\frac{\kappa}{2}\right )^{2}\, 
\left ( \frac{n_{s}^{2}}{s} + \frac{n_{t}^{2}}{t}
+\frac{n_{u}^{2}}{u}\right )\, .
\ee
Clearly, it is bi-linear in the polarisation vectors $\epsilon_{i}^{\mu}$ and displays a consistent pole structure, which are
necessary ingredients for it to be a graviton amplitude. Also the gauge invariance may be tested
straightforwardly using eqs.~\eqref{gfns} and~\eqref{gfntu},
\be
\label{gfM}
 M_{4}^{\text{tree}}\Bigr |_{\epsilon_{4}\to p_{4}} = 
 2\, (n_{s}+n_{t}+n_{u})\, \alpha(\{\epsilon_{i},p_{i}\})\, ,
 \ee
which now vanishes by virtue of the kinematical Jacobi identity~\eqref{kinematicJacobi}. We shall show
momentarily that the result~\eqref{4gravitonamp} is equivalent to \eqn{KLT2}. In order to do so,
let us express the gluon amplitude~\eqref{fullcolour4gluon} in a minimal colour and kinematical basis.
Going to the DDM basis of section~\ref{sec:1.9} amounts to eliminating $c_{t}$ via
$c_{t}=-c_{s}-c_{u}$~as
\bal{
c_{s} &= f^{a_{1}a_{2}e}f^{ea_{3}a_{4}}= 
\raisebox{0.3cm}{
\begin{tikzpicture}[baseline={(current bounding box.center)}, scale = 0.8]
\begin{feynman}
\coordinate (1) at (-1.2,0);
\coordinate (2) at (-0.5,0.7);
\coordinate (e1) at (-0.5,0);
\coordinate (e2) at (0.5,0);
\coordinate (3) at (0.5,0.7);
\coordinate (4) at (1.2,0);
\draw [gluon] (e1) -- (e2);
\draw [gluon] (1) -- (e1);
\draw [gluon] (2) -- (e1);
\draw [gluon] (3) -- (e2);
\draw [gluon] (4) -- (e2);
  \draw (1) node [left] {$1$};
    \draw (2) node [above] {$2$};
      \draw (3) node [above] {$3$};
        \draw (4) node [right] {$4$};
        \draw [fill] (e1) circle (.08);
       \draw [fill] (e2) circle (.08);
                   \end{feynman}
\end{tikzpicture}}\, , \\
-c_{u} &= f^{a_{1}a_{3}e}f^{ea_{2}a_{4}}= 
\raisebox{0.3cm}{
\begin{tikzpicture}[baseline={(current bounding box.center)}, scale = 0.8]
\begin{feynman}
\coordinate (1) at (-1.2,0);
\coordinate (2) at (-0.5,0.7);
\coordinate (e1) at (-0.5,0);
\coordinate (e2) at (0.5,0);
\coordinate (3) at (0.5,0.7);
\coordinate (4) at (1.2,0);
\draw [gluon] (e1) -- (e2);
\draw [gluon] (1) -- (e1);
\draw [gluon] (2) -- (e1);
\draw [gluon] (3) -- (e2);
\draw [gluon] (4) -- (e2);
  \draw (1) node [left] {$1$};
    \draw (2) node [above] {$3$};
      \draw (3) node [above] {$2$};
        \draw (4) node [right] {$4$};
        \draw [fill] (e1) circle (.08);
       \draw [fill] (e2) circle (.08);
                   \end{feynman}
\end{tikzpicture}}\, .
}
The amplitude then takes the form
\bal{
\label{fullcolour4gluonDDM}
\frac{\i}{g^{2}}\mathbf{A}_{4}^{\text{tree}}&= c_{s}\left ( \frac{n_{s}}{s} - \frac{n_{t}}{t}\right )
- c_{u}\left ( \frac{n_{t}}{t} - \frac{n_{u}}{u}\right ) \\
&= c_{s}\, A_{4}^{\text{tree}}(1,2,3,4) - c_{u}\, A_{4}^{\text{tree}}(1,3,2,4) \,,
}
hence the two colour-ordered amplitudes are to be identified as 
\be
A_{4}^{\text{tree}}(1,2,3,4)= \frac{n_{s}}{s} - \frac{n_{t}}{t}\, , \qquad
 A_{4}^{\text{tree}}(1,3,2,4) =\frac{n_{t}}{t} - \frac{n_{u}}{u}\, .
\ee
Their gauge invariance follows from eqs.~\eqref{gfns} and~\eqref{gfntu}, 
e.g.\
$$A_{4}^{\text{tree}}(1,2,3,4)\Bigl|_{\epsilon_{4}\to p_{4}}= \left(\frac{s}{s}-\frac{t}{t} \right)\, \alpha=0\, .$$
This must be the case, as we argued before for the gauge invariance of the colour-ordered amplitudes.
Due to the kinematical Jacobi identity, the above representation of the amplitude is not yet
minimal. By eliminating (in analogy to $c_{t}$) now $n_{t}$ as well via $n_{t}=-n_{s}-n_{u}$
in \eqn{fullcolour4gluonDDM} we find
the relation
\be
\begin{pmatrix}
A_{4}^{\text{tree}}(1,2,3,4)\cr  A_{4}^{\text{tree}}(1,3,2,4) 
\end{pmatrix}
=
\begin{pmatrix}
s^{-1}+t^{-1} & t^{-1}\cr -t^{-1} & -u^{-1}-t^{-1}
\end{pmatrix}
\,
\begin{pmatrix}
n_{s}\cr  n_{u} 
\end{pmatrix}\, .
\ee
From this expression we learn that the two colour-ordered amplitudes 
$A_{4}^{\text{tree}}(1,2,3,4)$ and  $A_{4}^{\text{tree}}(1,3,2,4)$ \emph{cannot}
be independent of each other: while they are gauge invariant, the kinematical numerators
$n_{s}$ and $n_{u}$ are not, and hence the $2\times 2$ matrix relating them is not invertible.
The linear dependance reads 
\be
\label{BCJ4}
s\, A_{4}^{\text{tree}}(1,2,3,4) = u \, A_{4}^{\text{tree}}(1,3,2,4) \, . 
\ee
This is the Bern-Carrasco-Johannson (BCJ) relation for a four-point amplitude. The general relation
reads, cf.~\eqref{BCJid},
\be \label{eq:BCJn}
\sum_{i=2}^{n-1}p_{1}\cdot (p_{2} + \ldots + p_{i}) \, A^{\text{tree}}_{n}(2,\ldots, i,1, i+1,\ldots,n) =0 \,,
\ee
which for $n=4$ reduces to \eqn{BCJ4}. 
As a matter of fact, using this kinematical $2\times 2$ matrix the coloured amplitude may be written as
\be
\mathbf{A}_{4}^{\text{tree}}= -\i g^{2}
\begin{pmatrix}
c_{s} & -c_{u}
\end{pmatrix}
\begin{pmatrix}
s^{-1}+t^{-1} & t^{-1}\cr -t^{-1} & -u^{-1}-t^{-1}
\end{pmatrix}
\,
\begin{pmatrix}
n_{s}\cr  n_{u} 
\end{pmatrix}\, .
\ee

Returning to the conjectured form of the four-graviton amplitude~\eqref{4gravitonamp}, we can use
there the relations $n_{t}=-n_{u}-n_{s}$ and $n_{u}=t A_{4}^{\text{tree}}(1,2,3,4) + n_{s}u/s$.
One finds
\be
\label{M4final}
M_{4}^{\text{tree}}(1,2,3,4)= - \i \frac{st}{u} \Bigl [ A_{4}^{\text{tree}}(1,2,3,4) \Bigr ]^{2}\, .
\ee
This very compact result may be transformed into the one we derived in \eqn{KLT2} upon using
the BCJ relation~\eqref{BCJ4}. Cyclically permute $1\to 2\to 3 \to 4\to 1$ in \eqref{BCJ4}
(under which $s\leftrightarrow t$ but $u$ is inert) to reach 
\be
t\, A_{4}^{\text{tree}}(2,3,4,1) = u \, A_{4}^{\text{tree}}(2,4,3,1)  \,.
\ee
With the cyclicity of the colour-ordered amplitudes, one has $A_{4}^{\text{tree}}(1,2,4,3)=
\frac{t}{u} A_{4}^{\text{tree}}(1,2,3,4)$, which inserted in \eqn{KLT2} yields \eqn{M4final}.

\begin{exer}{Kinematical Jacobi identity}
\label{Ex:2.kin_Jacobi}
Prove the kinematical Jacobi identity~\eqref{kinematicJacobi} for the coloured tree-level four-gluon amplitude in $D$ dimensions given in eq.~\eqref{fullcolour4gluon}. 
For the solution see \hyperref[Sol:2.kin_Jacobi]{chapter~5}.
\end{exer}

\subsection{The double-copy relation}

The general statement of the duality between colour and kinematics is as follows. A general coloured
$n$-gluon amplitude may be written as
\be
\label{AnCK}
\mathbf{A}_{n}^{\text{tree}} = - \i g^{n-2} \sum_{i} \frac{c_{i} \, n_{i}}{\prod_{\alpha_{i}} D_{\alpha_{i}}}\, ,
\ee
where the sum is over all $n$-point diagrams with a trivalent vertex structure. 
Here $c_{i}$ denote the colour factors (made up of the structure constants and possibly of generators),
$n_{i}$ the numerators (made up of the momenta and polarisation vectors), and the $D_{\alpha_{i}}=p_{\alpha_{i}}^{2}-m_{\alpha_{i}}^{2}$ are the inverse propagators.
We note that any graph may be made tri-valent upon inserting the identity $1= D_{\alpha_{i}}/D_{\alpha_{i}}$
in order to lift  the four-gluon vertices to a sum of three  $s$-$t$-$u$ channel diagrams as dictated by the colour structure. Due to the Jacobi identity, the colour factors obey algebraic relations
of the form $c_{i}-c_{j}=c_{k}$. Colour-kinematic duality now asserts that it is always possible to
find a representation of the amplitude~\eqref{AnCK} in which also the kinematical numerators obey
an analogous identity $n_{i}-n_{j}=n_{k}$. This may be reached by possibly adding overall
zeros to the amplitude.

The $n$-graviton scattering amplitude is then obtained upon replacing colour by kinematics
$c_{i}\to n_{i}$,
\be
M_{n}^{\text{tree}} = - \i \left (\frac{k}{2}\right) ^{n-2} \sum_{i} \frac{n_{i}^{2}}{\prod_{\alpha_{i}} D_{\alpha_{i}}}\, .
\ee
In general it is non-trivial to find numerators which satisfy the colour-kinematics duality. A possible route
is to start out from an ansatz, which one then constrains to match the amplitude and to obey the duality. 
At tree level the duality has been proven~\cite{ch2_Bjerrum-Bohr:2010pnr,ch2_Bern:2010yg,ch2_Bjerrum-Bohr:2016axv}, while at the Lagrangian level a full understanding
is still lacking~\cite{ch2_Bern:2010yg,ch2_Tolotti:2013caa,ch2_Bonezzi:2022bse}. A comprehensive review of the double-copy relation is given in~\cite{ch2_Bern:2019prr}.

The $n$-point generalisation of the squaring relation~\eqref{M4final} is known as
the  Kawai-Lewellen-Tye (KLT) relation and takes the form~\cite{ch2_Kawai:1985xq} 
\be  \label{eq:MtoA2}
M_{n}^{\text{tree}} = \sum_{\sigma,\rho\in S_{n-3}} A_{n}^{\text{tree}}(1,\sigma,n-1,n)
\, S[\sigma|\rho]\, A_{n}^{\text{tree}}(1,\rho,n,n-1)\, .
\ee
Here $\sigma$ and $\rho$ range over the $(n-3)!$ permutations of the elements $\{2,\ldots,n-2\}$.
The KLT  kernel $S[\sigma|\rho]$ are the entries of an $(n-3)!\times (n-3)!$ matrix of kinematic polynomials. 
A closed form expression reads~\cite{ch2_Bjerrum-Bohr:2010pnr,ch2_Bjerrum-Bohr:2010kyi}
\be
\label{KLTkernel}
S[\sigma|\rho] = \prod_{i=2}^{n-2} \biggl[ 
2 \, p_{1}\cdot p_{\sigma_{i}} + \sum_{j=2}^{i} 2 \, p_{\sigma_{i}}\cdot p_{\sigma_{j}}\, \theta(\sigma_{j},
\sigma_{i})_{\rho}
\biggr] \,,
\ee
where $\theta(\sigma_{j}, \sigma_{i})_{\rho}=1$ when $\sigma_{j}$ is before $\sigma_{i}$ in the permutation $\rho$,
and zero otherwise. For $n=4$ the matrix degenerates to a scalar and reproduces \eqn{KLT2}.

\begin{exer}{Five-point KLT relation}
\label{Ex:2.KLT}
Prove the five-point KLT relation
\begin{align} \label{eq:KLT5pt}
M_{5}^{\text{tree}}(1,2,3,4,5) = s_{12} \, s_{34} \, A_5^{\text{tree}}(1,2,3,4,5) \, A^{\text{tree}}_5(1,2,5,3,4) + (2\leftrightarrow 3) \,.
\end{align}
For the solution see \hyperref[Sol:2.KLT]{chapter~5}.
\end{exer}

\chapter{Loop integrands and amplitudes}
\label{ch:loopamps}

\abstract{
In this chapter we study the structure of loop-level scattering amplitudes. The
appearance of integrals over internal loop momenta gives rise to a new set of
functions that go beyond the rational functions of spinor products seen at
tree-level. We will use the unitarity of scattering amplitudes to show that
discontinuities in loop amplitudes can be determined from tree-level
information as a result of factorisation when loop momentum dependent
propagators go on-shell. We then show that generalised discontinuities can be
used to break loop amplitudes further into small tree-level building blocks. We
then turn our attention to a general method for one-loop dimensionally
regulated amplitudes in which a basis of functions is determined as well as a
technique to determine their coefficients from on-shell data.
}

\section{Introduction to loop amplitudes}
\label{sec:3.1}

Perturbative predictions for scattering amplitudes allow us to explore the
quantum nature of fundamental interactions. Explicit computations within
quantum field theory, in particular using the method of Feynman diagrams,
quickly lead to an explosion of both analytic and algebraic complexity.

Loop-level amplitudes involve integration of internal---virtual---momenta,
which takes us beyond the simple rational functions we have encountered at tree
level. In Chapter~\ref{ch:trees} we have seen that the analysis of the poles of
tree-level amplitudes led to factorisation when the poles vanish. Equivalently
we may say amplitudes factorise when the internal momenta go on-shell. This
factorisation was observed when considering the soft and collinear limits of
amplitudes and also, after analytic continuation to complex momenta, on the
residues in the BCFW construction. As we will see, the integrals over the
virtual momenta give rise to functions with branch cuts, such as logarithms.
These branch cuts lead to discontinuities, which are a new feature of
loop-level amplitudes.  At the level of the integrand, which is a rational
function of the internal and external momenta, we may associate discontinuities
with poles dependent on the virtual (or loop) momentum. Analysing the integrand
at points where these poles vanish will again lead to factorisation into
simpler objects. Our main aim for this chapter is to turn these words into a
concrete computational method in which we may re-use on-shell tree-level
amplitudes to directly obtain information about loop amplitudes.

There is, however, another major new feature of loop amplitudes. Integrals over
virtual momenta can lead to divergences, which have to be regulated.
Divergences at large values of the loop momentum are known as ultraviolet (UV),
while divergences at small values of the loop momentum are known as infrared
(IR). In these lecture notes we will follow the procedure of dimensional regularisation,
which regulates both IR and UV regions using an analytical continuation of the
space-time dimension to $D=4-2\eps$, where $\eps$ is a small parameter. We
postpone a more detailed discussion of the dimensional regularisation until
chapter~\ref{ch:loopints} (see section~\ref{sec:DimReg}), where we consider the
evaluation of the loop integrals.  UV and IR divergences must cancel for
physical predictions. UV divergences are removed through the procedure of
renormalisation, which is covered in standard field theory textbooks (e.g.\ \cite{ch3_Sterman:1993hfp,ch3_Peskin:1995ev,ch3_Schwartz:2014sze}). The
cancellation of IR singularities is more complicated and in general beyond the
scope of these lecture notes. We have seen that IR singularities also appear in
tree-level amplitudes in soft and collinear limits, and it is these divergences
that must cancel the IR divergences in the virtual amplitudes. This
cancellation only happens at the level of cross-sections, where the squared
amplitude is integrated over an inclusive phase space. The topic is worthy of
study in its own right, and the interested reader may like to explore the
review~\cite{ch3_Agarwal:2021ais}.

\begin{figure}[b]
  \begin{subfigure}{0.4\textwidth}
    \centering
    \includegraphics[width=0.5\textwidth]{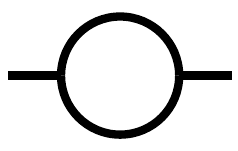}
    \label{fig:topo-bub}
    \caption{bubble}
  \end{subfigure}
  \hfill
  \begin{subfigure}{0.4\textwidth}
    \hspace*{0.17\textwidth}
    \includegraphics[width=0.5\textwidth]{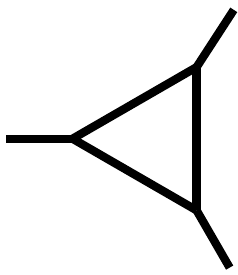}
    \label{fig:topo-tri}
    \caption{triangle}
  \end{subfigure}
  \\
  \begin{subfigure}{0.4\textwidth}
    \centering
    \includegraphics[width=0.5\textwidth]{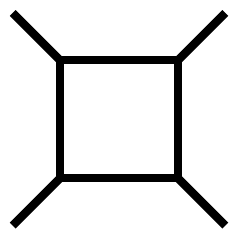}
    \label{fig:topo-box}
    \caption{box}
  \end{subfigure}
  \hfill
  \begin{subfigure}{0.4\textwidth}
    \centering
    \includegraphics[width=0.74\textwidth]{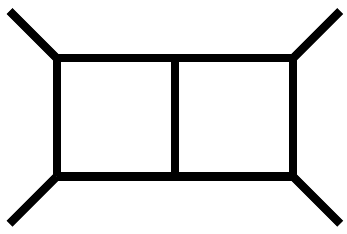}
    \label{fig:topo-dblbox}
    \caption{two-loop double box}
  \end{subfigure}
  \caption{Example loop topology graphs.}
  \label{fig:topo}
\end{figure}

We begin this discussion with some general observations on the structure of
loop amplitudes. We will consider loop amplitudes in Yang-Mills theory (YM) in
which the colour structure has been stripped off as discussed in
section~\ref{sec:1.10}.  An amplitude with $n$ external legs at $L$ loops may
be written in terms of a set of Feynman integrals, $F$, together with rational
coefficients $c$. The amplitude will depend on the external momenta of each leg
as well as their helicity, and mass (we may consider YM coupled to matter). We
will write the arguments of the amplitudes as a list of integers which
represent these properties.  The external momenta will be denoted $p_i$ with $i
= 1,\ldots,n$.  As in the previous chapters, we take them to be all outgoing,
and hence they satisfy momentum conservation in the form $\sum_{i=1}^n p_i =
0$. When analytically continuing the dimension we must rescale the coupling to
make sure that we are still expanding in a dimensionless quantity. This scale
is arbitrary and we will represent it with the symbol $\mu_{\rm R}$. To write
down a general expression for the amplitude we must introduce a number of
conventions. Let us first write the expression and then proceed with the
explanation of the normalisations and symbols:
\begin{align} \begin{aligned}
  A_n^{(L),[D]}(1,\ldots,n) = \ &
    \left(g_{\rm YM} \,\mu_{\rm R}^{(4-D)/2}\right)^{n-2}
    \left(\frac{\i \, \alpha_{\rm YM}\,\mu_{\rm R}^{(4-D)}}{(4\pi)^{(D-2)/2}}\right)^L \\&
  \sum_T c_T^{[D]} (1,\cdots,n) \, F_{T}^{(L),[D]}(p_1,\cdots,p_{n-1}) \,.
  \end{aligned}
  \label{eq:loopamp-general}
\end{align}
Here the expansion is in the coupling $\alpha_{\rm YM} = g_{\rm YM}^2/(4\pi)$.
The linear combination sums over a set of loop \textit{topologies}, $T$, which
are defined by the set of propagators and loop-momentum dependent numerators.
They may also potentially contain propagators with higher powers. The
coefficients $c_T^{[D]}$ depend on the momenta, helicities and masses of the
external legs, and on the masses of the internal particles. On the other hand, the
integrals $F_{T}^{(L),[D]}$ only depend on the $n-1$ independent external
momenta and on the masses of the internal and external particles (which are
suppressed in the notation above for conciseness).  The factors of $\i$ and
$\pi$ are due to the normalisation of the integrals, which is given below.
Example graphs of possible loop topologies are shown in figure~\ref{fig:topo}.
The structure of the amplitude is not specific to YM theory apart from the
couplings.  For a useful separation of coefficients and integrals, we need to
identify a (linearly) independent basis of integrals, which defines the sum over
topologies $T$. A precise definition of this basis at one loop is one of the
main aims of this chapter. In addition, we will show how on-shell techniques can
be used to directly extract the coefficients of the basis integrals. The
couplings and dependence on $\mu_{\rm R}$ can be easily restored at the end of a
calculation through dimensional analysis, and so we set $g_{\rm YM}=1$ and $\mu_{\rm R}=1$
for the remainder of this chapter. Furthermore, the factors of
${\alpha_{\rm YM}\mu_{\rm R}^{(4-D)}}/{(4\pi)^{(D-2)/2}}$ will also be suppressed,
since they may be restored at the end of the computation as well.

\begin{figure}[t]
\centering
\includegraphics[width=0.5\textwidth]{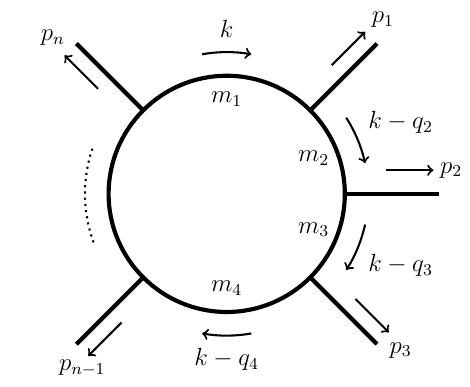}
\caption{The generic one-loop integral.}
\label{fig:onelooppic}
\end{figure}

At one loop all topologies take the form of an $n$-gon with a potential numerator function $N$,
\begin{equation}
  F_n^{(1),[D]}[N] = \int_k \frac{N}{\prod_{a=1}^{n}\bigl[-(k-q_a)^2+m_{a}^2-\i0\bigr] } \,,
  \label{eq:oneloopintegraldef}
\end{equation}
where $q_a = \sum_{b=1}^{a-1} p_b$ (with $q_1 = 0$) are the momenta flowing in
each propagator, which also have a mass $m_{a}$. We have also introduced a
short hand for the integration measure:
\begin{equation}
  \int_k \coloneqq \int \frac{\d^D k}{\i \pi^{D/2}} \,.
\end{equation}
Note the different loop integration measure with respect to the $\d^D k/(2\pi)^D$ in the Feynman rules. This difference is responsible for the factors of $\i$ and $\pi$ in \eqn{eq:loopamp-general}, and will be motivated in section~\ref{sec:introduction}.
The configuration of momenta and propagators of \eqn{eq:oneloopintegraldef} is shown graphically in figure~\ref{fig:onelooppic}. Cases in which the numerator
is one, $F_n^{(1),[D]}[1]$, are referred to as \emph{scalar integrals}. When
using an integer $n$ to represent the topology we are already indicating that
it is a one-loop integral, and so the loop-order superscript will be dropped for
the remainder of this chapter. If no numerator is specified it should
considered to be a scalar integral and so $F_n^{[D]} \equiv F_n^{[D]}[1] \equiv
F_n^{(1),[D]}[1]$. A small imaginary part $\i0$ follows from the Feynman
prescription for the propagators that was introduced in the Feynman rules.
This $\i0$ prescription will mainly play a role only when evaluating the
integrals, and so we will drop it from the propagator expressions in cases where
it is not necessary. We will follow the standard convention to refer to the
simple one-loop topologies according to the polygon that represents the number of
propagators, e.g.\ bubble for two propagators, triangle for three propagators,
box for four propagators, and so on (see figure ~\ref{fig:topo} again). An integral with one propagator is
referred to as a tadpole integral.

The coefficients of the integrals in \eqn{eq:loopamp-general} can be expanded around the physical space-time dimension $D=4$, as
\begin{equation}
  c_T^{[D]} = c_T^{(0)} + \eps \, c_T^{(1)} + \eps^2 \, c_T^{(2)} + \ldots
  \label{eq:coeffexp}
\end{equation}
The order at which we must expand the coefficients to ensure the correct result
for the amplitudes as $\eps\to 0$ will depend on the overall divergences present
in the loop integrals.

We may also consider the \emph{integrand} of an amplitude $A_n^{(L),[D]}(1,\ldots,n)$, which we denote $I_n^{(L),[D]}(1,\ldots,n)$, and is a rational function satisfying
\begin{equation}
  A_n^{(L),[D]}(1,\ldots,n) = \int \prod_{l=1}^L \frac{\d^D k_l}{\i\pi^{D/2}} \, I_n^{(L),[D]}(k, 1,\ldots,n) \,.
\end{equation}
The amplitude integrands are not uniquely defined, as they may differ by terms which integrate to zero, thus giving rise to the same amplitude. The simplest choice is to use the Feynman-diagram expansion to define the integrand.
If we make the assumption that a colour-ordered one-loop amplitude has a single
ordering of the external legs for the topology with $n$ propagators (this is the case for the
leading-colour approximation in Yang-Mills theory), we can be more explicit and
write
\begin{equation} \label{eq:integrand_ordered_1L}
  I_n^{(1),[D]}(k,1,\ldots,n) = \frac{{N}(k,1,\ldots,n)}{\prod_{a=1}^{n}\bigl[-(k-q_a)^2+m_{a}^2-\i0\bigr]} \,.
\end{equation}
In the same way that we can try to find a basis of Feynman integrals for the
amplitude, we may also ask if there exists a basis of loop-momentum dependent
numerator functions $\{f_x\}$ such that the integrand numerator in \eqn{eq:integrand_ordered_1L} can be written as
\begin{equation}
  N(k,1,\ldots,n) = \sum_x c_x(1,\ldots,n) \,f_x(k,p_1,\ldots,p_{n-1}) \,,
\end{equation}
where the coefficients $c_x$ are rational functions of the external kinematics,
and $f_x$ are independent scalar products dependent on the loop momentum.
Again, the sum over $x$ and the definition of ``independent'' here are not yet
defined but we can motivate the construction with a simple example. If we
consider a one-loop integrand $I_n^{(1),[D]}(k,1,2,3)$ in which
${N}(k,1,2,3) = k\cdot q_{2}$, then we can express the numerator in terms of
the difference of two propagators in order to rewrite everything in terms of scalar integral
functions:
\begin{eqnarray}
  \begin{aligned}
  {N}(k,1,2,3)
  &= \frac{1}{2}\left[ \bigl(-(k-q_2)^2+m_2^2\bigr) - \bigl(-k^2+m_1^2\bigr) + q_2^2-m_2^2+m_1^2 \right] \\
  &= \frac{1}{2}\left\{1,-1,q_2^2-m_2^2+m_1^2\right\} \cdot \left\{-(k-q_2)^2+m_2^2, -k^2+m_1^2, 1\right\}^{\top} \,.
  \end{aligned}
\end{eqnarray}
So in this case the functions $f_x$ are the inverse propagators and $1$.

\begin{figure}[t!]
  \centering
  \includegraphics[width=0.8\textwidth]{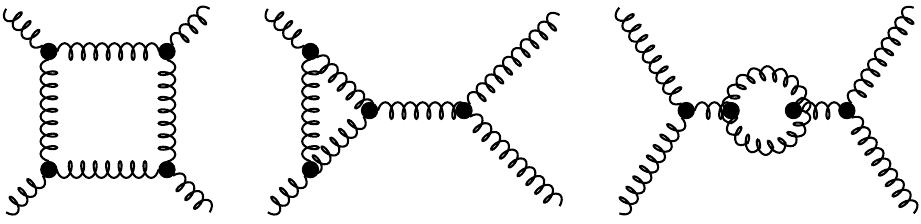}
  \caption{Sample diagrams contributing to the four-gluon scattering amplitude at one-loop order.}
  \label{fig:4gluon1L}
\end{figure}

Throughout this chapter we will use the example of four-gluon scattering to
illustrate general methods for loop integrands and amplitudes. Sample diagrams
for this process are shown in figure~\ref{fig:4gluon1L}. In this case the most
complicated topology is the box graph which, following the structure of
the three-gluon vertex, has a maximum of four powers of the loop momentum
in the numerator function. Graphs containing vertex corrections or those with
bubble insertions will contain triangle and bubble tensor integrals
respectively. As we will see, writing the loop-momentum dependence of the numerator in terms of
the propagators in each graph will allow us to find a basis of scalar Feynman
integrals, so that the particular form of \eqn{eq:loopamp-general} relevant for four-gluon scattering becomes
\begin{align}
\begin{aligned}
  A_4^{(1),[D]}&(1,2,3,4) =
  \i \, c_{{\rm box}}^{[D]} \, F_{4}^{[D]}[1](p_1,p_2,p_3) \\&
  + \i \, c_{{\rm tri},1}^{[D]} \, F_{3}^{[D]}[1](p_1,p_2)
  + \i \, c_{{\rm tri},2}^{[D]} \, F_{3}^{[D]}[1](p_1,p_{23}) \\&
  + \i \, c_{{\rm tri},3}^{[D]} \, F_{3}^{[D]}[1](p_{12},p_3)
  + \i \, c_{{\rm tri},4}^{[D]} \, F_{3}^{[D]}[1](p_2,p_3) \\&
  + \i \, c_{{\rm bub},1}^{[D]} \, F_{2}^{[D]}[1](p_{12})
  + \i \, c_{{\rm bub},2}^{[D]} \, F_{2}^{[D]}[1](p_{23}) \,.
 \end{aligned}
\end{align}
where we have used the notation,
\begin{equation}
  p_{i_1 \cdots i_n} = p_{i_1} + p_{i_2} + \cdots + p_{i_n},
\end{equation}
which will also be used for the invariants,
\begin{equation}
  s_{i_1 \cdots i_n} = p_{i_1\cdots i_n}^2.
\end{equation}
The fact that only scalar integrals appear remains to be proven of course. We
will also set about the task of extracting the coefficients of these scalar
integrals and the task of generalising to the $n$-point case.

The chapter is organised as follows. We start in section~\ref{sec:3.2} by
demonstrating that the unitarity of the $S$-matrix leads to deep insights into
the structure of loop amplitudes. This will lead us to consider the
discontinuities of loop amplitudes, and show that they may be computed from the
product of tree-level amplitudes. Using the example of four-gluon scattering, we will demonstrate that this results in an extremely efficient technique to
identify simple integral structure in the amplitude. We will then explore
generalised discontinuities of loop amplitudes in section~\ref{sec:3.3}, and use
the ``Cutkosky rules'' to make a direct connection with the pole structure of the
integrand. The factorisation on these poles leads to the generalised unitarity
method, which allows for the computation of the coefficients of the
scalar box integrals. Section~\ref{sec:3.4} lays the ground work for the
general treatment of any loop amplitude, as we use tensor reduction and
integrand-level analysis of transverse spaces to identify relations between
Feynman integrals. Section~\ref{sec:3.5} is dedicated to the derivation of the
complete decomposition of a general one-loop amplitude into a basis of scalar
integrals, and how their coefficients may be extracted from products of
tree-level amplitudes via generalised unitarity. In section~\ref{sec:rat4g} we
put all of this technology to work to complete the computation of the one-loop four-gluon scattering amplitude
in dimensional regularisation. Finally we give some outlook and extensions of
the ideas presented here, and consider efficient computations using rational
parametrisations of the external kinematics in section~\ref{sec:3.7} and
extensions to two-loop integrands in section~\ref{sec:3.8}.

Further information on the topics presented in this chapter can be found in a number of comprehensive reviews, for example see~\cite{ch3_Dixon:1996wi,ch3_Dixon:2013uaa,ch3_Ellis:2011cr}.

\begin{svgraybox}
  \textbf{Ultraviolet power counting.}
  Before we get started with the main topics of this chapter it is useful to recall how we can quantify UV divergences.
  The divergences at large values of the loop momentum can be estimated at the integrand level by considering the
  scaling behaviour. For example, using general polar coordinates, one-loop scalar integrals become
  \begin{equation} \label{eq:UVpowercounting}
    F_n^{[D]}[1] \overset{|k|\to\infty}{\to} \int \frac{(-1)^n |k|^{D-1} \d |k| \, \d\Omega}{\i\pi^{D/2}} \frac{1}{k^{2n}} \,,
  \end{equation}
  from which we see a divergence if $n\leq D/2$. If $n=2$ in $D=4-2\eps$, we have that
  \begin{equation}
    F_{2}^{[4-2\eps]}[1] \overset{|k|\to\infty}{\to} \int \frac{\d |k| \, \d\Omega}{\i\pi^{2-\eps}} \frac{1}{|k|}\,,
    \label{eq:UVcountingbubble}
  \end{equation}
  and we see a logarithmic divergence. Infrared divergences, such as the soft
  and collinear configurations discussed in chapter~\ref{ch:trees}, are more difficult to
  classify but can also be regulated by the analytic continuation of the
  dimension. We will discuss this further in chapter~\ref{ch:loopints}, where we focus on the
  integration over loop momenta. 
\end{svgraybox}

\section{Unitarity and cut construction \label{sec:3.2}}

The unitarity of the $S$-matrix provides some of the most fundamental
constraints on the analytic form of on-shell amplitudes. 
The initial steps
follow the discussion of the optical theorem and Cutkosky analysis of the
discontinuities of Feynman integrals~\cite{ch3_Cutkosky:1960sp} that may be familiar to many readers.

The scattering amplitudes associated with an $S$-matrix  $S = \1 + \i T$ are
determined through the transition matrix $T$. Transition matrix elements
$\langle F | T | I \rangle$ are a measure of the probability of a initial state $I$
evolving into a final state $F$. 

The unitarity condition of the $S$-matrix
\begin{equation}
  S^\dagger S = \1
  \end{equation}
implies a non-linear constraint on the transition matrix:
\begin{align}
 - \i \bigl(T-T^\dagger \bigr) =  T^\dagger T  \,.
  \label{eq:Sunitarity}
  \end{align}
The asymptotic states obey the completeness relations
\begin{equation} 
  \sum_{n} \int \prod_{j=1}^n \frac{\d^3 \vec{k}_j}{(2\pi)^3 \, 2 E_j} \ |\{k\}_n \rangle\langle \{k\}_n | = \1 \,,
\end{equation}
where $|\{k\}_n  \rangle$ indicate multi-particle states $|\{k\}_j\rangle
\coloneqq |k_1,k_2, \ldots,k_n\rangle$, $k_i^{\mu} = ( E_i, \vec{k}_i)$, and
$E_i^2 = |\vec{k}_i|^2 + m_i^2$, $m_i$ being the mass of the $i$th particle. Therefore,
when we contract \eqn{eq:Sunitarity} with the initial and final states
and insert a complete set of states in the product $T T^\dagger$, we determine
that
\begin{equation}
  -\i\langle F| \bigl(T-T^\dagger\bigr)|I\rangle =
  \sum_{n} \int \prod_{j=1}^n \frac{\d^3 \vec{k}_j}{(2\pi)^3 \, 2 E_j} \,\langle F| T^\dagger|\{k\}_n\rangle \,\langle \{k\}_n |T|I\rangle \,.
  \label{eq:unitarity-Tmatexp}
\end{equation}

The scattering amplitude $A(I\to F)$ is extracted from the transition matrix
element stripped of the overall momentum-conserving delta function. If we represent
this amplitude by a picture,
\begin{align}
  A(I \to F) = \raisebox{-0.4cm}{\includegraphics[width=1.5cm]{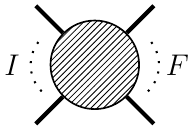}} \,,
\end{align}
we can show the relation \eqref{eq:unitarity-Tmatexp} graphically, as
\begin{align} \begin{aligned}
  -2\i \left(
       \raisebox{-0.4cm}{\includegraphics[width=1.5cm]{ch3_loopamps/figures/ampIFall.pdf}}
     - \raisebox{-0.4cm}{\includegraphics[width=1.5cm]{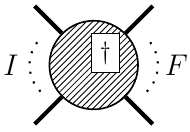}}
  \right)
  &= \sum_{n=2}^{\infty} \int \d\Phi_n(P, \{k\}_n) \raisebox{-0.4cm}{\includegraphics[width=3cm]{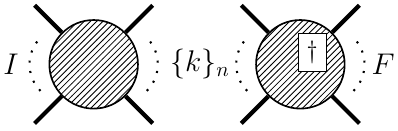}}\,,
  \end{aligned}
  \label{eq:unitarity-pictorial}
\end{align}
where the integration measure now includes the on-shell delta function that ensures an on-shell final state phase-space integral,
\begin{align}
  \d\Phi_n(P, \{k\}_n) = (2\pi)^4 \delta^{(4)}\left( P-\sum_n k_n \right) \prod_{i=1}^n \frac{\d^3 \vec{k}_i}{(2\pi)^3 \, E_i}.
  \label{eq:phase_space}
\end{align}
The momentum $P$ above represents the total incoming momentum $P = \sum_i p_i$.

The LHS of \eqn{eq:unitarity-pictorial} may be shown to be
proportional to the discontinuity of the amplitude across the real $P^2$ axis.
While we will not see any specific examples until chapter~\ref{ch:loopints}, Feynman integrals
will in general contain branch cuts. The simplest function of this type is the
logarithm $\log(x)$,\footnote{We saw the logarithm appear in the UV limit of of
the one-loop bubble function $F_{2}^{[4-2\eps]}[1]$ in \eqn{eq:UVcountingbubble}.} which
has a branch cut across the (negative) real $x$ axis. For a generic function
$f(x)$, the discontinuity across the real $x$ axis is defined as ${\rm
Disc}_{x} f(x) \coloneqq  f(x+\i0) - f(x-\i0)$. For the logarithm this gives
${\rm Disc}_{x} \log(x) = 2\pi \i \, \Theta(-x)$, where $\Theta(-x)$ is the
Heaviside step function (for further details see exercise~\ref{Ex:Discontinuities}).
Rational functions, such as the tree-level amplitudes we have encountered up
until now, do not contain branch cuts. We can take the unitarity constraint to
imply that scattering amplitudes do contain branch cuts. Following this
argument, it is possible to show that
\begin{align} \begin{aligned}
  -2\i \left(
       \raisebox{-0.4cm}{\includegraphics[width=1.5cm]{ch3_loopamps/figures/ampIFall.pdf}}
     - \raisebox{-0.4cm}{\includegraphics[width=1.5cm]{ch3_loopamps/figures/ampFIall.pdf}}
  \right)
  &= {\rm Disc}_{P^2}\left( \raisebox{-0.4cm}{\includegraphics[width=1.5cm]{ch3_loopamps/figures/ampIFall.pdf}} \right) \\
  \end{aligned}
  \label{eq:amp-disc}
\end{align}
where the discontinuity is across the branch cut in the invariant $P^2$, \break ${\rm
Disc}_{P^2} \,A(\cdots,P^2,\cdots) =  A(\ldots,P^2+\i0,\ldots) -
A(\ldots,P^2-\i0,\ldots)$. This is referred to as the discontinuity in the $P^2$ channel.

We may now expanding the relation~\eqref{eq:unitarity-pictorial}
perturbatively, in a coupling $g$, which results in a set of extremely useful
equations for on-shell amplitudes. We can represent the perturbative expansion
of the amplitudes this pictorially by,
\begin{equation}
  \raisebox{-0.4cm}{\includegraphics[width=1.5cm]{ch3_loopamps/figures/ampIFall.pdf}} = g^{\rm LO} \left(
  \raisebox{-0.4cm}{\includegraphics[width=1.5cm]{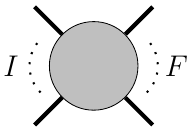}} +
  g^2 \raisebox{-0.4cm}{\includegraphics[width=1.5cm]{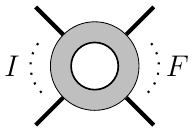}} +
  g^4 \raisebox{-0.4cm}{\includegraphics[width=1.5cm]{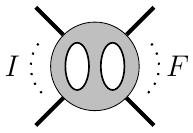}} + \ldots  \right) \,,
  \label{eq:pertexp-pictorial}
\end{equation}
where $g^{\rm LO}$ is the leading-order coupling (e.g.\ $g^{\rm LO} = g^{n-2}$
for a $n$-gluon amplitude in YM theory).  By substituting this expansion into
the unitarity relation~\eqref{eq:unitarity-pictorial} we find equations order
by order in the coupling $g$. Explicitly, up to third order we find
\begin{align}
  {\rm Disc}_{P^2}\left( \raisebox{-0.4cm}{\includegraphics[width=1.5cm]{ch3_loopamps/figures/ampIF0.pdf}} \right) &= 0 \,, \\
  {\rm Disc}_{P^2}\left( \raisebox{-0.4cm}{\includegraphics[width=1.5cm]{ch3_loopamps/figures/ampIF1.pdf}} \right) &=
         \int \d\Phi_2 \,\raisebox{-0.55cm}{\includegraphics[width=3cm]{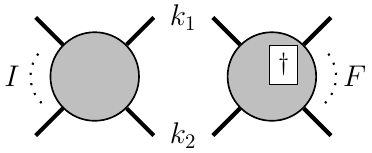}} \,, \\
  {\rm Disc}_{P^2}\left( \raisebox{-0.4cm}{\includegraphics[width=1.5cm]{ch3_loopamps/figures/ampIF2.pdf}} \right) &=
  \int \d\Phi_2 \,\left( \raisebox{-0.55cm}{\includegraphics[width=3cm]{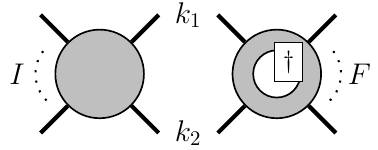}}
                    + \raisebox{-0.55cm}{\includegraphics[width=3cm]{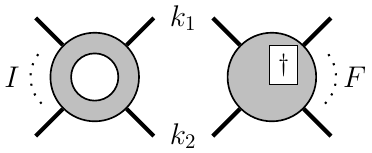}} \right)\nonumber\\& \phantom{= {}}
       + \int \d\Phi_3 \,\raisebox{-0.55cm}{\includegraphics[width=3cm]{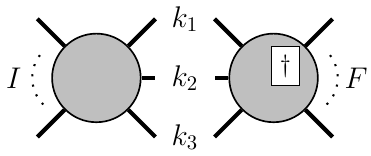}} \,.
\end{align}
The first of these equations confirms that rational tree-level amplitudes do
not contain branch cuts. At one-loop order we find that the product of two
\textit{on-shell} tree-level amplitudes is directly related to the
discontinuity in the channel $P^2$ of the one-loop amplitude.

Re-using the on-shell tree-level amplitudes we have found in chapters~\ref{ch:intro} and~\ref{ch:trees} inside the factorised loop integrand is an extremely efficient way of
computing the discontinuity of a loop amplitude. It avoids some of the large
intermediate algebraic steps that would be found following the expansion of the
loop amplitude into Feynman diagrams. The act of putting an internal propagator
on-shell, as we have done through the insertion of a complete set of states, is
referred to a the \emph{cut} of a loop amplitude.

However, we must still find a way to upgrade the discontinuity of the amplitude to the
full amplitude. One method to do this is to perform a dispersion integral. To
express this concretely, let us specify that the amplitude depends on $r$
invariants $s_1 = P_1^2,\ldots,s_r = P_r^2$. Then, we have that 
\begin{equation}
  A^{(1)}(s_1,\cdots,s_r) = \sum_{i=1}^r \int \frac{\d s'_i}{s_i'-s_i} \, {\rm Disc}_{s_i} \,A^{(1)}.
  \label{eq:dispersion}
\end{equation}
This story has been known for a long time and traces back to the work of
Cutkosky~\cite{ch3_Cutkosky:1960sp} and the days of the analytic $S$-matrix~\cite{ch3_Eden:1966dnq}.
The modern unitarity method (Bern, Dixon, Dunbar, Kosower~\cite{ch3_Bern:1994zx})
transformed the approach into a powerful computational tool by the cut
constraints with knowledge of a \textit{basis} of loop integrals, and the
spinor-helicity method for the compact representation of on-shell tree
amplitudes. Rather than computing the dispersion relation, one uses the
unitarity cut constraints to project out information about the rational
integral coefficients from a representation of the amplitudes such as the one
shown in \eqn{eq:loopamp-general}. It is this procedure that we refer to as
\textit{cut construction} of a scattering amplitude.

The on-shell phase space $\d\Phi_n$ contains Dirac $\delta$ functions which
ensure the intermediate particles are on-shell. We recall in fact that
\eqn{eq:phase_space} can be rewritten in a manifestly Lorentz-invariant form as
\begin{align}
\d\Phi_n(P, \{k\}_n) = (2\pi)^4 \delta^{(4)}\left( P-\sum_n k_n \right) \prod_{i=1}^n \frac{\d^4 k_i}{(2\pi)^4 } (2\pi) \delta^{(+)}\bigl(k_i^2-m_i^2 \bigr) \,,
\end{align} 
where
\begin{align}
 \delta^{(+)}\bigl(k_i^2-m_i^2 \bigr) \coloneqq  \delta \bigl(k_i^2-m_i^2 \bigr) \, \Theta\bigl(k_i^0\bigr)\,,
\end{align}
with the Heaviside step function $\Theta$ ensuring the positivity of the energy.

Since it will become a fundamental part of our amplitude analysis, it is useful to introduce some
notation for the action of imposing these on-shell constraints on internal
particles of a loop diagram. The operation of computing the discontinuity
across a two-particle factorisation or \emph{cut} will be represented as $\mathcal{C}_{L|R}$, where
the indices $L$ and $R$ will be the list of external particles entering the left and right side of the cut respectively,
\begin{align}
  & \mathcal{C}_{i_1\ldots i_m|i_{m+1}\ldots i_n}\left( A_n^{(1),[D]} \right) \coloneqq {\rm Disc}_{s_{i_1\ldots i_m}}\left( A_n^{(1),[D]} \right)
     \nonumber\\
    &= \int \d\Phi_2 \sum_{h_i=\pm}
       \i A_{m+1}^{(0)}\bigl(-l_1^{-h_1}, i_1,\ldots,i_m, l_2^{h_2} \bigr)
    \, \i A_{n-m+2}^{(0)}\bigl(-l_2^{-h_2}, i_{m+1},\ldots,i_n, l_1^{h_1} \bigr) \,,
  \label{eq:Cnotationdef}
\end{align}
where
\begin{align} 
\label{eq:onshellPSmeasure}
\begin{aligned}
  \d\Phi_2 & =
  \frac{\d^4 l_1}{(2\pi)^4}
  \frac{\d^4 l_2}{(2\pi)^4}
     (2\pi)^4 \delta^{(4)}(l_1-l_2-p_{i_1\ldots i_m}) \\
     & \phantom{= {}} \times  (2\pi) \delta^{(+)}(l_1^2-m_1^2) \, (2\pi) \delta^{(+)}(l_2^2-m_2^2) \,.
\end{aligned}
\end{align}
The $\delta$ functions ensure momentum conservation, and that the internal
momenta $l_1$ and $l_2$ are at their on-shell values $m_1$ and $m_2$.
Notice that two factors of $\i$ appear in the factorisation into tree
amplitudes. This appears for exactly the same reason as it did in BCFW
recursion, where each factorised on-shell gluon propagator contributed a factor of
$\i$ (see below \eqn{eq:multiparticlepole}). For fermion propagators this factor would change as discussed in section~\ref{sec:2.1} and exercise~\ref{Ex:3.1}.  

We will also use the operation $\mathcal{C}_{L|R}$ to act on the integrand of the amplitude,
and therefore only represent the product of tree-level amplitudes,
\begin{align}
\begin{aligned}
   &\mathcal{C}_{i_1\ldots i_m|i_{m+1}\ldots i_n}\left( I_n^{(1),[D]} \right)
    = \\
   & \sum_{h_i=\pm} \i A_{m+2}^{(0)} \bigl(-l_1^{-h_1}, i_1,\ldots,i_m, l_2^{h_2} \bigr) \, \i A_{n-m+2}^{(0)}\bigl(-l_2^{-h_2}, i_{m+1},\ldots,i_n, l_1^{h_1} \bigr) \,,
 \end{aligned}
  \label{eq:Cnotationdef}
\end{align}
where the on-shell conditions for $l_1$, $l_2$ are understood to be imposed.

\begin{example}{Example: The $s_{12}$-channel cut of the $gg\to gg$ MHV scattering amplitude}

  Let us consider the leading-colour\footnote{This is the coefficient of the single-trace term in the colour ordered one-loop amplitude given in \eqn{leading-colour-one-loop}, denoted there as $A^{(1)}_{n;1}.$} four-gluon MHV amplitude in
pure Yang-Mills theory, $A^{(1)}(1^-,2^-,3^+,4^+)$. We begin with the familiar
Parke-Taylor formula~\eqref{ParkeTaylor1} for the tree amplitudes (this time setting the coupling to 1),
\begin{equation}
  A^{(0)}(1^-,2^-,3^+,4^+) = \frac{\i \spA12^3}{\spA23\spA34\spA41} \,.
  \label{eq:4gtree}
\end{equation}
Note that by using the four-dimensional tree-level amplitudes inside the cut we are only resolving the first term in the expansion of the integral coefficient expressed in \eqn{eq:coeffexp}.
The $s_{12}$-channel is associated with the invariant $s_{12} = (p_1+p_2)^2$, and the discontinuity
is obtained from the following product of two tree amplitudes summed over all possible
helicity states,
\begin{equation}
  \mathcal{C}_{12|34}\left( I_4^{(1)}(1^-,2^-,3^+,4^+) \right) = \sum_{h_i=\pm} \i A^{(0)}(-l_1^{-h_1},1^-,2^-,l_2^{h_2}) \, \i A^{(0)}(-l_2^{-h_2},3^+,4^+,l_1^{h_1}) \,,
  \label{eq:4gschannelintegrand}
\end{equation}
where, as imposed by the Lorentz invariant phase-space measure $\d\Phi_2$,
$l_2=l_1-p_{12}$ and $l_i^2=0$. Since all tree amplitudes with all-like
helicities or those with a single positive (or negative) helicity vanish, only
a single term contributes to the cut. In order to keep the notation as compact
as possible, we will use $\mathcal{C}_{12|34}$ to refer to
$\mathcal{C}_{12|34}\bigl(I_4^{(1)}(1^-,2^-,3^+,4^+)\bigr)$ 
for the remainder of this section. Replacing the tree-level amplitudes with their spinor-bracket forms leads to
\begin{align}
  \mathcal{C}_{12|34} & = \i \, A^{(0)}(-l_1^{+},1^-,2^-,l_2^{+}) \, \i \, A^{(0)}(-l_2^{-},3^+,4^+,l_1^{-}) \nonumber\\
  &= \frac{\spA12^3}{\spA2{l_2}\spA{l_2}{(-l_1)}\spA{(-l_1)}1} \frac{\spA{l_1}{(-l_2)}^3}{\spA{(-l_2)}3\spA34\spA4{l_1}} \nonumber\\
  \label{eq:C12|34YM}
  &= \frac{\spA12^3}{\spA2{l_2}\spA{l_2}{l_1}\spA{l_1}1} \frac{\spA{l_1}{l_2}^3}{\spA{l_2}3\spA34\spA4{l_1}} \,.
\end{align}
Above we have used the phase convention~\eqref{negpspin} for the spinors of the loop momenta, namely $|(-l_i)\ran = \i |l_i\ran$.
We can now apply some spinor identities to recast the integrand into a form that
can be identified with one-loop integral topologies. For example,
\begin{equation}
  \spA{l_1}1 = \frac{2 \, l_1\cdot p_1}{\spB1{l_1}} = -\frac{(l_1-p_1)^2}{\spB1{l_1}} \,.
\end{equation}
Hence we find
\begin{align}
  \mathcal{C}_{12|34} &= \frac{\spA12^3}{\spA34} \frac{\spA{l_1}{l_2}^2 \spB2{l_2}\spB{l_1}1\spB{l_2}3\spB4{l_1}}{(l_2+p_2)^2 (l_1-p_1)^2 (l_2-p_3)^2(l_1+p_4)^2} \nonumber\\
  &= \frac{\spA12^3}{\spA34} \frac{\spBB{1}{l_1 l_2}{2}\spBB{3}{l_2 l_1}{4}}{(l_2+p_2)^2 (l_1-p_1)^2 (l_2-p_3)^2(l_1+p_4)^2} \,.
\end{align}
The numerator can also be reduced using the on-shell kinematics,
\begin{equation}
  \spBB{1}{l_1 l_2}{2} = \spBB{1}{l_1 (l_1-p_{12})}{2} = -\spAB{1}{l_1}{1}\spB12 = (l_1-p_1)^2 \spB12 \,.
\end{equation}
This leads us to the simple result
\begin{align}
  \mathcal{C}_{12|34} &= \frac{\spA12^3}{\spA34} \frac{\spB12\spB34}{(l_1-p_1)^2 (l_1+p_4)^2} \nonumber\\
  &= -\frac{\spA12^3}{\spA23\spA34\spA41} \frac{s_{12}s_{23}}{(l_1-p_1)^2 (l_1+p_4)^2} \,.
\end{align}
We can now identify the integrand of the double cut of the one-loop scalar box integral,
\begin{align}
  F_4^{[D]}(p_1, p_2, p_3) \equiv F_4^{[D]}(s_{12}, s_{23}) = \int_k \frac{1}{k^2 (k-p_1)^2  (k-p_{12})^2 (k+p_4)^2 } \,.
\end{align}
The discontinuity of this function in the $s_{12}$ channel is given by 
\begin{align}
  {\rm Disc}_{s_{12}} F_4^{[D]}(s_{12}, s_{23}) = \int \d\Phi_2 \, \frac{1}{(l_1-p_1)^2 (l_1+p_4)^2 } \,,
\end{align}
where the cut propagators have been replaced by on-shell delta functions. We
will justify this step in the next section. Our final
result for the discontinuity for the one-loop amplitude is then
\begin{align}
  {\rm Disc}_{s_{12}} A^{(1)}(1^-,2^-,3^+,4^+) = - s_{12} s_{23} A^{(0)}(1^-,2^-,3^+,4^+) \, {\rm Disc}_{s_{12}} F_4^{[D]}(s_{12}, s_{23}) \,.
\end{align}
\end{example}

This result deserves a few remarks. The on-shell approach has uncovered some
dramatic simplifications and the final cut contains only one scalar integral
function. As we will see, this is not the most general structure we can
encounter, and we will need to work harder to find a complete function basis. It
should also be noted that we have only identified the leading term in the
$\eps$ expansion~\eqref{eq:coeffexp} of the integral coefficient, since the tree amplitudes were
evaluated in four dimensions. These contributions to loop amplitudes are
usually referred to as \textit{cut-constructible}.

\begin{svgraybox}
\textbf{Which Feynman diagrams have we calculated?} We can try to put this in the context of the Feynman diagram expansion. In an
axial gauge without ghosts, one can check that there are 39 diagrams
contributing to the full colour four-gluon one-loop amplitude, of which 17
contribute at leading colour. In the $s_{12}$-channel cut only 9 of the 17
ordered diagrams contribute.  What we should take from this is that a direct
Feynman diagram computation is not at all prohibitive here with only a modest
number of diagrams contributing.  However, one of these diagrams is the most
complicated tensor integral we can find for massless four-particle kinematics,
and will contain four powers of loop momentum in the numerator. The number of
three- and four-point vertices also means each diagram will expand to a large
number of terms.  The use of compact on-shell trees has allowed us to avoid
a lot of this complexity. As we have highlighted, some contributions have been
dropped but we have obtained a lot of information about the amplitude.
\end{svgraybox}

We may now try to complete the cut-constructible part of the four-gluon
scattering amplitude by considering the cut in the other independent invariant.

\newpage 

\begin{example}{Example: The $s_{23}$-channel cut of $gg\to gg$ MHV scattering amplitude}

The $s_{23}$-channel cut associated with the invariant $s_{23} = (p_2+p_3)^2$ of our example
$A^{(1)}(1^-,2^-,3^+,4^+)$ is slightly more complicated, since the sum over internal
helicities contains more non-zero elements. The cut integrand can be written as
\begin{align}
  \mathcal{C}_{23|41}&\left(I^{(1)}(1^-,2^-,3^+,4^+)\right) = \nonumber\\&
  \sum_{h_i=\pm} \i \, A^{(0)}\bigl(-l_1^{-h_1},2^-,3^+,l_2^{h_2}\bigr) \, \i \, A^{(0)}\bigl(-l_2^{-h_2},4^+,1^-,l_1^{h_1}\bigr) \nonumber\\
  = \ & \i \, A^{(0)}(-l_1^+,2^-,3^+,l_2^-) \, \i \, A^{(0)}(-l_2^+,4^+,1^-,l_1^-) \nonumber\\
  &+ \i \, A^{(0)}(-l_1^-,2^-,3^+,l_2^+) \, \i \, A^{(0)}(-l_2^-,4^+,1^-,l_1^+) \nonumber\\
  = \ &  \frac{\spA2{l_2}^4}{\spA{l_1}2\spA23\spA3{l_2}\spA{l_2}{l_1}} \frac{\spA1{l_1}^3}{\spA{l_2}2\spA41\spA{l_1}{l_2}} \nonumber\\
  &+ \frac{\spA{l_1}2^3}{\spA23\spA3{l_2}\spA{l_2}{l_1}} \frac{\spA{l_2}1^4}{\spA{l_2}2\spA41\spA1{l_1}\spA{l_1}{l_2}} \,.
  \label{eq:4gtchannelintegrand}
\end{align}
While not the most complicated expression, it is not as easy to express this in
terms of a basis of cut scalar integral functions as it was in the case of the
$s_{12}$-channel.
The aim is to reduce the complexity of the dependency on expressions
involving the loop momentum, and to identify the integral topologies that we
expect to find. This means identifying loop-momentum dependent propagators of
the form $(l+p)^2$. Performing the spinor algebra would be difficult if there
was no target to aim for, so we can also remind ourselves of the $s_{12}$-channel cut
result, which identified a simple scalar box integral as defined in Section~\ref{sec:3.1}. 
The $s_{23}$-channel cut should also be sensitive to the same function and so we can try to expose this term.
Let us look at the expression again, putting everything over a common
denominator (as before we use the short hand $\mathcal{C}_{23|41}$ to refer to $\mathcal{C}_{23|41}\left(I^{(1)}(1^-,2^-,3^+,4^+)\right)$ in this section),
\begin{align}
  \mathcal{C}_{23|41} =& \frac{\spA{l_1}1^4\spA{l_2}2^4 + \spA{l_1}2^4\spA{l_2}{1}^4}{\spA{l_1}{l_2}^2\spA{l_1}1\spA{l_1}{2}\spA{l_2}3\spA{l_2}4\spA23\spA14} \,.
\end{align}
The $s_{12}$-channel cut contained the spinor bracket $\spA{l_1}{l_2}$ in the numerator and so, following the motivation to expose a similar box structure, we can apply a Schouten identity,
\begin{equation}
  \spA{l_1}1\spA{l_2}2 - \spA{l_1}{l_2}\spA12 - \spA{l_1}2\spA{l_2}1 = 0 \,,
\end{equation}
which will produce a term very similar to the $s_{12}$-channel integrand. We can then write
\begin{align}
  \mathcal{C}_{23|41} = \ & \mathcal{C}_{23|41}^{\rm box} + \mathcal{C}_{23|41}' \,,
  \label{eq:tchannelcut1}
\end{align}
where
\begin{align} 
  \mathcal{C}_{23|41}^{\rm box} & =  \frac{\spA{l_1}{l_2}^2\spA12^4}{\spA{l_1}1\spA{l_1}{2}\spA{l_2}3\spA{l_2}4\spA23\spA14}\nonumber\\
  & = - \bigl(-\i A^{(0)}(1^-,2^-,3^+,4^+) \bigr) s_{12} s_{23} \frac{1}{(l_1-p_2)^2 (l_1+p_1)^2} \,,
  \label{eq:tchannelcut2}
\end{align}
and---after a reasonable amount of spinor manipulation---one finds
\begin{align}
  \mathcal{C}_{23|41}' = -\i A^{(0)}(1^-,2^-,3^+,4^+) &\frac{- 2}{s_{12}^2s_{23}^2}\left( \trmgamtwo1{\slashed{l}_2}32^2 + s_{12}s_{23}\trmgamtwo1{\slashed{l}_2}32 + 2 s_{12}^2s_{23}^2 \right)\nonumber\\&\times
  \left( 1 + \frac{\trmgamtwo1{\slashed{l}_1}42}{(l_1+p_1)^2 s_{12}} - \frac{\trmgamtwo2{\slashed{l}_1}31}{(l_1-p_2)^2 s_{12}} \right) \,,
  \label{eq:tchannelcut3}
\end{align}
where we have used the notation $\trm{a}{b}{c}{d} = {\rm tr}\left(
(1-\gamma_5)\slashed{a}\slashed{b}\slashed{c}\slashed{d} \right)/2 =
\spAB{a}{bcd}{a}$. The second part, $\mathcal{C}_{23|41}'$, contains three different propagator factors. After expanding
we can identify them as cut bubble and triangle configurations with loop-momentum dependence remaining in the numerator. The numerators in this case are up
to \emph{rank} three in the loop momentum, where rank refers to the power of loop momentum appearing the numerator. Simplification will require further
reduction techniques, which we will introduce in section~\ref{sec:3.4}, and will be used to identify a basis of integral functions. 
\end{example}

\begin{exer}{The four-gluon amplitude in $\mathcal{N}=4$ super-symmetric Yang-Mills theory}
\label{Ex:3.1}
Supersymmetry is an additional symmetry between particles of different spins.
This can relate fermions and scalars or fermions and gauge bosons, and the
precise type of supersymmetry requires us to specify how the degrees of freedom (d.o.f.)
are connected. Maximally supersymmetric Yang-Mills theory or $\mathcal{N}=4$
super-symmetric Yang-mills (sYM) theory has the maximum number of connections between
gluon, gluinos (adjoint-representation fermions) and scalars (also in the
adjoint representation). Connecting all degrees of freedom requires 2 gluon
degrees of freedom, i.e.\ positive and negative helicity, 4 gluinos flavours
also with positive and negative helicity, and 3 complex scalar degrees of
freedom (or equivalently 6 real scalars). The consequences of this additional
symmetry are remarkable cancellations and appearance of hidden structures that
are still an active field of research. For the purposes of this exercise, all
that we need to know about the theory are its particle content and the
tree-level amplitudes needed inside the double cuts.

\begin{table}[t]
\centering
\begin{tabular}[h]{c|ccccc}
  particle &$g^+$ & $\Lambda^+$ & $\phi$ & $\Lambda^-$ & $g^-$ \\
  \hline \\
  d.o.f.\ & 1 & 4 & 6 & 4 & 1
\end{tabular}
\caption{The particle content of $\mathcal{N}=4$ sYM theory and their degrees of freedom (d.o.f.). Here
$g^\pm$ represent positive- and negative-helicity gluons, $\Lambda^\pm$
represent positive- and negative-helicity gluinos, and $\phi$ represent real scalars.}
\label{fig:nis4part}
\end{table}

The particle content of $\mathcal{N}=4$ sYM theory is summarised in fig.~\ref{fig:nis4part}.
In addition to the Parke-Taylor MHV formula~\eqref{ParkeTaylor1} for gluons we also have
\begin{align}
\begin{aligned}
  A^{(0)}(1_\Lambda^-,2^-,3^+,4_\Lambda^+) &= \frac{-\i \, \spA12^3 \spA24}{\spA12\spA23\spA34\spA41} \,, \\
  A^{(0)}(1_\Lambda^-,2^+,3^-,4_\Lambda^+) &= \frac{-\i \, \spA13^3 \spA34}{\spA12\spA23\spA34\spA41} \,, \\
  A^{(0)}(1_\Lambda^+,2^-,3^+,4_\Lambda^-) &= \frac{-\i \, \spA12 \spA24^3}{\spA12\spA23\spA34\spA41} \,, \\
  A^{(0)}(1_\Lambda^+,2^+,3^-,4_\Lambda^-) &= \frac{-\i \, \spA13 \spA34^3}{\spA12\spA23\spA34\spA41} \,, \\
  A^{(0)}(1_\phi,2^-,3^+,4_\phi) &= \frac{\i \, \spA12^2\spA24^2}{\spA12\spA23\spA34\spA41} \,, \\
  A^{(0)}(1_\phi,2^+,3^-,4_\phi) &= \frac{\i \, \spA13^2\spA34^2}{\spA12\spA23\spA34\spA41} \,,
\end{aligned}
\end{align}
where we omit the particle subscripts for gluons.
All other amplitudes with two like-helicity gluons are zero.

Use these tree-level amplitudes to show that the cut four-gluon one-loop integrands $I^{(1)}_{\mathcal{N}=4}(1^-,2^-,3^+,4^+)$ in $\mathcal{N}=4$ sYM theory are given by
\begin{align}
  \mathcal{C}_{12|34}\left( I^{(1)}_{\mathcal{N}=4}(1^-,2^-,3^+,4^+) \right) &= \mathcal{C}_{12|34}\left( I^{(1)}(1^-,2^-,3^+,4^+) \right) \,, \\
  \mathcal{C}_{23|41}\left( I^{(1)}_{\mathcal{N}=4}(1^-,2^-,3^+,4^+) \right) &= \mathcal{C}_{23|41}^{\rm box}\left( I^{(1)}(1^-,2^-,3^+,4^+) \right) \,,
\end{align}
where on the RHSs are the cut integrands in YM theory computed above.
In contrast to YM theory, in $\mathcal{N}=4$ sYM theory both cuts match the one-loop scalar box integral~\cite{ch3_Bern:1994zx}. In other words, the term $\mathcal{C}'_{23|41}$ containing different propagator factors in \eqn{eq:tchannelcut1} is absent from the $s_{23}$-channel cut in sYM theory. After summing over the two independent cuts we thus find that
\begin{align}
\begin{aligned}
  A^{(1),\mathcal{N}=4}(1^-,2^-,3^+,4^+) = \ & - A^{(0)}(1^-,2^-,3^+,4^+) \, s_{12} s_{23} \, F^{[D]}_4(s_{12},s_{23}) \\ &+ \text{terms missed by cuts in 4D} \,.
\end{aligned}
\end{align}
Note that we have upgraded the integral into $D$-dimensions in order to regulate divergences.
Hint: the gluino's contribution to the cuts comes with a negative sign as a result of the Feynman rule for the closed fermion loops.
For the solution see \hyperref[Sol:3.1]{chapter~5}.
\end{exer}

We finish this section with a few remarks.
\begin{itemize}
  \item The unitarity cuts allowed us to extract information about the rational
    coefficients of one-loop integrals from the product of on-shell tree
    amplitudes.
  \smallskip
  \item While in simple cases such as the $s_{12}$-channel MHV four-gluon cut or
    maximally super-symmetric theories spinor manipulations were sufficient to
    identify an integral basis, additional work will be
    required to identify a basis of integrals in general. We will return to this point in Section~\ref{sec:3.4}.
  \smallskip
  \item The double cuts project out information on multiple coefficients and
    integral structures at the same time. If there were an operation that could
    project out one integral coefficient at a time, this would avoid difficult
    kinematic manipulations. This would be particularly important for
    amplitudes with more external legs, where the algebra can quickly get out of
    hand. We will explore this line of thought in the next section.
\end{itemize}

\section{Generalised unitarity \label{sec:3.3}}

The name ``generalised unitarity'' refers to the action of putting more than two
propagators inside the loop amplitude on-shell. In fact the name is something
of a misnomer, since the connection with the unitarity of the $S$-matrix is now
lost. A better term could be \textit{generalised discontinuities}, which relates
to the work of Cutkosky~\cite{ch3_Cutkosky:1960sp}.

Let us consider a one dimensional integral over the real line,
\begin{equation}
  f\bigl(p^2\bigr) = \int \d k \frac{1}{k^2-p^2} \,,
\end{equation}
where $p$ is a real number.\footnote{Note that this integral diverges over the
full range $(-\infty,\infty)$, the argument presented still follows if a large-$k$ cut-off regulator is introduced.} This function has a discontinuity
\begin{align}
  {\rm Disc}_{p^2}\left(f\bigl(p^2\bigr)\right)
  &= f\bigl(p^2+\i0\bigr)-f\bigl(p^2-\i0 \bigr) \nonumber\\
  &= \int \d k \left( \frac{1}{k^2-p^2-\i 0} - \frac{1}{k^2-p^2+\i 0} \right) \,.
\end{align}
The two terms on the RHS of this relation can be expanded into principle values and $\delta$ functions,
\begin{align}
  \frac{1}{k^2-p^2 \pm \i 0} = \mp \i \pi \, \delta\bigl(k^2-p^2\bigr) + \mathcal{P}\left( \frac{1}{k^2-p^2} \right) \,,
\end{align}
of which only the $\delta$-function contributions remain,
\begin{equation}
  {\rm Disc}_{p^2}\left(f(p^2)\right) = \int \d k \, 2\pi \i \, \delta(k^2-p^2) \,.
\end{equation}
Following this argument one can show that a multiple discontinuity (or
multiple cut) of an amplitude can be obtained by replacing
\begin{equation} \label{eq:cut_delta}
  \frac{1}{k^2-m^2+\i0} \rightarrow -2\pi\i \, \delta^{(+)}\left( k^2 - m^2 \right) \,,
\end{equation}
for a subset of the propagators in the integrand of a loop diagram. The act of replacing a
propagator by a $\delta$ function as in \eqn{eq:cut_delta} is referred to as
\emph{cutting} that propagator.  The integrand will then \textit{factorise}
into on-shell tree amplitudes.

\begin{important}{Multiple cuts of scattering amplitudes}
  By systematically putting loop propagators on-shell using the above
  \eqn{eq:cut_delta}, we can break up the loop amplitude into manageable pieces
  each of which isolates a particular subset of Feynman integral topologies. A
  \textit{maximal cut} of a scattering amplitude is the contribution in which
  the highest number of propagators are put on-shell. Our
  one-loop four-gluon example has at most four propagators from the box
  configuration and so the maximal cut is a \textit{quadruple cut}. We may thus
  use the factorised product of tree-level amplitudes to obtain information
  about the coefficient of the box integrals. We may then proceed to release
  cut constraints and use \textit{triple cuts} which will identify both
  triangle and box topologies. Since we have previously identified the box
  configurations, the triangle integral coefficients can now be uniquely
  identified. We may then proceed with the \textit{double cuts} that relate to
  the discontinuity of the one-loop amplitude and so on until the complete
  function is determined. This top-down approach can be taken at higher loop
  order as well. Cuts may be taken in four (using four-dimensional tree-level amplitudes in the
  factorisation) or $D=4-2\eps$ dimensions.
\end{important}

\begin{example}{Example: Quadruple cut of $gg\to gg$ MHV scattering amplitude}

The maximal cut of the ordered four gluon amplitude isolates a single Feynman
diagram by putting four propagators on-shell.

If we can find a solution to the system of equations which places all four propagators on-shell,
then the four-dimensional part of the loop integration will be completely
fixed. As with the double cuts, we will remain in four dimensions for the time
being, and come back to the issue of dimensional regularisation later. Let us
denote this quadruple cut operation as $\mathcal{C}_{1|2|3|4}$, and represent
the action on the four gluon amplitude using the following graphical notation:
  \begin{equation}
    \mathcal{C}_{1|2|3|4} \left(  \raisebox{-1.3cm}{\includegraphics[width=2.5cm]{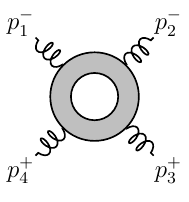}} \right)
    \phantom{xx} = 
    \raisebox{-1.5cm}{\includegraphics[width=3.5cm]{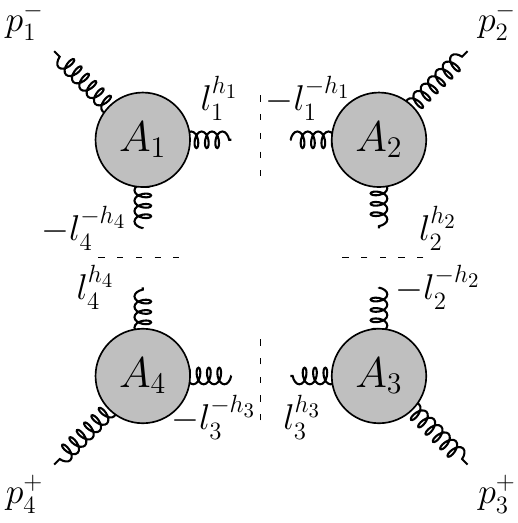}}
    \label{eq:quadcut4g}
  \end{equation}
Each cut leads to a factorised product of trees where the sum over polarisation
states is implicit. The momenta in each of the four propagators, $l_i$, have
been put on-shell by solving the conditions $l_i^2 = 0$. To find an explicit
solution we can use a basis constructed from the spinors of the external
momenta such as
\begin{equation}
  l_1^\mu = \alpha_1 \, p_1^\mu + \alpha_2 \, p_2^\mu +  \alpha_3 \, \frac{1}{2}\spAB{1}{\gamma^\mu}{2} + \alpha_4 \, \frac{1}{2} \spAB{2}{\gamma^\mu}{1} \,.
  \label{eq:quadcutmombasis}
\end{equation}
Using momentum conservation, we re-write the four on-shell constraints as
\begin{align} \begin{aligned}
  l_1^2 &= 0 \,, \\
  l_2^2 &= (l_1-p_2)^2 = 0 \overset{l_1^2=0}{=} 2 \, l_1\cdot p_2 = 0 \,, \\
  l_3^2 &= (l_1-p_2-p_3)^2 = 0 \overset{
        \begin{subarray}{l}
          l_1^2=0\\
          l_2\cdot p_2=0
        \end{subarray}
       }{\Rightarrow} 2 \, l_1\cdot p_3 = s_{23} \,, \\
  l_4^2 &= (l_1+p_1)^2 = 0 \overset{l_1^2=0}{=} 2 \, l_1\cdot p_1 = 0 \,,
  \end{aligned}
  \label{eq:quadcuteqs1}
\end{align}
which then are easily translated into conditions on the coefficients $\alpha_i$,\footnote{The identities used perform the spinor-helicity algebra are given Section~\ref{sec:1.7}, in particular the Fierz identity in \eqn{eq:Fierz}. The reader may also refer to exercise~\ref{Ex:1.6} where the identity is proven.}
\begin{align} 
\label{eq:quadcuteqs2}
\begin{aligned}
  &\alpha_1\alpha_2 - \alpha_3\alpha_4 = 0 \,, \\
  &\alpha_1 s_{12} = 0 \,, \\
  &\alpha_1 s_{13} + \alpha_2 s_{23} + \alpha_3 \spAB132 + \alpha_4 \spAB231 = s_{23} \,, \\
  &\alpha_2 s_{12} = 0 \,.
 \end{aligned}
\end{align}
These must hold for generic external kinematics, i.e., for $s_{12} \neq 0 $ and $s_{23} \neq 0$ ($s_{13}=-s_{12}-s_{23}$ because of momentum conservation). 
The second and fourth equations in the system~\eqref{eq:quadcuteqs2} allow us to simplify the first constraint, which
becomes $\alpha_3 \alpha_4 = 0$, and so we see that there are exactly two
solutions to the quadruple cut on-shell conditions:
\begin{align}
  \vec{\alpha}^{(1)} &= \Bigl\{ 0, 0, \tfrac{\spA23}{\spA13} , 0 \Bigr\} \,, \\
  \vec{\alpha}^{(2)} &= \Bigl\{ 0, 0, 0, \tfrac{\spB23}{\spB13} \Bigr\} \,,
  \label{eq:quadcutsol}
\end{align}
where we introduced the short-hand notation $\vec{\alpha} = \{\alpha_1,\alpha_2,\alpha_3,\alpha_4\}$.
These solutions deserve a few remarks. We see that the two solutions are
complex, and in fact are complex conjugates of each other. In order to extract
the value of the quadruple cut we will sum and average over the two solutions
as well as the sum over helicity in the factorised product of trees. For now we
simply state that this is the correct method to obtain the scalar integral
coefficient, though we will return to prove this later in
section~\ref{ssec:4dintegrandbasis}. The fact that the loop momenta are complex
means we must analytically continue the factorised tree-level amplitudes for
complex momenta as well.\footnote{The on-shell delta functions in the cut
integrals should also be reinterpreted as residue computations.} This step is
quite familiar to us, since we have already encountered analytic continuation
of tree amplitudes in the context of BCFW recursion. However, we should not
underestimate the importance of having a well defined analytic continuation. In
fact this feature was one of the main obstacles in the analytic $S$-matrix
program of the~1960's.

Turning back to our example, all that remains to do is to substitute the
on-shell solutions into the tree-amplitude expressions. Since these tree-level
amplitudes will be of the form of Parke-Taylor MHV amplitudes, it is first
convenient to write explicit spinor solutions for the the loop momenta $l_i$.
There is a flexibility in how to do this because of the little group symmetry,
but a simple choice using $z = {\spA23}/{\spA13}$ is
\begin{align} \begin{aligned}
  |l_1^{(1)} \r> &= |1\r> \,, & |l_1^{(1)}] &= z \, |2] \,, \\
  |l_2^{(1)} \r> &= z\, |1\r> - |2\r> \,, \qquad \quad & |l_2^{(1)}] &= |2] \,, \\
  |l_3^{(1)} \r> &= z\, |1\r> - |2\r> \,, & |l_3^{(1)}] &= \frac{s_{23}}{z \, s_{12}}\bigl(|1] + z \, |2]\bigr) \,, \\
  |l_4^{(1)} \r> &= |1\r> \,, & |l_4^{(1)}] &= |1] + z |2] \,,
  \end{aligned}
  \label{eq:quadcutsolspinors}
\end{align}
for the first solution, while the second is obtained by complex conjugation (i.e.\ replacing $|\r>
\leftrightarrow |]$ which also means $z\leftrightarrow z^\dagger$). Other choices of spinor normalisation will not affect the
final answer.

\begin{figure}[h]
  \centering
  \begin{tabular}[h]{cc}
    \includegraphics[width=4cm]{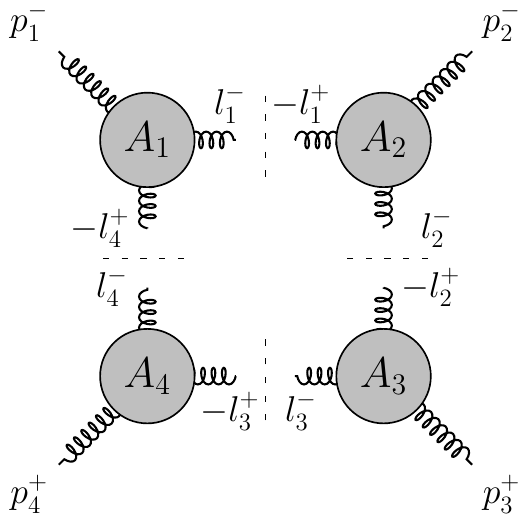} &
    \includegraphics[width=4cm]{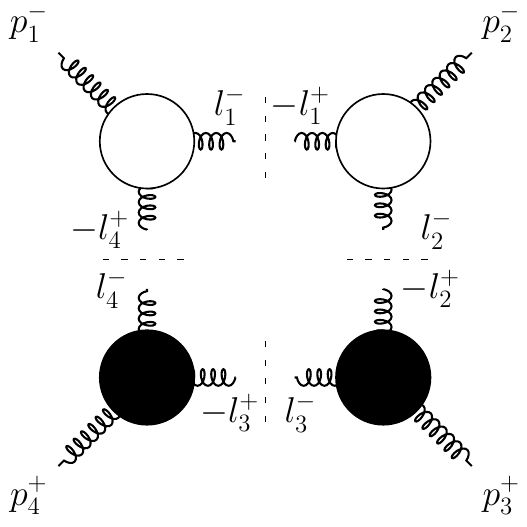}
  \end{tabular}
  \caption{An example helicity configuration contributing to the quadruple cut of the four gluon MHV amplitude. In the right panel the configuration is shown with MHV-type vertices shaded in white and \MHVb vertices shaded in black.}
  \label{fig:4gcutexhel}
\end{figure}

We are now ready to start substituting the on-shell solutions into the
expressions for the tree amplitudes. Let us start with the configuration of
internal helicities shown in figure \ref{fig:4gcutexhel}. Each of the
three-point amplitudes is either MHV or $\overline{\rm MHV}$. The first
amplitude at the top left of the cut evaluated on the first on-shell solution
is
\begin{equation}
  A_1  = \i \, \frac{\spA1{l_1}^3}{\spA{l_1}{(-l_4)}\spA{(-l_4)}{l_1}} \,,
  \label{eq:4ggucutexhel1}
\end{equation}
from which it is simple to see $A_1|_{l_i^{(1)}} = 0$, since the spinor solution
in \eqn{eq:quadcutsolspinors} shows the angle spinor for $l_1$ is
proportional to the angle spinor of $p_1$. One may worry about the remaining
spinor products in the denominator since they can also be shown to vanish on
the solution $l_1^{(1)}$, but the overall dimension of the three-point amplitude
ensures that $A_1|_{l_i^{(1)}}$ is indeed zero. There is a similar story for the
solution $l_1^{(2)}$, where we see that, since $|l_1^{(2)}\r> \propto |2\r>$, then
$A_2|_{l_1^{(2)}} = 0$. As a result we find no contribution from this helicity
configuration. This is a general feature of quadruple cuts for massless theories,
and we can use the fact that three-point amplitudes contain only angle or square
brackets to conclude:
\begin{important}{Three-point vertex rule for unitarity cuts.}
  Unitarity cuts of one-loop amplitudes do not support adjacent MHV (or \MHVb) three-point vertices.
\end{important}
A popular and convenient graphical notation is to shade the three-point
vertices to indicate whether they are of either MHV (white) or \MHVb~(black),
as shown in the right panel of Figure~\ref{fig:4gcutexhel}, which demonstrates that
this internal helicity configuration vanishes since we have highlighted
adjacent MHV amplitudes. Applying this rule to the full helicity
sum leads us to find only two non-vanishing contributions, one for each of the
two solutions:
\begin{align}
  \mathcal{C}_{1|2|3|4} \left(  \raisebox{-1.3cm}{\includegraphics[width=2.5cm]{ch3_loopamps/figures/4g1Lampblob_MHV.pdf}} \right) \Bigg|_{l_i^{(1)}}
    \phantom{xx} &=
    \raisebox{-1.7cm}{\includegraphics[width=3.5cm]{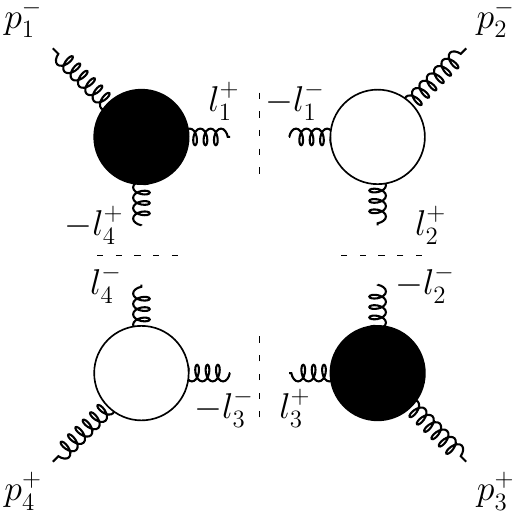}} \Bigg|_{l_i^{(1)}} \,, \\
  \mathcal{C}_{1|2|3|4} \left(  \raisebox{-1.3cm}{\includegraphics[width=2.5cm]{ch3_loopamps/figures/4g1Lampblob_MHV.pdf}} \right) \Bigg|_{l_i^{(2)}}
    \phantom{xx} &=
    \raisebox{-1.7cm}{\includegraphics[width=3.5cm]{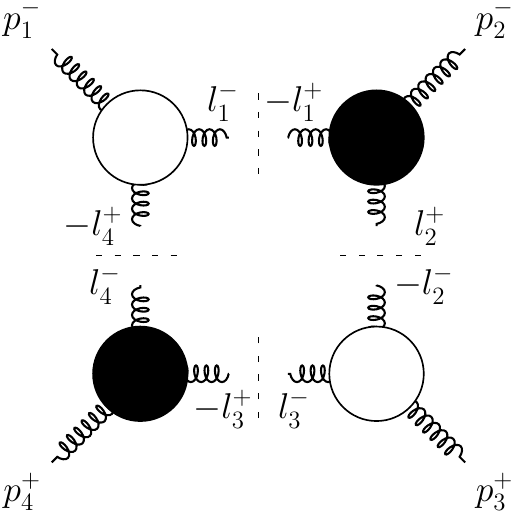}} \Bigg|_{l_i^{(2)}} \,.
  \label{eq:4gcutnonvanishinghels}
\end{align}

We can now complete the computation of the quadruple cut:
\begin{align}
    \raisebox{-1.5cm}{\includegraphics[width=3cm]{ch3_loopamps/figures/4g1L_gucut_hel1.pdf}} \Bigg|_{l_i^{(1)}}
    =
    \frac{\spB{l_1}{l_4}^3}{\spB{l_4}{1}\spB{1}{l_1}}
    \frac{\spA{l_1}{2}^3}{\spA{2}{l_2}\spA{l_2}{l_1}}
    \frac{\spB{3}{l_3}^3}{\spB{l_3}{l_2}\spB{l_2}{3}}
    \frac{\spA{l_3}{l_4}^3}{\spA{l_4}{4}\spA{4}{l_3}} \bigg|_{l_i^{(1)}}\,.
  \label{eq:4gcutresidue1a}
\end{align}
Substituting \eqn{eq:quadcutsolspinors} for $l_1, l_2$ and $l_4$ gives
\begin{align}
    \raisebox{-1.5cm}{\includegraphics[width=3cm]{ch3_loopamps/figures/4g1L_gucut_hel1.pdf}} \bigg|_{l_i^{(1)}}
    \hspace{-0.3cm} =
    \frac{(z\spB21)^3}{(z\spB21)(z\spB12)}
    \frac{\spA12^3}{(z\spA21)(-\spA21)}
    \frac{\spB{3}{l_3}^3}{\spB{l_3}{2}\spB23}
    \frac{\spA{l_3}{1}^3}{\spA14\spA{4}{l_3}} \Bigg|_{l_i^{(1)}}\,.
  \label{eq:4gcutresidue1b}
\end{align}
The substitution of $l_3$ would require some spinor manipulation at first sight but the spinor products can be combined together to make ``sandwiches'' of the $l_3$ momentum,
after which where we can evaluate using momentum conservation:
\begin{equation}
  \frac{\spB{3}{l_3}^3}{\spB{l_3}{2}}\frac{\spA{l_3}{1}^3}{\spA{4}{l_3}} \bigg|_{l_i^{(1)}} = \frac{\spAB1{l_3}3^3}{\spAB4{l_3}2} \bigg|_{l_i^{(1)} } = \frac{\spA12^3\spB23^3}{\spA43\spB32} = \frac{\spA12^3\spB23^2}{\spA34} \,.
  \label{eq:l3sub1}
\end{equation}
Putting everything together, we find the final result is simply\footnote{As is always the case with spinor algebra there are many paths to reach the same final result. Here we have attempted to be very explicit but the reader may prefer alternative derivations. For example, by expanding in the spinor basis for $p_1$ and $p_2$ we broke some symmetry in the original configuration. One can apply some more algebra to show that $|l_3^{(1)}\r> = |3\r>, |l_3^{(1)}] = \tfrac{\spA14}{\spA13} |4]$ for example, in which we would get a simple result without first combining into spinor sandwiches.}
\begin{align}
    \raisebox{-2cm}{\includegraphics[width=4cm]{ch3_loopamps/figures/4g1L_gucut_hel2.pdf}} \Bigg|_{l_i^{(2)}} 
    = -s_{12} s_{23} \bigl(-\i A^{(0)}(1^-,2^-,3^+,4^+) \bigr) \,.
\end{align}
A similar computation for the other non-zero cut configuration leads to
\begin{align}
    \raisebox{-2cm}{\includegraphics[width=4cm]{ch3_loopamps/figures/4g1L_gucut_hel2.pdf}} \Bigg|_{l_i^{(2)}} 
    = -s_{12} s_{23} \bigl(-\i A^{(0)}(1^-,2^-,3^+,4^+) \bigr) \,.
\end{align}
The coefficient of the scalar box integral is the average of the two solutions, and so we recover the result observed from the double cuts, that is
\begin{equation}
  A^{(1)}(1^-,2^-,3^+,4^+) =  \i \, c_{0;1|2|3|4}(1^-,2^-,3^+,4^+) \, F_4^{[D]}(s_{12},s_{23}) + \text{subtopologies}\,,
\end{equation}
where
\begin{align}
  c_{0;1|2|3|4}(1^-,2^-,3^+,4^+) &= \frac{1}{2} \sum_{s=1}^2 \mathcal{C}_{1|2|3|4}\left( I^{(1)}(1^-,2^-,3^+,4^+) \right) \Bigl|_{l_i^{(s)}} \nonumber \\
  &= -s_{12} s_{23} \bigl(-\i \, A^{(0)}(1^-,2^-,3^+,4^+)\bigr) \,.
  \label{eq:4gquadcutfinal}
\end{align}

\end{example}

\begin{exer}{Quadruple cuts of five-gluon MHV scattering amplitudes}
\label{Ex:3.2}
\vspace{-0.6cm}
\begin{enumerate}[a)]
\item  Follow the method of quadruple cuts for the one-loop five-gluon amplitudes to show that
\begin{equation}
  c_{0;1|2|3|45}(1^-,2^-,3^+,4^+,5^+) = \frac{\i}{2} s_{12} s_{23} A^{(0)}(1^-,2^-,3^+,4^+,5^+) \,.
\end{equation}
\item A more complicated example is required to show that we will not always find that box coefficients are proportional to tree-level amplitudes. Using the same technique, show that
\begin{align}
\begin{aligned}
   c_{0;1|23|4|5}(1^+,2^+,3^-,4^+,5^-) = & \, \frac{\i}{2}  s_{45} s_{15} \, A^{(0)}(1^+,2^+,3^-,4^+,5^-) \\
  & \times \left[ \left( \frac{\spA34 \spA15}{\spA14 \spA35} \right)^4 + \left( \frac{\spA13 \spA45}{\spA14 \spA35} \right)^4 \right]  \,.
 \end{aligned}
\end{align}
\end{enumerate}
For the solution see \hyperref[Sol:3.2]{chapter~5}.
\end{exer}

\section{Reduction methods \label{sec:3.4}}

Through the concepts of unitarity and generalised unitarity cuts we have been
able to understand better the meaning of equation \eqref{eq:loopamp-general}.
While performing two-particle cuts, we saw that the cut-constructible part of
the four-gluon MHV amplitude could be written in terms of a scalar box integral
and sub-topologies written in terms of triangle and bubble integrals with some
non-trivial numerator function. In this section will we show how to reduce this
loop dependent tensor numerators to basis integral functions. We will then see how we can extend these ideas to find a basis of integrand level structures. 

\subsection{Tensor reduction}
\label{sec:TensorReduction}

This approach to the computation of one-loop amplitudes due to Passarino and
Veltman~\cite{ch3_Passarino:1978jh} revolutionised the field of precision theoretical predictions for high
energy experiments. The method is remarkably and elegantly simple. We will
restrict ourselves to massless propagators as before although the method is
equally applicable in the general case. Consider a tensor integral such as
\begin{equation}
  F_n^{[D]}(p_1,\ldots,p_{n-1})[k^\mu] = \int_k \frac{k^\mu}{\prod_{a=1}^{n}[-(k-q_a)^2]} \,,
  \label{eq:tensorintexample}
\end{equation}
where $q_{a} = \sum_{b=1}^{a-1} p_b$ as before. Feynman's $\i 0$ prescription is irrelevant for the purpose of this section, hence we omit it. After integration the
integral can only depend on the independent external momenta, and so the vector can be described by a linear combination of $n-1$ external
momenta, as
\begin{equation}
  F^{[D]}_{n}[k^\mu] = \sum_{i=1}^{n-1} a_{n,i} \, p_i^\mu \,,
  \label{eq:testintexampledecomp}
\end{equation}
where we have dropped the momentum argument on the LHS for a more
compact notation. The coefficients $a_{n,i}$, referred to as \emph{form factors}, can then be determined by
constructing a linear system of equations through contractions of eqs.~\eqref{eq:tensorintexample} and~\eqref{eq:testintexampledecomp} with the basis
vectors $p_i$. For illustrative purposes it is useful to take a specific
example, say $n=2$ with $p_1^2\neq 0$. We then find one equation,
\begin{align}
  \int_k \frac{k\cdot p_1}{k^2 (k-p_1)^2} = a_{2,1} \, p_1^2 \,.
\end{align}
By rewriting the scalar product in the numerator in terms of inverse propagators through
 $2 \, k\cdot p_1 = k^2 - (k-p_1)^2 + p_1^2$, we can expand the LHS into three scalar integrals,
\begin{align}
  \frac{1}{2} \int_k \frac{1}{(k-p_1)^2}  - \frac{1}{2} \int_k \frac{1}{k^2} + \frac{p_1^2}{2} \int_k \frac{1}{k^2 (k-p_1)^2} = a_{2,1} \, p_1^2 \,.
\end{align}
The first two scalar integrals on the LHS have the topology of a tadpole. Since they do not depend on any external scale, they are zero in dimensional regularisation,\footnote{We will come back to this non-trivial aspect in chapter~\ref{ch:loopints}.} and so we arrive at the well known result
\begin{align}
  a_{2,1} = \frac{1}{2} F^{[D]}_2(p_1)[1] \,.
  \label{eq:PVbubR1sol}
\end{align}
For a general tensor we can decompose into bases of external momenta and the metric tensor, for example,
\begin{align}
  F^{[D]}_{2}[k^{\mu_1}k^{\mu_2}] &=  \int_k \frac{k^{\mu_1} k^{\mu_2}}{k^2 (k-p_1)^2} = a_{2,00} \, \eta^{\mu_1\mu_2} + a_{2,11} \, p_1^{\mu_1} p_1^{\mu_2}  \,,\label{eq:PVbubR2dec}\\
  F^{[D]}_{2}[k^{\mu_1}k^{\mu_2} k^{\mu_3}] &=  \int_k \frac{k^{\mu_1} k^{\mu_2}k^{\mu_3}}{k^2 (k-p_1)^2} \nonumber\\
  & = a_{2,001} \left( \eta^{\mu_1\mu_2} p_1^{\mu_3} + \eta^{\mu_2\mu_3} p_1^{\mu_1} + \eta^{\mu_3\mu_1} p_1^{\mu_2} \right) \nonumber\\
  & \phantom{= {}} + a_{2,111} \, p_1^{\mu_2} p_1^{\mu_2} p_1^{\mu_3} \,.
\label{eq:PVbubR3dec}
\end{align}
Note that the final example is a rank-three two-point function, which would not
appear in a conventional renormalisable gauge theory, which permits a maximum
tensor rank of $n$ for a $n$-point one-loop function. This follows from the
restrictions on the mass dimension of the operators that represent the
interactions leading to a general counting of one power of momentum per
three-point vertex. This is not the case for gravity theories (see section~\ref{sec:1.4}) or effective field theories.

Explicit solutions for the form factors are easy to find with an automated computer algebra
system, although for higher-point integrals can quickly generate large
expressions and complicated denominators from the determinant of the linear
system of equations. These are known as \emph{Gram determinants}, since they are
related to the Gram matrix computed from the independent external momenta, that is, the matrix of entries
$[G_n]_{ij} = p_i \cdot p_j$ with $i,j=1,\ldots,n-1$.

There are many references in which well organised analytic solutions are presented, see~\cite{ch3_Ellis:2011cr} and references therein for a summary. 
Many of these have seen extensive use in high energy physics applications. We leave the complete solution to the bubble system as an exercise.

\begin{exer}{Tensor decomposition of the bubble integral}
\label{Ex:3.3}
\vspace{-0.6cm}
\begin{enumerate}[a)]
\item Prove that the form factors in the decomposition of the rank-two bubble integral in \eqn{eq:PVbubR2dec} are given by
\begin{align} \label{eq:rank2bubsol}
\begin{aligned}
  a_{2,00} &= -\frac{p_1^2}{4(D-1)} \, F_2^{[D]}(p_1)[1] \,, \\
  a_{2,11} &= \frac{D}{4(D-1)} \, F_2^{[D]}(p_1)[1] \,.
\end{aligned}
\end{align}

\item Prove that the form factors in the decomposition of the rank-three bubble integral in \eqn{eq:PVbubR3dec} are given by
\begin{align} \label{eq:rank3bubsol}
\begin{aligned}
  a_{2,001} &= -\frac{p_1^2}{8(D-1)} \, F_2^{[D]}(p_1)[1] \,, \\
  a_{2,111} &= \frac{D+2}{8(D-1)} \, F_2^{[D]}(p_1)[1] \,.
\end{aligned}
\end{align}
\end{enumerate}
For the solution see \hyperref[Sol:3.3]{chapter~5}.
\end{exer}

\begin{example}{Example: Reducing the $gg\to gg$ $s_{23}$-channel cut to scalar integrals}

At the end of section~\label{sec:3.2} (eqs.~\eqref{eq:tchannelcut1}--\eqref{eq:tchannelcut3})
we reached an expression for the $s_{23}$-channel double cut of the four-gluon MHV
amplitudes written in terms of cut Feynman integrals. The box contribution was
already in a reduced form, while the triangle and bubble sub-topologies had
non-trivial dependence in the numerator. Since we will use many different
generalised cuts, we use the notation from section~\ref{sec:3.3} for the
quadruple cut $\mathcal{C}_{1|2|3|4}$, also for the double cuts
$\mathcal{C}_{I|J}$, triple cuts $\mathcal{C}_{I|J|K}$, and so on. The
$s_{12}$-channel cut of a four-particle process is therefore represented as
$\mathcal{C}_{12|34}$, and $\mathcal{C}_{23|41}$ represents the $s_{23}$-channel cut.
Written explicitly in terms of the integral functions $F_n^{[D]}$ the result is
\begin{align} \label{eq:4gcut23recall}
  \mathcal{C}_{23|41}&\left( A^{(1)}(1^-,2^-,3^+,4^+) \right)
    = \mathcal{C}^{\rm box}_{23|41} + \mathcal{C}'_{23|41} \,,
\end{align}
where\footnote{Note that we have changed the double cut to apply to the amplitude rather than the integrand in this section, which affects the factors of $\i$.}
\begin{align}
 & \mathcal{C}^{\rm box}_{23|41} = - A^{(0)}(1^-,2^-,3^+,4^+) \, s_{12} s_{23} \, \mathcal{C}_{23|41}\left( F_4^{[D]}(p_2,p_3,p_4) \right) \,, \\
 \label{eq:4gcut23recallCprime}
 &  \mathcal{C}'_{23|41} = - 2 A^{(0)}(1^-,2^-,3^+,4^+) \, \frac{1}{s_{12}^2s_{23}^2} \times \nonumber\\
  & \quad \mathcal{C}_{23|41} \left(
    F_2^{[D]}(p_{23})[\mathcal{N}_2 ]
  + \frac{1}{s_{12}} F_3^{[D]}(p_{23},p_4)[\mathcal{N}_{3,a}]
  + \frac{1}{s_{12}} F_3^{[D]}(p_2,p_3)[\mathcal{N}_{3,b}]
  \right)\,. 
\end{align}
Here the non-trivial numerators are given by
\begin{align}
  \mathcal{N}_2 &= \trmgamtwo1{\slashed{l}_2}32^2 + s_{12}s_{23} \, \trmgamtwo1{\slashed{l}_2}32 + 2 s_{12}^2s_{23}^2 \,, \\
  \mathcal{N}_{3,a} &= \trmgamtwo1{\slashed{l}_1}42 \left(\trmgamtwo1{\slashed{l}_2}32^2 + s_{12}s_{23} \, \trmgamtwo1{\slashed{l}_2}32 + 2 s_{12}^2s_{23}^2 \right) \,, \\
  \mathcal{N}_{3,b} &= \trmgamtwo2{\slashed{l}_1}31 \left(\trmgamtwo1{\slashed{l}_2}32^2 + s_{12}s_{23} \, \trmgamtwo1{\slashed{l}_2}32 + 2 s_{12}^2s_{23}^2 \right) \,,
\end{align}
where $l_1=k$ and $l_2=k-p_{23}$. We can also represent this equation
graphically, which helps to keep track of the integral topologies. We draw the
$s_{23}$-channel cut of the box integral as
\begin{equation}
  \mathcal{C}_{23|41}\left( \raisebox{-0.9cm}{\includegraphics[width=1.8cm]{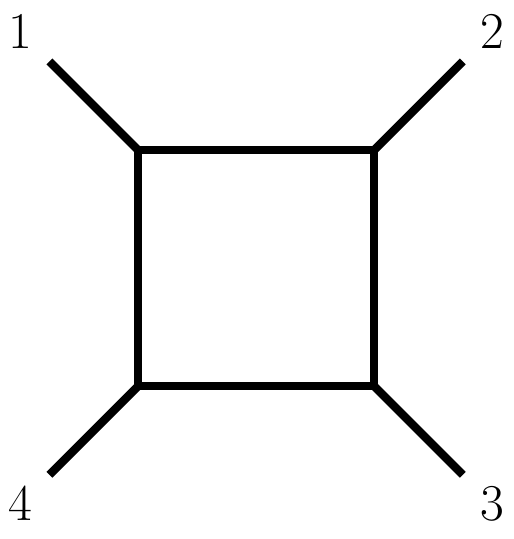}} \right)
  \ = \
  \raisebox{-0.9cm}{\includegraphics[width=1.8cm]{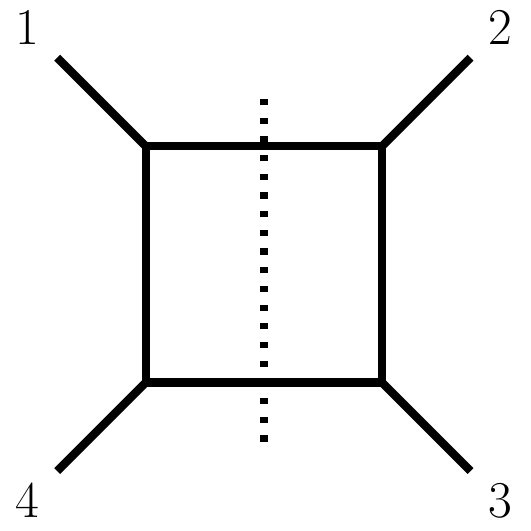}} \,,
\end{equation}
and so the $s_{23}$-channel cut can be represented as
\begin{align}
    &\mathcal{C}_{23|41} \left(  \raisebox{-0.9cm}{\includegraphics[width=1.8cm]{ch3_loopamps/figures/4g1Lampblob_MHV.pdf}} \right)
    \ = \
    - \raisebox{-0.9cm}{\includegraphics[width=1.8cm]{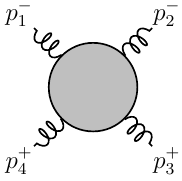}} \left[\rule{0cm}{1cm}\right. 
    s_{12} s_{23} \raisebox{-0.9cm}{\includegraphics[width=1.8cm]{ch3_loopamps/figures/boxint-tcut.pdf}} \nonumber\\&
    + \frac{2}{s_{12}^2s_{23}^2} \left(
    \raisebox{-0.4cm}{\includegraphics[width=1.8cm]{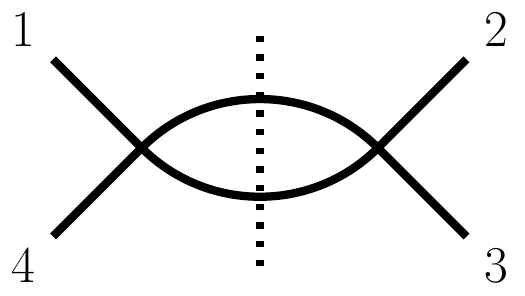}}[\mathcal{N}_2]
    + \raisebox{-0.8cm}{\includegraphics[width=1.8cm]{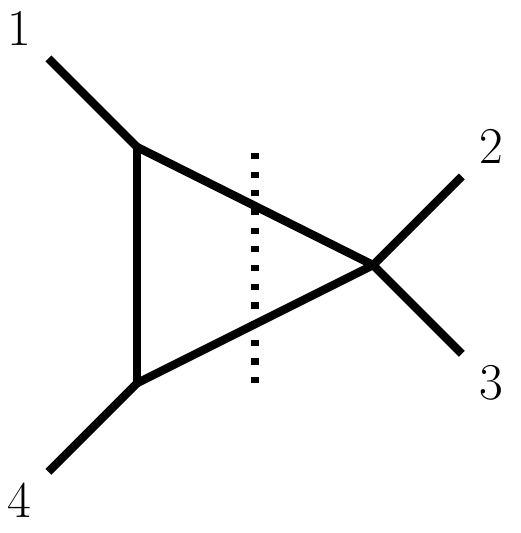}}[\mathcal{N}_{3,a}]
    + \raisebox{-0.8cm}{\includegraphics[width=1.8cm]{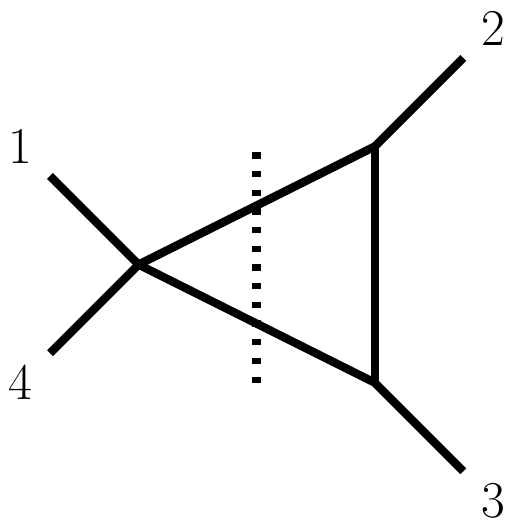}}[\mathcal{N}_{3,b}]
    \right)
    \left.\rule{0cm}{1cm}\right] \,.
\end{align}

As we can see, there is still a bit of work to do to simplify this expression. Let us start with the bubble configuration, which has rank-one and and rank-two tensor numerators,
\begin{align} \begin{aligned}
  F_2^{[D]}&(p_{23})[\mathcal{N}_2] = \\
   & \trmgamtwo1{\gamma_{\mu_1}}32 \, \trmgamtwo1{\gamma_{\mu_2}}32 \, F_2^{[D]}(p_{23})\bigl[(k-p_{23})^{\mu_1}(k-p_{23})^{\mu_2}\bigr] \\
  & + s_{12} s_{23} \, \trmgamtwo1{\gamma_{\mu_1}}32 \, F_2^{[D]}(p_{23})\bigl[(k-p_{23})^{\mu_1}\bigr] + 2 \, s_{12}^2s_{23}^2 \, F_2^{[D]}(p_{23})[1] \,.
  \end{aligned}
  \label{eq:c23bubred}
\end{align}
We have already reduced the rank-one integral so we may substitute the form-factor decomposition~\eqref{eq:PVbubR1sol},
\begin{align} \begin{aligned}
  s_{12} s_{23} \, \trmgamtwo{1}{\gamma^{\mu}}{3}{2}
  F_2^{[D]}(p_{23})\bigl[(k-p_{23})^{\mu}\bigr] & = s_{12} s_{23} \left( -\frac{1}{2}\trmgamtwo1{\slashed{p}_{23}}32 \right) F^{[D]}_2(p_{23})[1] \\
  & = -\frac{1}{2} s_{12}^2 s_{23}^2 \, F^{[D]}_2(p_{23})[1] \,.
  \end{aligned}
\end{align}
The rank-two integral will involve several steps of algebra but follows exactly
the same strategy. From \eqn{eq:PVbubR2dec} with the form factors in ~\eqn{eq:rank2bubsol} we may substitute into the
rank-two integral above, obtaining
\begin{align}
  &\trmgamtwo1{\gamma_{\mu_1}}32 \, \trmgamtwo1{\gamma_{\mu_2}}32 \, F_2^{[D]}(p_{23})\bigl[(k-p_{23})^{\mu_1}(k-p_{23})^{\mu_2} \bigr] \nonumber\\
  & \qquad \qquad = a_{2,00} \, \trmgamtwo1{\gamma_{\mu_1}}32 \, \trmgamtwo1{\gamma^{\mu_1}}32 + a_{2,11} \, \trmgamtwo1{\slashed{p}_{23}}32{}^2 \nonumber\\
  & \qquad \qquad = a_{2,11} \, s_{12}^2 s_{23}^2 \nonumber\\
  & \qquad \qquad = \frac{D}{4 (D-1)} F^{[D]}_2(p_{23})[1] \,.
\end{align}
The triangle tensor integrals look troubling at first sight, since we must use reduction for up to rank-three integrals,
\begin{align}
  F^{[D]}_{3}(p_1,p_2)&[k^{\mu_1}] =  a_{3,1} \, p_1^{\mu_1} + a_{3,2} \, p_2^{\mu_1} \,, \label{eq:PVtriR1dec}\\
  F^{[D]}_{3}(p_1,p_2)&[k^{\mu_1}k^{\mu_2}] = \nonumber\\
  & a_{3,00} \, \eta^{\mu_1\mu_2} + a_{2,11} \, p_1^{\mu_1} p_1^{\mu_2}
  + a_{3,12} \left( p_1^{\mu_1} p_2^{\mu_2} + p_1^{\mu_1} p_2^{\mu_2} \right)
  + a_{3,22} \, p_2^{\mu_1} p_2^{\mu_2} \,, \label{eq:PVtriR2dec}\\
  F^{[D]}_{3}(p_1,p_2)&[k^{\mu_1}k^{\mu_2} k^{\mu_3}] = \nonumber\\&
    a_{3,001} \left( \eta^{\mu_1\mu_2} p_1^{\mu_3} + \eta^{\mu_2\mu_3} p_1^{\mu_1} + \eta^{\mu_3\mu_1} p_1^{\mu_2}\right) \nonumber\\&
  + a_{3,002} \left( \eta^{\mu_1\mu_2} p_2^{\mu_3} + \eta^{\mu_2\mu_3} p_2^{\mu_1} + \eta^{\mu_3\mu_1} p_2^{\mu_2}\right) \nonumber\\&
  + a_{3,111} \, p_1^{\mu_1} p_1^{\mu_2} p_1^{\mu_3}
  + a_{3,222} \, p_2^{\mu_1} p_2^{\mu_2} p_2^{\mu_3} \nonumber\\&
  + a_{3,112} \left(p_1^{\mu_1} p_1^{\mu_2} p_2^{\mu_3} + p_1^{\mu_2} p_1^{\mu_3} p_2^{\mu_1} + p_1^{\mu_3} p_1^{\mu_1} p_2^{\mu_2} \right) \nonumber\\&
  + a_{3,122} \left(p_1^{\mu_1} p_2^{\mu_2} p_2^{\mu_3} + p_1^{\mu_2} p_2^{\mu_3} p_2^{\mu_1} + p_1^{\mu_3} p_1^{\mu_2} p_2^{\mu_2} \right) \,.
\label{eq:PVtriR3dec}
\end{align}
Let us start with the easiest rank-one part of the integral $F_3^{[D]}(p_{23},p_{234})$,
\begin{align}
  F_3^{[D]}(p_{23},p_{234})&\bigl[ 2 \, \trmgamtwo1{\slashed{l}_1}42 \, s_{12}^2s_{23}^2 \bigr] \nonumber\\
  &= 2  \, s_{12}^2s_{23}^2 \left( a_{3,1} \, \trmgamtwo1{\slashed{p}_{23}}42 + a_{3,2} \, \trmgamtwo1{\slashed{p}_{234}}42\right) \nonumber\\
 &= 0 \,.
\end{align}
In fact, if we follow through with all the form-factor substitutions we will
see that all tensor triangles reduce to zero. So with a little bit of extra
work with tensor reduction we have found a compact final answer:
\begin{align}
  C'_{23|41} = \ & - 2 \, A^{(0)}(1^-,2^-,3^+,4^+) \frac{1}{s_{12}^2s_{23}^2} \nonumber\\
  & \ \times \mathcal{C}_{23|41}\left(F_2^{[D]}(p_{23})\bigl[\trm\1{l_2}32^2 + s_{12}s_{23} \, \trm1{l_2}32 + 2 \, s_{12}^2s_{23}^2 \bigr]\right)
   \nonumber\\
  = \ & -2 \, A^{(0)}(1^-,2^-,3^+,4^+) \left( \frac{D}{4 (D-1)} -\frac{1}{2} + 2 \right) \mathcal{C}_{23|41}\left(F_2^{[D]}(p_{23})[1]\right) \nonumber\\
  = \ & - A^{(0)}(1^-,2^-,3^+,4^+) \frac{7D-6}{2(D-1)} \, \mathcal{C}_{23|41}\left(F_2^{[D]}(p_{23})[1]\right) \,.
\end{align}
\end{example}

\subsection{Transverse spaces and transverse integration \label{ssec:transverse}}

The development of integrand-reduction techniques led to many efficient
methods for the processing of tensor integrals. Here we would like to
highlight the advantages of decomposing the loop momenta into transverse
components. This feature has been exploited in a number of situations
including---but not exclusively---the loop-momentum basis of Van
Neerven-Vermaseren~\cite{ch3_vanNeerven:1983vr}, Baikov integral
representations~\cite{ch3_Baikov:1996rk},
and the adaptive integrand-reduction method~\cite{ch3_Mastrolia:2016dhn}.

We start by considering an $n$-point, $L$-loop Feynman integral in $D=4-2\eps$
dimensions which depend on $n-1$ independent momenta,
\begin{equation}
  F_n^{(L),[4-2\eps]}(p_1,\cdots,p_{n-1}) = \int_{k_1}\cdots\int_{k_L}  \,f(\{k\}, \{p\}) \,.
\end{equation}
We continue to assume the external momenta $\{p\}$ are in four
dimensions. This implies that the independent external momenta span a space of dimension $m=n-1$ if $n\le 4$, or $m=4$ if $n > 4$. For a one-loop
box integral $m=3$, for a pentagon $m=4$, and a hexagon would also have $m=4$.
So we may write the loop momenta as a decomposition of an $m$-dimensional and
$4-m-2\eps$ dimensional space,
\begin{equation}
  k_i^\mu = k_i^{\mu,[m]} + k_i^{\mu,[4-2\eps-m]} \,,
\end{equation}
which are orthogonal, 
\begin{equation}
  k_i^{\mu,[m]}\cdot k_j^{\mu,[4-m-2\eps]} = 0 \,.
\end{equation}
If $m<4$ (i.e., if $n \le 4$), we may even consider three transverse spaces of dimensions $m$, $4-m$ and $-2\eps$,
\begin{equation}
  k_i^\mu = k_i^{\mu,[m]} + k_i^{\mu,[4-m]} + k_i^{\mu,[-2\eps]} \,,
\end{equation}
all orthogonal to each other.
Since the various indices become cumbersome at this point it is convenient to introduce some notation:
\begin{align*}
  &k_i^{\mu,[m]} = k^\mu_{i,\parallel} \text{ spans the physical space of the loop integral},\\
  &k_i^{\mu,[4-m]} = k^\mu_{i,\perp} \text{ spans the spurious space of the loop integral},\\
  &k_i^{\mu,[-2\eps]} \text{ spans the extra-dimensional space of the loop integral},\\
  &k_i^{\mu,[4-m-2\eps]} \text{ spans both spurious and extra-dimensional space of the loop integral}.
\end{align*}
We use the term \textit{physical} to indicate the space after integration, i.e.\
the space of independent external momenta that we used for tensor reduction in section~\ref{sec:3.4}.
The term \textit{spurious} space is used to indicate terms that are non-zero at
the level of the integrand but which will vanish after integration.
For each of these spaces a spanning set of momenta can be found. The construction
of such a basis is often referred to as the Van Neerven-Vermaseren
basis~\cite{ch3_vanNeerven:1983vr}. For the physical space $k^\mu_{i,\parallel}$ the
\textit{independent} external momenta can be used. For the spurious space $k^\mu_{i,\perp}$ we may find an orthogonal basis of
four-dimensional vectors. Traditionally these are denoted $\omega_i^\mu$ and depend on the vectors $p_i^\mu$ spanning the physical space.
In the Van Neerven-Vermaseren construction they are orthonormal, i.e.\ $\omega_i\cdot \omega_j = N_i \delta_{ij}$ with $N_i=1$.
In a given implementation it can be beneficial to let $N_i \neq 1$ in order to
avoid square roots in the kinematic invariants.

Let us give an explicit example to make things clearer. Consider a one-loop
triangle configuration with massless external momenta $p_1, p_2$ and $p_3$.
This means the spurious space has dimension two and the physical space may be
spanned by $\{p_1, p_2\}$. General four-dimensional vectors may be written as
\begin{equation}
  \omega_i^\mu(p_1,p_2) = \alpha_{1i} \, p_1^\mu + \alpha_{2i} \, p_2^\mu + \alpha_{3i} \, \frac{1}{2} \spAB{1}{\gamma^\mu 2 X}{1} + \alpha_{4i} \, \frac{1}{2} \spAB{2}{\gamma^\mu 1 X}{2} \,,
  \label{eq:spuriousvectorexampeEQ1}
\end{equation}
where $i=1,2$, and we have introduced an arbitrary reference momentum $X$ to ensure the coefficients $\alpha$ are free of any spinor phase.

The spurious vectors satisfy the conditions $\omega_i\cdot p_j = 0$ and $\omega_i\cdot \omega_j = \delta_{ij}$ if
\begin{align}
  \alpha_{1i} = \alpha_{2i} = 0 \,, \qquad
  \alpha_{31}\alpha_{42} + \alpha_{41}\alpha_{32} = 0 \,, \qquad
  - \frac{1}{2} s_{12} {\rm tr}(\slashed1 \slashed{X} \slashed2 \slashed{X}) \alpha_{3i} \alpha_{4i} = 1 \,,
  \label{eq:spuriousvectorexampeEQ2}
\end{align}
for $i=1,2$, and so we find the explicit representation
\begin{align} \label{eq:triangle_omegas}
\begin{aligned}
  \omega_1^\mu(p_1,p_2) &= \frac{\sqrt{2}}{\sqrt{s_{12}{\rm tr}_(\slashed1 \slashed{X} \slashed2 \slashed{X})}}\bigl(\spAB{1}{\gamma^\mu 2 X}{1}  - \spAB{2}{\gamma^\mu 1 X }{2}\bigr) \,, \\
  \omega_2^\mu(p_1,p_2) &= \frac{\i\sqrt{2}}{\sqrt{s_{12}{\rm tr}(\slashed1 \slashed{X} \slashed2 \slashed{X})}}\bigl(\spAB{1}{\gamma^\mu 2 X}{1}  + \spAB{2}{\gamma^\mu 1 X}{2}\bigr) \,.
\end{aligned} 
\end{align}
In the context of a full amplitude computation, the momentum $X$ could be one
of the other independent external momenta. It is also clear in this expression that
the complicated prefactor results from the assertion that the vectors should be
orthonormal, and can be avoided by releasing the condition without affecting any
final amplitude level results. Numerators for amplitudes in the transverse
space (which are Lorentz scalars) may now be expressed in terms of scalar
products $k_1\cdot \omega_i$,
\begin{equation}
  k^\mu_{1,\perp} =  (k_1\cdot \omega_1) \, \omega_1^\mu + (k_1\cdot \omega_2) \, \omega_2^\mu \,.
\end{equation}

We could try to span the extra-dimensional space with explicit vectors, though
we would be forced to introduce a fixed embedding dimension larger than four to
do it. As a result we would lose all the convenient four-dimensional spinor-helicity methods
that have been working well so far. Instead, we simply identify the independent
scalar products appearing, which depend only on the number of loops. At one-loop
there would only be a single extra-dimensional scalar product,
\begin{equation}
  k_1^{[-2\eps]}\cdot k_1^{[-2\eps]} \eqqcolon -\mu_{11}\,.
  \label{eq:extradimSP1}
\end{equation}
The definition of the extra-dimensional scalar product $\mu_{11}$
includes a sign so this appears as an effective mass term in the
propagator,
\begin{equation}
  k_1^2 = k_{1,\parallel}^2 + k_{1,\perp}^2 - \mu_{11}\,.
\end{equation}
At higher loops more extra-dimensional scales will be introduced, which may be
labelled $\mu_{ij} \coloneqq - k_i^{[-2\eps]} \cdot k_j^{[-2\eps]}$. At
two-loops there would be three such scales and $L(L+1)/2$ for the general $L$-loop case.

We finish this section by demonstrating a useful application of the
decomposition into transverse spaces: \emph{transverse integration}. If the numerator
for a given tensor integral lives in a transverse space, we may provide a
general tensor decomposition using only the metric tensor in the transverse
space. For example, returning to the one-loop triangle, consider:
\begin{align}
  \int_k \frac{k\cdot\omega_i}{k^2 (k-p_1)^2 (k-p_{12})^2}  
  &= \omega_{i,\mu} \int_k \frac{k^\mu}{k^2 (k-p_1)^2 (k-p_{12})^2} \nonumber \\
    &= \omega_{i,\mu}  \left(a_{3,1} \, p_1^\mu + a_{3,2} \, p_2^{\mu}\right) \nonumber \\
    & = 0\,,
\end{align}
where we have used the Passarino-Veltman reduction from the previous section.
Following the same logic it should be clear that for any odd power, $r$, of the
numerator $(k\cdot\omega_i)^r$, the integral will vanish. Even powers do not
vanish, but we may expand the tensor integral using the metric tensor
$\eta^{\mu\nu, [m]}$ for $m=2$, which satisfies
\begin{equation}
  \eta_{\ \ \mu}^{\mu, [m]} = m\,.
\end{equation}
The tensor decomposition can be written in terms of a form factor
$\tilde{a}_{3,00}^{[2]}$ which multiplies the metric tensor of the two-dimensional
spurious space. For instance, for $r=2$, we have that
\begin{align}
  \int_k \frac{(k\cdot \omega_i)^2}{k^2 (k-p_1)^2 (k-p_{12})^2} 
  &= \omega_{i,\mu_1} \, \omega_{i,\mu_2} \int_k \frac{k_\perp^{\mu_1} k_\perp^{\mu_2}}{k^2 (k-p_1)^2 (k-p_{12})^2} \nonumber\\
  &= \omega_{i,\mu_1} \, \omega_{i,\mu_2} \, \tilde{a}_{3,00}^{[2]} \, \eta^{\mu_1\mu_2, [2]} \nonumber\\
  &= \omega_i^2 \, \tilde{a}_{3,00}^{[2]} \,.
  \label{eq:triwsqtransverseint}
\end{align}
Projecting using the tensor $\eta_{\mu_1\mu_2}^{[2]}$ leads to
\begin{align}
  \tilde{a}_{3,00}^{[2]} \, \eta^{\mu_1\mu_2, [2]} \, \eta_{\mu_1\mu_2}^{[2]} &= 2 \, \tilde{a}_{3,00}^{[2]} \\
   &= \int_k \frac{k_\perp^2}{k^2 (k-p_1)^2 (k-p_{12})^2} \\
   &= \int_k \frac{k^2-k_{\parallel}^2+\mu_{11}}{k^2 (k-p_1)^2 (k-p_{12})^2}\,.
\end{align}
The term $k_{\parallel}^2$ may be written in terms of the propagators by decomposing $k^{\mu}_{\parallel}$ as
\begin{align}
k_{\parallel}^\mu = \frac{k_{\parallel}\cdot p_2}{p_1\cdot p_2} \, p_1^\mu + \frac{k_{\parallel}\cdot p_1}{p_1\cdot p_2} \, p_2^\mu \,.
\end{align}
This in fact implies that 
\begin{align}
   k_{\parallel}^2 = \frac{2 \, (k_{\parallel}\cdot p_1) (k_{\parallel}\cdot p_2)}{p_1\cdot p_2}\,,
\end{align}
which can be expressed in terms of the propagators using
\begin{align}
  2 \, k_{\parallel}\cdot p_1 &= k^2 - (k-p_1)^2 \,, \\
  2 \, k_{\parallel}\cdot p_2 &= (k-p_1)^2 - (k-p_{12})^2 + s_{12} \,.
\end{align}
In this way we write the original rank-two tensor integrals in terms of the
usual Feynman integrals but including a triangle with $\mu_{11}$ in the
numerator. We will see later that there are methods to integrate it directly.
We may also observe a very similar but alternative derivation of the same integral:
\begin{align}
  \int_k \frac{(k\cdot\omega_i)^2}{k^2 (k-p_1)^2 (k-p_{12})^2} 
  &= \omega_{i,\mu_1} \omega_{i,\mu_2} \int_k \frac{k_\perp^{\mu_1} k_\perp^{\mu_2}}{k^2 (k-p_1)^2 (k-p_{12})^2} \\
  &= \omega_{i,\mu_1} \omega_{i,\mu_2} \int_k \frac{k^{\mu_1, [2-2\eps]} k^{\mu_2, [2-2\eps]}}{k^2 (k-p_1)^2 (k-p_{12})^2} \\
  &= \omega_{i,\mu_1} \omega_{i,\mu_2} \, \tilde{a}^{[2-2\eps]}_{3,00} \, \eta^{\mu_1\mu_2, [2-2\eps]}\,.
\end{align}
This time the form factor $\tilde{a}^{[2-2\eps]}_{3,00}$ multiplies the metric
tensor of the $(2-2\eps)$-dimensional spurious and extra-dimensional spaces. Contracting this expression with the same metric tensor gives
\begin{align}
\begin{aligned}
   \tilde{a}^{[2-2\eps]}_{3,00} \, \eta^{\mu_1\mu_2, [2-2\eps]} \, \eta_{\mu_1\mu_2}^{[2-2\eps]} &= 2(1-\eps) \, \tilde{a}^{[2-2\eps]}_{3,00} \\
   &= \int_k \frac{k^2-k_{\parallel}^2}{k^2 (k-p_1)^2 (k-p_{12})^2}\,,
 \end{aligned}
\end{align}
which does not include the $\mu_{11}$ integral. We recall that $k_{\parallel}^2$ can be written in terms of propagators as discussed above. Putting everything together we obtain
\begin{align}
\begin{aligned}
\int_k \frac{(k\cdot\omega_i)^2}{k^2 (k-p_1)^2 (k-p_{12})^2}
&= \omega_{i}^2 \, \frac{1}{2} \int_k \frac{k^2-k_{\parallel}^2+\mu_{11}}{k^2 (k-p_1)^2 (k-p_{12})^2} \\
&= \omega_{i}^2 \, \frac{1}{2(1-\eps)} \int_k \frac{k^2-k_{\parallel}^2}{k^2 (k-p_1)^2 (k-p_{12})^2}\,.
\end{aligned}
\end{align}
Comparing both results we find that
\begin{equation}
  \int_k \frac{\mu_{11}}{k^2 (k-p_1)^2 (k-p_{12})^2} = -\frac{\eps}{1-\eps} \int_k \frac{k^2-k_{\parallel}^2}{k^2 (k-p_1)^2 (k-p_{12})^2}\,.
  \label{eq:mu11triformula}
\end{equation}
This exercise demonstrates a number of ways to move around the space of loop
integrals and integrands, and gives us enough technology to describe a complete
procedure for the computation of one-loop amplitudes.

\begin{exer}{Spurious loop-momentum space for the box integral}
\label{Ex:3.4}
Consider a one-loop box configuration with massless external momenta $p_1$, $p_2$, $p_3$, and $p_4$ ($p_i^2 = 0$, $p_1+p_2+p_3+p_4=0$). 

\begin{enumerate}[a)]
\item Determine the dimension of the physical and of the spurious space. Construct a basis of the latter using the spinors associated with the external momenta.
\smallskip
\item Show that the vector spanning the spurious space is proportional to
\begin{align} \label{eq:omega4pt}
\omega^{\mu}(p_1,p_2,p_3) = \epsilon^{\mu \nu \rho \sigma} p_{1\mu} p_{2\rho} p_{3\sigma} \,. 
\end{align}
\end{enumerate}
\noindent
Note that $\omega^{\mu}$ in \eqn{eq:omega4pt} spans the one-loop box spurious space also for massive or off-shell external momenta ($p_i^2 \neq 0$). In other words, $\omega(p_1,p_2,p_3) \cdot p_i = 0$ for any $i=1,\ldots,4$.
For the solution see \hyperref[Sol:3.4]{chapter~5}.
\end{exer}

\section{General integral and integrand bases for one-loop amplitudes \label{sec:3.5}}

\subsection{The one-loop integral basis}

At the beginning of this chapter we already introduced the idea that any
loop amplitude may be decomposed into an analytic part (a Feynman integral) and
an algebraic (or rational) coefficient. Using our notation for one-loop amplitudes (while continuing to suppress couplings, $\mu_{\rm R}$ and $4\pi$ prefactors) this reads
\begin{align}
  A_n^{(1),[D]}(1,\ldots,n) =
  \sum_{m,N} \i \, c_{m,N}^{[D]} (1,\cdots,n) \, F_{m}^{[D]}(p_1,\cdots,p_{n-1})[N] \,,
  \label{eq:loopamp-general-repeat}
\end{align}
where we must finalise the sum of propagator multiplicity $m$ and numerators
$N$ which form an integral basis.  In the examples of tensor integral reduction
we have seen how the large numbers of tensor integrals which appear in the
amplitude can be reduced onto a small number of integrals resulting in large
simplifications. If we collect the results from sections~\ref{sec:3.2} and~\ref{sec:3.4} we can summarise the information we have gathered about the four-gluon MHV amplitude in $D=4-2\eps$ dimensions as
\begin{align}
  A^{(1),[4-2\eps]}(1^-,&2^-,3^+,4^+) = A^{(0)}(1^-,2^-,3^+,4^+) \nonumber\\&
     \times \bigg(-s_{12}s_{23} F^{[4-2\eps]}_4(p_{1},p_{2},p_{3})[1] - \frac{11}{3} F^{[4-2\eps]}_2(p_{23})[1] \bigg)
     \nonumber\\&
     + \text{terms missed in 4D} \,.
   \label{eq:4g1loop}
\end{align} 
Here we have used the notation from section~\ref{sec:3.4}, where the integrals
are labelled by the external momenta flowing in the propagators and the
numerator dependence, in both cases here just simple scalar integrals.

At one loop the tensor-reduction method is sufficient to completely classify the
basis of all one-loop integrals for an arbitrary number of external momenta
with arbitrary kinematics. In order for it to be clear where we are going as we
derive this basis, let us begin by quoting the final result.

\begin{important}{A general formula for one-loop amplitudes}
A one-loop amplitude in dimensional regularisation with four-dimensional external states can be written as
\begin{align}
  A^{(1),[4-2\eps]}_n&(1,\ldots,n) = \nonumber\\&
    \sum_{1\leq i_1<i_2<i_3<i_4\leq n} \i \, c_{0;i_1|i_2|i_3|i_4} \, F^{[4-2\eps]}_4(p_{i_1,i_2-1},p_{i_2,i_3-1},p_{i_3,i_4-1})[1]\nonumber\\&
  + \sum_{1\leq i_1<i_2<i_3\leq n} \i \, c_{0;i_1|i_2|i_3} \, F^{[4-2\eps]}_3(p_{i_1,i_2-1},p_{i_2,i_3-1})[1]\nonumber\\&
  + \sum_{1\leq i_1<i_2\leq n} \i \, c_{0;i_1|i_2} \, F^{[4-2\eps]}_2(p_{i_1,i_2-1})[1]\nonumber\\&
  + \sum_{1\leq i_1\leq n} \i \, c_{0;i_1} F^{[4-2\eps]}_1(p_{i_1})[1]\nonumber\\&
  + R(1,\ldots,n) + \mathcal{O}(\eps) \,,
  \label{eq:oneloop4dbasis}  
\end{align}
where $R$ is a rational function of the external kinematics. There is quite a
lot going on here so let us unpack it. Firstly we see that result is quoted
in $4-2\eps$ dimensions and given only up to terms of $\mathcal{O}(\eps^0)$. No
integral functions with more than four propagators are required and only scalar
numerators appear. The rational coefficients, $c$, do not depend on $\eps$ and
may be obtained using four-dimensional unitarity cuts. The only term missed by
the four-dimensional cuts is a rational term. We will need to work a little
harder before we can justify that the remaining contribution is simply a
rational function, and for now we leave it as statement without proof. Let us also
clarify that the term \textit{rational function of the external kinematics} is
used to indicate a rational function of spinor products and so is a tree-like
function. We note that, for massless theories, bubbles on external legs and tadpoles
vanish in dimensional regularisation, so the basis simplifies.
\end{important}

The fact that no scalar integrals with more than four propagators are required
in $D=4-2\eps$ dimensions can be seen by considering the linear dependence of
the internal and external momenta.  The argument relies on our assumption that
the external momenta live in four dimensions. As a result, in a $D=4$
dimensional loop-momentum space, there can only ever be four independent
propagators and therefore the pentagon integral can be written entirely in
terms of box integrals. Following the exercise below we can see that the linear
dependence of the momenta can be related to the vanishing of the associated
Gram matrix. If the internal momentum is in $D=4-2\eps$ dimensions there is one
additional degree of freedom that means the pentagon integral is also
independent, but all integrals with a higher number of propagators will be
completely reducible. Using the basis choice in \eqn{eq:oneloop4dbasis} the
contribution of the pentagon-type integral has been moved to the terms of
$\mathcal{O}(\eps)$, in order to make the property that the pentagon vanishes
in four dimensions manifest. We will see later exactly how this can be
achieved. This fact was also demonstrated implicitly in exercise~\ref{Ex:3.2}, where the four-dimensional quadruple cuts of the five-gluon amplitude identified only the box
coefficients and did not detect any pentagon integral function.

The origin of rational terms, $R$, comes from terms in the integral
coefficients of higher order in $\eps$ multiplied by potential divergences in the
integrals resulting in terms like $\tfrac{\eps}{\eps}$ in the expansion.
Through an integrand-level analysis using the transverse decomposition one can
find an explicit representation of these terms, as we will show later.

\begin{exer}{Reducibility of the pentagon in four dimensions}
\label{Ex:3.5}
\vspace{-0.6cm}
\begin{enumerate}[a)]

\item Prove that the massless scalar triangle integral $F_3^{[D]}$ defined by \eqn{eq:oneloopintegraldef} with $n=3$, $m_a=0$, and $N=1$ is reducible in $D=2$ dimensions. For simplicity assume that $p_2^2=0=p_3^2$. Hint: introduce a two-dimensional parametrisation of the loop momentum, and use it to derive a relation among the inverse propagators.

\smallskip
\item The \emph{Gram matrix} $G$ of a set of momenta $q_1,\ldots,q_n$ is the matrix of entries 
\begin{align} \label{eq:gram_def}
\left[G\left(q_1,q_2,\ldots, q_n\right)\right]_{ij} = q_i \cdot q_j \,,
\end{align}
for $i,j=1,\ldots,n$. If the momenta are linearly dependent, their Gram matrix
has vanishing determinant. Prove that the relation among the inverse
propagators found at the previous step is equivalent to the vanishing of a Gram
determinant.

\smallskip
\item Use a Gram-determinant condition to prove that the massless pentagon integral $F_5^{[D]}$ defined by \eqn{eq:oneloopintegraldef} with $n=5$, $m_a=0$, and $N=1$ is reducible in $D=4$ dimensions. Parametrise the kinematics in terms of independent invariants assuming that $p_i^2=0$ for all $i=1,\ldots,5$.
\end{enumerate}
For the solution see \hyperref[Sol:3.5]{chapter~5} and the \texttt{Mathematica} notebook \break \href{https://scattering-amplitudes.mpp.mpg.de/scattering-amplitudes-in-qft/Exercises/}{\texttt{Ex3.5\_Reducibility.wl}}~\cite{website-ch3}.
\end{exer}

\subsection{A one-loop integrand basis in four dimensions \label{ssec:4dintegrandbasis}}

The integrand-reduction method is often referred to simply as the \emph{OPP method},
following the initials of the authors who introduced the method: Ossola,
Papadopoulos and Pittau~\cite{ch3_Ossola:2006us}. We have already made the first steps necessary to
follow this method in section~\ref{ssec:transverse}, where we
discussed how to provide general parametrisations for the loop momenta at the
integrand level using transverse spaces. The easiest place to start is with
terms with the maximal number of propagators, often referred to as the maximal
cut. At one loop this means the box configurations.

Throughout our general discussion on the one-loop integrands we do not have a
particular multiplicity of external legs in mind. As a result we will use the
notation $1|2|3|4$ for a general box configuration in which the momenta at the
four vertices $p_1, p_2, p_3, p_4$ are considered to be combinations of external legs. A similar notation also
applies for triangle and bubble configurations.

\subsubsection{The box integrand in four dimensions \label{sssec:4dbox}}

\begin{figure}[t]
  \centering
  \includegraphics[width=0.5\textwidth]{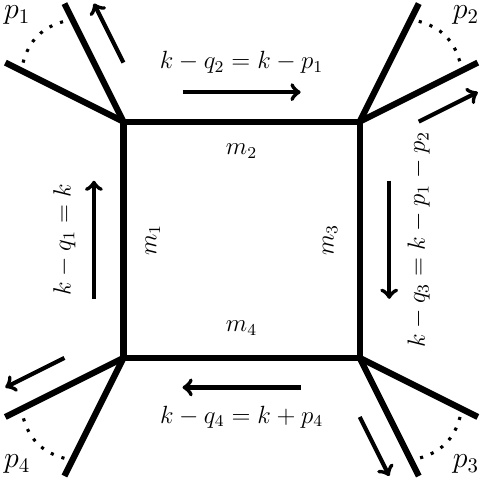}
  \caption{A general box configuration labelled according to the conventions of section~\ref{sssec:4dbox}. The arrows denote the direction of the momenta.}
  \label{fig:generalbox}
\end{figure}

The box configuration has three independent external momenta, and so the loop
momentum has a spurious space of dimension one. Let us consider a general
configuration $1|2|3|4$ with four masses and arbitrary momenta entering each vertex. The
inverse propagators are labelled as $D_i = (k-q_i)^2 - m_i^2$. In terms of the external momenta $p_i$ they are given by

\begin{align} 
\begin{alignedat}{2}
  & D_1 = k^2-m_1^2\,, && D_3 = (k-p_{12})^2 - m_3^2 \,, \\
  & D_2 = (k-p_1)^2 - m_2^2\,, \qquad \qquad && D_4 = (k+p_4)^2 - m_4^2 \,.
\end{alignedat}
\label{eq:boxprops}
\end{align}
This configuration is shown graphically in figure~\ref{fig:generalbox}.
Discarding the extra-dimensional terms in the loop momenta, and denoting the spanning vectors for the physical
space as $\vec{v}^\mu = \{p_1^\mu,p_2^\mu,p_3^\mu\}$, we may write
\begin{align}
  k^\mu &= k^\mu_{\parallel} + (k\cdot\omega) \, \omega^\mu,
  \label{eq:momparam} \\
  k^\mu_{\parallel} &= \vec{\alpha} \cdot \vec{v}^\mu,
  \label{eq:kphysbasis}
\end{align}
where $\vec{\alpha} = \{\alpha_1, \alpha_2, \alpha_3\}$ and $\omega \cdot p_i =
0$.\footnote{We use the inner product symbol $\cdot$ for all spaces, so that  $p\cdot q = p^\mu q_\mu$ and $\vec{\alpha}\cdot \vec{v}^\mu = \alpha_i v_i^\mu$ with summation over repeated indices implicit.} As we showed in the triangle example in section~\ref{ssec:transverse}, the
coefficients $\vec{\alpha}$ may be written in terms of external invariants
$q_i\cdot q_j$ and Lorentz-scalar products $k\cdot q_i$, where as before $q_i =
\sum_{l=1}^{i-1} p_l$. The latter can be written as
the difference of two inverse propagators, e.g.\ $k\cdot p_1 = (D_1 - D_2  +
p_1^2 + m_1^2 - m_2^2)/2$. This tells us that the loop-momentum dependence of
$\vec{\alpha}$ can be written completely in terms of the four inverse
propagators.

Our aim is to define an irreducible numerator, $\Delta_{1|2|3|4}$, that
parametrises all possible loop-momentum dependence in the numerator of the box
topology. In other words, we are interested in taking the quadruple cut of the
numerator function. Working at the integrand level, we re-use the notation
$\mathcal{C}_{1|2|3|4}$ used for the amplitudes in section~\ref{sec:3.3} to
indicate the inverse propagators are being set to zero:
$\mathcal{C}_{1|2|3|4}(f) \coloneqq f |_{D_i\to 0}$ for all $i=1,\ldots,4$. As
we have just ascertained that the loop-momentum dependence of $\vec{\alpha}$ is
entirely in terms of inverse propagators, the relevant part of the
loop-momentum parametrisation in \eqn{eq:momparam} depends on only one
loop-momentum dependent scalar product, $k\cdot\omega$, which we refer to as
the \emph{irreducible scalar product} (ISP). All terms in $\vec{\alpha}$
proportional to inverse propagators fall into sub-topologies which will be
dealt with later. For a renormalisable gauge theory\footnote{For other
theories, effective theories or gravity the only change would be to increase
the upper limit on the sum.} the dependence of the numerator must then have the
form
\begin{equation}
  \Delta_{1|2|3|4}(k\cdot\omega) = \sum_{i=0}^4 c_{i;1|2|3|4} \, (k\cdot\omega)^i \,.
  \label{eq:box4dintegrandparam}
\end{equation}
There is however one more constraint on the loop-momentum dependence of the numerator that comes from the
condition $D_1= 0$:
\begin{equation}
  \mathcal{C}_{1|2|3|4} \left( k^2 -m_1^2 \right) = \mathcal{C}_{1|2|3|4} \left( k_{\parallel}^2  \right) + (k\cdot\omega)^2 \, \omega^2 - m_1^2  = 0 \,.
  \label{eq:boxksqcondition}
\end{equation}
Since $\mathcal{C}_{1|2|3|4} \bigl( k_{\parallel}^2 \bigr)$ is a function of the external invariants
only---the inverse propagators are set to zero---it is a constant as far as the loop-momentum dependence goes. 
The condition~\eqref{eq:boxksqcondition} therefore states that monomials with more than one power of $k\cdot\omega$ are \textit{reducible}.
We may thus write the numerator of this box configuration as
\begin{equation}
  \Delta_{1|2|3|4}(k\cdot\omega) = c_{0;1|2|3|4} + c_{1;1|2|3|4} \, (k\cdot\omega) \,,
  \label{eq:box4dintegrandparam}
\end{equation}
where the rational coefficients $c_0$ and $c_1$ can be determined for an
arbitrary process from the cuts of Feynman diagrams or, as we have described
earlier in this chapter, from the product of tree-level amplitudes. To put
this more concretely, we can use the explicit loop-momentum solutions to \eqn{eq:boxksqcondition},
\begin{equation}
  k^{\pm,\mu} = \mathcal{C}_{1|2|3|4}\left(k_{\parallel}^\mu\right) \pm \sqrt{\frac{m_1^2-\mathcal{C}_{1|2|3|4}\bigl(k_{\parallel}^2\bigr)}{\omega^2}} \, \omega^\mu \,,
\end{equation}
to write down an expression for the quadruple cut of an arbitrary one-loop amplitude:
\begin{align}
  \mathcal{C}_{1|2|3|4} \left(A_n^{(1),[4-2\eps]}\right) 
  & = \int_k \Delta_{1|2|3|4} (k \cdot \omega) \, \left(\prod_{i=1}^4 (-2\pi\i) \, \delta^{(+)}(D_i) \right)   \nonumber\\
  & = \int_k \biggl(I^{(1)} \prod_{i=1}^4 D_i \biggr) \,\left(\prod_{i=1}^4 (-2\pi\i) \, \delta^{(+)}(D_i) \right) \,,
  \label{eq:quadcutrelation1}
\end{align}
where $I^{(1)}$ represents the integrand of the one-loop amplitude as
introduced previously. Identifying the integrands subject to the delta function
constraints leads us to the algebraic relation
\begin{align}
  \Delta_{1|2|3|4} (k \cdot \omega) \bigg|_{D_i = 0} = \biggl(I^{(1)} \prod_{i=1}^4 D_i \biggr) \bigg|_{D_i = 0}.
  \label{eq:quadcutrelation2}
\end{align}
If we had explicitly performed the integration over the delta functions in
\eqn{eq:quadcutrelation1} we would also obtain a Jacobian factor but it would
cancel on both sides of the relation.  The integrand $I^{(1)}$ could be
obtained by simply taking the subset of Feynman diagrams which have the same
four propagators and substituting the on-shell values of the loop momenta
$k^\pm$. In a physical gauge,\footnote{In exercise~\ref{Ex:1.7} we showed in
the light-like axial gauge that the numerator of the propagator can be written
as the product of polarisation vectors in the on-shell limit times a factor of
$\i$. Incidentally, this is the origin of the factors of $\i$ which accompany
the tree-level amplitudes in the factorisation on the cuts, see e.g.\
\eqn{eq:quadcuteq}.} it is easy to see that this subset
of diagrams may also be written as the product of tree-level amplitudes summed
over the internal helicity configurations:
\begin{align}
  \Biggr( I^{(1)} \prod_{i=1}^4 D_i \Biggl) \Biggl|_{k^\pm} & = \nonumber\\
  & \hspace{-2.15cm}  \sum_{h_i=\pm} \Bigg[ \i \, A^{(0)}\Bigl((-k^\pm)^{-h_1},p_1,(k^\pm-p_1)^{h_2} \Bigr) \,
  \i \, A^{(0)}\Bigl((-k^\pm+p_1)^{-h_2},p_2,(k^\pm-p_{12})^{h_3}\Bigr) \, \nonumber\\
  & \hspace{-2.15cm} \phantom{ \sum_{h_i=\pm} \Bigg[} \, \i \, A^{(0)}\Bigl((-k^\pm+p_{12})^{-h_3},p_3,(k^\pm+p_4)^{h_4}\Bigr) \,
  \i \, A^{(0)}\Bigl((-k^\pm-p_4)^{-h_4},p_4,(k^\pm)^{h_1}\Bigr) \Bigg] \nonumber\\
  & = \Delta_{1|2|3|4}(k^\pm\cdot\omega) \,.
  \label{eq:quadcuteq}
\end{align}
Using the on-shell values for the ISP,
\begin{equation}
  k^\pm\cdot \omega = \pm \sqrt{\frac{m_1^2-\mathcal{C}_{1|2|3|4}\bigl(k_{\parallel}^2\bigr)}{\omega^2}} \,,
  \label{eq:onshellISPsol}
\end{equation}
we can invert \eqn{eq:quadcuteq} to find solutions for the rational coefficients $c_{i;1|2|3|4}$ of the
parametrisation of $\Delta_{1|2|3|4}$ in \eqn{eq:box4dintegrandparam}. We obtain
\begin{align}
  c_{0;1|2|3|4} &= \frac{1}{2}\left( \overline{I}^{(1)}(k^+) +  \overline{I}^{(1)}(k^-) \right), \\
  c_{1;1|2|3|4} &= \frac{1}{2}\sqrt{\frac{\omega^2}{m_1^2-\mathcal{C}_{1|2|3|4}\bigl( k_{\parallel}^2 \bigr)}} \left( \overline{I}^{(1)}(k^+) -  \overline{I}^{(1)}(k^-) \right) \,,
\end{align}
where we introduced the short-hand notation
\begin{align}
\overline{I}^{(1)}(k^\pm) \coloneqq \Biggr( I^{(1)} \prod_{i=1}^4 D_i \Biggl) \Biggl|_{k^\pm} \,.
\end{align}
This proves the averaging prescription that we already applied during generalised unitarity cut computations in section~\ref{sec:3.3}. 
Having determined both coefficients from the on-shell cut, we see that the contribution to the amplitude is simply
\begin{equation}
  \int_k \frac{\Delta_{1|2|3|4}(k)}{(-D_1) (-D_2) (-D_3) (-D_4)}  = c_{0;1|2|3|4} \, F^{[4-2\eps]}_4(p_1,p_2,p_3)[1] \,,
  \label{eq:Delta4integrated}
\end{equation}
as the second, spurious, element integrates to zero following the arguments
presented earlier in section~\ref{ssec:transverse} regarding transverse integration. Notice that the minus signs on the inverse propagators have been put in to match the conventions for the Feynamn integrals in \eqn{eq:oneloopintegraldef}.

We have now demonstrated two important facts: a general basis for the
(four-dimensional) box part of any one-loop amplitude is simply the scalar box
integral, and its coefficient may be extracted by a purely algebraic
procedure using generalised unitarity cuts.

\subsubsection{The triangle integrand in four dimensions}

The procedure for determining the remaining parts of the amplitude and
establishing a complete integral basis is to reduce the number of cut
propagators to the triangle (then bubble, then tadpole) contributions.

We begin in exactly the same fashion as the box configuration by defining the inverse propagators according to
\begin{align}  \label{eq:triprops}
D_1 = k^2-m_1^2 \,, \qquad
D_2 = (k-p_1)^2 - m_2^2 \,, \qquad
D_3 = (k-p_{12})^2 - m_3^2 \,,
\end{align}
and parametrise the loop momenta using two orthogonal spurious vectors $\omega_i$,
\begin{align} 
\begin{aligned}
  k^\mu &= k_{\parallel}^\mu + (k\cdot\omega_1) \, \omega_1^{\mu} + (k\cdot\omega_2) \, \omega_2^{\mu} \,, \\
  k_{\parallel}^\mu &= \vec{\alpha}\cdot \vec{v}^\mu \,,
\end{aligned}
  \label{eq:momparamtri}
\end{align}
where $\vec{v}^{\mu} = \{p_1^{\mu},p_2^{\mu}\}$ with $\omega_i^{\mu} p_{j\, \mu} = 0$ for $i,j=1,2$, and $\vec{\alpha} =
\{\alpha_1, \alpha_2\}$. We have seen the explicit construction of the physical
and spurious spaces in section~\ref{ssec:transverse} so it is clear that
$\vec{\alpha}$ only depends on the inverse propagators, and hence on the triple cut
have no loop-momentum dependence. We therefore have two ISPs with which to
parametrise our triangle integrand: $k\cdot\omega_1$ and $k\cdot\omega_2$. For a
renormalisable gauge theory the maximum tensor rank is three and so a general
parametrisation is
\begin{equation}
  \Delta_{1|2|3}(k\cdot\omega_1, k\cdot\omega_2) = \sum_{i,j} c_{ij}  \,(k\cdot\omega_1)^i \, (k\cdot\omega_2)^j \,,
  \label{eq:triparam1}
\end{equation}
with $i+j \leq3$. This parametrisation is subject to the constraint
\begin{equation}
  \mathcal{C}_{1|2|3} \left( k^2 - m_1^2 \right) = \mathcal{C}_{1|2|3} \left( k_{\parallel}^2 \right) - m_1^2 + (k\cdot\omega_1)^2 \omega_1^2 + (k\cdot\omega_2)^2 \omega_2^2 = 0 \,.
  \label{eq:triksqcondition}
\end{equation}
It is slightly more difficult to apply this constraint to find a general
parametrisation since we now have multivariate polynomials. One could attempt
to deploy the mathematical technology to deal with such problems, introducing a
polynomial ordering and performing polynomial division with respect to a
Gr\"{o}bner basis,\footnote{Many computer algebra systems now come equipped with
decent implementations of polynomial division algorithms, so this can be a
practical method.} but it is easier to analyse this case by hand and
we will find an extremely convenient choice. Firstly we expand
\eqn{eq:triparam1} explicitly,
\begin{align}
  \Delta_{1|2|3}&(k\cdot\omega_1, k\cdot\omega_2) = c_{00}
  + c_{10} (k\cdot\omega_1) 
  + c_{01} (k\cdot\omega_2) \nonumber\\& 
  + c_{20} (k\cdot\omega_1)^2
  + c_{11} (k\cdot\omega_1) (k\cdot\omega_2) 
  + c_{02} (k\cdot\omega_2)^2 \nonumber\\& 
  + c_{30} (k\cdot\omega_1)^3
  + c_{21} (k\cdot\omega_1)^2 (k\cdot\omega_2)
  + c_{12} (k\cdot\omega_1) (k\cdot\omega_2)^2
  + c_{03} (k\cdot\omega_2)^3 \,.
  \label{eq:triparam2}
\end{align}
Removing all monomials with even powers of $k\cdot\omega_2$ would be a valid choice
and eliminate the implicit dependence on three of the monomials. The resulting
integrand parametrisation would still contain an integral quadratic in
$k\cdot\omega_1$, which we have seen does not vanish after integration. In the example of transverse integration from \eqn{eq:triwsqtransverseint} we saw that the integral of
$(k\cdot\omega_1)^2$ is the same as $(k\cdot\omega_2)^2$ up to a normalisation.
Therefore, in order to obtain a simple result after (transverse) integration, we choose
\begin{align} 
\begin{aligned}
  \Delta_{1|2|3}&(k\cdot\omega_1, k\cdot\omega_2) = c_{0;1|2|3}
  + c_{1;1|2|3} (k\cdot\omega_1) 
  + c_{2;1|2|3} (k\cdot\omega_2) \\& 
  + c_{3;1|2|3} \left((k\cdot\omega_1)^2 - \frac{\omega_1^2}{\omega_2^2} (k\cdot\omega_2)^2 \right)
  + c_{4;1|2|3} (k\cdot\omega_1) (k\cdot\omega_2) \\&
  + c_{5;1|2|3} (k\cdot\omega_1)^3
  + c_{6;1|2|3} (k\cdot\omega_1)^2 (k\cdot\omega_2) \,,
\end{aligned}
\label{eq:triparam}
\end{align}
where we constructed a spurious integral from the linear combination of $(k\cdot\omega_1)^2$ and $(k\cdot\omega_2)^2$.
As a result, after integration, all terms except the scalar integral coefficient vanish:
\begin{equation}
  \int_k \frac{\Delta_{1|2|3}(k\cdot\omega_1, k\cdot\omega_2)}{(-D_1) (-D_2) (-D_3)}  = c_{0;1|2|3} \, F^{[4-2\eps]}_3(p_1,p_2)[1] \,.
  \label{eq:Delta3integrated}
\end{equation}
The extraction of the rational coefficients can be performed algebraically by
using the information computed using the quadruple cuts as input. Following the
box example as a reference we write
\begin{align}
  \mathcal{C}_{1|2|3} & \left( A^{(1),[4-2\eps]}_n \right) \nonumber\\&
  = \int_k \left( -\Delta_{1|2|3}(k\cdot\omega_1, k\cdot\omega_2)  + \sum_X \frac{\Delta_{1|2|3|X}(k\cdot\omega)}{D_X} \right) \left(\prod_{i=1}^3 (-2\pi\i)\, \delta^{(+)}(D_i)\right) \nonumber\\&
  = \int_k I^{(1)}(k) \left(\prod_{i=1}^3 D_i \, (-2\pi\i)\, \delta^{(+)}(D_i) \right) \,,
\end{align}
which implies that
\begin{align}
  -\Delta_{1|2|3}(k\cdot\omega_1, k\cdot\omega_2) \bigg|_{D_i = 0} = \left( I^{(1)}(k) \prod_{i=1}^3 D_i - \sum_X \frac{\Delta_{1|2|3|X}(k\cdot\omega)}{D_X} \right) \bigg|_{D_i = 0} \,.
  \label{eq:triplecuteq}
\end{align}
The sum over $X$ for the box numerators indicates that we must include all
boxes that share the three propagators $\{D_1,D_2,D_3\}$, with $D_X$ being the
fourth propagator which completes the box.  Denoting the box configurations in
the subtraction as $1|2|3|X$ is a slight abuse of notation, and we give an
explicit example below to clarify.

The integrand factorises into tree amplitudes as before,
\begin{align} \begin{aligned}
  \left( I^{(1)}(k) \prod_{i=1}^3 D_i \right) \bigg|_{D_i = 0} = \sum_{h_i=\pm} \bigg[ &
   \i \, A^{(0)}\Bigl((-k)^{-h_1},p_1,(k-p_1)^{h_2} \Bigr) \\
   &  \i \, A^{(0)}\Bigl((-k+p_1)^{-h_2},p_2,(k-p_{12})^{h_3}\Bigr) \\
   &  \i \, A^{(0)}\Bigl((-k+p_{12})^{-h_3},p_3,(k)^{h_1}\Bigr) \bigg] \bigg|_{D_i = 0} \,,
\end{aligned} \end{align}
but on this occasion solving the on-shell constraints will lead to a family of
loop momenta parametrised by a single variable. The choice of this
parametrisation is not an immediate issue since it is more important to realise
that the triple cut condition in \eqn{eq:triplecuteq} must be valid
for any value of the free variable, which is sufficient to find the conditions
necessary to fix the rational coefficients $c_{i;1|2|3}$.

\begin{example}{Example: Subtraction terms for a six-point amplitude}
  Since the sum over $X$ and the cut notation $1|2|3|X$ in \eqn{eq:triplecuteq}
  are schematic, it is helpful to see an explicit example. Consider the
  application to an amplitude with six external legs where we are computing the
  triple cut $12|34|56$. The sum over $X$ for the box subtraction terms would
  then indicate the following set:
  \begin{equation}
    \bigl\{ 1|2|34|56 \,, 12|3|4|56 \,, 12|34|5|6 \bigr\}\,.
  \end{equation}
\end{example}

\subsubsection{The bubble integrand in four dimensions}

At this point the integrand reduction strategy of Ossola, Papadopoulos and
Pittau should be clear: continue to reduce the on-shell constraints on the
propagators until all the integral basis coefficients have been determined.

The equation for the irreducible numerator in the case of bubble
configurations, $\Delta_{1|2}$, is constructed from three irreducible scalar
products: $k\cdot\omega_1$, $k\cdot\omega_2$ and $k\cdot\omega_3$. Note that
$\omega_1$ and $\omega_2$ are not the same spurious vectors for the triangle
configuration. We will consider a configuration with inverse propagators
\begin{align}
  D_1 = k^2-m_1^2 \,, \qquad \qquad D_2 = (k-p_1)^2 - m_2^2 \,,
  \label{eq:bubbleprops}
\end{align}
and parametrise the loop momenta using the three orthogonal spurious vectors $\omega_i$ as
\begin{align}   \label{eq:momparambubble}
  k^\mu & = k^{\mu}_{\parallel} + (k\cdot\omega_1) \, \omega_1^\mu + (k\cdot\omega_2) \, \omega_2^\mu + (k\cdot\omega_3) \, \omega_3^\mu \,, \\
  k^{\mu}_{\parallel} & = \vec{\alpha} \cdot \vec{v}^\mu \,,
\end{align}
where $\vec{v}^\mu = \{p_1^\mu\}$ and $\vec{\alpha}=\{\alpha_1\}$. The remainder of the derivation we leave as an exercise, although perhaps the result is already clear by analogy to the triple cut case.

\begin{exer}{Parametrising the bubble integrand}
\label{Ex:3.6}
\vspace{-0.6cm}
\begin{enumerate}[a)]
\item Show that the irreducible numerator of a general bubble configuration can be written as
\begin{align} \label{eq:bubble_num_param}
\begin{aligned}
  \Delta_{1|2}&(k\cdot\omega_1, k\cdot\omega_2, k\cdot\omega_3) = c_{0;1|2} \\
 & + c_{1;1|2} (k\cdot\omega_1) 
  + c_{2;1|2} (k\cdot\omega_2) 
  + c_{3;1|2} (k\cdot\omega_3) \\ 
&  + c_{4;1|2} (k\cdot\omega_1) (k\cdot\omega_2) 
  + c_{5;1|2} (k\cdot\omega_1) (k\cdot\omega_3) 
  + c_{6;1|2} (k\cdot\omega_2) (k\cdot\omega_3) \\  
 & + c_{7;1|2} \left((k\cdot\omega_1)^2 - \frac{\omega_1^2}{\omega_3^2} (k\cdot\omega_3)^2 \right)
  + c_{8;1|2} \left((k\cdot\omega_2)^2 - \frac{\omega_2^2}{\omega_3^2} (k\cdot\omega_3)^2 \right) \,,
\end{aligned}
\end{align}
which results in
\begin{align}
  \int_k \frac{\Delta_{1|2}(k\cdot\omega_1, k\cdot\omega_2, k\cdot\omega_3)}{(-D_1) (-D_2)}  = c_{0;1|2} \, F^{[4-2\eps]}_2(p_1)[1] \,.
\end{align}
\item Show that the irreducible numerator can be determined on the double cut from
\begin{align}
\begin{aligned}
  \Delta_{1|2}&(k\cdot\omega_1, k\cdot\omega_2, k\cdot\omega_3) \bigg|_{D_i = 0} =
     \bigg( I^{(1)}(k) \prod_{i=1}^2 D_i \\&
     + \sum_X \frac{\Delta_{1|2|X}(k\cdot\omega_1^{X}, k\cdot\omega_2^{X})}{D_X} - \sum_{Y,Z} \frac{\Delta_{1|2|Y|Z}(k\cdot\omega^{YZ})}{D_Y D_Z} \bigg) \bigg|_{D_i = 0} \,,
\end{aligned}
  \label{eq:doublecuteq}
\end{align}
where the sum on $X$ indicates all triangle configurations, and the sum on $Y,Z$ indicates all box configurations. Again the sign on the triangle irreducible numerator $\Delta_{1|2|X}$ matches our sign conventions on the Feynman integrals.
\end{enumerate}
For the solution see \hyperref[Sol:3.6]{chapter~5}.
\end{exer}

For massless theories such as Yang-Mills theory we have now completed an
algebraic approach for the determination of the four-dimensional, or cut
constructible, part of the one-loop integrand. In massive theories there are
still contributions for integrals which only depend on the mass: tadpoles and
bubbles on external massive legs. One can continue to apply the integrand
reduction procedure to these cases and a full analysis can, for example, be
found in reference~\cite{ch3_Ellis:2011cr}. There is a subtlety in the method
though since the unrenormalised amplitude diverges when applying bubble cuts on
external lines and so we cannot factorise into the product of tree amplitudes
without additional regulators. The issue is connected with wave-function
renormalisation and the interested reader can find further information in the
literature~\cite{ch3_Ellis:2008ir,ch3_Britto:2011cr,ch3_Badger:2017gta}.

\subsection{$D$-dimensional integrands and rational terms}

The results of section~\ref{ssec:4dintegrandbasis} gave us confidence in the
proposed general one-loop formula in \eqn{eq:oneloop4dbasis}. The fact
that we found only scalar integrals after removing the spurious terms was
consistent with the tensor integral reduction method introduced in section~\ref{sec:3.4}, so we can consider this result as a confirmation of the latter. The
integrand-level matching of cut diagrams to irreducible numerators enabled the
\textit{algebraic} determination of the integral coefficients, leaving the
remaining integration of the basis integrals as a separate problem. That
problem is not to be underestimated of course and is the subject of the next
chapter.

The aim for the rest of this section is to extend our analysis to $D=4-2\eps$
dimensional integrands and amplitudes. This means that we can no longer avoid
the contribution of the pentagon which now becomes the starting point for the
top down integrand reduction approach of Ossola, Papdopoulos and Pittau. 

\subsubsection{The pentagon integrand}

Since there are four independent external momenta in the pentagon configuration
there is no spurious space. In exercise~\ref{Ex:3.5} we showed that the pentagon
configuration was completely reducible in four dimensions. In $D=4-2\eps$, we
need to clarify this point and to extend our four dimensional analysis of the
integrand. The decomposition of the transverse space gives us the starting
point. For concreteness we can specify the propagators,
\begin{align}
\begin{alignedat}{3}
& D_1 = k^2-m_1^2\,, && D_2 = (k-p_1)^2 - m_2^2\,, \qquad && D_3 = (k-p_{12})^2 - m_3^2 \,, \\
& D_4 = (k-p_{123})^2 - m_4^2\,, \qquad && D_5 = (k+p_5)^2 - m_5^2\,, && \\
\end{alignedat}
\label{eq:boxprops}
\end{align}
the spanning set of external momenta (which are four dimensional),
\begin{align}
\vec{v}^{\mu} = \{ p_1^{\mu}, p_2^{\mu}, p_3^{\mu}, p_4^{\mu} \} \,,
\end{align}
and the parametrisation of the loop momentum,
\begin{equation}
  k^\mu = \vec{\alpha}\cdot \vec{v}^\mu + k^{\mu, [-2\eps]} \,.
\end{equation}
Note that in this case there is no spurious space, as the external momenta $\vec{v}^{\mu}$ are sufficient to span the entire four-dimensional space.
As before, the coefficients $\vec{\alpha}$ of the spanning vectors of the physical
space are functions of the propagators and the external invariants. The
on-shell condition $D_1=0$ gives the key constraint for the determination of
the integrand basis,
\begin{equation}
  k^2 - m_1^2 = \vec{\alpha}\cdot G \cdot \vec{\alpha}^{\top} - \mu_{11} -m_1^2 = 0 \,,
  \label{eq:pentmucondition}
\end{equation}
where we have explicitly used the Gram matrix $G_{ij} = \vec{v}_i^{\mu}
\vec{v}_{j\,\mu}$. After applying the four on-shell conditions
$\{D_2,D_3,D_4,D_5\}=0$, the only ISP in this equation is $\mu_{11}$, and we
see that on the quintuple cut this ISP will be a constant expression written in
terms of external invariants. We may therefore parametrise the irreducible
numerator as
\begin{equation}
  \Delta_{1|2|3|4|5}(\mu_{11}) = \sum_{i=0}^2 c_{i;1|2|3|4|5} \, \mu_{11}^i \,,
\end{equation}
subject to the constraint in \eqn{eq:pentmucondition}. A minimal solution
to this would seem to take just the scalar pentagon as a basis integrand,
however this would not be consistent with the complete reduction of the
pentagon in four dimensions. The next-to-minimal choice is a single power of $\mu_{11}$,
\begin{equation}
  \Delta_{1|2|3|4|5}(\mu_{11}) = c_{1;1|2|3|4|5} \, \mu_{11} \,,
\end{equation}
which vanishes explicitly in the $D\to 4$ limit as we had previously argued. We
will not explicitly perform the integration but simply state that it is
possible to show the integral vanishes up to $\mathcal{O}(\eps)$: 
\begin{equation}
  \int_k \frac{\Delta_{1|2|3|4|5}(\mu_{11})}{D_1 D_2 D_3 D_4 D_5}  = \mathcal{O}(\eps) \,. 
\end{equation}

\subsubsection{Extending the box, triangle and bubble integrand basis to $D=4-2\eps$ dimensions}

The integrand reduction procedure is now rather easy to extend into
$D$-dimensions, since all we have to do is track the additional dependence on
the extra-dimensional ISP~$\mu_{11}$. The box irreducible numerator then becomes a polynomial in two ISPs,
\begin{equation}
  \Delta_{1|2|3|4}(k\cdot\omega, \mu_{11}) = \sum_{i,j} c_i \, (k\cdot\omega)^i \mu_{11}^j \,,
\end{equation}
where $i+2j<4$, and
\begin{equation}
  \mathcal{C}_{1|2|3|4} \left( k^2 - m_1^2 \right) = \mathcal{C}_{1|2|3|4} \left( k_{\parallel}^2  \right) - m_1^2 + (k\cdot\omega)^2 \omega^2 - \mu_{11} = 0\,.
  \label{eq:boxksqconditionD}
\end{equation}
The simplest solution is to eliminate $\mu_{11}$ from the parametrisation
completely, leaving five monomials in the ISP $k\cdot\omega$. Two of these
monomials would not vanish after integration however and again, as in the
pentagon case, the $D\to 4$ limit would not match our four-dimensional
analysis. Instead we choose the two terms from the four-dimensional
parametrisation and three more monomials proportional to $\mu_{11}$, 
\begin{align}
  \Delta_{1|2|3|4}(k\cdot\omega, \mu_{11}) & =  c_{0;1|2|3|4}
  + c_{1;1|2|3|4} \, (k\cdot\omega)
  + c_{2;1|2|3|4} \, \mu_{11} \nonumber\\
  & \phantom{= {}} + c_{3;1|2|3|4} \, (k\cdot\omega) \, \mu_{11}
  + c_{4;1|2|3|4} \, \mu_{11}^2 \,.
  \label{eq:boxparamD}
\end{align}
The result integrating this expression turns out to be incredibly simple but,
as before with the pentagon, requires some additional integration technology.
In $4-2\eps$ dimensions it turns out that the integral with $\mu_{11}$ in the
denominator vanishes up to $\mathcal{O}(\eps)$ while the $\mu_{11}^2$ integral
gives rise to a finite, and rational, contribution,
\begin{align}
  \int_k &\frac{\Delta_{1|2|3|4}(k\cdot\omega, \mu_{11})}{(-D_1) (-D_2) (-D_3) (-D_4)} \nonumber \\
  & = c_{0;1|2|3|4} \, F^{[4-2\eps]}_4(p_1,p_2,p_3)[1]
    + c_{2;1|2|3|4} \, F^{[4-2\eps]}_4(p_1,p_2,p_3)[\mu_{11}] \nonumber\\
  & \phantom{= {}} + c_{4;1|2|3|4} \, F^{[4-2\eps]}_4(p_1,p_2,p_3)[\mu_{11}^2] \nonumber\\&
    = 
    c_{0;1|2|3|4} \, F^{[4-2\eps]}_4(p_1,p_2,p_3)[1] - \frac{1}{6} c_{4;1|2|3|4} + \mathcal{O}(\eps) \,.
\end{align}
The final steps are to repeat the analysis for the triangle and bubble, and so we can quote the results for the triangle irreducible numerator,
\begin{align}
  \Delta_{1|2|3}&(k\cdot\omega_1, k\cdot\omega_2, \mu_{11}) = c_{0;1|2|3}
  + c_{1;1|2|3} \, (k\cdot\omega_1) 
  + c_{2;1|2|3} \, (k\cdot\omega_2) \nonumber\\& 
  + c_{3;1|2|3} \, \left((k\cdot\omega_1)^2 - \frac{\omega_1^2}{\omega_2^2} (k\cdot\omega_2)^2 \right)
  + c_{4;1|2|3} \, (k\cdot\omega_1) (k\cdot\omega_2) \nonumber\\& 
  + c_{5;1|2|3} \, (k\cdot\omega_1)^3
  + c_{6;1|2|3} \, (k\cdot\omega_1)^2 (k\cdot\omega_2)\nonumber\\&
  + c_{7;1|2|3} \, \mu_{11} 
  + c_{8;1|2|3} \, (k\cdot\omega_1) \, \mu_{11} 
  + c_{9;1|2|3} \, (k\cdot\omega_2) \, \mu_{11} \,,
  \label{eq:triparamD}
\end{align}
and its integrated form,
\begin{align}
  \int_k &\frac{\Delta_{1|2|3}(k\cdot\omega_1, k\cdot\omega_2, \mu_{11})}{(-D_1) (-D_2) (-D_3)} \nonumber\\& =
      c_{0;1|2|3} \, F^{[4-2\eps]}_3(p_1,p_2)[1]
    + c_{7;1|2|3} \, F^{[4-2\eps]}_3(p_1,p_2)[\mu_{11}] \nonumber\\&
    = 
    c_{0;1|2|3} \, F^{[4-2\eps]}_3(p_1,p_2)[1] - \frac{1}{2} c_{7;1|2|3} + \mathcal{O}(\eps) \,,
\end{align}
and also the bubble irreducible numerator
\begin{align}
  \Delta_{1|2}&(k\cdot\omega_1, k\cdot\omega_2, k\cdot\omega_3, \mu_{11}) = c_{0;1|2}
  + c_{1;1|2} \, (k\cdot\omega_1) 
  + c_{2;1|2} \, (k\cdot\omega_2) \nonumber\\& 
  + c_{3;1|2} \, (k\cdot\omega_3) 
  + c_{4;1|2} \, \left((k\cdot\omega_1)^2 - \frac{\omega_1^2}{\omega_3^2} (k\cdot\omega_3)^2 \right)
  + c_{5;1|2} \, \left((k\cdot\omega_2)^2 - \frac{\omega_2^2}{\omega_3^2} (k\cdot\omega_3)^2 \right) \nonumber\\&
  + c_{6;1|2} \, (k\cdot\omega_1) (k\cdot\omega_2) 
  + c_{7;1|2} \, (k\cdot\omega_1) (k\cdot\omega_3) 
  + c_{8;1|2} \, (k\cdot\omega_2) (k\cdot\omega_3) 
  + c_{9;1|2} \, \mu_{11} \,,
  \label{eq:bubparamD}
\end{align}
and its integrated form,
\begin{align}
  \int_k &\frac{\Delta_{1|2}(k\cdot\omega_1, k\cdot\omega_2, k\cdot\omega_3, \mu_{11})}{(-D_1) (-D_2)} \nonumber\\
  & \quad =
      c_{0;1|2} \, F^{[4-2\eps]}_2(p_1)[1]
    + c_{9;1|2} \, F^{[4-2\eps]}_2(p_1)[\mu_{11}] \nonumber\\
    & \quad = 
    c_{0;1|2} \, F^{[4-2\eps]}_2(p_1)[1] - \frac{p_1^2-m_1^2-m_2^2}{6} \, c_{9;1|2} + \mathcal{O}(\eps) \,.
\end{align}
When integrating the irreducible numerators we have used the following results for integrals in $D=4-2\eps$ dimensions:
\begin{align}
  F^{[4-2\eps]}_5(p_1,p_2,p_3,p_4)[\mu_{11}] &= \mathcal{O}(\eps) \,, \\
  F^{[4-2\eps]}_4(p_1,p_2,p_3)[\mu_{11}] &= \mathcal{O}(\eps) \,, \\
  F^{[4-2\eps]}_4(p_1,p_2,p_3)[\mu_{11}^2] &= -\frac{1}{6} + \mathcal{O}(\eps) \,, \\
  F^{[4-2\eps]}_3(p_1,p_2)[\mu_{11}] &= -\frac{1}{2} + \mathcal{O}(\eps) \,, \\
  F^{[4-2\eps]}_2(p_1)[\mu_{11}] &= -\frac{p_1^2-m_1^2-m_2^2}{6} + \mathcal{O}(\eps) \,.
\end{align}
Explicitly proving these results requires technology for loop integration that
we have not yet introduced. One interesting observation~\cite{ch3_Bern:1995db} is that the
$\mu_{11}$ numerators give rise to dimension-shifted integrals:
\begin{equation}
  F^{[4-2\eps]}_n(p_1,\ldots,p_{n-1})[\mu_{11}^r] = \left(\prod_{s=0}^{r-1}(s-\eps) \right) \, F^{[4+2r-2\eps]}_n(p_1,\ldots,p_{n-1})[1] \,.
  \label{eq:dimshift}
\end{equation}
The fact that the relation is proportional to $\eps$ shows that we are only
interested in the poles of the dimension-shifted integrals. Shifting the
dimension up improves the infrared behaviour and so all the possible poles in
the dimension-shifted integrals are of UV origin. The vanishing of the pentagon
and box integrals with the $\mu_{11}$ numerator can then be understood since the
integrand-level power-counting argument shows that the $(6-2\eps)$-dimensional scalar
integrals are UV finite.

\begin{exer}{Dimension-shifting relation at one loop}
\label{Ex:3.DimShift}
Prove the dimension-shifting relation~\eqref{eq:dimshift}~\cite{ch3_Bern:1995db}. Assume that the external momenta $p_i$ are four dimensional, and decompose the loop momentum into a four- and an extra-dimensional parts (see section~\ref{ssec:transverse}). The key of the proof is that the integrand of the integral on the LHS of \eqn{eq:dimshift} depends on the loop momentum only through its four-dimensional part and $\mu_{11}$.
Switch to radial and angular coordinates in the extra-dimensional subspace, carry out the angular integration, and absorb the factor of $\mu_{11}^r$ in the numerator into the radial part of a $(2r-2\eps)$-dimensional loop-integration measure. Use the following Gamma-function identity to simplify the ratio of the angular integrals,
\begin{align} \label{eq:gamma_ratio}
\frac{\Gamma(r-\eps)}{\Gamma(-\eps)} = \prod_{s=0}^{r-1} (s-\eps) \,.
\end{align}
We will prove the above in the solution of exercise~\ref{Ex:ExpGamma}. 
Finally, putting together the $(2r-2\eps)$-dimensional and the four-dimensional loop-integration measures gives the RHS of \eqn{eq:dimshift}.
For the solution see \hyperref[Sol:3.DimShift]{chapter~5}.

\end{exer}

\subsection{Final expressions for one-loop amplitudes in $D$-dimensions}

We have now completed the analysis at one loop. We have used a general integrand
parametrisation to prove the basis used in \eqn{eq:oneloop4dbasis} but
have now identified the connection between the rational terms and the
extra-dimensional terms missed by the $4D$ cuts. We can therefore give an
explicit formula for the rational term $R$:
\begin{align}
  A^{(1),[4-2\eps]}_n&(1,\ldots,n) = \nonumber\\&
    \sum_{1\leq i_1<i_2<i_3<i_4\leq n} \i \, c_{0;i_1|i_2|i_3|i_4} \, F^{[4-2\eps]}_4(p_{i_1,i_2-1},p_{i_2,i_3-1},p_{i_3,i_4-1})[1]\nonumber\\&
  + \sum_{1\leq i_1<i_2<i_3\leq n} \i \, c_{0;i_1|i_2|i_3} \, F^{[4-2\eps]}_3(p_{i_1,i_2-1},p_{i_3,i_3-1})[1]\nonumber\\&
  + \sum_{1\leq i_1<i_2\leq n} \i \, c_{0;i_1|i_2} \, F^{[4-2\eps]}_2(p_{i_1,i_2-1})[1]\nonumber\\&
  + \sum_{1\leq i_1\leq n} \i \, c_{0;i_1} \, F^{[4-2\eps]}_1(p_{i_1})[1]\nonumber\\&
  + R(1,\ldots,n) + \mathcal{O}(\eps) \,,\\
  R(1,\ldots,&n) = 
  -\frac{1}{6} \sum_{1\leq i_1<i_2<i_3<i_4\leq n} \i \, c_{4;i_1|i_2|i_3|i_4} \nonumber\\&
  -\frac{1}{2} \sum_{1\leq i_1<i_2<i_3\leq n} \i \, c_{7;i_1|i_2|i_3} \nonumber\\&
  -\frac{1}{6} \sum_{1\leq i_1<i_2\leq n}  \left(p_{i_1,i_2-1}^2 - m_{i_1}^2 - m_{i_2}^2 \right) \, \i \, c_{9;i_1|i_2} \,.
\end{align}
We have also demonstrated that the coefficients of the integral basis can be
extracted from products of tree-level amplitudes via generalised unitarity
cuts using a completely algebraic method.

\begin{important}{Automated approaches to one-loop amplitude computations}
  The coefficients of the integral basis presented above may now be extracted
  by solving the quadruple, triple and bubble cut conditions in
  eqs.~\eqref{eq:quadcutrelation2}, \eqref{eq:triplecuteq} and
  \eqref{eq:doublecuteq} and/or their $D$-dimensional equivalents.
  The coefficients of the scalar loop integrals are completely determined from
  factorised products of on-shell tree amplitudes in four dimensions.
  These coefficients can be determined numerically by inverting the cut conditions,
  a technique that allows large intermediate expressions to be sidestepped.
  The method is relatively easy to automate for high-multiplicity processes and has been used for
  precise phenomenological studies at high energy colliders with around five
  final-state particles~\cite{ch3_Berger:2008sj,ch3_Giele:2008bc,ch3_Hirschi:2011pa,ch3_Bevilacqua:2011xh,ch3_Badger:2012pg,ch3_GoSam:2014iqq,ch3_Actis:2016mpe,ch3_Buccioni:2019sur}.
\end{important}

There are still a couple of loose ends however. Aside from the fact that we did
not prove the results of integration that led to the rational terms, we have
also not explicitly demonstrated how the $D$-dimensional basis coefficients can
be extracted from tree-level amplitudes. Whichever approach is taken, some
information from tree-amplitudes in dimensions other than four must be used.
Numerically, this information can be efficiently extracted using recursion
relations for fixed integer values of the ``spin dimension'' $d_{\rm s} = \eta^\mu_{\phantom{\mu} \mu}$ for the numerator algebra
~\cite{ch3_Giele:2008ve}. Alternatively one can use explicit spinor-helicity
constructions in higher dimensions~\cite{ch3_Bern:2010qa}. 
Both approaches are completely general but the dramatic simplicity of amplitudes in four dimensions
uncovered by the spinor-helicity formalism is lost. At one loop an alternative
approach is also available in which we may exploit the fact that the extra-dimensional
dependence of the loop integrand is equivalent to a shift in the mass of the
propagating particles. In section~\ref{sec:rat4g} we will describe the steps
required to directly compute the rational terms of the four-gluon amplitude
using tree amplitudes in four dimensions but with massive internal scalar
particles.

\subsection{The direct extraction method}

Let us return to the triple cut equation that we used to determine the triangle
integrand coefficients, \eqn{eq:triplecuteq}. We did not give an explicit
solution to the system of equations but instead remarked that it could be
sampled numerically and inverted to find the coefficients $c_{i;1|2|3}$. In
this section we consider an analytic solution which is able to extract only the
information that remains after integration. The application to the
four-dimensional cut-constructible terms was presented by
Forde~\cite{ch3_Forde:2007mi} and later extended to include the rational
terms~\cite{ch3_Badger:2008cm}.

Solving the on-shell conditions $D_i^2=0$ requires a solution to the ISP
constraint given in \eqn{eq:triksqcondition}. The aim here is to find a particular
parametrisation for the on-shell solution which allows us to extract unknown
coefficients in the irreducible numerator $\Delta_{1|2|3}$. Let us start by introducing a short hand for the on-shell value for
the square of the physical loop momentum,
\begin{equation}
  \mathcal{C}_{1|2|3}\bigl(k_{\parallel}^2\bigr) = \vec{\alpha} \cdot G \cdot \vec{\alpha}^{\top} \bigl|_{D_i=0} \,.
\end{equation}
The condition $\mathcal{C}_{1|2|3}(D_1) = 0$ leads to a family of solutions:
\begin{equation}
  0 = \mathcal{C}_{1|2|3}(D_1) = \mathcal{C}_{1|2|3}\bigl(k_{\parallel}^2 + (k\cdot\omega_1)^2 \omega_1^2 + (k\cdot\omega_2)^2 \omega_2^2 - m_1^2 \bigr) \,.
\end{equation}
This family of solutions can be parametrised by a single variable, $\theta$, as
\begin{align}
  \mathcal{C}_{1|2|3} (k\cdot\omega_1) = \sqrt{\gamma} \cos\theta \,, \qquad \qquad
  \mathcal{C}_{1|2|3} (k\cdot\omega_2) = \i \sqrt{\gamma} \sin\theta \,.
  \label{eq:kwparam}
\end{align}
Introducing a light-like complex vector $\chi^\mu$ to write $\omega_1 = \chi +
\chi^\dagger$ and $\omega_2 = \chi - \chi^\dagger$ ($\dagger$ indicates complex conjugation), we find 
\begin{align}
\gamma = \frac{m_1^2 - \mathcal{C}_{1|2|3}\bigl(k_{\parallel}^2 \bigr)}{2 \, \chi\cdot\chi^\dagger} \,.
\end{align}
Efficient numerical solutions for the coefficients of $\Delta_{1|2|3}$ can be obtained by using values of
$\theta$ distributed equally on a circle, which is related to the method of
discrete Fourier projections~\cite{ch3_Berger:2008sj,ch3_Mastrolia:2008jb}.

\begin{exer}{Projecting out the triangle coefficients}
\label{Ex:3.7}
Expand the sine and cosine in \eqn{eq:kwparam} into exponentials to write
\begin{equation} \label{eq:Delta123exp}
  \Delta_{1|2|3}(\theta) = \sum_{k=-3}^3 d_{k;1|2|3} \, \mathrm{e}^{\i k \theta} \,,
\end{equation}
where $d_{0;1|2|3} = c_{0;1|2|3}$. Using seven discrete values for $\theta$,
\begin{equation} \label{eq:thetak}
  \theta_k = \frac{2\pi k}{7}\,, \qquad k=-3,\ldots,3\,,
\end{equation}
show that
\begin{equation} \label{eq:d123fourier}
  d_{k;1|2|3} = \frac{1}{7} \sum_{l=-3}^{3} \mathrm{e}^{-\i k \theta_l} \, \Delta_{1|2|3}(\theta_l) \,.
\end{equation}
This discrete Fourier projection is easy to generalise to higher-rank numerators, try it for a maximum tensor rank of four.
For the solution see \hyperref[Sol:3.7]{chapter~5}.
\end{exer}

An analytic solution for $c_{0;1|2|3}$ is complicated by the appearance of the
box terms on the RHS of \eqn{eq:triplecuteq}. It would be
useful if the values of $\theta$ used for the extraction of the scalar triangle
coefficient from the product of trees made the box subtraction terms as simple
as possible. With this in mind, we choose to reparametrise our solution again in terms of a single, complex, parameter $t$,
\begin{align}
  t = \sqrt{\gamma} \, \mathrm{e}^{\i \theta} \,.
\end{align}
Using this parametrisation we can re-write the box subtraction term indicated in \eqn{eq:triplecuteq} with an additional propagator,
\begin{equation}
  D_X = (k-p_X)^2 - m_X^2 \,,
  \label{eq:boxsubprop}
\end{equation}
where have used a symbol $p_X$ to represent the momentum flowing in the propagator. The spurious direction for this box can also be neatly written using the vector $\chi$,
\begin{equation}
  \omega = \chi \, (\chi^\dagger\cdot p_X) - \chi^\dagger \, (\chi\cdot p_X) \,.
\end{equation}
The box subtraction term then can be written as
\begin{align}
\frac{\Delta_{1|2|3|X}(k\cdot\omega)}{D_X} \bigg|_{D_i=0} = \frac{\mathcal{C}_{1|2|3}\bigl(\Delta_{1|2|3|X}(k\cdot\omega)\bigr) }{ \mathcal{C}_{1|2|3}(D_X) } \,,
\end{align}
with
\begin{align}
\begin{aligned}
  & \mathcal{C}_{1|2|3}\bigl(\Delta_{1|2|3|X}(k\cdot\omega)\bigr) =
  c_{0;1|2|3|X} \\
  & \qquad \quad + c_{1;1|2|3|X}\bigl[ \mathcal{C}_{1|2|3}(k\cdot\omega_1) (\omega_1\cdot\omega) + \mathcal{C}_{1|2|3}(k\cdot\omega_2) (\omega_2\cdot\omega) \bigr] \,,\\
 & \mathcal{C}_{1|2|3}(D_X) =
  P_X^2 + m_1^2 - m_X^2
  - 2 \, \mathcal{C}_{1|2|3}(k_{\parallel}\cdot p_X) \\
  & \qquad \quad + \mathcal{C}_{1|2|3}(k\cdot\omega_1) (\omega_1\cdot p_X)
  + \mathcal{C}_{1|2|3}(k\cdot\omega_2) (\omega_2\cdot p_X) \,,
\end{aligned}
\end{align}
where we used the triangle parametrisation of the loop momentum in \eqn{eq:momparamtri}.
Recalling that $\omega_1 = \chi+\chi^{\dagger}$ and  $\omega_2 = \chi-\chi^{\dagger}$ we can write
$\omega_1\cdot\omega = - (\chi\cdot\chi^\dagger) (\omega_2\cdot p_X)$ and
$\omega_2\cdot\omega = - (\chi\cdot\chi^\dagger) (\omega_1\cdot p_X)$, hence
\begin{align}
\begin{aligned}
&\mathcal{C}_{1|2|3}\bigl(\Delta_{1|2|3|X}(k\cdot\omega)\bigr) =
  c_{0;1|2|3|X} \\
  &- (\chi\cdot\chi^\dagger) \, c_{1;1|2|3|X}\bigl[ \mathcal{C}_{1|2|3}(k\cdot\omega_1) (\omega_2\cdot p_X) + \mathcal{C}_{1|2|3}(k\cdot\omega_2) (\omega_1\cdot p_X) \bigr] \,.
\end{aligned}
\end{align}
We now substitute $k\cdot\omega_i$ using the parametrisation in $t$,
\begin{align} 
  \mathcal{C}_{1|2|3}(k\cdot\omega_1) = \frac{1}{2}\left( t + \frac{\gamma}{t} \right) \,, \quad \qquad
  \mathcal{C}_{1|2|3}(k\cdot\omega_2) = \frac{1}{2}\left( t - \frac{\gamma}{t} \right) \,,
\end{align}
and observe that
\begin{align} \label{eq:tlim_boxnum}
\begin{aligned}
  &\frac{\Delta_{1|2|3|X}(k\cdot\omega)}{D_X} \bigg|_{D_i=0} \overset{t\to \infty}{\to} - (\chi\cdot\chi^\dagger) \, c_{1;1|2|3|X} \,, \\
  &\frac{\Delta_{1|2|3|X}(k\cdot\omega)}{D_X} \bigg|_{D_i=0} \overset{t\to 0}{\to} + (\chi\cdot\chi^\dagger) \, c_{1;1|2|3|X} \,.
\end{aligned}
\end{align}
Therefore, the sum of the box subtraction terms cancel between the two extreme
values of the loop momenta. The triangle's irreducible numerator becomes a
simple polynomial in $t$ in the same limits:
\begin{align}
  &\Delta_{1|2|3} \overset{t\to \infty}{\to} c_{0;1|2|3} + t \, c_{1;1|2|3} + \cdots \,, \\
  &\Delta_{1|2|3} \overset{t\to 0}{\to} c_{0;1|2|3} + \frac{1}{t} c_{2;1|2|3} + \cdots \,.
\end{align}
In keeping with the literature, we define an operation to extract the
components of the Laurent polynomial at infinity, ${\rm Inf}$. The ${\rm Inf}$
operation keeps all terms in a rational function, say $f(x)$, that  do not vanish in the $x \to \infty$
limit,
\begin{equation} \label{eq:InfDef}
  {\rm Inf}_x[f(x)] = \sum_{i=0}^m c_i \, x^i \,,
\end{equation}
where $c_i$ are some numerical values. Since we are considering the integrands of scattering amplitudes the maximum
exponent $m$ will always be finite. The coefficient of the $i^{\rm th}$ term in
the series is denoted ${\rm Inf}_x[f(x)]_{x^i}$. We may therefore write
\begin{equation}
  {\rm Inf}_t \left[\Delta_{1|2|3}\right]_{t^0} = {\rm Inf}_{1/t} \left[\Delta_{1|2|3}\right]_{t^0} = c_{0;1|2|3} \,.
\end{equation}
From \eqn{eq:tlim_boxnum} it follows that the box subtraction terms cancel in the sum,
\begin{align}
{\rm Inf}_t \left[ \frac{\Delta_{1|2|3|X}(k\cdot\omega)}{D_X} \bigg|_{D_i=0} \right]_{t^0} + {\rm Inf}_{1/t} \left[ \frac{\Delta_{1|2|3|X}(k\cdot\omega)}{D_X} \bigg|_{D_i=0} \right]_{t^0} = 0 \,.
\end{align}
As a result, the coefficient of the scalar triangle can be extracted directly from the product of on-shell trees, as
\begin{equation} \label{eq:direct_extraction_triangle}
  c_{0;1|2|3} = -\frac{1}{2} \left\{
        {\rm Inf}_t \left[\left( I^{(1)} \prod_{i=1}^3 D_i \right) \bigg|_{D_i = 0} \right]_{t^0}
  + {\rm Inf}_{1/t} \left[\left( I^{(1)} \prod_{i=1}^3 D_i \right) \bigg|_{D_i = 0} \right]_{t^0}\right\} \,.
\end{equation}

There is an obvious route from here, as we can apply the same method for the extraction bubble
coefficients from the double cut. The analysis is unfortunately not so smooth
since, while the box subtraction terms cancel out as described above, some of the
triangle subtraction terms remain.

There are a number of steps to complete: 1) we must find a suitable basis for
the on-shell loop momentum, 2) we must find a suitable basis for the spurious
vectors, and 3) we must substitute and expand the on-shell loop momentum into both sides of
the double cut equation, \eqn{eq:doublecuteq}.

We consider a bubble configuration with a momentum $P$. We switch the notation
slightly to avoid too many subscripts and focus on one generic triangle
subtraction term which we label with momenta $P$, $Q$ and $R$ where $P = -Q-R$. The double cut
notation is therefore $\mathcal{C}_{P|QR}$.  Since the physical space has only
one dimension we are missing an additional direction with which we can span the
loop momentum space. This forces us to introduce an arbitrary light-like
direction $n^\mu$ such that
\begin{equation}
  P^{\flat,\mu} = P - \frac{S}{2 \, P\cdot n} \, n^\mu \,,
  \label{eq:Pflatdef}
\end{equation}
with $P^2 = S$, is a second massless direction with which we may construct a spanning basis for the loop momentum,
\begin{equation} \label{eq:kPn}
  k^\mu = \alpha_1 \, P^{\flat,\mu} + \alpha_2 \, n^\mu
  + \alpha_3 \, \frac{1}{2}\spAB{P^\flat}{\gamma^\mu}{n} \, \Phi_{\rm bub} + \alpha_4 \, \frac{1}{2}\spAB{n}{\gamma^\mu}{P^\flat} \, \Phi_{\rm bub}^{-1} \,.
\end{equation}
Here $\Phi_{\rm bub}$ is an arbitrary factor which ensures that the coefficients $\alpha_i$ are free of spinor phases. We may give an explicit expression using one of the other linearly independent momenta, say $X$, as $\Phi_{\rm bub}=\spAB{n}{X}{P^{\flat}}$, as we did in \eqn{eq:spuriousvectorexampeEQ1}. The arbitrary factor $\Phi_{\rm bub}$ will however cancel out of the results, and we will thus leave it symbolic.
The on-shell constraints $k^2 = m_1^2$ and $2\,k \cdot P = S + m_1^2 - m_2^2 = \hat{S}$
have a two-parameter family of solutions. We parametrise it in terms of
parameters, $t$ and $y$, as
\begin{align}
\begin{aligned}
  \alpha_1 &= y \,, \qquad \quad &
  \alpha_2 &= \frac{\hat{S}- S \, y}{2 \, n \cdot P} \,, \\
  \alpha_3 &= t \,, &
  \alpha_4 &= \frac{y \, (\hat{S} - S \, y) - m_1^2}{2 \, t \, (n\cdot P)} \,.
\end{aligned}
  \label{eq:dblcutonshelsol}
\end{align}
We can represent the spurious vectors in the same basis,
\begin{align} \label{eq:spurious_omega_bubble}
  \omega_{1,{\rm bub}}^\mu &= \frac{1}{2}\spAB{P^\flat}{\gamma^\mu}{n} \, \Phi_{\rm bub} + \frac{1}{2}\spAB{n}{\gamma^\mu}{P^\flat} \, \Phi_{\rm bub}^{-1} \,, \\ 
  \omega_{2,{\rm bub}}^\mu &= \frac{1}{2}\spAB{P^\flat}{\gamma^\mu}{n} \, \Phi_{\rm bub} - \frac{1}{2}\spAB{n}{\gamma^\mu}{P^\flat} \, \Phi_{\rm bub}^{-1} \,,\\
  \omega_{3,{\rm bub}}^\mu &= P^{\flat,\mu} - \frac{S}{2 \, P\cdot n} \, n^\mu \,,
\end{align}
where, again, the phase factor $\Phi_{\rm bub}$ ensures that all summands are free of spinor phases.
Now we evaluate the spurious ISPs,
\begin{align}
  \mathcal{C}_{P|QR}(k\cdot\omega_{1,{\rm bub}})  &= - P\cdot n \left(t + \frac{y \, (\hat{S} - S \, y) - m_1^2}{2 \, t \, (P\cdot n)}\right) \,,\\
  \mathcal{C}_{P|QR}(k\cdot\omega_{2,{\rm bub}})  &= - P\cdot n \left(-t + \frac{y \, (\hat{S} - S \, y) - m_1^2}{2 \, t \, (P\cdot n)}\right) \,,\\
  \mathcal{C}_{P|QR}(k\cdot\omega_{3,{\rm bub}})  &= \frac{1}{2}\left( \hat{S}- 2 \, S \, y \right) \,,
\end{align}
and substitute them into the LHS of \eqn{eq:doublecuteq}, which
becomes a Laurent polynomial in $y$ and $t$. Note that the arbitrary phase factor $\Phi_{\rm bub}$ cancels out in the ISPs. One can then show that the direct
extraction of the scalar bubble coefficient can be obtained~using
\begin{equation}
  c_{0;P|QR} = \Delta_{P|QR}\bigl|_{t^0,y^0} + \frac{\hat{S}}{2 \, S} \, \Delta_{P|QR} \bigl|_{t^0,y^1}+ \frac{1}{3}\left( \frac{\hat{S}^2}{S^2} - \frac{m_1^2}{S} \right)\Delta_{P|QR}\bigl|_{t^0,y^2} \,.
  \label{eq:directbubform}
\end{equation}
To complete the bubble-extraction formula we must evaluate the RHS of
\eqn{eq:doublecuteq} at the same on-shell solution, and so we need to find a
representation of the spurious vectors in the box and triangle coefficients. Each
triangle subtraction term will depend on one additional momentum, say $Q$, while we have
two momenta $Q$ and $R$ for each box. For the case where both momenta $Q$ and
$R$ are massive ($Q^2 = T, R^2 = U$), we need to construct more light-like
projections in order to span the spurious space. A convenient way to do this is to consider linear combinations of $P^{\mu}$ and $Q^{\mu}$,
\begin{align}
  \check{P}^{\mu} = \frac{\gamma \, (\gamma \, P^\mu - S \, Q^\mu)}{\gamma^2 - S \, T} \,, \qquad \quad
  \check{Q}^{\mu} =  \frac{\gamma \, (\gamma \, Q^\mu - T \, P^\mu)}{\gamma^2 - S \, T} \,.
\end{align}
Requiring that $\check{P}^{\mu}$ and $\check{Q}^{\mu}$ are light-like gives two possible projections: $\gamma_{\pm} = P \cdot Q \pm \sqrt{(P\cdot Q)^2 -
S \, T}$. The argument of the square root is (minus) the Gram determinant $\mathrm{det} \, G(P,Q)$ (see \eqn{eq:gram_def}). The spurious
vectors are then simple to write down:
\begin{align} \label{eq:spurious_vectors_tri_box}
  \omega_{1,{\rm tri}}^\mu &= \frac{1}{2}\spAB{\check{P}}{\gamma^\mu}{\check{Q}} \, \Phi_{\rm tri} + \frac{1}{2}\spAB{\check{Q}}{\gamma^\mu}{\check{P}} \, \Phi_{\rm tri}^{-1} \,,\\ 
  \omega_{2,{\rm tri}}^\mu &= \frac{1}{2}\spAB{\check{P}}{\gamma^\mu}{\check{Q}}\, \Phi_{\rm tri} - \frac{1}{2}\spAB{\check{Q}}{\gamma^\mu}{\check{P}}\, \Phi_{\rm tri}^{-1} \,,\\ 
  \omega_{\rm box}^\mu &= \frac{1}{2}\spAB{\check{P}}{\gamma^\mu}{\check{Q}} \spAB{\check{Q}}{R}{\check{P}} - \frac{1}{2}\spAB{\check{Q}}{\gamma^\mu}{\check{P}} \spAB{\check{P}}{R}{\check{Q}} \,.
\end{align}
As above, $\Phi_{\rm tri}$ is an arbitrary factor which makes the triangle spurious vectors free of spinor phases. For instance, one may write $\Phi_{\rm tri}=\spAB{\check{Q}}{X}{\check{P}}$, where $X$ is an arbitrary momentum linearly independent of $Q$ and $P$. In $\omega_{\rm box}^{\mu}$, on the other hand, the phase factor is explicit, and is chosen so as to make $\omega_{\rm box}^{\mu}$ orthogonal to $R^{\mu}$.
As we have seen for the triangle coefficient, the idea is to consider the
behaviour of the integrand at large values for the loop momenta where the
additional uncut propagators suppress as many contributions as possible. In
order to write down the procedure concisely we introduce an operation
$\mathcal{P}$ which, acting on a function of $y$ and $t$, gives
\begin{equation} \label{eq:Pdef}
  \mathcal{P}\bigl(f(y,t) \bigr) = f \bigl|_{t^0,y^0} + \frac{\hat{S}}{2 \, S}f \bigl|_{t^0,y^1}+ \frac{1}{3}\left( \frac{\hat{S}^2}{S^2} - \frac{m_1^2}{S} \right)f\bigl|_{t^0,y^2} \,,
\end{equation}
and combine it with the limit of $y,t\to \infty$. From \eqn{eq:directbubform} we see that
$\mathcal{P}\bigl(\Delta_{1|2}(y,t) \bigr) = c_{0;1|2}$. The first term from the RHS of \eqn{eq:doublecuteq} can be extracted from the product of
two tree-level amplitudes,
\begin{align}
  \mathcal{P} {\rm Inf}_y {\rm Inf}_t \bigg[ I^{(1)}\bigl(k(y,t) \bigr) \prod_{i=1}^2 D_i \, \Bigl|_{D_1=D_2=0} \bigg] \,.
\end{align}
Using the definitions above for the on-shell loop momenta and the box spurious
vector, one can show that the box subtraction terms vanish in the limit $y,t\to
\infty$. The triangle subtraction term does not vanish but it is simple to
obtain an explicitly solution in terms of the spurious triangle coefficients
$c_{i|1|2|Q}$. While the procedure to extract the relevant contributions is
simple, the result is not particularly compact, especially for the higher
tensor rank coefficients. Therefore, we present the result here up to the
rank-one coefficients:
\begin{align}
\begin{aligned}
  & \mathcal{P} {\rm Inf}_y {\rm Inf}_t \bigg[ \frac{\Delta_{P|Q|R}(k\cdot\omega_{1,{\rm tri}}, k\cdot\omega_{2,{\rm tri}})}{-(k-P-Q)^2+m_3^2} \, \biggl|_{D_1=D_2=0} \bigg] \\
  & \hspace{0.5cm} =  ( c_{1;P|Q|R} + c_{2;P|Q|R} ) \frac{\spA{P^\flat}{\check{P}} \spB{\check{Q}}{n} \, \Phi_{\rm tri} }{2 \, \spAB{P^{\flat}}{Q}{n} } \\
  & \hspace{0.5cm} \phantom{{} = {}} + ( c_{1;P|Q|R} - c_{2;P|Q|R} ) \frac{ \spA{P^\flat}{\check{Q}} \spB{\check{P}}{n} \, \Phi_{\rm tri}^{-1} }{ 2 \, \spAB{P^{\flat}}{Q}{n} } + \ldots
\end{aligned}
\end{align}
The combination of both double cut and triangle subtraction terms gives the
general formula for the bubble coefficient, where we must remember to average
over the two projections for the triangle subtractions:
\begin{align} \label{eq:bubble_direct_extraction}
\begin{aligned}
  c_{0;P|QR} & =
  \mathcal{P} {\rm Inf}_y {\rm Inf}_t \bigg[ I^{(1)}\bigl(k(y,t)\bigr) \prod_{i=1}^2 D_i \bigg] \\
  & \phantom{ = {}} - \frac{1}{2} \sum_{Q} \sum_{\gamma=\gamma_\pm} \mathcal{P} {\rm Inf}_y {\rm Inf}_t \bigg[ \frac{\Delta_{1|2|Q}(k\cdot\omega_{1,{\rm tri}}, k\cdot\omega_{2,{\rm tri}})}{-(k-P-Q)^2+m_3^2} \bigg] \,.
\end{aligned}
\end{align}
Practical applications of this formula require a bit of practice, as many spinor identities are required to simplify the various projected momenta.

\begin{exer}{Rank-one triangle reduction with direct extraction}
\label{Ex:3.DirectExtraction}
To verify the analysis it is useful to consider a simple example,
\begin{align}
  F_3^{[D]}(P,Q)[k\cdot Z] = \int_k \frac{k\cdot Z}{
  \bigl[-k^2+m_1^2\bigr]\bigl[-(k-P)^2+m_2^2 \bigr] \bigl[-(k-P-Q)^2+m_3^2 \bigr]} \,,
\end{align}
where $Z^{\mu}$ is an arbitrary momentum, $P^2=S$ and $Q^2=T$. We denote the third momentum $R=-P-Q$.

\begin{enumerate}[a)]
\item Use the Passarino-Veltman method to show that the bubble coefficient of the $P$ channel, i.e., the coefficient of $F_2^{[D]}(P)[1]$, is given by
\begin{align}
  c_{0;P|QR} = \frac{(P\cdot Q) (P\cdot Z) - S \, (Q\cdot Z)}{2 \, \bigl( (P\cdot Q)^2 - S \, T \bigr)} \,.
\end{align}

\item To obtain the same result with the direct extraction method it is recommended to use a computer algebra system. We provide the intermediate steps to guide you through the process.
First, compute the triple-cut coefficients:
\begin{align}
c_{1;P|Q|R} = - \frac{Z\cdot \omega_{1,{\rm tri}}}{2 \, \check{P}\cdot \check{Q}} \,, \qquad \quad
c_{2;P|Q|R} = \frac{Z\cdot \omega_{2,{\rm tri}}}{2 \, \check{P}\cdot \check{Q}} \,.
\end{align}
The higher-rank coefficients vanish, and $c_{0;P|Q|R}$ is irrelevant for our purposes.
Then compute the double-cut part of the bubble,
\begin{align}
  \mathcal{P} {\rm Inf}_y {\rm Inf}_t \left[  \mathcal{C}_{P|Q} \left( \frac{k\cdot Z}{-(k-P-Q)^2+m_3^2} \right)  \right]
  = \frac{1}{2} \frac{\spAB{P^\flat}{Z}{n}}{\spAB{P^\flat}{Q}{n}} \,,
\end{align}
and finally put together the triangle subtraction term,
\begin{align}
  \mathcal{P} {\rm Inf}_y {\rm Inf}_t \left[ \mathcal{C}_{P|Q} \left(\frac{\Delta_{P|Q|R}(k\cdot\omega_{1,{\rm tri}}, k\cdot\omega_{2,{\rm tri}})}{-(k-P-Q)^2+m_3^2} \right) \right]
 = - \frac{\langle P^{\flat}|\check{P} Z \check{Q}|n] + (\check{P} \leftrightarrow \check{Q})}{4 \, (\check{P}\cdot \check{Q}) \, \spAB{P^{\flat}}{Q}{n} } \,,
\end{align}
which after summation over the two projections gives
\begin{align}
\begin{aligned}
 \frac{1}{2} \sum_{\gamma=\gamma_\pm} \mathcal{P} {\rm Inf}_y {\rm Inf}_t & \left[ \mathcal{C}_{P|Q}\left( \frac{\Delta_{P|Q|R}(k\cdot\omega_{1,{\rm tri}}, k\cdot\omega_{2,{\rm tri}})}{-(k-P-Q)^2+m_3^2} \right) \right] \\
 & = - \frac{(P\cdot Q) (P\cdot Z) - S \, (Q\cdot Z)}{2 \, \bigl( (P\cdot Q)^2 - S\,T \bigr)} + \frac{1}{2} \frac{\spAB{P^\flat}{Z}{n}}{\spAB{P^\flat}{Q}{n}} \,.
\end{aligned}
\end{align}
It is now easy to verify that by putting together the double cut and the triangle subtraction as in \eqn{eq:bubble_direct_extraction} we recover the Passarino-Veltman result.
\end{enumerate}

\noindent
For the solution see \hyperref[Sol:3.DirectExtraction]{chapter~5} and the \texttt{Mathematica} notebook \break \href{https://scattering-amplitudes.mpp.mpg.de/scattering-amplitudes-in-qft/Exercises/}{\texttt{Ex3.9\_DirectExtraction.wl}}~\cite{website-ch3}.
\end{exer}

The extension of this method to the $D$-dimensional cuts and the rational terms
is straightforward, since we may use it to first perform the four-dimensional
analysis including an mass shift in the propagators $m_i^2\to m_i^2-\mu_{11}$, and then consider the
$\mu_{11}\to \infty$ limit to directly probe the rational terms. An explicit demonstration of this technique is the final task for this chapter. 

\section{Project: Rational terms of the four-gluon amplitude \label{sec:rat4g}}

We would like to complete the computation of the four-gluon adjacent helicity MHV amplitude that we
have done in part throughout this chapter. This requires us to fix the rational
term, and the complete computation we will follow requires a substantial amount
of algebra to perform the direct extraction of the $D$-dimensional monomials in
the integrand. Alternative methods can also work nicely in this case, for
example fixing the ambiguity through requiring universal factorisation in
collinear limits or simply automating a Feynman-diagram computation, since the
four-point massless kinematics are relatively simple. In this section we will
outline the necessary steps, and leave the algebra as an extended exercise or
computer algebra project for the interested reader.

The first observation we make is to see that the extra-dimensional components
of the internal gluon lines are identical to those obtained by using a massive
internal scalar with the mass $\mu^2 = \mu_{11}$. The tree-level helicity
amplitudes we need in the cut can easily be derived using the methods described
in chapter~\ref{ch:trees}. One slight subtlety is that the three-point amplitude for two scalar fields ($S$) and one gluon depends on an arbitrary reference direction, which we will call $\xi$ in
this section. The results we need are given in eqs.~\eqref{ssp} and~\eqref{ssm}, which we repeat here for convenience (setting the coupling to $1$):
\begin{align}
  A^{(0)}_3(1_S, 2_g^+, 3_S; \xi) &= \i \frac{\spAB{\xi}{3}{2}}{\spA{\xi}{2}} \,, \\
  A^{(0)}_3(1_S, 2_g^-, 3_S; \xi) &= \i \frac{\spAB{2}{3}{\xi}}{\spB{2}{\xi}} \,.
\end{align}
When using a product of these amplitudes inside a cut one can make convenient
choices of the reference vector to simplify the spinor algebra. The two
independent four-point amplitudes were obtained through BCFW recursion given in \eqn{gpgpss}  and exercise~\ref{Ex:2.helicity}:
\begin{align}
  A^{(0)}_4(1_S, 2_g^+, 3_g^+, 4_S) &= \i \frac{\mu^2 \spB23}{\spA23 \spAB212} \,, \\
  A^{(0)}_4(1_S, 2_g^-, 3_g^+, 4_S) &= \i \frac{\spAB213^2}{s_{23} \spAB212} \,.
\end{align}

Before performing the full computation for the MHV amplitude, the simplest case
is the all-plus helicity amplitude, which vanishes at tree level. Due to
additional symmetries this helicity sector turns out to be even simpler than
the other vanishing tree-level amplitudes with a single minus helicity. As a
warm-up exercise we can perform the quadruple cut $1|2|3|4$. Choosing a spinorial basis with momenta $p_1$ and $p_4$,
\begin{align}
  k^\mu &= \vec{\alpha} \cdot \vec{v}^\mu \,, \\
  \vec{v}^\mu &= \left\{p_1^\mu, p_4^\mu, \tfrac{1}{2}\spAB1{\gamma^\mu}4, \tfrac{1}{2}\spAB4{\gamma^\mu}1\right\} \,,
\end{align}
with $\vec{\alpha}=\{\alpha_1,\alpha_2,\alpha_3,\alpha_4\}$, leads to two on-shell solutions $k_{\pm}^{\mu}$, with 
\begin{align}
  \vec{\alpha}_{\pm} &= \left\{0, 0,  \frac{\spB12}{\spB42} X_\pm,  \frac{\spA12}{\spA42} X_\mp \right\} \,, \\
  X_\pm & = \frac{1}{2}\left( 1 \pm \sqrt{1-\frac{4 \, \mu^2 \, s_{13}}{s_{12} \, s_{23}}}\right) \,.
\end{align}
The product of tree-level amplitudes can now be evaluated as follows:
\begin{align}
  & 2 \, \i \, A^{(0)}_3\Bigl((-k)_S, 1^+_g, (k-p_1)_S; p_2\Bigr) \, \i \, A^{(0)}_3\Bigl((-k+p_1)_S, 2^+_g, (k-p_{12})_S; p_1\Bigr) \nonumber\\
  & \hspace{0.9cm} \times \i \, A^{(0)}_3\Bigl((-k+p_{12})_S, 3^+_g, (k+p_4)_S; p_4\Bigr) \, \i \, A^{(0)}_3\Bigl((-k-p_4)_S, 4^+_g, k_S; p_3\Bigr) \Bigl|_{k=k_{\pm}} \nonumber\\
  &\hspace{3cm} = 2 \, \frac{\spAB2k1}{\spA21} \frac{\spAB1k2}{\spA12} \frac{\spAB4k3}{\spA43} \frac{\spAB3k4}{\spA34} \biggl|_{k=k_{\pm}} \nonumber\\
  &\hspace{3cm} = 2 \, \alpha_3^2 \alpha_4^2 \, \frac{\spB12\spB34 s_{23}^2}{\spA12\spA34} \biggl|_{\vec{\alpha}=\vec{\alpha}_{\pm}} \nonumber\\
  &\hspace{3cm} = 2 \, \frac{\mu^4 s_{12} s_{34}}{\spA12^2 \spA34^2}\nonumber\\
  &\hspace{3cm} = -2 \, \frac{\mu^4 s_{12} s_{23}}{\spA12\spA23\spA34\spA41} \,.
\end{align}
The overall factor of $2$ must be included to match the complex scalar degrees of
freedom with the extra-dimensional components of the gluon polarisation sum.
In general we should average over the two on-shell solutions but in this case
it turns out both give the same simple result, which only contains one of the
five possible irreducible numerator monomials in $\Delta_{1|2|3|4}$. From here we can read directly the coefficient of $\mu^4$ which contributes to
the rational term,
\begin{align}
  c_{4;1|2|3|4}(1^+,2^+,3^+,4^+) &= -2 \, \frac{s_{12} s_{23}}{\spA12\spA23\spA34\spA41} \,.
\end{align}
The quadruple cuts for the other helicity configurations are simple to compute with the same on-shell loop momentum solution. Explicitly for the adjacent MHV configuration one can find,
\begin{align}
  c_{4;1|2|3|4}(1^-,2^-,3^+,4^+) &= -2 \bigl(-\i A^{(0)}_4(1^-,2^-,3^+,4^+) \bigr) \frac{s_{23}}{s_{12}} \,,
\end{align}
Note that for higher-multiplicity amplitudes additional
uncut propagators would appear with non-trivial $\mu^2$ dependence in the
denominator. An additional limit of $\mu\to \infty$ is then necessary before
extracting the $\mu^4$ term.

The symmetry of the all-plus configuration leads to some dramatic and
unexpected cancellations so that the quadruple cut contribution actually fixes
the full amplitude. The other amplitudes require a bit more work. There are
four independent triangle contributions: $1|2|34, 1|23|4, 12|3|4$ and $2|3|41$.
Each has one possible contribution to the rational term which, in the case of
the $1|2|34$ configuration, is given by
\begin{align}
\begin{aligned}
  & c_{7;1|2|3}(1,2,3,4) = {\rm Inf}_{\mu^2} {\rm Inf}_t \bigg[
    -2 \, \i \, A^{(0)}_3\Bigl((-k)_S, 1_g, (k-p_1)_S; p_2\Bigr) \\
    & \hspace{0.4cm} \times \i \, A^{(0)}_3\Bigl((-k+p_1)_S, 2_g, (k-p_{12})_S; p_1\Bigr) \,
    \i \, A^{(0)}_4\Bigl((-k+p_{12})_S, 3_g, 4_g, k_S\Bigr)\bigg]_{t^0,\mu^2} \,,
\end{aligned}
\end{align}
with similar formulae for the other permutations. The final results for the adjacent MHV amplitude are
\begin{align}
  c_{7;1|2|34}(1^-,2^-,3^+,4^+) &= 0 \,, \\
  c_{7;1|23|4}(1^-,2^-,3^+,4^+) &= 0 \,, \\
  c_{7;12|3|4}(1^-,2^-,3^+,4^+) &= 0 \,, \\
  c_{7;2|3|41}(1^-,2^-,3^+,4^+) &= 0 \,.
\end{align}
Finally we turn to the bubble contributions, of which there are two configurations: $12|34$ and $23|41$. The relevant coefficient for the rational term is
\begin{align}
\begin{aligned}
  & c_{9;12|34}(1,2,3,4) =\mathcal{P} {\rm Inf}_{\mu^2}  {\rm Inf}_y {\rm Inf}_t \Bigg[
        - \frac{\Delta_{1|2|34}}{-(k-p_1)^2+\mu^2} - \frac{\Delta_{12|3|4}}{-(k+p_4)^2+\mu^2} \\
    & \hspace{1cm} +  2 \, \i \, A^{(0)}_4\bigl((-k)_S, 1_g, 2_g, (k-p_{12})_S\bigr) \, \i \, A^{(0)}_4\bigl((-k+p_{12})_S, 3_g, 4_g, k_S\bigr) \Bigg] \,,
\end{aligned}
\end{align}
and similarly for $23|41$. The reference vector $n^\mu$ used to form the double-cut loop-momentum basis can be chosen to simplify the algebra. If we choose it
to be $p_2^\mu$ for the $12|34$ cut then the $1^{\rm st}$ triangle subtraction will give
zero. Furthermore, since the triangle contribution $12|3|4$ has massless
external legs $3$ and $4$, there is only one value for $\gamma$ in the
light-like projection. The final results for the adjacent MHV amplitude are nice and compact:
\begin{align}
  c_{9;12|34}(1^-,2^-,3^+,4^+) &= 0 \,, \\ 
  c_{9;23|41}(1^-,2^-,3^+,4^+) &= -2 \bigl(-\i \, A^{(0)}_4(1^-,2^-,3^+,4^+)\bigr) \frac{2 \, s_{12} - 3 \, s_{23}}{3 \, s_{12} s_{23}} \,.
\end{align}
We are now finally ready to assemble the full amplitude. Together with the values for the integrals evaluated up to $\mathcal{O}\bigl(\eps^0\bigr)$, we have
\begin{align}
  R(1^-,2^-,3^+,4^+)
  &= -\frac{1}{6} \i \, c_{4;1|2|3|4}(1^-,2^-,3^+,4^+) - \frac{s_{23}}{6} \i \, c_{9;23|41}(1^-,2^-,3^+,4^+) \nonumber\\
  &= \frac{2}{9} \, A^{(0)}_4(1^-,2^-,3^+,4^+) \,.
\end{align}
Combined with the cut-constructible terms from \eqn{eq:4g1loop}, the only task
remaining is to evaluate the basis integrals, which brings us neatly to the
subject of the next chapter. Our final result, reinstating the correct prefactors, is
\begin{align}
\label{eq:full_split_helicity_four_gluon_one_loop_integrand}
\begin{aligned}
  &A^{(1),[4-2\eps]}_4(1^-,2^-,3^+,4^+) = \frac{\alpha_{\rm YM} \, \mu_{\rm R}^{2\eps}}{(4\pi)^{2-\eps}} \, A^{(0)}(1^-,2^-,3^+,4^+) \\
  & \hspace{0.4cm} \times \bigg[-s_{12}s_{23}F_4^{[4-2\eps]}(p_1, p_2, p_3)[1] - \frac{11}{3} F_2^{[4-2\eps]}(p_{23})[1]  + \frac{2}{9} \bigg] + \mathcal{O}(\eps) \,.
\end{aligned}
\end{align}
This result, along with the other independent helicity configurations and partonic channels, can be found in the following references~\cite{ch3_Bern:1990ux,ch3_Bern:1994fz}.

\section{Outlook: Rational representations of the external kinematics \label{sec:3.7}}

Having completed the exercises in this section it becomes clear that analytic
computations using the spinor-helicity formalism have limitations, especially
when working with pen and paper. Computer algebra systems have always been
essential for research in this area, and we can think of designing new systems
tuned to alleviate bottlenecks in the current state-of-the-art calculations.
The major flaw of the spinor-helicity formalism that we have encountered is the
redundancy of representations, which stems from the lack of manifest momentum
conservation and Schouten identities. This quickly becomes an annoyance as one
performs many spinor-helicity manipulations. A recent idea, perhaps introduced
with other motivations about manifest amplitude symmetries in mind, is that of
\emph{momentum twistors}. Introduced by Hodges~\cite{ch3_Hodges:2009hk}, these
differ from Penrose's twistor formalism by addressing \textit{dual conformal
invariance} rather than the usual conformal invariance. We will not attempt a
full review of the formalism here but try to explain the entry-level concepts
that can lead to very practical methods for amplitude calculations. One
important recent development has been the combination of rational kinematic
parametrisations with modular arithmetic over finite (prime) fields. This
technique enables multiple numerical evaluations modulo a (large) prime number
to be used to obtain fully analytic expressions for the coefficients in an
integral or integrand basis. A full description of this method is beyond the
scope of these lecture notes but we encourage the readers to follow some recent literature
and implementations~\cite{ch3_Peraro:2016wsq,ch3_Peraro:2019svx}.

The spinor-helicity formalism makes the on-shell condition for any massless
momenta manifest. A convenient way
to make the momentum conservation for an $n$-particle system manifest is to
introduce \emph{dual momentum variables} $y_i$ as
\begin{align}
p_i^{\mu} \eqqcolon y_{i+1}^{\mu} - y_i^{\mu} \,,
\end{align}
with $y_{n+1}=y_1$. It is easy to verify that, with this parametrisation, $ \sum_{i=1}^n p_i = 0$.
The dual momenta can then be used to form an positive-helicity
spinor $\mu_i$,
\begin{align}
|\mu_i]  \coloneqq \slashed{y_i} | i \ran \,.
\end{align}
Since we may span any
positive-helicity spinor $|i]$ in a basis of two independent positive-helicity
spinors, say $|\mu_i]$ and $|\mu_{i+1}]$, we may write,
\begin{equation}
  |i] = \alpha_i \, |\mu_i] + \beta_i \, |\mu_{i+1}] \,.
\end{equation}
By projecting out the coefficients and using the properties of the dual momenta one can show that
\begin{equation} \label{eq:LambdaTildeFromMT}
  |i] = \frac{\spA{i}{i+1}\,|\mu_{i-1}] + \spA{i+1}{i-1} \,|\mu_{i}] + \spA{i-1}{i}\,|\mu_{i+1}]}{\spA{i-1}{i}\spA{i}{i+1}} \,.
\end{equation}
The power of this formalism can then be appreciated by observing that, for
any random parametrisation of the $4\times n$ components in $|i\ran$ and
$|\mu_{i}]$, both momentum conservation and on-shellness will be manifest. The object $Z_i =
(|i\ran, |\mu_{i}])^{\top}$ is called a momentum twistor, and the system of $n$ momentum
twistors has Poincar\'{e} symmetry, meaning that only $3n-10$ components are
independent. We refer to~\cite{ch3_Elvang:2013cua} for further reading on this topic.

\begin{exer}{Momentum-twistor parametrisations}
\label{Ex:3.9}
Consider the kinematics of a massless $2\to 2$ scattering process. In the momentum-twistor formalism, it is described by a $4\times 4$ matrix $Z = \bigl( Z_1 \, Z_2 \, Z_3 \, Z_4 \bigr)$ of momentum twistors $Z_i = (|i\ran, |\mu_{i}])^{\top}$. Thanks to Poincar\'e symmetry, only $3\times4-10=2$ entries of $Z$ are independent. 
In order to obtain a parametrisation of $Z$ in terms of the minimal number of independent variables, one needs to make use of the full Poincar\'e group. In particular, one uses the little group invariance $(\lambda_i,\tilde{\lambda}_i) \equiv (\mathrm{e}^{\i \varphi_i} \lambda_i, \mathrm{e}^{-\i \varphi_i} \tilde{\lambda}_i)$ to fix some of the components by choosing explicit phases $\varphi_i$. As a result, the helicity scaling of the expressions is obscured. This is not a problem, as we can always divide all quantities by an arbitrary phase factor, and use the momentum-twistor parametrisation for the phase-free ratios. For the four-point case, a minimal parametrisation in terms of two independent variables $x$ and $y$ may be chosen as
\begin{align} \label{eq:Z4pt}
Z =
  \begin{pmatrix}
    1 & 0 & \frac{1}{y} & -y \\
    0 & 1 & 1 & 1 \\
    0 & 0 & -\frac{x}{y} & 0 \\
    0 & 0 & 0 & x \\
  \end{pmatrix} \,.
\end{align}
Using this parametrisation, calculate
\begin{enumerate}[a)]
  \item the positive helicity spinors $\tla_i$,
  \item the components of the four-momenta $p_i^\mu$,
  \item the invariants $s_{ij}$ and spinor products $\spA ij$, $\spB ij$,
\end{enumerate}
in terms of the free parameters $x$ and $y$. For what value of $x$ and $y$ do we
recover a standard $2\to2$ phase-space parametrisation in terms of energies and angles?
In the four-particle case the MHV and $\overline{\rm MHV}$ helicity configurations coincide. 
Use both the spinor-helicity formalism and the momentum-twistor parametrisation above to show that the MHV and $\overline{\rm MHV}$ Parke-Taylor formulae in eqs.~\eqref{ParkeTaylor1} and~\eqref{ParkeTaylor1MHVbar} are equivalent for $n=4$.
For the solution see \hyperref[Sol:3.9]{chapter~5}.
\end{exer}

\section{Outlook: Multi-loop amplitude methods \label{sec:3.8}}

The purely algebraic method outlined in this chapter is extremely powerful and
has led to the development of fully automated numerical programs~\cite{ch3_Berger:2008sj,ch3_Giele:2008bc,ch3_Hirschi:2011pa,ch3_Bevilacqua:2011xh,ch3_Badger:2012pg,ch3_GoSam:2014iqq,ch3_Actis:2016mpe,ch3_Buccioni:2019sur}. There are
however an increasingly large number of observables that require more accurate
perturbative predictions.

Many of the techniques presented here generalise in a straightforward way to
higher-loop cases. There is however a new feature that presents a substantial
additional challenge in identifying a suitable basis of integral functions. The
integrand-reduction procedure has been extended to the multi-loop case with
some explicit results obtained for amplitudes, or parts of amplitudes, at two
and three loops. Owing to the larger number of irreducible scalar products,
additional technology is required to reduce the amplitude to a basis of loop
integrals. Nevertheless the integrand can in principle be constructed from the
products of tree-level amplitudes by following the one-loop methodology.

\begin{figure}[h]
  \centering
  \includegraphics[width=0.4\textwidth]{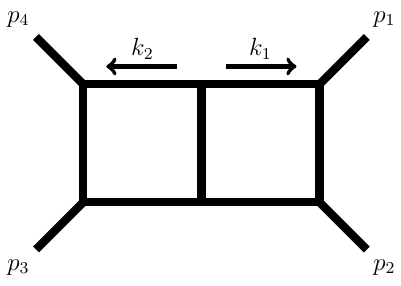}
  \caption{A two-loop double box configuration.}
  \label{fig:2Ldblbox}
\end{figure}

We can run through a simple example to get a sense of the new features.
Consider a two-loop double box with seven massless propagators and massless external
legs as shown in figure~\ref{fig:2Ldblbox}. The propagators can be written in
terms of scalar products which are linear in the loop momenta $k_i\cdot p_j$, and three
scalar products quadratic in the loop momenta: $k_1^2$, $k_2^2$ and $k_1 \cdot k_2$. The
linear scalar products can be written as the difference of two inverse
propagators. The list of all possible scalar products $k_i \cdot \vec{v}_{j}$, where $\vec{v}^{\mu}$ is a
spanning set of momenta such as $\vec{v}^{\mu} = \{p_1^{\mu},p_2^{\mu},p_4^{\mu},\omega^{\mu}\}$, can then be
separated into the groups of 1) \emph{reducible scalar products}, that can be
written as the difference of propagators (plus functions of the external kinematic invariants), and 2) a set of four
irreducible scalar products (ISPs), for example $k_1\cdot p_4$, $k_2\cdot p_1$, $k_1\cdot\omega$,
and $k_2\cdot\omega$. Being more explicit, we decompose the loop momenta into transverse spaces,
\begin{align}
  k_i^\mu = k^\mu_{i,\parallel} + (k_i\cdot\omega)\,\omega^\mu \,.
\end{align}
The constraints on the scalar products $k_i\cdot p_j$ fix $k^\mu_{i,\parallel}$, which becomes a function of the two ISPs $k_1\cdot p_4$ and $k_2\cdot p_1$.
The remaining on-shell constraints for $k_1^2=0$, $k_2^2=0$ and $k_1 \cdot k_2=0$ can now be recast into conditions on all four ISPs,
\begin{align}
  &f_{11}(k_1\cdot p_4, k_2\cdot p_1) + (k_1\cdot\omega)^2 = 0 \,, \\
  &f_{22}(k_1\cdot p_4, k_2\cdot p_1) + (k_2\cdot\omega)^2 = 0 \,, \\
  &f_{12}(k_1\cdot p_4, k_2\cdot p_1) + (k_1\cdot\omega)(k_2\cdot\omega) = 0 \,,
\end{align}
where $f_{ij} = \mathcal{C}_{\rm double-box}(k_{1,\parallel}\cdot k_{2,\parallel})/\omega^2$ with $\mathcal{C}_{\rm double-box}$ indicating the hepta-cut.
The irreducible numerator $\Delta_{\rm double-box}(k_1\cdot p_4, k_2\cdot p_1,
k_1\cdot\omega, k_2\cdot\omega)$ may then be constructed by forming a general
polynomial of rank four\footnote{For QCD the maximum rank appearing in Feynman
diagrams in the Feynman gauge is four, in different theories a higher degree polynomial maybe necessary.} in the four ISPs and then removing the (multi-variate)
constraints through polynomial division. This operation involves the
computation of a Gr\"{o}bner basis, which can be computationally expensive but for
the construction of a general basis is unlikely to present a
bottleneck. In this case a suitable basis can be found to be~\cite{ch3_Badger:2012dp}
\begin{align}
& \Delta_{\rm double-box}(k_1\cdot p_4, k_2\cdot p_1, k_1\cdot\omega, k_2\cdot\omega) =
    c_0 \nonumber\\
    & \hspace{0.4cm}
    + c_1 (k_1\cdot p_4)
    + c_2 (k_2\cdot p_1)
    + c_3 (k_1\cdot p_4)^2
    + c_4 (k_2\cdot p_1)^2
    + c_5 (k_1\cdot p_4)(k_2\cdot p_1)\nonumber\\
    & \hspace{0.4cm}
    + c_6 (k_1\cdot p_4)^3
    + c_7 (k_1\cdot p_4)(k_2\cdot p_1)^2
    + c_8 (k_1\cdot p_4)(k_2\cdot p_1)^2
    + c_9 (k_2\cdot p_1)^3\nonumber\\
    & \hspace{0.4cm}
    + c_{10} (k_1\cdot p_4)^4
    + c_{11} (k_1\cdot p_4)^3 (k_2\cdot p_1)
    + c_{12} (k_1\cdot p_4) (k_2\cdot p_1)^3
    + c_{13} (k_2\cdot p_1)^4 \nonumber\\
    & \hspace{0.4cm}
    + \text{spurious terms} \,.
\end{align}
The additional tensor integrals can be reduced to a smaller basis of integrals
by the use of integration-by-parts identities, which will be a main topic in the
next chapter. In this case it turns out that only two of these integrals can
actually be considered independent. While the determination of
the integrand basis can be useful, for current state-of-the art problems the
integration-by-parts system presents a considerable computational bottleneck.

\chapter{Loop integration techniques and special functions}
\label{ch:loopints}

\abstract{
In this chapter we introduce methods for evaluating Feynman loop integrals.
We introduce basic methods such as Feynman and Mellin parametrisations, and present
a number of one-loop examples. Working in dimensional regularisation, we discuss
ultraviolet and infrared divergences. 
We then introduce special functions encountered in loop calculations and discuss their properties.
Focusing on their defining differential equations, we show how the symbol method is a useful
tool for keeping track of functional identities.
We then connect back to Feynman integrals by showing how differential equations can be effectively 
used to read off the special functions appearing in them. 
In particular, we discuss residue-based methods that streamline such computations.
}

\section{Introduction to loop integrals}
\label{sec:4.1}

Feynman integrals play a crucial role in quantum field theory, as they often arise when seeking to make perturbative predictions. As such, it is important to understand how to evaluate them or at least have some knowledge of their behaviour.
One example of where Feynman integrals appear is in the study of correlation functions in position space. At the loop level, these integrals depend on the positions of the operators. Another example is the computation of anomalous dimensions of composite local operators, which are of particular interest due to their dependence on the coupling constant.
In momentum space, Feynman integrals also appear in the calculation of scattering amplitudes and other on-shell processes.
Feynman integrals also have applications in other areas, such as gravitational wave physics and  cosmology. While the specific types of integrals may vary in these different scenarios, methods known from particle physics are often applicable. In particular, the differential equation method discussed in this chapter has already proven to be useful in these areas.

One key point is that we can gain insight into these loop integrals by examining the properties of their rational integrands. We have a lot of knowledge about these rational functions and how to analyse them, such as through recursion relations or generalised unitarity.
The challenge is to understand what happens when we perform the $D-$dimensional, or four-dimensional, integration over internal loop momenta. This transforms the rational functions into special functions, such as logarithms, polylogarithms, and their generalisations.
It is interesting to consider how the properties of these special functions come from the integrand and how we can utilise this understanding. 
We will discuss their properties and how to best think about them.
We will explore the connection between Feynman integrals and differential equations. We will learn how to apply the differential equation method and how to use hints from the integrands to simplify the procedure. 

This chapter is organised as follows. Section~\ref{sec:introduction} quickly recalls prerequisites from quantum field theory and establishes conventions. For background material, we refer to standard textbooks, such as~\cite{ch4Peskin:1995ev}. 
For a general and more extensive introduction to Feynman integrals, we refer to the useful book~\cite{ch4Smirnov:2012gma} and the very comprehensive monograph~\cite{ch4Weinzierl:2022eaz}. Both of these references contain many further specialised topics that go beyond the scope of these lecture notes.
In section~\ref{ref:subsection:propertiesofspecialfunctions} we discuss relevant special functions from a differential equation viewpoint that facilitates seeing the connection to Feynman integrals.
In section~\ref{sec:DEFeynmanintegrals} we then discuss the differential equations method for Feynman integrals.
This chapter is complementary to the lecture notes~\cite{ch4Henn:2014qga}.

 We cover the following topics: conventions, Feynman parametrisation, Mellin-Barnes representation.
 We introduce two one-loop examples of Feynman integrals that will be useful throughout this chapter: the two-dimensional massive bubble integral, and the four-dimensional massless box integrals, that are relevant to the scattering processes discussed in the rest of these lecture notes.

 \section{Conventions and basic methods} 
\label{sec:introduction}

 \subsection{Conventions for Minkowski-space integrals}
 \label{sec:ConventionsCh4}

Unless otherwise stated, integrals are defined in Minkowski
space with ``mostly-minus'' metric, i.e.\ $\eta_{\mu\nu}=\text{diag}(+,-,-,-)$ in four dimensions.
When discussing quantities in general dimension $D$, likewise we take the metric to be $\eta_{\mu\nu}=\text{diag}(+,-,\ldots,-,-)$.
Up to overall factors, the  momentum-space Feynman propagator in $D$ dimensions for a particle of mass $m$ and with momentum $p$ reads 
\begin{align}
\frac{\i}{p^2 -m^2 + \i 0}\,.
\end{align}
Here the Feynman prescription ``$\i 0$'' stands for a small, positive imaginary part that moves the poles of the propagator off the real axis. It is important for causality.

\smallskip

{\it Kinematics}. Momentum-space Feynman integrals depend on external momenta and other parameters such as masses of particles. 
The external momenta are usually denoted by $p_{i}$, $i=1,\ldots, n$. The Feynman integrand depends on these momenta, and in addition on loop momenta, that are integrated over. The result of the integration depends on the external momenta via Lorentz invariants, such as $p_i \cdot p_j = \eta_{\mu \nu} p^{\mu}_{i} p^{\nu}_{j}$, where $\eta_{\mu \nu}$ is the metric tensor, for example.

\smallskip

At loop level, one encounters integrals over loop momenta, with the
integrand being given by propagators and possibly numerator 
factors. Let us begin by discussing the loop integrals with trivial
numerators first.

In order to explain how to define integrals in $D$-dimensional Minkowski space,
we begin with the following example:
\begin{align}\label{example_propagator}
 \int \frac{\d^{D}k}{\i \pi^{D/2}}  \frac{1}{(-k^2 + m^2- \i 0)^a} =  \int \frac{\d^{D-1}\vec{k} }{\i \pi^{D/2}}  \int  \frac{\d k_{0}}{(-k_{0}^2 + \vec{k}^2 + m^2-\i 0)^a}    \,,
\end{align}
where $k = (k_0, \vec{k})$.
We will see presently what range of the parameters $D$ and $a$ is allowed for the integral to converge.
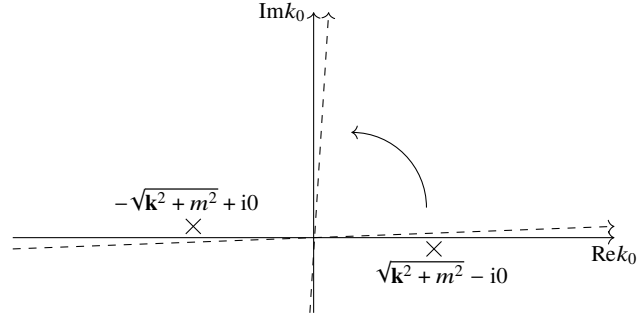
\begin{figure}[t]
\centering
\begin{tikzpicture}[baseline={(current bounding box.center)}, scale =1.0]
  \draw[->] (-4,0) -- (4,0) node[below] {${\rm Re}{k_0}$};
  \draw[->] (0,-1) -- (0,3) node[left] {${\rm Im}{k_0}$};
    \draw[dashed,->] (-4,-0.15) -- (4,+0.15);
    \draw[dashed,->] (-0.05,-1.0) -- (+0.2,3);
    \draw[] (1.5,0.1-0.15) -- (1.7,-0.1-0.15);
    \draw[] (1.5,-0.1-0.15) -- (1.7,0.1-0.15) node[below=0.14cm] {$\sqrt{{\bf k}^2+m^2}- \i 0$};
    \draw[] (-1.5,0.1+0.15) -- (-1.7,-0.1+0.15);
    \draw[] (-1.5,-0.1+0.15) -- (-1.7,0.1+0.15) node[above] {$-\sqrt{{\bf k}^2+m^2}+ \i 0$};
    \draw[->] (1.5,0.4) arc (0:90:1cm);
      \end{tikzpicture}
   \caption{Wick rotation: the $k_{0}$ integration contour is rotated from being parallel to the horizontal axis, to being parallel to the vertical axis, avoiding the propagator poles (which have a small imaginary part due to the Feynman $\i 0$ prescription).
} 
\label{fig:feyn0}
\end{figure}
Consider the integration over $k_{0}$. We see that there are two poles in the complex $k_{0}$ plane, at $k_{0}^\pm = \pm \sqrt{ \vec{k}^2+m^2 } \mp \i 0$.\footnote{Note that $\i 0$ is understood as a positive infinitesimal quantity. 
For this reason, we have absorbed a positive factor of 
$2\sqrt{ \vec{k}^2+m^2}$ into $\i 0$, and we have dropped terms of order $(\i 0)^{2}$.} We can rotate the contour of integration for $k_{0}$ in the complex plane (Wick rotation) so that the integration contour becomes  parallel to the imaginary axis, $k_{0} = \i k_{0,E}$.  This is done in a way that avoids crossing the propagator poles, see Fig.~\ref{fig:feyn0}.
Defining a Euclidean $D$-dimensional vector $k_E = (k_{0,E},\vec{k})$ we
arrive at
\begin{align}
 \int \frac{\d^{D}k_{E}}{ \pi^{D/2}}  \frac{1}{\bigl(k_{E}^2 + m^2-\i 0 \bigr)^a}  \,.
\end{align}
This is now in the form of a Euclidean-space integral, and we could drop the $\i 0$ prescription.
For integer $D$, this integral can be carried out using the following three steps.
First, we write the propagator in the Schwinger parametrisation, 
\begin{align}\label{schwinger}
\frac{1}{x^a} = \frac{1}{\Gamma(a)} \int_0^{\infty} \d\alpha\, \alpha^{a-1} \mathrm{e}^{-\alpha x}\,.
\end{align}
This formula can also be interpreted as the definition of the $\Gamma$ function, which we will encounter frequently in the context of Feynman integrals.
Note that the RHS of \eqn{schwinger} is well-defined for $a>0$. 
Second, we use Gaussian integral formula,
\begin{align}
\label{eq:GaussVolume}
\int_{-\infty}^{\infty} \d y \,\mathrm{e}^{-A y^2} = \sqrt{ \frac{{\pi}}{{A}}} \,,
\end{align}
to carry out the $D$-dimensional loop integration over $k$ (assuming integer $D$).
Third, we use \eqn{schwinger} again, to obtain
\begin{align}\label{single_propagator}
 \int \frac{\d^{D}k}{\i \pi^{D/2}}  \frac{1}{(-k^2 + m^2-\i 0)^a} = \frac{\Gamma(a-D/2)}{\Gamma(a)} \frac{1}{(m^2-\i 0)^{a-D/2}} \,.
\end{align}
Note that the dependence on $m^2$ in \eqn{single_propagator} could have been determined in advance by dimensional analysis.
Another simple consistency check can be performed by differentiating w.r.t.~$m^2$, which gives a recursion relation in $a$.

\begin{svgraybox}
{\bf Useful convention choices.} 
We follow the conventions of~\cite{ch4Smirnov:2012gma}, which helps to sort out many trivial factors of $-1, \i,$ and $\pi$.
\begin{itemize}
\item Choice of loop integration measure. Our choice of ${\d^{D}k}/({\i \pi^{D/2}})$ has the following desirable features. Firstly,
 the factor of $\i$ disappears after Wick rotation, and secondly the factors of $\pi$ compensate for the ``volume factor'' from the $D$-dimensional Gaussian integration~\eqref{eq:GaussVolume}. 
Experience shows that our definition of loop measure is natural from the viewpoint of transcendental weight properties to be discussed later, and in particular the occurrences of $\pi$ that appear after integration have a less trivial origin. 

\item Choice of ``effective coupling''. Note that the above choice of measure differs from the factor of $(2 \pi)^{-D}$ that Feynman rules give per loop, so we recommend splitting this factor up when defining the loop expansion. Often one organises the perturbative expansion in terms of an ``effective'' coupling, such as $g_{\rm YM}^2/\pi^2$ in four-dimensional Yang-Mills theory, to absorb factors of $\pi$. In QCD, $\alpha_s = g_{\rm QCD}^2/(4 \pi)$ is commonly used.

\item Choice of propagator factors. We included a minus sign in the propagator on the LHS of~\eqn{single_propagator}. This avoids minus signs on the RHS, and has the effect that, for certain integrals, the RHS is positive definite for certain values of the parameters.  Equation~\eqref{single_propagator} is a case in point. 
\end{itemize}
\end{svgraybox}

When dealing with massless particles, we may also encounter the situation where we need to evaluate the RHS of \eqn{single_propagator} for $m=0$. In this case, the only answer consistent with scaling is zero (for $a -D/2\neq 0$). We therefore set all scaleless integrals in dimensional regularisation to zero.

\subsection{Divergences and dimensional regularisation}
\label{sec:DimReg}
The derivation of \eqn{single_propagator} assumed integer $D$ (and $a$).
Moreover, when considering the convergence conditions for the different computational steps, we find the conditions $a>0$ and $a-D/2>0$.
We can also see this by inspecting the arguments of the $\Gamma$ functions in the final formula~\eqref{single_propagator}.
So, for example, the integral is well-defined for $a=3, D=4$.
It will be important in the following to extend the range of validity to non-integer values of $a$ and $D$, and beyond the range indicated by the inequalities.
But what is meant by integration for fractional dimension $D$?
Since we know the answer only for integer $D$, the analytic continuation is not unique. 
Therefore we need to make a choice. 
We can do so by taking the RHS of \eqn{single_propagator} as the {\it definition} for the integral in $D$ dimensions.
As we will see, all other, more complicated, integrals can be related to this one.

One of the main motivations for defining integrals for non-integer $D$ is that in quantum field theory one frequently encounters divergences.
Ultraviolet (UV) divergences are well known from textbooks. 
They are related to renormalisation of wavefunctions, masses and the coupling in QFT, and as such play an important role in making the theory well-defined. Beyond that, they can also lead to coupling-dependent scaling dimensions of operators in QFT, which are relevant for example in strong interaction physics, for example, or in describing critical phenomena in condensed matter physics. 
While in principle ultraviolet divergences could be dealt with by introducing certain cutoffs, it is both practically and conceptually very convenient to regularise them dimensionally, i.e.\ by setting $D=4-2\eps$, for $\eps>0$  (see the discussion on power counting in chapter~\ref{ch:loopamps}), and by considering the Laurent expansion as $\eps$ is small.

Another type of divergences are infrared (IR) ones. These can occur when on-shell processes involving massless particles are considered. 
One way of thinking about this is to start from UV-finite momentum-space correlation functions in general kinematics, and then to specialise them to 
on-shell kinematics, for example by setting $p_i^2=0$ in the case of external massless particles. 
In general, this leads to a new type of divergence. The most common case corresponds to the following regions of loop momenta: \emph{soft} (all components of the loop momentum are small) and \emph{collinear} (a loop momentum becomes collinear to an external on-shell momentum).
The behaviour of loop integrands in these configurations is closely related to the properties of tree-level amplitudes discussed in section~\ref{sec:2.1}.
Such infrared divergences can also be treated within dimensional regularisation, but with with $\eps<0$.\footnote{In practice, one may have situations where infrared and ultraviolet divergences are present at the same time. Thanks to analytic continuation, these divergences can be treated consistently within dimensional regularisation.}

 \subsection{Statement of the general problem}
\label{subsec:genproblem}

The main goal is the computation of Feynman integrals, represented by the function $F$, which depend on various parameters such as momenta $p_i$ and masses $m_j$, and on the number of space-time dimensions $D$:
 \begin{align}
F(p_{i}; m_{j}; D ) = \int \d^{D}{k_1} \ldots \, \d^{D}{k_L} \, I(p_{i}; k_{j}; m_{k}) \,. 
\end{align}
 These integrals are defined in $D$ dimensions, where $D=4-2 \eps$ in dimensional regularisation. 
 The method we discuss can be applied to a range of theories and models, though the complexity of the result increases with the number of parameters considered.

As an example, consider a scattering process involving incoming and outgoing particles, for which we want to compute the corresponding Feynman integrals. To approach this problem, we will start with special functions that are known to appear in certain calculations and then generalise from there. For one-loop calculations in four dimensions, it has been observed that apart from logarithms, only a class of functions called dilogarithms are needed. We will discuss the latter in more detail in section~\ref{ref:subsection:propertiesofspecialfunctions}.
Consider a Feynman integral $F$ that depends on $D=4-2 \epsilon$ and has a small $\eps$ expansion 
\begin{align}
F(D) = \sum_{j = j_{0}}^{j_{\rm max}} \eps^{j} F^{(j)} + {\mathcal{O}}\bigl(\eps^{j_{\rm max}+1} \bigr) \,,
\end{align}
where we omitted the dependence on the kinematics.
Since we are ultimately interested in finite results for four-dimensional observables, we can typically truncate the expansion at a certain order $j_{\rm max}$ and discard the higher order in $\eps$ terms. 

For example, in the case of one-loop amplitudes, the leading term is a double pole ($j_{0}=-2$), and one might neglect evanescent terms---that is, terms which vanish in $D=4$ dimensions---by setting $j_{\rm max}=0$. 
In this case, it is known that only logarithms and dilogarithms are needed to express the answer.

\begin{figure}[t]
\centering
\includegraphics[width=0.5\textwidth]{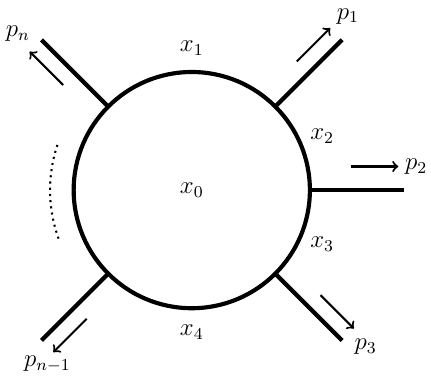}
   \caption{Generic one-loop $n$-point Feynman diagram and dual coordinates $x$. The latter denote the different dual regions. 
   }
\label{fig:feyn1dualcoordinates}
\end{figure}

Let us consider a generic one-loop scalar $n$-point Feynman integral, as in chapter~\ref{ch:loopamps}, 
%
\begin{align}
\begin{aligned}
F^{1-{\rm loop}}_{n} = & \int \frac{\d^{D}k}{\i \pi^{D/2}} \frac{1}{\bigl[-k^2+m_1^2\bigr]^{a_1}}  \\
& \times \frac{1}{ \bigl[- (k+p_1 )^2 + m_2^2\bigr]^{a_2} \ldots \bigl[- (k+p_1 + \ldots + p_{n-1})^2 + m_n^2\bigr]^{a_n} }\,,
\end{aligned}
\end{align}
see fig.~\ref{fig:onelooppic}, where the external momenta $p_{i}$ may or may not satisfy on-shell conditions.
It is convenient to introduce \emph{dual}, or \emph{region coordinates} $x_i$. 
Each dual coordinate labels one of the regions that the Feynman diagram separates the plane into, as in fig.~\ref{fig:feyn1dualcoordinates}.\footnote{For this purpose, we view the external legs as extending to infinity.} The momentum flowing in each of the edges of the graph is then given by the difference of the coordinates of the adjacent dual regions,
\begin{align} \label{eq:dual_coordinates}
k = x_{1} - x_{0}\,, \quad p_{j} = x_{j+1} - x_{j} \,,
\end{align}
with the identification $x_{j+n} \equiv x_{j}$. 
Then the integral above takes the simple form
\begin{align}\label{oneloop_dual}
F^{1-{\rm loop}}_{n} = \int \frac{\d^{D}x_{0}}{\i \pi^{D/2}} \prod_{j=1}^{n} \frac{1}{ \bigl(- x_{0j}^2 + m_{j}^2 \bigr)^{a_{j} } } \,,
\end{align}
where $x_{ij} \coloneqq x_{i} - x_{j}$.
Translation invariance in the dual space corresponds to the freedom of redefining the loop integration variables $k$ in the initial
integral. Momentum conservation implies that the external momenta form a closed polygon in dual space, with the vertices being the dual coordinates $x_{i}$ and the edges being the momenta $p_{i}$.

 \subsection{Feynman parametrisation}
\label{subsec:Feynmanparam}

It is often convenient to exchange the integration over space-time for parametric integrals. 
Formulae for doing so for general Feynman integrals are given in ref.~\cite{ch4Smirnov:2012gma}. 

The idea of Feynman parametrisation is to exchange the space-time integration for a certain number of auxiliary integrations (over Feynman parameters).
This can be done systematically. 
Let us show how it is done explicitly at one loop. 
The starting point is the general one-loop integral given in \eqn{oneloop_dual}.
The goal is to relate this to the case in \eqn{single_propagator},
by introducing auxiliary integration parameters. 

This method is closely related to the Schwinger formula~\eqref{schwinger} encountered earlier.
The Feynman trick is based on the following identity:
\begin{align}\label{Schwinger2}
\frac{1}{X_1^{a_1} X_2^{a_2 }} = \frac{\Gamma(a_1 + a_2 )}{\Gamma(a_1 ) \Gamma(a_2 )}  \int_0^\infty \frac{\d\alpha_1 \d\alpha_2}{{\rm GL}(1)} \frac{ {\alpha_1}^{a_1-1} {\alpha_2}^{a_2-1}}{ (\alpha_1 X_1 + \alpha_2 X_2 )^{a_1 + a_2} } \,.
\end{align}
The RHS of this formula requires some explanation. The form on the RHS is invariant under general linear (GL) transformations, i.e.\ arbitrary rescalings of $\alpha_1$ and $\alpha_2$. The measure ${\d\alpha_1 \d\alpha_2}/{{\rm GL}(1)}$ means that one ``mods out'' by such transformations, so that in effect the integration is only one-dimensional. For example, one could fix $\alpha_2 = 1$, upon which the integration measure becomes $\int_0^\infty \d\alpha_1$. Another common choice is to set $\alpha_1 + \alpha_2 = 1$.  In other words, modding out by the ${\rm GL}(1)$ transformations amounts to inserting a Dirac $\delta$ function---e.g.\ $\delta(\alpha_1+\alpha_2-1)$ or $\delta(\alpha_2-1)$---under the integral sign in \eqn{Schwinger2}.

Feynman integrals typically have many propagators (corresponding to the number of edges), so we need 
a generalisation of \eqn{Schwinger2} to an arbitrary number $n$ of denominator factors:
\begin{align}\label{Feynman_general}
\frac{1}{ \prod_{i=1}^{n}  X_i^{a_i} }  = \frac{\Gamma(a_1 + \ldots + a_n) }{\prod_{i=1}^{n} \Gamma(a_i )}  \int_0^\infty \frac{ \prod_{i=1}^{n}  \d\alpha_i  }{{\rm GL}(1)}    
\frac{  \prod_{i=1}^{n}    \alpha_i^{a_i-1}  } { (\sum_{i=1}^{n} \alpha_i X_i)^{a_1 + \ldots + a_n } }\,.
\end{align}
This can be shown by mathematical induction.

Applying \eqn{Feynman_general} to the one-loop formula~\eqref{oneloop_dual} yields an integral over a single factor in the integrand.
By performing a change of variables in the integration variables $k$, this can be brought into the form of \eqn{single_propagator},
and hence the space-time integration can be performed (see exercise~\ref{Ex:MasslessBubble} for an example).
The result is
\begin{align}\label{oneloop_master}
F_{n} =  
\frac{\Gamma(a-D/2)}{\prod_{i=1}^{n} \Gamma(a_{i})}  
\int_{0}^{\infty}  \frac{   \prod_{i=1}^{n} 
 \d\alpha_{i} \alpha_{i}^{a_{i}-1} }{ {\rm GL}(1)} 
\frac{U^{a-D} }{ (V + U \sum_{i=1}^{n} m_{i}^2 \alpha_i  -\i 0)^{a-D/2}} \,,
\end{align}
where $U = \sum_{i=1}^{n} \alpha_{i}$ and $V = \sum_{i<j} \alpha_i \alpha_j (-x_{ij}^2)$.
These polynomials, called \emph{Symanzik polynomials}, have a graph-theoretical interpretation, see e.g.~\cite{ch4Smirnov:2012gma}. 
Consider the graph corresponding to the propagators forming the loop, where an edge corresponding to a denominator $X_i$ has label $\alpha_i$. 
Then, consider all ways of removing a minimal number of lines so that the graph becomes a tree. To each such term, associate the product of $\alpha_i$ factors of the removed factors. Summing over all such terms gives $U$. Similarly, to define $V$, one considers all ways of removing factors to obtain two trees, and again takes the products of the $\alpha_i$, but this time weighted by (minus) the momentum squared flowing through these lines, which yields $-\alpha_i \alpha_j x_{ij}^2$ at one loop.

Depending on the situation, different choices of fixing the  ${\rm GL}(1)$ invariance in \eqn{oneloop_master} may be particularly convenient. One may fix $U=1$, for example, or alternatively one may set one of the Feynman parameters $\alpha_i$ to $1$.

\begin{figure}[h]
  \begin{subfigure}{0.5\textwidth}
    \centering
    \includegraphics[width=0.7\textwidth]{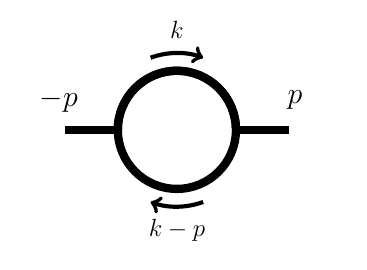}
     \caption{}
    \label{fig:feynbubbleandtriangle-bubble}
    \quad\quad\quad\quad\quad\quad
  \end{subfigure}
  \begin{subfigure}{0.4\textwidth}
    \centering
    \vspace*{-4mm}
    \hspace*{-6mm}\includegraphics[width=0.85\textwidth]{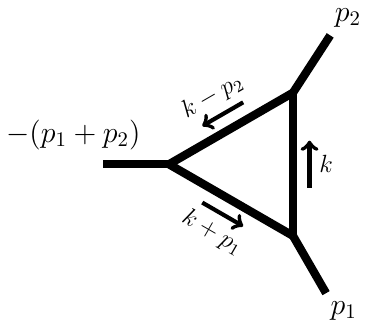}
    \caption{}
    \label{fig:feynbubbleandtriangle-triangle}
  \end{subfigure}
  \caption{Bubble and triangle Feynman integrals discussed in the main text. The arrows denote the direction of the momenta.}
  \label{fig:feynbubbleandtriangle}
\end{figure}

\newpage

\begin{exer}{The massless bubble integral}
\label{Ex:MasslessBubble}
Consider the bubble integral, cf. fig.~\ref{fig:feynbubbleandtriangle-bubble}, with massless propagators but generic propagator powers:
\begin{align} \label{eq:massless_bubble}
F_{2}\left( a_1,a_2 ; D\right) = \int \frac{\d^D k}{\i \pi^{\frac{D}{2}}} \frac{1}{\left(-k^2  - \i 0\right)^{a_1} \left(-(k-p)^2  - \i 0 \right)^{a_2} } \,.
\end{align}

\begin{enumerate}[a)]

\item Use the identity~\eqref{Schwinger2} to write down the Feynman parameterisation. Verify that it matches the general formula~\eqref{oneloop_master}, and read off the Symanzik polynomials.

\smallskip
\item Show that the integral evaluates to
\begin{align} \label{eq:Ibubblechapter4}
F_{2}\left( a_1,a_2 ; D \right) = B\left(a_1, a_2 \right) \, \bigl(-p^2 - \i 0 \bigr)^{\frac{D}{2}-a_1-a_2} \,,
\end{align}
with
\begin{align} \label{eq:Bbubble}
B\left(a_1,a_2 ; D\right) = \frac{\Gamma\left(a_1+a_2-\frac{D}{2}\right) \Gamma\left(\frac{D}{2}-a_1\right) \Gamma\left(\frac{D}{2}-a_2\right)}{\Gamma\left(a_1\right) \Gamma\left(a_2\right) \Gamma\left(D-a_1-a_2\right)} \,.
\end{align}

\end{enumerate}
For the solution see \hyperref[Sol:MasslessBubble]{chapter~5}.
\end{exer}

\begin{example}{Example: Infrared-divergent massless triangle integral}

As an example of the one-loop Feynman parameter formula~\eqref{oneloop_master}, let us consider the massless on-shell triangle diagram of fig.~ \ref{fig:feynbubbleandtriangle-triangle},
\begin{align}\label{F3examplemomentumspace}
F_3 =  \int \frac{\d^{D}k}{\i \pi^{D/2}} \frac{1}{[-k^2 -\i 0] [-(k+p_{1})^2-\i 0] [- (k-p_{2})^2 -\i 0]} \,.
\end{align}
In order to use \eqn{oneloop_master}, we need to match the kinematics to the general notation used there.
Note that \eqn{F3examplemomentumspace} is a special case of \eqn{oneloop_dual} with $n=3$ and the following choices:
zero particle masses $m_1=m_2=m_3=0$, and unit propagator exponents $a_1 = a_2= a_3 =1$.
Moreover, we consider the massless on-shell conditions $p_{1}^2= p_{2}^2 = 0$, 
so that the integral depends on the dimensionful variable $s = (p_{1}+p_{2})^2$, and on $D$.
Translating this to dual coordinates, we have $x_{12}^2= 0, x_{23}^2=0, x_{13}^2 =s$.
\begin{exer}{Feynman parametrisation}
\label{Ex:4.FeynPar}
Draw the triangle diagram in \eqn{F3examplemomentumspace} using both the momentum-space and the dual-space labelling, and verify the above identification of variables.
Use this to write down the Feynman parametrisation for $F_{3}$.
For the solution see \hyperref[Sol:4.FeynPar]{chapter~5}.
\end{exer}

Having established this notation, we can readily employ our main one-loop formula~\ref{oneloop_master}.
Setting $D=4-2 \eps$, it gives
\begin{align}\label{F3triangleFeynman}
F_3 =  \int_0^{\infty} \frac{ \d\alpha_1 \d\alpha_2 \d\alpha_3}{{\rm GL}(1)} \frac{\Gamma(1+\eps)}{(-s \, \alpha_1 \alpha_3 -\i 0)^{1+\eps} (\alpha_1 + \alpha_2 + \alpha_3 )^{1-2\eps} } \, .
\end{align}
For simplicity, let us consider the so-called Euclidean kinematic region, where $s<0$. In this case, we see that the denominator on the RHS is positive, and therefore the $\i 0$ prescription is not needed. 
Later we may be interested in analytically continuing the integral to other kinematic regions. Noticing that the $\alpha$-parameter polynomial multiplying $-s$ is positive, we can conveniently record the information on the correct analytic continuation prescription  by giving a small imaginary part to $s$: $s \to s + \i 0$.
In the present case, the dependence on $s$ is actually trivial: it is dictated by the overall dimensionality of the integral.
This implies that $F_3$ depends on $s$ as $(-s-\i 0)^{-1-\eps}$.

Let us comment on the convergence properties of \eqn{F3triangleFeynman}.
The expression is valid for $\eps<0$. 
This is consistent with our expectations, since this integral is UV-finite (see the power counting in \eqn{eq:UVpowercounting}) but has IR divergences.
The integral would actually be finite for $p_1^2 \neq 0, p_2^2 \neq 0$.
For on-shell kinematics $p_1^2=p_2^2=0$, one can see there are problematic regions of loop momentum in \eqn{F3examplemomentumspace}
that lead to divergences when integrating in $D=4$ dimensions. 
The soft region occurs when all components of $k$ in \eqn{F3examplemomentumspace} are small. 
On top of this, there are two collinear regions, where $k \sim p_1$ and $k\sim p_2$, respectively. 
One may convince oneself by power counting (see e.g.~\cite{ch4Agarwal:2021ais}  
for more details) that these regions lead to logarithmic divergences ($1/\eps$ in dimensional regularisation).
Moreover, each collinear region overlaps with the soft region, so that the divergences can appear simultaneously. 
We therefore expect the leading term of $F_3$ as $\eps \to 0$ to be a double pole $1/\eps^2$.

Let us now verify this explicitly.
In order to carry out the $\alpha$ integrals we introduce the following useful formula,
\begin{align}\label{alpha_integrals}
\int_0^\infty \frac{\prod_{i=1}^{n} \d\alpha_{i}\,
\alpha_{i}^{b_i-1}}{{\rm GL}(1)} \biggl(\sum_{i=1}^{n} \alpha_i \biggr)^{-b} = \frac{ \prod_{i=1}^{n} \Gamma(b_i )} {\Gamma( b)} \,,\quad {\rm with} \quad b=\sum_{i=1}^{n} b_i\,.
\end{align}
Applying this to \eqn{F3triangleFeynman}, for $b_1=-\eps, b_2 = 1, b_3=-\eps$,  we find
\begin{align}
F_{3} = (-s-\i 0)^{-1-\eps} \frac{\Gamma(1+\eps) \Gamma^2(-\eps)}{\Gamma(1-2 \eps)} \,.
\end{align}
We wish to expand this formula for small $\eps$. To do so, we need to familiarise us with the properties of the $\Gamma$ function.

\begin{svgraybox}
{\bf Important properties of the $\Gamma$ function.}
In the calculation above we encountered a first special function, the $\Gamma$ function.
It is defined as 
\begin{align}
\Gamma(x) = \int_0^\infty \d t \, t^{x-1} \mathrm{e}^{-t} \,.
\end{align}
This formula converges for $x >0$. To define $\Gamma(x)$ for complex $x$, one uses analytic continuation.
Here we collect several important properties of the $\Gamma$ function.
It satisfies the recurrence
\begin{align} \label{eq:gammarecurrence}
\Gamma(x+1) = x \, \Gamma(x) \,.
\end{align}
It has the expansion
\begin{align} \label{eq:logGammaTaylor}
\log \Gamma(1+x) = - \gamma_{\text{E}} \, x + \sum_{n=2}^{\infty} \frac{(-1)^n \, x^n}{n} \zeta_n \,,
\end{align}
for $|x|<1$.
Here, Euler's constant is
\begin{align} \label{eq:EulerConstant}
\gamma_{\text{E}} = - \Gamma'(1) = 0.57721\ldots \,,
\end{align}
and Riemann's zeta constants appear,
\begin{align} \label{eq:zetan}
\zeta_n = \sum_{k=1}^{\infty} \frac{1}{k^n}  \,, \qquad \forall \, n\ge 2 \,.
\end{align}
For even $n$, these evaluate to powers of $\pi$, e.g.\ 
$\zeta_2 = \pi^2/6$ and $\zeta_{4} = \pi^4/90$.
\end{svgraybox}
Using the expansion~\eqref{eq:logGammaTaylor}, as well as \eqn{eq:gammarecurrence}, we find
\begin{align}
\mathrm{e}^{\eps \gamma_{\rm E}} F_{3} =  (-s)^{-1-\eps} \left[ \frac{1}{\eps^2} - \frac{\pi^2}{12} - \frac{7}{3} \zeta_3 \eps - \frac{47 \pi^4}{1440} \eps^2 + \cO(\eps^3)\right]\,.
\end{align}
Here we multiplied $F_{3}$ by a factor of $\mathrm{e}^{\eps \gamma_{\rm E}}$ (in general, one takes one
such factor per loop order),
in order to avoid the explicit appearance of $\gamma_{\rm E}$ in the expansion.
\end{example}

\begin{exer}{Taylor series of the Log-Gamma function}
\label{Ex:TaylorLogGamma}
In this guided exercise we prove the Taylor series of the Log-Gamma function in \eqn{eq:logGammaTaylor}. 
The Taylor series of $\log \Gamma(1+x)$ around $x=0$ is given by
\begin{align} \label{eq:loggammataylorgen}
\log \Gamma(1+x) = \sum_{n=1}^{\infty} \frac{x^n}{n!} \left( \frac{\d^n}{\d x^n} \log \Gamma(1+x) \right)\biggl|_{x=0} \,.
\end{align}
The first-order derivative by definition gives Euler's constant $\euler$~\eqref{eq:EulerConstant}.
In order to compute the higher-order derivatives, we derive a series representation for the \emph{digamma function} $\psi(x)$, i.e.\ the logarithmic derivative of the gamma function,
\begin{align} \label{eq:psidef}
\psi(x) \coloneqq \frac{\d \log \Gamma(x)}{\d x} = \frac{\Gamma'(x)}{\Gamma(x)} \,.
\end{align}

\begin{enumerate}[a)]

\item Prove the following recurrence relation for the digamma function,
\begin{align} \label{eq:psixn}
\psi(x+n) = \psi(x) + \sum_{k=0}^{n-1} \frac{1}{x+k} \,, \qquad n \in \mathbb{N}\,.
\end{align}

\item Prove the following series representation of the digamma function,
\begin{align} \label{eq:psiseries}
\psi(x) = - \gamma_{\text{E}} - \sum_{k=0}^{\infty} \left(\frac{1}{x+k}-\frac{1}{1+k} \right) \,.
\end{align}
Hint: study the limit of $\psi(x+n)-\psi(1+n)$ for $n\to \infty$ using Stirling's formula,
\begin{align} \label{eq:stirling}
\Gamma(x+1) = \sqrt{2 \pi x} \, x^x \, \text{e}^{-x} \left[1 + \mathcal{O}\bigl(1/x\bigr) \right] \,.
\end{align}

\item Use \eqn{eq:psiseries} to prove the Taylor series in \eqn{eq:logGammaTaylor}.

\end{enumerate}
For the solution see \hyperref[Sol:TaylorLogGamma]{chapter~5}.
\end{exer}

\begin{example}{Example: Ultraviolet divergent bubble integral}

An important example throughout this chapter will be the one-loop massive bubble integral.
The integral is defined as
\begin{align}
F_{2}(s,m^2;D) = \int \frac{\d^{D}k}{\i \pi^{D/2}} \frac{1}{(-k^2+m^2 - \i 0) [-(k-p)^2+m^2- \i 0]} \,.
\label{defbubble}
\end{align}
The integrated function depends on the external momentum $p$ via the Lorentz invariant $s=p^2$.
This is a special case of \eqn{oneloop_dual} with $n=2$, with uniform internal masses $m_1=m_2=m$, with unit propagator powers $a_1=a_2=1$, and with the single kinematic variable $x_{12}^2=p^2$.
In slight abuse of notation, we denote this integral by the same letter $F_2$ as we did for the massless bubble integral above.

Applying \eqn{oneloop_master}, we find
\begin{align}\label{eq:bubbleFeynman} 
F_{2}(s,m^2;D) = \Gamma\left(2-\frac{D}{2}\right)  \, \int_0^\infty \frac{\d \alpha_1 \d\alpha_2}{{\rm GL}(1)} 
\frac{(\alpha_1+\alpha_2)^{2-D} }{[\alpha_1 \alpha_2 (-s) + (\alpha_1 + \alpha_2)^2 m^2 -\i 0]^{2-D/2}  } \,.
\end{align} 
Just as in the triangle example above, we see that the integrand in this formula is positive definite in the Euclidean region $s<0, m>0$, and that we can absorb the $\i 0$ prescription into $s$.

We see that $\Gamma(2-D/2)$ is divergent for $D \to 4$, and requires $D<4$ for convergence. The parameter integral is instead well-defined for $D=4$. Therefore we can compute the integral for $D=4-2\eps$, with $\eps>0$. In the limit $\eps\to0$, we find
\begin{align}\label{BleadingpoleUV}
 F_{2}(s,m^2;D) = 
 \frac{1}{\eps}  
 + 
 {\mathcal{O}}\bigl(\eps^0\bigr)
  \,.
\end{align}
This divergence is of ultraviolet origin. As we discussed in chapter~\ref{ch:loopamps}, we can understand it by doing power counting in the momentum-space representation~\eqref{defbubble}. Consider the integrand for large loop momentum $k$. Switching to radial coordinates, the integration measure becomes $\d^{D}k = r^{D-2} \d r \, \d\Omega$, where $r$ is the radial direction, and $\Omega$ represents the angular integrations. At large $r$, the integrand goes as $\d r/r^{D-4}$. This converges for $D<4$, but leads to a logarithmic divergence at $D=4$. This is exactly what \eqn{BleadingpoleUV} encodes.
With the same power counting, we see that the integral in \eqn{defbubble} is finite in $D=2$. 
\end{example}

\begin{exer}{Finite two-dimensional bubble integral}
\label{Ex:4.2Dbubble}
Starting from the Feynman parametrisation in eq. (\ref{eq:bubbleFeynman}), carry out the remaining integration for $D=2$, for the kinematic region $s<0, m^2>0$, to find
\begin{align}\label{resbubbleexplicit}
F_{2}\bigl(s,m^2;D=2\bigr) =  \frac{2}{s \, \sqrt{1-4 m^2/s}} \log \left(  \frac{\sqrt{1-4 m^2/s}-1}{\sqrt{1-4 m^2/s}+1} \right) \,.
\end{align}
Hint: employ the change of variables $-s/m^2 = (1-x)^2/x$, with $0<x<1$.
For the solution see \hyperref[Sol:4.2Dbubble]{chapter~5}.
\end{exer}
We will also be interested in the dimensionally-regularised version of \eqn{defbubble}, i.e.\ the deformation where $D=2-2 \epsilon$. 
This is interesting for several reasons. Firstly, we have seen in chapter~3 that integrals in dimensions $D$ and $D\pm 2$ are related by certain recurrence relations, see \eqn{eq:dimshift} at one-loop order~\cite{ch4Lee:2012cn}.
Secondly, this integral for $D=2-2\eps$ will serve as a main example for understanding integration techniques in this chapter.

Before closing this section, let us mention the $L$-loop generalisation of \eqn{oneloop_master}.
\begin{svgraybox}
{\bf Feynman parametrisation for a scalar $L$-loop Feynman diagram.}
Consider now an $L$-loop scalar Feynman integral with $n$ denominator factors.
The graph may be planar or non-planar. As before, we label the $i$-th factor (which may have a generic mass $m_i^2$ and is raised to a power $a_i$) by the Feynman parameter $\alpha_i$.
Then, the generalisation of \eqn{oneloop_master} is given by
\begin{align}\label{FeynmanLloopgeneralization}
\frac{\Gamma(a-L D/2)}{\prod_{i=1}^{n} \Gamma(a_{i})}  
\int_{0}^{\infty}  \frac{   \prod_{i=1}^{n} 
 \d\alpha_{i} \, \alpha_{i}^{a_{i}-1} }{ {\rm GL}(1)} 
\frac{U^{a-L D} }{ (V + U \sum_{i=1}^{n} m_{i}^2 \alpha_i  -\i 0)^{a-L D/2}} \,.
\end{align}
Here $a= \sum_i a_i$, and the Symanzik polynomials $U$ and $V$ have the same graph theoretical definition mentioned above.
They have been implemented in various useful computer programs, for example~\cite{ch4Lee:2012cn}. 
\end{svgraybox}

\subsection{Summary}
In this section, we introduced conventions and notations for Feynman integrals.
The integrals are initially defined as space-time integrals, but other representations are also useful.
We showed how Feynman  
representations are obtained.
We discussed a number of sample one-loop integrals, and showed how ultraviolet and infrared divergences
are treated in dimensional regularisation. 
We also saw first examples of special functions appearing in the integrated answers, namely the $\Gamma$ function and the logarithm.
In the next sections, we introduce the Mellin-Barnes method, which will allow us to go beyond the cases treated so far, and see first examples of polylogarithms.
After that we discuss more systematically special functions appearing in Feynman integrals,
and propose a useful way for thinking about them in terms of their defining differential equations.

\section{Mellin-Barnes techniques}
\label{section:mellin}

In the previous section we saw how to derive parameter-integral formulae for Feynman integrals.
For a triangle diagram we derived the complete analytic answer by carrying out the parameter integrals,
and we did the same for the finite two-dimensional massive bubble integral.
In general, it is difficult to carry out the Feynman parameter integrals directly (see however interesting work in this direction, together with powerful algorithms~\cite{ch4Panzer:2014caa}).

Another useful representation 
trades the Feynman parameter integrals
for Mellin-Barnes integrals, as we describe 
presently. 
The resulting Mellin-Barnes representations makes certain properties of the integrals easier to see as compared to the Feynman parameter integrals. In particular, useful algorithms have been developed to resolve singularities in $\eps$ and to provide representations of the terms in the Laurent expansion of the Feynman integrals.
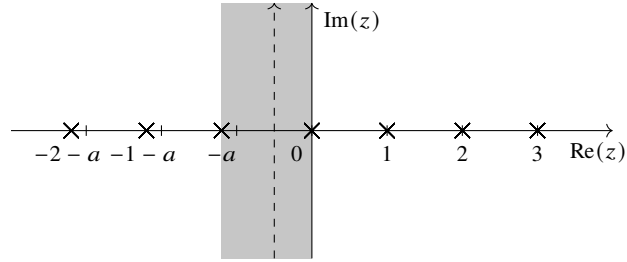
\begin{figure}[t]
\centering
\begin{tikzpicture}[baseline={(current bounding box.center)}]

  \newcommand \aval {1.2};
   
  \fill[gray!45] (-\aval,-1.7) rectangle (0,1.7);
    
  \draw [->] (-4,0) -- (4,0) node [below left, xshift=0.3cm, yshift=-0.05cm]  {$\text{Re}(z)$};
  \draw [->] (0,-1.7) -- (0,1.7) node [below right, xshift=0.05cm] {$\text{Im}(z)$};
  
  \foreach \n in {-3,-2,-1,0,1,2,3}{ 
    \draw (\n,-2pt) -- (\n,2pt); 
  }
    
    \draw[dashed,->] (-\aval/2+0.1, -1.7) -- (-\aval/2+0.1, 1.7);
  
  \draw[mark=x, mark size=4pt] plot (0,0) node[below, xshift=-0.2cm, yshift=-0.1cm] {$0$};
  \draw[mark=x, mark size=4pt] plot (1,0) node[below, yshift=-0.1cm] {$1$};
  \draw[mark=x, mark size=4pt] plot (2,0) node[below, yshift=-0.1cm] {$2$};
  \draw[mark=x, mark size=4pt] plot (3,0) node[below, yshift=-0.1cm] {$3$};

   \draw[mark=x, mark size=4pt] plot (-\aval,0) node[below, yshift=-0.175cm] {$-a$};
   \draw[mark=x, mark size=4pt] plot (-1-\aval,0) node[below, yshift=-0.1cm] {$-1-a \, \,$};
   \draw[mark=x, mark size=4pt] plot (-2-\aval,0) node[below, yshift=-0.1cm] {$-2-a \, \,$};

\end{tikzpicture}
\caption{Integration contour for Mellin-Barnes representation in eq.~(\ref{MB_basic}). The integration contour (dashed line) goes parallel to the vertical axis, with ${\rm Re}(z)=c$, with $-a<c<0$, i.e.\ to the right of the poles of $\Gamma(-z)$, and to the left of the poles of  $\Gamma(z+a)$ (shaded~area).}
\label{fig:mb1}
\end{figure}
The key formula is the following,
\begin{align}\label{MB_basic}
\frac{1}{(x+y)^{a}} = \frac{1}{\Gamma(a)} \, \int_{c - \i \infty}^{c+\i \infty} \frac{\d z}{2 \pi \i} \, \Gamma(-z) \Gamma(z+a) \, x^z y^{-a-z} \,,
\end{align}
where the integration contour is parallel to the imaginary axis, with real part $c$ in the interval $-a<c<0$.
See Fig.~\ref{fig:mb1}. In general, the integration contour is chosen such that the poles of $\Gamma$ functions of the type $\Gamma(z+\ldots )$ lie to its left, and the poles of  $\Gamma(-z+\ldots)$ lie to its right.

One can verify the validity of \eqn{MB_basic} by checking that the series expansions of its LHS and RHS agree.
Let us see this in detail.
Assume $x<y$. Then the LHS of \eqn{MB_basic} has the following series representation,
\begin{align}\label{MB_check}
\frac{1}{(x+y)^{a}} = y^{-a} \sum_{n\ge 0} \, (-1)^{n+1} \frac{ \Gamma(n+a)}{\Gamma(n+1)\Gamma(a)} \left( \frac{x}{y} \right)^n \,,\qquad x<y\,.
\end{align}
Let us see how this arises from the RHS of \eqn{MB_basic}. If $x<y$ we can close the integration contour in \eqn{MB_basic} on the right,
because the contribution from the semicircle at infinity vanishes.
By complex analysis, we get a contribution from (minus) all poles of $\Gamma(-z)$ situated at $z_{n} = n$, with $n = 0, 1, \ldots$.
Taking into account that the corresponding residues are $\text{Res}[\Gamma(-z), \, z=n]=(-1)^{n+1}/n!$ (see exercise~\ref{Ex:ExpGamma}),
one readily reproduces \eqn{MB_check}.
One may verify similarly the validity of \eqn{MB_basic} for $y<x$. In this case, one closes the integration contour on the left.

Equation~\eqref{MB_basic} can be used to factorise expressions, e.g.\ the denominator factors appearing 
in Feynman parametrisation.
Once factorised, \eqn{alpha_integrals} 
allows one to carry out the Feynman parameter integrals. 
In some sense, the Mellin-Barnes representation can therefore be considered
the inverse of the Feynman parametrisation.
Of course, this means that one is just trading one kind of integral representation for another.
However, the Mellin-Barnes representation is very flexible, and has a number of useful features, as
we will see shortly.

\begin{exer}{Laurent expansion of the Gamma function}
\label{Ex:ExpGamma}
The Gamma function $\Gamma(z)$ is holomorphic in the whole complex plane except for the non-positive integers, $z=0,-1,-2,\ldots$, where it has simple poles. 

\begin{enumerate}[a)]
\item Compute the Laurent expansion of $\Gamma(z)$ around $z=0$ up to order $z$,
\begin{align} \label{eq:GammaExp0}
\Gamma(z) = \frac{1}{z} - \gamma_{\text{E}} + \frac{z}{2}\left(\euler^2 + \zeta_2\right) + \mathcal{O}\left(z^2 \right)\,.
\end{align}

\item Using eq.~\eqref{eq:GammaExp0}, show that the Laurent expansion of $\Gamma(z)$ around $z=-n$, with $n\in\mathbb{N}_0$, is given by
\begin{align} \label{eq:GammaExpn}
\Gamma(z) = \frac{(-1)^n}{n!} \left\{ \frac{1}{z+n} + H_n-\euler + \frac{1}{2}(z+n) \left[\left(H_n - \euler\right)^2 + \zeta_2 + H_{n,2} \right] \right\} + \ldots \,,
\end{align}
where the ellipsis denotes terms of order $(z+n)^2$ or higher. Here, $H_{n,r}$ is the $n$-th harmonic number of order $r$,
\begin{align}  \label{eq:harmonicnumbers}
H_{n,r} \coloneqq \sum_{k=1}^n \frac{1}{k^r} \,, \qquad \qquad H_{n} \coloneqq H_{n,1} \,.
\end{align}

\end{enumerate}
For the solution see \hyperref[Sol:ExpGamma]{chapter~5}.
\end{exer}

\subsection{Mellin-Barnes representation of the one-loop box integral}

\begin{figure}[t]
  \centering
  \includegraphics[width=0.4\textwidth]{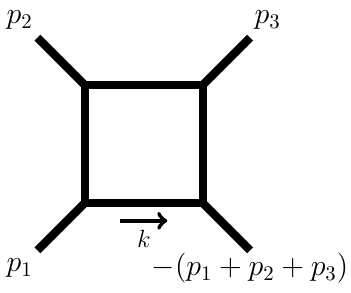}
  \caption{Massless one-loop four-point Feynman integral considered in the main text.}
  \label{fig:mb2}
\end{figure}

Let us apply the above procedure to the massless one-loop box integral,
cf.~fig.~\ref{fig:mb2}:
\begin{align}
F_{4} = \int \frac{\d^{D}k}{\i \pi^{D/2}} \frac{1}{k^2 (k+p_1)^2 (k+p_1 + p_2)^2 (k+p_1 + p_2 + p_3 )^2} \,,
\end{align}
with $D=4-2\eps$.
The $\i 0$ prescription is understood.
The external momenta are taken to be on-shell and massless, i.e.\ $p_i^2=0$.
It is a function of $s=(p_1 + p_2 )^2$, $t= (p_2 + p_3)^2 $, and $\eps$.
Power counting shows that this integral is ultraviolet finite, but it has soft and collinear divergences.
Therefore we expect the leading term to be a double pole in $\eps$, just as for the massless triangle integral computed above.

\begin{figure}[t]
\includegraphics[scale=0.7]{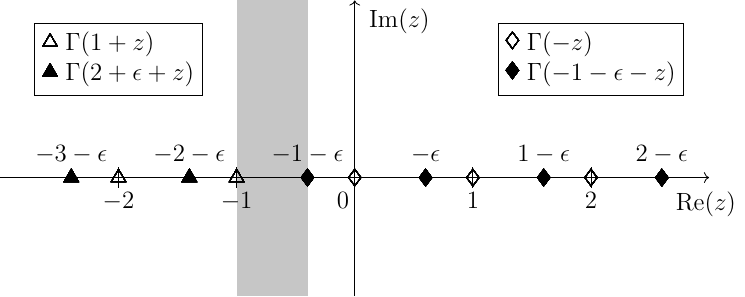}
\centering
\caption{Poles of the $\Gamma$ functions involved in the Mellin-Barnes parameterisation of the one-loop box integral~(\ref{eq:MBboxM}) assuming $-1<\eps<0$. For the integration in eq.~(\ref{MBbox}) to be well defined, the real part of $z$ must lie in the shaded area, between the right-most pole of the $\Gamma$ functions of the type $\Gamma(z+\ldots)$ and the left-most pole of those of the type $\Gamma(-z+\ldots)$.
}
\label{fig:box_MB}
\end{figure}

We start by writing down a Feynman parametrisation, using \eqn{oneloop_master},
\begin{align}\label{eq:F4Feynman}
F_{4} = \int_0^{\infty} \frac{\prod_{i=1}^{4} \d\alpha_{i}}{{\rm GL}(1)} \frac{\Gamma(2+\eps)}{[ \alpha_1 \alpha_3 (-s) + \alpha_2 \alpha_4 (-t)   ]^{2+\eps} \bigl(\sum_{i=1}^{4} \alpha_i\bigr)^{2 \eps}} \,.
\end{align}
Here we absorbed the $\i 0$ prescription into $s$ and $t$. In the following we take the kinematics to be in the Euclidean region $s<0,t<0$.
We can factorise the first factor of the integrand of \eqn{eq:F4Feynman} at the cost of introducing one Mellin-Barnes parameter integral, using \eqn{MB_basic}.
Then, the integral over the $\alpha$ parameters can be done with the help of \eqn{alpha_integrals}.
We find
\begin{align}\label{MBbox}
F_{4} = \int \frac{\d z}{2 \pi \i} \, M(s,t,z;\eps)\,,
\end{align}
with
\begin{align} \label{eq:MBboxM}
M(s,t,z;\eps) =   (-s)^z (-t)^{-2-\eps-z}  \Gamma(-z) \Gamma(2+\eps+z) \frac{\Gamma^2(1+z) \Gamma^2(-1-\eps-z)}{\Gamma(-2\eps)} \,.
\end{align}
For the integrations leading to this expression to be well defined, 
the real part of the arguments of each $\Gamma$ function must be positive. The pole structure of the relevant $\Gamma$ functions is shown in figure~\ref{fig:box_MB}.
We see that this implies in particular that $\eps<0$, 
which is expected since the integral is infrared divergent.
We can choose~e.g.
\begin{align}
 \Re(z) = -\frac{3}{4} \,, \quad  \eps = -\frac{1}{2}\,.
\end{align}
We will now explain how to analytically continue to $\eps \to 0$.

\subsection{Resolution of singularities in $\eps$}

Here we follow ref.~\cite{ch4Czakon:2005rk} and references therein.
We saw that the integral in \eqn{MBbox} is ill-defined for $\eps = 0$.
This can be traced back to the presence of the Gamma functions
$\Gamma(1+z)$ and $\Gamma(-1-\eps-z)$. The contour for the $z$ integration has to pass between the poles of these Gamma functions, which is only possible for $\eps<0$. In other words, as $\eps$ goes to $0$, the shaded area in fig.~\ref{fig:box_MB} is pinched between the right-most pole of $\Gamma(1+z)$ and the left-most pole of $\Gamma(-1-\eps-z)$. Before we can take the limit $\eps \to 0$, we must therefore deform the integration contour for $z$, so that it does not become pinched when taking the limit.
Let us deform the contour to the right.
This leads to a contribution of the residue at $z=-1-\eps$. In other words, 
\begin{align}
F_{4} = - \oint_{z=-1-\eps}  \frac{\d z}{2 \pi \i} \,  M(s,t,z;\eps) +  \int_{  \Re(z) = c }  \frac{\d z}{2 \pi \i} \, M(s,t,z;\eps) \,,
\end{align}
where $-1-\eps < c < 0$.
The value of this residue is
\begin{align} \label{eq:MBboxA}
A = \frac{ \left(-s\right)^{-{\eps}} }{s \, t} 
\frac{ \Gamma^2(-{\eps}) \Gamma ({\eps}+1) }{ \Gamma (-2
   {\eps})}
  \left[ 2 \, \psi(-{\eps})-\psi({\eps}+1)+\log\left(\frac{s}{t}\right)+\gamma_{\rm E} \right]\,,
\end{align}
where $\psi(z)$ is the digamma function, defined in \eqn{eq:psidef} of exercise~\ref{Ex:TaylorLogGamma}.

In the second term, we can safely Taylor expand in $\eps$.
We see that it is of $\cO(\eps)$, due to the presence of the factor $\Gamma(-2\eps)$ in the denominator. 
Here we keep only this leading term,
\begin{align} \label{eq:MBboxB}
B = -2 \eps \int_{  \Re(z) = c}  \frac{\d z}{2 \pi \i} \, \frac{1}{t^2} \left(\frac{s}{t}\right)^z  \Gamma(-z) \Gamma(2+z) {\Gamma^2(1+z) \Gamma^2(-1-z)}
  + \cO(\eps^2) \,,
\end{align}
where $-1<c<0$.
Therefore, remembering that the residue $A$ in \eqn{eq:MBboxA} contributes with a minus sign, we find
\begin{align}\label{eq:answerboxchapter4}
\mathrm{e}^{\eps \gamma_{\rm E}} F_{4} =  \frac{(-t)^{-\eps}}{s\,t} \left[  \frac{4}{\eps^2}  - \frac{2}{\eps}  \log \frac{s}{t} -  \frac{4 \pi^2}{3} + \cO(\eps)   \right] \,.
\end{align}
In exercise~\ref{Ex:4.BoxMellinBarnes} we compute also the $\cO(\eps)$ term.

\begin{svgraybox}
{\bf The full $--++$ helicity QCD amplitude.}
In chapter~\ref{ch:loopamps}, the one-loop four-gluon amplitude in the $++--$ helicity configuration was given in \eqn{eq:full_split_helicity_four_gluon_one_loop_integrand} in terms of box and bubble Feynman integrals.
Let us denote the ratio of the one-loop and the tree amplitude by
\begin{align}
 \mathrm{e}^{-\eps \gamma_{\rm E}}  \frac{\alpha_{\rm YM}}{(4 \pi)^{2-\eps}} M_{--++}^{(1)} \coloneqq  \frac{ A^{(1),[4-2\eps]}(1^-,2^-,3^+,4^+) }{A^{(0)}(1^-,2^-,3^+,4^+) }\,. 
\end{align}
Using the results for the integrals from eqs.~\eqref{eq:answerboxchapter4} and~\eqref{eq:Ibubblechapter4},
we find
\begin{align}\label{eq:finalresultQCDamplitude}
 M_{--++}^{(1)} =
&  - \frac{4}{\eps^2} +\frac{1}{\eps} \left[ -\frac{11}{3} +2  \log\left(- \frac{s}{\mu_{\rm R}^2} \right) + 2  \log\left(-\frac{t}{\mu_{\rm R}^2}\right) \right]   \\
&\hspace{-1cm}- 2  \log\left(-\frac{s}{\mu_{\rm R}^2}\right) \log\left(-\frac{t}{\mu_{\rm R}^2}\right) + \frac{4}{3}\pi^2  + \frac{11}{3} \log\left(-\frac{t}{\mu_{\rm R}^2}\right) - \frac{64}{9}  + {\cal O}(\eps) \nonumber
\,.
\end{align}
Here we have reinstated the dimensional regularisation scale $\mu_{\rm R}^2$.
We can rewrite this in the following instructive form,
\begin{align} \label{eq:finalresultQCDamplitude2}
\begin{aligned}
M_{--++}^{(1)}  =
& - \frac{2}{\eps^2}  \left(-\frac{s}{\mu_{\rm R}^2}\right)^{-\eps} - \frac{2}{\eps^2}  \left(-\frac{t}{\mu_{\rm R}^2}\right)^{-\eps} + \log^2 \left(\frac{s}{t}\right) +  \frac{4 \pi^2}{3}  \\
& \hspace{1cm}- \frac{11}{3 \eps}  + \frac{11}{3} \log\left(-\frac{t}{\mu_{\rm R}^2}\right) - \frac{64}{9} + {\cal O}(\eps) \,.
\end{aligned}
\end{align}
The special form of the poles in $\eps$ in \eqn{eq:finalresultQCDamplitude2} is related to the structure of ultraviolet and infrared divergences in Yang-Mills theories. 
It is due to the fact that ultraviolet and infrared effects come from separate regions. 
For an introduction to infrared divergences in Yang-Mills theories, see e.g.\ ref.~\cite{ch4Agarwal:2021ais}. 
\end{svgraybox}

\begin{exer}{Massless one-loop box with Mellin-Barnes parametrisation}
\label{Ex:4.BoxMellinBarnes}
Compute the order-$\eps$ term of the function $B$ in \eqn{eq:MBboxB}. Putting the latter together with the Laurent expansion of the residue $A$ in \eqn{eq:MBboxA} gives the analytic expression of the massless one-loop box integral $F_4$ up to order $\eps$:
\begin{align}  \label{eq:F4final}
\begin{aligned}
F_4 = \ & \frac{\text{e}^{-\epsilon \gamma_{\text{E}}} (-s)^{-\epsilon}}{s \, t} \biggl\{ \frac{4}{\eps^2} - \frac{2}{\eps} \log(x) -\frac{4 \pi^2}{3} +
  \epsilon \biggl[  2 \, \text{Li}_3\left(-\frac{1}{x}\right) + 
 2 \log(x) \, \text{Li}_2\left(-\frac{1}{x}\right)    \\
& + \log^3(x) + \frac{7 \pi^2}{6} \log(x) - \bigl(\log(x)^2+ \pi^2 \bigr) \log(1 + x)  - \frac{34}{3} \zeta_3 \biggr] + \mathcal{O}\bigl(\epsilon^2\bigr) \biggr\} \,,
\end{aligned}
\end{align}
where $x=t/s>0$. In this result we see for the first time the \emph{polylogarithm} $\mathrm{Li}_n(x)$, a special function which arises frequently in the computation of Feynman integrals. 
For $|x|<1$, the $n$-th polylogarithm ${\rm Li}_{n}(x)$ is defined as the power series
\begin{align}\label{def_polylogarithm_series}
{\rm Li}_{n}(x) \coloneqq \sum_{k\ge1 }\frac{x^k}{k^n} \,,\quad \quad {n=1,2,\ldots} \,.
\end{align}
The definition can be extended to the rest the complex plane by analytic continuation, e.g.\ by viewing the polylogarithms as solutions to differential equations. We will take this viewpoint in section~\ref{ref:subsection:propertiesofspecialfunctions}.
Note that the first polylogarithm is just a logarithm: ${\rm Li}_{1}(x) = - \log(1-x)$. The second polylogarithm, $\mathrm{Li}_2$, is typically referred to as \emph{dilogarithm}. At unit argument, the polylogarithms with $n\ge 2$ evaluate to Riemann's zeta constants: $\mathrm{Li}_n(1) = \zeta_n$.
For the solution see \hyperref[Sol:4.BoxMellinBarnes]{chapter~5}.

\end{exer}

\section{Special functions, differential equations, and transcendental~weight}
\label{ref:subsection:propertiesofspecialfunctions}

 \subsection{A first look at special functions in Feynman integrals}

In the previous section, we have already seen a few examples of special functions appearing in Feynman integrals,
namely the logarithm and the polylogarithm. We have also encountered special numbers: powers of $\pi$, as well as other transcendental constants such as $\zeta_3$.
The latter appear on their own, or arise as special values of the special functions, as we see presently.

These transcendental numbers and functions are ubiquitous in quantum field theory. For example, they may appear in anomalous dimensions of local operators, in the $\beta$ function governing the renormalisation group flow, or in scattering amplitudes. 
From a structural viewpoint it is very interesting to ask: what transcendental numbers may arise in a given computation?
Some of the techniques discussed later in this chapter came together from insights into this and related questions.

A first useful concept is the notion of \emph{transcendental weight}, or ``transcendentality''.
Roughly speaking, it describes the complexity of an expression. Rational numbers are assigned weight zero, while $\pi$ is assigned weight one, and more generally $\zeta_n$ is assigned weight $n$.
Likewise, the logarithm is assigned weight one, while the polylogarithm ${\rm Li}_n$ is assigned weight $n$.
The first interest in this definition came from two observations. Firstly, in the special ${\mathcal{N}}=4$ super Yang-Mills theory, quantities appear to always have a fixed, uniform weight. Secondly, for certain anomalous dimensions in QCD, which are not uniform in weight, the highest-weight piece agrees with the one computed in ${\mathcal{N}}=4$ super Yang-Mills theory~\cite{ch4Kotikov:2002ab}. These first observations stimulated more research that eventually led to a better understanding of transcendental weight, which allows one to predict which Feynman integrals have the maximal weight property. This insight is  useful for computing Feynman integrals, as we will discuss below.

We have seen a definition of the polylogarithm in \eqn{def_polylogarithm_series}. 
There are many examples of special functions in physics, and usually there exist several equivalent definitions.
The same is the case here. 
In many cases, a definition in terms of a defining differential equation is convenient.
We will follow this approach in this section, and will discover that it is very useful in the context of Feynman integrals.
Therefore let us first review the functions we encountered so far from this perspective, which is closely related to integral representations, and discuss some of their key~properties.

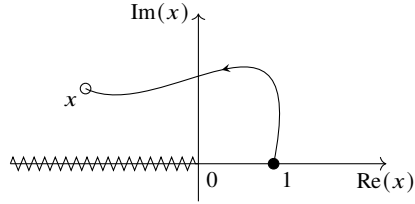
\begin{figure}[t]
\centering
\begin{tikzpicture}[baseline={(current bounding box.center)}, scale =1.0,
  decoration={markings,  mark= at position 0.5 with {\arrow{stealth}}}]
  \draw[->] (-2.5,0) -- (2.5,0) node[below] {${\rm Re}(x)$};
  \draw[->] (0,-0.5) -- (0,2) node[left] {${\rm Im}(x)$};
  \node[below right] at (0,0) {$0$};
  \filldraw (1,0) circle (2pt) node[below right] {$1$};
    \draw[decorate,decoration={zigzag,segment length=4pt}] (-2.5,0) -- (0,0);
    \draw[postaction={decorate}] (1,0) .. controls (1.5,2.5) and (-0.5,0.5) .. (-1.5,1);
      \draw (-1.5,1) circle (2pt) node[below left] {$x$};
      \end{tikzpicture}
   \caption{Integration contour to extend the definition of the logarithm to the complex plane with the branch cut along the negative real axis removed, as indicated by the zig-zag line.}
  \label{fig:logarithm}
\end{figure}

The logarithm can be defined as a single integral: 
\begin{align}\label{eq:logxreal}
\log x = \int_{1}^{x} \frac{\d t}{t} \,.
\end{align}
This equation is defined for real positive $x$. 
To extend the definition to complex argument, one places a branch cut along the negative real axis, and defines the answer in the cut complex plane by analytic continuation, i.e.\ by integrating along a contour from the base point $x=1$ to the argument $x \in \mathbb{C} \setminus \{x<0\}$, as shown in fig.~\ref{fig:logarithm}.

From \eqn{eq:logxreal} we can simply read off the derivative,
\begin{align}\label{eq:difflog}
\partial_x \log x &= \frac{1}{x} \,,
\end{align}
and we have $\log 1 = 0$.
Dilogarithms can be defined in a similar way, but with two integrals instead of one:
\begin{align} \label{eq:Li2_integral}
{\rm Li}_{2}(x) = - \int_{0}^{x} \frac{\d t}{t} \log(1-t) \,.
\end{align}
One may verify that this agrees with the series representation~\eqref{def_polylogarithm_series} by Taylor expanding the integrand in $t$.
From this we can read off the derivatives,
\begin{align}
\partial_x {\rm Li}_{2}(x) &= -\frac{1}{x} \log(1-x) \,,\\
\partial_x {\rm Li}_{2}(1-x) &= \frac{1}{1-x} \log(x)  \label{diffeqLi2}  \,,
\end{align}
as well as the special value  ${\rm Li}_{2}(0) = 0$. Like the logarithm, the dilogarithm ${\rm Li}_2(x)$ is a multi-valued function. Its branch points are at $x=1$ and infinity. Following the convention of the logarithm, the branch cut is along the positive real axis between $x=1$ and infinity (see exercise~\ref{Ex:Discontinuities}).
For more information about the dilogarithm, see~\cite{ch4Zagier:2007knq}.

We have seen that this, as well as the trilogarithm encountered above, are part of a larger class of polylogarithms, defined in terms of series in \eqn{def_polylogarithm_series}.
 In the following, it will be useful to think of these functions in terms of iterated integrals.
 To establish the connection, we note that 
\begin{align}
x \, \partial_x {\rm Li}_{n}(x) &=  {\rm Li}_{n-1}(x) \,,\quad {\rm for} \;\; n>2 \,,
\end{align}
which follows straightforwardly from \eqn{def_polylogarithm_series}. 
Therefore we can write
\begin{align}
 {\rm Li}_{n}(x) = \int_0^x  {\rm Li}_{n-1}(y) \frac{\d y}{y}  \,,\quad {\rm for} \;\; n>2 \,.
\end{align}
All polylogarithms ${\rm Li}_{n}(x)$ are multi-valued functions, with a branch cut along the positive real axis between $x=1$ and infinity.
Note that we can think of all those functions as iterated integrals over certain logarithmic integration kernels: $\d x/x$ and $\d x/(x-1)$.
This leads to another way to think about transcendental weight: it corresponds to the number of integrations in such an iterated-integral representation.

\begin{exer}{Discontinuities}
\label{Ex:Discontinuities}
The discontinuity of a univariate function $f(x)$ across the real $x$ axis is defined~as
\begin{align}
\mathrm{Disc}_x \left[f(x)\right] \coloneqq \lim_{\eta\to 0^+} \left[ f(x+\i \eta) - f(x- \i \eta) \right] \,.
\end{align}
\begin{enumerate}[a)]
\item Prove that the discontinuity of the logarithm is given by
\begin{align} \label{eq:disc_log}
\mathrm{Disc}_x \left[\log(x)\right] = 2 \pi \i \, \Theta(-x) \,,
\end{align}
where $\Theta$ denotes the Heaviside step function. 

\smallskip
\item The dilogarithm ${\rm Li}_2(x)$ has a branch cut along the real $x$ axis for $x>1$. Prove that the discontinuity is given by
\begin{align} \label{eq:disc_dilog}
\mathrm{Disc}_x \left[\text{Li}_2(x)\right] = 2 \pi \i \, \log(x) \, \Theta(x-1) \,.
\end{align}
Hint: use the identity
\begin{align} \label{eq:Li2identity}
{\rm Li}_2(x) = - {\rm Li}_2(1-x) - \log(1-x) \log(x) + \zeta_2 \,,
\end{align}
which we shall prove in exercise~\ref{Ex:4.W2basis}.

\end{enumerate}
For the solution see \hyperref[Sol:Discontinuities]{chapter~5}.
\end{exer}

\subsection{Special functions from differential equations: the dilogarithm}
\label{sec:sepcialfunctionsfromDE}

Let us now see how we can think of these functions conveniently from a differential equations approach. 
Say we are interested in the function ${\rm Li}_{2}(1-x)$, perhaps because we know that it can appear in a certain calculation.
Our goal is to find a defining set of differential equations for this function. Inspecting \eqn{diffeqLi2}, we see that $\log(x)$ appears in its derivative, so we consider this function also, as well as the constant $1$, which is required to write the derivative of $\log(x)$. 
Let us put these key functions into a vector,
\begin{align} \label{definitionfLi2}
\vec{f}(x)= \begin{pmatrix} {\rm Li}_{2}(1-x) \\ \log(x) \\ 1 \end{pmatrix}
\,.
 \end{align}
 A short calculation then shows that the following differential equation is satisfied,
  \begin{align} \label{exampleDELi2matrix2}
\partial_x \vec{f}(x)=  \left( \frac{A_0}{x}  +  \frac{A_1}{x-1} \right) \cdot \vec{f}(x)
\,.
 \end{align}
with the matrices
 \begin{align} \label{exampleDELi2matrix}
A_0 = {\begin{pmatrix}0 & 0 & 0 \\ 0 & 0 & 1 \\ 0 & 0 & 0 \end{pmatrix}}  \,, \quad  A_1 = {\begin{pmatrix}0 & -1 & 0 \\ 0 & 0 & 0 \\ 0 & 0 & 0 \end{pmatrix}}\,.
 \end{align}
The first-order differential equations~\eqref{exampleDELi2matrix2}, together with the boundary condition $\vec{f}(x=1) = (0,0,1)^{\top}$, uniquely fix the answer.

Equation~\eqref{exampleDELi2matrix2} encodes the singular points $x=0,1,\infty$ of the functions. As we will see later, the leading behaviour of $\vec{f}(x)$ is governed by the coefficient matrices of those singular points, which are $A_0,A_1,A_{\infty}=-A_0-A_1$, respectively.
The last point can be understood by changing variables to $y=1/x$, and inspecting the singularity at $y=0$.

Let us now see how this connects to the concept of transcendental weight.
Recall that, when referring to iterated integrals, the weight counts the number of integrations. So the rational constant has weight zero, a logarithm has weight one, the dilogarithm has weight two, and so on.
Looking at \eqn{definitionfLi2}, we see that the different components of $\vec{f}$ have different weight. 
In order to remedy this, we introduce a weight-counting factor $\eps$, to which we assign the weight $-1$~\cite{ch4Henn:2013pwa}.
For the moment, this is a purely formal definition. However, later we will see that this is natural in the context of dimensional regularisation.\footnote{In particular, poles $1/\eps$ in dimensional regularisation correspond to $\log \Lambda$ terms in cutoff regularisations, so it is natural that $1/\eps$ has the same weight as a logarithm.} 
With the weight-counting parameter $\eps$ at our disposal, we can define
\begin{align} \label{eq:defg}
\vec{g}(x;\eps)= \begin{pmatrix} \eps^2 {\rm Li}_{2}(1-x) \\ \eps \log(x) \\ 1 \end{pmatrix}
\,.
 \end{align}
This vector has uniform weight zero by definition. 
We find that it satisfies the following differential equations:
 \begin{align}\label{eqDEgLi2canonical0}
 \partial_x \vec{g}(x;\eps) = \eps \left( \frac{A_0}{x} + \frac{A_1}{x-1} \right) \cdot \vec{g}(x;\eps) \,.
 \end{align}
It is instructive to rewrite this in differential form, as
\begin{align}\label{eqDEgLi2canonical}
\d \, \vec{g}(x;\eps) = \eps \bigl[ A_0\, \d \log(x) + A_1\, \d \log(1-x) \bigr] \cdot \vec{g}(x;\eps) \,,
\end{align}
 where $\d= \d x \frac{\partial}{\partial x}$.
 We can see that the weights in this equation are consistent:
 $\d$ and $\eps$ have weight $-1$, $\log$ has weight $+1$, and the constant matrices $A_0$ and $A_1$ have weight zero. 
  Therefore $g$ has weight zero. 
  Since $g$ depends on $\eps$, this means that when expanding in a series around $\eps=0$, the weight of the coefficients increases with the order in $\eps$, starting with weight zero at order $\eps^0$.
  This is of course exactly what is expected from \eqn{eq:defg}.
  
 Let us now see how this arises from solving \eqn{eqDEgLi2canonical}. Plugging the ansatz
 \begin{align}
\vec{g}(x;\eps) = \sum_{k \ge 0} \eps^k \vec{g}^{(k)}(x) 
 \end{align}
into the DE, and looking at the different orders in $\eps$, we see that the equations decouple. The first few orders read 
\begin{align}\label{eq:DEgexpanded}
 & \partial_x \vec{g}^{(0)}(x) = 0 \,, \\
 &  \partial_x \vec{g}^{(1)}(x) = A(x) \cdot \vec{g}^{(0)}(x)\,,\\
 &  \partial_x \vec{g}^{(2)}(x) = A(x) \cdot \vec{g}^{(1)}(x) \,,
\end{align}
at order $\eps^{0}, \eps^{1}, \eps^{2}$, respectively, and so on.
Recalling the boundary condition $\vec{g}(1,\eps) = (0,0,1)^{\top}$,
the equations are readily solved, giving
 \begin{align}
 \vec{g}^{(0)}(x) = 
 \begin{pmatrix} 0 \\ 0 \\ 1 \end{pmatrix}\,,\quad
 \vec{g}^{(1)}(x) = 
 \begin{pmatrix} 0 \\ \log x \\ 0 \end{pmatrix}\,,\quad
\vec{g}^{(2)}(x) =
  \begin{pmatrix} {\rm Li}_{2}(1-x) \\ 0 \\ 0 \end{pmatrix}\,.
 \end{align}
 The higher-order equations read $ \partial_x \vec{g}^{(3)}(x) = A(x) \cdot \vec{g}^{(2)}(x)  =0 $, which lead to $\vec{g}^{(3)}(x) = (0,0,0)^{\top}$, and similarly at higher orders. In other words, the $\eps$ expansion stops at $\eps^2$. The reason is that, in this specific case, $A(x)$ is a nilpotent matrix, i.e., $A(x)^3 = 0$. 
This will be different for general Feynman integrals in $D=4-2\eps$ dimensions, but it will not limit the usefulness of the method.

\begin{important}{It is useful to think about special functions in terms of their defining differential equations.}
Just as in familiar textbook examples from quantum mechanics, differential equations turn out to be a useful
 way of {\it defining} classes of special functions. We shall see later in this chapter that this strategy is particularly effective for Feynman integrals.
\end{important}
 
\subsection{Comments on properties of the defining differential equations}
 \label{subsec:comments}

Let us make a number of important comments on the differential equations~\eqref{eqDEgLi2canonical} discussed in the last subsection.

\smallskip

1. {\it  Fuchsian nature of the singularities}. Equation~\eqref{eqDEgLi2canonical} has several special features. 
One of them is the nature of its singularities. We see that the matrices on its RHS have only simple poles at each singular point. 
It implies that the asymptotic solution near any singular limit, say $x\to 0$, can be expressed in terms of powers of $x$ and logarithms of $x$.
This type of singularity is called Fuchsian.
In contrast, consider differential equations with a non-Fuchsian singularity, e.g.
\begin{align} \label{eq:DEnonFuchsian}
\partial_x f(x) = \frac{a}{x^2} \, f(x) \,.
\end{align}
The solution to \eqn{eq:DEnonFuchsian} reads 
\begin{align}
f(x) = \mathrm{e}^{-a/x} f_0\,,
\end{align} 
for some boundary constant $f_0$. This has non-analytic behaviour at $x=0$, which is not expected from individual Feynman integrals.

We will see in section~\ref{sec:DEFeynmanintegrals} that the Fuchsian property is useful in several regards. Firstly, it may help in finding simple forms of the differential equations.
Secondly, analysing the behaviour of the equations near singular points provides crucial information for fixing integration constants based on physical principles, without additional calculations~\cite{ch4Henn:2020lye}.
Thirdly, the asymptotic expansion of an integral in a certain limit can be read off easily from the differential equations.

\smallskip

2. {\it  ``Gauge dependence'' of the differential equations}. 
The differential equations 
\begin{align}\label{eq:generalDEincommentsection}
\partial_x \vec{f}(x,\eps) = A(x,\eps) \cdot \vec{f}(x,\eps)
\end{align}
 are not unique in the following sense. 
Consider an invertible matrix $T(x,\eps)$, such that we can define the following change of basis,
\begin{align} \label{eq:gaugetransformation}
\vec{f} = T \cdot \vec{g} \,.
\end{align}
Then the new basis $\vec{g}$ satisfies similar differential equations, $\partial_x \vec{g} = B \cdot \vec{g}$, with a different matrix
\begin{align}\label{eq:defB}
B = T^{-1} \cdot A \cdot T - T^{-1} \cdot \partial_x T \,.
\end{align}
Note that the ``connection matrix'' $A$ transforms to $B$ as under a gauge transformation.

For this reason, even if a simple form of the differential equations such as \eqn{eqDEgLi2canonical} exists, this fact might be obscured if an unfortunate choice of basis is made. 
In particular, the Fuchsian property mentioned above may be obscured in this way.
However, a judicious basis choice can reveal the simplicity of the answer.

For example, consider the following matrix,
\begin{align}
T=  {\begin{pmatrix} 1+x & 0 & 1 \\ 1 & -x & 0 \\ 0 & 0 & 1  \end{pmatrix}} \,,\quad {\rm with} \quad T^{-1} = {\begin{pmatrix} \frac{1}{1+x} & 0 & -\frac{1}{1+x} \\ \frac{1}{x (1+x)} & -\frac{1}{x} & -\frac{1}{x (1+x)} \\ 0 & 0 & 1  \end{pmatrix}} \,.
\end{align}
Applying this to $A= A_0/x + A_1/(x-1)$ with~\eqref{exampleDELi2matrix}, we find that \eqn{eq:defB} evaluates~to
\begin{align}
B = {\begin{pmatrix} \frac{1-\eps-x}{(x-1)(x+1)} & \frac{\eps x}{(x-1)(x+1)} & 0 \\ \frac{1-\eps-x}{(x-1)(x+1)x} & \frac{1+\eps x-x^2}{(x-1)(x+1)x} & -\frac{\eps}{x^2}  \\ 0 & 0 & 0  \end{pmatrix}} \,.
\end{align}
This new form of the DE is far worse compared to the original one, for three reasons. First, the factorised $\eps$-dependence is lost. Second, there is a spurious divergence at $x=-1$. Third, the Fuchsian property at $x=0$ is no longer manifest, due to the $1/x^2$ term. This example underlines the importance of good guiding principles when dealing with this type of differential equations.

\smallskip

3.  {\it  Solution as a path-ordered exponential}. 
The analogy with gauge transformations mentioned in point~2 above allows us to write down the solution to the general differential equations~\eqref{eq:generalDEincommentsection}. 
The latter
is given by the following expression 
\begin{align}\label{eq:Wilsonlinesolution}
\vec{f}(x) = {\mathbb P} \exp\left[ \int_{\cal C} A(x') \d x' \right]  \cdot \vec{f}(x_0) \,.
\end{align}
Here ${\cal C}$ is a path connecting the \textit{base point} $x_{0}$ to the function argument $x$,
 ${\mathbb P}$ stands for path ordering along this path, and $f(x_{0})$ is the value of the function at the base~point.

In the formal expansion of the matrix exponential appearing in \eqn{eq:Wilsonlinesolution}, the arguments are evaluated at different points $x'$ along the path ${\cal C}$. In practice, one may consider the pull-back of the form $A(x') \d x'$ to the unit interval, parametrised by a parameter $t\in [0,1]$. 
The path-ordering then dictates that the matrices are ordered according to the ordering on the unit interval.

Equation~\eqref{eq:Wilsonlinesolution} may be familiar to readers. On the one hand, the path ordering also shows up as time-ordering in the evolution operator in quantum mechanics. On the other hand, somewhat more advanced, it shows up in gauge theory: given a gauge field $A$, the matrix  $ {\mathbb P} \exp\bigl[ \int_{\cal C} A(x') \d x' \bigr]$ represents a Wilson line connecting two points $x_0$ and $x$, along a contour ${\cal C}$. 

This second analogy also makes it clear that \eqn{eq:Wilsonlinesolution} enjoys a manifest {\it homotopy invariance}: two contours ${\cal C}$ and ${\cal C'}$ (connecting $x$ and $x_0$) give the same value, as long as they can be smoothly deformed into each other, without crossing poles of $A$.

For general $A$, the RHS of \eqn{eq:Wilsonlinesolution} is somewhat formal. However, in the cases considered  in these lectures, it can be made very explicit. Firstly, in the dilogarithm example of section~\ref{sec:sepcialfunctionsfromDE}, $A$ is nilpotent, and hence the path-ordered exponential has a finite number of terms (and hence number of iterated integrals) only. Secondly, in a later section we propose a method that achieves that $A \sim \eps$, which allows us to write the path-ordered exponential as a Taylor series in $\eps$. This is analogous to a perturbative expansion in the Yang-Mills coupling in gauge theory.

\smallskip

\begin{figure}[t]
\centering
\begin{tikzpicture}[baseline={(current bounding box.center)}, scale =1.0, 
   decoration={markings,  mark= at position 0.56 with {\arrow{stealth}}}]
  \draw[->] (-2.5,0) -- (2.5,0) node[below] {${\rm Re}(x)$};
  \draw[->] (0,-1) -- (0,2) node[left] {${\rm Im}(x)$};
  \node[below right] at (0,0) {$0$};
  \filldraw (1,0) circle (2pt) node[below right] {$1$};
    \draw[decorate,decoration={zigzag,segment length=4pt}] (-2.5,0) -- (0,0);
    \draw[postaction={decorate}] (1,0) .. controls (1.5,2.5) and (-0.5,0.5) .. (-1.5,1);
    \node at (1.15,1.2) {$\mathcal{C}$};
    \draw[postaction={decorate}]  (1,0) .. controls (-1.5,-1.5) and (-0.5,-0.5) .. (-1.5,1);
    \node at (-1.,-0.8) {$\mathcal{C}'$};
      \draw (-1.5,1) circle (2pt) node[below left] {$x$};
      \draw[mark=x, mark size=4pt,thick] plot (0,0);
      \draw[mark=x, mark size=4pt,thick] plot (1,0);
      \end{tikzpicture}
   \caption{
   Integration contours for solving the defining DE of the dilogarithm $\mathrm{Li}_2(1-x)$, eq.~(\ref{exampleDELi2matrix2}). The zig-zag line denotes the branch cut, and the crosses the poles of the connection matrix. The difference between integrating the DE along the two contours gives the discontinuity of the solution.}
  \label{fig:logarithm_disc}
\end{figure}
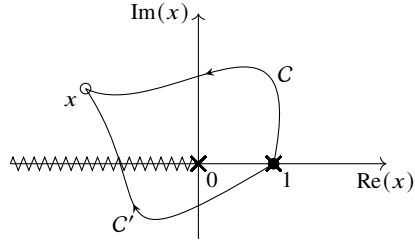

4. {\it Discontinuities} (see exercise~\ref{Ex:Discontinuities}) are also solutions to the differential equations. 
This is obvious from the form~\eqref{eq:Wilsonlinesolution} of the general solution.
Indeed, consider two contours ${\cal C}$ and ${\cal C'}$ that differ by a contour encircling a pole of $A$. 
Since both  ${\cal C}$ and ${\cal C'}$ are solutions to \eqn{eq:generalDEincommentsection}, so is their difference.
The latter corresponds to taking a discontinuity of $\vec{f}$.
For instance, in our example above (\eqn{exampleDELi2matrix2}), we may consider two contours $\mathcal{C}$ and $\mathcal{C}'$ as in figure~\ref{fig:logarithm_disc}, with $\mathcal{C}'$ crossing the branch cut starting from $x=0$. In this case, for $x\in \mathbb{C} \setminus \{x<0\}$, we find~that
\begin{align} \label{eq:disc_example}
\begin{aligned}
{\rm Disc}_{x=0} \, \vec{f}(x) & = \left\{ \mathbb{P} \, \mathrm{exp}\left[\int_{\mathcal{C}} A(x') \d x' \right] - \mathbb{P} \, \mathrm{exp}\left[\int_{\mathcal{C}'} A(x') \d x' \right] \right\} \cdot \vec{f}(x_0) \\
 & = 2 \pi \i \, \begin{pmatrix}  -\log(1-x) \\ 1 \\ 0 \end{pmatrix} \,,
\end{aligned}
\end{align}
which is equally a solution to \eqn{exampleDELi2matrix2}. This way of defining the discontinuity is more general than the one we have seen in exercise~\ref{Ex:Discontinuities} and chapter~\ref{ch:loopamps}---which was restricted to the real axis---and is more suitable to functions involving multiple branch points.\footnote{In fact, note that the connection matrix in the DE~\eqref{exampleDELi2matrix2} has a pole also at $x=1$. This branch point is ``hidden'' in the principal branch of $\mathrm{Li}_2(1-x)$, but comes to light when we cross the branch cut starting from $x=0$. Indeed, we see in \eqn{eq:disc_example} that the discontinuity of $\mathrm{Li}_2(1-x)$ has itself a branch point at $x=1$.} For this reason we indicate the branch point in the subscript of the discontinuity in \eqn{eq:disc_example}. 
This property of the discontinuity will be useful when analysing Feynman integrals.

\smallskip

5. {\it Relation to   Picard-Fuchs equation}. The issue raised in the previous point can be addressed in part in the following way. We can trade the first-order system of differential equations for higher-order equations for one of the integrals, called Picard-Fuchs equations~\cite{ch4Muller-Stach:2012tgj}. 
Let us illustrate this for the ${\rm Li}_{2}(1-x)$, which is the first component of ${\bf f}$ in \eqn{definitionfLi2}. Differentiating $\vec{f}$ multiple times with the help of \eqn{exampleDELi2matrix2}), one obtains a system of equations, from which one eliminates all functions except $\vec{f}_{1}$ and its derivatives.
In the present case we get 
\begin{align}
\partial_x x \partial_x (1-x) \partial_x {\rm Li}_{2}(1-x) = 0\,.
\end{align}
An advantage of this representation as compared to \eqn{eqDEgLi2canonical} is that it depends only on one function, ${\rm Li}_{2}(1-x)$, and not on the other elements in the vector ${\bf f}$. As such it does not suffer from the gauge dependence mentioned~above. 

\smallskip

6. {\it Dependence on the weight-counting parameter $\eps$}.
Recall that we assigned to $\eps$ transcendental weight $-1$.
We saw that when each $w$-fold iterated integral appearing in the basis $\vec{f}$ was multiplied by a factor $\eps^w$, then the differential equations had a simple, factorised dependence on $\eps$, cf.~\eqn{eqDEgLi2canonical}. Conversely, had we considered linear combinations of mixed weight, or had we not normalised the integrals appropriately, the dependence on $\eps$ would have been more complicated. This is important to bear in mind when applying the above philosophy to Feynman integrals. 
 
 \smallskip 
 
 7. {\it Uniform weight functions and pure functions}.
 It turns out that the simplicity of the differential equations considered in this chapter can be easily understood. For this it is useful to introduce the following concepts. A {\it uniform weight} function is a linear combination of functions of some uniform weight $w$, with coefficients that may be rational or algebraic functions. An example is 
 \begin{align}
  \frac{1}{1+x} {\rm Li}_{2}(1-x)  + \frac{x}{1-x} \log^2(1+ x) \,.
 \end{align}
 Such a function does not satisfy particularly nice differential equations. The reason is that a derivative $\partial_x$ can act either on one of the prefactors, or on the transcendental functions. As a result, one obtains a function of mixed transcendental weight. In contrast, consider a {\it pure function}, which is a ${\mathbb Q}$-linear combination of functions of some uniform weight $w$. An example for $w=2$ is
 \begin{align}
 {\rm Li}_{2}(1-x)  +  \log^2(1+ x) \,.
 \end{align} 
 Its derivative is
 \begin{align}
 \frac{1}{1-x} \log x + \frac{2}{1+x} \log(1+x) \,,
 \end{align}
 which has weight one, i.e.\ one less than the original function.
It is built from two new pure functions, for which we could iterate the differential procedure. In this way one can construct a system of differential equations similar to the one considered above for any pure function. This also generalises naturally to multi-variable functions.

\subsection{Functional identities and symbol method}
 \label{sec:functionalidentitiesandsymbolmethod}
 
We will see that the key properties of the special functions can be encoded in so-called {\it symbols}. Roughly speaking, symbols preserve the information on the integration kernels but disregard the integration constants. 
In this context, the integration kernels are called \emph{letters}, and their ensemble \emph{alphabet}. 
For example, the alphabet associated with the DE~\eqref{eqDEgLi2canonical} has two letters: $\d \log(x)$ and $\d \log(1-x)$.\footnote{Whenever the alphabet contains only $\d \log$-type integration kernels, it is sufficient to keep track of their arguments. In the case of \eqn{eqDEgLi2canonical}, for instance, one would say that the alphabet is $\{x,1-x\}$.}
Concatenating different alphabet letters into words corresponds to specific iterated integrals. Leveraging the basic addition identity of the logarithm, the symbol technique allows one to detect function identities by simple algebra. 

Consider as an example the following function,
\begin{align}\label{defgexamplesymbol}
g(x) \coloneqq {\rm Li}_{2}(x) + {\rm Li}_{2}\left(\frac{1}{x}\right) + \frac{1}{2} \log^2(-x) + \frac{\pi^2}{6} \,. 
\end{align}
Let us consider this for $x<0$, such that we stay away from branch cuts, and all summands are real-valued.
We now wish to show that $g(x) =0$.
It is instructive to do this in the most elementary way, namely to show that $g'(x) =0$, and that the identity is true at some value of $x$.
Using \eqn{diffeqLi2} we have
\begin{align}
x \, \partial_x g(x) = - \log(1-x) + \log\left(1- \frac{1}{x} \right) + \log(-x)  = 0 \,.
\end{align}
In the last step, we have assumed $x<0$, so that we can use $\log(a b) = \log(a) + \log(b)$ for $a,b>0$.
Moreover, using ${\rm Li}_{2}(-1) = -\pi^2/12$, one verifies that $g(-1) = 0$. This completes the proof of the dilogarithm inversion identity $g(x)=0$.

The symbol method can streamline finding avatars of such identities, i.e.\ identities up to possibly integration constants.
It leverages the fact that we are dealing with iterated integrals, whose integration kernels satisfy the basic logarithm identity
\begin{align} \label{eq:dlogid}
\d \log(a b) = \d \log(a) + \d \log(b)\,.
\end{align}
Unlike the analogous identity for the logarithm, \eqn{eq:dlogid} holds for any non-vanishing $a$ and $b$, as $\d \log c = 0$ for any constant $c$.
Let us now see this in practice, first giving an intuitive explanation, and then a formal definition. Given an iterated integral, say 
\begin{align}
 {\rm Li}_{2}(x) 
 =&  -  \int_0^x \d\log(y) \int_0^y \d\log(1-z) \,, \label{Li2dlog} 
 \end{align}
we read off its logarithmic integration kernels, which are $\d\log(1-z)$ and $\d\log(y)$, respectively, and record their arguments in the symbol, denoted by square brackets,
 \begin{align}\label{eq:Li2symbolexample}
{\cal S} \bigl( {\rm Li}_{2}(x) \bigr) =& - [1-x,x] \,.
 \end{align}
 An alternative notation in the literature is $- (1-x) \otimes x$. We prefer the bracket notation to make it clear that the minus sign in \eqn{eq:Li2symbolexample} multiplies the symbol, rather than being one of its entries.
 
 Note that the order of integration kernels in the symbol $[\ldots ]$ is opposite to that in the integral representation,~\eqn{Li2dlog}.
 Readers might find confusing why the entries of the symbol in \eqn{eq:Li2symbolexample} depend on $x$, while in \eqn{Li2dlog} they depend on the (mute) integration variables. To clarify this, we find it best to give the following formal definition of the symbol.
 
 \smallskip
 
 {\it Recursive symbol definition for iterated integrals}. Let $f^{(w)}$ be a uniform 
  weight-$w$ function whose derivative is given by 
 \begin{align}\label{defsymbol1}
\d f^{(w)} = \sum_i c_i \, f_{i}^{(w-1)} \d\log \alpha_i\,,
\end{align}
where $c_{i}$ are kinematic-independent constants, $f_{i}^{(w-1)}$ are uniform weight-$(w-1)$ functions, and $\alpha_i$ are algebraic expressions depending on the kinematic variables.
 Then we define the symbol ${\cal S}$ of $f^{(w)}$ iteratively to be 
\begin{align}\label{defsymbol2}
\mathcal{S}\bigl(f^{(w)}\bigr) = \sum_i c_i  \bigl[ \mathcal{S}\bigl(f_{i}^{(w-1)} \bigr) , \alpha_i \bigr] \,.
 \end{align}
The iterative definition starts at weight $0$ with the ``empty'' symbol $[] \coloneqq \mathcal{S}(1)$. Applying this definition to the logarithm gives
\begin{align} \label{eq:symbol_log}
\mathcal{S}(\log x ) = [x] \,,
\end{align}
while for the dilogarithm we readily recover \eqn{eq:Li2symbolexample}.

\begin{exer}{The symbol of a transcendental function}
\label{Ex:4.Symbol}
Compute the symbol of $\log(x) \log (1-x)$.
For the solution see \hyperref[Sol:4.Symbol]{chapter~5}.
\end{exer}

Note that \eqn{defsymbol1}, and hence the definition of the symbol~\eqref{defsymbol2}, does not know about integration constants. 
In particular, constants such as $\pi^2/6$ in \eqn{defgexamplesymbol} are set to zero by the symbol map.
Moreover, since the definition is recursive, at weight $w$ the symbol in principle misses $w$ integration constants.
Nevertheless, the symbol is very useful: it provides a shortcut to discovering that an identity between transcendental functions exists. 
Let us now see how this works.

\smallskip

{\it Basic symbol properties}. It follows from the basic identity~\eqref{eq:dlogid} that the symbol satisfies
\begin{align} \label{eq:symbol_properties} \begin{aligned}
& [\ldots , a \, b , \ldots ] = [\ldots, a , \ldots ] +  [\ldots, b , \ldots ]\,, \\
& [\ldots , x^c , \ldots ] =c \, [\ldots, x , \ldots ]\,.
\end{aligned} \end{align}
Moreover, $[\ldots , c , \ldots ] = 0$ if $c$ is a constant.

Let us see how this works on $g(x)$ of \eqn{defgexamplesymbol}. We have
\begin{align}
{\cal S}\bigl({\rm Li}_{2}(x) \bigr) = \ & - [1-x,x] \,, \\
{\cal S}\left({\rm Li}_{2}\left(\frac{1}{x}\right)\right) = \ & - \left[1-\frac{1}{x},\frac{1}{x} \right] =  \left[- \frac{1-x}{x},x\right]  = [1-x,x]-[x,x]  \,, \\
{\cal S}\left(\frac{1}{2} \log^2(-x) \right) = \ &  [-x,-x] = [x,x]  \,, \\
{\cal S}\left(\frac{\pi^2}{6}\right) = \ & 0\,.
\end{align}
From this we readily conclude that the symbol identity ${\cal S}\bigl( g(x) \bigr) = 0$ holds.
What this means is the following.

\begin{important}{Symbols allow us to effortlessly find ``avatars'' of identities between transcendental functions.}
Finding a symbol identity implies a corresponding functional identity, with integration constants yet to be fixed. This can be done systematically. The ``beyond-the-symbol terms'' may in general involve constants times lower-weight functions.
\end{important}

\smallskip

{\it Connection between first and last entries to discontinuities and differentiation, respectively}.
By definition~\eqref{defsymbol1}, the last entry tells us how a symbol behaves under differentiation.
Interestingly, the first entry also has an important meaning: it is related to discontinuities.
Both these properties can be understood by thinking about symbols as iterated integrals.

For a logarithm $\log x$, the symbol is just $[x]$. Taking the discontinuity (normalised by $1/(2 \pi \i)$) across the negative real axis corresponds to replacing $[x] \to 1$.
Higher-weight functions are more interesting. For example, taking a discontinuity of $-[x,1-x]$, which is the symbol of ${\rm Li}_{2}(1-x)$, yields $-[1-x] = {\mathcal S}\bigl(- \log (1-x) \bigr)$.
For more details, see exercise~\ref{Ex:Discontinuities}.
In summary, we have that
\begin{align} \begin{aligned}
\d \, [a_1,\ldots,a_{n-1},a_n] & = \d \log(a_n) \times [a_1,\ldots,a_{n-1}] \,, \\
\mathrm{Disc} [a_1,a_2,\ldots,a_n] & = \mathrm{Disc}(\log a_1) \times [a_2,\ldots,a_n] \,.
\end{aligned} \end{align}
We thus see that, on top of providing a short-cut to finding functional relations, the symbol also encodes manifestly the branch-cut structure and the derivatives of the~functions.

\smallskip

{\it Function basis at weight two}. 
The symbol method is very useful when looking for simplifications, or when comparing results of different calculations.
In fact, for a given symbol alphabet (i.e., a given set of integration kernels), and up to a given weight, it is possible to classify the full space of possible functions.

Let us discuss this in more detail for the alphabet consisting of $\{x, 1-x \}$ that we are already familiar with. At weight two, there are four symbols we can build from these letters, namely 
\begin{align}\label{eq:symbolbasisweight2}
\bigl\{ [x,x] , [x,1-x], [1-x,x], [1-x,1-x] \bigr\} \,.
\end{align}
Once we write down a basis for this space, we can then rewrite any other weight-two function with those integration kernels in terms of that basis.

\begin{exer}{Symbol basis and weight-two identities}
\label{Ex:4.W2basis}
\vspace{-0.6cm}
\begin{enumerate}[a)]
\item Verify that 
\begin{align} \label{eq:funcbasisW2}
\bigl\{ \log^2(x), \, \log^2 (1-x), \, \log(x) \log (1-x), \, {\rm Li}_{2}(1-x)\bigr\}
\end{align}
provides a basis for \eqn{eq:symbolbasisweight2}.
\smallskip
\item Compute the symbols of 
\begin{align} \label{eq:dilogs}
{\rm Li}_{2}(x) \,, \  {\rm Li}_{2}(1/x) \,, \  {\rm Li}_{2}\left(1/(1-x)\right) \,, \  {\rm Li}_{2}\left(x/(x-1)\right) \,, \ {\rm Li}_{2}\left((x-1)/x\right)  \,, 
\end{align}
and show that they ``live'' in the space spanned by the symbols in \eqn{eq:symbolbasisweight2}.  
\smallskip
\item Derive the identities (at symbol level) for rewriting the functions in point b in terms of the basis given in point a.
\end{enumerate}
For the solution see \hyperref[Sol:4.W2basis]{chapter~5}.
\end{exer}

\begin{important}{Knowledge of the symbol alphabet dramatically restricts the answer.}
If the symbol alphabet is known (or conjectured) for a given scattering amplitude, this places strong constraints on the answer. 
Combined with additional information, such as for example the behaviour of the amplitudes in certain limits, this can sometimes be used to entirely ``\emph{bootstrap}'' the answer, i.e.\ to obtain it without actually performing a Feynman-diagram calculation.
For more information, see~\cite{ch4Henn:2020omi} and references therein. 
\end{important}

So far we have discussed how to obtain the symbol of a given function, and used this for finding identities.
A related application can be to find simplifications. This is relevant if the symbols of individual terms in an expression are more complicated than the symbol of their sum. It may even be that individual terms contain spurious symbol letters, i.e.\ letters that cancel in the sum. In such cases the symbol is a good starting point for finding a simplified answer.
Given the simplified symbol, the task is then to come up with a (simple) function representation.
At weight two, it turns out that only dilogarithms and products of logarithms are needed, for suitable arguments. 
It is easy to make an ansatz for such arguments: given that the symbol of ${\rm Li}_{2}(z)$ contains both $z$ and $1-z$, these two expressions should be part of the symbol alphabet. We can see this explicitly for the example considered above. For example, both $z=x/(x-1)$ and $1-z = 1/(1-x)$ have factors within the $\{x,1-x\}$ alphabet, and therefore $z$ is a suitable dilogarithm argument for this alphabet. Conversely, $z=-x$ would lead to a new letter $1-z=1+x$. For further reading, 
cf.~\cite{ch4Goncharov:2010jf}.

\smallskip

{\it Multi-variable example}.
The definitions~\eqref{defsymbol1} and~\eqref{defsymbol2}) apply also to the multi-variable case. To illustrate this, let us consider the following function, which appears in the six-dimensional one-loop box integral (or, equivalently, it appears in the finite part of the corresponding four-dimensional box),
\begin{align}\label{examplef1}
f_{1}(u,v) = \frac{\pi^2}{6} - {\rm Li}_{2} \left( \frac{1-v}{u}\right)  - {\rm Li}_{2} \left( \frac{1-u}{v}\right)+  {\rm Li}_{2} \left( \frac{(1-u)(1-v)}{u v}\right) \,.
\end{align}
Looking at the symbols of the individual summands, one notices that the following symbol letters appear,
\begin{align}\label{alphabet1massbox}
\{ u, v, 1-u, 1-v, 1-u-v \} \,.
\end{align}
However, the full symbol is much simpler,
\begin{align}\label{examplef1symbol}
{\cal S}(f_{1}(u,v)) = [u,1-u]+[v,1-v]-[u,v]-[v,u] \,.
\end{align} 
It only requires four of the five symbol letters. Moreover, the first entry is either $u$ or $v$, which tells us that $f_1$ has branch cuts only along the negative real $u$ and $v$ axes. In contrast, the individual terms in \eqn{examplef1} have a (spurious) cut also at $u+v=1$.
All of this tells us that a simpler function representation exists.
Readers who worked through the exercise above might be able to guess one, e.g.
\begin{align}\label{examplef1b}
f_{2}(u,v) = \frac{\pi^2}{6} - {\rm Li}_{2} \left(1-u \right)  - {\rm Li}_{2} \left(1-v \right) - \log u \log v \,.
\end{align}
The full identity can be verified as was done for \eqn{defgexamplesymbol}.
In \eqn{examplef1b}, real-valuedness for $u>0,v>0$ is manifest.

\begin{exer}{Simplifying functions using the symbol}
\label{Ex:4.Symbolf2}
Prove that the symbol of $f_1(u,v)$ is given by \eqn{examplef1symbol}, and verify that ${\cal S}(f_{2}(u,v))  = {\cal S}(f_{1}(u,v))$.
For the solution see \hyperref[Sol:4.Symbolf2]{chapter~5}.
\end{exer}

For further interesting applications of the symbol method, interested readers can find how a twenty-page expression for a six-particle amplitude in ${\cal N}=4$ super Yang-Mills theory was famously simplified to just a few lines~\cite{ch4Goncharov:2010jf}, applications to Higgs boson amplitudes~\cite{ch4Duhr:2012fh}, and an example for simplifying functions appearing in the anomalous magnetic moment~\cite{ch4Henn:2020omi}.

\subsection{What differential equations do Feynman integrals satisfy?}

In the previous subsection, we analysed defining differential equations for the logarithm and dilogarithm. 
These functions are sufficient to describe one-loop Feynman integrals in four dimensions.
We have already seen that at higher orders in the dimensional regulator, further functions, such as the trilogarithm ${\rm Li}_{3}$, make an appearance,
and more complicated functions are expected at higher loops.
Furthermore, Feynman integrals depend in general on multiple kinematic or mass variables, so a generalisation to this case is needed as well.
It turns out that there are natural extensions in both directions.

How could the most general differential equations that Feynman integrals satisfy look like? Inspired by the dilogarithm toy example above, 
we start by making a few observations that are helpful in guiding us.
\begin{itemize}
\item One important guiding principle when looking for suitable more general differential equations 
are the general properties that are expected for Feynman integrals. 
A property is the behaviour in asymptotic limits, which implies that the differential equations are Fuchsian. 
Let us consider some $N$-vector of functions $\vec{f}(x)$ (generalising \eqn{definitionfLi2}) that satisfies a set of differential equations of the form\footnote{For simplicity of notation, we suppress the dependence on $\eps$ for the moment.}
\begin{align}
\partial_x \vec{f}(x) = A(x) \cdot \vec{f}(x) \,,
\end{align}
for some $N\times N$ matrix $A(x)$.
$A(x)$ will in general have singularities at certain locations $x_k$.
In view of the gauge dependence discussed in the preceding subsection, the exact form of $A(x)$ depends on the basis choice for $\vec{f}$.
For this reason, $A(x)$ may have higher poles at any of the $x_k$. 
However, the Fuchsian property guarantees that for each singular point $x_k$, a gauge transformation exists such that $A(x)$ has only a single pole $1/(x-x_{k})$ as $x \to x_{k}$. 
We will assume in the following that this is possible to achieve simultaneously for all singular points, although mathematical counterexamples exist.\footnote{See~\cite{ch4Henn:2014qga,ch4Lee:2014ioa} and references therein.}
\smallskip
\item As far as we are aware, in all cases known in the literature, the special functions needed to express Feynman integrals in are iterated integrals (defined over a certain set of integration kernels). In line with the previous point, we assume that the latter can be chosen such that they make the Fuchsian property manifest.
The simplest examples of such integration kernels are $\d x/(x-x_k) = \d \log(x-x_k)$ for a single variable $x$; in the case of multiple variables $\vec{x}$, it could be $\d\log \alpha(\vec{x})$, for some algebraic function $\alpha$. However, the literature knows also elliptic integration kernels (which locally behave as $\d x/x$).
\smallskip
\item We use the fact that iterated integrals have a natural notion of transcendental weight. While a Feynman integral could have terms of mixed weight, we can imagine a ``gauge transformation'' that disentangles such admixtures, so that each term is a {\it pure function} of uniform weight. If we then further normalise such pure functions by a weight-counting parameter $\eps$, one may expect $\eps$-factorised differential equations, as e.g.~\eqn{eqDEgLi2canonical}.
\end{itemize}
These considerations lead us to natural generalisations of the dilogarithm example.

\smallskip

{\it Generalisation of the differential equations to multiple singular points}.
The simplest form of $A(x)$ that achieves the above properties is the following:
\begin{align}
A(x) = \sum_{k} \frac{A_{k}}{x-x_{k}} \,,
\end{align}
with constant rational matrices $A_k$.
The associated class of special functions, sometimes called \emph{multiple polylogarithms}, Goncharov polylogarithms, or hyperlogarithms, are important in the Feynman integrals literature.
Since they go beyond the scope of these lecture notes, we refer interested readers to~\cite{ch4Duhr:2019wtr} and references therein.
Being iterated integrals with logarithmic integration kernels, these functions admit the notion of transcendental weight
discussed above. Let us therefore normalise the functions of weight $w$ with $\eps^{w}$, and arrange them into the vector $\vec{f}(x;\eps)$.
This leads to a natural generalisation of \eqn{eqDEgLi2canonical}, namely
\begin{align}
\label{eq:canonicalDEmanysingularpoints}
\partial_x \vec{f}(x;\eps) = \eps \left[ \sum_k \frac{A_{k}}{x-x_{k}} \right] \cdot \vec{f}(x;\eps) \,,
\end{align}
where $\vec{f}$ is a vector with $N$ components, and $A_{k}$ are constant $N \times N$ matrices.

\smallskip

{\it Examples}. All presently known four-point box integrals satisfy this equation with $x_{k} = \{0,1\}$. 
The number $N$ depends on the specific Feynman integrals, and is $3$ for a one-loop box integral, $8$ for a two-loop double-box integral, 
and for non-planar three-loop integrals the number can be in the hundreds, see~\cite{ch4Henn:2020lye}.

\smallskip

{\it Generalisation to multiple variables}.
It is instructive to rewrite \eqn{eq:canonicalDEmanysingularpoints} in differential form, using $\d= \d x \, \partial_x$, similarly to \eqn{eqDEgLi2canonical}.
Then it becomes
\begin{align}
\label{eq:canonicalDEmanysingularpoints2}
\d\, \vec{f}(x;\eps) = \eps \left[ \sum_k {A_{k}} \d \log({x-x_{k}}) \right] \cdot  \vec{f}(x;\eps) \,.
\end{align}
This form is suitable for generalisation to multiple variables. 
Indeed, if in \eqn{eq:canonicalDEmanysingularpoints2} the positions $x_{k}$ depend on some other variables, 
then one may consider partial derivatives in those variables as well.
However, there are also Feynman integrals with more complicated dependence on the arguments.
For example, in the case of a bubble diagram with an internal mass $m$,
we found the following logarithm, see \eqn{resbubbleexplicit},
\begin{align}\label{eq:exampleHiggs} 
\log \left( \frac{ \sqrt{1-4 m^2/s } -1 }{ \sqrt{1-4 m^2/s } +1} \right) \,.
\end{align}
Although it is possible to perform a change of variables that removes the square root and allows one to treat this integral in terms
of the differential equations written above (see exercise~\ref{Ex:4.2Dbubble}), it is a harbinger of more general structures.
Equation~\eqref{eq:exampleHiggs} motivates a further generalisation where one keeps the $\d \log(\ldots)$ structure of the integration kernels,
but allows for more general arguments than $x-x_k$. In particular, it is natural to allow arbitrary algebraic expressions.
Let us therefore denote by $x$ a set of variables, and let $\alpha(x)$ be a set of algebraic expressions.
Then a generalisation of \eqn{eq:canonicalDEmanysingularpoints2} is
\begin{align}
\label{eq:canonicalDEmanysingularpointsmultiplevariables}
\d \, \vec{f}(\vec{x};\eps) = \eps \, \sum_k A_{k}  \, \d  \log[ \alpha_k(\vec{x})]  \cdot \vec{f}(\vec{x};\eps) \,.
\end{align}
This is what is called \textit{canonical form} of the differential equations in the case of logarithmic integration kernels.
See Table~\ref{tab:examples:alphabets} for examples of integration kernels $\{ \alpha_k \}$ of one-loop integrals.

\begin{table}[t]
\centering
\renewcommand{\arraystretch}{2.0}
\begin{tabular}{|>{\centering\arraybackslash}p{4.8 cm}|>{\centering\arraybackslash}p{1.6 cm}|>{\centering\arraybackslash}p{5.0 cm}|}
\hline
\textbf{One-loop Feynman integral}  & \textbf{Variables} $\vec{x}$ & \textbf{Integration kernels} $\alpha_{k}(\vec{x})$ \\
\hline
On-shell box integral \hspace{2cm}(cf.\ Exercise~\ref{Ex:4.BoxDE})& $s, t$ & $\{ s, t, s+t \}$ \\
\hline
Box-integral with one off-shell leg \hspace{1cm} (cf.\ eq.~(\ref{alphabet1massbox}) and Exercise~\ref{Ex:4.Symbolf2}) )& $s, t, p^2$ & $ \{ s, t, s+t, s-p^2, t-p^2, s+t-p^2 \}$ \\
\hline
Bubble integral with internal mass \hspace{1cm} (cf.\ Exercise~\ref{Ex:4.BubbleDE}) & $s, m^2$ & $\left\{ m^2, s-4 \, m^2 ,  \frac{ \sqrt{1-4 \, m^2/s } -1 }{ \sqrt{1-4 \, m^2/s } +1} \right\}$ \\
\hline
\end{tabular}
\caption{Examples of sets of variables $\vec{x}$ and integration kernels $\alpha_{k}(\vec{x})$ appearing in the canonical differential equations~(\ref{eq:canonicalDEmanysingularpointsmultiplevariables}) for several one-loop Feynman integrals.}
\label{tab:examples:alphabets}
\end{table}

\smallskip

{\it Generalisation beyond $\d\log$ integation kernels}.
Equation~\eqref{eq:canonicalDEmanysingularpointsmultiplevariables} covers a large class of cases.
As we hinted at above, even more general cases exist, where the connection matrix is not written as a sum of logarithms:
\begin{align}
\label{eq:canonicalDEmanysingularpointsmultiplevariables3}
\d \, \vec{f}(\vec{x};\eps) = \eps \, \d \tilde{A}(\vec{x}) \cdot \vec{f}(\vec{x};\eps) \,.
\end{align}
Here the assumed iterative structure of the special functions is realised by having $\eps$ as a book-keeping variable for the complexity. 
It is not yet understood what the most general form of the connection matrix $\tilde{A}$ is.
The Fuchsian property restricts the form of the integration kernels. 
Say $\tilde{A}(\vec{x})$ is singular at $\vec{x}_{0}$. Parametrising the limit $\vec{x} = \vec{x}_{0} + \tau \, \vec{y}$ for generic $\vec{y}$,
we have the requirement
\begin{align}
 \stackrel[{\tau \to 0}]{}{\lim}  \tilde{A}(\vec{x}_{0} + \tau \, \vec{y} ) \sim B \log(\tau) \,,
\end{align}
for some matrix $B$. In other words $\tilde{A}(\vec{x})$ {\it locally} behaves as a logarithm.
This however leaves open the possibility that {\it globally} $\tilde{A}(\vec{x})$ is more complicated.

The first example of such integration kernels occurs in the so-called sunrise integral, see \cite{ch4Bourjaily:2022bwx} and references therein.
The special functions one finds are multiple elliptic polylogarithms and generalisations thereof.
It is an active topic of research how to best think about these functions,
in particular in terms of canonical differential equations~\eqref{eq:canonicalDEmanysingularpointsmultiplevariables3} with specific differential forms.
An open question is what form such equations take, generalising \eqn{eq:canonicalDEmanysingularpointsmultiplevariables},
but being more specific than the very general form~\eqref{eq:canonicalDEmanysingularpointsmultiplevariables3}.

\begin{important}{Canonical differential equations are a useful language for describing the invariant information content of Feynman integrals.}
Providing the alphabet of integration kernels, together with the corresponding constant matrices, i.e.\ the sets $\alpha_k$ and $A_k$ of \eqn{eq:canonicalDEmanysingularpointsmultiplevariables}, is arguably the neatest way of encoding what the Feynman integral actually is. This is also very flexible, thanks to the homotopy invariance of the solution~\eqref{eq:Wilsonlinesolution}. The latter is completely analytic, and can be a good starting point for the numerical evaluation as well, see e.g.~\cite{ch4Chicherin:2021dyp}.
\end{important}

We conclude this survey of special functions relevant for Feynman integrals. Readers may wonder how these ``thought experiments'' are actually operationalised for Feynman integrals. This question will be answered in section~\ref{sec:DEFeynmanintegrals}. Here we just satisfy a first curiosity: what is the weight-counting parameter for Feynman diagrams, and does it have anything to do with the $\eps$ in dimensional regularisation? And if so, how could this possibly make any sense? The answer is yes, and it actually turns out that the dimensional regularisation parameter can naturally be thought of as having transcendental weight $-1$. The reason is that a pole $1/\eps$ in dimensional regularisation could equivalently be described by $\log \Lambda$, where $\Lambda$ is some cutoff. For this reason it is natural to identify $\eps$ from our toy example above with the dimensional regularisation parameter!

\section{Differential equations for Feynman integrals}
\label{sec:DEFeynmanintegrals}

In this section we explain the differential equations method for computing Feynman integrals.
The main steps are to obtain the relevant differential equations, and then to put them into a convenient form that makes it easy to solve for them---the canonical form we encountered in the previous section. For the first step algorithms have existed for a long time, and we follow here the strategy introduced by~\cite{ch4Laporta:2000dsw}.
Novel ideas stemming from the experience with loop integrals in supersymmetric theories have streamlined the second step.

The upcoming subsections will delve into the method's details, but let us anticipate here the main steps:
\begin{enumerate}
\item Define a ``family''  of Feynman graphs of interest (cf.~section~\ref{sec:integral_family}).
\item Write down the integration by parts identities (cf.~section~\ref{sec:integral_family}).
\item Find a basis of so-called master integrals (cf.~section~\ref{sec:integral_family}).
\item Set up differential equations for the basis integrals (cf.~section~\ref{subsec:obtainingDE}).
\item Perform consistency checks (cf.~section~\ref{sec:dimensionalanalysis}).
\item Choose a good integral basis (cf.~sections~\ref{sec:canonical} and~\ref{sec:dlog}).
\item Transform the differential equations to canonical form (cf.~section~\ref{sec:canonical}).
\item Fix the boundary values (cf.~section~\ref{subsec:solvingDE}).
\item Solve the differential equations (cf.~section~\ref{subsec:solvingDE}).
\end{enumerate}

 \subsection{Organisation of the calculation in terms of integral families}
 \label{sec:integral_family}

Let us illustrate how to organise a general calculation using the massive bubble integrals (cf.\ figure~\ref{fig:feynbubbleandtriangle-bubble}) as an example.
As explained in chapter~\ref{ch:loopamps}, it is sufficient to consider scalar integrals.
Given a Feynman diagram, it turns out to be useful to define an \emph{integral family}, where the propagators are raised to arbitrary (integer) powers:
\begin{align}\label{deffamily}
G_{a_1, a_2}(s,m^2;D) = \int \frac{\d^{D}k}{\i \pi^{D/2}} \frac{1}{[-k^2+m^2]^{a_1} [-(k+p)^2+m^2]^{a_2}} \,,
\end{align}
where we omitted the $\i 0$ prescription, and we recall that $s=p^2$.
It turns out that integrals with different values of propagator powers $(a_1, a_2)$ satisfy linear relations.
One can define a (finite-dimensional) basis in this space. The basis elements are called \emph{master integrals}.

\smallskip

 {\it Integration-by-parts identities in momentum space}.
We have that
\begin{align}
\int \frac{\d^{D}k}{\i \pi^{D/2}}  \frac{\partial}{\partial k^{\mu}} \left\{ v^\mu \frac{1}{[-k^2+m^2]^{a_1} [-(k+p)^2+m^2]^{a_2}} \right\} = 0\,,
\end{align}
for any four-vector $v^\mu$.
This follows simply from integrating by parts the total derivative. The boundary terms at infinity vanish, at least for some range of $a_1, a_2, D$, and, by analytic continuation, everywhere.
Writing this equation for $v^\mu=k^\mu$
yields the following integration-by-parts (IBP) relation:
\begin{align}\label{IBP1}
\begin{aligned}
0 = \ & (D-2 a_1 - a_2 ) \, G_{a_1 , a_2} - a_2 \, G_{a_1 - 1,a_2 + 1}  \\
 & +2 m^2 a_1 \, G_{a_1+1,a_2} + (2 m^2-s) \, a_2 \, G_{a_1, a_2 + 1} \,.
 \end{aligned}
\end{align}
A second relation follows from $v=k+p$ or, equivalently, from noticing that 
\begin{align}\label{IBPsymmetry}
G_{a_1, a_2} = G_{a_2, a_1}\,,
\end{align} 
by symmetry.

\smallskip

 {\it Master integrals and basis choice}.
 From the IBP relation (\ref{IBP1}) and its symmetric version it follows that there are two master integrals (MIs). 
In practice, one generates a system of identities for a range of values $(a_1, a_2)$, and then performs a Gauss elimination procedure (with some ranking, e.g.\ preferring integrals with lower indices $a_i$, or with lower $a_1+a_2$)~\cite{ch4Laporta:2000dsw}.

The master integrals can be chosen for example as
\begin{align}\label{choicemaster}
\{ G_{1,1} , G_{0,1} \} \,.
\end{align}
The statement that these two integrals are master integrals, or basis integrals, means that for any $a_1,a_2$ there exist some $c_1, c_2$, such that
\begin{align}
G_{a_1,a_2} = c_1 \, G_{1,1} + c_2 \, G_{0,1} \,.
\end{align}
The $c_i$ are rational functions in $m^2, s, D$.

For example, we have (setting $D=2-2\epsilon$ without loss of generality)
\begin{align}\label{IBPexamples}
& G_{2,0} = \frac{\epsilon}{m^2} \, G_{0,1} \,, \\
& G_{2,1} = \frac{\epsilon}{m^2 (4 m^2-s)} \, G_{0,1} + \frac{ 1+2 \eps}{4m^2-s} \, G_{1,1} \,. 
\end{align}
A number of comments are due.
\begin{itemize}
\item There exist several computer algebra implementations and publicly available codes for generating and solving IBP relations. 
See exercises~\ref{Ex:4.BubbleDE} and~\ref{Ex:4.BoxDE} for examples.
\smallskip
\item The number of master integrals can be determined in various ways~\cite{ch4Lee:2013hzt,ch4Bitoun:2017nre}. 
Note however that what is counted exactly, i.e.\ what is meant by the number of master integrals, may differ depending on the reference.
In general it is advisable to compute this number, and then compare with the result obtained from analysing the IBP relations.
\smallskip
\item It is useful to organise master integrals according to their number of propagators. One speaks of integral ``sectors''. One useful feature is that integral sectors correspond to certain blocks in the differential equations satisfied by the integrals.
For example, the ``tadpole'' integrals form a subsector within the bubble integral family. We will see this explicitly in subsection~\ref{subsec:obtainingDE}.

\smallskip
\item The choice of master integrals is important, and can significantly impact how easy or complicated a calculation is. In section~\ref{sec:dlog} we introduce a method for choosing the master integrals wisely, motivated by the properties of transcendental functions discussed in section~\ref{ref:subsection:propertiesofspecialfunctions}.
\end{itemize}

\begin{exer}{The massless two-loop kite integral}
\label{Ex:4.Kite}
Consider the following massless two-loop Feynman integral, 
\begin{equation} \label{eq:kite_def}
   F_{\text{kite}}\bigl(s; D\bigr) = \, \vcenter{\hbox{\includegraphics[width=3.cm]{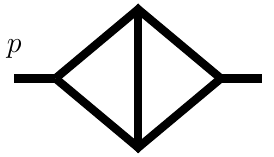}}}  \,.    
\end{equation}
All propagators are massless, and $s=p^2$.
Define the corresponding integral family. Write down the integration by parts identities for one of the triangle sub-integrals, and use them to express $F_{\text{kite}}\left(s; D\right)$ in terms of one-loop bubble integrals. Use the formula~\eqref{eq:Ibubblechapter4} to rewrite the latter in terms of $\Gamma$ functions, and show that
\begin{align} \label{eq:kite_result}
 F_{\text{kite}}\left(s; 4-2\eps \right) = \frac{6 \, \zeta_3}{-s} + \mathcal{O}\left(\eps\right) \,.
\end{align} 
For the solution see \hyperref[Sol:4.Kite]{chapter~5}.
\end{exer}

  \subsection{Obtaining the differential equations}
  \label{subsec:obtainingDE}
We know that for the bubble integral family~\eqref{deffamily} there are two master integrals, which can be chosen as in \eqn{choicemaster}.
We wish to know the derivatives of these integrals, as this would amount to knowing the derivative of any integral in the~family.

For any integral of the form~\eqref{deffamily}, it is straightforward to compute the derivative w.r.t.~$m^2$.
For example, we have
\begin{align}
\partial_{m^2} G_{a_1,a_2} = -a_1 G_{a_1 + 1,a_2} - a_2 G_{a_1, a_2 + 1 } \,.
\end{align}
Applying this for the two master integrals, we simply have
\begin{align}
\partial_{m^2} \begin{pmatrix}
G_{0,1}\\
G_{1,1} 
\end{pmatrix}
=
\begin{pmatrix}
-G_{2,0}\\
-2 \, G_{2,1} 
\end{pmatrix}\,.
\end{align}
Then, using the IBP relations~\eqref{IBPexamples}, we find
\begin{align}\label{DEexample1}
\partial_{m^2} \begin{pmatrix}
G_{0,1} \\
G_{1,1}
\end{pmatrix}
=
\left[
\begin{pmatrix}
0 & 0 \\
0 & \frac{-2}{4 m^2-s} 
\end{pmatrix}
+
\epsilon
\begin{pmatrix}
\frac{-1}{m^2} & 0 \\
\frac{-2}{m^2(4m^2-s)} & \frac{-4}{4 m^2-s} 
\end{pmatrix}
\right] \cdot
\begin{pmatrix}
G_{0,1} \\
G_{1,1}
\end{pmatrix}
\,.
\end{align}
Similarly, one can obtain the differential equations w.r.t.~$s$ by using $ \partial_s = 1/(2 s) p^{\mu} \partial_{p^{\mu}}$.
One finds
\begin{align}\label{DEexample1s}
\partial_{s} \begin{pmatrix}
G_{0,1} \\
G_{1,1}
\end{pmatrix}
=
\left[
\begin{pmatrix}
0 & 0 \\
0 & \frac{s-2 m^2}{s(4 m^2-s)} 
\end{pmatrix}
+
\epsilon
\begin{pmatrix}
0 & 0 \\
\frac{2}{s(4m^2-s)} & \frac{1}{4 m^2-s} \end{pmatrix}
\right] \cdot
\begin{pmatrix}
G_{0,1} \\
G_{1,1}
\end{pmatrix}
\,.
\end{align}

\subsection{Dimensional analysis and integrability check}
\label{sec:dimensionalanalysis}

There are two simple consistency checks we can perform of the differential equation matrices just obtained. 

\smallskip

{\it 1. Scaling relation}.
The integral $G_{a_1,a_2}(s,m^2;D)$ has overall (mass) dimension $D-2 a_1 - 2 a_2$. In other words, one can write
\begin{align}\label{eq:dimension}
G_{a_1,a_2}(s,m^2;D) = m^{D-2 a_1 -2 a_2} \, g(s/m^2;D)\,,
\end{align}
for some function $g$.
This implies the differential equation 
(dilatation relation)
\begin{align}
\left[ s \, \partial_s + m^2 \partial_{m^2} \right] G_{a_1,a_2} = (D/2-a_1-a_2) \, G_{a_1,a_2} \,.
\end{align}
Indeed, applying this differential operator to the massive bubble example, and using eqs.~\eqref{DEexample1} and~\eqref{DEexample1s}, we find
\begin{align}\label{eq:bubblerescaling}
 \left[ s \, \partial_s + m^2 \partial_{m^2} \right] 
 \begin{pmatrix}
G_{0,1} \\
G_{1,1}
\end{pmatrix} =
\begin{pmatrix}
-\eps & 0 \\
0 & -1-\eps 
\end{pmatrix} \cdot
\begin{pmatrix}
G_{0,1} \\
G_{1,1}
\end{pmatrix}\,.
\end{align}
Equation~\eqref{eq:bubblerescaling} is as expected. It is a diagonal matrix, with the diagonal entries corresponding to the scaling dimensions, measured in units of the dimension of $m^2$, cf.\ \eqn{eq:dimension}. The latter can be verified by dimensional analysis of the original definition in terms of loop integrals.

We remark that one could modify the definition of the master integrals, by simply rescaling them with a dimensional prefactor, to set their overall scaling dimension to zero. In our case $(m^2)^{\eps}$ and $(m^2)^{1+\eps}$ would achieve this. This would allow us to talk about single-variable differential equations in the variable $s/m^2$, as in \eqn{eq:dimension}.
However, in general we prefer not to include fractional terms such as $(m^2)^{\eps}$ in the definition, as this may obscure physical properties, e.g.\ when considering a limit $m\to 0$ or $m\to \infty$. Moreover, as we shall see, within the setup proposed here, dealing with multiple variables is not substantially more complicated as acompared to one variable.

\smallskip

{\it 2. Integrability conditions}.
A second check follows from the commutativity of partial derivatives, in our case $\partial_s \partial_{m^2}  -\partial_{m^2} \partial_{s} = 0$.
Applying this to our basis of master integrals, we get
\begin{align}\label{eq:integrability}
\left( \partial_{s} A_{m^2} - \partial_{m^2} A_{s} + A_{m^2} \cdot A_{s} - A_{s} \cdot A_{m^2} \right) \cdot \begin{pmatrix}
G_{0,1} \\
G_{1,1}
\end{pmatrix} =0 \,.
\end{align}  
One can verify indeed that the matrix appearing in \eqn{eq:integrability} vanishes identically.

\smallskip 

We close this subsection with a comment. Whenever one of the two checks discussed here fails, e.g.\ when one gets a non-vanishing matrix on the RHS of \eqn{eq:integrability}, this most likely points to some mistake in a calculation or implementation step. However, note that \eqn{eq:integrability} also admits solutions for non-vanishing matrices on the right-hand side if the master integrals are not linearly independent. It can indeed happen in practice that there are ``hidden'' IBP relations (that would e.g.\ be discovered by considering a larger set of IBP relations). In this case these checks may give hints for such missing relations. Note however that the converse is not true: successful scaling and integrability tests do not guarantee that one has found all IBP relations.

\subsection{Canonical differential equations}
\label{sec:canonical}

The differential equations~\eqref{DEexample1} and~\eqref{DEexample1s} are already rather simple,
however by comparing to \eqn{eq:canonicalDEmanysingularpointsmultiplevariables} we see that they are not yet quite in canonical form.
In particular, they contain a $\eps^0$ term.
We will see in section~\ref{sec:dlog} how to directly find canonical differential equations but, for now,
let us proceed in a more pedestrian way.
We may attempt to ``integrate out'' the unwanted $\eps^0$ term, by changing basis from
\begin{align}
\vec{g} = 
 \begin{pmatrix}
G_{0,1}\\
G_{1,1}
\end{pmatrix}
\end{align}
to
\begin{align}
\vec{f} = T \cdot \vec{g}\,,
\end{align}
for some suitable invertible matrix $T$.
The differential equations for the new basis $\vec{f}$ are governed by \eqn{eq:defB}.
Demanding that this matrix is free of $\eps^0$ terms leads us to the following auxiliary problem:
\begin{align}
\partial_{m^2} T
=
- T  \cdot 
\begin{pmatrix}
0 & 0 \\
0 & \frac{-2}{4 m^2-s} 
\end{pmatrix}
\,,
\qquad 
\partial_{s} T
=
- T \cdot
\begin{pmatrix}
0 & 0 \\
0 & \frac{-2 m^2+s}{s(4 m^2-s)} 
\end{pmatrix}
 \,.
\end{align} 
This leads to the transformation matrix
\begin{align}
T = 
\begin{pmatrix}
1 & 0 \\
0 &\sqrt{(-s) (4m^2-s)}  
\end{pmatrix}
 \,,
\end{align}
and hence to the following new basis\footnote{In section~\ref{sec:dlog} we will see that this basis can be motivated in an entirely different way, without the need to analyse differential equations.}
\begin{align}\label{eq:UTbasisbubble}
\vec{f} = \begin{pmatrix}
G_{0,1}\\
\sqrt{(-s) (4m^2-s)} \, G_{1,1}
\end{pmatrix}\,.
\end{align}
Assuming $s<0,m^2>0$, one finds
\begin{align}\label{DEexample2}
\partial_{m^2} \, \vec{f} 
=
\epsilon \,
\begin{pmatrix}
\frac{-1}{m^2} & 0 \\
\frac{-2}{m^2 \sqrt{1-4 m^2/s}}  & \frac{-4 }{4 m^2-s}  
\end{pmatrix}
\cdot \vec{f} 
\,.
\end{align}
There is a similar equation for $\partial_s$.

So, structurally we have two differential equations 
\begin{align}
\partial_{m^2} \, \vec{f} = \eps\, A_{m^2} \cdot \vec{f} \,, \qquad
\partial_s \, \vec{f} = \eps\, A_{s} \cdot \vec{f} \,.
\end{align}
The two partial derivative equations can be combined in a single equation using the total differential $\d = \d s \, \partial_s + \d{m^2} \, \partial_{m^2}$.
Then we have
\begin{align}\label{canonical1}
\d \, \vec{f} =  \eps\,  (\d \tilde{A}) \cdot \vec{f} \,,
\end{align}
provided that $\tilde{A}$ satisfies
\begin{align}
\partial_{m^2} \, \tilde{A} = A_{m^2}  \,, \qquad
\partial_s \, \tilde{A} = A_{s}  \,.
\end{align}
We find the following $\tilde{A}$ solves these equations,
\begin{align}\label{canonical2}
\tilde{A} =   \begin{pmatrix}
-\log m^2 & 0 \\
-2 \log\left( \frac{\sqrt{1-4 m^2/s}-1}{ \sqrt{1-4 m^2/s} +1 } \right)  & - \log( 4 m^2-s )    
\end{pmatrix}  \,.
\end{align}
Equations~\eqref{canonical1} and~\eqref{canonical2} are an example of {\it canonical} differential equations for Feynman integrals~\cite{ch4Henn:2013pwa}.
The specific form~\eqref{canonical2} is an instance of the general case~\eqref{eq:canonicalDEmanysingularpointsmultiplevariables}, with $N=2$.
There are three alphabet letters, namely
\begin{align}\label{eq:alphabetbubble}
    \left\{ m^2, \frac{\sqrt{1-4 m^2/s}-1}{ \sqrt{1-4 m^2/s} +1 }, 4m^2-s \right\} \,.
\end{align}

\subsection{Solving the differential equations}
\label{subsec:solvingDE}

{\it General solution to the differential equations}.
Here we discuss how to solve the canonical differential equations.
We had already seen in section~\ref{sec:sepcialfunctionsfromDE} that the general solution takes the form of a path-ordered exponential, cf.~\eqref{eq:Wilsonlinesolution}. Adopting this equation to the present case, we have 
\begin{align}\label{eq:Wilsonlinesolutionbubble}
\vec{f}(\vec{x};\eps)  = {\mathbb P}\exp \left[\eps \int_{\cal C} \d \tilde{A}(\vec{x}') \right]  \cdot \vec{f}(\vec{x}_0;\eps) \,,
\end{align}
with $\tilde{A}$ from \eqn{canonical2}, and where $\vec{x}$ refers collectively to the set of kinematic variables $\vec{x}=(s,m^2)$, and $\vec{x}_0$ is an arbitrary base point for the integral, which takes the value $\vec{f}(\vec{x}_0;\eps)$ there. This corresponds to the fact that the system of first-order equations uniquely fixes the answer up to a boundary condition. We will discuss this presently.

There are simplifications thanks to the fact that the matrix in the exponent on the RHS of \eqn{eq:Wilsonlinesolutionbubble} is proportional to $\eps$. Therefore we can expand the exponential perturbatively in $\eps$. Moreover, due to the fact that the matrix on the RHS (cf.~\eqn{canonical2}) contains only logarithmic integration kernels, 
the answer are iterated integrals with the alphabet of \eqn{eq:alphabetbubble}.
In fact, we shall see presently that the answer up to the finite part is written in terms of much simpler functions. 
But this is not essential. The main message is that the class of special functions at our disposal is large enough to express the general solution to \eqn{canonical1} with \eqn{canonical2}.

\smallskip
In \eqn{eq:Wilsonlinesolutionbubble}, $\vec{f}(\vec{x}_0;\eps)$ is a boundary vector at a given base point $\vec{x}_{0}$.
As such, \eqn{eq:Wilsonlinesolutionbubble} expresses the general solution to the differential equations.
In most cases, one is interested in the specific solution that corresponds to the Feynman integrals at hand.
This means that it is necessary to provide a boundary condition.
In other words, for a Feynman integral depending on multiple variables $\vec{x}$, one needs to know its value at one specific point $\vec{x}_{0}$.

\begin{important}{Fixing the boundary conditions from physical consistency conditions.}
One might naively think that a completely separate calculation is needed for this.
However, experience shows that one can obtain the boundary information from physical consistency conditions.
This approach is well known in the literature, but turns out to be especially easy within the canonical differential equations approach, 
which moreover offers additional insights~\cite{ch4Henn:2020lye}.
As a result, in most calculations this allows one to fix all integration constants, up to an overall normalisation.
\end{important}

The key is to consider the behaviour near singular points (or rather singular kinematic subvarieties) of the differential equations.
The singular points are  easily identified from the alphabet~\eqref{canonical2}. They correspond to kinematic configurations where alphabet letters tend to zero or infinity. In our case, this corresponds to $s=0, s=m^2, s=\infty, m^2=0, m^2 =\infty$.

In the present case, it turns out that $s=0$ is a suitable boundary configuration. The reason is that, physically, one knows that this limit is non-singular (due to the presence of the internal mass). In other words, we can simply set $p=0$ in \eqn{defbubble}. This reduces the bubble integral to a tadpole integral. 
However, since its normalisation factor in \eqn{eq:UTbasisbubble} vanishes, we do not even need to know its value. 
This fixes the boundary constant at $s=0$, up to the value of the tadpole integral. 
The calculation of the latter is elementary, with the result 
\begin{align}
G_{0,1}= \Gamma(\eps) \, \bigl(m^2\bigr)^{-\eps} \,,
\end{align}
which follows from \eqn{single_propagator} with $a=1$ and $D=2 -2 \eps$.
Therefore the boundary condition is
\begin{align}\label{boundarybubble1}
\vec{f}(s=0,m^2;D=2-2\eps) = 
 \begin{pmatrix}
\Gamma(\eps) (m^2)^{-\eps} \\
0
\end{pmatrix}\,.
\end{align}
This fixes the answer of the differential equation to all orders in $\eps$.

\smallskip

{\it Solution in terms of multiple polylogarithms}.
The alphabet \eqn{eq:alphabetbubble} can be rationalised using a simple change of variables.
Indeed, setting $s = -m^2 (1-x)^2/x$, and assuming $0<x<1$, \eqn{eq:alphabetbubble} becomes 
\begin{align}\label{eq:alphabetbubble2}
    \left\{ m^2, x , m^2 \frac{(1+x)^2}{x}  \right\} \,,
\end{align}
i.e.\ the alphabet, written in the independent variables $m^2$ and $x$ is simply
\begin{align}
    \left\{ m^2, x, 1+x \right\} \,.
\end{align}
This means that the answer can be written in terms of a special subclass of iterated integrals, called harmonic polylogarithms.\footnote{For more information on particular classes of special functions, and how to handle them efficiently, we refer interested readers to~\cite{ch4Duhr:2019wtr} and references therein.}
Moreover, the dependence on $m^2$ in the new alphabet becomes trivial, as it corresponds to the overall scale. We can therefore set $m^2=1$ without loss of generality, and solve the equations as a function of $x$ only. Equivalently, we could multiply all integrals by $(m^2)^{-\eps}$.

With this in mind, let us make the following final basis choice (the normalisation is motivated by \eqn{boundarybubble1}):
\begin{align}\label{eq:UTbasisbubblefinal}
\vec{f}(x;\eps) \coloneqq \frac{1}{(m^2)^{-\eps} \Gamma(\eps)}  \begin{pmatrix}
G_{0,1}\\
\sqrt{(-s) (4m^2-s)} \, G_{1,1}
\end{pmatrix}\,.
\end{align}
It satisfies the differential equations
\begin{align}\label{DEbubblefinalAtilde} 
\d \, \vec{f}(x;\eps) = \eps \, \d
 \begin{pmatrix}
0 & 0 \\
-2 \log x  & \log \frac{x}{(1+x)^2}  
\end{pmatrix} \cdot
 \vec{f}(x;\eps)\,,
\end{align}
with the boundary condition
\begin{align}\label{DEbubblefinalbdry}
\vec{f}(1;\eps)  =  \begin{pmatrix}
1\\
0
\end{pmatrix}\,.
\end{align}
More explicitly, \eqn{DEbubblefinalAtilde} is
\begin{align}\label{DEbubblefinalAtilde2} 
\partial_x \vec{f}(x;\eps) = \eps  \left[ \frac{1}{x} 
 \begin{pmatrix}
0 & 0 \\
-2   & 1   
\end{pmatrix} 
+
\frac{1}{1+x}
 \begin{pmatrix}
0 & 0 \\
0  & -2   
\end{pmatrix} 
\right] \cdot
 \vec{f}(x;\eps)\,.
\end{align}
We can now solve this equation, together with the boundary condition~\eqref{DEbubblefinalbdry}, order by order in $\eps$.
To do so, we set
\begin{align}
 \vec{f}(x;\eps) = \sum_{k \ge 0} \eps^k  \vec{f}^{(k)}(x) \,,
\end{align}
up to some order in $\eps$.
The key point is that the equations~\eqref{DEbubblefinalAtilde} decouple order by order in $\eps$,  when expressed in terms of $ \vec{f}^{(k)}(x)$.
For the first few orders, we straightforwardly find
\begin{align} 
\vec{f}^{(0)}(x) =  \begin{pmatrix}
1  \\
0     
\end{pmatrix} \,,
\end{align}
and 
\begin{align}\label{solbubbleweight1}
\vec{f}^{(1)}(x) =  \begin{pmatrix}
0  \\
-2 \log x     
\end{pmatrix} \,,
\end{align}
and 
\begin{align} \label{solbubbleweight2}
\vec{f}^{(2)}(x) =  \begin{pmatrix}
0  \\
4 \, {\rm Li}_{2}(-x) +4 \log x \log(1+x) -\log^2 x+\pi^2/3 
\end{pmatrix} \,.
\end{align} 
Recalling the definition~\eqref{eq:UTbasisbubblefinal},
this gives
\begin{align}
F_{2}\bigl(s,m^2;D=2- 2\eps\bigr) = 
\frac{\Gamma(1+\eps) (m^2)^{-\eps}}{\sqrt{(-s)(4 m^2-s)}} \left[   
- 2  \log\left( \frac{\sqrt{1-4 m^2/s}-1}{ \sqrt{1-4 m^2/s} +1 } \right) + {\cal O}(\eps) 
  \right] \,.
\end{align}
This agrees with \eqn{resbubbleexplicit}.
Furthermore, \eqn{solbubbleweight2} provides the next term in the $\eps$ expansion, and higher-order terms can be straightforwardly generated.

\newpage

\begin{exer}{The massive bubble integrals with the differential equations method}
\label{Ex:4.BubbleDE}
Write a \texttt{Mathematica} notebook for computing the massive bubble integrals~\eqref{deffamily} with $D=2-2\eps$ following step by step the discussion in section~\ref{sec:DEFeynmanintegrals}.
Use the package \textsc{LiteRed}~\cite{ch4Lee:2012cn} to perform the IBP reductions and differentiate the integrals. 
For the solution see the \texttt{Mathematica} notebook \href{https://scattering-amplitudes.mpp.mpg.de/scattering-amplitudes-in-qft/Exercises/}{\tt Ex4.12\_BubbleDE.wl}~\cite{ch4website-ch4}.

\end{exer}

\section{Feynman integrals of uniform transcendental weight}
\label{sec:dlog}

In the previous subsections, we have discussed special functions appearing in Feynman diagrams.
We have seen in the preceding chapters that they satisfy simple, canonical differential equations. 
However, we have also seen that the equations have a gauge degree of freedom, which corresponds to making a basis choice for the coupled system of $N$ equations. This freedom implies that the DE may be written in an equivalent, albeit unnecessarily complicated form. Therefore an important question is: how can we make sure that the equations we get for Feynman integrals will have the desired simple, canonical form?

We address this problem in this section. This builds on the observations of section~\ref{subsec:comments} that pure uniform weight functions satisfy canonical differential equations. In this section we present a conjecture that gives a criterium for when a Feynman integral evaluates to such functions. Taken together, this provides the information for how to choose a set of Feynman integrals that satisfy canonical differential equations.

\subsection{Connection to differential equations and (unitarity) cuts}
\label{sec:DEandCuts}

We have already seen in section~\ref{subsec:comments} that, in the context of the differential equations satisfied by Feynman integrals, it is natural to consider discontinuities.
It turns out that there is a useful connection to (generalised) unitarity methods.

This is based on the important observation that unitarity cuts of an integral satisfy the same differential
equations, albeit with different boundary conditions.
We can see this in the following way~\cite{ch4Anastasiou:2002yz}.
When we ``cut'' a propagator, we effectively replace $1/(-k^2+m^2)$ by $\delta(-k^2+m^2)$.
As was discussed in section~\ref{sec:3.3} of chapter~\ref{ch:loopamps},
the cut can be written as a difference of propagators with different $\i 0$ prescriptions.
However, this prescription does not affect the derivation of the IBP relations, and hence one gets the same differential equations.
Of course, the boundary values of the cut integrals differ from the original ones, and in particular one could put to zero integrals that obviously vanish on a cut because the relevant cut propagators are absent.

Consider the one-loop massive bubble integrals $G_{1,1}$ and $G_{1,0}$ of \eqn{defbubble} as an example.
Applying an $s$-channel unitarity cut, and setting\footnote{One key feature of the canonical differential equations is that their RHS vanishes for $\eps=0$.
It is therefore interesting to ask whether we can solve for the $\eps=0$ part of the differential equations. However, it is also interesting (albeit more involved) to study cut integrals for general $D$.} $D=2$ we obtain
\begin{align}\label{eq:B11cutintegral}
G_{1,1} \longrightarrow \int  \d^{2}k \, \delta\bigl(-k^2+m^2 \bigr) \, \delta\bigl(- (k+p)^2+m^2 \bigr)  = \frac{1}{\sqrt{(-s)(4 m^2-s)}} \,.
\end{align}
This is exactly the factor we introduced in \eqn{eq:UTbasisbubble} to remove the $\eps=0$ part of the differential equation.
Performing the same cut on the tadpole integral $G_{1,0}$ vanishes, simply because it is
missing one propagator that was cut. 
Indeed, by cutting all propagators---whose number happens to coincide here with the space-time dimension $D=2$---we are taking a so-called \emph{maximal cut}. This allows us to focus on a given integral sector.
In terms of the differential equations, this means that this cut describes a block of the differential equations,
whose size corresponds to the number of relevant master integrals.
In the present case, there is just one master integral with two propagators, so the block corresponds of one element
of the differential equations.

We can also learn something about the tadpole integral, by considering generalised unitarity cuts. 
The most obvious is to cut the one propagator that is present. It is instructive to do this:
\begin{align}\label{eq:B10cutintegral}
G_{1,0} \longrightarrow \int  \d^{2}k \, \delta\bigl(-k^2+m^2\bigr)   \,.
\end{align}
It is convenient to introduce two light-like vectors $p_1$ and $p_2$, with $p_1^2 = p_2^2 = 0$, such that $p=p_1 + p_2$, and $2 \, p_1 \cdot p_2 = s$.
This allows us to parametrise $k= \beta_1 p_1 + \beta_2 p_2$. 
Taking into account the Jacobian from the change of variables, and using the delta function to fix one integration, we find
\begin{align}
G_{1,0} \longrightarrow \int \d\beta_1 \d\beta_2 \, s \, \delta\bigl(-s \beta_1 \beta_2 + m^2\bigr) = \int \frac{\d\beta_2}{\beta_2} \,.
\end{align}
Unlike \eqn{eq:B11cutintegral}, here the integrations are not fully fixed.
Moreover, the first integration has produced a new singularity, $1/\beta_2$, that was not present initially.
This is a typical situation in generalised unitarity. 
We can define a further cut and localise the integral at one of its singularities, either $\beta_2= 0$ or $\beta_2 = \infty$.
The result is just $\pm 1$, which confirms the normalisation choice in \eqn{eq:UTbasisbubble}.

\subsection{Integrals with constant leading singularities and uniform weight conjecture}

The calculations above provided useful insight about diagonal blocks of the differential equations (in the present case, the diagonal elements, since the block size happened to be one). Can one also learn something about the off-diagonal terms with this method?
The answer---conjecturally---is yes. The idea is first to think of the above calculations not of changing the integrand, but instead of changing the integration contour.
The original integration is over two-dimensional Minkowski space. In the calculations above, we have effectively replaced this by certain residue calculations.
The idea then is to generalise this to arbitrary residues we can take for a given loop integrand.
In the case where those residues completely localise the loop integrations, one speaks specifically of \emph{leading singularities}. 
We know that leading singularities that correspond to maximal cuts inform us about diagonal blocks of the differential equations. 
The assumption is that the other residues ``know about'' the off-diagonal parts, even though we do not know the precise mapping.
However, for the present purposes, this precise map is not relevant: if we can normalise all leading singularities 
to constants, then their derivatives will obviously be trivial. 
This gives us a useful tool for obtaining differential equations whose RHS is proportional to $\eps$.

How does one see that all leading singularities are kinematic-independent constants?
Let us review the tadpole and bubble integrands from this viewpoint.
As explained, we now focus on the {\it integrand},
\begin{align}
    \omega_{1,0} = \frac{\d^2k}{-k^2+m^2} \propto \frac{\d\beta_1 \d\beta_2 \, s}{-s \beta_1 \beta_2 + m^2} \,,
\end{align}
where we have used the same loop parametrisation as above.
Integrating one variable at a time, we can write this in the following way,
\begin{align}
\omega_{1,0} \propto \d \log \bigl(-s \beta_1 \beta_2 + m^2 \bigr) \, \d\log \bigl(\beta_2 \bigr) \,.
\end{align}
Here $\d= \d\beta_1 \, \partial_{\beta_1} + \d\beta_2 \, \partial_{\beta_2}$. The differential forms satisfy $\d\beta_i \d\beta_j = - \d\beta_j \d\beta_i$, 
and hence e.g.~$\d\beta_2 \d\beta_2 =0$.
In other words, upon further changing variables, the form~reads
\begin{align} \label{eq:omega10dlog}
\omega_{1,0} \propto \frac{\d\tau_1}{\tau_1} \frac{\d\tau_2}{\tau_2} \,.
\end{align}
This makes it clear that any leading singularity of this integral evaluates to a constant, and hence that any of its derivatives vanishes, as desired.

Repeating this analysis for the bubble is slightly more interesting. We leave it as an exercise to readers.
The result can be written as
\begin{align} \label{eq:bubble_dlog} \begin{aligned}
    \omega_{1,1} = \ & \frac{\d^2k}{(-k^2+m^2)[-(k+p)^2+m^2]}\,, \\
     \propto &  \frac{1}{\sqrt{(-s)(4m^2-s)}} \, \d\log\left[\frac{-k^2+m^2}{(k-k_{\pm})^2} \right] \, \d\log\left[\frac{-(k+p)^2+m^2}{(k-k_{\pm})^2} \right] \,.
\end{aligned} \end{align}
Here $k_{\pm}$ is any of the two solutions to the cut equations $-k^2+m^2=-(k+p)^2+m^2=0$.
This expression makes it manifest that $\omega_{1,1}$ has only logarithmic singularities (i.e., of type $\d x/x$), and its maximal residue (i.e., its leading singularity) is 
\begin{align}\label{massiveleadingsing}
    \oint \omega_{1,1} \propto \frac{1}{\sqrt{(-s)(4m^2-s)}} \,.
\end{align}
Therefore we conclude that the integral $ {\sqrt{(-s)(4m^2-s)}} \, \int  \omega_{1,1}$ is a good basis integral that may lead to canonical differential equations. 
Indeed, note that this is consistent with \eqn{resbubbleexplicit}.

\begin{exer}{``$\d \log$'' form of the massive bubble integrand with $D=2$}
\label{Ex:4.BubbleCut}
Use the parameterisation introduced above ($k=\beta_1 p_1 + \beta_2 p_2$) to prove that the integrand of the massive bubble in $D=2$ dimensions can be expressed as a ``$\d \log$'' form. Show that the latter is equivalent to the momentum-space $\d \log$ form in \eqn{eq:bubble_dlog}. 
For the solution see \hyperref[Sol:4.BubbleCut]{chapter~5}.
\end{exer}

Interestingly, the question of which Feynman integrals evaluate to uniform weight functions was previously studied 
independently from the differential equations.
Understanding initially came from studies of scattering amplitudes in ${\mathcal N} =4$ super Yang-Mills theory, 
but it turned out that the observations made there were applicable more generally~\cite{ch4Arkani-Hamed:2010pyv,ch4Henn:2013pwa}. 
This led to the following conjecture. A Feynman integral integrates to a pure function if 
\begin{enumerate}
\item its integrand, and iterated residues thereof, only contains simple poles,
\item the maximal residues are normalised to constants.
\end{enumerate}
These two criteria are in one-to-one correspondence to the properties discussed above. 
The first requirement is intended to remove less than maximal weight functions, and therefore lead to integrals with uniform and maximal weight. 
See exercise~\ref{Ex:4.KiteLS} for an example of an integrand with a double pole, which is a sign for a weight-drop. 
As computed explicitly in exercise~\ref{Ex:4.Kite}), the answer has weight three, which is less than the maximal weight four expected for two-loop integrals in four dimensions.
The second requirement addresses another potential problem: if an integral is a sum of maximal weight functions with different algebraic prefactors, 
this spoils the desired simple structure under differentiation. Normalising all prefactors to kinematic-independent constants solves this issue.

\begin{important}{Integrand conjecture as a practical tool for finding canonical differential equations.}
The above integrand conjecture has proven preciously useful for choosing bases of Feynman integrals that satisfy canonical differential equations. What renders this method particularly powerful is that it can be used at the level of the loop integrand, independently of questions about IBP identities, and without knowledge of the differential equations.
\end{important}

\vspace{-0.6cm}
\begin{exer}{An integrand with double poles: the two-loop kite in $D=4$}
\label{Ex:4.KiteLS}
Compute the four-dimensional maximal cut of the two-loop kite integral defined in exercise~\ref{Ex:4.Kite}, and show that---on the maximal cut---its integrand has a double pole. Hint: introduce two auxiliary light-like momenta $p_1$ and $p_2$ ($p_i^2=0$) such that $p=p_1+p_2$, and use the spinors associated with them to construct a basis in which to expand the loop momenta.
For the solution see \hyperref[Sol:4.KiteLS]{chapter~5}.
\end{exer}

\vspace{-0.6cm}
\begin{exer}{Computing leading singularities with \textsc{DlogBasis}}
\label{Ex:4.dlogs}
The \texttt{Mathematica} package \textsc{DlogBasis}~\cite{ch4Henn:2020lye} provides a suite of tools for computing leading singularities and checking whether a given integrand can be cast into $\d \log$ form, based on the partial fractioning procedure we used to solve exercise~\ref{Ex:4.BubbleCut}. Use \textsc{DlogBasis} to do the following.
\begin{enumerate}[a)]
\item Verify the leading singularities of the massive tadpole and bubble integrals given in eqs.~\eqref{eq:omega10dlog} and~\eqref{massiveleadingsing}.
\item Verify that the integrand of the two-loop kite integral with $D=4$ studied in exercise~\ref{Ex:4.KiteLS} has a double pole.
\item Consider the integrands of the following massless box and triangle integrals,
\vspace{-0.2cm}
\begin{align}
 \int \omega^{\text{box}} = \vcenter{\hbox{\includegraphics[width=1.8cm]{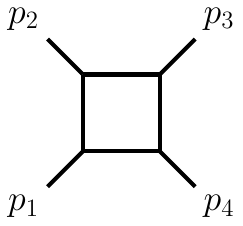}}} \,,  \ \
 \int \omega^{\text{tr.}}_{s} = \vcenter{\hbox{\includegraphics[width=1.8cm]{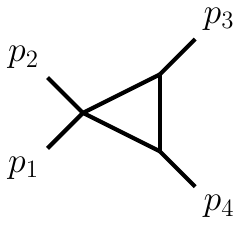}}} \,, \ \
 \int \omega^{\text{tr.}}_{t} = \vcenter{\hbox{\includegraphics[width=1.8cm]{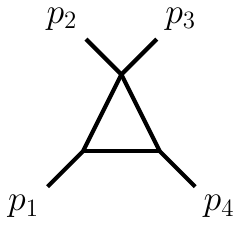}}} \,, 
\end{align}
\vspace{-0.1cm}
with $p_i^2 = 0$.
Show that their leading singularities in $D=4$ dimensions are
\begin{align}
\oint \omega^{\text{box}} \propto \frac{1}{s \, t} \,, \qquad \quad \oint \omega^{\text{tr.}}_s \propto \frac{1}{s} \,,  \qquad \quad
  \oint \omega^{\text{tr.}}_t \propto \frac{1}{t} \,,
\end{align}
where $s=2 \, p_1 \cdot p_2$ and $t = 2\, p_2 \cdot p_3$. Parametrise the integrands using \textsc{DlogBasis}' utilities to expand the loop momentum in a four-dimensional basis constructed from the spinors associated with $p_1$ and $p_2$.
\end{enumerate}
For the solution see the \texttt{Mathematica} notebook \break \href{https://scattering-amplitudes.mpp.mpg.de/scattering-amplitudes-in-qft/Exercises/}{\texttt{Ex4.15\_LeadingSingularities.wl}}~\cite{ch4website-ch4}. 

\end{exer}

\begin{exer}{The box integrals with the differential equations method}
\label{Ex:4.BoxDE}
Write a \texttt{Mathematica} notebook to compute the massless one-loop box integrals,
\begin{align} \label{eq:box_def}
G^{\mathrm{box}}_{a_1,a_2,a_3,a_4} = \int \frac{\d^D k}{\i \pi^{D/2}} \frac{1}{D_1^{a_1} D_2^{a_2} D_3^{a_3} D_4^{a_4} }\,,
\end{align}
where
\begin{align}\begin{alignedat}{2}
& D_1 =  -k^2 - \i 0\,,  & \quad & D_3 =  -(k+p_1+p_2)^2 - \i 0 \,, \\
& D_2 = -(k+p_1)^2 - \i 0 \,, & \qquad \quad & D_4 = - (k-p_4)^2 - \i 0\,,
\end{alignedat}\end{align}
with $p_i^2=0$ and $p_1+p_2+p_3+p_4=0$, using the method of DEs.
Parameterise the kinematics in terms of $s = 2 \, p_1 \cdot p_2$ and $t = 2 \, p_2 \cdot p_3$, and assume that $s<0$ and $t<0$. In this domain, called Euclidean region, the integrals are real valued, and we may thus neglect the $\i 0$'s.
Use the package \textsc{LiteRed}~\cite{ch4Lee:2012cn} to perform the IBP reductions and differentiate the integrals. 
\begin{enumerate}[a)]
\item Define the family and solve the IBP relations to find a basis of master integrals. 
\item Compute the DEs satisfied by the master integrals as functions of $s$ and $t$. Check the scaling relation and the integrability conditions.
\item Change basis of master integrals to
\begin{align} \label{eq:fdef}
\vec{f}(s,t;\eps) = c(\eps) \, \begin{pmatrix}  
   s \, t \, G^{\text{box}}_{1,1,1,1} \\
   s \, G^{\text{box}}_{1,1,1,0} \\ 
   t \, G^{\text{box}}_{1,1,0,1}
\end{pmatrix} \,,
\end{align}
where $c(\eps) = \eps^2 \, \text{e}^{\eps \gamma_{\text{E}}}$. From exercise~\ref{Ex:4.dlogs} we know that, for $D=4$, the integrals in $\vec{f}$ contain only simple poles at the integrand level, and have constant leading singularities.
Compute the transformation matrix and the DEs satisfied by $\vec{f}$. Verify that the latter are in canonical form.

\item Change variables from $(s,t)$ to $(s,x)$, with $x=t/s$.

\item Determine the weight-0 boundary values. Use the results of exercise~\ref{Ex:MasslessBubble} for the master integrals of bubble type. Fix the remaining value by imposing that the solution to the DEs is finite at $u=-s-t = 0$ ($x=-1$). Write a function which produces the symbol of the solution up to a given order in $\eps$.

\item Determine the boundary values at the basepoint $x_0 = 1$ order by order in $\eps$. 
Write a function which produces the analytic solution up to a given order in $\eps$.

\item Verify that the solution for the box integral $G^{\text{box}}_{1,1,1,1}$ agrees with the result obtained through the Mellin-Barnes method in \eqn{eq:F4final}.

\end{enumerate}
For the solution see \hyperref[Sol:4.BoxDE]{chapter~5} and the \texttt{Mathematica} notebook \href{https://scattering-amplitudes.mpp.mpg.de/scattering-amplitudes-in-qft/Exercises/}{\texttt{Ex4.16\_BoxDE.wl}}~\cite{ch4website-ch4}. 

\end{exer}

\chapter{Solutions to the exercises}
\label{ch:solutions}

\section*{Exercise \ref{Ex:1.1}: Manipulating spinor indices}
\label{Prob:1.1}
The sigma-matrix four-vector is defined as $(\bar{\sigma}^{\mu})^{\dot{\alpha} \alpha} \coloneqq (\1,-\vec\sigma)^{\top}$, where $\vec\sigma = (\hat\sigma_1,\hat\sigma_2, \hat\sigma_3)$ is the list of the Pauli matrices $\hat\sigma_i$.\footnote{For the sake of clarity, here we use $\hat{\sigma}_i$ (with $i=1,2,3$) for the $i$th Pauli matrix. This way, the symbol $\sigma$ unambiguously refers to the sigma-matrix four-vectors and their components. This distinction is necessary, as \eqn{eq:sigmalowermu} implies that $\sigma_i = - \hat\sigma_i$.} We rewrite $(\sigma^{\mu})_{\alpha \dot\alpha} = - \epsilon_{\dot\alpha \dot\beta} (\bar{\sigma}^{\mu})^{\dot\beta \beta} \epsilon_{\beta \alpha}$ in matrix notation, as
\begin{align}
\sigma^{\mu} = - \begin{pmatrix} 0 & 1 \\ -1 & 0 \\ \end{pmatrix} \cdot \bar{\sigma}^{\mu} \cdot \begin{pmatrix} 0 & 1 \\ -1 & 0 \\ \end{pmatrix} \,.
\end{align}
Substituting the explicit expressions for $\bar{\sigma}^{\mu}$ gives $\sigma^0 = \1$ and $\sigma^i = \hat\sigma_i$, hence $\sigma^{\mu} = (\1, \vec\sigma)^{\top}$. Multiplying the latter by the metric tensor proves the second identity,
\begin{align} \label{eq:sigmalowermu}
\sigma_{\mu} = \eta_{\mu \nu} \sigma^{\nu} = \begin{pmatrix} 1 & 0 & 0 & 0 \\ 0 & -1 & 0 & 0 \\ 0 & 0 & -1 & 0 \\ 0 & 0 & 0 & -1 \\ \end{pmatrix} \cdot \begin{pmatrix} \1 \\ \hat\sigma_1 \\ \hat\sigma_2 \\ \hat\sigma_3 \\ \end{pmatrix} = (\1, -\vec\sigma)^{\top} \,.
\end{align}

To prove the third identity, $\tr{\sigma^{\mu} \bar{\sigma}^{\nu}} = 2 \eta^{\mu\nu}$, we consider it for fixed values of $\mu$ and $\nu$. 
The facts that the Pauli matrices have vanishing trace and obey the anti-commutation relation $\{\hat\sigma_i,\hat\sigma_j \} = 2 \delta_{ij}$ imply that
\begin{align}
\begin{array}{ll}
\tr{\sigma^0 \bar{\sigma}^0} = \tr{\1} = 2 \,, & \quad \tr{\sigma^0 \bar{\sigma}^i} = -\tr{ \hat{\sigma}_i} = 0\,, \\
\tr{\sigma^i \bar{\sigma}^0} = \tr{\hat{\sigma}_i} = 0 \,, & \quad \tr{\sigma^i \bar{\sigma}^j} = -\tr{ \hat\sigma_i \hat\sigma_j } = - 2\delta_{ij}\,, \\
\end{array}
\end{align}
for $i,j=1,2,3$. Putting these together gives the third identity.

The Pauli matrices and the identity matrix form a basis of all $2\times2$ matrices. Any $2\times2$ matrix $M_{\alpha \dot\alpha}$ can thus be expressed as
\begin{align}
M_{\alpha \dot\alpha} = m_{\mu} (\sigma^{\mu})_{\alpha \dot\alpha} \,.
\end{align}
By contracting both sides by $(\bar\sigma^{\nu})^{\dot\alpha \beta}$ and computing the trace using the third identity, we can express the coefficients of the expansion in terms of $M$ as $m^{\mu} = \tr{M \bar{\sigma}^{\mu}}/2$.
Substituting this into the expansion of $M_{\alpha\dot\alpha}$ gives 
\begin{align}
2 \, M_{\alpha\dot\alpha} = M_{\beta\dot\beta} (\bar{\sigma}_{\mu})^{\dot\beta \beta} (\sigma^{\mu})_{\alpha\dot\alpha} \,.
\end{align}
Since this holds for any matrix $M$, it follows that
\begin{align}
(\sigma^{\mu})_{\alpha\dot\alpha}  (\bar{\sigma}_{\mu})^{\dot\beta \beta}  = 2 \, \delta^{\beta}_{\, \alpha} \delta^{\dot\beta}_{\, \beta} \,.
\end{align}
Contracting both sides with suitable Levi-Civita symbols gives the fourth identity.


\section*{Exercise \ref{Ex:1.2}: Massless Dirac equation and Weyl spinors}
\label{Prob:1.2}

\begin{enumerate}[a)]

\item Any Dirac spinor $\xi$ can be decomposed as $\xi = \xi_+ + \xi_-$, where $\xi_{\pm}$ satisfy the helicity relations
\begin{align}
P^{\pm} \xi_{\pm} =  \xi_{\pm} \,, \qquad \quad P_{\pm} \xi_{\mp} = 0 \,,
\end{align}
with $P_{\pm} = (\1 \pm \gamma^{5})/2$.
Using the Dirac representation of the $\gamma$ matrices in \eqn{eq:gamma_dirac} we have that
\begin{align}
P_{\pm} = \begin{pmatrix} \1_2 & \pm \1_2 \\  \pm \1_2 & \1_2 \end{pmatrix} \,.
\end{align}
The helicity relations then constrain the form of  $\xi_{\pm}$ to have only two independent components:
\begin{align}
 \xi_{+} = \left(\xi^0, \xi^1, \xi^0, \xi^1 \right)^{\top} \,, \qquad  \xi_{-} = \left(\xi^0, \xi^1, -\xi^0, -\xi^1 \right)^{\top} \,.
\end{align}
Indeed, $u_+$ and $v_-$ ($u_-$ and $u_+$) have the form of $\xi^+$ ($\xi^-$). We now focus on $\xi_+$. We change variables from $k^{\mu}$ to $k^{\pm}$ and $\mathrm{e}^{\i \phi}$, which have the benefit of implementing $k^2=0$. Then we have that 
\begin{align}
\gamma^{\mu} k_{\mu} = \begin{pmatrix} 
\frac{k^++k^-}{2}  & 0 & \frac{k^--k^+}{2} & - \mathrm{e}^{-\i \phi} \sqrt{k^+ k^-} \\
0 & \frac{k^++k^-}{2} & - \mathrm{e}^{\i \phi} \sqrt{k^+ k^-} & \frac{k^+-k^-}{2} \\
\frac{k^+-k^-}{2} & \mathrm{e}^{-\i \phi} \sqrt{k^+ k^-} & - \frac{k^++k^-}{2} & 0 \\
 \mathrm{e}^{\i \phi} \sqrt{k^+ k^-} & \frac{k^--k^+}{2} & 0 & -\frac{k^++k^-}{2} \\
 \end{pmatrix} \,.
\end{align}
Plugging the generic form of $\xi_{+}$ into the Dirac equation $\gamma^{\mu} k_{\mu} \xi^+ = 0$ gives one equation, which fixes $\xi^1$ in terms of $\xi^0$,
\begin{align}
\xi^1 = \mathrm{e}^{\i \phi} \sqrt{\frac{k^-}{k^+}} \, \xi^0 \,.
\end{align}
Since the equation is homogeneous, the overall normalisation is arbitrary. Choosing $\xi^0 = \sqrt{k^+}/\sqrt{2}$ gives the expressions for $\xi_+=u_+=v_-$ given above.

\smallskip
\item For any Dirac spinor $\xi$ we have
\begin{align}
\bar{\xi} P_{\pm} = \xi^{\dagger} \gamma^0 P_{\pm} = \xi^{\dagger} P_{\mp} \gamma^0 = \left(\gamma^0 P_{\pm} \xi \right)^{\dagger} \,,
\end{align}
where we used that $(\gamma^5)^{\dagger} = \gamma^5$ and $\{\gamma^5,\gamma^0\} = 0$. From this it follows that
\begin{align}
\bar{u}_{\pm} P_{\pm} = 0 \,, \qquad \bar{u}_{\pm} P_{\mp} = \bar{u}_{\pm} \,, \qquad \bar{v}_{\pm} P_{\pm} = \bar{v}_{\pm} \,, \qquad  
 \bar{v}_{\pm} P_{\mp} = 0 \,.
\end{align}

\smallskip
\item Through matrix multiplication we obtain the explicit expression of $U$,
\begin{align}
U = \frac{1}{\sqrt{2}} \begin{pmatrix} \1_{2} & -\1_{2} \\ \1_{2} & \1_{2} \\ \end{pmatrix} \,,
\end{align}
which is indeed a unitary matrix. The Dirac matrices in the chiral basis then are
\begin{align}
\gamma^{0}_{\text{ch}}=U \gamma^{0} U^{\dagger} =  \left ( \begin{matrix}0 & {\1}_{2\times 2}\cr {\1}_{2\times 2} & 0\end{matrix}\right )\,, \qquad 
\gamma^{i}_{\text{ch}}= U \gamma^{i} U^{\dagger} = \left ( \begin{matrix}0 & \vec\sigma^{i}\cr -\vec\sigma^{i} & 0\end{matrix}\right )\, ,
\end{align}
with $i=1,2,3$. Putting these together gives
\begin{align} \label{eq:gammaCh}
\gamma^{\mu}_{\text{ch}} = \begin{pmatrix} 0 & \sigma^{\mu} \\ \overline{\sigma}^{\mu} & 0 \\ \end{pmatrix} \,.
\end{align}
Similarly, we obtain the expression of $\gamma^5$, which in this basis is diagonal,
\begin{align} \label{eq:gamma5ch}
\gamma^5_{\text{ch}}=U \gamma^{5} U^{\dagger} = \begin{pmatrix}- \1_{2} & 0 \\ 0 & \1_{2} \end{pmatrix} \,.
\end{align}
Finally, the solutions to the Dirac equation in the chiral basis are given by
\begin{align}
U u_+ = \left( 0, 0, \sqrt{k^+}, \mathrm{e}^{\i \phi} \sqrt{k^-} \right)^{\top} \,, \qquad 
U u_- =  \left( \sqrt{k^-} \mathrm{e}^{-\i \phi}, -\sqrt{k^+}, 0, 0 \right)^{\top}\,,
\end{align}
and similarly for $v_{\pm}$.

\smallskip
\item The product of four Dirac matrices in the chiral representation~\eqref{eq:gammaCh} is given by
\begin{align}
\gamma^{\mu} \gamma^{\nu} \gamma^{\rho} \gamma^{\tau} 
         = \begin{pmatrix} \sigma^{\mu} \bar{\sigma}^{\nu} \sigma^{\rho} \bar{\sigma}^{\tau} & 0 \\ 0 & \bar{\sigma}^{\mu} \sigma^{\nu} \bar{\sigma}^{\rho}  \sigma^{\tau} \\ \end{pmatrix} \,.
\end{align}
Multiplying to the right by
\begin{align}
\frac{1}{2} \left( \1 - \gamma_5 \right) = \begin{pmatrix} \1 & 0 \\ 0 & 0 \\ \end{pmatrix} 
\end{align}
selects the top left entry,
\begin{align}
\frac{1}{2}  \gamma^{\mu} \gamma^{\nu} \gamma^{\rho} \gamma^{\tau}  \left( \1 - \gamma_5 \right)  = \begin{pmatrix} \sigma^{\mu} \bar{\sigma}^{\nu} \sigma^{\rho} \bar{\sigma}^{\tau} & 0 \\ 0 & 0 \\ \end{pmatrix} \,.
\end{align}
Taking the trace of both sides of this equation finally gives eq.~\eqref{eq:trSigma2gamma}. Note that this result does not depend on the representation of the Dirac matrices, as the unitarity matrices relating different representations drop out from the trace. Using $\1+\gamma_5$ instead gives a relation for $\Tr\left(\bar{\sigma}^{\mu} \sigma^{\nu} \bar{\sigma}^{\rho} \sigma^{\tau}\right)$.

\end{enumerate}


\section*{Exercise \ref{Ex:1.3}: SU($N_c$) identities}
\label{Sol:1.3}

\begin{enumerate}[a)]

\item The Jacobi identity for the generators~\eqref{eq:JacobiGenerators} can be proven directly by expanding all commutators. We recall that the commutator is bilinear.
The first term on the left-hand side gives
\begin{align}
\left[T^{a}, [ T^{b}, T^{c}]\right]  = T^a T^b T^c - T^a T^c T^b - T^b T^c T^a + T^c T^b T^a \,.
\end{align}
Summing both sides of this equation over the cyclic permutations of the indices ($\{a,b,c\}$, $\{b,c,a\}$, $\{c,a,b\}$) gives \eqn{eq:JacobiGenerators}.

\smallskip
\item We substitute the commutation relations~\eqref{I.15} into the Jacobi identity for the generators~\eqref{eq:JacobiGenerators}. The first term gives
\begin{align}
\left[T^{a}, [ T^{b}, T^{c}]\right] = - 2 f^{bce} f^{aeg} T^{g} \,.
\end{align}
By summing over the cyclic permutations of the indices and removing the overall constant factor we obtain
\begin{align} \label{eq:jacobiproof}
\left(f^{abe} f^{ceg} + f^{bce} f^{aeg} + f^{cae} f^{beg} \right) T^{g} = 0 \,.
\end{align}
We recall that repeated indices are summed over. Since the generators $T^{g}$ are linearly independent, their coefficients in \eqn{eq:jacobiproof} have to vanish separately. This gives the Jacobi identity~\eqref{Jacobif}.

\smallskip
\item Any $N_c \times N_c$ complex matrix $M$ can be decomposed into the identity $\1_{N_c}$ and the $su(N_c)$ generators $T^a$ (with $a=1,\ldots, N_c^2-1$),
\begin{align}
M = m_0 \, \1_{N_c} + m_a \, T^a \,.
\end{align}
As usual, the repeated indices are summed over. The coefficients of the expansion can be obtained by multiplying both sides by $\1_{N}$ and $T^a$, and taking the trace. Using the tracelessness of $T^a$ and $\Tr T^a T^b = \delta^{ab}$ we obtain that $m_0 = \Tr(M)/N_c$ and $m_a = \Tr (M T^a)$. We then rewrite the expansion as 
\begin{align}
\left[ \left(T^a\right)_{i_1}^{\ j_1} \left(T^a\right)_{i_2}^{\ j_2} - \delta_{i_1}^{\ j_2} \delta_{i_2}^{\ j_1}  + \frac{1}{N_c} \delta_{i_1}^{\ j_1} \delta_{i_2}^{\ j_2}  \right] M_{j_2}^{\ i_2} = 0\,.
\end{align}
Since this equation holds for any complex matrix $M$, it follows that the coefficient of $M_{j_2}^{\ i_2}$ vanishes. This yields the desired relation.

\end{enumerate}


\section*{Exercise \ref{Ex:1.4}: Casimir operators}
\label{Sol:1.4}

\begin{enumerate}[a)]

\item The commutator of $T^a T^a$ with the generators $T^b$ is given by
\begin{align}
\begin{aligned}
\left[ T^a T^a, T^b \right] & = T^a  \left[ T^a, T^b \right] + \left[ T^a, T^b \right] T^a \\
& = \i \sqrt{2} f^{abc} \left( T^{a} T^{c} + T^{c} T^{a} \right) \,,
\end{aligned}
\end{align}
which vanishes because of the anti-symmetry of $f^{abc}$. In the first line we used that $[A B,C] = A [B,C] + [A,C] B$, which can be proven by expanding all commutators, and in the second line we applied the commutation relations~\eqref{I.15}.

\smallskip
\item The Casimir invariant of the fundamental representation follows directly from the completeness relation~\eqref{eq:completenessSUN}, 
\begin{align}
\begin{aligned}
\left(T^a_F \right)_{i}^{\ k} \left(T^a_F \right)_{k}^{\ j} & = \delta_{i}^{\ j} \delta_{k}^{\ k} - \frac{1}{N_c} \delta_{i}^{\ k} \delta_{k}^{\ j} \\
& = \frac{N_c^2-1}{N_c} \left( \1_{N_c} \right)_{ij} \,,
\end{aligned}
\end{align}
from which we read off that $C_F = (N_c^2-1)/N_c$. For the adjoint representation, we use \eqn{eq:GeneratorsAdjoint} to express the generators in terms of structure constants,
\begin{align}
\left(T^a_A T^a_A \right)^{bc} = 2 f^{bak} f^{cak} \,.
\end{align}
We express one of the structure constants in terms of generators through \eqn{I.17},
\begin{align}
\left(T^a_A T^a_A \right)^{bc} = - \i 2 \sqrt{2} \Tr\left(T^b_F T^a_F T^k_F \right) f^{cak} \,,
\end{align}
where we also used the anti-symmetry of $f^{cak}$ to remove the commutator from the trace. Next, we move $f^{cak}$ into the trace and rewrite $ f^{cak} T^k_F $ in terms of a commutator,
\begin{align}
\left(T^a_A T^a_A\right)^{bc} = 2 \Tr \left( T^b_F T^a_F [T^a_F, T^c_F] \right) \,.
\end{align}
Applying the completeness relation~\eqref{eq:completenessSUN} finally gives
\begin{align}
\left(T^a_A T^a_A\right)^{bc} = 2 \, N_c \left( \1\right)^{bc} \,,
\end{align}
from which we see that $C_A = 2 \, N_c$. \\

Note that in many QCD contexts it is customary to normalise the generators so that $\mathrm{Tr}(T^a T^b) = \delta^{ab}/2$, as opposed to $\mathrm{Tr}(T^a T^b) = \delta^{ab}$ as we do here. This different normalisation results in $C_A = N_c$ and $C_F = (N_c^2-1)/(2 N_c)$.

\end{enumerate}


\section*{Exercise \ref{Ex:1.6}: Spinor identities}
\label{Sol:1.6}

The identities a) and b) follow straightforwardly from the definition of the bra-ket notation and from the expression of $\gamma^{\mu}$ in terms of Pauli matrices,
\begin{align}
\spBA{i}{\gamma^{\mu}}{j} & = \begin{pmatrix} 0 & \tilde{\lambda}_i \\ \end{pmatrix} \cdot \begin{pmatrix} 0 & \sigma^{\mu} \\ \bar{\sigma}^{\mu} & 0 \\ \end{pmatrix} \cdot
  \begin{pmatrix} \lambda_j \\ 0 \\ \end{pmatrix} = (\tilde{\lambda}_i)_{\dot\alpha} (\bar{\sigma}^{\mu})^{\dot{\alpha} \alpha} (\lambda_j)_{\alpha} \,, \\
\spAB{i}{\gamma^{\mu}}{j} & = \begin{pmatrix} \lambda_i & 0 \\ \end{pmatrix} \cdot \begin{pmatrix} 0 & \sigma^{\mu} \\ \bar{\sigma}^{\mu} & 0 \\ \end{pmatrix} \cdot
  \begin{pmatrix} 0 \\ \tilde{\lambda}_j \\ \end{pmatrix} = \lambda_i^{\alpha} (\sigma^{\mu})_{\alpha\dot{\alpha}} \tilde{\lambda}_j^{\dot\alpha}\,.
\end{align}
Setting $j=i$ in the previous identities and using that $(\sigma^{\mu})_{\beta\dot{\beta}} = \epsilon_{\beta\alpha} \epsilon_{\dot{\beta}\dot{\alpha}} (\bar{\sigma}^{\mu})^{\dot{\alpha}\alpha}$ gives the relation c),
\begin{align}
\spBA{i}{\gamma^{\mu}}{i} = \epsilon_{\dot{\alpha}\dot{\beta}}  \tilde{\lambda}_i^{\dot\beta} (\bar{\sigma}^{\mu})^{\dot{\alpha} \alpha} \epsilon_{\alpha \beta} \lambda_i^{\beta} = \lambda_i^{\beta} (\sigma^{\mu})_{\beta \dot{\beta}} \tilde{\lambda}_i^{\dot{\beta}} = \spAB{i}{\gamma^{\mu}}{i} \,.
\end{align}
We obtain the relation d) by substituting the identities $\tilde{\lambda}_i^{\dot{\alpha}} \lambda_i^{\alpha} = p_i^{\mu} (\bar{\sigma}_{\mu})^{\dot{\alpha}\alpha}$
and $\text{tr}\left(\sigma^{\mu} \bar{\sigma}^{\nu}\right) = 2 \eta^{\mu\nu}$ into b) with $j=i$,
\begin{align}
\spAB{i}{\gamma^{\mu}}{i} =  (\sigma^{\mu})_{\alpha\dot{\alpha}} (p_i)_{\nu} (\bar{\sigma}^{\nu})^{\dot\alpha \alpha} = 2 p_i^{\mu} \,.
\end{align}
In order to prove the Schouten identity, we recall that a spinor $\lambda_i$ is a two-dimensional object. We can therefore expand $\lambda_3$ in a basis constructed from $\lambda_1$ and $\lambda_2$,
\begin{align} \label{eq:lambda3exp}
\lambda_3^{\alpha} = c_1 \lambda_1^{\alpha} + c_2 \lambda_2^{\alpha} \,.
\end{align}
Contracting both sides of this equation by $\lambda_1$ and $\lambda_2$ gives a linear system of equations for the coefficients $c_1$ and $c_2$,
\begin{align}
\begin{cases} \spA 31 = c_2 \spA 21 \\
\spA 32 = c_1 \spA 12 \\
\end{cases}\,.
\end{align}
Substituting the solution of this system into Eq.~\eqref{eq:lambda3exp} and rearranging the terms gives the Schouten identity.
Finally, the identity a) and $(\bar{\sigma}^{\mu})^{\dot{\alpha} \beta} (\bar{\sigma}_{\mu})^{\dot{\beta} \alpha} = 2  \epsilon^{\dot{\alpha}\dot{\beta}}  \epsilon^{\beta \alpha}$ give the Fierz rearrangement, 
\begin{align}
\begin{aligned}
\spBA{i}{\gamma^{\mu}}{j} \spBA{k}{\gamma_{\mu}}{l} & = (\tilde{\lambda}_i)_{\dot{\alpha}} (\bar{\sigma}^{\mu})^{\dot{\alpha} \beta} (\lambda_j)_{\beta} 
  (\tilde{\lambda}_k)_{\dot{\beta}} (\bar{\sigma}_{\mu})^{\dot{\beta} \alpha} (\lambda_l)_{\alpha} \\
& = 2 (\tilde{\lambda}_i)_{\dot{\alpha}} \epsilon^{\dot{\alpha}\dot{\beta}} (\tilde{\lambda}_k)_{\dot{\beta}}
  (\lambda_j)_{\beta}  \epsilon^{\beta \alpha} (\lambda_l)_{\alpha} \\
& = 2  \spB ik \spA lj \,.
\end{aligned}
\end{align}


\section*{Exercise \ref{Ex:1.5}: Lorentz generators in the spinor-helicity formalism}
\label{Sol:1.5}

\begin{enumerate}[a)]

\item The Lorentz generators in the scalar representation are obtained by setting to zero the $x$-independent representation matrices $S^{\mu\nu}$ in \eqn{Mrepgen}:
\be
M^{\mu\nu} = \i\,  \left (x^{\mu}\,\frac{\partial}{\partial x_{\nu}} - 
x^{\nu}\,\frac{\partial}{\partial x_{\mu}} \right ) \, .
\ee
We act with $M^{\mu\nu}$ on a generic function $f(x)$, which we express in terms of its Fourier transform as 
$f(x) = \int \d^4 p \, \mathrm{e}^{\i p\cdot x} \tilde{f}(p)$. By integrating by parts and using $x^{\mu} = - \i \frac{\partial}{\partial p_{\mu}} \mathrm{e}^{\i p\cdot x}$ we obtain
\begin{align}
M^{\mu \nu} f(x) = \int \d^4 p \, \mathrm{e}^{\i p\cdot x} \tilde{M}^{\mu \nu} \tilde{f}(p) \,,
\end{align}
where
\begin{align}
\tilde{M}^{\mu \nu} = \i \left( p^{\mu} \frac{\partial}{\partial p_{\nu}} - p^{\nu} \frac{\partial}{\partial p_{\mu}} \right)
\end{align}
is the momentum-space realisation of the Lorentz generators. Indeed, one can verify that this form of the generators satisfies the commutation relations of the Poincar\'e algebra in eqs.~\eqref{I.4} and~\eqref{I.5}.

\smallskip
\item We begin with $m_{\alpha\beta}$. It is instructive to spell out the indices of $S^{\mu\nu}_{\rm L}$,
\begin{align}
\left(S^{\mu\nu}_{\rm L}\right)_{\alpha\beta} = \frac{\i}{4}  \epsilon_{\beta\gamma} \left[
   \left(\sigma^{\mu}\right)_{\alpha\dot\alpha} \left(\bar{\sigma}^{\nu}\right)^{\dot\alpha\gamma} -
   \left(\sigma^{\nu}\right)_{\alpha\dot\alpha} \left(\bar{\sigma}^{\mu}\right)^{\dot\alpha\gamma} \right] \,.
\end{align}
Contracting it with $\tilde{M}_{\mu\nu}$ and doing a little spinor algebra gives
\begin{align} \label{eq:mab_ex}
m_{\alpha\beta} = \frac{1}{2} \lambda_{\beta}  \left(\sigma^{\mu}\right)_{\alpha\dot\alpha} \tilde{\lambda}^{\dot\alpha} \frac{\partial}{\partial p^{\mu}} +
\frac{1}{2} \lambda_{\alpha} \tilde{\lambda}_{\dot \alpha}  \left(\bar{\sigma}^{\mu}\right)^{\dot\alpha\gamma} \epsilon_{\gamma\beta} \frac{\partial}{\partial p^{\mu}} \,.
\end{align}
We now need to express the derivatives with respect to $p^{\mu}$ in terms of derivatives with respect to $\lambda^{\alpha}$ and $\tilde{\lambda}^{\dot\alpha}$. For this purpose, we use the identity $p^{\mu} = \tilde{\lambda}^{\dot\alpha} \lambda^{\alpha} \left(\sigma^{\mu}\right)_{\alpha\dot\alpha}/2$ (see exercise~\ref{Ex:1.6}), which allows us to use the chain rule,
\begin{align} \label{eq:dp_1}
\frac{\partial}{\partial \lambda^{\alpha}} = \frac{\partial p^{\mu}}{\partial \lambda^{\alpha}} \frac{\partial}{\partial p^{\mu}} = \frac{1}{2} \left(\sigma^{\mu}\right)_{\alpha\dot\alpha} \tilde{\lambda}^{\dot\alpha} \frac{\partial}{\partial p^{\mu}} \,.
\end{align}
This takes care of the first term on the RHS of \eqn{eq:mab_ex}. For the second term, we do the same but with the equivalent identity $p^{\mu} = \lambda_{\alpha} \tilde{\lambda}_{\dot\alpha} \left(\bar{\sigma}^{\mu}\right)^{\dot\alpha\alpha}/2$. Using that $\frac{\partial \lambda_{\beta}}{\partial \lambda^{\alpha}} = \epsilon_{\beta\alpha}$ we obtain 
\begin{align} \label{eq:dp_2}
\frac{\partial}{\partial \lambda^{\alpha}} = \frac{1}{2} \tilde{\lambda}_{\dot\beta}  \left(\bar{\sigma}^{\mu}\right)^{\dot\beta \beta}   \epsilon_{\beta \alpha} \frac{\partial}{\partial p^{\mu}} \,.
\end{align}
Substituting eqs.~\eqref{eq:dp_1} and~\eqref{eq:dp_2} into \eqn{eq:mab_ex} finally gives the desired expression of $m_{\alpha \beta}$. The computation of $\overline{m}_{\dot\alpha \dot\beta}$ is analogous.

\smallskip
\item The $n$-particle generators are given by
\begin{equation}
\begin{aligned}
& m_{\alpha \beta} = \sum_{k=1}^n \left( \lambda_{k \alpha} \frac{\partial}{\partial \lambda_k^{\beta}} + \lambda_{k \beta} \frac{\partial}{\partial \lambda_k^{\alpha}}\right) \,, \quad 
\overline{m}_{\dot\alpha \dot\beta} = \sum_{k=1}^n \left( \tilde{\lambda}_{k  \dot\alpha} \frac{\partial}{\partial \tilde{\lambda}_k^{\dot\beta}} + \tilde{\lambda}_{k \dot\beta} \frac{\partial}{\partial \tilde{\lambda}_k^{\dot\alpha}}\right) \,, \\
& \tilde{M}^{\mu \nu} = \i \sum_{k=1}^n \left( p_k^{\mu} \frac{\partial}{\partial p_{k \nu}} - p_k^{\nu} \frac{\partial}{\partial p_{k \mu}} \right) \,.
\end{aligned}
\end{equation}
We act with $m_{\alpha \beta}$ and $\overline{m}_{\dot\alpha \dot\beta}$ on $\langle ij \rangle = \lambda_i^{\gamma} \lambda_{j\,\gamma}$ and $[ ij ] = \tilde{\lambda}_{i \, \dot\gamma} \lambda_{j}^{\dot\gamma}$. $\langle ij \rangle$ ($[ij]$) depends only on the $\lambda_i$ ($\tilde{\lambda}_i$) spinors, and is thus trivially annihilated by $\overline{m}_{\dot\alpha \dot\beta}$ ($m_{\alpha \beta}$). With a bit more of spinor algebra we can show that $\langle ij \rangle$ is annihilated also by $m_{\alpha \beta}$,
\begin{align}
\begin{aligned}
m_{\alpha \beta} \langle ij \rangle & = \sum_{k=1}^n \left[
   \delta_{ik} \delta^{\gamma}_{\, \beta} \lambda_{k \, \alpha} \lambda_{j \, \gamma} +
   \delta_{jk} \epsilon_{\gamma \beta} \lambda_{k \, \alpha} \lambda_i^{\gamma} + \left( \alpha \leftrightarrow \beta \right) \right] = \\
   & =  \lambda_{i \, \alpha} \lambda_{j \, \beta} -  \lambda_{i \, \beta} \lambda_{j \, \alpha} + \left( \alpha \leftrightarrow \beta \right) = 0 \,.
\end{aligned}
\end{align}
Similarly we can show that $\overline{m}_{\dot\alpha \dot\beta} [ij] =0 $. The Lorentz generators are first-order differential operators. As a result, any function of a Lorentz-invariant object is Lorentz invariant as well. We can thus immediately conclude that $s_{ij} = \vev{ij}\bev{ji}$ is annihilated by $m_{\alpha \beta}$ and $\overline{m}_{\dot\alpha \dot\beta}$. Alternatively, we can show that
\begin{align}
\tilde{M}_{\mu \nu} s_{ij} = 2 \i \left[ p_{i \mu} p_{j \nu} +  p_{i \nu} p_{j \mu}  - \left(\mu \leftrightarrow \nu \right) \right] = 0 \,.
\end{align}

\end{enumerate}


\section*{Exercise \ref{Ex:1.7}: Gluon polarisations}
\label{Sol:1.7}

\begin{enumerate}[a)]
\item In order to construct an explicit expression for the polarisation vectors we will write a general ansatz and apply constraints to fix all free coefficients. The polarisation vector $\epsilon^{\dot{\alpha}\alpha}_i$ is a four-dimensional object which satisfies constraints involving the corresponding external momentum $p_i^{\dot{\alpha} \alpha} = \tilde{\lambda}_i^{\dot{\alpha}} \lambda_i^{\alpha}$ and reference vector $r_i^{\dot{\alpha} \alpha} = \tilde{\mu}_i^{\dot{\alpha}} \mu^{\alpha}$. For generic kinematics, i.e.\ for $p_i \cdot r_i \neq 0$ (and thus $\spA{\lambda_i}{\mu_i} \neq 0$ and $\spB{\tilde{\lambda}_i}{\tilde{\mu}_i}\neq 0$), one can show that $\tilde{\lambda}_i^{\dot{\alpha}} \lambda_i^{\alpha}$, $\tilde{\mu}_i^{\dot{\alpha}} \mu_i^{\alpha}$,
 $\tilde{\lambda}_i^{\dot{\alpha}} \mu_i^{\alpha}$ and $\tilde{\mu}_i^{\dot{\alpha}} \lambda_i^{\alpha}$ are linearly independent, and thus form a basis in which we can expand $\epsilon^{\dot{\alpha}\alpha}_i$. Our ansatz for $\epsilon^{\dot{\alpha}\alpha}_i$ therefore is
\begin{align}
\epsilon^{\dot{\alpha}\alpha}_i = c_1 \, \tilde{\lambda}_i^{\dot{\alpha}} \lambda_i^{\alpha} + c_2 \, \tilde{\mu}_i^{\dot{\alpha}} \mu_i^{\alpha} +
  c_3 \, \tilde{\lambda}_i^{\dot{\alpha}} \mu_i^{\alpha} + c_4 \, \tilde{\mu}_i^{\dot{\alpha}} \lambda_i^{\alpha} \,.
\end{align}
The transversality and the gauge choice,
\begin{align}
\begin{aligned}
& \epsilon^{\dot{\alpha}\alpha}_i \left(p_i\right)_{\alpha \dot{\alpha}} = c_2 \spA{\mu_i}{\lambda_i} \spB{\tilde{\lambda}_i}{\tilde{\mu}_i} = 0 \,, \\
&  \epsilon^{\dot{\alpha}\alpha}_i \left(r_i\right)_{\alpha \dot{\alpha}} = c_1 \spA{\lambda_i}{\mu_i} \spB{\tilde{\mu}_i}{\tilde{\lambda}_i} = 0 \,,
\end{aligned}
\end{align}
imply that $c_1 = c_2 = 0$. The light-like condition,
\begin{align}
\epsilon^{\dot{\alpha}\alpha}_i \left(\epsilon_i\right)_{\alpha \dot{\alpha}}= 2 c_3 c_4 \spA{\lambda_i}{\mu_i} \spB{\tilde{\lambda}_i}{\tilde{\mu}_i} = 0 \,,
\end{align}
has two solutions: $c_3=0$ and $c_4 = 0$. We parametrise the two solutions as
\begin{align}
\epsilon_{A,i}^{\dot{\alpha}\alpha} = n_A \tilde{\lambda}_i^{\dot{\alpha}} \mu_i^{\alpha} \,, \qquad \quad
\epsilon_{B,i}^{\dot{\alpha}\alpha} = n_B \tilde{\mu}_i^{\dot{\alpha}} \lambda_i^{\alpha} \,.
\end{align}
Next, we normalise the two solutions such that $\epsilon_{A,i} \cdot \epsilon_{B,i}=-1$ and $\epsilon_{A,i}^* = \epsilon_{B,i}$. This implies that 
\begin{align} \label{eq:constraint_nAnB}
n_A n_B = \frac{-\sqrt{2}}{\spA{\lambda_i}{\mu_i}} \frac{\sqrt{2}}{\spB{\tilde{\lambda}_i}{\tilde{\mu}_i}} \,, \qquad \qquad n_A^*=n_B \,.
\end{align}
There is now some freedom in fixing $n_A$ and $n_B$, which we must use to ensure that the two solutions have the correct helicity scaling. We may parametrise $n_A = n \, \mathrm{e}^{\i \varphi}$ and $n_B = n \, \mathrm{e}^{-\i \varphi}$ with real $n$ and $\varphi$, and fix the phase $\varphi$ by requiring that the solutions are eigenvectors of the helicity operator. It is however simpler to follow a heuristic approach. Recalling that $\spA{\lambda_i}{\mu_i}^* = - \spB{\tilde{\lambda}_i}{\tilde{\mu}_i}$, we notice that a particularly simple solution to the constraints~\eqref{eq:constraint_nAnB} is given by $n_A = -\sqrt{2}/\spA{\lambda_i}{\mu_i}$ and $n_B =  \sqrt{2}/\spB{\tilde{\lambda}_i}{\tilde{\mu}_i}$. Following this guess, we have two fully determined vectors which satisfy all constraints of the polarisation vectors:
\begin{align}
\epsilon_{A,i}^{\dot{\alpha}\alpha} = -\sqrt{2} \, \frac{\tilde{\lambda}_i^{\dot{\alpha}} \mu_i^{\alpha}}{\spA{\lambda_i}{\mu_i}} \,, \qquad \quad
\epsilon_{B,i}^{\dot{\alpha}\alpha} = \sqrt{2} \, \frac{\tilde{\mu}_i^{\dot{\alpha}} \lambda_i^{\alpha}}{\spB{\tilde{\lambda}_i}{\tilde{\mu}_i}}  \,.
\end{align}
Finally, we need to check that $\epsilon_{A,i}^{\dot{\alpha}\alpha}$ and $\epsilon_{B,i}^{\dot{\alpha}\alpha}$ are indeed eigenvectors of the helicity generator $h$ in \eqn{eq:hel_gen}, which in this case takes the form
\begin{align}
h = \frac{1}{2} \left[
-\la_{i}^{\a}\frac{\partial}{\partial \la_{i}^{\a}} -\mu_{i}^{\a}\frac{\partial}{\partial \mu_{i}^{\a}}  + \tla_{i}^{\da}\frac{\partial}{\partial \tla_{i}^{\da}} +
 \tilde{\mu}_{i}^{\da}\frac{\partial}{\partial \tilde{\mu}_{i}^{\da}}
\right]\, .
\end{align}
The explicit computation yields that
\begin{align} 
h \epsilon_{A,i}^{\dot{\alpha}\alpha} = +\epsilon_{A,i}^{\dot{\alpha}\alpha}\,, \qquad \quad h \epsilon_{B,i}^{\dot{\alpha}\alpha} = -\epsilon_{B,i}^{\dot{\alpha}\alpha}\,.
\end{align}
We can therefore identify $\epsilon_{A,i}^{\dot{\alpha}\alpha}=\epsilon_{+,i}^{\dot{\alpha}\alpha}$ and $\epsilon_{B,i}^{\dot{\alpha}\alpha}=\epsilon_{-,i}^{\dot{\alpha}\alpha}$, which completes the derivation.

\smallskip

\item 
We rewrite the spinor expressions for the polarisation vectors as Lorentz vectors using the identities of exercise~\ref{Ex:1.6},
\begin{align}
\epsilon_{+,i}^{\mu} = -\frac{1}{\sqrt{2}} \frac{ \mu_i^{\alpha} (\sigma^{\mu})_{\alpha\dot{\alpha}} \tilde{\lambda}_i^{\dot{\alpha}}}{\spA{\lambda_i}{\mu_i}} \,, \qquad \quad
  \epsilon_{-,i}^{\mu} = \frac{1}{\sqrt{2}} \frac{ \lambda_i^{\alpha} (\sigma^{\mu})_{\alpha\dot{\alpha}} \tilde{\mu}_i^{\dot{\alpha}}}{\spB{\tilde{\lambda}_i}{\tilde{\mu}_i}} \,.
\end{align}
Plugging these expressions into the polarisation sum gives
\begin{align}
\sum_{h=\pm} \epsilon_{h,i}^{\mu} \epsilon_{h,i}^{* \nu} = -\frac{1}{2} \frac{  (\sigma^{\mu})_{\alpha\dot{\alpha}} \tilde{\lambda}_i^{\dot{\alpha}} \lambda_i^{\beta}
  (\sigma^{\nu})_{\beta\dot{\beta}}  \tilde{\mu}_i^{\dot{\beta}} \mu_i^{\alpha} +
 \left(\mu \leftrightarrow \nu \right)}{\spA{\lambda_i}{\mu} \spB{\tilde{\lambda}_i}{\tilde{\mu}_i}} \,.
\end{align}
Next, we rewrite the numerator in terms of traces as
\begin{align}
\sum_{h=\pm} \epsilon_{h,i}^{\mu} \epsilon_{h,i}^{* \nu} = -\frac{1}{2} \frac{\text{Tr}\left[\sigma^{\mu} p_i \sigma^{\nu} r_i \right]+
 \left(\mu \leftrightarrow \nu \right)}{\spA{\lambda_i}{\mu} \spB{\tilde{\lambda}_i}{\tilde{\mu}_i}} \,.
\end{align}
We then use the identity~\eqref{eq:trSigma2gamma} to rewrite the trace of Pauli matrices in terms of Dirac matrices. Finally, by using
\begin{align} \label{eq:gamma_traces}
\begin{aligned}
& \Tr\left(\gamma^{\mu} \gamma^{\nu} \gamma^{\rho} \gamma^{\tau}\right) = 4 \left(\eta^{\mu \nu} \eta^{\rho \tau} - \eta^{\mu \rho} \eta^{\nu \tau} + \eta^{\mu \tau} \eta^{\nu \rho} \right) \,, \\
& \Tr\left(\gamma^{\mu} \gamma^{\nu} \gamma^{\rho} \gamma^{\tau} \gamma_5 \right) = - 4 \, \i \, \epsilon^{\mu \nu \rho \tau} \,,
\end{aligned}
\end{align}
we obtain
\begin{align}
\sum_{h=\pm} \epsilon_{h,i}^{\mu} \epsilon_{h,i}^{* \nu}  = -\eta^{\mu\nu} + \frac{p_i^{\mu} r_i^{\nu}+p_i^{\nu} r_i^{\mu}}{p_i \cdot r_i} \,.
\end{align}

\end{enumerate}


\section*{Exercise \ref{Ex:1.8}: Colour-ordered Feynman rules}
\label{Sol:1.8}

We start from the full Feynman rule four-point vertex~\eqref{V4YM} contracted with dummy polarisation
vectors $\epsilon_{i}$,
\begin{align}
V_{4}&= -\i g^2\, f^{abe}\, f^{cde}\, \left[ \epsdot{1}{3}\epsdot{2}{4}- \epsdot{1}{2}\epsdot{3}{4} \right]
+\text{cyclic}\,,
\end{align}
and use $f^{abe}\, f^{cde}=-\Tr([T^{a},T^{b}]\, [T^{c},T^{d}]) / 2$, which is
obtained from eqs.~\eqref{I.17} and~\eqref{eq:completenessSUN}. Note that the $U(1)$ piece cancels out here. 
Expanding out the commutators in the traces and collecting terms of identical colour
ordering gives
\begin{align}
V_{4}&= \frac{\i g^{2}}{2} \Tr\left(T^{a}T^{b}T^{c}T^{d}\right)\, \bigl[ 2\epsdot{1}{2}\epsdot{3}{4}
-\epsdot{1}{3}\epsdot{2}{4}\nn\\ & \qquad -\epsdot{1}{4}\epsdot{2}{3} \bigr]+\text{cyclic}\, ,
\end{align}
which is the result quoted in eq.~\eqref{Feynrulveert}.


\section*{Exercise \ref{Ex:1.9}: Independent gluon partial amplitudes}
\label{Sol:1.9}

a) Taking parity and cyclicity into account we have the following independent four-gluon tree-level amplitudes:
\begin{align*}
A^{\text{tree}}_{4}(1^{+},2^{+},3^{+},4^{+}) \, ,\qquad & A^{\text{tree}}_{4}(1^{-},2^{+},3^{+},4^{+}) \, ,
\\
A^{\text{tree}}_{4}(1^{-},2^{-},3^{+},4^{+})\, , \qquad & A^{\text{tree}}_{4}(1^{-},2^{+},3^{-},4^{+})\, .
\end{align*}
The last two are related via the $U(1)$ decoupling theorem as 
\begin{align}
\begin{aligned}
A^{\text{tree}}_{4}(1^{-},2^{+},3^{-},4^{+}) &= - A^{\text{tree}}_{4}(1^{-},2^{+},4^{+},3^{-}) - A^{\text{tree}}_{4}(1^{-},4^{+},2^{+},3^{-}) \\
&= - A^{\text{tree}}_{4}(3^{-},1^{-},2^{+},4^{+}) - A^{\text{tree}}_{4}(3^{-},1^{-},4^{+},2^{+}) \, .
\end{aligned}
\end{align}
Hence only the three amplitudes $A^{\text{tree}}_{4}(1^{+},2^{+},3^{+},4^{+})$,
$A^{\text{tree}}_{4}(1^{-},2^{+},3^{+},4^{+})$ and $A^{\text{tree}}_{4}(1^{-},2^{-},3^{+},4^{+}) $
are independent. In fact the first two of this list vanish, so there is only one independent
four-gluon amplitude at tree-level to be computed.

\medskip\noindent
b) Moving on to the five-gluon case, we have the four 
cyclic and parity independent amplitudes
\begin{align*}
A^{\text{tree}}_{5}(1^{+},2^{+},3^{+},4^{+},5^{+}) \, ,\qquad & A^{\text{tree}}_{5}(1^{-},2^{+},3^{+},4^{+},5^{+}) \, ,
\\
A^{\text{tree}}_{5}(1^{-},2^{-},3^{+},4^{+},5^{+})\, , \qquad & A^{\text{tree}}_{4}(1^{-},2^{+},3^{-},4^{+},5^{+})\, .
\end{align*}
Looking at the following $U(1)$ decoupling relation we may again relate the last amplitude in the
above list to the third one
\begin{align}
\begin{aligned}
&A^{\text{tree}}_{5}(2^{+},3^{-},4^{+},5^{+},1^{-}) = \\ & =- A^{\text{tree}}_{5}(3^{-},2^{+},4^{+},5^{+},1^{-}) - A^{\text{tree}}_{5}(3^{-},4^{+},2^{+},5^{-},1^{-}) 
- A^{\text{tree}}_{5}(3^{-},4^{+},5^{+},2^{+},1^{-}) 
\\ 
&= - A^{\text{tree}}_{5}(1^{-},3^{-},2^{+},4^{+},5^{+}) - A^{\text{tree}}_{5}(1^{-},3^{-},4^{+},2^{+},5^{+}) -  A^{\text{tree}}_{5}(1^{-},3^{-},4^{+},5^{+},2^{+}) \,.
\end{aligned}
\end{align}
Hence also for the five-gluon case there are only three independent amplitudes:
$A^{\text{tree}}_{5}(1^{+},2^{+},3^{+},4^{+},5^{+})$,
$A^{\text{tree}}_{5}(1^{-},2^{+},3^{+},4^{+},5^{+})$, $A^{\text{tree}}_{5}(1^{-},2^{-},3^{+},4^{+},5^{+}) $. The first two in this list vanish, leaving us with one independent and
non-trivial five-gluon tree-level amplitude, of the MHV type.


\section*{Exercise \ref{Ex:MHV3}: The $\overline{\text{MHV}}_3$ amplitude}
\label{Sol:MHV3}

Using the three-point vertex in eq.~\eqref{Feynrulveert} we obtain
\begin{align} \label{eq:A3antiMHV1}
\begin{aligned}
A_3^{\text{tree}}\left(1^+, 2^+, 3^-\right) = & \frac{\i g}{\sqrt{2}} \bigl\{ (\epsilon_{+,1}\cdot \epsilon_{+,2})\, (p_1-p_2)\cdot \epsilon_{-,3} \\
& + (\epsilon_{+,2}\cdot \epsilon_{-,3}) \, (p_2-p_3) \cdot \epsilon_{+,1} +
(\epsilon_{-,3}\cdot \epsilon_{+,1})\, (p_3-p_1) \cdot \epsilon_{+,2} \bigr\} \,.
\end{aligned}
\end{align}
Choosing the same reference momentum for all polarisations, $r^{\dot\alpha \alpha} = \tilde{\mu}^{\dot\alpha} \mu^{\alpha}$, we have
\begin{align}
\epsilon_{+,1} \cdot \epsilon_{+,2} = 0 \,, \quad
\epsilon_{+,i} \cdot \epsilon_{-,j} = - \frac{\vev{\mu j}\bev{\mu i}}{\vev{i \mu}\bev{j\mu}}\, ,
\quad (p_{i}-p_{j})\cdot\epsilon_{+,k}= \sqrt{2}\, \frac{\bev{ki}\vev{\mu i}}{\vev{k\mu}}\,,
\end{align}
where we used that $p_i+p_j+p_k=0$.
Substituting these into \eqn{eq:A3antiMHV1} yields
\begin{align} \label{eq:A3antiMHV2}
\begin{aligned}
A_3^{\text{tree}}\left(1^+, 2^+, 3^-\right) & = \i g \frac{ \spA{\mu}{3} }{[3\mu] \spA{1}{\mu} \spA{2}{\mu} } \Bigl( [12][\mu2]\spA{2}{\mu} - [1\mu]  \overbrace{ \spA{\mu}{3}[32] }^{-\spA{\mu}{1}[12] }\Bigr) \\
& = \i g \frac{ \spA{\mu}{3} [12] }{[3\mu] \spA{1}{\mu} \spA{2}{\mu} } \Bigl(  \overbrace{ [\mu2]\spA{2}{\mu} + \spA{\mu}{1} [1\mu]}^{-[\mu3]\spA{3}{\mu}} \Bigr) \\
& =  -\i g \frac{\vev{\mu 3}^{2} \bev{12}}{\vev{1\mu}\vev{2\mu}} \, .
\end{aligned}
\end{align}
Since the left-handed spinors are collinear, we may set $\la_{2}=a\la_{1}$ and $\la_{3}=b\la_{1}$.
Momentum conservation $\la_{1}(\tla_{1} + a\tla_{2}+b\tla_{3})=0$ then implies that
$a = \bev{31}/\bev{23}$ and $b= \bev{12}/\bev{23}$.
Substituting these into eq.~\eqref{eq:A3antiMHV2} finally gives
\begin{align}
A_3^{\text{tree}}\left(1^+, 2^+, 3^-\right) = -\i g \frac{\bev{12}^{3}}{\bev{23}\bev{31}}\, .
\end{align}

\newpage

\section*{Exercise \ref{Ex:1.11}: Four-point quark-gluon scattering}
\label{Sol:1.11}
There are two colour-ordered diagrams contributing to $A_{\bar{q}qgg}^{\text{tree}}(1^{-}_{\bar q}, 2^{+}_{q}, 3^{-}, 4^{+})$:
\begin{align}
A_{\bar{q}qgg}^{\text{tree}}(1^{-}_{\bar q}, 2^{+}_{q}, 3^{-}, 4^{+}) = \underset{\textstyle\mathrm{(I)}}{\vcenter{\hbox{\includegraphics[scale=0.7]{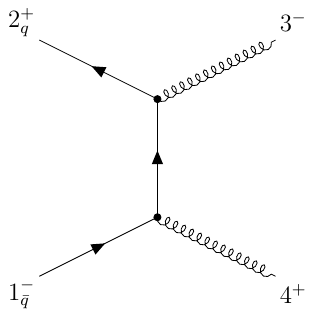}}}}
+ \underset{\textstyle\mathrm{(II)}}{\vcenter{\hbox{\includegraphics[scale=0.7]{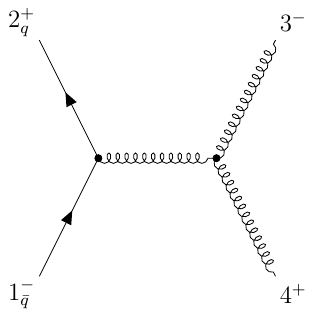}}}} \,.
\end{align}
The first graph (I) is proportional to
$[2| \slashed{\epsilon}_{3,-} (\slashed{p}_1+\slashed{p}_2) \slashed{\epsilon}_{4,+} |1\rangle\,,$
which vanishes for the reference-vector choice $\mu_{4}\, \tilde\mu_{4}=p_{1}$. This is so as
\begin{align}
\slashed{\epsilon}_{4,+} = - \frac{\sqrt{2}}{\vev{4 \mu_4}} \bigl(|4]\langle \mu_4| + | \mu_4 \rangle [4| \bigr) 
 \quad \Rightarrow \quad 
\slashed{\epsilon}_{4,+} |1\rangle = - \frac{\sqrt{2}}{\vev{4 \mu_4}} |4] \langle \mu_4 1 \rangle \stackrel{\mu_{4}=\la_{1}}{=}0 \,.
\end{align}
Evaluating the second graph (II) with the colour-ordered Feynman rules we obtain
\begin{align}
\label{graphII}
\text{(II)} = \frac{\i g^2}{2 q^2} [2|\gamma_{\mu}|1\rangle 
\Bigl [ \underbrace{ (\epsilon_{3,-}\cdot \epsilon_{4,+})\, p_{34}^{\mu}}_{(1)}
+ \underbrace{\epsilon_{4,+}^{\mu} \, (p_{4q}\cdot\epsilon_{3,-})}_{(2)} +
\underbrace{ \epsilon^{\mu}_{3,-}\, (p_{q3}\cdot
\epsilon_{4,+})}_{(3)}  \Bigr ]  \,,
\end{align}
where $q=p_1+p_2$ and $p_{ij} = p_i-p_j$.
The term $(2)$ vanishes for our choice $\mu_{4}\, \tilde\mu_{4}=p_{1}$,
\begin{align}
(2) \propto [2| \slashed{\epsilon}_{4,+} |1\rangle 
\stackrel{\mu_{4}\tilde\mu_{4}=p_{1}}{=} 0 \,.
\end{align}
For the term $(3)$, we note that $\slashed{\epsilon}_{3,-}
= \sqrt{2} \, (\, |3\rangle\, [\mu_{3}| + |\mu_{3}]\,\langle 3| \,)/[3\mu_{3}]$ 
to find
\begin{align}
(3) \propto
[2| \slashed{\epsilon}_{3}^{-}\, |1\rangle = \frac{\sqrt{2}}{[3\mu_{3}]}\,
\bev{2\mu_{3}}\, \vev{31}\,,
\end{align}
which is killed by the choice $\mu_3\tilde{\mu}_3 = p_2$.
Hence, for this choice of reference vectors only the term $(1)$ in \eqn{graphII} contributes. 
One has
\begin{align}
\epsilon_{3,-}\cdot\epsilon_{+,4}= - \frac{\vev{\mu_{4}3}\, \bev{\mu_{3}\,4}}{\vev{4\mu_{4}}
\, \bev{3\mu_{3}}}  \stackrel{\mu_{3}=\la_2, \mu_{4}=\la_1}{=}
-\frac{\vev{13}\, \bev{24}}{\vev{41}\, \bev{32}}\, ,
\end{align}
and, using momentum conservation,
\begin{align}
 [2|(\slashed{p}_{3}\, - \slashed{p}_{4}) |1\rangle = 2 \, \spB 23 \spA 31 \,.
\end{align}
Inserting these into the term $(1)$ of \eqn{graphII} and using $q^{2}=\vev{12}\bev{21}$ yields
\begin{align}
A^{\text{tree}}_{\bar q q gg}(1^-_{\bar{q}}, 2^+_q, 3^-, 4^+) &= - \i g^2 \frac{\vev{13}^{2}}{\vev{12}\vev{41}} \, \overbrace{\frac{\bev{24}\vev{43}}{\bev{21}\vev{43}}}^{-\bev{21}\vev{13}} = - \i g^2 \frac{\vev{13}^{3}\vev{23}}{\vev{12}\vev{23}\vev{34}\vev{41}}\,,
\end{align}
as claimed. 
The helicity count of our result is straightforward and correct:
\begin{align*}
&h_{1} [ A^{\text{tree}}_{\bar q q gg} ]=-\frac{1}{2}(3-1-1)=-\ft12 \,, \qquad
& h_{2}[A^{\text{tree}}_{\bar q q gg}]=-\ft12(1-1-1)=+\ft12\, , \\
& h_{3}[A^{\text{tree}}_{\bar q q gg} ]=-\ft12(4-1-1)=-1\, , \qquad
& h_{4}[A^{\text{tree}}_{\bar q q gg}]=-\ft12(0-1-1)=+1\,.
\end{align*}


\section*{Exercise \ref{Exer:2.1}: The vanishing splitting function $\mathrm{Split}^{\mathrm{tree}}_+(x,a^+,b^+)$}
\label{Sol:2.1}
We parametrise the collinear limit $5^+ \parallel 6^+$ by
\begin{align}
\lambda_5 = \sqrt{x} \, \lambda_P \,, \qquad \tilde{\lambda}_5 = \sqrt{x} \, \tilde{\lambda}_P \,, \qquad 
\lambda_6 = \sqrt{1-x} \, \lambda_P \,, \qquad \tilde{\lambda}_5 = \sqrt{1-x} \, \tilde{\lambda}_P \,,
\end{align} 
with $P = \lambda_P \tilde{\lambda}_P = p_5 + p_6$. Substituting this into the Parke-Taylor formula~\eqref{ParkeTaylor1} for $A_{6}^{\rm tree}(1^{-}, 2^{-}, 3^{+}, 4^{+}, 5^{+}, 6^{+})$ gives
\begin{align}
A_{6}^{\rm tree}(1^{-}, 2^{-}, 3^{+}, 4^{+}, 5^{+}, 6^{+}) \stackrel{5 \parallel 6}{\longrightarrow} \frac{g}{\sqrt{x(1-x)}\,\vev{56}}\, \frac{\i g^3 \vev{12}^{4}}{\vev{12}\vev{23}\vev{34}\vev{4P}\vev{P1}} \,.
\end{align}
Comparing this to the expected collinear behaviour from \eqn{eq:collinear_limit},
\begin{align} \begin{aligned}
A_{6}^{\rm tree}(1^{-}, 2^{-}, 3^{+}, 4^{+}, 5^{+}, 6^{+}) \stackrel{5 \parallel 6}{\longrightarrow} \, & {\rm Split}_{-}^{\rm tree}(x,5^{+},6^{+})\, 
A_{5}^{\rm tree}(1^{-}, 2^{-}, 3^{+}, 4^{+}, P^{+}) \\ 
& + {\rm Split}_{+}^{\rm tree}(x,5^{+},6^{+}) \, A_{5}^{\rm tree}(1^{-}, 2^{-}, 3^{+}, 4^{+}, P^{-}) \,, 
\end{aligned} \end{align}
and using \eqn{ParkeTaylor1} for the $5$-gluon amplitudes, we see that the term with ${\rm Split}_{+}^{\rm tree}(x,5^{+},6^{+})$ is absent. Since $ A_{5}^{\rm tree}(1^{-}, 2^{-}, 3^{+}, 4^{+}, P^{-}) \neq 0$, we deduce that
\begin{align}
{\rm Split}_{+}^{\rm tree}(x,5^{+},6^{+}) = 0 \,,
\end{align} 
as claimed.


\section*{Exercise \ref{Ex:2.2}: Soft functions in the spinor-helicity formalism}
\label{Sol:2.2}
The leading soft function for a positive-helicity gluon with colour-ordered neighbours $a$ and $b$ is given by eq.~\eqref{eq:S0explicit} with $n=a$ and $1=b$,
\begin{align} \label{eq:S0YMex1}
\mathcal{S}^{[0]}_{\text{YM}}\left(a,q^+,b\right) = \frac{g}{\sqrt{2}} \left(\frac{\epsilon_+ \cdot p_b}{p_b \cdot q} - \frac{\epsilon_+ \cdot p_a}{p_a \cdot q} \right)  \,.
\end{align}
Using \eqn{s1polarrel} for the polarisation vector with $\mu$ as reference spinor, we have that
\begin{align}
\frac{\epsilon_+ \cdot p_i}{p_i \cdot q} = \sqrt{2} \frac{\spA{\mu}{i}}{\spA{q}{i} \spA{\mu}{q}} \,.
\end{align}
Substituting this with $i=a,b$ into eq.~\eqref{eq:S0YMex1} and using the Schouten identity give
\begin{align}
\mathcal{S}^{[0]}_{\text{YM}}\left(a,q^+,b\right) =  g \, \frac{\spA ab}{\spA aq \spA qb} \,,
\end{align}
as claimed. We can obtain the negative-helicity soft function by acting with spacetime parity on the positive-helicity one. Parity exchanges $\lambda^{\alpha}$ and $\tilde{\lambda}^{\dot\alpha}$, which amounts to swapping $\spA ij$ with $\spB ji$.

We now turn to a positive-helicity graviton. The starting point is again eq.~\eqref{eq:S0explicit},
\begin{align} \label{eq:S0GMex1}
\mathcal{S}^{[0]}_{\text{GR}}\left(q^{++},1,\ldots,n\right) = \kappa \sum_{a=1}^n \frac{\epsilon^{++}_{\mu\nu} \, p_a^{\mu} p_a^{\nu}}{p_a \cdot q} \,.
\end{align}
We parametrise the graviton's polarisation vector by two copies of the gauge-field~one, 
\begin{align}
\epsilon_{++}^{\mu\nu}(q) = \epsilon_+^{\mu}(q,x) \, \epsilon_+^{\nu}(q,y) \,,
\end{align}
where we spelled out the arbitrary reference vectors $x$ and $y$. Substituting this into \eqn{eq:S0GMex1} and using \eqn{s1polarrel} for the polarisation vectors gives the desired result:
\begin{align}
\mathcal{S}^{[0]}_{\text{GR}}\left(q^{++},1,\ldots,n\right) = \kappa \sum_{a=1}^n \frac{\spA xa \spA ya \spB aq}{\spA xq \spA yq \spA aq} \,.
\end{align}
As above, the negative-helicity result can by obtained through parity conjugation.

\newpage

\section*{Exercise \ref{Ex:qqggg}: A $\bar{q}qggg$ amplitude from collinear and soft limits}
\label{Sol:qqggg}
Let us consider the collinear limit $3^{-} \parallel 4^{+}$ of the quark-gluon amplitude \break
$A_{\bar{q}qggg}^{\text{tree}}(1_{\bar q}^{-}, 2_{q}^{+}, 3^{-}, 4^{+}, 5^{+})$. We parametrise the limit with
$$
\la_{3}\to \sqrt{x}\, \la_{P}\, , \quad  \la_{4}\to\sqrt{1-x}\, \la_{P} \,,
$$
where $P=p_3+p_4$. The collinear factorisation theorem implies that
\begin{align} \label{5qqlimit}
\begin{aligned}
A_{\bar{q}qggg}^{\text{tree}}(1_{\bar q}^{-}, 2_{q}^{+}, 3^{-}, 4^{+}, 5^{+})  
\stackrel{3 \parallel 4}{\longrightarrow} \, &
\text{Split}^{\text{tree}}_{+}(x,3^{-},4^{+})\, A_{\bar{q}qgg}^{\text{tree}}
(1_{\bar q}^{-}, 2_{q}^{+}, P^{-}, 5^{+}) \\ & +  
\text{Split}^{\text{tree}}_{-}(x,3^{-},4^{+})\, 
A_{\bar{q}qgg}^{\text{tree}}(1_{\bar q}^{-}, 2_{q}^{+}, P^{+}, 5^{+}) \\
= \, & \frac{g \, x^{2}}{\sqrt{x (1-x)}\, \vev{34}}\, \frac{- \i g^2 \vev{1P}^{3}\vev{2P}}{\vev{12}\vev{2P}\vev{P5}
\vev{51}} \,,
\end{aligned}
\end{align}
where we inserted \eqn{splittingfunctions} for the splitting functions, and eqs.~\eqref{eq:Aqqgg1} and~\eqref{eq:Aqqgg2} for the amplitudes.
The limiting form of \eqn{5qqlimit} suggests that the amplitude before the limit takes the~form
\begin{align}
A_{\bar{q}qggg}^{\text{tree}}(1_{\bar q}^{-}, 2_{q}^{+}, 3^{-}, 4^{+}, 5^{+}) = - \i g^3 \frac{\vev{13}^{3}\vev{23}}
{\vev{12}\vev{23}\vev{34}\vev{45}\vev{51}} \, .
\end{align}
The form above leads us to conjecture the following $n$-particle generalisation:
\be
A_{\bar{q}qg\ldots g}^{\text{tree}}(1_{\bar q}^{-}, 2_{q}^{+}, 3^{-}, 4^{+}, \ldots, n^{+}) = - \i g^{n-2} \frac{\vev{13}^{3}\vev{23}}
{\vev{12}\vev{23}\vev{34}\ldots\vev{n1}}\, .
\label{qqnpoint}
\ee
By analogy with eq.~\eqn{5qqlimit}, we see that the conjectured form of the $n$-particle amplitude~\eqn{qqnpoint} 
is consistent with the collinear limits $3^{-} \parallel 4^{+}$ and $i^{+} \parallel (i+1)^{+}$ for $i=4,\ldots , n-1$. 
Let us also study two soft limits.
First we take $\la_{3}\to 0$. Then we immediately see that
\begin{align}
A_{\bar{q}qg\ldots g}^{\text{tree}}(1_{\bar q}^{-}, 2_{q}^{+}, 3^{-}, 4^{+},\ldots, n^{+}) \stackrel{3^{-}\to 0}{\longrightarrow}  0 \,.
\end{align}
Since the expected behaviour in the limit is
\begin{align}
A_{\bar{q}qg\ldots g}^{\text{tree}}(1_{\bar q}^{-}, 2_{q}^{+}, 3^{-}, 4^{+},\ldots, n^{+})  \stackrel{3^{-}\to 0}{\longrightarrow} \mathcal{S}^{[0]}(2,3^{-},4)\,  
A_{\bar{q}qg\ldots g}^{\text{tree}}(1_{\bar q}^{-}, 2_{q}^{+}, 4^{+}, \ldots, n^{+}) \, ,
\end{align}
and the relevant soft function is not zero, this implies that 
\begin{align}
A_{\bar{q}qg\ldots g}^{\text{tree}}(1_{\bar q}^{-}, 2_{q}^{+}, 3^{+}, \ldots, n^{+}) = 0 \,,
\end{align}
which is thus the conjectured $n$-particle generalisation of \eqn{eq:Aqqgg1}. 
Taking the soft limit $4^{+}\to 0$ (or any other positive-helicity gluon leg) on the other hand again allows us to check the self-consistency of \eqn{qqnpoint},
\begin{align}
A_{\bar{q}qg\ldots g}^{\text{tree}}(1_{\bar q}^{-}, 2_{q}^{+}, 3^{-}, 4^{+}, \ldots, n^{+}) \stackrel{4^{+}\to 0}{\longrightarrow} 
\underbrace{g \frac{\vev{35}}{\vev{34}\vev{45}}}_{\mathcal{S}^{[0]}(3,4^{+},5)}\, A_{\bar{q}qg\ldots g}^{\text{tree}}(1_{\bar q}^{-}, 2_{q}^{+}, 3^{-}, 
5^{+}, \ldots, n^{+}) \, .
\end{align}


\section*{Exercise \ref{Ex:6gNMHV}: The six-gluon split-helicity NMHV amplitude}
\label{Sol:6gNMHV}
We want to determine the NMHV six-gluon amplitude 
$A_{6}^{\text{tree}}(1^{+},2^{+},3^{+},4^{-},5^{-},6^{-})$. The $[6^-1^+\rangle$ shift leads to the BCFW recursion relation
\begin{align}
\begin{aligned}
& A_{6}^{\text{tree}}(1^{+}, 2^+, 3^+, 4^-, 5^-, 6^{-}) = \\
& \quad \sum_{i=3}^{5} \sum_{h=\pm} A_{i}^{\text{tree}}\left(\hat 1^+ , 2^+,\ldots , i-1, -{\hat P}^{-h}_{i}(z_{P_{i}})\right)\, 
\frac{\i}{P_{i}^{2}}
\, A_{8-i}^{\text{tree}}\left({\hat P}^{h}_{i}(z_{P_{i}}),i,\ldots , 5^-, \hat{6}^- \right)\,.
\end{aligned}
\end{align}
Since the all-plus/minus and single-plus/minus tree amplitudes vanish, only two contributions are non-zero. Diagrammatically, they are given by
\begin{align*}
\underset{\textstyle\mathrm{(I)}}{\vcenter{\hbox{\includegraphics[scale=0.7]{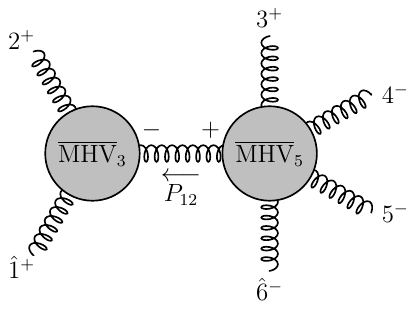}}}}
\quad \underset{\textstyle\mathrm{(II)}}{\vcenter{\hbox{\includegraphics[scale=0.7]{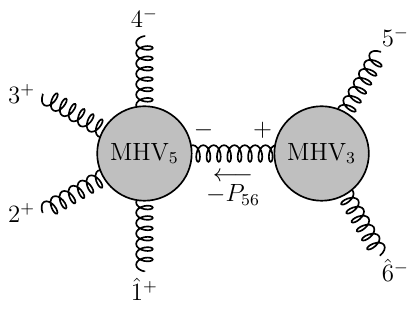}}}} 
\end{align*}
where we used the short-hand notation $P_{ij} = p_i+p_j$. The first diagram is given by
\begin{align}
\label{sh1}
\text{(I)} = \frac{- \i g \, \spB{\hat{1}}{2}^3}{\spB{2}{(-\hat P_{12})}\, \spB{(-\hat P_{12})}{\hat 1}}
\times \frac{\i}{\vev{12}\bev{21}} \times 
\frac{-\i g^3 \spB{\hat P_{12}}{3}^3}{\spB{3}{4}\,\spB{4}{5}\, \spB{5}{\hat{6}}\, \spB{\hat{6}}{ 
\hat P_{12}}} \, .
\end{align}
The corresponding $z$-pole is at $z_{\text{I}} = P_{12}^2/\spAB{6}{P_{12}}{1} = \spA 12 / \spA 62$. Hence we~have
\begin{align}
|\hat{1}\rangle = \frac{\spA{1}{6}}{\spA{6}{2}} |2\rangle \,, \quad |\hat 1] = |1] \,, \quad
|\hat{6}\rangle = |6\rangle \,, \quad |\hat{6}] = |6] + \frac{\spA{1}{2}}{\spA{6}{2}} |1] \,,
\end{align}
where we used the Schouten identity to simplify $|\hat{1}\rangle$, and, for $\hat{P}_{12} = \hat{p}_1 + p_2$,
\begin{align}
|\hat{P}_{12}\rangle = |2\rangle \,, \qquad  |\hat{P}_{12}] = |2] + \frac{\spA 61}{\spA 62} |1] \,.
\end{align}
By combining the above we obtain
\begin{align} \begin{aligned}
\bev{2 \hat{P}_{12}} &= \frac{\vev{61}}{\vev{62}}\, \bev{21}\, , \quad
\bev{\hat{P}_{12} \hat{1}} =\bev{21}\, , \quad  \bev{5\hat 6} = \frac{[5|P_{16}|2\rangle}{\spA 62}\, , \\
\bev{\hat{P}_{12} 3} &= \frac{\langle 6 | P_{12}|3]}{\vev{62}}\, , \quad
\bev{\hat 6 \hat{P}_{12}} = - \frac{P_{26}^{2}+P_{12}^{2}+P_{16}^{2}}{\vev{62}} \,.
\end{aligned} \end{align}
The expression for $\bev{\hat 6 \,\hat P_{12}}$ can be further simplified by noting that $P_{26}^{2}+P_{12}^{2}+P_{16}^{2} = (p_6+p_1+p_2)^2$.
Substituting all these into~\eqn{sh1}, with the sign convention~\eqref{negpspin}, gives our final expression for diagram (I):
\begin{align}
\text{(I)} = \i g^4 \frac{\spAB{6}{P_{12}}{3}^3}{\spA 61 \spA 12 \spB 34 \spB 45 \spBA{5}{P_{16}}{2}} \frac{1}{(p_6+p_1+p_2)^2}\,.
\end{align}
For the second diagram we start with
\begin{align} \label{eq:6gdiagII}
\text{(II)} =  \frac{\i g^3 \, \spA{4}{\hat P_{56}}^{3}}{\vev{\hat P_{56}\hat{1}}\vev{\hat{1}2}
\vev{23}\vev{34}}\times \frac{\i}{\vev{56}\bev{65}} \times \frac{\i g \, \vev{5\hat 6}^{3}}
{\vev{6 (-\hat{P}_{56})}\vev{(-\hat{P}_{56}) 5}} \,.
\end{align}
Now the shift parameter takes the value $z_{\text{II}} = \spB 65 / \spB 51$, which implies
\begin{align}
|\hat{1}\rangle = |1\rangle + \frac{\spB 56}{\spB 51} |6\rangle \,, \quad 
|\hat{6}\rangle = |6\rangle \,, \quad 
|\hat{P}_{56}\rangle = |5\rangle + \frac{\spB 16}{\spB 15} |6\rangle \,,
\end{align}
and hence
\begin{align} \begin{aligned}
\vev{4 \hat{P}_{56}} &= \frac{\langle 4 | P_{56}|1]}{\bev{51}}\, , \quad
\vev{\hat{P}_{56} \hat{1}} = \frac{(p_1+p_5+p_6)^2}{\bev{15}}\, ,\quad
\vev{5\hat{6}} = \vev{56}\, ,\\ 
\vev{6 \hat{P}_{56}}&= \vev{65}\, ,\quad
\vev{\hat{P}_{56} 5}= \frac{\bev{16}}{\bev{15}} \vev{65}\, , \quad
\vev{\hat{1}2}= \frac{[5|P_{16}|2\rangle}{\bev{51}}\, .
\end{aligned} \end{align}
Plugging these into \eqn{eq:6gdiagII} yields
\begin{align}
\text{(II)} = \i g^4 \frac{\spAB{4}{P_{56}}{1}^3}{ \vev{23} \vev{34} [16][65] \spBA{5}{P_{16}}{2} }  \frac{1}{(p_1+p_5+p_6)^2} \,.
\end{align}
Finally, by combining the two diagrams we obtain
\begin{align} \begin{aligned}
A_{6}^{\text{tree}}(1^{+},2^{+},3^{+}, {} & 4^{-},5^{-},6^{-}) = \i g^4 \biggl( \frac{\spAB{6}{P_{12}}{3}^3}{\spA 61 \spA 12 \spB 34 \spB 45 \spBA{5}{P_{16}}{2}} \frac{1}{(p_6+p_1+p_2)^2} \\
& + \frac{\spAB{4}{P_{56}}{1}^3}{ \vev{23} \vev{34} [16][65] \spBA{5}{P_{16}}{2} }  \frac{1}{(p_1+p_5+p_6)^2} \biggr) \,.
\end{aligned} \end{align}


\section*{Exercise \ref{Ex:6gSoft}: Soft limit of the six-gluon split-helicity amplitude}
\label{Sol:6gSoft}
In the soft limit $p_5 \to 0$ we have the reduced momentum-conservation condition $p_{1}+p_{2}+p_{3}+p_{4}+p_{6}=0$, which implies that $(p_6+p_1+p_2)^2 = \spA 34 \spB 43$ and $(p_1+p_5+p_6)^2 = \spA 16 \spB 61$. Using these in \eqn{eq:A6NMHV} and pulling out the pole term $(\bev{45}\bev{56})^{-1}$ gives
\begin{align} 
\begin{aligned}
A_{6}^{\text{tree}}&(1^{+}, 2^{+}, 3^{+}, 4^{-}, 5^{-}, 6^{-}) \stackrel{5^- \to 0}{\longrightarrow}
\\ & \quad 
\frac{\i g^4}{[5|P_{16}|2\rangle\, \bev{45}\bev{56}} \biggl(
\frac{\langle 6|P_{12}|3]^{3} \bev{56}}{\vev{61}\vev{12}\bev{34}^2 \vev{43}}
+ \frac{ \langle 4|P_{56}|1]^{3} \bev{54} }{\vev{23}\vev{34} \bev{16}^2 \vev{61}}\,
\biggr) \,.
\end{aligned} 
\end{align}
We use the reduced momentum conservation and the Dirac equation to simplify
\begin{align}
\langle 6|P_{12}|3] & = - \langle 6|(p_3+p_4+p_6)|3] \nonumber \\
& = - \spA{6}{4} \spB{4}{3}\,,
\end{align}
and the soft limit for $\langle 4|P_{56}|1] = \spA{4}{6} \spB{6}{1}$, obtaining
\begin{align} 
\begin{aligned}
A_{6}^{\text{tree}}&(1^{+}, 2^{+}, 3^{+}, 4^{-}, 5^{-}, 6^{-}) \stackrel{5^- \to 0}{\longrightarrow}
\\ & \quad 
 \frac{\i g^4}{[5|P_{16}|2\rangle\, \bev{45}\bev{56}}\, \frac{\vev{46}^{3}}{\vev{12}\vev{23}\vev{34}
\vev{61}} \bigl(\bev{34}\bev{56} \vev{23} +  \bev{16}\bev{45}\vev{12} \bigr) \,.
\end{aligned} 
\end{align}
The two terms in the parentheses may be simplified using a Schouten identity~as
\begin{align}
 \overbrace{\bev{34}\bev{56}}^{-\bev{45}\bev{36}-\bev{53}\bev{46}} \hspace{-0.5cm} \vev{23}
 + \bev{16}\bev{45} \vev{12} & = \bev{54}[6|\overbrace{P_{13}}^{-p_{4}}|2\rangle + \bev{35}\bev{46}\vev{23} \nonumber \\
& = \bev{46}[5|P_{34}|2\rangle \nonumber \\
& = - \bev{46}[5|P_{16}|2\rangle\, .
\end{align}
By plugging this into the above we find
\begin{align}
A_{6}^{\text{tree}}(1^{+}, 2^{+}, 3^{+}, 4^{-}, 5^{-}, 6^{-})  &\stackrel{5^{-}\to 0}{\longrightarrow}
- g \frac{\bev{46}}{\bev{45}\bev{56}}  \times \frac{\i g^3 \, \vev{46}^{3}}{\vev{12}\vev{23}\vev{34}
\vev{61}}  \,,
\end{align}
which indeed matches the expected factorisation,
\begin{align}
A_{6}^{\text{tree}}(1^{+}, 2^{+}, 3^{+}, 4^{-}, 5^{-}, 6^{-})  &\stackrel{5^{-}\to 0}{\longrightarrow} \mathcal{S}^{[0]}_{\text{YM}}(4,5^{-},6) \times A_{5}^{\text{tree}}(1^{+}, 2^{+}, 3^{+}, 4^{-}, 6^{-})\,,
\end{align}
with the soft function given in \eqn{eq:SoftYM_minus}.


\section*{Exercise \ref{Ex:2.helicity}: Mixed-helicity four-point scalar-gluon amplitude}
\label{Sol:2.helicity}
The $[4^- 1^+\rangle$ shift leads to the BCFW recursion
\begin{align} \begin{aligned}
A_{4}(1^{+},2_{\phi},3_{\bar\phi},4^{-}) & = \vcenter{\hbox{\includegraphics[scale=0.7]{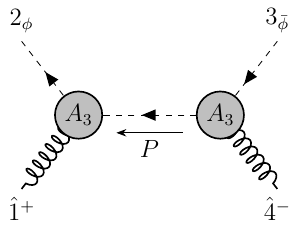}}} \\
& = A_{3}\left( (-\hat{P})_{\bar\phi}, \hat{1}^{+}, 2_{\phi} \right) \, \frac{\i}{P^2-m^2} \,
A_{3}\left(3_{\bar\phi},\hat{4}^{-}, \hat{P}_{\phi}\right) \\
& = \left( - \i \, \frac{ \spAB{r_1}{(-\hat{P})}{\hat{1}}}{\vev{r_1 \hat{1}}} \right) \frac{\i}{P^{2}-m^{2}} \left(-\i \,
\frac{\spAB{\hat{4}}{p_3}{r_4} }{\bev{\hat{4} r_4}} \right)
\,,
\end{aligned} \end{align}
where we used eqs.~\eqref{ssp} and~\eqref{ssm} for the three-point scalar-gluon amplitudes (with $g=1$), $P=p_1+p_2$, and $r_1$ ($r_4$) denotes the reference momentum of the gluon leg $1$ ($4$).
With the gauge choice $r_{1}=\hat{p}_4$ and $r_{4}=\hat{p}_1$ along with the identities
$|\hat 4\rangle=|4\rangle$ and $|\hat 1]=|1]$ for the $[4^- 1^+\rangle$ shift one has
\begin{align}
\vev{r_1 \hat{1}}&= \vev{41} \,, \quad 
\bev{\hat{4} \, r_4}= \bev{4 1} \,, \quad
\spAB{r_1}{\hat{P}}{\hat{1}} &= - \spAB{4}{p_3}{1} \,, \quad
\spAB{ \hat{4} }{p_3}{r_4}  &=  \spAB{4}{p_3}{1} \,.
\end{align}
Plugging these into the above yields the final compact result
\begin{align}
A_{4}(1^{+},2_{\phi},3_{\bar\phi},4^{-}) =  \i \,
\frac{\langle 4| p_{3}| 1]^{2}}{ (p_1+p_4)^2 \, [(p_1+p_2)^2-m^2] } \,.
\end{align}


\section*{Exercise \ref{Ex:2.confalg}: Conformal algebra}
\label{Sol:2.confalg}
The commutation relations with the dilatation operator $d$,
\begin{align}
\left[d,p^{\a\da} \right]=p^{\a\da}\, , \quad \left[d,k_{\a\da} \right]=-k_{\a\da}\, , \quad 
\left[d,m_{\a\b} \right]=0=\left[d,{\overline m}_{\da\db}\right] \,,
\end{align}
are manifest from dimensional analysis. We recall in fact that $d$ measures the mass dimension, i.e.~$[d,f] = [f] f$ where $[f]$ denotes the dimension of $f$ in units of mass, and that the helicity spinors $\lambda_i$ and $\tilde{\lambda}_i$ have mass dimension $1/2$.
It remains for us to compute the commutator $[k_{\a\da}, p^{\b\db}]$, which is given by
\begin{align}
\left[k_{\a\da},p^{\b\db} \right] = 
\left[\partial_{\a},\la^{\b}\tla^{\db} \right]\, \partial_{\da}+ \partial_{\a} \left[ \partial_{\da}, \la^{\b}\tla^{\db} \right]
= \delta^{\b}_{\ \a}\, \tla^{\db}
\, \partial_{\da} + \delta^{\db}_{\ \da}\, \la^{\b}\partial_{\a}
+ \delta^{\b}_{\ \a}\, \delta^{\db}_{\da} \,.
\end{align}
By using \eqn{symlapa} for a single particle with raised index,
\begin{align}
\la^{\beta}\,\partial_{\alpha} = \epsilon^{\beta\rho} \, 
\la_{(\rho} \partial_{\alpha)}+ \frac{1}{2} \underbrace{\epsilon^{\beta\rho}\, \epsilon_{\rho\alpha}}_{=\, \delta^{\beta}_{\ \alpha}} \, \la^{\gamma}\partial_{\gamma} \,,
\end{align}
and the analogous equation with dotted indices, we obtain
\begin{align}
\left[k_{\a\da},p^{\b\db} \right] & = \delta^{\b}_{\ \a}\, \epsilon^{\db\dot\rho} \underbrace{\tilde{\la}_{(\dot{\rho}} \partial_{\dot{\alpha})}}_{ = \, \overline{m}_{\dot{\rho}\dot{\alpha}} }
+ \, \delta^{\db}_{\ \da}\, \epsilon^{\b\rho} \underbrace{\la_{(\rho} \partial_{\alpha)}}_{= \, m_{\rho \alpha}} + \, \delta^{\b}_{\ \a}\delta^{\db}_{\ \da}
 \Bigl( \underbrace{ \frac{1}{2} \la^{\gamma}\partial_{\gamma}+ \frac{1}{2}\tla^{\dot\gamma}\partial_{\dot\gamma}
+ 1}_{= \, d} \Bigr) \,,
\end{align}
which concludes the proof of \eqn{confalgebra}.


\section*{Exercise \ref{Ex:2.invspecialconf}: Inversion and special conformal transformations}
\label{Sol:2.invspecialconf}
\noindent
a) Using the inversion transformation $I \, x^{\mu}= x^{\mu}/x^{2}$ and the translation
transformation $P^{\mu} \, x = x^{\mu }-a^{\mu}$ we 
have
\begin{align}
I \, P^{\mu} \, I \, x^{\mu}= I \, P^{\mu} \, \frac{x^{\mu}}{x^{2}} & =
 I \, \frac{x^{\mu}-a^{\mu}}{(x-a)^{2}} = 
\frac{\frac{x^{\mu}}{x^{2}}-a^{\mu}}{\left(\frac{x}{x^{2}}-a \right)^{2}}=
\frac{x^{\mu}-a^{\mu}x^{2}}{1-2 \, a\cdot x + a^{2}\, x^{2}} \,,
\end{align}
which equals the finite special conformal transformation in \eqn{finiteKaction}.
\medskip

\noindent
b) We begin by computing the Jacobian factor $|\partial x'/\partial x|$, i.e.\ the absolute value of the determinant of the matrix with entries $\partial x'^{\mu}/\partial x^{\nu}$ for $\mu,\nu=0,1,2,3$. It is convenient to decompose the special conformal transformation $x\to x'$ as in point~a):
\begin{align}
x^{\mu} \quad \overset{I}{\longrightarrow} \quad y^{\mu} \coloneqq \frac{x^{\mu}}{x^2} \quad \overset{P^{\mu}}{\longrightarrow} \quad z^{\mu} \coloneqq y^{\mu} - a^{\mu} 
\quad \overset{I}{\longrightarrow} \quad x'^{\mu} \coloneqq \frac{z^{\mu}}{z^2} \,.
\end{align}
The Jacobian factor for $x\to x'$ then factorises into the product of the Jacobian factors for the three separate transformations:
\begin{align}
\left| \frac{\partial x'}{\partial x} \right| = \left| \frac{\partial x'}{\partial z} \right| \, \left| \frac{\partial z}{\partial y} \right| \, \left| \frac{\partial y}{\partial x} \right| \,.
\end{align}
For the first inversion, $x^{\mu} \to y^{\mu}$, we have that
\begin{align}
\frac{\partial y^{\mu}}{\partial x^{\nu}} = \frac{1}{x^2}\left( \eta^{\mu}{}_{\nu} - 2 \, \frac{x^{\mu} x_{\nu}}{x^2} \right) \,,
\end{align}
so that the Jacobian factor takes the form
\begin{align}
\left| \frac{\partial y}{\partial x} \right| = \bigl(x^2\bigr)^{-4} \left| \mathrm{det}\left( \eta^{\mu}{}_{\nu} - 2 \, \frac{x^{\mu} x_{\nu}}{x^2} \right) \right| \,.
\end{align}
We use the representation of the determinant in terms of Levi-Civita symbols:
\begin{align}
\left| \frac{\partial y}{\partial x} \right| = \frac{\bigl(x^2\bigr)^{-4}}{4!} \left| 
  \epsilon_{\mu_1 \mu_2 \mu_3 \mu_4} \epsilon^{\nu_1 \nu_2 \nu_3 \nu_4}
  \left( \eta^{\mu_1}{}_{\nu_1} - 2 \, \frac{x^{\mu_1} x_{\nu_1}}{x^2} \right) \ldots  \left( \eta^{\mu_4}{}_{\nu_4} - 2 \, \frac{x^{\mu_4} x_{\nu_4}}{x^2} \right)  \right| \,.
\end{align}
The contractions involving two, one, or no factors of $\eta^{\mu_i}{}_{\nu_i}$ vanish because of the anti-symmetry of the Levi-Civita symbol.
The contractions with three factors of $\eta^{\mu_i}{}_{\nu_i}$ are equal. This leads us to
\begin{align}
\left| \frac{\partial y}{\partial x} \right| = \frac{\bigl(x^2\bigr)^{-4}}{4!} \left| 
  \epsilon_{\mu \nu \rho \sigma} \epsilon^{\mu \nu \rho \sigma} + 4 \, 
  \epsilon_{\mu \nu \rho \sigma_1} \epsilon^{\mu \nu \rho \sigma_2} \left( - 2 \, \frac{x^{\sigma_1} x_{\sigma_2}}{x^2} \right)
  \right| \,.
\end{align}
Using the identities $ \epsilon_{\mu \nu \rho \sigma} \epsilon^{\mu \nu \rho \sigma} = - 4!$ and $ \epsilon_{\mu \nu \rho \sigma_1} \epsilon^{\mu \nu \rho \sigma_2} = - 3! \, \eta^{\sigma_2}{}_{\sigma_1}$\footnote{We can derive this identity by tensor decomposition (see section~\ref{sec:TensorReduction}). We write $ \epsilon_{\mu \nu \rho \sigma_1} \epsilon^{\mu \nu \rho \sigma_2} = c \, \eta^{\sigma_2}{}_{\sigma_1}$, and fix $c=-3!$ by contracting both sides by $\eta^{\sigma_1}{}_{\sigma_2}$ and solving for $c$.} gives
\begin{align}
\left| \frac{\partial y}{\partial x} \right| = \bigl(x^2\bigr)^{-4}  \,.
\end{align}
Similarly, for the second inversion, $z \to x'$, we have 
\begin{align} \begin{aligned}
\left| \frac{\partial x'}{\partial z} \right| & = \bigl(z^2\bigr)^{-4}  \\
& = \left( \frac{1 - 2 \, a \cdot x + a^2 x^2}{x^2} \right)^{-4} \,.
\end{aligned} \end{align}
The Jacobian factor of the translation $y^{\mu} \to z^{\mu} = y^{\mu}-a^{\mu}$ is simply $1$, as the translation parameter $a^{\mu}$ does not depend on $y^{\mu}$. Putting the above together gives
\begin{align}
\left| \frac{\partial x'}{\partial x} \right| =  \left(1 - 2 \, a \cdot x + a^2 x^2\right)^{-4} \,.
\end{align}

\noindent
We now consider the transformation rule for the scalar field $\Phi$ in \eqn{eq:conformal_scalar},
\begin{align} \label{eq:conformal_scalar_2}
\Phi^{\prime}(x^{\prime}) =  \left(1 - 2 \, a \cdot x + a^2 x^2\right)^{\Delta} \Phi(x) \,,
\end{align}
and expand both sides in a Taylor series around $a^{\mu} = 0$. For the LHS we obtain
\begin{align} \label{eq:fprimeTaylor} \begin{aligned}
\Phi^{\prime}\left( x^{\prime} \right) & = \Phi^{\prime}(x) + a^{\mu} \left( \frac{\partial x^{\prime \, \nu}}{\partial a^{\mu}} \right) \biggl|_{a=0} \partial_{\nu} \Phi^{\prime}(x) + \mathcal{O}\bigl(a^2\bigr) \\
& = \Phi^{\prime}(x) + a^{\mu} \left( - \eta_{\mu}{}^{\nu} x^2 + 2 \, x_{\mu} \, x^{\nu} \right) \partial_{\nu} \Phi^{\prime}(x) + \mathcal{O}\bigl(a^2\bigr) \,.
\end{aligned} \end{align}
Since $\Phi^{\prime}(x) - \Phi(x) = \mathcal{O}(a)$, we can replace $\partial_{\nu} \Phi^{\prime}(x) $ by $\partial_{\nu} \Phi(x) $ in the above. Plugging this into \eqn{eq:conformal_scalar_2} and expanding also the RHS gives
\begin{align}
 \Phi^{\prime}(x) + a^{\mu} \left( - \eta_{\mu}{}^{\nu} x^2 + 2 \, x_{\mu} \, x^{\nu} \right) \partial_{\nu} \Phi^{\prime}(x)  = \Phi(x) - 2 \, \Delta \, (a \cdot x) \, \Phi(x) + \mathcal{O}\bigl(a^2\bigr) \,.
\end{align}
By comparing this to the defining equation of the generators~\eqref{eq:fxprime},
\begin{align}
\Phi^{\prime}\left( x \right ) = \left[ 1 - \i \, a^{\mu} \, K_{\mu} +  \mathcal{O}\bigl(a^2\bigr) \right] \, \Phi(x) \,,
\end{align}
we can read off the explicit form of the generator,
\begin{align}
K_{\mu} = \i \left[ x^2 \, \partial_{\mu} - 2 \, x_{\mu} \left( x^{\nu} \partial_{\nu} + \Delta \right) \right] \,,
\end{align}
as claimed.


\section*{Exercise \ref{Ex:2.kin_Jacobi}: Kinematical Jacobi identity}
\label{Sol:2.kin_Jacobi}
We start from the expression of $n_s$ given in eq.~\eqref{ns}.
We choose the reference momenta $r_i$ for the polarisation vectors $\epsilon_i$ so as to kill as many terms as possible. We recall that $\epsilon_i \cdot r_i = 0$. Choosing $r_1 = p_2$, $r_2 = p_1$, $r_3 = p_4$, and $r_4 = p_3$ yields
\begin{align}
\begin{aligned}
n_{s}= \phantom{ } & (p_1 \cdot p_2) \left[ (\epsilon_1\cdot \epsilon_2) (\epsilon_3\cdot \epsilon_4)   
  - (\epsilon_1\cdot \epsilon_3) (\epsilon_2\cdot \epsilon_4) 
  + (\epsilon_2\cdot \epsilon_3) (\epsilon_1\cdot \epsilon_4) \right] \\
  & + 2 \, (p_2 \cdot p_3)  (\epsilon_1\cdot \epsilon_2) (\epsilon_3\cdot \epsilon_4) \,.
\end{aligned}
\end{align}
The other factors are obtained from $n_s$ by replacing the particles' labels as
\begin{align}
n_t = n_s \bigl|_{1 \to 2, 2 \to 3, 3 \to 1} \,, \qquad \qquad n_u = n_s \bigl|_{1 \to 3, 2 \to 1, 3 \to 2} \,.
\end{align}
Adding the three factors gives
\begin{align}
\begin{aligned}
n_s + n_t + n_u = \phantom{ } & \bigl[ (\epsilon_1 \cdot \epsilon_4) (\epsilon_2 \cdot \epsilon_3) + 
  (\epsilon_1 \cdot \epsilon_3) (\epsilon_2 \cdot \epsilon_4) + (\epsilon_1 \cdot \epsilon_2) (\epsilon_3 \cdot \epsilon_4) \bigr]  \\
 & \times   ( p_1 \cdot p_2 +  p_2 \cdot p_3 + p_3 \cdot p_1 ) \,,
\end{aligned}
\end{align}
which vanishes because of momentum conservation.


\section*{Exercise \ref{Ex:2.KLT}: Five-point KLT relation}
\label{Sol:2.KLT}
The squaring relation~\eqref{eq:MtoA2} in the five-point case reads
\begin{align}  \label{eq:MtoA2_5pt}
M_{5}^{\text{tree}}(1,2,3,4,5) = \sum_{\sigma,\rho\in S_{2}} A_{5}^{\text{tree}}(1,\sigma,4,5)
\, S[\sigma|\rho] \, A_{5}^{\text{tree}}(1,\rho,5,4) \,,
\end{align}
where $S_2$ is the set of permutations of $\{2,3\}$, namely $S_2 = \bigr\{ \{2,3\}, \, \{3,2\} \bigl\}$. 
We recall that the KLT kernels $S[\sigma|\rho]$ are given by \eqn{KLTkernel} with $n=5$,
\begin{align}
S[\sigma|\rho] = \prod_{i=2}^{3} \biggl[ 
2 \, p_{1}\cdot p_{\sigma_{i}} + \sum_{j=2}^{i} 2 \, p_{\sigma_{i}}\cdot p_{\sigma_{j}}\, \theta(\sigma_{j},
\sigma_{i})_{\rho}
\biggr] \,,
\end{align}
where $\theta(\sigma_{j}, \sigma_{i})_{\rho}=1$ if $\sigma_{j}$ is before $\sigma_{i}$ in the permutation $\rho$,
and zero otherwise. We then have
\begin{align} \begin{aligned}
S\left[(2,3)|(2,3)\right] & = 2 \, p_1 \cdot p_2 \, \bigl[ 2 \, p_1 \cdot p_3 + 2 \, p_3 \cdot p_2 \, \overbrace{\theta(2,3)_{(2,3)}}^{= \, 1} \bigr] \\
& = s_{12} ( s_{13} + s_{23} ) \,,
\end{aligned} \end{align}
where $s_{ij} = 2 \, p_i \cdot p_j$, and similarly
\begin{align} \begin{aligned}
& S\left[(3,2)|(2,3)\right] = s_{12} \, s_{13} = S\left[(2,3)|(3,2)\right]  \,, \\ 
& S\left[(3,2)|(3,2)\right] = s_{13} (s_{12}+s_{23}) \,.
\end{aligned} \end{align}
Plugging the above into the squaring relation~\eqref{eq:MtoA2_5pt} gives
\begin{align} \label{eq:M5KLT1} \begin{aligned}
& M_{5}^{\text{tree}}(1,2,3,4,5) = \\
 & \quad s_{12} A_5^{\text{tree}}(1,2,3,4,5) \bigl[ s_{13} A^{\text{tree}}_5(1,3,2,5,4) + (s_{13}+s_{23}) A^{\text{tree}}_5(1,2,3,5,4) \bigr]  \\
 & \quad + s_{13} A_5^{\text{tree}}(1,3,2,4,5) \bigl[ s_{12} A_5^{\text{tree}}(1,2,3,5,4) + (s_{12}+s_{23}) A_5^{\text{tree}}(1,3,2,5,4) \bigr] \,.
\end{aligned} \end{align}
The terms in the square brackets can be simplified using the BCJ relations~\eqref{eq:BCJn}. For instance, for the first term we use
\begin{align} \begin{aligned}
& p_3 \cdot p_1 \, A^{\text{tree}}_5(1,3,2,5,4) + p_3 \cdot (p_1+p_2) \, A^{\text{tree}}_5(1,2,3,5,4) \\
& +  p_3 \cdot (p_1+p_2+p_5) \, A^{\text{tree}}_5(1,2,5,3,4) = 0 \,,
\end{aligned} \end{align}
which is obtained by replacing $1\to 3$, $2\to 1$, $3\to 2$, $4 \to 5$ and $5 \to 4$ in \eqn{eq:BCJn} with $n=5$. Substituting
\begin{align} \begin{aligned}
s_{13} \, A^{\text{tree}}_5(1,3,2,5,4) + (s_{13}+s_{23}) \, A^{\text{tree}}_5(1,2,3,5,4)  = s_{34} \, A^{\text{tree}}_5(1,2,5,3,4) \,, \\
s_{12} \, A_5^{\text{tree}}(1,2,3,5,4) + (s_{12}+s_{23}) \, A_5^{\text{tree}}(1,3,2,5,4) =  s_{24} \, A_5^{\text{tree}}(1,3,5,2,4) \,,
\end{aligned} \end{align}
into \eqn{eq:M5KLT1} finally gives
\begin{align} \begin{aligned}
M_{5}^{\text{tree}}(1,2,3,4,5) & = s_{12} \, s_{34} \, A_5^{\text{tree}}(1,2,3,4,5) \, A^{\text{tree}}_5(1,2,5,3,4) \\
 & \phantom{{} =} + s_{13} \, s_{24} \, A_5^{\text{tree}}(1,3,2,4,5) \, A_5^{\text{tree}}(1,3,5,2,4) \,,
\end{aligned} \end{align}
as claimed.


\section*{Exercise \ref{Ex:3.1}: The four-gluon amplitude in $\mathcal{N}=4$ super-symmetric Yang-Mills theory}
\label{Sol:3.1}
We begin with the $s_{12}$ channel. The cut integrand is given by the following product of tree amplitudes:
\begin{align}
\mathcal{C}^{\mathcal{N}=4}_{12|34} & \coloneqq \mathcal{C}_{12|34}\left(I^{(1)}_{\mathcal{N}=4}(1^-,2^-,3^+,4^+) \right) \nonumber \\
& \phantom{:} =  \sum_{h_1,h_2} \i \, A^{(0)}\left( (-l_1)^{-h_1},1^-,2^-,(l_2)^{h_2} \right) \, \i \, A^{(0)}\left( (-l_2)^{-h_2},3^+,4^+,(l_1)^{h_1} \right) \nonumber \\
& \phantom{\coloneqq {}} - 4 \, \sum_{h_1,h_2} A^{(0)}\left( (-l_1)^{-h_1}_{\Lambda},1^-,2^-,(l_2)^{h_2}_{\Lambda} \right) \, A^{(0)}\left( (-l_2)^{-h_2}_{\Lambda},3^+,4^+,(l_1)^{h_1}_{\Lambda} \right) \nonumber \\
& \phantom{\coloneqq {}} + 6 \, \i \, A^{(0)}\left( (-l_1)_{\phi},1^-,2^-,(l_2)_{\phi} \right) \, \i \, A^{(0)}\left( (-l_2)_{\phi},3^+,4^+,(l_1)_{\phi} \right) \,.
\end{align}
The constant factors multiplying the amplitudes deserve a few remarks. First, we have a factor counting each field's multiplicity in the $\mathcal{N}=4$ super-multiplet: $1$ gluon ($g$), $4$ gluinos ($\Lambda$), and $6$ scalars ($\phi$). Next, the factors of imaginary unit ``$\i$'' follow from the factorisation properties of tree-level amplitudes as discussed below \eqn{eq:multiparticlepole}. In particular, note that the factorisation of the fermion line does not require any factors of ``$\i$'', as opposed to gluons and scalars. Finally, the gluino's contribution comes with a further factor of $-1$ coming from the Feynman rule for the closed fermion loop. The only non-vanishing contribution comes from the product of purely gluonic amplitudes with $h_1=-$ and $h_2=+$,
\begin{align}
\mathcal{C}_{12|34}^{\mathcal{N}=4} = 
  \i \, A^{(0)}\left( (-l_1)^{+},1^-,2^-,(l_2)^{+} \right) \, \i \,
  A^{(0)}\left( (-l_2)^{-},3^+,4^+ ,(l_1)^{-} \right) \,.
\end{align}
This is the same as in the non-supersymmetric YM theory computed in section~\ref{sec:3.2} (see \eqn{eq:C12|34YM}), hence we can immediately see that 
\begin{align}
\mathcal{C}_{12|34}\left( I^{(1)}_{\mathcal{N}=4}(1^-,2^-,3^+,4^+) \right)=\mathcal{C}_{12|34}\left( I^{(1)}(1^-,2^-,3^+,4^+) \right) \,, 
\end{align}
as claimed.

In contrast to the $s_{12}$ channel, all fields contribute to the $s_{23}$-channel cut:
\begin{align}
\begin{aligned}
\mathcal{C}_{23|41}^{\mathcal{N}=4} & \coloneqq \mathcal{C}_{23|41}\left(I^{(1)}_{\mathcal{N}=4}(1^-,2^-,3^+,4^+) \right)  \\
& \phantom{:} =  \i \, A^{(0)}\left( (-l_1)^{+},2^-,3^+,(l_2)^{-} \right)
  \, \i \, A^{(0)}\left( (-l_2)^{+},4^+,1^-,(l_1)^{-} \right) \\
  & \phantom{{} \coloneqq}  + \i \, A^{(0)}\left( (-l_1)^{-},2^-,3^+,(l_2)^{+} \right)
  \, \i \, A^{(0)}\left( (-l_2)^{-},4^+,1^-,(l_1)^{+} \right) \\
  &  \phantom{{} \coloneqq}  - 4 \, A^{(0)}\left( (-l_1)^{-}_{\Lambda},2^-,3^+,(l_2)^{+}_{\Lambda} \right) \,
  A^{(0)}\left( (-l_2)^{-}_{\Lambda},4^+,1^-,(l_1)^{+}_{\Lambda} \right) \\ 
  &  \phantom{{} \coloneqq}  - 4 \, A^{(0)}\left( (-l_1)^{+}_{\Lambda},2^-,3^+,(l_2)^{-}_{\Lambda} \right) \,
  A^{(0)}\left( (-l_2)^{+}_{\Lambda},4^+,1^-,(l_1)^{-}_{\Lambda} \right) \\ 
  &  \phantom{{} \coloneqq}  + 6 \, \i \, A^{(0)}\left( (-l_1)_{\phi},2^-,3^+,(l_2)_{\phi} \right) \, \i \,
  A^{(0)}\left( (-l_2)_{\phi},4^+,1^-,(l_1)_{\phi} \right) \,.
\end{aligned}
\end{align}
The second and fourth terms can be obtained by swapping $1 \leftrightarrow 2$ and $3 \leftrightarrow 4$ in the first and the third ones, respectively. We put all terms over a common denominator:
\begin{align}
{\rm D} = \langle 1 l_1 \rangle \langle l_1 l_2 \rangle \langle l_2 4 \rangle \langle 4 1 \rangle \langle l_1 2 \rangle \langle 23 \rangle \langle 3 l_2 \rangle \langle l_2 l_1 \rangle \,.
\end{align}
Factoring it out we have
\begin{align}
\begin{aligned}
{\rm D} \, \mathcal{C}_{23|41}^{\mathcal{N}=4} = \ & \spA 1{l_1}^4  \spA 2{l_2}^4 + \spA 2{l_1}^4 \spA 1{l_2}^4 - 4 \, \spA {l_1}1 \spA{l_1}2^3 \spA{l_2}2 \spA{l_2}1^3 \\
 & - 4 \, \spA {l_1}2 \spA{l_1}1^3 \spA{l_2}1 \spA{l_2}2^3   + 6 \, \spA {l_1}1^2  \spA {l_2}1^2  \spA {l_1}2^2  \spA {l_2}2^2\,.
\end{aligned}
\end{align}
Here the ``magic'' of $\mathcal{N}=4$ super Yang-Mills theory comes into play: the five terms above conspire together to form the fourth power of a binomial, 
\begin{align}
\mathcal{C}_{23|41}^{\mathcal{N}=4} = \frac{\bigl(\spA 1{l_2} \spA 2 {l_1} - \spA 1{l_1} \spA 2{l_2}\bigr)^4}{{\rm D}} \,,
\end{align}
which can be further simplified using a Schouten identity, obtaining
\begin{align}
\mathcal{C}_{23|41}^{\mathcal{N}=4} &  = \frac{ \spA {l_1}{l_2}^4 \spA 12^4}{{\rm D}} \nonumber \\
& = \frac{  \spA {l_1}{l_2}^2 \spA 12^4}{\spA {l_1}1 \spA {l_1} 2 \spA {l_2}3 \spA {l_2}4 \spA 23 \spA 14} \,.
\end{align}
This matches $\mathcal{C}_{23|41}^{\text{box}}$ (see \eqn{eq:tchannelcut2}), and we can thus conclude that
\begin{align}
\mathcal{C}_{23|41}\left( I^{(1)}_{\mathcal{N}=4}(1^-,2^-,3^+,4^+) \right) =\mathcal{C}_{23|41}^{\text{box}}\left( I^{(1)}(1^-,2^-,3^+,4^+) \right) \,.
\end{align}


\section*{Exercise \ref{Ex:3.2}: Quadruple cuts of five-gluon MHV scattering amplitudes}
\label{Sol:3.2}
\begin{enumerate}[a)]
\item We parametrise the loop momentum $l_1$ using the spinors of the external momenta as in \eqn{eq:quadcutmombasis}. We then rewrite the quadruple cut equations,
\begin{align}
\begin{cases} l_1^2 = 0 \,, \\
l_2^2 = (l_1-p_2)^2 = 0  \,, \\
l_3^2 = (l_1-p_2-p_3)^2 = 0 \,, \\
l_4^2 = (l_1+p_1)^2 = 0 \,,
\end{cases}
\end{align}
in terms of the parameters $\alpha_i$, as
\begin{align}
\begin{cases} \alpha_1 s_{12} = 0 \,, \\
\alpha_2 s_{12} = 0 \,, \\
(\alpha_1 \alpha_2-\alpha_3 \alpha_4)s_{12} = 0 \,, \\
\alpha_1 s_{13} + \alpha_2 s_{23} + \alpha_3 \spA13 \spB32 + \alpha_4 \spA23 \spB31 = s_{23} \,.
\end{cases}
\end{align}
For generic kinematics this system has two solutions:
\begin{align}
\left(l_1^{(1)}\right)^{\mu} = \frac{\spA 23}{\spA 13} \frac{1}{2} \langle 1 | \gamma^{\mu} | 2 ] \,, \qquad 
\left(l_1^{(2)}\right)^{\mu} = \frac{\spB 23}{\spB 13} \frac{1}{2} \langle 2 | \gamma^{\mu} | 1 ] \,.
\end{align}
The spinors of the on-shell loop momenta on the first solution can be chosen as
\begin{align} \label{eq:l1solSpinors}
\begin{alignedat}{2}
 & |l_1^{(1)} \r> =  \frac{\spA23}{\spA13} |1\r> \,, \qquad \quad && |l_1^{(1)}] = |2] \,, \\
 & |l_2^{(1)} \r> = \frac{\spA21}{\spA13} |3\r> \,, && |l_2^{(1)}] = |2] \,, \\
 & |l_3^{(1)} \r> = |3\r>  \,, && |l_3^{(1)}] = \frac{\spA21}{\spA13} |2] - |3] \,, \\
 & |l_4^{(1)} \r> = |1\r> \,, && |l_4^{(1)}] = |1] + \frac{\spA23}{\spA13} |2] \,.
\end{alignedat}
\end{align}
The spinors for the second solution, $l_1^{(2)}$, are obtained by swapping $\langle \rangle \leftrightarrow []$ in the first one.
For each solution $l_1^{(s)}$, the quadruple cut is is obtained by summing over all internal helicity configurations $\vec h = (h_1,h_2,h_3,h_4)$ (with $h_i = \pm$) the product of four tree-level amplitudes, 
 \begin{equation}
    \mathcal{C}_{1|2|3|45}\left( I^{(1)}(1^-,2^-,3^+,4^+,5^+) \right) \Bigl|_{l_1^{(s)}} = \sum_{\vec{h}} 
    \raisebox{-2cm}{\includegraphics[width=4.3cm]{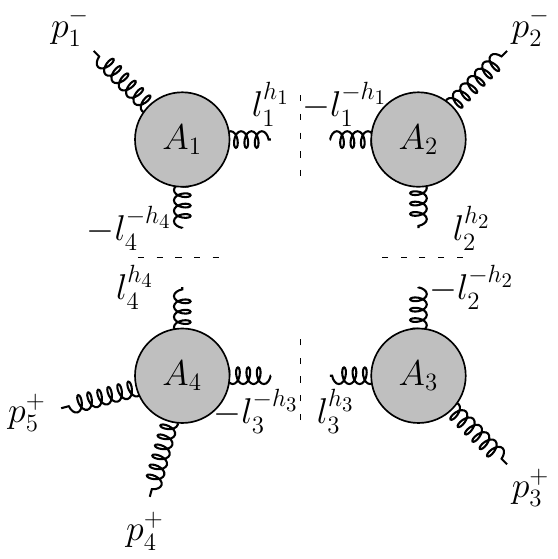}} \Biggl|_{l_1^{(s)}} \,.
    \label{eq:quadcut1_5g_generic}
  \end{equation}
Consider $A_4$. The only non-vanishing four-gluon tree-level amplitude with two positive-helicity gluons is the MHV one, namely $h_4=-h_3=-$. $A_3$ is thus $\overline{\text{MHV}}$, and $h_2=+$. Since MHV/$\overline{\text{MHV}}$ three-point vertices cannot be adjacent, $A_2$ must be MHV, and $A_1$ $\overline{\text{MHV}}$. This fixes the remaining helicity, $h_1=+$. The quadruple cuts therefore receive contribution from one helicity configuration only, which we represent using the black/white notation as
  \begin{equation}
    \mathcal{C}_{1|2|3|45}\left( I^{(1)}(1^-,2^-,3^+,4^+,5^+) \right)  = 
    \raisebox{-2cm}{\includegraphics[width=4.3cm]{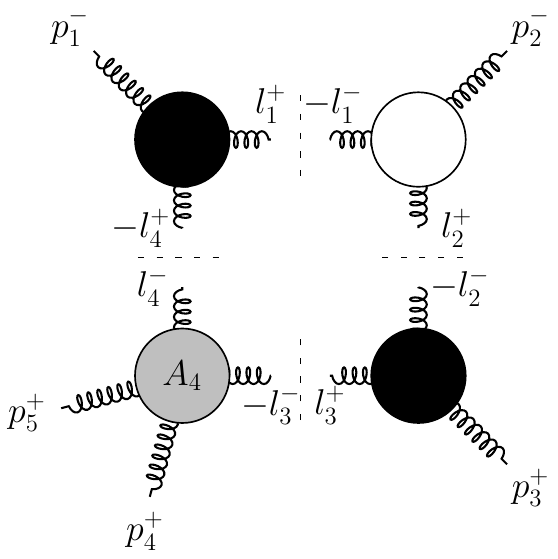}}   \,.
    \label{eq:quadcut1_5g_generic}
  \end{equation}
Recall that the trivalent vertices impose constraints on the momenta. The $\overline{\text{MHV}}$ vertex attached to $p_1$ and the MHV vertex attached to $p_2$ imply that $|l_1\rangle \propto |1\rangle$ and $|l_1] \propto |2]$, or equivalently that $l_1^{\mu} \propto \langle 1 | \gamma^{\mu} |2]$.
Only the solution $l_1^{(1)}$ is compatible with this constraint. Indeed, we can show explicitly that the contribution from the second solution vanishes, for instance
\begin{align}
\vcenter{\hbox{\includegraphics[width=2.2cm]{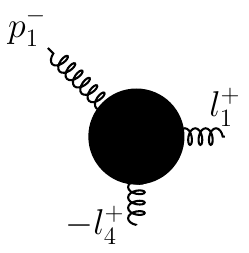}}} \Biggl|_{l_1^{(2)}} 
= \frac{[l_1 (-l_4)]^3}{[l_1(-l_4)] [(-l_4)1]} \biggr|_{l_1^{(2)}} = 0 \,,
\end{align}
where we used $|l_1^{(2)}] = \bigl([23]/[13] \bigr) |1]$ and $|l_4^{(2)}] = |1]$. We assign spinors to $-p$ according to the convention~\eqref{negpspin}, namely $|-p\rangle = \i |p\rangle$ and $|-p] = \i |p]$. We thus have that
\begin{align}
 \mathcal{C}_{1|2|3|45}\left( I^{(1)}(1^-,2^-,3^+,4^+,5^+) \right) \biggl|_{l_1^{(2)}}  = 0 \,.
\end{align}
On the first solution, the quadruple cut is given by
\begin{align}
 \mathcal{C}_{1|2|3|45} \bigl|_{l_1^{(1)}}  = 
\frac{\spB {l_1}{l_4}^3}{\spB 1{l_1} \spB {l_4}1} \frac{\spA{l_1}2^3}{\spA2 {l_2} \spA{l_2}{l_1}}
  \frac{\spB 3{l_3}^3}{\spB{l_3}{l_2} \spB{l_2}3} \frac{\spA {l_4}{l_3}^3}{\spA{l_3}4 \spA45 \spA 5{l_4} } \Biggl|_{l_1^{(1)}} \,,
\end{align}
where we omitted the argument of $\mathcal{C}$ for the sake of compactness. Plugging in the spinors from \eqn{eq:l1solSpinors} and simplifying gives
\begin{align}
 \mathcal{C}_{1|2|3|45}\left( I^{(1)}(1^-,2^-,3^+,4^+,5^+) \right) \biggl|_{l_1^{(1)}}  = \i \, s_{12} s_{34} \left( \frac{\i \, \spA12^3}{\spA23 \spA34 \spA45 \spA51} \right) \,,
\end{align}
where in the parentheses we recognise the tree-level amplitude. Averaging over the two cut solutions 
(as in \eqn{eq:4gquadcutfinal} for the four-gluon case) gives the four-dimensional coefficient of the scalar box integral,
\begin{align}
c_{0;1|2|3|45}(1^-,2^-,3^+,4^+,5^+)  & = \frac{1}{2} \sum_{s=1}^2  \mathcal{C}_{1|2|3|45}\left( I^{(1)}(1^-,2^-,3^+,4^+,5^+) \right) \biggl|_{l_1^{(s)}} \nonumber \\
& = \frac{\i}{2} s_{12} s_{34} A^{(0)}(1^-,2^-,3^+,4^+,5^+) \,,
\end{align}
as claimed.

\smallskip

\item The solution of the quadruple cut can be obtained as in part a) of this exercise. Alternatively, we can take a more direct route by exploiting the black/white formalism for the trivalent vertices.
On each of the two solutions $l_1^{(s)}$ the quadruple cut is given by
\begin{align}
 \mathcal{C}_{1|23|4|5} \left( I^{(1)}(1^+,2^+,3^-,4^+,5^-)  \right) \Bigl|_{l_1^{(s)}} = \sum_{\vec{h}} \vcenter{\hbox{\includegraphics[width=4.3cm]{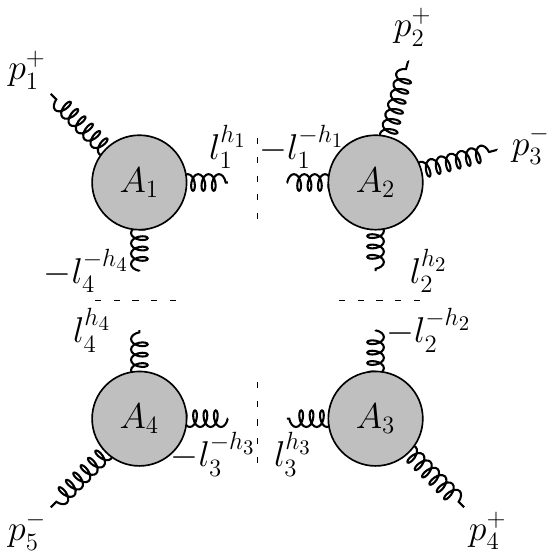}}} \Biggl|_{l_1^{(s)}}\,.
\end{align}
The only non-vanishing tree-level four-point amplitude is the MHV (or equivalently $\overline{\text{MHV}}$) one, so we have either $h_1=h_2=-$ or $h_1=h_2=+$. Specifying $h_1$ and $h_2$ and excluding adjacent black/white vertices fixes all the other helicities, so that the quadruple cut receives contribution from two helicity configurations:
\begin{align}
 \mathcal{C}_{1|23|4|5}^{(a)} = \hspace{-0.3cm} \vcenter{\hbox{\includegraphics[width=4.3cm]{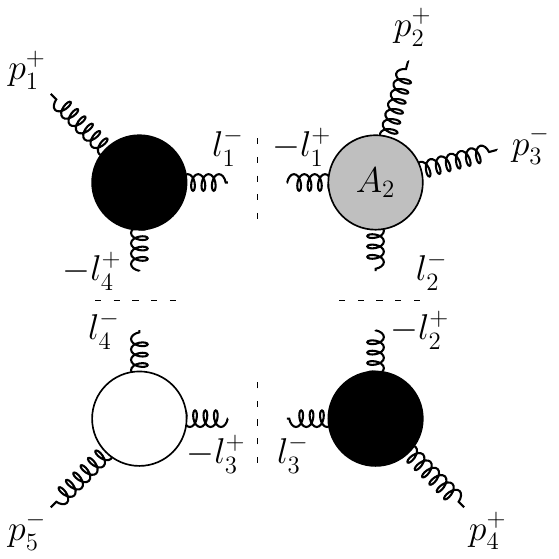}}} \hspace{-0.3cm} \,,
\ \
 \mathcal{C}_{1|23|4|5}^{(b)} = \hspace{-0.3cm} \vcenter{\hbox{\includegraphics[width=4.3cm]{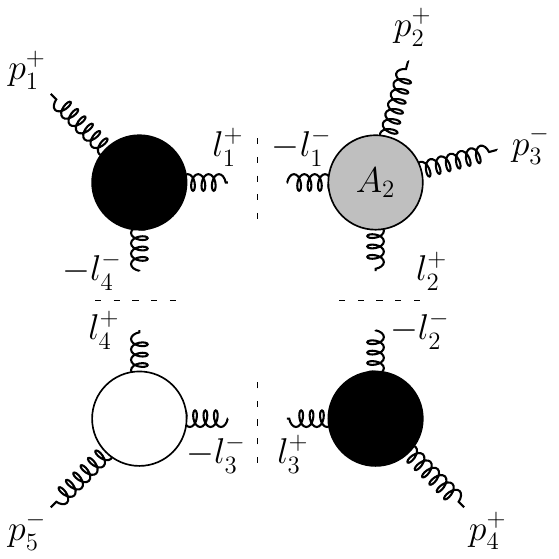}}} \hspace{-0.3cm} \,.
\end{align}
In both cases the trivalent vertices constrain $|l_4\rangle \propto |1\rangle$ and $|l_4] \propto |5]$. The two configurations are thus non-vanishing only on one solution of the quadruple cut, say $l_1^{(1)}$, which we parametrise starting from $l_4$ as
\begin{align}
\left( l_4^{(1)} \right)^{\mu} = a \, \frac{1}{2} \langle 1 | \gamma^{\mu} |5] \,.
\end{align}
The value of $a$ is fixed by requiring that $l_2^{(1)} = l_4^{(1)}+p_4+p_5$ is on shell (i.e.\ $\bigl(l_2^{(1)}\bigr)^2=0$), which gives $a=\spA45 / \spA 14$.
The spinors for the internal momenta on this solution can then be chosen as
\begin{align} \label{eq:l1solSpinors}
\begin{alignedat}{2}
 & |l_1^{(1)} \r> = |1\r> \,, && |l_1^{(1)}] = \frac{\spA45}{\spA14} |5] - |1] \,, \\
 & |l_2^{(1)} \r> = |4\r> \,, && |l_2^{(1)}] = |4] + \frac{\spA15}{\spA14} |5] \,, \\
 & |l_3^{(1)} \r> = |4\r>  \,, && |l_3^{(1)}] = \frac{\spA15}{\spA14} |5] \,, \\
 & |l_4^{(1)} \r> = |1\r> \,, \qquad \quad && |l_4^{(1)}] = \frac{\spA45}{\spA14} |5] \,.
\end{alignedat}
\end{align}
The first contribution to the quadruple cut is given by
\begin{align} 
\begin{aligned}
\mathcal{C}_{1|23|4|5}^{(a)} \Bigl|_{l_1^{(1)}} & = \frac{[l_4 1]^3}{[1l_1][l_1l_4]} \frac{\langle3 l_2\rangle^3}{\spA{l_2}{l_1} \spA{l_1}{2} \spA23} 
  \frac{[l_2 4]^3}{[4l_3][l_3l_2]} \frac{\spA5{l_4}^3}{\spA{l_4}{l_3} \spA{l_3}5}  \biggl|_{l_1^{(1)}} \\
  & = \i \, s_{45} s_{15} \left( \frac{\spA34 \spA15}{\spA14 \spA35} \right)^4 \left(\frac{\i \, \spA35^4}{\spA12 \spA23 \spA34 \spA45 \spA51} \right) \,,
\end{aligned}
\end{align}
where in the right-most parentheses of the second line we recognise the tree-level amplitude $A^{(0)}(1^+,2^+,3^-,4^+,5^-)$. The computation of the second term is analogous. Summing up the two contributions finally gives
\begin{align}
\mathcal{C}_{1|23|4|5} \Bigl|_{l_1^{(1)}} =  \i \, s_{45} s_{15} \, A^{(0)} \, 
  \left[ \left( \frac{\spA34 \spA15}{\spA14 \spA35} \right)^4 + \left( \frac{\spA13 \spA45}{\spA14 \spA35} \right)^4 \right]  \,,
\end{align}
where we omitted the argument of $\mathcal{C}_{1|23|4|5}$ and $A^{(0)}$ for compactness.
The second solution, $l_1^{(2)}$, is the complex conjugate of the first one. The quadruple cut vanishes on it by the argument above,
\begin{align}
\mathcal{C}_{1|23|4|5} \left(I^{(1)}(1^+,2^+,3^-,4^+,5^-)\right) \Bigl|_{l_1^{(2)}} =  0\,.
\end{align}
Finally, we obtain the coefficient of the scalar box function at order $\eps^0$ by averaging over the two solutions:
\begin{align}
c_{0;1|23|4|5}(1^+,2^+,3^-,4^+,5^-) =  \frac{\i}{2} s_{45} s_{15} \, A^{(0)} \, 
  \left[ \left( \frac{\spA34 \spA15}{\spA14 \spA35} \right)^4 + \left( \frac{\spA13 \spA45}{\spA14 \spA35} \right)^4 \right]  \,.
\end{align}

\end{enumerate}


\section*{Exercise \ref{Ex:3.3}: Tensor decomposition of the bubble integral}
\label{Sol:3.3}
\begin{enumerate}[a)]
\item We contract both sides of the form-factor decomposition in \eqn{eq:PVbubR2dec} by the basis tensors $\eta^{\mu_1 \mu_2}$ and $p_1^{\mu_1} p_1^{\mu_2}$, obtaining
\begin{align} \label{eq:PVbubR2sys}
\begin{cases}
F_2^{[D]} \left[k^2 \right] = a_{2,00} \, D + a_{2,11} \, p_1^2 \,, \\
F_2^{[D]}\left[(k\cdot p_1)^2 \right]  = a_{2,00} \, p_1^2 + a_{2,11} \, (p_1^2)^2 \,. \\
\end{cases}
\end{align}
For the sake of simplicity we omit the dependence of the bubble integrals on $p_1$, and we introduce the short-hand notations $D_1=k^2$ and $D_2 = (k-p_1)^2$ for the inverse propagators.
Solving the linear system~\eqref{eq:PVbubR2sys} for the form factors gives
\begin{align} \label{eq:PVbubR2sol1}
\begin{aligned}
a_{2,00} & = \frac{1}{D-1} \Bigl( F_2^{[D]}\left[k^2 \right] - \frac{1}{p_1^2} F_2^{[D]}\left[(k\cdot p_1)^2 \right] \Bigr) \,, \\
a_{2,11} & = \frac{1}{p_1^2 (D-1)} \Bigl( \frac{D}{p_1^2}  F_2^{[D]}\left[(k\cdot p_1)^2 \right] - F_2^{[D]}\left[k^2 \right]\Bigr) \,.
\end{aligned}
\end{align}
The contraction of the rank-2 bubble with $\eta^{\mu_1\mu_2}$ is given by a scaleless integral and thus vanishes in dimensional regularisation,
\begin{align} \label{eq:I2dotgfin}
F_2^{[D]}\left[ k^2 \right] = \int_k \frac{1}{(k-p_1)^2} = 0 \,.
\end{align}
The contraction with $p_1^{\mu_1} p_1^{\mu_2}$ is instead given by
\begin{align} \label{eq:I2dotp1p1}
F_2^{[D]}\left[ (k\cdot p_1)^2 \right] = 
  \frac{1}{4} \int_k \left( \frac{(p_1^2)^2}{D_1 D_2} + \frac{D_1}{D_2} + \frac{D_2}{D_1} - 2 + 2 \frac{p_1^2}{D_2} - 2 \frac{p_1^2}{D_1} \right) \,,
\end{align}
where we used that $2\, k\cdot p_1 = D_1 - D_2 + p_1^2$. All terms but the first one vanish in dimensional regularisation. To see this explicitly, consider for instance the second term. By shifting the loop momentum by $p_1$ we can rewrite it as a combination of manifestly scaleless integrals,
\begin{align}
\int_k \frac{k^2}{(k-p_1)^2} = \int_k 1 + p_1^2 \int_k \frac{1}{k^2} + 2 p_1^{\mu} \int_k \frac{k_{\mu}}{k^2} \,,
\end{align}
which vanish in dimensional regularisation (see section~\ref{sec:ConventionsCh4} in chapter~\ref{ch:loopints}).
Equation~\eqref{eq:I2dotp1p1} thus reduces to
\begin{align} \label{eq:I2dotp1p1fin}
F_2^{[D]}\left[ (k\cdot p_1)^2 \right] = 
  \frac{(p_1^2)^2}{4} F_2^{[D]}[1] \,.
\end{align}
Substituting eqs.~\eqref{eq:I2dotgfin} and~\eqref{eq:I2dotp1p1fin} into \eqn{eq:PVbubR2sol1} finally gives
\begin{align}
\begin{aligned}
a_{2,00} & = -\frac{p_1^2}{4 (D-1)} \, F_2^{[D]}\left[1 \right]  \,, \\
a_{2,11} & = \frac{D}{4 (D-1)} \, F_2^{[D]}\left[1 \right] \,.
\end{aligned}
\end{align}

\item We proceed as we did in part a). For compactness, we define
\begin{align} \begin{aligned}
& T_{1}^{\mu_1 \mu_2 \mu_3} = \eta^{\mu_1\mu_2} p_1^{\mu_3} + \eta^{\mu_2\mu_3} p_1^{\mu_1} + \eta^{\mu_3\mu_1} p_1^{\mu_2} \,, \\
& T_{2}^{\mu_1 \mu_2 \mu_3} = p_1^{\mu_1} p_1^{\mu_2} p_1^{\mu_3} \,.
\end{aligned} \end{align}
Note that $F_2^{[D]}\left[k^{\mu_1} k^{\mu_2} k^{\mu_3} \right]$ is symmetric under permutations of the Lorentz indices. While $T_2$ enjoys this symmetry, the three separate terms of $T_1$ do not. That is why they appear together in $T_1$ rather than with distinct form factors. The symmetry property would in fact constrain the latter to be equal. We then contract both sides of the tensor decomposition~\eqref{eq:PVbubR3dec} with the basis tensors, and solve the ensuing $2\times 2$ linear system for the form factors,
Using the following contractions,
\begin{align}\begin{aligned}
& F_2^{[D]}\left[k^{\mu_1} k^{\mu_2} k^{\mu_3} \right] \left(T_1\right)_{\mu_1 \mu_2 \mu_3} = 0 \,, \\
& F_2^{[D]}\left[k^{\mu_1} k^{\mu_2} k^{\mu_3} \right] \left(T_2\right)_{\mu_1 \mu_2 \mu_3} = \frac{(p_1^2)^3}{8} F_2^{[D]}\left[1\right]\,, \\
& T_1^{\mu_1\mu_2\mu_3} \left(T_1\right)_{\mu_1\mu_2\mu_3} = 3 p_1^2 (D+2) \,, \\
& T_1^{\mu_1\mu_2\mu_3} \left(T_2\right)_{\mu_1\mu_2\mu_3} = 3 (p_1^2)^2 \,, \\
& T_2^{\mu_1\mu_2\mu_3} \left(T_2\right)_{\mu_1\mu_2\mu_3} = (p_1^2)^3 \,,
\end{aligned} \end{align}
we obtain
\begin{align} \begin{aligned}
a_{2,001} & = - \frac{p_1^2}{8 (D-1)} \, F_2^{[D]}\left[1\right] \,, \\
a_{2,111} & = \frac{D+2}{8(D-1)} \, F_2^{[D]} \left[1\right] \,.
\end{aligned} \end{align}

\end{enumerate}

\section*{Exercise \ref{Ex:3.4}: Spurious loop-momentum space for the box integral}
\label{Sol:3.4}
\begin{enumerate}[a)]
\item The physical space is 3-dimensional, and may be spanned by $\{p_1,p_2,p_3\}$. The spurious space is 1-dimensional. In order to construct a vector $\omega$ spanning the spurious space, we start from a generic ansatz made from the spinors associated with $p_1$ and $p_2$,
\begin{align}
\omega^{\mu} = \alpha_1 p_1^{\mu} + \alpha_2 p_2^{\mu} + \alpha_3 \frac{1}{2} \spAB{1}{\gamma^{\mu}}{2} +  \alpha_4 \frac{1}{2} \spAB{2}{\gamma^{\mu}}{1} \,,
\end{align}
and constrain it by imposing the orthogonality to the external momenta and the normalisation ($\omega^2 = 1$).
While $\omega \cdot p_1 = 0$ and $\omega \cdot p_2 = 0$ fix $\alpha_1 = \alpha_2 = 0$, the orthogonality to $p_3$ and the normalisation imply
\begin{align}
\alpha_3 \spA{1}{3} \spB{3}{2} + \alpha_4 \spA{2}{3} \spB{3}{1} = 0 \,, \qquad \quad \alpha_3 \alpha_4 \, s_{12} = -1 \,, 
\end{align} 
where $s_{ij} = (p_i + p_j)^2$. The solution is given by
\begin{align} \label{eq:omega4ptSpinors}
\omega^{\mu} = \frac{1}{2 \sqrt{s_{12} s_{23} s_{13}}} \bigl[ \spAB{1}{\gamma^{\mu}}{2} \spA{2}{3} \spB{3}{1} - 
   \spAB{2}{\gamma^{\mu}}{1}  \spA{1}{3} \spB{3}{2} \bigr] \,.
\end{align}

\item We rewrite the spinor chains in \eqn{eq:omega4ptSpinors} in terms of traces of Pauli matrices,
\begin{align}
\omega^{\mu} = \frac{1}{2 \sqrt{s_{12} s_{23} s_{13}}} \bigl[ \text{Tr}\left( \sigma^{\mu} \bar{\sigma}^{\rho} \sigma^{\tau} \bar{\sigma}^{\nu} \right) -
    \text{Tr}\left( \sigma^{\mu} \bar{\sigma}^{\nu} \sigma^{\tau} \bar{\sigma}^{\rho} \right) \bigr] p_{1 \nu} p_{2 \rho} p_{3 \tau} \,.
\end{align}
We trade the Pauli matrices for Dirac matrices through eq.~\eqref{eq:trSigma2gamma}. The terms free of $\gamma_5$ cancel out thanks to the cyclicity of the trace and the identity $\text{Tr}\left( \gamma^{\mu} \gamma^{\nu} \gamma^{\rho} \gamma^{\tau} \right) = \text{Tr}\left( \gamma^{\tau} \gamma^{\rho} \gamma^{\nu} \gamma^{\mu} \right)$. We rewrite the traces with $\gamma_5$ in terms of the Levi-Civita symbol using $\text{Tr}\left(\gamma^{\mu} \gamma^{\nu} \gamma^{\rho} \gamma^{\tau} \gamma_5\right) = -4 \i \eps^{\mu\nu\rho\tau}$, obtaining
\begin{align}
\omega^{\mu} = \frac{2 \i}{ \sqrt{s_{12} s_{23} s_{13}}}  \eps^{\mu \nu \rho \tau} p_{1\nu} p_{2\rho} p_{3\tau} \,.
\end{align}

\end{enumerate}


\section*{Exercise \ref{Ex:3.5}: Reducibility of the pentagon in four dimensions}
\label{Sol:3.5}
\begin{enumerate}[a)]
\item We rewrite the triangle integral as
\begin{align}
F^{[D]}_3(p_1,p_2) = \int_k \frac{1}{(-D_1) \, (-D_2) \, (-D_3)} \,,
\end{align}
with inverse propagators
\begin{align}
D_1 = k^2 \,, \quad D_2 =(k-p_1)^2 \,, \quad D_3 = (k-p_1-p_2)^2 \,.
\end{align}
The $\i 0$ is irrelevant here, and we thus omit it.
We introduce a two-dimensional parametrisation of the loop momentum $k^{\mu}$ by expanding it in a basis formed by two independent external momenta, say $p_1^{\mu}$ and $p_2^{\mu}$, as
\begin{align} \label{eq:k2D}
k^{\mu} = a_1 \, p_1^{\mu} + a_2 \, p_2^{\mu} \,.
\end{align}
Since there are only two degrees of freedom, parametrised by $a_1$ and $a_2$, the three inverse propagators of the triangle integral cannot be algebraically independent (over the field of the rational functions of $s$). In order to find the relation among them, we express them in terms of $a_1$ and $a_2$,
\begin{align} \label{eq:D2a}
\begin{aligned}
& D_1 = s \, a_1 (a_1-a_2) \,, \\ 
& D_2 = s \, (1-a_1) (1+a_2-a_1) \,, \\ 
& D_3 = s \, (1 - a_1) (a_2 - a_1) \,,
\end{aligned}
\end{align}
with $s=p_1^2$.
We then solve two of these equations to express $a_1$ and $a_2$ in terms of inverse propagators. Choosing $D_1$ and $D_3$ we obtain
\begin{align}
a_1 = \frac{D_1}{D_1-D_3} \,, \quad a_2 = \frac{(D_1-D_3)^2 - s \, D_1}{s \, (D_3-D_1)} \,.
\end{align}
Plugging these into the expression of $D_2$ in \eqn{eq:D2a} gives a relation among the three inverse propagators,
\begin{align} \label{eq:triangle_relation_2D}
1 = \frac{1}{s} \left( D_1+D_2-D_3 - \frac{D_1 D_2}{D_3} \right) \,.
\end{align}
Inserting this into the numerator of the triangle integral, expanding, and removing the scaleless integrals which integrate to zero finally gives
\begin{align} \begin{aligned}
F^{[2-2\eps]}_{3}(p_1,p_2) & = \frac{1}{s} \int_k \frac{1}{k^2 (k-p_1)^2} + \text{terms missed in $2D$} \\
& = \frac{1}{s} \, F_2^{[2-2\eps]}(p_1)  + \text{terms missed in $2D$}\,.
\end{aligned}
\end{align}
Up to terms which are missed by the two-dimensional analysis, the triangle integral in $D=2-2\eps$ dimensions can be expressed in terms of a bubble integral, and is thus reducible.

\smallskip

\item In $D=2$ dimensions, any three momenta are linearly dependent. The Gram matrix $G(k,p_1,p_2)$ therefore has vanishing determinant,
\begin{align}
-\frac{1}{4} s^2 k^2 - s \, (k\cdot p_1) (k \cdot p_2) - s \, (k \cdot p_2)^2 = 0 \,,
\end{align}
which can be verified using a two-dimensional parametrisation of $k^{\mu}$ such as \eqn{eq:k2D}. In order to convert it into a relation among the inverse propagators, we express the scalar products of the loop momentum in terms of inverse propagators,
\begin{align} 
k^2 = D_1 \,, \qquad k\cdot p_1 = \frac{D_1-D_2+s}{2} \,, \qquad k\cdot p_2 = \frac{D_2-D_3-s}{2} \,.
\end{align}
Expressing the determinant of $G(k,p_1,p_2)$ in terms of inverse propagators gives the relation~\eqref{eq:triangle_relation_2D}.

\smallskip

\item The steps are the same as for the previous point, but the algebraic manipulations are cumbersome. We implement them in the \texttt{Mathematica} notebook \href{https://scattering-amplitudes.mpp.mpg.de/scattering-amplitudes-in-qft/Exercises/}{\texttt{Ex3.5\_Reducibility.wl}}~\cite{ch5-website}.
In $D=4$ dimensions the following Gram determinant vanishes:
\begin{align} \label{eq:pentagon_Gramdet}
\text{det} \, G\left(k,p_1,p_2,p_3,p_4\right) = 0\,.
\end{align}
We aim to rewrite this in terms of the inverse propagators of the pentagon:
\begin{align} \label{eq:pentagon_Ds}
\begin{array}{ll}
D_1 = k^2 \,, &  D_4 = (k-p_1-p_2-p_3)^2 \,,  \\
D_2 = (k-p_1)^2 \,, \qquad \quad & D_5 = (k-p_1-p_2-p_3-p_4)^2 \,. \\
D_3 = (k-p_1-p_2)^2 \,, \qquad \qquad & \\
\end{array}
\end{align}

The first step is to parametrise the kinematics in terms of independent invariants $s_{ij} = (p_i+p_j)^2$. It is instructive to count the latter for a generic number of particles $n$.
There are $n(n+1)/2$ distinct scalar products $p_i\cdot p_j$ with $i,j=1,\ldots,n$. Momentum conservation gives $n$ constraints, as we may contract $\sum_{i=1}^n p_i^{\mu} = 0$ by $p_j^{\mu}$ for any $j=1,\ldots,n$. Moreover, we have 
$n$ on-shell constraints: $p_i^2=0$ for $i=1,\ldots,n$. We are thus left with $\frac{n(n+1)}{2} -2 n = n (n-3)/2 $ independent invariants. For $n=4$ that gives $2$ independent invariants---the familiar $s$ and $t$ Mandelstam invariants---while for $n=5$ we have $5$. It is convenient to choose them as $\vec{s} \coloneqq \{s_{12},s_{23},s_{34},s_{45},s_{51}\}$. We now need to express all scalar products $p_i\cdot p_j$ in terms of $\vec{s}$. We may do so by solving the linear system of equations obtained from momentum conservation as discussed above:
\begin{align}
\sum_{i=1}^5 p_i \cdot p_j = 0 \,, \forall \, j=1,\ldots, 5\,.
\end{align}
We rewrite the latter in terms of $s_{ij}$'s and solve. We do this in the \texttt{Mathematica} notebook, obtaining---for example---that
$p_1 \cdot p_4 = (s_{23} - s_{45} - s_{51})/2$.

We now turn our attention to the scalar products involving the loop momentum: $k^2$, and $k\cdot p_i$ for $i=1,\ldots,4$ ($k\cdot p_5$ is related to the others by momentum conservation). Having parametrised the kinematics in terms of independent invariants $\vec{s}$, we may solve the system~\eqref{eq:pentagon_Ds} to express them in terms of inverse propagators and $\vec{s}$.
For example, we obtain that $k \cdot p_3 = (D_3 - D_4 + s_{45} -s_{12})/2$.
Using this result, we can rewrite the Gram-determinant condition~\eqref{eq:pentagon_Gramdet} as
\begin{align}
1 = \sum_{i=1}^5 A_i D_i + \sum_{i\le j=1}^5 B_{ij} D_i D_j \,,
\end{align}
where $A_i$ and $B_{ij}$ are rational functions of the invariants $\vec{s}$. Plugging this into the numerator of the pentagon integral and expanding finally gives the reduction into integrals with fewer propagators, up to terms missed in $D=4$.

\end{enumerate}


\section*{Exercise \ref{Ex:3.6}: Parametrising the bubble integrand}
\label{Sol:3.6}
a) We parametrise the loop momentum as in \eqn{eq:momparambubble}. We recall that $p_1 \cdot \omega_i = 0$ and $\omega_i \cdot \omega_j = \delta_{ij} \, \omega_i^2$. The coefficient $\alpha_1$ can be expressed in terms of propagators and external invariants by noticing that $\alpha_1 = k\cdot p_1/p_1^2$, and rewriting $k\cdot p_1$ in terms of inverse propagators~\eqref{eq:bubbleprops}. This gives
\begin{align} \label{eq:alpha1toD}
\alpha_1 = \frac{D_1-D_2 +p_1^2 + m_1^2-m_2^2}{2 \, p_1^2}\,.
\end{align}
We thus see that $\alpha_1$ does not depend on the loop momentum on the bubble cut $D_1=D_2=0$. As a result, the loop-momentum parametrisation of the bubble numerator $\Delta_{1|2}$ depends only on three ISPs: $k\cdot \omega_i$ for $i=1,2,3$. The maximum tensor rank for a renormalisable gauge theory is two, hence a general parametrisation~is
\begin{align} \label{eq:Delta12}
\begin{aligned}
  \Delta_{1|2}&(k \cdot \omega_1, k \cdot \omega_2, k \cdot \omega_3) = c_{000} \\
 & + c_{100} (k \cdot \omega_1) 
  + c_{010} (k \cdot \omega_2) + c_{001} (k \cdot \omega_3)  \\
 & + c_{110} (k \cdot \omega_1) (k \cdot \omega_2) 
  + c_{101} (k \cdot \omega_1) (k \cdot \omega_3) 
  + c_{011} (k \cdot \omega_2) (k \cdot \omega_3) \\
  & + c_{200} (k \cdot \omega_1)^2 
  + c_{020} (k \cdot \omega_2)^2 
  + c_{002} (k \cdot \omega_3)^2  \,.
\end{aligned}
\end{align}
The cut condition $D_1=0$ implies one more constraint on the loop-momentum dependence:
\begin{align} \label{eq:C12k}
\mathcal{C}_{1|2}\left(k_{\parallel}^2\right) + \sum_{i=1}^3 (k\cdot \omega_i)^2 \omega_i^2 -m_1^2 = 0 \,.
\end{align}
Since $\mathcal{C}_{1|2}\bigl(k_{\parallel}^2\bigl) = (m_1^2 - m_2^2 + p_1^2)^2/(4 \, p_1^2)$ does not depend on the loop momentum on the cut, we may use \eqn{eq:C12k} to eliminate, say, $(k\cdot \omega_3)^2$ from the numerator~\eqref{eq:Delta12}. It is however more convenient to implement the constraint~\eqref{eq:C12k} so as to maximise the number of terms which integrate to zero.
The terms in the second and third line on the RHS of \eqn{eq:Delta12} contain odd powers of $k\cdot \omega_i$, and thus vanish upon integration.
Using transverse integration one can show that
\begin{align}
\int_k \frac{(k \cdot \omega_i)^2}{D_1 D_2} = \frac{\omega_i^2}{\omega_j^2} \int_k \frac{(k \cdot \omega_j)^2}{D_1 D_2} \,.
\end{align}
We can then use the constraint~\eqref{eq:C12k} to group $(k\cdot \omega_1)^2$, $(k\cdot \omega_2)^2$ and $(k\cdot \omega_3)^2$ into two terms which vanish upon integration. This can be achieved for instance as
\begin{align}
\begin{aligned}
  \Delta_{1|2}&(k \cdot \omega_1, k \cdot \omega_2, k \cdot \omega_3) = c_{0;1|2} \\
 & + c_{1;1|2} (k \cdot \omega_1) 
  + c_{2;1|2} (k \cdot \omega_2) + c_{3;1|2} (k \cdot \omega_3)  \\
 & + c_{4;1|2} (k \cdot \omega_1) (k \cdot \omega_2) 
  + c_{5;1|2} (k \cdot \omega_1) (k \cdot \omega_3) 
  + c_{6;1|2} (k \cdot \omega_2) (k \cdot \omega_3) \\
  & + c_{7;1|2} \left[ (k \cdot \omega_1)^2  - \frac{\omega_1^2}{\omega_3^2} (k \cdot \omega_3)^2  \right]
  + c_{8;1|2} \left[ (k \cdot \omega_2)^2  - \frac{\omega_2^2}{\omega_3^2} (k \cdot \omega_3)^2  \right] \,,
\end{aligned}
\end{align}
such that only the term with coefficient $c_{0;1|2}$ survives upon integration, as claimed.

\smallskip
\noindent
b) The bubble cut of the one-loop amplitude $A^{(1),[4-2\eps]}_n$ is by definition given by
\begin{align} \label{eq:C12lhs}
\mathcal{C}_{1|2} \left(A^{(1),[4-2\eps]}_n\right) = \int_k \left[ I^{(1)}(k) \prod_{i=1}^2 \left( D_i \, (-2 \pi \i) \, \delta^{(+)}\left(D_i\right) \right) \right] \,,
\end{align}
where $I^{(1)}(k)$ denotes the integrand of $A^{(1),[4-2\eps]}_n$. We parametrise the latter in terms of boxes, triangles, and bubbles.
The terms which survive on the bubble cut $1|2$ are
\begin{align}
& \mathcal{C}_{1|2} \left(A^{(1),[4-2\eps]}_n\right) = \int_k \biggl[ \frac{\Delta_{1|2}\left(k\cdot \omega_1, k\cdot \omega_2, k\cdot \omega_3\right)}{(-D_1)(-D_2)} + \sum_X \frac{\Delta_{1|2|X}\left( k \cdot \omega^X_1, k\cdot \omega^X_2 \right)}{(-D_1)(-D_2)(-D_X)}  \nonumber \\
& \hspace{1cm} + \sum_{Y,Z} \frac{\Delta_{1|2|Y|Z}\left( k \cdot \omega^{YZ} \right)}{(-D_1)(-D_2)(-D_Y) (-D_Z)} + \ldots \biggr]  \prod_{i=1}^2 \left( D_i \, (-2 \pi \i ) \delta^{(+)}(D_i)\right)   \nonumber \\
& \phantom{\mathcal{C}_{1|2} \left(A^{(1)}\right)} = \int_k \biggl[ \Delta_{1|2}\left(k\cdot \omega_1, k\cdot \omega_2, k\cdot \omega_3\right) + \sum_X \frac{\Delta_{1|2|X}\left( k \cdot \omega^X_1, k\cdot \omega^X_2 \right)}{-D_X} \nonumber \\
&  \phantom{\mathcal{C}_{1|2} \left(A^{(1)}\right) = \int_k \biggl[} + \sum_{Y,Z} \frac{\Delta_{1|2|Y|Z}\left( k \cdot \omega^{YZ} \right)}{D_Y D_Z} \biggr] \prod_{i=1}^2 \left( (-2 \pi \i ) \delta^{(+)}(D_i)\right) \,. \label{eq:C12rhs}
\end{align}
Here, the ellipsis denotes terms which vanish on the cut. The sum over $X$ runs over all triangle configurations which share the propagators $1/(-D_1)$ and $1/(-D_2)$, $\omega_1^X$ and $\omega_2^X$ are the vectors spanning the corresponding spurious loop-momentum space, and $1/(-D_X)$ is the propagator which completes the triangle. Similarly, the sum over $Y,Z$ runs over all box configurations sharing the propagators $1/(-D_1)$ and $1/(-D_2)$, $\omega^{YZ}$ spans their spurious-loop momentum space, and $1/(-D_Y)$ and $1/(-D_Z)$ are the inverse propagators which complete the box. Equating eqs.~\eqref{eq:C12lhs} and~\eqref{eq:C12rhs} and solving for $\Delta_{1|2}$ gives
\begin{align}
\begin{aligned}
  \Delta_{1|2}&\left(k\cdot \omega_1, k\cdot \omega_2, k\cdot \omega_3\right) \biggl|_{D_i=0} = \biggl(
    I^{(1)}(k) \prod_{i=1}^2 D_i \\
  &  + \sum_X \frac{\Delta_{1|2|X}\left( k \cdot \omega^X_1, k\cdot \omega^X_2 \right)}{D_X}
- \sum_{Y,Z} \frac{\Delta_{1|2|Y|Z}\left( k \cdot \omega^{YZ} \right)}{D_Y D_Z} \biggr) \biggl|_{D_i=0} \,,
\end{aligned}
\end{align}
as claimed.


\section*{Exercise \ref{Ex:3.DimShift}: Dimension-shifting relation at one loop}
\label{Sol:3.DimShift}
We decompose the loop momentum into a four- and a $(-2\eps)$-dimensional parts as
\begin{align}
k_1^{\mu} = k_1^{[4], \, \mu} + k_1^{[-2\eps], \, \mu} \,,
\end{align}
with $k_1^{[-2\eps]} \cdot k_1^{[4]} = 0 = k_1^{[-2\eps]} \cdot p_i$ and $ k_1^{[-2\eps]} \cdot k_1^{[-2\eps]} = - \mu_{11}$. Note that $\mu_{11} > 0$.
The loop-integration measure factorises as $\d^D k_1 = \d^4 k_1^{[4]} \, \d^{-2\eps} k_1^{[-2\eps]}$.
We rewrite the integral on the LHS of \eqn{eq:dimshift} as
\begin{align} \label{eq:dimshift_eq1}
F^{[4-2\eps]}_n(p_1,\ldots,p_{n-1})[\mu_{11}^r] = \int \frac{\d^D k_1}{\i \pi^{D/2}} \, \mu_{11}^r \,  \mathcal{F}_n\left( k_1^{[4]}, \mu_{11} \right) \,.
\end{align}
The integrand $\mathcal{F}_n$ depends on the loop momentum only through its four-dimensional components $k_1^{[4], \, \mu}$ and $\mu_{11}$. In other words, the integrand does not depend on the angular coordinates of the $(-2\eps)$-dimensional subspace.
We thus introduce angular and radial coordinates as
\begin{align}
\d^{-2\eps} k_1^{[-2\eps]} = \frac{1}{2} \, \d \Omega_{-2\eps} \, \mu_{11}^{-1-\eps} \d \mu_{11} \,,
\end{align}
and carry out the $(-2\eps)$-dimensional angular integration in \eqn{eq:dimshift_eq1}. We recall that the surface area of a unit-radius sphere in $m$-dimensions is given by
\begin{align} \label{eq:area-sphere}
\Omega_{m} \coloneqq \int \d \Omega_{m} = \frac{2 \, \pi^{m/2}}{\Gamma\left(\frac{m}{2}\right)} \,.
\end{align}
We obtain
\begin{align} \label{eq:dimshift_eq2}
F^{[4-2\eps]}_n(p_1,\ldots,p_{n-1})[\mu_{11}^r] = \frac{\Omega_{-2\eps}}{2} \int \frac{\d^4 k_1^{[4]}}{\i \pi^{2-\eps}} \int_{0}^{\infty} \d \mu_{11} \, \mu_{11}^{r-1-\eps} \, \mathcal{F}_n\left( k_1^{[4]}, \mu_{11} \right) \,.
\end{align}
We view the remaining $\mu_{11}$ integration as the radial integration in a $(2r-2\eps)$-dimensional subspace. The loop-integration measure in the latter is in fact given~by
\begin{align}
\d^{2r - 2\eps} k_1^{[2r - 2\eps]} = \frac{1}{2} \, \d \Omega_{2r-2\eps} \, \mu_{11}^{r-1-\eps} \d \mu_{11} \,.
\end{align}
Exploiting again the independence of the integrand on the angular coordinates, we rewrite \eqn{eq:dimshift_eq2} as
\begin{align} \label{eq:dimshift_eq3}
F^{[4-2\eps]}_n(p_1,\ldots,p_{n-1})[\mu_{11}^r] = \frac{\pi^r \Omega_{-2\eps}}{\Omega_{2 r-2\eps}} 
  \int \frac{ \d^{4} k_1^{[4]} \,  \d^{2r-2\eps} k_1^{[2r-2\eps]} }{\i \pi^{2+r-\eps}}  \, \mathcal{F}_n\left( k_1^{[4]}, \mu_{11} \right) \,.
\end{align}
Using \eqn{eq:area-sphere} for the prefactor gives
\begin{align}
\frac{\pi^r \Omega_{-2\eps}}{\Omega_{2 r-2\eps}} = \frac{\Gamma(r-\eps)}{\Gamma(-\eps)}  \,,
\end{align}
which we simplify using \eqn{eq:gamma_ratio}. The loop integration on the RHS of \eqn{eq:dimshift_eq3} matches the scalar one-loop integral (i.e., with numerator $1$) with loop momentum in $D=4+2r - 2\eps$ dimensions and four-dimensional external momenta, namely
\begin{align}
F_n^{[4-2\eps]}(p_1,\ldots,p_{n-1})[\mu_{11}^r] = \left( \prod_{s=0}^{r-1}(s-\eps) \right) \, F_{n}^{[4+2 r - 2\eps]}(p_1,\ldots,p_{n-1})[1] \,,
\end{align}
as claimed.


\section*{Exercise \ref{Ex:3.7}: Projecting out the triangle coefficients}
\label{Sol:3.7}

The solution follows from the theory of discrete Fourier transform. Let $N$ be a positive integer.
The functions
\begin{align}
\left\{\text{e}^{\frac{2 \pi \i}{N} k l}\,, \ l = 0,\ldots, N-1 \right\} \,,
\end{align}
with $k\in \mathbb{Z}$, form an orthogonal basis of the space of complex-valued functions on the set of the $N^{\text{th}}$ roots of unity, $\{ \text{e}^{2\pi\i l/N}, \, l=0,\ldots,N-1\}$. In other words, they satisfy the orthogonality condition
\begin{align} \label{eq:orthogonality}
\sum_{l=0}^{N-1} \text{e}^{\frac{2\pi\i}{N} (n-k) l} = \delta_{n,k} \, N \,.
\end{align}
This is straightforward for $n=k$. For $n\neq k$, \eqn{eq:orthogonality} follows from the identity
\begin{align}
\sum_{l=0}^{N-1} z^l = \frac{1-z^N}{1-z}  
\end{align}
with $z = \text{e}^{2\pi\i (n-k)/N}$, and hence $z^N = 1$.

We can then use the orthogonality condition to project out the triangle coefficients $d_{k;1|2|3}$. Using eqs.~\eqref{eq:Delta123exp} and~\eqref{eq:thetak} we have that
\begin{align}
\sum_{l=-3}^3 \text{e}^{-\i k \theta_l} \Delta_{1|2|3}(\theta_l) = \sum_{n=-3}^3 d_{n;1|2|3} \, \text{e}^{- 3 (n-k) \frac{2\pi\i}{7}} \sum_{l=0}^6 \text{e}^{\frac{2\pi \i}{7} (n-k) l} \,.
\end{align}
Substituting \eqn{eq:orthogonality} with $N=7$, and solving for $d_{k;1|2|3}$ gives
\begin{align}
d_{k;1|2|3} = \frac{1}{7} \sum_{l=-3}^3 \text{e}^{-\i k \theta_l} \Delta_{1|2|3}(\theta_l)\,,
\end{align}
as claimed.

\smallskip

We now consider a rank-$4$ four-dimensional triangle numerator $\Delta_{1|2|3}^{(4)}(k\cdot \omega_1, k\cdot \omega_2)$. 
We parametrise the family of solutions to the triple cut by the angle $\theta$ as in \eqn{eq:kwparam}.
Expanding sine and cosine into exponentials gives
\begin{align}
\Delta_{1|2|3}^{(4)}(\theta) = \sum_{k=-4}^4 d_{k;1|2|3}^{(4)} \, \text{e}^{\i k \theta} \,.
\end{align}
The coefficients $d_{k;1|2|3}^{(4)}$ can then be projected out using the $9^{\text{th}}$ roots of unity $\text{e}^{\i \theta_l'}$, with $\theta_l' = 2\pi \, l/9$, for $l=-4,\ldots,4$. By using the orthogonality condition~\eqref{eq:orthogonality} with $N=9$ we obtain
 \begin{align}
d_{k;1|2|3}^{(4)} = \frac{1}{9} \sum_{l=-4}^4 \text{e}^{-\i k \theta_l'} \Delta_{1|2|3}^{(4)}\left(\theta_l'\right)\,.
\end{align}

\section*{Exercise \ref{Ex:3.DirectExtraction}: Rank-one triangle reduction with direct extraction}
\label{Sol:3.DirectExtraction}
a) After integration, the tensor integral $F_3^{[D]}(P,Q)[k^{\mu}]$ can only be a function of $P^{\mu}$ and $Q^{\mu}$. We thus expand it as
\begin{align}
F_3^{[D]}(P,Q)[k^{\mu}] = c_1 \, P^{\mu} + c_2 \, Q^{\mu} \,.
\end{align}
Contracting both sides by $P^{\mu}$ and $Q^{\mu}$, and solving for the coefficients gives
\begin{align} \label{eq:triangle_c1c2}
\begin{aligned}
& c_1 =  \frac{1}{(P\cdot Q)^2-S \, T} \left[ (P\cdot Q) \, F_3^{[D]}(P,Q)[k\cdot Q] - T \, F_3^{[D]}(P,Q)[k\cdot P] \right] \,, \\
& c_2 =  \frac{1}{(P\cdot Q)^2-S \, T} \left[ (P\cdot Q) \,  F_3^{[D]}(P,Q)[k\cdot P] - S \, F_3^{[D]}(P,Q)[k\cdot Q]  \right] \,.
\end{aligned}
\end{align}
Next, we need to rewrite the integrals above in terms of scalar integrals. To this end, we express
the scalar products $k \cdot P$ and $k \cdot Q$ in terms of inverse propagators $D_i$,~as
\begin{align}
k \cdot P = \frac{1}{2} \bigl(D_1-D_2 + 
\hat{S} \bigr) \,, \qquad \quad
k \cdot Q = \frac{1}{2} \bigl( D_2 - D_3 + \hat{T} \bigr) \,,
\end{align}
where 
\begin{align}
D_1 = k^2 - m_1^2 \,, \quad D_2 = (k-P)^2 - m_2^2 \,, \quad D_3 = (k-P-Q)^2-m_3^2\,,
\end{align}
and 
\begin{align}
\hat{S} \coloneqq S + m_1^2 - m_2^2 \,, \qquad \hat{T} \coloneqq T + m_2^2 - m_3^2 + 2 \, (Q\cdot P) \,.
\end{align}
We thus have that
\begin{align}
F_3^{[D]} (P,Q)[k \cdot Q] & = \frac{1}{2} \left[ \int_k \frac{1}{D_1 \, D_2} - \int_k \frac{1}{D_1 \, D_3} - \int_k \frac{\hat{T}}{D_1 \, D_2 \, D_3}  \right] \nonumber \\
& = \frac{1}{2} \left[ F_2^{[D]}(P) - F_2^{[D]}(P+Q) + \hat{T}  \, F_3^{[D]}(P,Q) \right] \,,
\end{align}
while $F_3^{[D]} (P,Q)[k \cdot P] $ contains no $P$-channel bubble. We recall that $F_n^{[D]}(\cdots) \equiv F_n^{[D]}(\cdots)[1]$. Putting the above together gives
\begin{align}
F_3^{[D]}(P,Q)[k\cdot Z] & = c_1 \, (P\cdot Z) + c_2 \, (Q\cdot Z) \nonumber \\
& = \frac{1}{2} \frac{(P\cdot Q) (P\cdot Z) -  (Q\cdot Z) \, S}{(P\cdot Q)^2-S \, T} \, F_2^{[D]}(P) + \ldots \,,
\end{align}
where the ellipsis denotes terms which do not involve $P$-channel bubbles.
Finally, we can read off that the coefficient of the $P$-channel scalar bubble integral is given~by
\begin{align} \label{eq:c0PQ_res}
c_{0;P|QR} = \frac{(P\cdot Q) (P\cdot Z) -  (Q\cdot Z) \, S}{2\bigl( (P\cdot Q)^2-S \, T \bigr) } \,,
\end{align}
as claimed.

\medskip
\noindent
b) We outline here the main steps of the solution, while the computations are performed in the \texttt{Mathematica} notebook \href{https://scattering-amplitudes.mpp.mpg.de/scattering-amplitudes-in-qft/Exercises/}{\texttt{Ex3.9\_DirectExtraction.wl}}~\cite{ch5-website}. Since we are considering a triangle integral, all quadruple cuts vanish. The coefficients of the box numerator are thus zero. We parametrise the triangle numerator $\Delta_{P|Q|R}$ as in \eqn{eq:triparam2},
\begin{align} \begin{aligned}
& \Delta_{P|Q|R}(k\cdot \omega_{1,{\rm tri}}, k\cdot \omega_{2,{\rm tri}}) = c_{0;P|Q|R} + c_{1;P|Q|R} \, (k\cdot \omega_{1,{\rm tri}}) + c_{2;P|Q|R} \, (k\cdot \omega_{2,{\rm tri}}) \\
& \hspace{0.5cm} +  c_{3;P|Q|R} \left( (k\cdot \omega_{1,{\rm tri}})^2 - \frac{\omega_{1,{\rm tri}}^2}{\omega_{2,{\rm tri}}^2} (k\cdot \omega_{2,{\rm tri}})^2 \right) + c_{4;P|Q|R} \,  (k\cdot \omega_{1,{\rm tri}})  (k\cdot \omega_{2,{\rm tri}}) \\
& \hspace{0.5cm} + c_{5;P|Q|R} \,  (k\cdot \omega_{1,{\rm tri}})^3 + c_{6;P|Q|R} \,  (k\cdot \omega_{1,{\rm tri}})^2  (k\cdot \omega_{2,{\rm tri}}) \,,
\end{aligned} \end{align}
with the spurious vectors as in \eqn{eq:spurious_vectors_tri_box},
 \begin{align}
  \omega_{1,{\rm tri}}^\mu &= \frac{1}{2}\spAB{\check{P}}{\gamma^\mu}{\check{Q}} \, \Phi_{\rm tri} + \frac{1}{2}\spAB{\check{Q}}{\gamma^\mu}{\check{P}} \, \Phi_{\rm tri}^{-1}\,,\\ 
  \omega_{2,{\rm tri}}^\mu &= \frac{1}{2}\spAB{\check{P}}{\gamma^\mu}{\check{Q}}  \, \Phi_{\rm tri} - \frac{1}{2}\spAB{\check{Q}}{\gamma^\mu}{\check{P}} \, \Phi_{\rm tri}^{-1} \,.
\end{align}
Here, $\Phi_{\rm tri}$ is an arbitrary factor which makes the summands phase-free. E.g.\ we may choose $\Phi_{\rm tri}=\spAB{\check{Q}}{Z}{\check{P}}$, but its expression is irrelevant as it cancels out from the result.
Moreover, we have the light-like projections
\begin{align} \label{eq:ex_PxQx}
\check{P}^{\mu} = \frac{\gamma \, \left(\gamma P^{\mu} - S \, Q^{\mu} \right)}{\gamma^2 - S \, T} \,, \qquad \quad \check{Q}^{\mu} = \frac{\gamma \, \left(\gamma Q^{\mu} - T \, P^{\mu}\right) }{\gamma^2 - S \, T}  \,,
\end{align}
with two projections $\gamma_{\pm} = (P\cdot Q) \pm \sqrt{ (P\cdot Q)^2 -S \, T}$. We parametrise the loop momentum on the triple cut $P|Q|R$ ($D_1 = D_2 = D_3 = 0$) in terms of $t$ as discussed in section~\ref{sec:3.5}:
\begin{align} \label{eq:ex_PQRk}
\mathcal{C}_{P|Q|R}\left(k^{\mu}\right) = \beta_1 \, P^{\mu}+ \beta_2 \, Q^{\mu} + \frac{1}{2} \left(t+\frac{\gamma_{\rm tri}}{t}\right) \, \omega_{1, {\rm tri}}^{\mu}  + \frac{1}{2} \left(t-\frac{\gamma_{\rm tri}}{t}\right) \, \omega_{2, {\rm tri}}^{\mu} \,,
\end{align}
with
\begin{align}
\beta_1 = \frac{\hat{S} \, T- \hat{T} \, (P\cdot Q)}{2 \, \bigl(S \, T- (P \cdot Q)^2 \bigr)} \,, \qquad \quad
\beta_2 = \frac{S \, \hat{T}- \hat{S} \, (P\cdot Q)}{2 \, \bigl(S \, T- (P \cdot Q)^2 \bigr)} \,.
\end{align}
We use $\gamma_{\rm tri}$ in \eqn{eq:ex_PQRk} to distinguish it from the $\gamma$ used in \eqn{eq:ex_PxQx}. Its value is fixed by the constraint $D_1 = 0$, and we omit it here for conciseness.
We now determine the triangle coefficients $c_{i;P|Q|R}$ by solving
\begin{align} \label{eq:ex_triangle_match}
\mathcal{C}_{P|Q|R}\left( \Delta_{P|Q|R}(k\cdot \omega_{1,{\rm tri}}, k\cdot \omega_{2,{\rm tri}}) \right) = \mathcal{C}_{P|Q|R} \left(k \cdot Z \right) \,.
\end{align}
We recall that the box subtraction terms are zero in this case.
In section~\ref{sec:3.5} we have seen how to extract directly $c_{0;P|Q|R}$ using the operation ``${\rm Inf}$'' (see \eqn{eq:direct_extraction_triangle}). Here however we need all triangle coefficients. The two sides of \eqn{eq:ex_triangle_match} are Laurent polynomials in $t$, with the loop-momentum parametrisation~\eqref{eq:ex_PQRk}. As the equation holds for any value of $t$, we may solve it separately order by order in $t$. This gives enough constraints to fix all triangle coefficients. We~find
\begin{align}
c_{1;P|Q|R} = - \frac{Z\cdot \omega_{1,{\rm tri}}}{2 \, \check{P}\cdot \check{Q}} \,, \qquad \quad
c_{2;P|Q|R} = \frac{Z\cdot \omega_{2,{\rm tri}}}{2 \, \check{P}\cdot \check{Q}} \,.
\end{align}
The coefficient of the scalar triangle integral, $c_{0;P|Q|R}$, will not contribute to the bubble coefficient, and we thus omit it here. The higher-rank coefficients, $c_{i;P|Q|R}$ with $i=3,\ldots,6$ all vanish, as we could have guessed from start by noticing that the example integral we are studying has a rank-one numerator.

\smallskip

We can now move on to the bubble coefficients. We parametrise the bubble numerator as in \eqn{eq:bubble_num_param},
\begin{align} 
\begin{aligned}
 & \Delta_{P|QR}(k\cdot\omega_{1,{\rm bub}}, k\cdot\omega_{2,{\rm bub}}, k\cdot\omega_{3,{\rm bub}}) = c_{0;P|QR} \\
 & \hspace{0.2cm} + c_{1;P|QR} (k\cdot\omega_{1,{\rm bub}}) 
  + c_{2;P|QR} (k\cdot\omega_{2,{\rm bub}}) 
  + c_{3;P|QR} (k\cdot\omega_{3,{\rm bub}}) \\ 
& \hspace{0.2cm} + c_{4;P|QR} (k\cdot\omega_{1,{\rm bub}}) (k\cdot\omega_{2,{\rm bub}}) 
  + c_{5;P|QR} (k\cdot\omega_{1,{\rm bub}}) (k\cdot\omega_{3,{\rm bub}}) \\
& \hspace{0.2cm} + c_{6;P|QR} (k\cdot\omega_{2,{\rm bub}}) (k\cdot\omega_{3,{\rm bub}})   
  + c_{7;P|QR} \left((k\cdot\omega_{1,{\rm bub}})^2 - \frac{\omega_{1,{\rm bub}}^2}{\omega_{3,{\rm bub}}^2} (k\cdot\omega_{3,{\rm bub}})^2 \right) \\
& \hspace{0.2cm} + c_{8;P|QR} \left((k\cdot\omega_{2,{\rm bub}})^2 - \frac{\omega_{2,{\rm bub}}^2}{\omega_{3,{\rm bub}}^2} (k\cdot\omega_{3,{\rm bub}})^2 \right) \,,
\end{aligned}
\end{align}
with the spurious vectors as in \eqn{eq:spurious_omega_bubble},
\begin{align}
  \omega_{1,{\rm bub}}^\mu &= \frac{1}{2}\spAB{P^\flat}{\gamma^\mu}{n} \, \Phi_{\rm bub} + \frac{1}{2}\spAB{n}{\gamma^\mu}{P^\flat} \, \Phi_{\rm bub}^{-1} \,, \\ 
  \omega_{2,{\rm bub}}^\mu &= \frac{1}{2}\spAB{P^\flat}{\gamma^\mu}{n} \, \Phi_{\rm bub} - \frac{1}{2}\spAB{n}{\gamma^\mu}{P^\flat} \,  \Phi_{\rm bub}^{-1} \,,\\
  \omega_{3,{\rm bub}}^\mu &= P^{\flat,\mu} - \frac{S}{2 \, P\cdot n} \, n^\mu \,,
\end{align}
where $n^{\mu}$ is an arbitrary light-like momentum, and
\begin{align}
P^{\flat, \, \mu} = P^{\mu} - \frac{S}{2 \, P\cdot n} \, n^{\mu} \,.
\end{align}
We may choose the phase factor e.g.\ as $\Phi_{\rm bub} = \spAB{n}{Z}{P^\flat}$ but---just like $\Phi_{\rm tri}$---this will not appear in the result.
We parametrise the loop momentum on the double cut $P|QR$ ($D_1 = D_2 = 0$) in terms of $t$ and $y$ as in \eqn{eq:kPn}:
\begin{align} \label{eq:ex_PQk}
\mathcal{C}_{P|QR}\left(k^{\mu}\right) = \alpha_1 \, P^{\flat, \, \mu}+ \alpha_2 \, n^{\mu} + \alpha_3 \, \frac{1}{2} \, \spAB{P^{\flat}}{\gamma^{\mu}}{n} \, \Phi_{\rm bub} + \alpha_4 \, \frac{1}{2} \spAB{n}{\gamma^{\mu}}{P^{\flat}} \, \Phi_{\rm bub}^{-1} \,,
\end{align}
with
\begin{align}
  \alpha_1 = y \,, \qquad 
  \alpha_2 = \frac{\hat{S}- S \, y}{2 \, n \cdot P} \,, \qquad
  \alpha_3 = t \,, \qquad
  \alpha_4 = \frac{y \, (\hat{S} - S \, y) - m_1^2}{2 \, t \, (n\cdot P)} \,.
\end{align}
The bubble coefficients $c_{i;P|QR}$ are fixed through \eqn{eq:doublecuteq} with the box subtraction term set to zero:
\begin{align}
\mathcal{C}_{P|QR} \bigl( \Delta_{P|QR}(\{k\cdot\omega_{i,{\rm bub}}\}) \bigr) = \mathcal{C}_{P|QR} \left( - \frac{k\cdot Z}{D_3} + \frac{\Delta_{P|Q|R}( \{k\cdot\omega_{i,{\rm tri}}\} )}{D_3}  \right)  \,.
\end{align}
We can extract the coefficient $c_{0;P|Q}$ directly using \eqn{eq:bubble_direct_extraction}, as
\begin{align} \label{eq:c0PQ_extraction}
c_{0;P|QR} = \mathcal{P} \mathrm{Inf}_y \mathrm{Inf}_t \left[ \mathcal{C}_{P|QR}\left( - \frac{k\cdot Z}{D_3} \right) - \frac{1}{2} \sum_{\gamma=\gamma_{\pm}} \mathcal{C}_{P|QR}\left( \frac{\Delta_{P|Q|R}( \{k\cdot\omega_{i,{\rm tri}}\} )}{-D_3} \right) \right] \,,
\end{align}
with the operator $\mathcal{P}$ defined in \eqn{eq:Pdef},
\begin{align}
  \mathcal{P}\bigl(f(y,t) \bigr) = f \bigl|_{t^0,y^0} + \frac{\hat{S}}{2 \, S}f \bigl|_{t^0,y^1}+ \frac{1}{3}\left( \frac{\hat{S}^2}{S^2} - \frac{m_1^2}{S} \right)f\bigl|_{t^0,y^2} \,,
\end{align}
while $\mathrm{Inf}_x$ expands a rational function around $x =\infty$ and keeps only the terms that do not vanish in the limit (see \eqn{eq:InfDef}).
We obtain
\begin{align} \label{eq:ex_tr_cut}
 \mathcal{P} \mathrm{Inf}_y \mathrm{Inf}_t \left[  \mathcal{C}_{P|QR}\left( -  \frac{k\cdot Z}{D_3} \right) \right] = \frac{1}{2} \, \frac{\spAB{P^{\flat}}{Z}{n}}{\spAB{P^{\flat}}{Q}{n}} \,,
\end{align}
and
\begin{align} \label{eq:ex_tr_sub}
 \mathcal{P} \mathrm{Inf}_y \mathrm{Inf}_t \left[ \mathcal{C}_{P|QR}\left( \frac{\Delta_{P|Q|R}( \{k\cdot\omega_{i,{\rm tri}}\} )}{ - D_3} \right) \right]  = - \frac{\langle P^{\flat}|\check{P} Z \check{Q}|n] + \langle P^{\flat}|\check{Q} Z \check{P}|n] }{4 \, (\check{P}\cdot \check{Q}) \, \spAB{P^{\flat}}{Q}{n} } \,.
\end{align}
We may simplify the RHS of \eqn{eq:ex_tr_sub} by rewriting
\begin{align} 
\langle P^{\flat}|\check{P} Z \check{Q}|n] & = \spAB{\check{Q}}{Z}{\check{P}} \, \spA{\check{P}}{P^{\flat}} [n\check{Q}] \nonumber \\
& = \mathrm{Tr}\left(\sigma^{\mu} \bar{\sigma}^{\nu} \sigma^{\rho} \bar{\sigma}^{\tau} \right) \, Z_{\mu} \check{P}_{\nu} V_{\rho} \check{Q}_{\tau} \,,
\end{align}
where we introduced the short-hand
$ V^{\mu} = \spAB{P^{\flat}}{\gamma^{\mu}}{n}/2$. Using the identity~\eqref{eq:trSigma2gamma} we can trade the $\sigma$-matrix trace for a $\gamma$-matrix trace, obtaining
\begin{align}
\langle P^{\flat}|\check{P} Z \check{Q}|n] = \frac{1}{2} \text{Tr}\left( \slashed{Z} \check{\slashed{P}} \slashed{V} \check{\slashed{Q}} (1-\gamma_5) \right) \,.
\end{align}
The trace with $\gamma_5$ is fully anti-symmetric in the four momenta, and thus cancels out between $\langle P^{\flat}|\check{P} Z \check{Q}|n]$ and $\langle P^{\flat}|\check{Q} Z \check{P}|n]$. The remaining traces can be expressed in terms of scalar products using the familiar rule for the Dirac matrices (see \eqn{eq:gamma_traces}). We then obtain
\begin{align}
\begin{aligned}
 & \mathcal{P} \mathrm{Inf}_y \mathrm{Inf}_t \left[ \mathcal{C}_{P|QR}\left( \frac{\Delta_{P|Q|R}( \{k\cdot\omega_{i,{\rm tri}}\} )}{-D_3} \right) \right]  \\
 & \hspace{1.5cm} =  
 \frac{1}{2} \frac{(S \, T + \gamma^2) (P\cdot Z) - 2 \, \gamma \, S \, (Q\cdot Z)}{2 \, \gamma \, S \, T -  (S\, T +\gamma^2) (P\cdot Q)} +
 \frac{1}{2} \frac{\spAB{P^\flat}{Z}{n}}{\spAB{P^\flat}{Q}{n}}  \,.
 \end{aligned}
\end{align}
Averaging over the two projections $\gamma_{\pm}$ then gives
\begin{align}
\begin{aligned}
 & \frac{1}{2} \sum_{\gamma=\gamma_\pm} \mathcal{P} {\rm Inf}_y {\rm Inf}_t  \left[ \mathcal{C}_{P|QR}\left( \frac{\Delta_{P|Q|R}(\{k\cdot\omega_{i,{\rm tri}}\})}{-D_3} \right) \right] \\
 &  \hspace{1.5cm} = - \frac{(P\cdot Q) (P\cdot Z) - S \, (Q\cdot Z)}{2 \, \bigl( (P\cdot Q)^2 - S\,T \bigr)} + \frac{1}{2} \frac{\spAB{P^\flat}{Z}{n}}{\spAB{P^\flat}{Q}{n}} \,.
\end{aligned}
\end{align}
Substituting this and \eqn{eq:ex_tr_cut} into \eqn{eq:c0PQ_extraction} finally gives
\begin{align}
c_{0;P|QR} =  \frac{(P\cdot Q) (P\cdot Z) - S \, (Q\cdot Z)}{2 \, \bigl( (P\cdot Q)^2 - S\,T \bigr)}  \,,
\end{align}
in agreement with the result of the Passarino-Veltman reduction given in \eqn{eq:c0PQ_res}.

\section*{Exercise \ref{Ex:3.9}: Momentum-twistor parametrisations}
\label{Sol:3.9}
The matrix $Z$ in \eqn{eq:Z4pt} has the form  
\begin{align}
Z = \left( Z_1 \, Z_2 \, Z_3 \, Z_4 \right) \,, \quad \text{with} \quad Z_i = \begin{pmatrix} \lambda_{i\alpha} \\ \mu_i^{\dot{\alpha}} \end{pmatrix}  \,.
\end{align}
We can thus read off $\lambda_{i\alpha}$ and compute all $\spA ij$ through $\spA ij = - \lambda_{i \alpha} \epsilon^{\alpha \beta} \lambda_{j \beta}$. Our conventions for $\epsilon^{\alpha \beta}$ are given in exercise~\ref{Prob:1.1}. For instance, we have that
\begin{align}
\spA 12 = - ( 1 \ 0 ) \cdot \begin{pmatrix} 0 & -1 \\ 1 & 0 \\ \end{pmatrix} \cdot \begin{pmatrix} 0 \\ 1 \\ \end{pmatrix} = 1 \,.
\end{align}
Repeating this for all $\spA ij$ we get
\begin{align} \label{eq:spAs}
\spA 12 = \spA 13 = \spA 14 = 1 \,, \quad \spA 23 = - \frac{1}{y} \,,  \quad  \spA 24 = y \,,  \quad \spA 34 = \frac{1+y^2}{y} \,.
\end{align}
From these we can see explicitly that the helicity information is obscured, as some $\spA ij$ are set to constants.
Next, we compute the $\tilde{\lambda}_i$ through \eqn{eq:LambdaTildeFromMT}. E.g.\ we~have
\begin{align} \label{eq:lambdat1ex}
\tilde{\lambda}_1^{\dot{\alpha}} = \frac{\spA 12 \, \mu_4^{\dot{\alpha}} + \spA 24 \, \mu_1^{\dot{\alpha}}  + \spA 41 \, \mu_2^{\dot{\alpha}}}{\spA 41 \spA 12} \,.
\end{align}
From \eqn{eq:Z4pt} we read off $\mu_1^{\dot\alpha} = \mu_2^{\dot\alpha} = (0,0)^{\top}$ and $\mu_4^{\dot\alpha} = (0, x)^{\top}$. Substituting this and \eqn{eq:spAs} into \eqn{eq:lambdat1ex} gives $\tilde{\lambda}_1^{\dot{\alpha}} = -\mu_4^{\dot{\alpha}} $. The other $\tilde{\lambda}_i$ are obtained similarly:
\begin{align} \label{eq:lambdatildes}
\tilde{\lambda}_1^{\dot{\alpha}} = \begin{pmatrix} 0\\ -x \end{pmatrix} \,, \quad 
\tilde{\lambda}_2^{\dot{\alpha}} = \begin{pmatrix} x \\ 0 \end{pmatrix} \,, \quad
\tilde{\lambda}_3^{\dot{\alpha}} = \frac{x \, y}{1+y^2} \begin{pmatrix} -y \\ 1 \end{pmatrix} \,, \quad
\tilde{\lambda}_4^{\dot{\alpha}} = \frac{-x}{1+y^2} \begin{pmatrix} 1\\  y \end{pmatrix} \,.
\end{align}
These allow us to determine all $\spB ij$ through $\spB ij = - \tilde{\lambda}_i^{\dot\alpha} \epsilon_{\dot\alpha \dot\beta} \tilde{\lambda}_j^{\dot\alpha}$,
\begin{align} \label{eq:spBs}
\begin{aligned}
\spB 12 &= -x^2 \,,          &  \spB 13 &=\frac{x^2 y^2}{1+y^2} \,,             &  \spB 14&=\frac{x^2}{1+y^2} \,, \\
\spB 23 &= - \frac{x^2 y}{1+y^2} \,,        &  \spB 24 &=\frac{x^2 y}{1+y^2}\,,  &  \spB 34 & = -\frac{x^2 y}{1+y^2} \,.
\end{aligned}
\end{align}
We calculate $s_{ij}$ from $\spA ij$ and $\spB ij$ through $s_{ij} = \spA ij \spB ji$. Thanks to momentum conservation, only two are independent. We choose
\begin{align} \label{eq:sijs}
s_{12} = x^2 \,, \qquad s_{23} = - \frac{x^2}{1+y^2} \,.
\end{align}
The others are determined from these as $s_{13}=s_{24}=-s_{12}-s_{23}$, $s_{14}=s_{23}$, and $s_{34}=s_{12}$.
We obtain the momenta $p_i^{\mu}$ from $\lambda_{i\alpha}$ and $\tilde{\lambda}_i^{\dot\alpha}$ through $p_i^{\mu} = - \lambda_{i\alpha} \epsilon^{\alpha\beta} (\sigma^{\mu})_{\beta\dot\beta} \tilde{\lambda}_i^{\dot\beta}$. See exercise~\ref{Prob:1.1} for our conventions on $\sigma^{\mu}$. We obtain
\begin{align} \label{eq:explmom}
p_1^{\mu} = \frac{x}{2} \begin{pmatrix} -1 \\ 0 \\ 0 \\ - 1\end{pmatrix} \,, \qquad p_2^{\mu} = \frac{x}{2} \begin{pmatrix} - 1 \\ 0 \\ 0 \\ 1 \end{pmatrix} \,, \qquad
p_3^{\mu} = \frac{x}{2} \begin{pmatrix} 1 \\ \frac{2 y}{1 + y^2} \\ 0 \\ \frac{1-y^2}{1+y^2} \end{pmatrix}\,,
\end{align}
and $p_4 = -p_1-p_2-p_3$. This parametrisation describes two incoming particles with momenta $-p_1$ and $-p_2$ traveling along the $z$ axis with energy $E = x/2$ in their center-of-mass frame. The outgoing particles, with momenta $p_3$ and $p_4$, lie on the $xz$-plane. The angle $\theta$ between the three-momentum $\vec{p}_3$ and the $z$ axis is related to $y$ through $y = \tan(\theta/2)$.

\smallskip

Let us consider the tree-level four-gluon amplitude $A^{(0)}_4(1^-,2^+,3^-, \allowbreak 4^+)$. Using the Parke-Taylor formulae~\eqref{ParkeTaylor1} and~\eqref{ParkeTaylor1MHVbar} (with $g=1$) it may be written either~as
\begin{align}
A^{(0)}_{4, \mathrm{MHV}}(1^-,2^+,3^-,4^+) = \i \, \frac{\langle 13 \rangle^4 }{\spA 12 \spA 23 \spA 34 \spA41} \,,
\end{align}
or as
\begin{align}
A^{(0)}_{4, \overline{\mathrm{MHV}}}(1^-,2^+,3^-,4^+) = \i \, \frac{[ 24]^4 }{[12] [23] [34] [41]} \,.
\end{align}
Using the momentum-twistor parametrisation in eqs.~\eqref{eq:spAs} and~\eqref{eq:spBs} it is straightforward to see that both expressions evaluate to
\begin{align}
A^{(0)}_4(1^-,2^+,3^-,4^+) = \i \, \frac{y^2}{1+y^2} \,.
\end{align}
Showing this with the spinor-helicity formalism alone requires some gymnastics with momentum conservation. For instance, we may proceed as 
\begin{align}
\frac{A^{(0)}_{4,  \overline{\mathrm{MHV}}}(1^-,2^+,3^-,4^+)}{A^{(0)}_{4, \mathrm{MHV}}(1^-,2^+,3^-,4^+)} & = \frac{[24]^4 \spA 12 \spA 23 \spA 34 \spA41}{\langle 13 \rangle^4 [12] [23] [34] [41] } \nonumber \\
& = \frac{\overbrace{\spA12 [24]}^{- \spA13 [34]} \, \overbrace{\spA32 [24]}^{- \spA31 [14]} \, \overbrace{\spA34 [42]}^{-\spA 31 [12]} \, \overbrace{[24] \spA41}^{-[23] \spA31}}{\langle 13 \rangle^4 [12] [23] [34] [41] } \nonumber \\
& = 1 \,.
\end{align}
The proof for the adjacent MHV configuration $A^{(0)}_4(1^-,2^-,3^+, 4^+)$ is analogous.


\section*{Exercise \ref{Ex:MasslessBubble}: The massless bubble integral}
\label{Sol:MasslessBubble}
a) Applying the Feynman trick~\eqref{Schwinger2} to the bubble integral~\eqref{eq:massless_bubble} gives
\begin{align}
F_2 = \frac{\Gamma(a_1+a_2)}{\Gamma(a_1) \Gamma(a_2)} \int \frac{\d\alpha_1 \d\alpha_2}{\text{GL}(1)} \int \frac{\d^D k}{\i \pi^{\frac{D}{2}}} \frac{\alpha_1^{a_1-1} \alpha_2^{a_2-1} (\alpha_1+\alpha_2)^{-a_1-a_2}}{ \left[-M^2 - \i 0 \right]^{a_1+a_2}} \,,
\end{align}
where
\begin{align} 
M^2 = k^2 - \frac{2 \, \alpha_2}{\alpha_1+\alpha_2} \, p \cdot k + \frac{\alpha_2}{\alpha_1+\alpha_2} p^2 \,.
\end{align}
We complete the square in $M^2$,
\begin{align}
M^2 = \biggl( k - \frac{\alpha_2}{\alpha_1+\alpha_2} p \biggr)^2 + p^2 \frac{\alpha_1 \alpha_2}{(\alpha_1+\alpha_2)^2} \,,
\end{align}
and shift the loop momentum as $k \to k-\alpha_2/(\alpha_1+\alpha_2) p$. This gives 
\begin{align}
F_2 = \frac{\Gamma(a_1+a_2)}{\Gamma(a_1) \Gamma(a_2)} \int \frac{\d\alpha_1 \d\alpha_2}{\text{GL}(1)} \int \frac{\d^D k}{\i \pi^{\frac{D}{2}}} \frac{\alpha_1^{a_1-1} \alpha_2^{a_2-1} (\alpha_1+\alpha_2)^{-a_1-a_2}}{ \left[-k^2 - \frac{\alpha_1\alpha_2}{(\alpha_1+\alpha_2)^2} p^2 - \i 0 \right]^{a_1+a_2}} \,.
\end{align}
We can now carry out the integration in $k$ using the formula~\eqref{single_propagator}, obtaining
\begin{align} \label{eq:ex4.1_step}
F_2 = \frac{\Gamma\left(a_1+a_2-\frac{D}{2}\right)}{\Gamma(a_1) \Gamma(a_2)} \int \frac{\d\alpha_1 \d\alpha_2}{\text{GL}(1)} \alpha_1^{a_1-1} \alpha_2^{a_2-1} \frac{(\alpha_1+\alpha_2)^{a_1+a_2-D}}{\left( - \alpha_1 \alpha_2 p^2 - \i 0\right)^{a_1+a_2-\frac{D}{2}}} \,.
\end{align}
This formula is the Feynman parameterisation for the massless bubble integral. It matches the one-loop master formula~\eqref{oneloop_master}, with  $U=\alpha_1+\alpha_2$ and $V=-\alpha_1 \alpha_2 p^2$. 

\smallskip
\noindent
b) We use the $\text{GL}(1)$ invariance to fix $\alpha_1+\alpha_2=1$, namely we insert $\delta(\alpha_1+\alpha_2-1)$ under the integral sign in \eqn{eq:ex4.1_step}, and we absorb the $\i 0$ prescription into a small positive imaginary part of $p^2$. We can carry out the remaining integration in terms of Gamma functions, obtaining
\begin{align}
F_2 & = \bigl(-p^2-\i 0\bigr)^{\frac{D}{2}-a_1-a_2} \frac{\Gamma\left(a_1+a_2-\frac{D}{2}\right)}{\Gamma(a_1) \Gamma(a_2)} \int_{0}^{1} \d\alpha_1 \, \alpha_1^{\frac{D}{2}-a_2-1} (1-\alpha_1)^{\frac{D}{2}-a_1-1} \nonumber \\ 
& =  \bigl(-p^2-\i 0\bigr)^{\frac{D}{2}-a_1-a_2} \frac{\Gamma\left(a_1+a_2-\frac{D}{2}\right) \Gamma\left(\frac{D}{2}-a_1\right) \Gamma\left(\frac{D}{2}-a_2\right) }{\Gamma(a_1) \Gamma(a_2)  \Gamma(D - a_1 - a_2)}  \,,
\end{align}
as claimed.


\section*{Exercise \ref{Ex:4.FeynPar}: Feynman parametrisation}
\label{Sol:4.FeynPar}
\begin{figure}[t]
\includegraphics[scale=0.7]{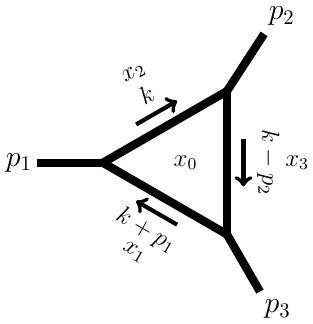}
\centering
\caption{Diagram of the triangle Feynman integral $F_3$ defined in eq.~(\ref{F3examplemomentumspace}). We write next to each internal edge the corresponding momentum. The arrows denote the directions of the momenta. The edges of the graph divide the space into four regions, which we label by the dual coordinates $x_i$.} 
\label{fig:triangle}
\end{figure}

We draw the diagram of the triangle Feynman integral $F_3$~\eqref{F3examplemomentumspace} in figure~\ref{fig:triangle}, with both momentum-space and dual-space labelling. We assign the dual coordinate $x_0$ to the region inside the loop, and relate the other dual coordinates to the external momenta according to \eqn{eq:dual_coordinates}:
\begin{align}
p_1 = x_2 - x_1 \,, \quad \quad p_2 = x_3 - x_2 \,, \quad \quad p_3 = x_1 - x_3 \,.
\end{align}
Note that momentum conservation ($p_1+p_2+p_3 = 0$) is automatically satisfied in terms of the dual coordinates.
The loop momenta are then given by
\begin{align} \label{eq:props2x}
k = x_0 - x_2 \,, \quad \quad k+p_1 = x_0 - x_1 \,, \quad \quad k-p_2 = x_0 - x_3 \,,
\end{align}
so that the integral takes the form of \eqn{oneloop_dual},
\begin{align} \label{eq:F3dual}
F_3 = \int \frac{\d^D x_0}{\i \pi^{D/2}} \prod_{j=1}^3 \frac{1}{- x^2_{0j} - \i 0} \,.
\end{align}
The relation between the loop momentum $k$ and the dual variables in \eqn{eq:props2x} differs from that given in \eqn{eq:dual_coordinates}, $k=x_1-x_0$. We emphasise that this is just a convention, as we are free to redefine the loop integration variables. The dual regions, on the other hand, are invariant. 
This means that once we assign coordinates $x_i$ to the dual regions, the integral takes the 
form of \eqn{eq:F3dual} (in agreement with the general formula~\eqref{oneloop_dual}),
regardless of the loop-momentum labelling we started from.

The kinematic constraints $p_1^2 = p_2^2 = 0$ and $p_3^2 = s$ imply that
\begin{align}
x^2_{12} = p_1^2 = 0 \,, \qquad x^2_{23} = p_2^2 = 0 \,, \qquad x^2_{13} = p_3^2 = s \,,
\end{align}
as claimed. 
The Symanzik polynomials are given by
\begin{align}
U = \alpha_1 + \alpha_2 + \alpha_3 \,, \qquad V = - s \, \alpha_1 \alpha_3 \,.
\end{align}
Substituting the above into \eqn{eq:F3dual} with $D=4-2 \eps$ 
gives the following Feynman parameterisation:
\begin{align}
F_3 = \Gamma(1+\eps) \int_0^{\infty} \frac{\d \alpha_1 \d \alpha_2 \d \alpha_3}{\text{GL}(1)} 
  \frac{1}{(\alpha_1+\alpha_2+\alpha_3)^{1-2\eps} (- s \, \alpha_1 \alpha_3 - \i 0)^{1+\eps} } \,.
\end{align}


\section*{Exercise \ref{Ex:TaylorLogGamma}: Taylor series of the Log-Gamma function}
\label{Sol:TaylorLogGamma}
\begin{enumerate}[a)]
\item The recurrence relation of the digamma function follows from that of the $\Gamma$ function~\eqref{eq:gammarecurrence}. 
Differentiating the latter, and dividing both sides by $x \Gamma(x)$ gives
\begin{align}
\frac{ \Gamma'(x+1)}{\Gamma(x+1)} = \frac{1}{x} + \frac{\Gamma'(x)}{\Gamma(x)} \,,
\end{align}
where we have used $x \Gamma(x) = \Gamma(x+1)$.
Comparing to the definition of the digamma function in \eqn{eq:psidef}, we can rewrite this as
\begin{align} \label{eq:psirecurrence}
\psi(x+1) = \frac{1}{x} + \psi(x) \,.
\end{align}
We apply \eqn{eq:psirecurrence} recursively starting from $\psi(x+n)$ with $n\in\mathbb{N}$,
\begin{align}
\begin{aligned}
\psi(x+n) & = \frac{1}{x+n-1} + \psi(x+n-1) \\
& = \frac{1}{x+n-1} + \frac{1}{x+n-2} + \psi(x+n-2)  \\
& = \sum_{s=1}^n \frac{1}{x+n-s} + \psi(x) \,.
\end{aligned} 
\end{align}
Changing the summation index to $k=n-s$ in the last line gives \eqn{eq:psixn}.

\smallskip 

\item Consider the difference $\psi(x+n) - \psi(1+n)$. Using \eqn{eq:psixn} we can rewrite it as
\begin{align} \label{eq:psidiff}
\psi(x+n) - \psi(1+n) = \sum_{k=0}^{n-1} \left(\frac{1}{x+k} - \frac{1}{1+k}\right) + \psi(x) + \gamma_{\text{E}} \,,
\end{align}
where we recall that $\psi(1) = - \gamma_{\text{E}}$. In order to study the limit $n \to \infty$ we use Stirling's formula~\eqref{eq:stirling}, which implies the following approximation for $\psi(x)$,
\begin{align}
\psi(1+x) = \frac{1}{2 x} + \log(x) + \mathcal{O}\left(\frac{1}{x^2}\right) \,.
\end{align}
It follows that 
\begin{align}
\lim_{n\to \infty} \left[ \psi(x+n) - \psi(1+n) \right] = 0 \,.
\end{align}
Taking the limit $n\to \infty$ of both sides of \eqn{eq:psidiff} gives \eqn{eq:psiseries}.

\smallskip

\item 
The series series representation~\eqref{eq:psiseries} of the digamma function allows us the compute the higher-order derivatives in the Taylor expansion~\eqref{eq:loggammataylorgen},
\begin{align} \label{eq:dndloggamma}
\begin{aligned}
\frac{\d^n}{\d x^n} \log \Gamma(1+x) \biggl|_{x=0}  & = \frac{\d^{n-1}}{\d x^{n-1}} \psi(1+x) \biggl|_{x=0} \\
& = (-1)^n \, (n-1)! \, \sum_{k=0}^{\infty} \frac{1}{(1+k)^{n}} \,,
\end{aligned}
\end{align}
for $n\ge 2$.
By changing the summation index to $k'=k+1$, we recognise in the last line the definition of the Riemann zeta constant $\zeta_n$ given in \eqn{eq:zetan}. Substituting \eqn{eq:dndloggamma} into \eqn{eq:loggammataylorgen} and simplifying finally give \eqn{eq:logGammaTaylor}.

\end{enumerate}


\section*{Exercise \ref{Ex:4.2Dbubble}: Finite two-dimensional bubble integral}
\label{Sol:4.2Dbubble}
We start from the Feynman parameterisation in \eqn{eq:bubbleFeynman}. We set $D=2$, fix the $\text{GL}(1)$ freedom such that $\alpha_1+\alpha_2=1$, and absorb the $\i 0$ prescription in a small positive imaginary part of $s$. We obtain
\begin{align}
F_2\bigl(s,m^2; D=2\bigr) = \int_0^1 \frac{\d \alpha_1}{-s \, \alpha_1 (1-\alpha_1 ) + m^2} \,.
\end{align}
In order to carry out the integration, we factor the denominator and decompose the integrand into partial fractions w.r.t.~$\alpha_1$:
\begin{align}
F_2\bigl(s,m^2; D=2\bigr) = \frac{1}{s \, (\alpha_1^+-\alpha_1^-)} \int_0^1 \d \alpha_1 \left( \frac{1}{\alpha_1 - \alpha_1^+} - \frac{1}{\alpha_1 - \alpha_1^-} \right) \,,
\end{align}
where
\begin{align}
\alpha_1^{\pm} = \frac{1}{2} \left( 1 \pm \sqrt{\Delta} \right) \,, \qquad \Delta = 1 - 4 \, \frac{m^2}{s} \,. 
\end{align}
For $s<0$ and $m^2>0$, we have that $\alpha_1^+ >1$ and $\alpha_1^-<0$. The integration then yields
\begin{align} \label{eq:Bintermediate}
F_2\bigl(s,m^2;D=2\bigr) = \frac{1}{s \, \sqrt{\Delta}} \log \left[ \frac{\alpha_1^- (\alpha_1^+-1)}{\alpha_1^+ (\alpha_1^--1)} \right] \,.
\end{align}
We may simplify the expression by changing variables to $s$ and $x$ through 
\begin{align} \label{eq:m2x}
m^2 = -s \frac{x}{(1-x)^2} \,,
\end{align}
with $0<x<1$. The discriminant $\Delta$ in fact becomes a perfect square, and $\sqrt{\Delta}$ a rational function. The choice of the branch of the square root is arbitrary. We choose
\begin{align} \label{eq:branch_choice}
\sqrt{\Delta} = \frac{1+x}{1-x} \,,
\end{align}
which is positive for $0<x<1$. Equation~\eqref{eq:Bintermediate} then simplifies to
\begin{align} \label{eq:Bintermediate2}
F_2\bigl(s,m^2;D=2\bigr) = \frac{2}{s} \frac{1-x}{1+x} \log(x) \,.
\end{align}

Equation~\eqref{eq:Bintermediate2} is very simple, but hides a symmetry property. We said above that the choice of the branch of $\sqrt{\Delta}$ is arbitrary. In other words, $F_2$ must be invariant under $\sqrt{\Delta} \to - \sqrt{\Delta}$. Let us work out how $x$ changes under this transformation. Solving \eqn{eq:m2x} for $x$ gives two solutions. We choose the one such that $0<x<1$ for $s<0$ and $m^2>0$, which is compatible with \eqn{eq:branch_choice}:
\begin{align}
x = 1 - \frac{1}{2} \frac{s}{m^2} \left(1-\sqrt{\Delta}\right) \,.
\end{align}
One may then verify that $1/x = x \bigl|_{\sqrt{\Delta} \to - \sqrt{\Delta}} $.
Therefore, when changing the sign of $\sqrt{\Delta}$, both the logarithm in \eqn{eq:Bintermediate2} and its coefficient gain a factor of $-1$, so that $F_2$ is indeed invariant. This property is very common in Feynman integrals involving square roots. A particularly convenient way to make it manifest is to rewrite the argument of the logarithm in the form 
\begin{align}
\log\left( \frac{\sqrt{\Delta}-a}{\sqrt{\Delta}+a} \right) \,,
\end{align}
for some rational function $a$. In the triangle case, dimensional analysis tells us that $a$ must be a constant. Indeed, one may verify with $a=1$ we recover $\log(x)$. 
Our final expression for the two-dimensional bubble integral therefore is
\begin{align}
F_2\bigl(s,m^2; D=2\bigr) =  \frac{2}{s \, \sqrt{\Delta}} \log \left(  \frac{\sqrt{\Delta}-1}{\sqrt{\Delta}+1} \right) \,.
\end{align}


\section*{Exercise \ref{Ex:ExpGamma}: Laurent expansion of the Gamma function}
\label{Sol:ExpGamma}
\begin{enumerate}[a)]

\item Since $\Gamma(z)$ has a simple pole at $z=0$, $\Gamma(z+1) = z \Gamma(z)$ admits a Taylor expansion around $z=0$,
\begin{align} \label{eq:Gammazp1exp}
\Gamma(z+1) = 1 - z \, \euler + \frac{z^2}{2} \Gamma''(1) + \mathcal{O}\bigl(z^3\bigr) \,.
\end{align}
In order to evaluate the second derivative of $\Gamma(z)$, we relate it to the digamma function through \eqn{eq:psidef}. Then we have that
\begin{align} \label{eq:gammapp1}
\Gamma''(1) = \psi'(1) + \euler^2 \,.
\end{align}
Finally, we can evaluate $\psi'(1)$ using the series representation of the digamma function~\eqref{eq:psiseries}, obtaining
\begin{align} \label{eq:psip1}
\psi'(1) = \sum_{k=0}^{\infty} \frac{1}{(k+1)^2} = \zeta_2 \,.
\end{align}
Substituting eqs.~\eqref{eq:psip1} and~\eqref{eq:gammapp1} into \eqn{eq:Gammazp1exp}, and dividing by $z$ both sides of the equation gives the desired Laurent expansion~\eqref{eq:GammaExp0}.

\smallskip

\item In order to exploit the Laurent expansion around $z=0$ computed in the previous part, we apply the recurrence relation~\eqref{eq:gammarecurrence} iteratively until we get
\begin{align} \label{eq:gammaznrec}
\Gamma(z) =  \frac{\Gamma(z+n)}{z(z+1) \ldots (z+n-1) }  \,.
\end{align}
The Laurent expansion of $\Gamma(z+n)$ around $z=-n$ is then obtained from \eqn{eq:GammaExp0} by replacing $z$ with $z+n$.
The remaining factors are regular at $z=-n$, and their Taylor expansion is given in terms of harmonic numbers~\eqref{eq:harmonicnumbers} by
\begin{align} \label{eq:factorsexp}
\prod_{k=0}^{n-1} \frac{1}{z+k}  =  \frac{(-1)^n}{n!} \left[ 1 + (z+n) \, H_n + \frac{(z+n)^2}{2} \left( H_{n,2} + H_n^2 \right) + \mathcal{O}\left((z+n)^3\right) \right] \,.
\end{align}
Substituting \eqn{eq:GammaExp0} with $z \to z+n$ and \eqn{eq:factorsexp} into \eqn{eq:gammaznrec}, and expanding up to order $(z+n)$ gives \eqn{eq:GammaExpn}.

\end{enumerate}


\section*{Exercise \ref{Ex:4.BoxMellinBarnes}: Massless one-loop box with Mellin-Barnes parametrisation}
\label{Sol:4.BoxMellinBarnes}
\begin{figure}[t]
\includegraphics[scale=0.7]{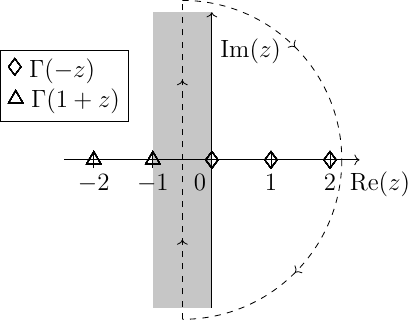}
\centering
\caption{Pole structure of the $\Gamma$ functions in eq.~(\ref{eq:MBboxB1}), and integration contour (dashed line).}
\label{fig:box_MB_ex}
\end{figure}

We begin by rewriting the function $B$ in \eqn{eq:MBboxB} as
\begin{align} \label{eq:B2B1}
B = - \frac{2 \, \eps}{t^2} \, B^{(1)} + \mathcal{O}\bigl(\eps^2\bigr) \,,
\end{align}
where
\begin{align} \label{eq:B1def}
B^{(1)} = \int_{\Re(z)=c'} \frac{\d z}{2\pi \i} \, x^{-z} \, \Gamma(-z)^3 \, \frac{\Gamma(1+z)^3}{1+z} \,.
\end{align}
We applied the recurrence relation~\eqref{eq:gammarecurrence} twice---with $n=-1-z$ and $n=1+z$---to simplify the expression w.r.t.~\eqn{eq:MBboxB}. The pole structure of $B^{(1)}$ is depicted in figure~\ref{fig:box_MB_ex}. From the latter we see that $-1 < c' <0$. E.g., we may set $c'=-1/2$.
In order to carry out the integration we close the contour at infinity. Assuming that $x>1$, we close the contour to the right, as shown in figure~\ref{fig:box_MB_ex}. The contribution from the semi-circle at infinity vanishes, and the integral is given by
\begin{align} \label{eq:MBboxB1}
B^{(1)} = - \sum_{n=0}^{\infty}  \text{Res}\left[ x^{-z} \, \Gamma(-z)^3 \, \frac{\Gamma(1+z)^3}{1+z} , \, z=n \right] \,.
\end{align}
The minus sign comes from the clockwise direction of the loop. To compute the residues, we make use of the Laurent expansions computed in exercise~\ref{Ex:ExpGamma}. The factor of $\Gamma(-z)^3$ entails a triple pole at $z=n$ (with $n=0,1,\ldots$), so that we need the expansion of all functions involved around $z=n$ up to the third order. We obtain the Laurent expansion of $\Gamma(-z)$ around $z=n$ by replacing $z\to -z$ in \eqn{eq:GammaExpn},
\begin{align} \label{eq:ExpGammamz}
\Gamma(-z) = \frac{(-1)^n}{n!} \left\{ \frac{-1}{z-n} + H_n-\euler - \frac{z-n}{2} \left[\left(H_n - \euler\right)^2 + \zeta_2 + H_{n,2} \right] \right\} + \ldots \,.\end{align}
In order to compute the Taylor expansion of $\Gamma(1+z)$ we leverage what we learnt about the digamma function in exercise~\ref{Ex:TaylorLogGamma}.
The first derivative is given by
\begin{align}
\frac{\d}{\d z} \Gamma(1+z) \biggl|_{z=n} = \Gamma(1+n) \, \psi(1+n) = n! \, (H_n - \gamma_{\text{E}}) \,,
\end{align}
where we used the recurrence relation~\eqref{eq:psixn}. We recall the definition of the harmonic numbers in \eqn{eq:harmonicnumbers}. For the second derivative we use again \eqn{eq:psixn} to write $\psi'(1+n) = \psi'(1) - H_{n,2}$, and \eqn{eq:psiseries} to evaluate $\psi'(1) = \zeta_2$. We thus obtain
\begin{align}
\frac{\d^2}{\d z^2} \Gamma(1+z) \biggl|_{z=n} =& n! \, \bigl[ \psi(1+n)^2 + \psi'(1+n) \bigl] \nonumber \\
=& n! \, \bigl[ (H_n-\gamma_{\text{E}})^2 + \zeta_2 - H_{n,2} \bigr] \,.
\end{align}
Putting the above together gives the Taylor expansion of $\Gamma(1+z)$ around $z=n$:
\begin{align} \label{eq:ExpGamma1pz}
\Gamma(1+z) = n! \left\{ 1 + (z-n) \bigl( H_n - \gamma_{\text{E}}\bigr) + \frac{(z-n)^2}{2} \bigl[ (H_n-\gamma_{\text{E}})^2+\zeta_2-H_{n,2} \bigr] 
  + \ldots \right\}\,.
\end{align}
Attentive readers may notice that the expansion of $\Gamma(1+z) \Gamma(-z)$ is much simpler:
\begin{align}
\Gamma(-z) \, \Gamma(1+z) = - (-1)^n \, \biggl[ \frac{1}{z-n} + \zeta_2 \, (z-n) \biggr] + \mathcal{O}\bigl((z-n)^2\bigr) \,.
\end{align}
We could have arrived directly at this result through Euler's reflection formula,
\begin{align}
\Gamma(-z) \, \Gamma(1+z) = -\frac{\pi}{\sin(\pi z)} \,.
\end{align}
The Taylor expansions of the other functions in \eqn{eq:B1def} are straightforward.
Substituting them into \eqn{eq:MBboxB1} and taking the residue gives
\begin{align} \label{eq:MBboxB1b}
B^{(1)}  = \sum_{n=0}^{\infty} \frac{(-x)^{-n}}{(1+n)^3} + \log(x) \sum_{n=0}^{\infty} \frac{(-x)^{-n}}{(1+n)^2}  +
  \frac{1}{2} \bigl(\pi^2+\log^2(x)\bigr) \sum_{n=0}^{\infty} \frac{(-x)^{-n}}{1+n} \,.
\end{align}
The series can be summed in terms of polylogarithms through their definition~\eqref{def_polylogarithm_series}:
\begin{align}
\sum_{n=0}^{\infty}  \frac{(-x)^{-n}}{(1+n)^k} = - x \, \mathrm{Li}_k\left(-\frac{1}{x}\right) \,,
\end{align}
for $k=1,2,\ldots$ and $x>1$. Plugging this into \eqn{eq:MBboxB1b}, and the latter into \eqn{eq:B2B1} gives our final expression for $B$:
\begin{align} \label{eq:MBboxBfinal}
B = \frac{2 \, \eps}{s\, t} \, \biggl[ \mathrm{Li}_3\left(-\frac{1}{x}\right) + \log(x) \, \mathrm{Li}_2\left(-\frac{1}{x}\right) -
  \frac{1}{2}\bigl(\pi^2 + \log^2(x)\bigr) \log\left(1+\frac{1}{x}\right) \biggr] + \mathcal{O}\bigl(\eps^2\bigr) \,.
\end{align}
The expression of the massless one-loop box $F_4$ in \eqn{eq:F4final} is obtained by subtracting from $B$ in \eqn{eq:MBboxBfinal} the Laurent expansion of the residue $A$ in \eqn{eq:MBboxA}.


\section*{Exercise \ref{Ex:Discontinuities}: Discontinuities}
\label{Sol:Discontinuities}
\begin{enumerate}[a)]

\item The logarithm of a complex variable $z = |z| \text{e}^{\i \, \text{arg}(z)}$ is defined as
\begin{align}
\log(z) = \log|z| + \i \, \text{arg}(z) \,,
\end{align}
where $|z|$ is the absolute value of $z$, and the argument of $z$ ($\text{arg}(z)$) is the counterclockwise angle from the positive real axis to the line connecting $z$ with the origin. $\log|z|$ is a continuous function of $z$, hence the discontinuity of $\log(z)$ originates from $\text{arg}(z)$. As $z$ approaches the negative real axis from above (below), $\text{arg}(z)$ approaches $\pi$ ($-\pi$). In other words, we have that
\begin{align}
 \lim_{\eta\to 0^+}\text{arg} (x \pm \i \eta) = \pm \pi \, \Theta(-x)  \,,
\end{align}
for $x \in \mathbb{R}$.
The discontinuity of the logarithm across the real axis is thus given by
\begin{align}
\begin{aligned}
\text{Disc}_x \left[ \log (x) \right] & = \lim_{\eta\to 0^+} \i \left[ \text{arg}(x+\i \eta) -  \text{arg}(x-\i \eta) \right] \\
 & = 2\i \pi \, \Theta(-x) \,.
 \end{aligned}
\end{align}

\item We rewrite the identity~\eqref{eq:Li2identity} here for convenience:
\begin{align} \label{eq:Li2identitySol}
{\rm Li}_2(x) = - {\rm Li}_2(1-x) - \log(1-x) \log(x) + \zeta_2 \,.
\end{align}
This equation is well defined for $0<x<1$. 
Focusing on the RHS, however, we see that all functions are well defined for $x>0$ except for $\log(1-x)$.
We can thus make use of \eqn{eq:Li2identitySol} to reduce the analytic continuation of ${\rm Li}_2(x)$ to $x>1$ 
to that of $\log(1-x)$. For $x>1$ we have that
\begin{align}
{\rm Li}_2(x + \i \eta) - {\rm Li}_2(x - \i \eta) = \log(x) \left[ \log(1-x+\i \eta ) - \log(1-x-\i \eta) \right] + \mathcal{O}(\eta) \,.
\end{align}
Hence, the discontinuity of the dilogarithm follows from that of the logarithm, as
\begin{align}
\text{Disc}_x \left[ {\rm Li}_2 (x) \right] & = \log (x) \, \text{Disc}_x \left[ \log(1-x) \right]  \nonumber \\
& = 2 \pi \i \, \log(x) \, \Theta(x-1) \,.
\end{align}

\end{enumerate}


\section*{Exercise \ref{Ex:4.Symbol}: The symbol of a transcendental function}
\label{Sol:4.Symbol}
We make use of the recursive definition of the symbol~\eqref{defsymbol2}. 
The differential of $\log(x) \log (1-x)$ is given by
\begin{align}
\d \left[ \log(x) \log (1-x) \right] =  \log (1-x) \, \d \log(x) +  \log (x) \, \d \log (1-x)  \,.
\end{align}
Through \eqn{defsymbol2} we then have that 
\begin{align}
\mathcal{S}\bigl( \log(x) \log (1-x) \bigr) = [ S(\log(1-x)) , x ] + [ S(\log(x)), 1-x] \,.
\end{align}
Since we already know that $S(\log a) = [a]$ (\eqn{eq:symbol_log}), the final result is
\begin{align} \label{eq:sloglog1}
\mathcal{S}\bigl( \log(x) \log (1-x) \bigr) = [x, 1-x] + [1-x, x] \,.
\end{align} 
Alternatively, one may replace each function in the product by its symbol,
\begin{align} \label{eq:sloglog2}
\mathcal{S}\bigl( \log(x) \log (1-x) \bigr) = [x] \times [1-x]  \,.
\end{align}
By comparing \eqn{eq:sloglog1} and \eqn{eq:sloglog2} we see that
\begin{align}
[x] \times [1-x] = [x,1-x] + [1-x, x] \,.
\end{align}
This is a very important property of the symbol called \emph{shuffle product}. It allows us to express the product of two symbols of weights $n_1$ and $n_2$ as a linear combination of symbols of weight $n_1+n_2$. For example, at weight three we have
\begin{align}
[a] \times [b,c] = [a,b,c] + [b,a,c] + [b,c,a] \,.
\end{align}
We refer the interested readers to ref.~\cite{ch5-Duhr:2019tlz}.


\section*{Exercise \ref{Ex:4.W2basis}: Symbol basis and weight-two identities}
\label{Sol:4.W2basis}
\noindent
a) The symbol method turns finding relations among special functions into a linear algebra problem. The first step is to put the symbols of \eqn{eq:symbolbasisweight2} and the functions of \eqn{eq:funcbasisW2} in two vectors:
\begin{align}
& \vec{b} = \bigl( [x,x] , [x,1-x], [1-x,x], [1-x,1-x] \bigr)^{\top} \,, \\
& \vec{g} = \bigl( \log^2(x),  \log^2 (1-x),  \log(x) \log (1-x),  {\rm Li}_{2}(1-x) \bigr)^{\top} \,.
\end{align}
The elements of $\vec{b}$ are a basis of all weight-two symbols in the alphabet $\{x, 1-x\}$. We can thus express the symbol of $\vec{g}$ in the basis $\vec{b}$, as
\begin{align}
\mathcal{S}(\vec{g}) = \begin{pmatrix} 
2 \, [x,x] \\
2 \, [1-x,1-x] \\
[x,1-x]+[1-x,x] \\
- [x,1-x] \\
\end{pmatrix} = M \cdot \vec{b} \,, \qquad \text{with} \quad 
M = \begin{pmatrix}
2 & 0 & 0 & 0 \\
0 & 0 & 0 & 2 \\
0 & 1 & 1 & 0 \\
0 & -1 & 0 & 0 \\
\end{pmatrix} \,.
\end{align}
Since the matrix $M$ has non-zero determinant, we can invert it to express the weight-two symbol basis $\vec{b}$ in terms of the symbols of the functions in $\vec{g}$,
\begin{align} \label{eq:b2g}
\vec{b} = M^{-1} \cdot \mathcal{S}(\vec{g}) \,, \qquad \text{with} \quad 
M^{-1} = \begin{pmatrix}
\frac{1}{2} & 0 & 0 & 0 \\
0 & 0 & 0 & -1 \\
0 & 0 & 1 & 1 \\
0 & \frac{1}{2} & 0 & 0 \\
\end{pmatrix} \,.
\end{align}
Therefore, $\vec{g}$ is a basis of the weight-two symbols in the alphabet $\{x,1-x\}$ as well.

\smallskip
\noindent
b) Let us start from ${\rm Li}_{2}\left(x/(x-1)\right)$. Using \eqn{eq:Li2symbolexample}, its symbol is given by
\begin{align}
\mathcal{S}\biggl[{\rm Li}_{2}\left(\frac{x}{x-1}\right)\biggr] = - \left[1-\frac{x}{x-1}, \frac{x}{x-1} \right] \,.
\end{align}
The properties~\eqref{eq:symbol_properties} allow us to express the latter in terms of the letters $\{x,1-x\}$:
\begin{align}
\mathcal{S}\biggl[{\rm Li}_{2}\left(\frac{x}{x-1}\right)\biggr] = [1-x, x ] - [1-x, 1-x] \,.
\end{align}
We can then express the symbol of ${\rm Li}_{2}(x/(x-1))$ in the basis $\vec{b}$, as
\begin{align}
\mathcal{S}\biggl[{\rm Li}_{2}\left(\frac{x}{x-1}\right)\biggr] = (0, 0, 1, -1) \cdot \vec{b} \,.
\end{align}
In this sense, ${\rm Li}_{2}(x/(x-1))$ `lives' in the space spanned by \eqn{eq:symbolbasisweight2}. 
We do the same for the other dilogarithms in \eqn{eq:dilogs}:
\begin{align}
& \mathcal{S}\bigl[ {\rm Li}_{2}(x) \bigr] = - [1-x,x] = (0, 0, -1, 0) \cdot \vec{b} \,, \\ 
& \mathcal{S}\biggl[ {\rm Li}_{2}\left(\frac{1}{x}\right)\biggr]  = [1 - x, x] - [x, x] = (-1, 0, 1, 0) \cdot \vec{b} \,, \\
& \mathcal{S}\biggl[  {\rm Li}_{2}\left(\frac{1}{1-x}\right)\biggr]  = -[1 - x, 1 - x] + [x, 1 - x] = (0, 1, 0, -1) \cdot \vec{b} \,, \\ 
& \mathcal{S}\biggl[  {\rm Li}_{2}\left(\frac{x-1}{x} \right)\biggr]  = [x, 1 - x] - [x, x] = (-1, 1, 0, 0) \cdot \vec{b} \,. 
\end{align}

\smallskip
\noindent
c) Having expanded the symbols of the dilogarithms in \eqn{eq:dilogs} in the basis $\vec{b}$, 
we can change basis from $\vec{b}$ to $\mathcal{S}(\vec{g})$ as in \eqn{eq:b2g}. For example, we have
\begin{align} \begin{aligned}
\mathcal{S}\biggl[{\rm Li}_{2}\left(\frac{x}{x-1}\right)\biggr] & = (0, 0, 1, -1) \cdot M^{-1} \cdot S(\vec{g}) \\
& = \mathcal{S}\biggl[-\frac{1}{2} \log^2 (1-x) + \log(x) \log (1-x) + {\rm Li}_{2}(1-x) \biggl] \,.
\end{aligned} \end{align}
Doing the same for the other dilogarithms in \eqn{eq:dilogs} gives the following identities:
\begin{align}
& \mathcal{S}\biggl[  {\rm Li}_{2}(x) + {\rm Li}_{2}(1 - x) + \log(x) \log(1 - x)  \biggr] = 0 \,, \\
& \mathcal{S}\biggl[ {\rm Li}_{2}\left(\frac{1}{x}\right) - {\rm Li}_{2}(1 - x)  - \log(x) \log(1 - x) + \frac{1}{2} \log^2(x) \biggr] = 0 \,, \\
& \mathcal{S}\biggl[ {\rm Li}_{2}\left(\frac{1}{1-x}\right) + {\rm Li}_2(1 - x)  + \frac{1}{2} \log^2(1 - x) \biggr] = 0 \,, \\
& \mathcal{S}\biggl[  {\rm Li}_{2}\left(\frac{x-1}{x} \right) + {\rm Li}_2(1 - x) + \frac{1}{2} \log^2(x) \biggr] = 0 \,.
\end{align}
The terms which are missed by the symbol may be fixed as done for \eqn{defgexamplesymbol}.


\section*{Exercise \ref{Ex:4.Symbolf2}: Simplifying functions using the symbol}
\label{Sol:4.Symbolf2}
We compute the symbol of $f_1(u,v)$ in \eqn{examplef1}. The general strategy is the following:
we use the rule in \eqn{eq:Li2symbolexample} for the dilogarithms; next, we put all letters over a common denominator and factor them; finally, we use the symbol properties~\eqref{eq:symbol_properties} to expand the symbol until all letters are irreducible factors.
This gives
\begin{align}
& \mathcal{S}\biggl[ {\rm Li}_2\left(\frac{1-v}{u}\right) \biggr] = [u+v-1,u] - [u+v-1, 1-v] + [u,1-v] - [u,u] \,, \\
& \begin{aligned}
\mathcal{S}\biggl[ {\rm Li}_2\left(\frac{(1-u)(1-v)}{u v}\right) \biggr] = \ & \bigl( [u+v-1,u]-[u+v-1,1-u]+[u,1-u] \\
&  +[u,1-v]-[u,u]-[u,v] \bigr) + \bigl(u \leftrightarrow v \bigr) \,.
\end{aligned} \end{align}
${\rm Li}_2((1-u)/v)$ is obtained from ${\rm Li}_2((1-v)/u)$ by exchanging $u \leftrightarrow v$, and so is its symbol. The symbol of $\pi^2/6$ vanishes.
Putting the above together gives
\begin{align} \label{eq:Sf1}
\mathcal{S}\bigl[ f_1(u,v) \bigr] = [u, 1 - u] + [v, 1 - v] - [u, v] - [v, u] \,,
\end{align}
as claimed. Note that the letter $u+v-1$ drops out in the sum. In other words, $u+v=1$ is a branch point for the separate terms in the expression of $f_1(u,v)$ given in \eqn{examplef1}, but not for $f_1(u,v)$.

On the RHS of \eqn{eq:Sf1} we recognise in $ [u, 1 - u] $ ($[v, 1 - v]$) the symbol of $-{\rm Li}_2(u)$ ($-{\rm Li}_2(v)$), 
and in $[u, v] + [v, u]$ the symbol of $\log u \log v$ (see exercise~\ref{Ex:4.Symbol}). 
An alternative and simpler expression for $f_1(u,v)$ is thus given by
\begin{align}
\mathcal{S}\bigl[f_1(u,v)\bigr] =\mathcal{S}\bigl[  -{\rm Li}_2(u) -{\rm Li}_2(v) - \log u \log v \bigr] \,,
\end{align}
which matches---at symbol level---the expression of $f_2(u,v)$ in \eqn{examplef1b}.


\section*{Exercise \ref{Ex:4.Kite}: The massless two-loop kite integral}
\label{Sol:4.Kite}
We define the integral family as
\begin{align} \label{eq:kitefamily}
G^{\text{kite}}_{a_1,a_2,a_3,a_4,a_5} \coloneqq \int \frac{\d^D k_1}{\i \pi^{D/2}} \frac{\d^D k_2}{\i \pi^{D/2}}
   \frac{1}{D_1^{a_1} D_2^{a_2} D_3^{a_3} D_4^{a_4} D_5^{a_5}} \,,
\end{align}
where the inverse propagators $D_i$ are given by\footnote{In chapter~\ref{ch:loopints} we write the inverse propagators as $D_a = -(k-q_a)^2 + m_a^2 - \i 0 $, which is natural for the loop integration (see section~\ref{sec:ConventionsCh4}), as opposed to $D_a = +(k-q_a)^2 - m_a^2 + \i 0$. We used the latter in chapter~\ref{ch:loopamps} as it is more convenient for the unitarity methods.}
\begin{align} \label{eq:D2sp}
\begin{array}{lll}
D_1 = - k_1^2 \,,  & D_2 = -(k_1-p)^2 \,, \quad \quad & D_3 = -k_2^2 \,, \\
D_4 = -(k_2+p)^2\,, \quad \quad & D_5 = -(k_1+k_2)^2\,. & \\
\end{array}
\end{align}
Feynman's $\i 0$ prescription is dropped, as it plays no part here. The desired integral is
\begin{align}
F_{\text{kite}}\bigl(s; D\bigr) = G^{\text{kite}}_{1,1,1,1,1} \,.
\end{align}
The IBP relations for the triangle sub-integral with loop momentum $k_2$ are given by
\begin{align}
\int \frac{\d^D k_1}{\i \pi^{D/2}} \frac{\d^D k_2}{\i \pi^{D/2}} \frac{\partial}{\partial k_2^{\mu}}  
   \frac{ q^{\mu} }{D_1 D_2 D_3 D_4 D_5}  = 0 \,,
\end{align}
for any momentum $q$. Upon differentiating we rewrite the scalar products in terms of inverse propagators---and thus of integrals of the family~\eqref{eq:kitefamily}---by inverting the system of equations~\eqref{eq:D2sp} (e.g.\ $k_1 \cdot p =( D_2 - D_1 + s)/2$). 
There are three independent choices for $q$: $k_1$, $k_2$, and $p$. We need to find a linear combination of these such that the resulting IBP relation contains only $F_{\text{kite}}$ and bubble-type integrals. Using $q=k_1+k_2$ gives
\begin{align}
(D-4) \, G^{\text{kite}}_{1, 1, 1, 1, 1} & - G^{\text{kite}}_{1, 1, 1, 2, 0} - 
 G^{\text{kite}}_{1, 1, 2, 1, 0} + G^{\text{kite}}_{0, 1, 2, 1, 1} + G^{\text{kite}}_{1, 0, 1, 2, 1}  = 0 \,.
 \end{align}
The graph symmetries imply that 
\begin{align}
G^{\text{kite}}_{1, 1, 1, 2, 0} = G^{\text{kite}}_{1, 1, 2, 1, 0} \,, \qquad \quad  G^{\text{kite}}_{1, 0, 1, 2, 1} = G^{\text{kite}}_{0, 1, 2, 1, 1} \,.
\end{align}
$G^{\text{kite}}_{1, 1, 2, 1, 0}$ is a product of bubble integrals. Using \eqn{eq:Ibubblechapter4} for the latter gives 
\begin{align}
G^{\text{kite}}_{1, 1, 2, 1, 0} = B(1,1) \, B(1,2) \, (-s)^{D-5} \,.
\end{align}
$G^{\text{kite}}_{0, 1, 2, 1, 1}$ instead has a bubble sub-integral. By using \eqn{eq:Ibubblechapter4} iteratively we obtain
\begin{align}
\begin{aligned}
G^{\text{kite}}_{0, 1, 2, 1, 1} & = \ \vcenter{\hbox{\includegraphics[width=2.5cm]{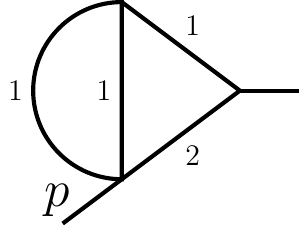}}}  \\
& = B(1,1) \times \vcenter{\hbox{\includegraphics[width=2.cm]{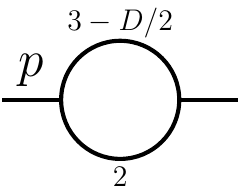}}}  \\
& = B(1,1) \, B\left(3-\frac{D}{2},2\right) \, (-s)^{D-5} \,,
\end{aligned}
\end{align}
where the numbers next to the propagators in the graphs are their exponents. 
Putting the above together gives
\begin{align}
G^{\text{kite}}_{1, 1, 1, 1, 1} = \frac{2}{D-4} \, B(1,1) \left[ B(1,2) -B\left(3-\frac{D}{2}, 2\right) \right] (-s)^{D-5} \,,
\end{align}
which can be expressed in terms of gamma functions through \eqn{eq:Bbubble}. Setting $D=4-2\eps$ and expanding the $\Gamma$ functions around $\eps=0$ through \eqn{eq:logGammaTaylor} finally gives \eqn{eq:kite_result}.


\section*{Exercise \ref{Ex:4.BubbleCut}: ``$\d \log$'' form of the massive bubble integrand with $D=2$}
\label{Sol:4.BubbleCut}
We start from the integrand $\omega_{1,1}$ in the first line of \eqn{eq:bubble_dlog}. We use the parameterisation $k = \beta_1 p_1 + \beta_2 p_2$, with $p=p_1+p_2$, $p_1^2=p_2^2=0$ and $2 \, p_1 \cdot p_2 = s$.
We change integration variables from $k$ to $\beta_1$ and $\beta_2$.
The propagators are given~by
\begin{align} \label{eq:prop_beta} \begin{aligned}
& -k^2 + m^2 = -s \, \bigl( \beta_1 \beta_2 - x \bigr) \,, \\
& -(k+p)^2 + m^2 = -s \, \bigl[ (1+\beta_1)(1+\beta_2) - x \bigr] \,,
\end{aligned} \end{align}
where we introduced the short-hand notation $x=m^2/s$.
The Jacobian factor is a function of $s$, and dimensional analysis tells us that $J\propto s$. The constant prefactor is irrelevant here.\footnote{For an example of how to compute the Jacobian factor, see the solution to exercise~\ref{Ex:4.KiteLS}.}
The integrand then reads
\begin{align} \label{eq:ometa11beta}
\omega_{1,1} \propto \frac{1}{s} \, \frac{\d \beta_1 \d \beta_2}{(\beta_1 \beta_2 - x) \bigl[ (1+\beta_1)(1+\beta_2)-x\bigr]} \,.
\end{align}
Our goal is to rewrite $\omega_{1,1}$ in a ``$\d \log$'' form. First, we decompose $\omega_{1,1}$ into partial fractions w.r.t.~$\beta_2$:
\begin{align}
s \, \omega_{1,1} \propto \frac{\d \beta_1}{\beta_1^2 + \beta_1 + x} \, \frac{\d \beta_2}{\beta_2 - x/\beta_1} - 
  \frac{\d \beta_1}{\beta_1^2 + \beta_1 + x} \, \frac{\d \beta_2}{\beta_2 + 1 - x/(1+\beta_1)} \,.
\end{align}
We then push all $\beta_2$-dependent factors into $\d \log$ factors, obtaining
\begin{align} \begin{aligned}
s \, \omega_{1,1} & \propto \frac{\d \beta_1}{\beta_1^2 + \beta_1 + x} \, \d \log\left(\beta_2 - \frac{x}{\beta_1} \right) - 
  \frac{\d \beta_1}{\beta_1^2 + \beta_1 + x} \, \d \log\left(1+\beta_2 - \frac{x}{1+\beta_1} \right)  \\
 & \propto  \frac{\d \beta_1}{\beta_1^2 + \beta_1 + x} \, \d \log \left( \frac{\beta_1 \beta_2 - x}{(1 + \beta_1)(1 + \beta_2) - x} \right) \,.
\end{aligned} \end{align}
In order to do the same w.r.t.~$\beta_1$, we need to factor the polynomial in the denominator:
\begin{align} \label{eq:beta1pm}
\beta_1^2 + \beta_1 + x = \bigl(\beta_1 - \beta_1^+ \bigr) \bigl(\beta_1 - \beta_1^- \bigr) \,, \qquad \beta_1^{\pm} = \frac{1}{2} \Bigl(-1 \pm \sqrt{1-4 x} \Bigr) \,.
\end{align}
Partial fractioning w.r.t.~$\beta_1$ and rewriting all $\beta_1$-dependent factors into $\d \log$'s yields
\begin{align} \label{eq:omega11dlogbeta}
\omega_{1,1} \propto \frac{1}{s \, \sqrt{1-4 x}} \, \d\log\left( \frac{\beta_1 - \beta_1^+}{\beta_1 - \beta_1^-} \right) \, \d \log \left( \frac{\beta_1 \beta_2 - x}{(1 + \beta_1)(1 + \beta_2) - x} \right) \,.
\end{align}
Note that there is a lot of freedom in the expression of the $\d \log$ form. For instance, we might have started with a partial fraction decomposition w.r.t.~$\beta_1$, and we would have obtained a different---yet equivalent---expression. While the expression of the $\d \log$ form may vary, all singularities are always manifestly of the type $\d x/x$, and the leading singularities stay the same (up to the irrelevant sign). 

\smallskip

We now consider the momentum-space $\d \log$ form in \eqn{eq:bubble_dlog}, namely
\begin{align} \label{eq:omega11mom}
\omega_{1,1} \propto \frac{1}{s \, \sqrt{1-4x }} \, \d \log(\tau_1) \, \d \log(\tau_2) \,,
\end{align}
where
\begin{align} \label{eq:taus}
\tau_1 = \frac{-k^2+m^2}{(k-k_{\pm})^2} \,, \qquad \qquad \tau_2 = \frac{-(k+p)^2+m^2}{(k-k_{\pm})^2} \,.
\end{align}
We rewrite the latter in terms of $\beta_1$ and $\beta_2$, and show that it matches \eqn{eq:ometa11beta}.
We recall that $\d = \d\beta_1 \, \partial_{\beta_1} + \d\beta_2 \, \partial_{\beta_2}$ and $\d \beta_2 \, \d \beta_1 = - \d \beta_1 \, \d \beta_2$, which imply that
\begin{align} \label{eq:dk2dbeta}
\d \log(\tau_1) \, \d \log(\tau_2) = \left( \frac{\partial \tau_1}{\partial \beta_1} \frac{\partial \tau_2}{\partial \beta_2} - 
   \frac{\partial \tau_1}{\partial \beta_2} \frac{\partial \tau_2}{\partial \beta_1} \right) \frac{\d \beta_1 \, \d \beta_2}{\tau_1 \, \tau_2} \,.
\end{align}
In \eqn{eq:taus}, $k_{\pm}$ denotes either of the two solutions to the cut equations.
We choose $k_+ = \beta_1^+ \, p_1 + \beta_2^+ \, p_2$, where $\beta_{1}^+$ is given by \eqn{eq:beta1pm} and $\beta_2^+=x/\beta_1^+$. We then have
\begin{align} \label{eq:kmkp}
(k-k_+)^2  = \frac{s}{2} \, \Bigl[2x+\beta_1 + \beta_2 + 2 \beta_1 \beta_2 + (\beta_1 - \beta_2) \sqrt{1 - 4 x} \Bigr]\,.
\end{align}
We substitute eqs.~\eqref{eq:kmkp} and~\eqref{eq:prop_beta} into \eqn{eq:taus}, and the latter into \eqn{eq:dk2dbeta}. Simplifying the result---possibly with a computer-algebra system---gives 
\begin{align}
\frac{1}{s \, \sqrt{1-4x }} \, \d \log(\tau_1) \, \d \log(\tau_2) = \frac{1}{s} \, \frac{\d\beta_1 \, \d \beta_2}{(\beta_1 \beta_2 - x) \bigl[(1 + \beta_1)(1 + \beta_2) - x \bigr] } \,,
\end{align}
which matches the expression of $\omega_{1,1}$ in \eqn{eq:ometa11beta}, as claimed.


\section*{Exercise \ref{Ex:4.KiteLS}: An integrand with double poles: the two-loop kite in $D=4$}
\label{Sol:4.KiteLS}
We complement $p_1$ and $p_2$ with two momenta constructed from their spinors,
\begin{align}
p_3^{\mu} = \frac{1}{2} \spAB{1}{\gamma^{\mu}}{2} \,, \quad p_4^{\mu} = \frac{1}{2} \spAB{2}{\gamma^{\mu}}{1} \,,
\end{align}
to construct a four-dimensional basis. We have that $p_1 \cdot p_2  = - p_3 \cdot p_4 = s/2$,
while all other scalar products $p_i \cdot p_j$ vanish. We expand the loop momenta as
\begin{align}
k_1^{\mu} = \sum_{i=1}^4 a_i \, p_i^{\mu} \,, \qquad k_2^{\mu} = \sum_{i=1}^4 b_i \, p_i^{\mu} \,,
\end{align}
and change integration variables from $k_1^{\mu}$ and $k_2^{\mu}$ to $a_i$ and $b_i$. 
The inverse propagators defined in \eqn{eq:D2sp} are given by
\begin{align}\label{eq:D2ab}
\begin{aligned}
& \begin{array}{ll}
D_1 = \left(a_3 a_4 - a_1 a_2\right) s \,, \qquad \qquad & D_2 = \left(a_3 a_4 - a_1 a_2 + a_1 + a_2  -1 \right) s \,, \\
D_3 = \left(b_3 b_4 - b_1 b_2\right) s \,, \qquad \qquad & D_4 = \left(b_3 b_4 - b_1 b_2 - b_1 - b_2  -1 \right) s \,, \\
\end{array} \\
& \, D_5 = \left(a_3 a_4 - a_1 a_2 + b_3 b_4  - b_1 b_2 + a_3 b_4 + a_4 b_3 - a_1 b_2 - a_2 b_1 \right) s   \,.
\end{aligned}
\end{align}
The Jacobian $|J_1|$ of the change of variables $\{k_1^{\mu}\} \to \{a_i\}$ is the determinant of the $4 \times 4$ matrix with entries
\begin{align}
\left(J_1\right)^{\mu}_i = \frac{\partial k_1^{\mu}}{\partial a_i} \,, \quad \mu=0,\ldots,3 \,, \quad i=1,\ldots, 4\,.
\end{align}
Dimensional analysis tells us that $|J_1| \propto s^2$. It is instructive to compute it explicitly as well.
To do so, it is convenient to first consider
\begin{align}
\left(J_1\right)^{\mu}_i \eta_{\mu \nu} \left(J_1\right)^{\nu}_j = p_i \cdot p_j \,.
\end{align}
Taking the determinant on both sides gives
\begin{align}
|J_1|^2 = - \frac{s^4}{16}\,,
\end{align}
where the minus sign comes from the determinant of the metric tensor. The Jacobian for  $\{k_2^{\mu}\} \to \{b_i\}$ is similar.
The maximal cut is then given by
\begin{align}
F_{\text{kite}}^{\text{max cut}} \propto s^4 \int \d a_1 \d a_2 \d a_3 \d a_4 \d b_1 \d b_2 \d b_3 \d b_4 \prod_{i=1}^5 \delta\left(D_i \right) \,,
\end{align}
where the inverse propagators are expressed in terms of $a_i$ and $b_i$ through \eqn{eq:D2ab}, and the overall constant is neglected. We use the delta functions of $D_1$ and $D_3$ to fix $a_1=a_3 a_4/a_2$ and $b_1 = b_3 b_4/b_2$,
\begin{align} \begin{aligned}
& F_{\text{kite}}^{\text{max cut}} \propto \frac{1}{s} \int \frac{\d a_2 \d a_3 \d a_4 \d b_2 \d b_3 \d b_4}{a_2 b_2} \\
  & \phantom{I_{\text{kite}}} \delta\left(1-a_2 - \frac{a_3 a_4}{a_2} \right)  \delta\left(1+b_2 + \frac{b_3 b_4}{b_2} \right) 
   \delta\left( \frac{(a_2 b_3 - b_2 a_3) (a_2 b_4 - b_2 a_4)}{a_2 b_2}\right) \,.
\end{aligned}
\end{align}
We then use the first two delta functions to fix $a_3 = a_2(1-a_2)/a_4$ and $b_3 =-b_2(1+b_2)/b_4$, and the remaining one to fix $b_2 = a_2 b_4/a_4$. We obtain
\begin{align}
F_{\text{kite}}^{\text{max cut}} \propto \frac{1}{s} \int \frac{\d a_2 \d a_4 \d b_4}{a_4 (a_4+b_4)} \,.
\end{align}
New simple poles have appeared, and the integrand has a double pole at $a_2 \to \infty$. We can make this manifest by the change of variable $a_2 \to 1/\tilde{a}_2$, which maps the hidden double pole at $a_2 \to \infty$ into a manifest double pole at $\tilde{a}_2 = 0$,
\begin{align}
F_{\text{kite}}^{\text{max cut}} \propto \frac{1}{s} \int \frac{\d \tilde{a}_2 \d a_4 \d b_4}{\tilde{a}_2^2 a_4 (a_4+b_4)} \,.
\end{align}


\section*{Exercise \ref{Ex:4.BoxDE}: The box integrals with the differential equations method}
\label{Sol:4.BoxDE}
\begin{figure}[t]
\begin{minipage}{0.27\linewidth}
\centering
\includegraphics[width=\textwidth]{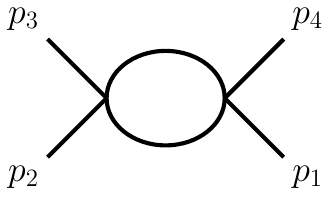}
$I^{\text{box}}_{0,1,0,1}$
\end{minipage}%
\hfill
\begin{minipage}{0.27\linewidth}
\centering
\includegraphics[width=\textwidth]{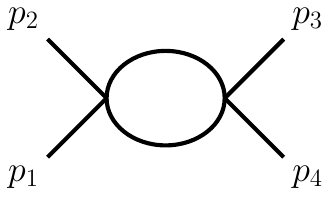}
$I^{\text{box}}_{1,0,1,0}$
\end{minipage}%
\hfill
\begin{minipage}{0.27\linewidth}
\centering
\includegraphics[width=\textwidth]{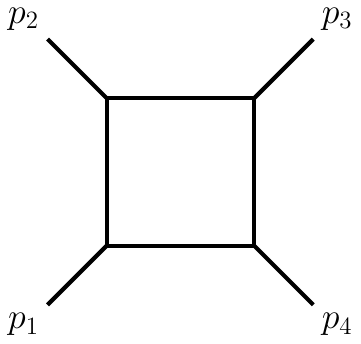}
$I^{\text{box}}_{1,1,1,1}$
\end{minipage}
\caption{Master integrals of the box integral family in eq.~(\ref{eq:gdef}).}
\label{fig:box_masters}
\end{figure}

\noindent
a) We define and analyse the box integral family using \textsc{LiteRed}~\cite{ch5-Lee:2012cn}. There are $3$ master integrals. \textsc{LiteRed}'s algorithm selects them as the $t$- and $s$-channel bubbles, and the box (see figure~\ref{fig:box_masters}). 
We denote them by $\vec{g}$,
\begin{align} \label{eq:gdef}
\vec{g}(s,t;\eps) = \begin{pmatrix} 
I^{\text{box}}_{0,1,0,1} \\ 
I^{\text{box}}_{1,0,1,0} \\
I^{\text{box}}_{1,1,1,1}
\end{pmatrix} \,.
\end{align}

\smallskip
\noindent
b) We differentiate $\vec{g}$ w.r.t.\ $s$ and $t$, and IBP-reduce the result. We obtain
\begin{align}
\begin{cases}
\partial_s \vec{g}(s,t;\eps)  = A_s(s,t;\eps)  \cdot   \vec{g}(s,t;\eps) \,, \\
\partial_t \vec{g}(s,t;\eps)  = A_t(s,t;\eps)  \cdot   \vec{g}(s,t;\eps) \,, \\
\end{cases}
\end{align}
where
\begin{align}
A_s = \begin{pmatrix}
0 & 0 & 0 \\
0 & -\frac{\eps}{s} & 0 \\ 
\frac{2 (2 \eps-1)}{s t (s + t)} &
\frac{2 (1 - 2 \eps)}{s^2 (s + t)} & 
  - \frac{s + t + \eps t}{s (s + t)} \\
\end{pmatrix} \,, \qquad
A_t = \begin{pmatrix}
-\frac{\eps}{t} & 0 & 0 \\
0 & 0 & 0 \\ 
\frac{2 (1 - 2 \eps)}{t^2 (s + t)} & 
 \frac{2 (2 \eps -1 )}{s t (s + t)} & 
 - \frac{s + t + \eps s}{t (s + t)} \\
\end{pmatrix} \,.
\end{align}
We verify the integrability conditions,
\begin{align}
A_s \cdot A_t - A_t \cdot A_s+ \partial_t A_s - \partial_s A_t = 0 \,,
\end{align}
and the scaling relation,
\begin{align}
s \, A_s + t \, A_t = \text{diag}( - \eps, - \eps , -2 -\eps ) \,.
\end{align}
The diagonal entries on the RHS of the scaling relation match the power counting of the integrals in $\vec{g}$ in units of $s$.

\smallskip
\noindent
c) We express $\vec{f}$ in terms of $\vec{g}$ as $\vec{f} = T^{-1} \cdot \vec{g}$ by IBP-reducing the integrals in \eqn{eq:fdef}.\footnote{We use the inverse of $T$ to match the convention of section~\ref{subsec:comments} for the gauge transformation.} We obtain 
\begin{align}
T^{-1} = c(\eps) \, \begin{pmatrix}
  0 & 0 & s t \\
  0 & \frac{2\eps-1}{\eps} & 0 \\
  \frac{2\eps-1}{\eps} & 0 & 0 \\
\end{pmatrix} \,.
\end{align}
The new basis $\vec{f}$ satisfies a system of DEs in canonical form,
\begin{align} \label{eq:box_canDE}
\begin{cases}
\partial_s \vec{f}(s,t;\eps)  = \eps \, B_s(s,t)  \cdot   \vec{f}(s,t;\eps) \,, \\
\partial_t \vec{f}(s,t;\eps)  = \eps \, B_t(s,t)  \cdot   \vec{f}(s,t;\eps) \,, \\
\end{cases}
\end{align}
where
\begin{align} \label{eq:boxBmatrices}
B_s = \begin{pmatrix}
\frac{1}{s + t}-\frac{1}{s} & \frac{2}{s + t}-\frac{2}{s} & \frac{2}{s + t} \\
0 & - \frac{1}{s} & 0 \\
0 & 0 & 0 \\
\end{pmatrix} \,, \qquad
B_t = \begin{pmatrix}
\frac{1}{s + t}-\frac{1}{t} & \frac{2}{s + t} & \frac{2}{s + t}-\frac{2}{t} \\
0 & 0 & 0 \\
0 & 0 & -\frac{1}{t} \\
\end{pmatrix} \,.
\end{align}
We compute $B_s$ and $B_t$ as in point b, or through the gauge transformation
\begin{align}
\eps \, B_s = T^{-1} \cdot A_s \cdot T -  T^{-1} \cdot  \partial_s T \,,
\end{align}
and similarly for $t$. From \eqn{eq:boxBmatrices} we see that the symbol alphabet of this family is $\{s, t, s+t\}$.
Thanks to the factorisation of $\eps$ on the RHS of the canonical DEs~\eqref{eq:box_canDE}, the integrability conditions split into
\begin{align}
B_s \cdot B_t - B_t \cdot B_s = 0 \,, \qquad \qquad \partial_s B_t - \partial_t B_s = 0 \,.
\end{align}
The scaling relation is given by
\begin{align}
s \, B_s + t \, B_t = -  \1_{3} \,.
\end{align}

\smallskip
\noindent
d) Viewed as a function of $s$ and $x=t/s$, $\vec{f}$ satisfies the canonical DEs
\begin{align}
\begin{cases}
\partial_s \vec{f}(s,x;\eps)  = \eps \, C_s(s,x)  \cdot   \vec{f}(s,x;\eps) \,, \\
\partial_x \vec{f}(s,x;\eps)  = \eps \, C_x(s,x)  \cdot   \vec{f}(s,x;\eps) \,, \\
\end{cases}
\end{align}
where $C_s$ and $C_x$ are related to $B_s$ and $B_t$ in \eqref{eq:boxBmatrices} through the chain rule,
\begin{align}
C_x  = s \, B_t \Bigl|_{t=x s} \,, \qquad \qquad C_s = B_s + \frac{t}{s} \, B_t \Bigl|_{t=x s} \,.
\end{align}
We observe that $C_x$  is a function of $x$ only, and $C_s$ of $s$. Thanks to this separation of variables, we can straightforwardly rewrite the canonical DEs in differential form,
\begin{align} \label{eq:box_canDEs_dlog}
\d \, \vec{f}(s,x;\eps) = \eps \, \bigl[ \d \, \tilde{C}(s,x) \bigr] \cdot \vec{f}(s,x;\eps) \,.
\end{align}
The connection matrix $\tilde{C}$ is given by a linear combination of logarithms of the alphabet letters $\alpha_1 = s$, $\alpha_2 = x$ and $\alpha_3 = 1+x$, 
\begin{align}
\tilde{C}(s,x) = \sum_{k=1}^3 C_k \, \log \alpha_k(s,x) \,,
\end{align}
with constant matrix coefficients,
\begin{align}
C_1 = - \1_{3} \,, \qquad 
C_2 = \begin{pmatrix}
  -1 & 0 & -2 \\
  0 & 0 & 0 \\
  0 & 0 & -1 \\
\end{pmatrix} \,, \qquad
C_3 = \begin{pmatrix}
  1 & 2 & 2 \\
  0 & 0 & 0 \\
  0 & 0 & 0 \\
\end{pmatrix}\,.
\end{align}

\smallskip
\noindent
e) We expand the pure integrals $\vec{f}$ as a Taylor series around $\eps=0$,\footnote{The factor $c(\eps)$ in the definition of the pure integrals~\eqref{eq:fdef} is chosen such that they are finite at $\eps=0$.}
\begin{align} \label{eq:f_eps_exp}
\vec{f}(s,x;\eps) = \sum_{w\ge 0} \eps^w \, \vec{f}^{(w)}(s,x) \,,
\end{align}
and define the weight-$w$ boundary values at $s=-1$ and $x=1$ as $\vec{b}^{(w)} = \vec{f}^{(w)}(-1,1)$.
Using \eqn{eq:Ibubblechapter4} for the bubble-type integrals we obtain
\begin{align} \label{eq:f23analytic} \begin{aligned}
& f_2(s,x;\eps) = -1+ \eps \, \log(-s) - \frac{\eps^2}{12} \left(6 \log^2(-s)-\pi^2\right) \\
& \phantom{f_2(s,x;\eps) = \, } + \frac{\eps^3}{12} \left(2 \, \log^3(-s)-\pi^2 \log(-s)+28 \, \zeta_3 \right) + \mathcal{O}\bigl(\eps^4\bigr) \,. \\
\end{aligned} \end{align}
The expression for $f_3(s,x;\eps)$ is obtained by trading $s$ for $t = s x$ in $f_2(s,x;\eps)$.
We leave $b_1^{(w)}$ as free parameters. The weight-$0$ boundary values are thus given by $ \vec{b}^{(0)} = (b_1^{(0)}, -1, -1)^{\top}$. We can now solve the canonical DEs in terms of symbols. In order to do so, we note that the canonical DEs~\eqref{eq:box_canDEs_dlog} imply the following DEs for the coefficients of the $\eps$ expansion,
\begin{align} \label{eq:box_canDEs_iteration}
\d \, f_i^{(w)}(s,x) =  \sum_{k=1}^3 \left[ \sum_{j=1}^3 (C_{k})_{ij} \, f_j^{(w-1)}(s,x) \right] \, \d \log\alpha_k(s,x) \,.
\end{align}
The iteration starts from $\vec{f}^{(0)} = \vec{b}^{(0)}$. We spelled out all indices in \eqn{eq:box_canDEs_iteration} to facilitate the comparison against the recursive definition of the symbol given by eqs.~\eqref{defsymbol1} and~\eqref{defsymbol2}. From this, we find the following recursive formula for the symbol of the solution to the canonical DEs:
\begin{align}
\mathcal{S} \Bigl( f_i^{(w)} \Bigr) = \sum_{k=1}^3 \sum_{j=1}^3 (C_{k})_{ij} \, \biggl[ \mathcal{S} \Bigl( f_j^{(w-1)} \Bigr) , \, \alpha_k \biggr] \,,
\end{align}
starting at weight $w=0$ with $\mathcal{S} \bigl( f_i^{(0)} \bigr) = b^{(0)}_i []$ (we recall that $[]$ denotes the empty symbol).
With a slight abuse of notation, we may rewrite this in a more compact form as
\begin{align}
\mathcal{S}\bigl( \vec{f}^{(w)} \bigr) = \sum_{k=1}^3 C_{k} \cdot \left[ \mathcal{S}\bigl(  \vec{f}^{(w-1)} \bigr) , \, \alpha_k \right] \,.
\end{align}
At transcendental weight 1 we thus have that
\begin{align}
\mathcal{S}\bigl(  \vec{f}^{(1)} \bigr)  = C_1 \cdot \vec{b}^{(0)} \, [s] + C_2 \cdot \vec{b}^{(0)} \, [x] +  C_3 \cdot \vec{b}^{(0)} \, [1 + x]  \,.
\end{align}
Since $[1+x]$ is the symbol of $\log(1+x)$, $\vec{f}^{(1)}(x)$ would diverge at $x=-1$ unless the coefficients of $[1+x]$ vanish. The finiteness at $x=-1$ thus implies
\begin{align}
C_3 \cdot \vec{b}^{(0)} = 0 \,,
\end{align}
which fixes $b_1^{(0)} = 4$. We now have everything we need to write down the symbol of the solution up to any order in $\eps$. For $f_1$, for instance, we obtain 
\begin{align} \begin{aligned}
& \mathcal{S}\left( f_1 \right) = 4 \, [] - 2 \, \eps \bigl( 2\, [s] + [x] \bigr)
  + 2 \, \eps^2 \bigl( 2 \, [s,s] + [s,x] + [x,s] \bigr) \\
& \, - 2 \eps^3 \Bigl( 2 \, [s,s,s] + [s,s,x] + [s,x,s] + [x,s,s] - [x,x,x]+[x,x,1+x] \Bigr) + \mathcal{O}\bigl(\eps^4\bigr) \,.
\end{aligned} \end{align}

\smallskip
\noindent
f) The dependence on $s$ is given by an overall factor of $(-s)^{-\eps}$,\footnote{The minus sign in front of $s$ ensures the positivity in the Euclidean region, where $s<0$.} which is fixed by dimensional analysis. We thus define
\begin{align}
\vec{f}(s,x;\eps) = (-s)^{-\eps} \, \vec{h}(x;\eps) \,,
\end{align}
where $\vec{h}(x;\eps)$ does not depend on $s$. We expand $\vec{h}(x;\eps)$ around $\eps=0$ as in \eqref{eq:f_eps_exp}. The coefficients of the expansion $\vec{h}^{(w)}(x)$ satisfy the recursive DEs
\begin{align} \label{eq:box_canDEs_iteration_h}
\partial_x \, \vec{h}^{(w)}(x) = \left[ \frac{C_2}{x} + \frac{C_3}{1+x} \right] \cdot \vec{h}^{(w-1)}(x) \,.
\end{align}
Equation~\eqref{eq:f23analytic} implies that
\begin{align} \begin{aligned}
& h_2(x;\eps) = -1 + \eps^2 \frac{\pi^2}{12} + \eps^3 \frac{7}{3} \zeta_3 + \mathcal{O}\bigl(\eps^4 \bigr) \,, \\
& h_3(x;\eps) = -1 + \eps \, \log(x) + \frac{\eps^2}{12} \left[ \pi^2 - 6 \, \log^2(x) \right] \\
& \phantom{h_3(x;\eps) =  \,} + \frac{\eps^3}{12} \left[ 2 \, \log^3(x)- \pi^2 \log(x) + 28 \, \zeta_3 \right] +  \mathcal{O}\bigl(\eps^4 \bigr) \,,
\end{aligned} \end{align}
which give us the boundary values $\vec{e}^{(w)} = \vec{h}^{(w)}(1)$ for the second and third integral. We have determined above that $e_1^{(0)}=4$, and we leave the remaining $e_1^{(w)}$'s as free parameters.
Integrating both sides of \eqn{eq:box_canDEs_iteration_h} gives
\begin{align} \label{eq:box_recursive_sol}
\vec{h}^{(w)}(x) =  \int_1^x \frac{\d x'}{x'} \, C_{2} \cdot \vec{h}^{(w-1)}(x') + \int_1^x \frac{\d x'}{1+x'} \, C_{3} \cdot \vec{h}^{(w-1)}(x') + \vec{e}^{(w)} \,,
\end{align}
starting from $\vec{h}^{(0)} = \vec{e}^{(0)}$. 
For arbitrary values of the undetermined $e_1^{(w)}$'s, the second integral on the RHS of \eqn{eq:box_recursive_sol} diverges at $x=-1$. We can thus fix the remaining boundary values by requiring that
\begin{align}
\lim_{x\to -1} C_3 \cdot \vec{h}^{(w)}(x) = 0 \,,
\end{align}
order by order in $\eps$.
For instance, at weight one we have that
\begin{align}
C_3 \cdot \vec{h}^{(1)}(x) = \left( e_1^{(1)} , \, 0, \, 0 \right)^{\top} \,.
\end{align}
The finiteness at $x=-1$ thus fixes $e_1^{(1)} = 0$. Iterating this up to weight 3 yields
\begin{align} \label{eq:boxDE_result} \begin{aligned}
& h_1 (x;\eps) =  4 + \eps \bigl[ - 2 \, \log(x) \bigr] + \eps^2 \biggl[ - \frac{4 \pi^2}{3} \biggr] + \eps^3 \biggl[ 
2 \, \text{Li}_3(-x) - 2 \, \log(x) \, \text{Li}_2(-x) \\
& \ + \frac{1}{3} \log^3(x)  - \log(x)^2 \log(1 + x) + \frac{7 \pi^2}{6} \log(x)   - \pi^2 \log(1 + x)  - \frac{34}{3} \zeta_3
\biggr] + \mathcal{O}\bigl(\eps^4\bigr) \,.
\end{aligned} \end{align}

\smallskip
\noindent
g) Equation~\eqref{eq:F4final} is related to \eqn{eq:boxDE_result} through $h_1 = \eps^2 s t \mathrm{e}^{\eps \gamma_{\text{E}}} (-s)^{\eps} F_4$. Up to transcendental weight two the equality is manifest. At weight three we find agreement after applying the identities
\begin{align} \label{eq:LiIdsmx} \begin{aligned}
& \mathrm{Li}_2\left(-\frac{1}{x}\right) =  - \mathrm{Li}_2(-x)-\frac{1}{2} \log^2(x)-\zeta_2 \,, \\
& \mathrm{Li}_3\left(-\frac{1}{x}\right) =  \mathrm{Li}_3(-x)+\frac{1}{3!} \log^3(x)+\zeta_2 \log(x) \,,
\end{aligned} \end{align}
for $x>0$, which we may prove by the symbol method, as discussed in chapter~\ref{sec:functionalidentitiesandsymbolmethod}.

\appendix

\chapter{Conventions and useful formulae}
\label{sec:Conventions}

\begin{itemize}
\item Index and metric conventions:
\begin{align*}
\eta_{\mu\nu} &= \text{diag}(+,-,-,-)\, ,  \qquad  p_{\mu}p^{\mu}=p_{0}^{2}-{\vec p}^{2} \, ,
\nn\\
\epsilon_{12}&=\epsilon_{\dot 1\dot 2}=\epsilon^{21}=\epsilon^{\dot 2\dot 1}=+1\, , \quad
\epsilon_{21}=\epsilon_{\dot 2\dot 1}=\epsilon^{12}=\epsilon^{\dot 1\dot 2}=-1\, ,\nn \\
(\bar\sigma^{\mu})^{\da\a}&=(\1,-\vec\sigma) \, , \quad
(\sigma^{\mu})_{\a\da}=\epsilon_{\a\b}
\, \epsilon_{\da\db}\, (\bar\sigma^{\mu})^{\db\b}= (\1,\vec\sigma) \, ,\nn \\
(\bar\sigma_{\mu})^{\da\a}&=(\1,\vec\sigma) \, , \quad 
(\sigma_{\mu})_{\a\da}=(\1,-\vec\sigma)\, , \nn\\
\sigma_{1}&=   
\begin{pmatrix} 
      0 & 1 \\
      1 & 0 \\
\end{pmatrix}\, ,
\quad 
\sigma_{2}=   
\begin{pmatrix} 
      0 & -\i \\
      \i & 0 \\
\end{pmatrix}\, ,
\quad 
\sigma_{3}=   
\begin{pmatrix} 
      1 &  0 \\
      0 & -1 \\
\end{pmatrix}\, , \nn\\
\chi^{\a}&= \epsilon^{\a\b}\,\chi_{\b}\, ,\qquad
\tilde\chi^{\da}= \epsilon^{\da\db}\,\tilde\chi_{\b}\, , \nn \\
(\chi_{\a})^{\ast} & =\tilde\chi_{\da}\, , \qquad
(\chi^{\a}\, \psi^{\b})^{\dagger}= (\psi^{\b})^{\dagger} \, (\chi^{\a})^{\dagger}\, .
\end{align*}
\item (Anti)-symmetrisation:
\begin{align*}
A_{(\mu}\, B_{\nu)} \coloneqq \frac{1}{2} \, ( A_{\mu} B_{\nu} + A_{\nu} B_{\mu})\, , \qquad
A_{[\mu}\, B_{\nu]} \coloneqq \frac{1}{2} \, ( A_{\mu} B_{\nu} - A_{\nu} B_{\mu})\, .
\end{align*}
\item Spinor helicity relations: 
\begin{align*}
p^{\a\da}&=\lambda^{\a}\, \tilde\lambda^{\da} \, ,  \qquad
p_{\da\a}=\epsilon_{\da\db}\, \epsilon_{\a\b}\, \lambda^{\b}\, \tilde\lambda^{\db}\, ,
\quad \la_{\a}=\epsilon_{\a\b}\,\la^{\b}\, ,
\quad \tla_{\da}=\epsilon_{\da\db}\,\tla^{\db}\, ,\nn\\
u_{+}(p)&= v_{-}(p)=    
\begin{pmatrix} 
      \la_{\alpha} \\
      0 \\
   \end{pmatrix}
   \eqqcolon |p\rangle\, ,\quad
u_{-}(p)= v_{+}(p)=    
\begin{pmatrix} 
      0 \\
      \tla^{\da} \\
   \end{pmatrix}
   \eqqcolon |p]\, , \nn\\
\bar u_{+}(p)&= \bar v_{-}(p) = (0\,\,\, \tla_{\da}) \eqqcolon [p| \, , \quad
   \bar u_{-}(p)= \bar v_{+}(p)  = (\la^{\a}\,\,\, 0) \eqqcolon \langle p| \, ,\nn \\
   \vev{\la_{i}\, \la_{j}} & \coloneqq \la_{i}^{\a}\, \la_{j\, \a}\, , \qquad
   \bev{\tla_{i}\, \tla_{j}} \coloneqq \tla_{i\,\da}\, \tla_{j}^{\da}\, .
   \end{align*}
Using the chiral representation of the Dirac matrices, we have
\begin{align*}
\slashed{p}  & = p_{\mu}\, \gamma^{\mu} =    \begin{pmatrix} 
      0 & p_{\a\da} \\
      p^{\da\a} & 0 \\
   \end{pmatrix}\,, \quad
   \quad
   p_{\a\da} \coloneqq p_{\mu}\, (\sigma^{\mu})_{\a\da}\, , \nn \\
   p^{\da\a}&= \epsilon^{\a\b}\, \epsilon^{\da\db}\,  p_{\b\db}= p_{\mu}\, (\bar\sigma^{\mu})^{\da\a}\,.
\end{align*}
We furthermore note 
\begin{align*}
[i|\gamma^{\mu}| j\rangle &= \langle j | \gamma^{\mu}|i]\, ,\quad 
\langle p | \gamma^{\mu} | p] = \la^{\a}\, \sigma^{\mu}_{\a\da}\, \tla^{\da}= 2\, p^{\mu}\, , 
\nn\\ 
[i|\gamma^{\mu}| j\rangle \, \langle l|\gamma_{\mu}|k] &= 2\, [ik]\, \langle lj \rangle\, .
\end{align*}
\item The generators of the Lorentz group in the spinor representation are given by
\begin{align*}
(\sigma^{\mu\nu})_\a{}^\b &= \frac{1}{4} \, \Bigl( (\sigma^{\mu})_{\a\da}\,  
({\bar\sigma}^{\nu})^{\da\b} - (\sigma^{\nu})_{\a\da}\,  
({\bar\sigma}^{\mu})^{\da\b} \Bigr) \, , \nn \\
({\bar\sigma}^{\mu\nu})^\a{}_\db &= \frac{1}{4} \, \Bigl( ({\bar\sigma}^{\mu})^{\da\a}\,  
({\sigma}^{\nu})_{\a\db} - ({\bar\sigma}^{\nu})^{\da\a}\,  
({\sigma}^{\mu})^{\a\db} \Bigr) \, .
\end{align*}
\item Complex conjugation properties:
\begin{align*}
(\la^{\a})^{\ast}= \tla^{\da}\, , \qquad \vev{ij}^{\ast}=
(\la_{i}^{\a}\, \la_{j\, \a})^{\ast}=
(\tla_{i}^{\da}\, \tla_{j\, \da})^{\ast}=
-\bev{ij}\, .
\end{align*}
\item Useful trace identities:
\begin{align*}
\Tr[\slashed{a}\,\slashed{b}\,\slashed{c}\,\slashed{d}]&=4\,[\, ( a\cdot b)(c\cdot d)-
( a\cdot c)(b\cdot d)+( a\cdot d)(b\cdot c)\, ]\, , \nn \\ 
\Tr[\slashed{a}\,\slashed{b}\, ] &= 4\,  ( a\cdot b)\, . 
\label{traceids}
\end{align*}
\item Expressions for the $\text{SU}(N_c)$ generators in the
fundamental representation:
\begin{itemize}
\item $\text{SU}(2)$: \quad $T^{a}=\ft{1}{\sqrt{2}}\, \sigma^{a}$, \quad $\sigma^{a}$: Pauli matrices $a=1,2,3$.\\
\item $\text{SU}(3)$: \quad $T^{a}=\ft{1}{\sqrt{2}}\, \lambda^{a}$, \quad $\lambda^{a}$: Gell-Mann matrices $a=1,\ldots,8$,\\
\begin{align*}
\begin{aligned}
& \lambda^{a}= \begin{pmatrix} \sigma^{a} & \\ & 0\\ \end{pmatrix}\,, \quad a=1,2,3\,, \\
& \lambda^{4}= \begin{pmatrix}  0 &  & 1 \\ & 0 & \\ 1 & & 0 \\ \end{pmatrix}\,, \qquad
\lambda^{5}= \begin{pmatrix}  0 &  & -\i\\ & 0 & \\ \i& & 0 \\ \end{pmatrix}\,, \qquad
\lambda^{6}= \begin{pmatrix}  0 &  &  \\ & 0 &1 \\  & 1 & 0 \\ \end{pmatrix}\,,  \\
& \lambda^{7}= \begin{pmatrix}  0 &  &  \\ & 0 &-\i\\  & \i& 0 \\ \end{pmatrix}\,, \qquad
\lambda^{8}= \frac{1}{\sqrt{3}}\begin{pmatrix}  1 &  &  \\ & 1 & \\  &  & -2 \\ \end{pmatrix}\,.
\end{aligned}
\end{align*}
The omitted entries in the above are $0$.
\smallskip
\item $\text{SU}(N_c)$: Explicit constructions for $T^{a}$ exist in terms of the 't~Hooft twist matrices, see~\cite{ch6-'tHooft:1981sz}.
\end{itemize}

\end{itemize}




\end{document}